北京大學

# Post-Doctoral Concluding Report

## ENERGY-VIEWPOINT-BASED ELECTROMAGNETIC MODAL ANALYSIS

Ren-Zun LIAN

Peking University (PKU)

# ENERGY-VIEWPOINT-BASED ELECTROMAGNETIC MODAL ANALYSIS

# 基于能量观点的电磁模式分析

Post-Doctoral Name: Ren-Zun LIAN

rzlian@vip.163.com

Cooperative Supervisor Name: Ming-Yao XIA

Post-Doctoral Station: Electronic Science and Technology

Discipline: Electromagnetic Field and Microwave Technology

Post-Doctoral Start Date: September 11, 2019

Post-Doctoral End Date: August 15, 2021

School of Electronics Engineering and Computer Science @ Peking University (PKU)

China • Beijing

August, 2021

# 版权声明

The objective is to introduce the energy point of view into the study of electromagnetic fields. The law of conservation of energy suggests to define energy functions for expressing the energy supplied to a field or delivered by it.

—— R. M. Fano, L. J. Chu (朱兰成) & R. B. Adler

The central purposes of this Post-Doctoral Concluding Report are the following two: (1) to reveal the core position of Energy Viewpoint in the realm of electromagnetic modal analysis; (2) to show how to do the energy-viewpoint-based modal analysis for various electromagnetic structures.

The major conclusions related to this report are that: Energy Conservation Law governs the energy utilization processes of various electromagnetic structures, and its energy source term sustains the steady energy utilization processes; the whole modal space of an electromagnetic structure is spanned by a series of Energy-Decoupled Modes (DMs), which don't have net energy exchange in any integral period; the DMs can be effectively constructed by orthogonalizing Energy Source Operator, which is just the operator form of the energy source term.

This report is not only a systematical integration but also a further sublimation for the author's Doctoral Dissertation "Research on the Work-Energy Principle Based Characteristic Mode Theory for Scattering Systems"[arXiv:1907.11787] and the author's Post-Doctoral Research Report "Research on the Power Transport Theorem Based Decoupling Mode Theory for Transceiving Systems"[arXiv:2103.01853], which can be downloaded from the following links.

Doctoral Dissertation: https://arxiv.org/abs/1907.11787

Post-Doctoral Research Report: https://arxiv.org/abs/2103.01853

Ren-Zun LIAN

Peking University (PKU) • Yan Yuan

# Abstract

Modal analysis method is one of the important study topics in computational physics and mathematical physics, and has many valuable engineering applications. The central purposes of this Post-Doctoral Concluding Report are the following two: (1) to reveal the core position of energy viewpoint in the realm of electromagnetic modal analysis; (2) to show how to do the energy-viewpoint-based modal analysis for various electromagnetic structures.

For a linear electromagnetic structure, it has many physically realizable steadily working modes, and all of the modes constitute a linear space — modal space. Energy Conservation Law implies that: a non-zero energy source is indispensable for sustaining the steady energy utilization of a mode (except self-oscillating modes). If the energy sources of mode α and mode β don't supply energies to mode β and mode α respectively (where mode α is different from mode β), then mode α and mode β are energy-decoupled. To construct a set of complete Energy-Decoupled Modes (DMs), which can span whole modal space, is valuable for analyzing and designing the objective electromagnetic structure.

For a certain electromagnetic structure, its different working manners (such as scattering, transmitting, receiving, power-guiding/transferring, energy-dissipating, and self-oscillating manners, etc.) involve different energy utilization processes (such as work-energy transformation, power transportation, energy dissipation, and self-oscillation processes, etc.), so the different manners need different kinds of energy sources to sustain their steady workings. From the mathematical point of view, the construction process for DMs is to formulate the operator expression of energy source first, and then to orthogonalize the Energy Source Operator. Thus generally speaking, the different working manners of a certain objective electromagnetic structure have different DM sets.

This report derives/reviews five different manifestation forms of Energy Conservation Law — Power Transport Theorem (PTT) form, Partial-Structure-Oriented Work-Energy Theorem (PS-WET) form, Entire-Structure-Oriented Work-Energy Theorem (ES-WET) form, Poynting's Theorem (PtT) form, and Lorentz's Reciprocity Theorem (LRT) form, where

i) PTT governs the energy utilization process in power transportation manner, and its energy source term is formulated as INPUT POWER OPERATOR (IPO);
ii) PS-WET governs the energy utilization process in partial-structure-oriented work-energy transformation manner, and its energy source term is formulated as PARTIAL-STRUCTURE-ORIENTED DRIVING POWER OPERATOR (PS-DPO);
iii) ES-WET governs the energy utilization process in entire-structure-oriented work-energy transformation manner, and its energy source term is formulated as ENTIRE-STRUCTURE-ORIENTED DRIVING POWER OPERATOR (ES-DPO);
iv) PtT governs the energy utilization processes in energy dissipation manner and self-oscillation manner, and its energy source term is formulated as POYNTING'S FLUX OPERATOR (PtFO);
v) LRT governs the energy coupling manner between the fields of two different working modes.

At the same time, this report also proves some beautiful equivalence relations existing among the different manifestation forms.

Based on the different manifestation forms of ENERGY CONSERVATION LAW and the corresponding ENERGY SOURCE OPERATORS, this report constructs DMs for a series of typical electromagnetic structures as below.

a) Under PTT framework and employing orthogonalizing IPO method, this report constructs the DMs for wave-port-fed antennas, waveguides, and some combined systems (such as waveguide-antenna and antenna-antenna cascaded systems).
b) Under PS-WET framework and employing orthogonalizing PS-DPO method, this report constructs the energy-decoupled CHARACTERISTIC MODES (CMs) for lumped-port-driven/local-near-field-driven antennas and waveguides, where the antennas and waveguides usually include passive loads.
c) Under ES-WET framework and employing orthogonalizing ES-DPO method, this report reveals that the conventional CMs work at scattering manner rather than transmitting manner, and then explains why the conventional CHARACTERISTIC MODE THEORY (CMT) fails to analyze many classical antennas working at transmitting manner; this report proves that the conventional CMs are energy-decoupled and don't contain scatterer-environment and scatterer-driver interaction informations; this report generalizes the conventional CMs to the environment-dependent/driver-dependent CMs with scatterer-environment/scatterer-driver interaction information.

d) Under PtT framework and employing orthogonalizing PtFO method, this report also constructs the DMs for energy-dissipating structures and self-oscillating structures, where the self-oscillating structures are source-free.

e) Employing LRT, this report derives some beautiful energy-decoupling features satisfied by the DMs and CMs.

To quantify the modal energy features of the obtained DMs and CMs, this report reviews some traditional physical quantities (such as modal significance), and re-defines some classical physical quantities (such as modal input impedance and admittance), and generalizes some conventional physical quantities (for example: Q-factor → Θ-factor), and also introduces some novel physical quantities (such as energy transporting and transferring coefficients).

In summary, the core position of energy viewpoint in the realm of electromagnetic modal analysis embodies in that many seemingly different modal analysis theories can actually be unified in an universal framework — ENERGY CONSERVATION LAW framework; under the framework, the energy-viewpoint-based modal analysis theories can effectively construct DMs/CMs by employing an universal method — orthogonalizing ENERGY SOURCE OPERATOR method, and the obtained DMs/CMs don't have net energy exchange in any integral period.

# 摘要

模式分析方法是计算物理学和数学物理学的重要研究课题，并且有极高的工程应用价值。此“博士后结题报告”的中心目的有两个：（1）揭示能量观点在电磁模式分析领域所处的核心地位；（2）展示如何基于能量观点为各种电磁结构做模式分析。

对于一个线性的电磁结构，其存在很多物理上可实现的稳定工作模式，且所有模式构成一个线性空间——模式空间。能量守恒定律（energy conservation law）指出：非零的能量源对于维持模式的稳定工作是必不可少的（自振荡模式除外）。若模式 α/模式 β 的能量源无法将能量提供给模式 β/模式 α（此处模式 α 不同于模式 β），则模式 α 和模式 β 是能量去耦的。在模式空间中构造一组完备的能量去耦模式（energy-decoupled modes, DMs）对于分析和设计目标电磁结构极有价值。

对于一个确定的电磁结构，其不同的工作方式（如散射、发射、接收、功率引导/传输、能量耗散以及自振荡等方式）涉及不同的能量使用过程（如功能转换、功率输运、能量耗散以及自振荡等过程），所以不同的工作方式需要不同类型的能量源去维持其稳定工作。从比较数学的视角来看，对 DMs 的构造过程就是一个先推导出能量源的算子表示，再去正交化该能量源算子（energy source operator）的过程。因此一般来讲，同一个目标电磁结构的不同工作方式具有不同的 DM 集合。

此报告推导/回顾了能量守恒定律的五种表现形式——功率输运定理（power transport theorem, PTT）形式、面向部分结构的功能定理（partial-structure-oriented work-energy theorem, PS-WET）形式、面向完整结构的功能定理（entire-structure-oriented work-energy theorem, ES-WET）形式、Poynting 定理（Poynting’s theorem, PtT）形式和 Lorentz 互易定理（Lorentz’s reciprocity theorem, LRT）形式，其中

i) PTT 支配着“功率输运方式”下的能量使用过程，其能量源项表现为输入功率算子（input power operator, IPO）；
ii) PS-WET 支配着“面向部分结构之功能转换方式”下的能量使用过程，其能量源项表现为面向部分结构的驱动功率算子（partial-structure-oriented driving power operator, PS-DPO）；
iii) ES-WET 支配着“面向完整结构之功能转换方式”下的能量使用过程，其能量源项表现为面向完整结构的驱动功率算子（entire-structure-oriented driving power operator, ES-DPO）；
iv) PtT 支配着“能量耗散方式”和“自振荡方式”下的能量使用过程，其能量源项

表现为 Poynting 通量算子（Poynting’s flux operator, PtFO）；

v) LRT 支配着两个不同工作模式的场之间的能量耦合方式。

同时，此报告还证明了一些存在于上述不同表现形式之间的等价关系。

基于能量守恒定律的不同表现形式以及相应的能量源算子，此报告为一系列典型的电磁结构构造了 DMs 如下。

a) 基于 PTT 框架和正交化 IPO 方法，此报告为波端口馈电的天线、波导和组合系统（如“波导-天线”和“天线-天线”级联系统）构造了 DMs。
b) 基于 PS-WET 框架和正交化 PS-DPO 方法，此报告为集总端口驱动/近场驱动的天线和波导构造了能量去耦的特征模式（energy-decoupled characteristic modes, energy-decoupled CMs），这些天线和波导通常包含无源加载的子结构。
c) 基于 ES-WET 框架和正交化 ES-DPO 方法，此报告揭示了这样的事实“传统的 CMs 工作于散射状态而非发射状态”，进而解释了为什么“传统的特征模式理论（characteristic mode theory, CMT）对很多工作于发射状态的天线失效”；此报告证明了“传统的 CMs 是能量去耦的，并且不包含‘散射体-环境’和‘散射体-驱动器’之间的相互作用信息”；此报告将传统的 CMs 进一步推广到那些包含有‘散射体-环境’和‘散射体-驱动器’之间相互作用信息的 CMs。
d) 基于 PtT 框架和正交化 PtFO 方法，此报告还为能量耗散结构和自振荡结构构造了 DMs，这里所说的自振荡结构是无源的。
e) 借助于 LRT，此报告还推导出了上述 DMs 和 CMs 所满足的一系列能量去耦特性。

为了定量地反应上述所得 DMs 和 CMs 在能量使用方面的特性，此报告回顾了一些常用的物理量（如模式显著性“modal significance, MS”），重新定义了一些经典的物理量（如模式输入阻抗与导纳），并对一些传统的物理量做了适当推广（如：Q 因子→Θ 因子），而且还引入了一些新的物理量（如能量输运和传输系数）。

综上所述，能量观点在电磁模式分析领域中所处的核心地位体现在，很多看似不同的电磁模式分析理论实际上可以被统一在一个通用的框架——能量守恒定律框架——之下；于该框架之下，基于能量观点的模式分析理论可以利用一个通用的方法——正交化能量源算子方法——为各种电磁结构有效地构造一组在任何完整周期内都不发生净能量交换的 DMs/CMs。

# SYMBOL AND ABBREVIATION LISTS

In the following discussions, we use some necessary mathematical symbols to represent the operators and physical quantities involved in this report. Now, the symbols and their meanings are listed as below for facilitating readers' references.

| SYMBOLS | MEANINGS |
|---|---|
| $\mathcal{F} / \mathcal{C} / \mathcal{P}$ | time-domain field/current/power |
| $\boldsymbol{F}$ / $\boldsymbol{C}$ / $P$ | frequency-domain field/current/power |
| $P_{\text{DRIV}}$ / $P_{\text{driv}}$ | entire/partial-structure-oriented driving power |
| $\mathbb{C} / \mathbb{P} / \mathbb{p}$ | matrix forms of current/power/power |
| $\mathbb{T} / \mathbb{t}$ | transformation matrix from independent currents to all/other currents |
| $\nabla \times / \nabla \cdot$ | curl/divergence operation |
| $*$ | convolution integral operation |
| $\dagger$ | conjugate transpose operation |
| $< \boldsymbol{f}, \boldsymbol{g} >_{\Omega}$ | inner product defined as $\int_{\Omega} \boldsymbol{f}^{\dagger} \cdot \boldsymbol{g} d\Omega$ |
| $\mu_0 / \varepsilon_0$ | free-space permeability/permittivity |
| $\eta_0 / k_0$ | free-space wave impedance/number |
| $\mu_{\text{r}} / \varepsilon_{\text{r}}$ | relative permeability/permittivity of material |
| $\boldsymbol{\mu} / \boldsymbol{\varepsilon} / \boldsymbol{\sigma}$ | material permeability/permittivity/conductivity tensors |
| $\mathbf{I}$ | unit dyad |
| $\text{V}$ | three-dimensional region |
| $\partial\text{V} / \text{int}\,\text{V} / \text{ext}\,\text{V}$ | boundary/interior/exterior of $\text{V}$ |
| $\partial\text{V}_{-} / \partial\text{V}_{+}$ | inner/outer surface of $\partial\text{V}$ |
| $\boldsymbol{n}_{\partial\text{V}}^{-} / \boldsymbol{n}_{\partial\text{V}}^{+}$ | inner/outer normal direction of $\partial\text{V}$ |
| $\boldsymbol{z}$ | direction of Z-axis |
| $\text{E}_3$ | three-dimensional Euclidean space |
| $\text{S}_{\infty}$ | outer boundary of $\text{E}_3$ |
| $\varnothing$ | null set |
| $\Theta$ | electric-magnetic energy-decoupling factor (in DMT) or field-current energy-decoupling factor (in CMT) |
| $\theta_m / \lambda_m$ | Θ-factor of the *m*-th energy-decoupled mode / characteristic mode |
| $Z / R / X$ | input impedance/resistance/reactance of wave-port-fed electromagnetic structure |
| $Y / G / B$ | input admittance/conductance/susceptance of wave-port-fed electromagnetic structure |
| $\lambda_z$ | waveguide wavelength along Z-axis |
| $T / f / \omega$ | time period / time frequency / angular frequency of time-harmonic field |
| $j$ | unit of imaginary numbers |
| $e^{j\omega t}$ | time factor of frequency-domain time-harmonic field |
| $G_0(\boldsymbol{r},\boldsymbol{r}')$ | equals to $e^{-jk_0\|\boldsymbol{r}-\boldsymbol{r}'\|} / 4\pi \|\boldsymbol{r}-\boldsymbol{r}'\|$ |
| $\mathbf{G}^{CF}(\boldsymbol{r},\boldsymbol{r}')$ | dyadic Green's function used to transform current into field |
| $\mathcal{L}_0(\boldsymbol{X})$ | equals to $[1+(1/k_0^2)\nabla\nabla\cdot]\int_\Omega G_0(\boldsymbol{r},\boldsymbol{r}')\boldsymbol{X}(\boldsymbol{r}')d\Omega'$ |
| $\mathcal{K}_0(\boldsymbol{X})$ | equals to $\nabla\times\int_\Omega G_0(\boldsymbol{r},\boldsymbol{r}')\boldsymbol{X}(\boldsymbol{r}')d\Omega'$ |
| $\mathcal{E}_0(\boldsymbol{J},\boldsymbol{M})$ | equals to $-j\omega\mu_0\mathcal{L}_0(\boldsymbol{J})-\mathcal{K}_0(\boldsymbol{M})$ |
| $\mathcal{H}_0(\boldsymbol{J},\boldsymbol{M})$ | equals to $+\mathcal{K}_0(\boldsymbol{J})-j\omega\varepsilon_0\mathcal{L}_0(\boldsymbol{M})$ |
| $\mathcal{F}(\boldsymbol{J},\boldsymbol{M})$ | equals to $\mathbf{G}^{JF}*\boldsymbol{J}+\mathbf{G}^{MF}*\boldsymbol{M}$ |
| $\text{P.V.}\mathcal{F}(\boldsymbol{J},\boldsymbol{M})$ | principal of operator $\mathcal{F}$ |
| $\delta_{mn}$ | Kronecker's delta symbol |

For the convenience of expressions, this report utilizes some abbreviations frequently. Now, we list the abbreviations and their full names in the following table for facilitating readers' references.

| ABBREVIATIONS | FULL NAMES |
|---|---|
| EM | electromagnetic |
| SLT | Sturm-Liouville theory |
| SM | scattering matrix |
| IE | integral equation |
| WET/WEP | work-energy theorem/principle |
| ES-WET | entire-structure-oriented WET |
| PS-WET | partial-structure-oriented WET |
| PtT | Poynting's theorem |
| PTT | power transport theorem |
| EMT | eigen-mode theory |
| CMT | characteristic mode theory |
| DMT | decoupling mode theory |
| SL-EMT | Sturm-Liouville EMT |
| SM-CMT | SM-based CMT |
| IE-CMT | IE-based CMT |
| ES-WET-CMT | ES-WET-based CMT |
| PS-WET-CMT | PS-WET-based CMT |
| PtT-DMT | PtT-based DMT |
| PTT-DMT | PTT-based DMT |
| SLO | Sturm-Liouville operator |
| PMO | perturbation matrix operator |
| IMO | impedance matrix operator |
| DPO | driving power operator |
| ES-DPO | entire-structure-oriented DPO |
| PS-DPO | partial-structure-oriented DPO |
| PtFO | Poynting's flux operator |
| IPO | input power operaor |
| CM | characteristic mode |
| DM | energy-decoupled mode |
| TE mode | transverse electric mode |
| TM mode | transverse magnetic mode |
| TEM mode | transverse electromagnetic mode |
| DVE | dependent variable elimination |
| SDC | solution domain compression |
| IVM | intermediate variable method |
| Θ-factor | electric-magnetic energy-decoupling factor (in DMT) or field-current energy-decoupling factor (in CMT) |
| Q-factor | quality factor |
| MS | modal significance |
| TC | transporting/transferring coefficient |
| PMCHWT operator | Poggio-Miller-Chang-Harrington-Wu-Tsai-based IMO |
| pmchwt operator | modified PMCHWT operator |
| JM-formed operator | J-M interaction form of energy source operator |
| EH-formed operator | E-H interaction form of energy source operator |
| JE-formed operator | J-E interaction form of energy source operator |
| HM-formed operator | H-M interaction form of energy source operator |
| DoJ | definition of J (equivalent surface electric current) |
| DoM | definition of M (equivalent surface magnetic current) |
| JE-DoJ | JE-formed operator with DoJ-based DVE |
| HM-DoM | HM-formed operator with DoM-based DVE |
| WPT | wireless power transfer |
| RCS | radar cross section |
| MoM | method of moments |

# CONTENTS

# CHAPTER 1 INTRODUCTION

**CHAPTER MOTIVATION:** This chapter is devoted to exposing the fact that: the conventional characteristic mode theory (CMT) fails to analyze some classical transmitting antennas, such as horn antenna and Yagi-Uda antenna. Based on this expose, this chapter introduces the following main topics focused on by this Post-Doctoral Concluding Report.

TOPIC 1. How to explain the failure of the conventional CMT-based modal analysis method for transmitting antennas?

TOPIC 2. How to establish an effective modal analysis method for transmitting antennas?

TOPIC 3. How to further generalize the transmitting-antenna-oriented modal analysis method to other electromagnetic (EM) structures?

Based on the above motivation, this chapter is organized as follows: research background and significance on EM modal analysis (Sec. 1.1) → research history and status related to EM modal analysis (Sec. 1.2) → major problem and challenge in the realm of EM modal analysis (Sec. 1.3) → main innovations and contributions of this report (Sec. 1.4) → research outline and roadmap of this report (Sec. 1.6).

## 1.1 Research Background and Significance on EM Modal Analysis

Energy is one of important information carriers. In EM engineering, it has had a long history to transmit, receive, and guide information by loading the information into EM energy. Just by controlling the space-time distribution of the energy, EM device (alternatively called EM structure/system) finally realizes the control for the information. The different space-time distributions of energy correspond to the different working modes (or simply called modes) of device. For a linear device, its all physically realizable modes constitute a linear space[1] — modal space.

In the modal space, some modes can exchange energies with each other, but some modes can not. The modes without energy exchange are called energy-decoupled modes (DMs). **Because of the absence of energy coupling, the DMs can work independently, and then can carry independent informations.** Due to the energy-decoupling and information-independence features, the DMs have many important engineering applications[2,3]. Thus, the DM-oriented modal analysis for EM structures has become one of the research hot spots in EM theory and engineering.

## 1.2 Research History and Status Related to EM Modal Analysis

In mathematical physics, some DM-oriented methods have been established, and they can be collectively referred to as modal analysis theory. The studies for modal analysis theory have had a relatively long history, and they can be dated back to some famous physicists and mathematicians, such as Sauveur, Bernoulli family (mainly John and Daniel), Euler, d'Alembert, Lagrange, Laplace, Fourier, Sturm, and Liouville *et al*.

### 1.2.1 Eigen-Mode Analysis

The most classical modal analysis theory used in electromagnetism is electromagnetic eigen-mode theory (EMT), and it has been used to construct the eigen-modes of closed EM structures, such as wave-guiding transmission lines and wave-oscillating resonance cavities etc., for many years.

Under famous Sturm-Liouville theory (SLT) framework, **electromagnetic EMT focuses on constructing a set of eigen-modes propagating or oscillating in a region with perfectly electric wall**, by solving Sturm-Liouville equation. The operator — Sturm-Liouville operator (SLO) — contained in Sturm-Liouville equation can be viewed as the generating operator of the eigen-modes. Some detailed discussions for electromagnetic EMT can be found in Refs. [2,3], and some detailed discussions on the mathematical foundation of SLT can be found in Ref. [4].

In fact, it is not difficult to prove that **the classical electromagnetic eigen-modes of "closed" metallic waveguides and cavities are energy-decoupled**[2-Sec.5.2].

### 1.2.2 From Eigen-Mode Analysis to SM-Based CM Analysis

In the 1960s, a seminal work on generalizing modal analysis theory from closed EM structures to open EM structures was done by Garbacz *et al.*[5~7] by employing the idea of multipole expansion and the theory of scattering matrix (SM), and the generalized theory is now called SM-based characteristic mode theory (CMT).

**The SM-based CMT (SM-CMT) aims to constructing a set of far-field-decoupled modes** — SM-based characteristic modes (CMs) — for open EM structures by orthogonalizing perturbation matrix operator (PMO). A detailed discussion on the lossless-structure-oriented SM-CMT can be found in Ref. [6], and some simplified discussions on the lossless-structure-oriented SM-CMT and on generalizing SM-CMT from lossless structures to lossy structures can be found in Ref. [8-Sec.2.2].

In addition, it is easy to prove that: **for lossless open structures, the far-field-decoupled SM-based CMs are *usually* also energy-decoupled; for lossy open structures, the energy-decoupling feature of SM-based CMs cannot be guaranteed, though the SM-based CMs are far-field-decoupled**.

### 1.2.3 From SM-Based CM Analysis to IE-Based CM Analysis

Following the seminal works of Garbacz *et al.*, Harrington *et al.*[9~13], under an alternative integral equation (IE) framework, constructed "another" kind of CMs — IE-based CMs — for open EM structures, by orthogonalizing impedance matrix operator (IMO).

In fact, it is not difficult to prove that[14,15]: **for lossless open structures, the IE-based CMs are not only far-field-decoupled but also energy-decoupled; for lossy open structures, the IE-based CMs[12-Sec.II] are energy-decoupled, but cannot guarantee far-field-decoupling feature**. Thus, for lossless open EM structures, the IE-based CMs are equivalent (but not necessarily identical) to the SM-based CMs in the sense of decoupling modal far fields; for lossy open EM structures, the IE-based CMs and the SM-based CMs are usually not equivalent to each other in the senses of both far-field decoupling and energy decoupling.

From SM-CMT to IE-CMT, it is not only a transformation for the theoretical framework of CMT — from SM framework to IE framework, but also a transformation for modal generating operator — from PMO to IMO, as shown in Tab. 1-1[8-Chap.2],[14].

Table 1-1 Evolutions of CMT and comparisons from the aspects of theoretical framework, modal generating operator, and modal core physical feature

| | **Theoretical Framework** | **Modal Generating Operator** | **Modal Core Physical Feature** |
|---|---|---|---|
| **SM-CMT[5~7]** | SM | PMO | far-field decoupling |
| | ↓ | | |
| **IE-CMT[9~13]** | IE | IMO | not clarified by its founders |
| | ↓ | | |
| **ES-WET-CMT[8,14,15]** | ES-WET | ES-DPO | energy decoupling |

The transformations significantly simplify the calculation process for CMs, because to obtain the IMO is much easier than to obtain the PMO. Due to its merits, IE-CMT has been used to analyze some kinds of transmitting antennas, such as patch antenna[16], monopole antenna[17], dipole antenna[18], meta-surface antenna[19], MIMO antenna[20], plasmonic nanoantenna[21], and dielectric resonator antenna[22] etc. Some typical IE-CMT-based antenna applications had been comprehensively summarized in Refs. [23~25].

After half a century (from 1970 to 2019) progress, IE-CMT has had a great development in many aspects. But, at the same time, some problems are also exposed[8-Sec.1.3],[14,15,26],[27-Sec.1.2.3], for example:

PROBLEM I: When objective EM structure is placed in a non-free-space environment, if the background Green's function is employed to establish the IE-based CM calculation formula, the obtained CMs seemingly don't satisfy a reasonable work-energy transformation relation. Then, how to reasonablely calculate the CMs of an objective EM structure, which is placed in non-free space?

PROBLEM II: When the objective EM structure is magneto-dielectric (i.e., both magnetic and dielectric), the CMs generated by symmetric IMO and the CMs generated by asymmetric IMO are not necessarily consistent with each other. Then, which IMO is the reasonable one for generating CMs?

PROBLEM III: When the objective EM structure is lossy, the far-field-decoupling feature and energy-decoupling feature cannot be simultaneously satisfied by IE-based CMs usually. Then, which feature is the core/indispensable physical feature of CMs, and why?

PROBLEM IV: When the objective EM structure is material (especially magneto-dielectric case) or metal-material composite, the physical meaning of IE-CMT-based characteristic values has not been interpreted successfully. Then, how to provide a unified physical interpretation for the characteristic values of metallic, material, and composite structures?

PROBLEM V: When the objective EM structure is material or composite, the modal generating operator formulated by surface currents will output some unwanted and spurious modes usually. Then, what is the reason leading to the unwanted and spurious modes, and how to suppress the modes?

PROBLEM VI: When the objective EM structure includes some metallic parts, the modal characteristic fields cannot satisfy the boundary conditions on the metallic boundaries. Then, how to explain this phenomenon?

In fact, the above PROBLEMS are closely related to the theoretical foundation of IE-CMT, and their solving will further promote the development of IE-CMT.

### 1.2.4 From IE-Based CM Analysis to ES-WET-Based CM Analysis

To resolve the PROBLEMS mentioned above, Lian *et al.*[8,14,15] re-established the IE-CMT under an alternative framework — ENTIRE-STRUCTURE-ORIENTED WORK-ENERGY THEOREM/PRINCIPLE (ES-WET/ES-WEP) framework, and constructed the IE-based CMs by orthogonalizing an alternative modal generating operator — ENTIRE-STRUCTURE-ORIENTED DRIVING POWER OPERATOR (ES-DPO). In fact, the ES-WET-based CMT (ES-WET-CMT) realizes the second transformation for the theoretical framework of CMT — from IE framework to ES-WET framework, and then the second transformation for the generating operator of CMs — from IMO to ES-DPO, as shown in Tab. 1-1[8-Chap.2],[14]. Taking the material structure shown in Fig. 1-1 as a typical example, the ES-WET-based resolutions for the above-listed PROBLEMS I~VI are simply summarized in this sub-section.

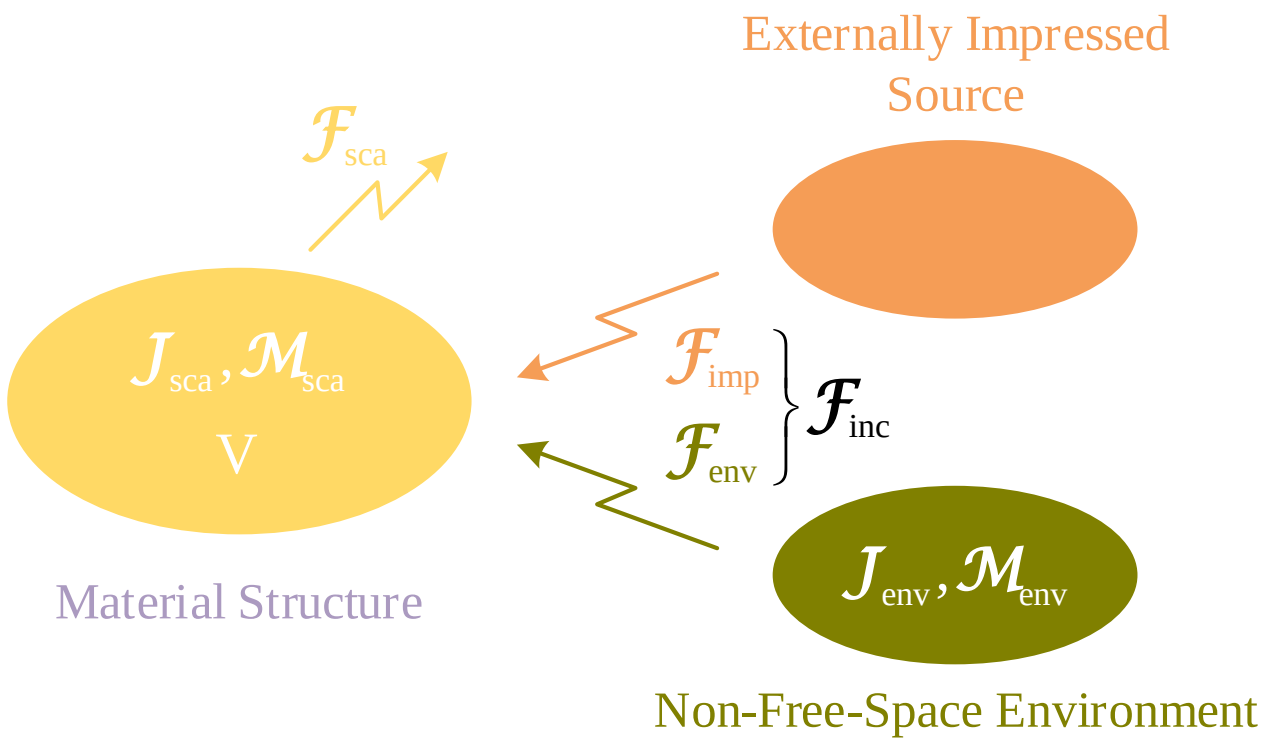

Figure 1-1 External-field-excited material structure placed in non-free-space environment.

In Fig. 1-1, the material structure is placed in a non-free-space environment, and excited by an externally impressed field $\mathcal{F}_{\rm imp}$, where $\mathcal{F}_{\rm imp}$ is the abbreviated form of electric and magnetic fields $(\mathcal{E}_{\rm imp}, \mathcal{H}_{\rm imp})$. The region occupied by the material structure is denoted as V. The magnetic permeability, dielectric permittivity, and electric conductivity of the material structure are $\boldsymbol{\mu}$, $\boldsymbol{\varepsilon}$, and $\boldsymbol{\sigma}$ respectively, and the material parameters are time-independent symmetrical dyads. Due to the existence of $\mathcal{F}_{\rm imp}$, currents $\mathcal{C}_{\rm sca}$ and $\mathcal{C}_{\rm env}$ are induced on V and environment respectively, and then second fields $\mathcal{F}_{\rm sca}$ and $\mathcal{F}_{\rm env}$ are generated by $\mathcal{C}_{\rm sca}$ and $\mathcal{C}_{\rm env}$ correspondingly, where $\mathcal{C}_{\rm sca/env}$ is the abbreviated form of electric and magnetic currents $(\mathcal{J}_{\rm sca/env}, \mathcal{M}_{\rm sca/env})$. For the convenience of the following discussions, the summation of $\mathcal{F}_{\rm imp}$, $\mathcal{F}_{\rm sca}$, and $\mathcal{F}_{\rm env}$ is called total field and denoted as $\mathcal{F}$, i.e., $\mathcal{F} = \mathcal{F}_{\rm imp} + \mathcal{F}_{\rm sca} + \mathcal{F}_{\rm env}$.

**ES-WET-Based Resolution for PROBLEM I**

Because of the time-domain Maxwell's equations $\nabla\times\mathcal{H}=\boldsymbol{\sigma}\cdot\mathcal{E}+\frac{\partial}{\partial t}\boldsymbol{\varepsilon}\cdot\mathcal{E}$ and $\nabla\times\mathcal{E}=-\frac{\partial}{\partial t}\boldsymbol{\mu}\cdot\mathcal{H}$ satisfied on V, there exists the following time-domain POYNTING's THEOREM (PtT)[28]

$$\oint\!\!\!\oint_{\partial\mathrm{V}}\left(\mathcal{E}\times\mathcal{H}\right)\cdot\boldsymbol{n}_{\partial\mathrm{V}}^{-}dS=\left\langle\boldsymbol{\sigma}\cdot\mathcal{E},\mathcal{E}\right\rangle_{\mathrm{V}}+\frac{d}{dt}\left[\frac{1}{2}\left\langle\mathcal{H},\boldsymbol{\mu}\cdot\mathcal{H}\right\rangle_{\mathrm{V}}+\frac{1}{2}\left\langle\boldsymbol{\varepsilon}\cdot\mathcal{E},\mathcal{E}\right\rangle_{\mathrm{V}}\right] \tag{1-1}$$

In PtT (1-1), $\partial\mathrm{V}$ is the boundary of V; $\boldsymbol{n}_{\partial\mathrm{V}}^{-}$ is the inner normal direction of $\partial\mathrm{V}$; the inner product is defined as that $<\boldsymbol{f},\boldsymbol{g}>_{\Omega}=\int_{\Omega}\boldsymbol{f}^{\dagger}\cdot\boldsymbol{g}d\Omega$, where superscript "†" is the conjugate transpose operation for a scalar/vector/matrix.

As proved in Refs. [14] and [27-Sec.1.2.4.4], the above PtT (1-1) can be equivalently transformed into the following alternative form

$$\begin{aligned}&<\boldsymbol{J}_{\mathrm{sca}},\overbrace{\mathcal{E}_{\mathrm{imp}}+\mathcal{E}_{\mathrm{env}}}^{\mathcal{E}_{\mathrm{inc}}}>_{\mathrm{V}}+<\mathcal{M}_{\mathrm{sca}},\overbrace{\mathcal{H}_{\mathrm{imp}}+\mathcal{H}_{\mathrm{env}}}^{\mathcal{H}_{\mathrm{inc}}}>_{\mathrm{V}}\\ =\;&\oint\!\!\!\oint_{\mathrm{S}_{\infty}}\left(\mathcal{E}_{\mathrm{sca}}\times\mathcal{H}_{\mathrm{sca}}\right)\cdot\boldsymbol{n}_{\mathrm{S}_{\infty}}^{+}dS+\left\langle\boldsymbol{\sigma}\cdot\mathcal{E},\mathcal{E}\right\rangle_{\mathrm{V}}\\ &+\frac{d}{dt}\left[(1/2)\left\langle\mathcal{H}_{\mathrm{sca}},\mu_{0}\mathcal{H}_{\mathrm{sca}}\right\rangle_{\mathrm{E}_{3}}+(1/2)\left\langle\varepsilon_{0}\mathcal{E}_{\mathrm{sca}},\mathcal{E}_{\mathrm{sca}}\right\rangle_{\mathrm{E}_{3}}\right]\\ &+\frac{d}{dt}\left[(1/2)\left\langle\mathcal{H},\Delta\boldsymbol{\mu}\cdot\mathcal{H}\right\rangle_{\mathrm{V}}+(1/2)\left\langle\Delta\boldsymbol{\varepsilon}\cdot\mathcal{E},\mathcal{E}\right\rangle_{\mathrm{V}}\right]\end{aligned} \tag{1-2}$$

In Eq. (1-2), $\mathcal{F}_{\mathrm{inc}}$ is the summation of $\mathcal{F}_{\mathrm{imp}}$ and $\mathcal{F}_{\mathrm{env}}$, and called externally incident field; $\mathrm{E}_3$ is the whole three-dimensional Euclidean space; $\mathrm{S}_{\infty}$ is the outer boundary of $\mathrm{E}_3$, and it is a closed spherical surface with infinite radius; $\boldsymbol{n}_{\mathrm{S}_{\infty}}^{+}$ is the outer normal direction of $\mathrm{S}_{\infty}$; $\Delta\boldsymbol{\mu}=\boldsymbol{\mu}-\mathbf{I}\mu_0$, and $\Delta\boldsymbol{\varepsilon}=\boldsymbol{\varepsilon}-\mathbf{I}\varepsilon_0$, where $\mathbf{I}$ is two-order unit dyad.

Integrating Eq. (1-2) on time interval $t_0\sim t_0+\Delta t$, the following equation is immediately obtained[8-Sec.2.4.2],[14].

$$\mathcal{W}_{\mathrm{DRIV}}=\mathcal{E}_{\mathrm{rad}}+\mathcal{E}_{\mathrm{dis}}+\Delta\mathcal{E}_{\mathrm{field}}+\Delta\mathcal{E}_{\mathrm{matter}} \tag{1-3}$$

where the left-hand side term is $\mathcal{W}_{\mathrm{DRIV}}=\int_{t_0}^{t_0+\Delta t}[<\boldsymbol{J}_{\mathrm{sca}},\mathcal{E}_{\mathrm{inc}}>_{\mathrm{V}}+<\mathcal{M}_{\mathrm{sca}},\mathcal{H}_{\mathrm{inc}}>_{\mathrm{V}}]dt$, and the terms in the right-hand side can be similarly interpreted. Obviously, Eq. (1-3) has a very clear physical interpretation: **in time interval $t_0\sim t_0+\Delta t$, the work $\mathcal{W}_{\mathrm{DRIV}}$ done by incident fields ($\mathcal{E}_{\mathrm{inc}},\mathcal{H}_{\mathrm{inc}}$) on induced currents ($\boldsymbol{J}_{\mathrm{sca}},\mathcal{M}_{\mathrm{sca}}$) is transformed into four parts — the radiated energy $\mathcal{E}_{\mathrm{rad}}$ passing through $\mathrm{S}_{\infty}$, the Joule heating energy $\mathcal{E}_{\mathrm{dis}}$ dissipated in V, the increment of the magnetic and electric field energies $\mathcal{E}_{\mathrm{field}}$ stored in $\mathrm{E}_3$, and the increment of the magnetization and polarization energies $\mathcal{E}_{\mathrm{matter}}$ stored in V**. Thus, Eq. (1-3) is a quantitative expression for the transformation between work and energy, and it is very similar to the WORK-ENERGY THEOREM in mechanism[29-Sec.6.2]. Then, Eq. (1-3) and its time-differential version (1-2) are collectively referred to as the ENTIRE-STRUCTURE-ORIENTED WORK-ENERGY THEOREM (ES-WET) in electromagnetism, where **to use modifier "entire-structure-oriented" is because $\mathcal{F}_{\text{inc}}$ directly acts on the entire EM structure rather than only on a partial structure** (for more details, please see Sec. 2.2 and Chaps. 4&5).

In fact, the above ES-WET provides an effective and reasonable resolution for the PROBLEM I mentioned in Sec. 1.2.3, and the resolution is that[8-Sec.3.4.1],[14]: when we focus on calculating the CMs of an objective EM structure placed in non-free-space environment, all external fields — externally impressed field and externally environmental field — can be treated as a whole — externally incident field, and this treatment can guarantee not only reasonable work-energy transformation relation but also environment-independent feature for CMs.

**ES-WET-Based Resolution for PROBLEM II**

**Work term $\mathcal{W}_{\text{DRIV}}$ is the source to drive the work-energy transformation**[8,14], so it is called entire-structure-oriented driving work, and the associated power $\mathcal{P}_{\text{DRIV}}$ is called entire-structure-oriented driving power. Equation (1-2) implies that the driving power has operator expression $\mathcal{P}_{\text{DRIV}} = <\mathcal{J}_{\text{sca}}, \mathcal{E}_{\text{inc}}>_{\text{V}} + <\mathcal{M}_{\text{sca}}, \mathcal{H}_{\text{inc}}>_{\text{V}}$, and the operator is accordingly called ENTIRE-STRUCTURE-ORIENTED DRIVING POWER OPERATOR (ES-DPO). The ES-DPO has two different frequency-domain versions as follows[8-Sec.4.2.1],[14]:

$$P_{\text{DRIV}} = (1/2)\langle \boldsymbol{J}_{\text{sca}}, \boldsymbol{E}_{\text{inc}}\rangle_{\text{V}} + (1/2)\langle \boldsymbol{M}_{\text{sca}}, \boldsymbol{H}_{\text{inc}}\rangle_{\text{V}} \tag{1-4}$$

$$\tilde{P}_{\text{DRIV}} = (1/2)\langle \boldsymbol{J}_{\text{sca}}, \boldsymbol{E}_{\text{inc}}\rangle_{\text{V}} + (1/2)\langle \boldsymbol{H}_{\text{inc}}, \boldsymbol{M}_{\text{sca}}\rangle_{\text{V}} \tag{1-5}$$

where coefficient 1/2 originates from the time average for the power-type quadratic quantity of time-harmonic EM field[28]. Obviously, $P_{\text{DRIV}}$ is equal to neither $\tilde{P}_{\text{DRIV}}$ nor $\tilde{P}_{\text{DRIV}}^{\dagger}$, if both $\boldsymbol{J}_{\text{sca}}$ and $\boldsymbol{M}_{\text{sca}}$ are not zero, and this is just the reason to use a "~" to distinguish them from each other.

In fact, the different frequency-domain ES-DPOs will generate different modal sets[8-Sec.4.2.2],[14]. Specifically, the CMs derived from orthogonalizing $P_{\text{DRIV}}$ satisfy the following orthogonality

$$P_{\text{DRIV}}^{m}\delta_{mn} = (1/2)\langle \boldsymbol{J}_{\text{sca}}^{m}, \boldsymbol{E}_{\text{inc}}^{n}\rangle_{\text{V}} + (1/2)\langle \boldsymbol{M}_{\text{sca}}^{m}, \boldsymbol{H}_{\text{inc}}^{n}\rangle_{\text{V}} \tag{1-6}$$

and the modes derived from orthogonalizing $\tilde{P}_{\text{DRIV}}$ satisfy the following orthogonality

$$\tilde{P}_{\text{DRIV}}^{m}\delta_{mn} = (1/2)\langle \boldsymbol{J}_{\text{sca}}^{m}, \boldsymbol{E}_{\text{inc}}^{n}\rangle_{\text{V}} + (1/2)\langle \boldsymbol{H}_{\text{inc}}^{m}, \boldsymbol{M}_{\text{sca}}^{n}\rangle_{\text{V}} \tag{1-7}$$

where $\delta_{mn}$ is Kronecker's delta symbol, and $P_{\text{DRIV}}^{m}$ and $\tilde{P}_{\text{DRIV}}^{m}$ are the corresponding modal powers. Obviously, the CMs satisfying relation (1-6) are completely decoupled, i.e., the action by the $n$-th modal fields $(\boldsymbol{E}_{\text{inc}}^{n}, \boldsymbol{H}_{\text{inc}}^{n})$ on the $m$-th modal currents $(\boldsymbol{J}_{\text{sca}}^{m}, \boldsymbol{M}_{\text{sca}}^{m})$ is zero if $m \neq n$. But, the modes satisfying relation (1-7) are not decoupled, as exhibited by the intertwining orthogonality (1-7) between $(\boldsymbol{E}_{\text{inc}}^{n}, \boldsymbol{M}_{\text{sca}}^{n})$ and $(\boldsymbol{J}_{\text{sca}}^{m}, \boldsymbol{H}_{\text{inc}}^{m})$ instead of between $(\boldsymbol{E}_{\text{inc}}^{n}, \boldsymbol{H}_{\text{inc}}^{n})$ and $(\boldsymbol{J}_{\text{sca}}^{m}, \boldsymbol{M}_{\text{sca}}^{m})$. Due to this, ES-WET-CMT selects $P_{\text{DRIV}}$ as CM generating operator rather than $\tilde{P}_{\text{DRIV}}$ [8-Sec.4.2.2],[14]. In addition, the real part of $P_{\text{DRIV}}^{m}$ is usually normalized to 1[9~13], i.e., $\operatorname{Re} P_{\text{DRIV}}^{m} = 1$, and then orthogonality (1-6) becomes $(1+\lambda_m)\delta_{mn} = (1/2) < \boldsymbol{J}_{\text{sca}}^{m}, \boldsymbol{E}_{\text{inc}}^{n} >_{\text{V}} + (1/2) < \boldsymbol{M}_{\text{sca}}^{m}, \boldsymbol{H}_{\text{inc}}^{n} >_{\text{V}}$, where $\lambda_m$ is the corresponding characteristic value. The physical reason to normalize CMs such that $\operatorname{Re} P_{\text{DRIV}}^{m} = 1$ was explained in Refs. [14] and [27-Sec.1.2.4.7].

In fact, ES-DPO $P_{\text{DRIV}}$ is equivalent to the symmetric IMO in the sense of generating CMs[8-Sec.2.4],[14,30,31], but not equivalent to the asymmetric IMO, and this provides a ES-WET-based resolution for the PROBLEM II mentioned in Sec. 1.2.3.

**ES-WET-Based Resolution for PROBLEM III**

In fact, the CMs satisfying frequency-domain power-decoupling feature (1-6) also satisfy the following time-domain energy-decoupling feature (or alternatively called time-averaged power-decoupling feature)[8-Sec.2.4.2],[14]

$$(1/T)\int_{t_0}^{t_0+T}\left[\left\langle \boldsymbol{J}_{\text{sca}}^{m}, \boldsymbol{\mathcal{E}}_{\text{inc}}^{n}\right\rangle_{\text{V}} + \left\langle \boldsymbol{\mathcal{M}}_{\text{sca}}^{m}, \boldsymbol{\mathcal{H}}_{\text{inc}}^{n}\right\rangle_{\text{V}}\right]dt = \delta_{mn} \tag{1-8}$$

and the following decoupling feature[14]

$$(1/2)\left\langle \frac{1}{\eta_0}\cdot \boldsymbol{E}_{\text{sca}}^{m}, \boldsymbol{E}_{\text{sca}}^{n}\right\rangle_{\text{S}_\infty} + (1/2)\left\langle \boldsymbol{\sigma}\cdot \boldsymbol{E}^{m}, \boldsymbol{E}^{n}\right\rangle_{\text{V}} = \delta_{mn} \tag{1-9}$$

where $\eta_0 = \sqrt{\mu_0/\varepsilon_0}$ is the free-space wave impedance ($\mu_0$ and $\varepsilon_0$ are free-space permeability and permittivity), and $T$ is the time period of the time-harmonic EM field.

The time-domain energy-decoupling feature (1-8) has a very clear physical interpretation: **in any integral period, the $n$-th modal fields $(\boldsymbol{\mathcal{E}}_{\text{inc}}^{n}, \boldsymbol{\mathcal{H}}_{\text{inc}}^{n})$ don't supply net energy to the $m$-th modal currents $(\boldsymbol{J}_{\text{sca}}^{m}, \boldsymbol{\mathcal{M}}_{\text{sca}}^{m})$, if $m \neq n$**. At the same time, the frequency-domain decoupling feature (1-9) clearly exhibits the fact that: when the material structure is lossy ($\boldsymbol{\sigma} \neq 0$), the modal far fields may not be orthogonal[14]. Then, it clearly reveals the core physical features of CMs — energy decoupling (rather than far-field orthogonality).

This is just the ES-WET-based resolution for the PROBLEM III mentioned in Sec. 1.2.3, and some more detailed physical explanations for this resolution can be found in Refs. [14] and [27-Sec.1.2.4.5].

**ES-WET-Based Resolution for PROBLEM IV**

Using the modal decomposition proposed in Refs. [8-Sec.3.3] and [32], any working mode $\boldsymbol{C}_{\text{sca}}$ can be decomposed into three energy-decoupled fundamental components as $\boldsymbol{C}_{\text{sca}} = \boldsymbol{C}_{\text{sca}}^{\text{ind}} + \boldsymbol{C}_{\text{sca}}^{\text{res}} + \boldsymbol{C}_{\text{sca}}^{\text{cap}}$, and then the incident field $\boldsymbol{F}_{\text{inc}}$ distributing on $\text{V}$ can be correspondingly decomposed as $\boldsymbol{F}_{\text{inc}} = \boldsymbol{F}_{\text{inc}}^{\text{ind}} + \boldsymbol{F}_{\text{inc}}^{\text{res}} + \boldsymbol{F}_{\text{inc}}^{\text{cap}}$. Based on the decompositions, Ref. [14] introduced the concept of Θ-factor as follows:

$$\Theta(\boldsymbol{C}_{\text{sca}}) = \frac{\text{Im}\left\{\frac{1}{2}\langle \boldsymbol{J}_{\text{sca}}^{\text{ind}}, \boldsymbol{E}_{\text{inc}}^{\text{ind}} \rangle_{\text{V}} + \frac{1}{2}\langle \boldsymbol{M}_{\text{sca}}^{\text{ind}}, \boldsymbol{H}_{\text{inc}}^{\text{ind}} \rangle_{\text{V}}\right\} - \text{Im}\left\{\frac{1}{2}\langle \boldsymbol{J}_{\text{sca}}^{\text{cap}}, \boldsymbol{E}_{\text{inc}}^{\text{cap}} \rangle_{\text{V}} + \frac{1}{2}\langle \boldsymbol{M}_{\text{sca}}^{\text{cap}}, \boldsymbol{H}_{\text{inc}}^{\text{cap}} \rangle_{\text{V}}\right\}}{\text{Re}\left\{\frac{1}{2}\langle \boldsymbol{J}_{\text{sca}}, \boldsymbol{E}_{\text{inc}} \rangle_{\text{V}} + \frac{1}{2}\langle \boldsymbol{M}_{\text{sca}}, \boldsymbol{H}_{\text{inc}} \rangle_{\text{V}}\right\}} \tag{1-10}$$

As explained in Ref. [14], **the above $\Theta(\boldsymbol{C}_{\text{sca}})$ quantitatively characterizes the mismatching degree between the phase of field $\boldsymbol{F}_{\text{inc}}$ and the phase of current $\boldsymbol{C}_{\text{sca}}$**, so it is called modal field-current phase-mismatching factor. In fact, the Θ-factor can also be viewed as a generalized version of the classical Q-fact (i.e., quality factor).

For any single CM $\boldsymbol{C}_{\text{sca}}^{m}$, there exists a very simple relation between Θ-factor and characteristic value as follows[14]:

$$\Theta\left(\boldsymbol{C}_{\text{sca}}^{m}\right) = |\lambda_m| \tag{1-11}$$

and this relation clearly reveals **the physical meaning of $|\lambda_m|$ — field-current phase-mismatching degree of the *m*-th CM**. In fact, this is just the reason why the CMs with smaller $|\lambda_m|$ are more desired usually.

In addition, the above physical interpretation for the characteristic values of material structures is also applicable to the characteristic values of metallic and composite structures, and then the PROBLEM IV mentioned in Sec. 1.2.3 is successfully resolved under ES-WET framework.

**ES-WET-Based Resolution for PROBLEM V**

Based on ES-WET (1-3), driving power $P_{\text{DRIV}}$ can be decomposed into two terms — dissipated power $P_{\text{dis}}$ and non-dissipated power $P_{\text{non-dis}}$ — as follows[27-Sec.1.2.4.8]:

$$P_{\text{DRIV}} = \underbrace{\left(P_{\text{DRIV}} - P_{\text{dis}}\right)}_{P_{\text{non-dis}}} + P_{\text{dis}} \tag{1-12}$$

If the equivalent surface currents on $\partial\text{V}$ are denoted as $(\boldsymbol{J}_{\text{equ}}, \boldsymbol{M}_{\text{equ}})$, which are defined as $\boldsymbol{J}_{\text{equ}} = \boldsymbol{n}_{\partial V}^{-} \times \boldsymbol{H}$ and $\boldsymbol{M}_{\text{equ}} = \boldsymbol{E} \times \boldsymbol{n}_{\partial V}^{-}$, then the $P_{\text{non-dis}}$ has the following operator expression[8-Sec.6.4.1],[27-Sec.1.2.4.8],[33]

$$P_{\text{non-dis}} = \underbrace{-\frac{1}{2}\left\langle \boldsymbol{J}_{\text{equ}}, \text{P.V.}\,\mathcal{E}_0\left(\boldsymbol{J}_{\text{equ}}, \boldsymbol{M}_{\text{equ}}\right)\right\rangle_{\partial V} - \frac{1}{2}\left\langle \boldsymbol{M}_{\text{equ}}, \text{P.V.}\,\mathcal{H}_0\left(\boldsymbol{J}_{\text{equ}}, \boldsymbol{M}_{\text{equ}}\right)\right\rangle_{\partial V}}_{\mathcal{P}_{\text{non-dis}}\left(\boldsymbol{J}_{\text{equ}}, \boldsymbol{M}_{\text{equ}}\right)} \tag{1-13}$$

but however the operator expression for $P_{\text{dis}}$ has the following multiple choices[27-Sec.1.2.4.8],[33]

$$P_{\text{dis}} = \begin{cases} \underbrace{\text{Re}\left\{-(1/2)\left\langle \boldsymbol{J}_{\text{equ}}, \boldsymbol{n}_{\partial V}^{-} \times \boldsymbol{M}_{\text{equ}}\right\rangle_{\partial V}\right\}}_{\mathcal{P}_{\text{dis}}^{\text{JM}}\left(\boldsymbol{J}_{\text{equ}}, \boldsymbol{M}_{\text{equ}}\right)} \\ \underbrace{\text{Re}\left\{-\left\langle \boldsymbol{J}_{\text{equ}}, \text{P.V.}\,\mathcal{E}_{\text{m}}\left(\boldsymbol{J}_{\text{equ}}, \boldsymbol{M}_{\text{equ}}\right)\right\rangle_{\partial V}\right\}}_{\mathcal{P}_{\text{dis}}^{\text{JE}}\left(\boldsymbol{J}_{\text{equ}}, \boldsymbol{M}_{\text{equ}}\right)} \\ \underbrace{\text{Re}\left\{-\left\langle \boldsymbol{M}_{\text{equ}}, \text{P.V.}\,\mathcal{H}_{\text{m}}\left(\boldsymbol{J}_{\text{equ}}, \boldsymbol{M}_{\text{equ}}\right)\right\rangle_{\partial V}\right\}}_{\mathcal{P}_{\text{dis}}^{\text{MH}}\left(\boldsymbol{J}_{\text{equ}}, \boldsymbol{M}_{\text{equ}}\right)} \\ \underbrace{\text{Re}\left\{-(1/2)\left\langle \boldsymbol{J}_{\text{equ}}, \text{P.V.}\,\mathcal{E}_{\text{m}}\left(\boldsymbol{J}_{\text{equ}}, \boldsymbol{M}_{\text{equ}}\right)\right\rangle_{\partial V} - (1/2)\left\langle \boldsymbol{M}_{\text{equ}}, \text{P.V.}\,\mathcal{H}_{\text{m}}\left(\boldsymbol{J}_{\text{equ}}, \boldsymbol{M}_{\text{equ}}\right)\right\rangle_{\partial V}\right\}}_{\mathcal{P}_{\text{dis}}^{\text{pmchwt}}\left(\boldsymbol{J}_{\text{equ}}, \boldsymbol{M}_{\text{equ}}\right)} \\ \underbrace{-(1/2)\left\langle \boldsymbol{J}_{\text{equ}}, \text{P.V.}\,\mathcal{E}_{\text{m}}\left(\boldsymbol{J}_{\text{equ}}, \boldsymbol{M}_{\text{equ}}\right)\right\rangle_{\partial V} - (1/2)\left\langle \boldsymbol{M}_{\text{equ}}, \text{P.V.}\,\mathcal{H}_{\text{m}}\left(\boldsymbol{J}_{\text{equ}}, \boldsymbol{M}_{\text{equ}}\right)\right\rangle_{\partial V}}_{\mathcal{P}_{\text{dis}}^{\text{PMCHWT}}\left(\boldsymbol{J}_{\text{equ}}, \boldsymbol{M}_{\text{equ}}\right)} \end{cases} \tag{1-14}$$

Here, operator $\mathcal{F}_{0/\text{m}}(\boldsymbol{J}, \boldsymbol{M})$ is defined as that $\mathcal{F}_{0/\text{m}}(\boldsymbol{J}, \boldsymbol{M}) = \mathbf{G}_{0/\text{m}}^{JF} * \boldsymbol{J} + \mathbf{G}_{0/\text{m}}^{MF} * \boldsymbol{M}$, where $\mathcal{F}_{0/\text{m}} = \mathcal{E}_{0/\text{m}} / \mathcal{H}_{0/\text{m}}$ and correspondingly $F = E / H$, and $\mathbf{G}_{0/\text{m}}^{JF}$ and $\mathbf{G}_{0/\text{m}}^{MF}$ are the dyadic Green's functions with parameters $(\mu_0, \varepsilon_0) / (\boldsymbol{\mu}, \boldsymbol{\varepsilon}, \boldsymbol{\sigma})$, and "$*$" is the convolution integral operator; $\text{P.V.}\,\mathcal{F}_{0/\text{m}}$ denotes the principal value of operator $\mathcal{F}_{0/\text{m}}$. Thus, the operator expression for $P_{\text{DRIV}}$ has the following multiple choices

$$P_{\text{DRIV}} = \mathcal{P}_{\text{non-dis}}\left(\boldsymbol{J}_{\text{equ}}, \boldsymbol{M}_{\text{equ}}\right) + \begin{cases} \mathcal{P}_{\text{dis}}^{\text{JM}}\left(\boldsymbol{J}_{\text{equ}}, \boldsymbol{M}_{\text{equ}}\right) \\ \mathcal{P}_{\text{dis}}^{\text{JE}}\left(\boldsymbol{J}_{\text{equ}}, \boldsymbol{M}_{\text{equ}}\right) \\ \mathcal{P}_{\text{dis}}^{\text{MH}}\left(\boldsymbol{J}_{\text{equ}}, \boldsymbol{M}_{\text{equ}}\right) \\ \mathcal{P}_{\text{dis}}^{\text{pmchwt}}\left(\boldsymbol{J}_{\text{equ}}, \boldsymbol{M}_{\text{equ}}\right) \\ \mathcal{P}_{\text{dis}}^{\text{PMCHWT}}\left(\boldsymbol{J}_{\text{equ}}, \boldsymbol{M}_{\text{equ}}\right) \end{cases} \tag{1-15}$$

The unwanted and spurious modes mentioned in the previous PROBLEM V are originated from the following reasons.

Reason 1: The different operator expressions in Eq. (1-15) have different numerical performances. **When an inappropriate one is used to calculate CMs, some unwanted modes will be resulted.**

Reason 2: The $\boldsymbol{J}_{\text{equ}}$ and $\boldsymbol{M}_{\text{equ}}$ are involved in ES-DPO, and they are not independent of each other. **When the dependence relation between $\boldsymbol{J}_{\text{equ}}$ and $\boldsymbol{M}_{\text{equ}}$ is overlooked, some spurious modes will be resulted.**

For details, please see Refs. [8-Chap.6], [27-Sec.1.2.4.8], and [14,33,34].

Taking the homogeneous isotropic material structures (with relative permeability $\mu_{\text{r}}$, relative permittivity $\varepsilon_{\text{r}}$, and conductivity $\sigma$) as example, Ref. [27-Sec.1.2.4.8] proposed an effective scheme to suppress the unwanted modes, and the scheme is to express ES-DPO as follows:

$$P_{\text{DRIV}} = \mathcal{P}_{\text{non-dis}}\left(\boldsymbol{J}_{\text{equ}},\boldsymbol{M}_{\text{equ}}\right) + \begin{cases} 0 & \text{, if } \sigma = 0 \\ \mathcal{P}_{\text{dis}}^{\text{JM}}\left(\boldsymbol{J}_{\text{equ}},\boldsymbol{M}_{\text{equ}}\right) \text{ or } \mathcal{P}_{\text{dis}}^{\text{MH}}\left(\boldsymbol{J}_{\text{equ}},\boldsymbol{M}_{\text{equ}}\right) & \text{, if } \sigma \neq 0 \text{ and } \mu_{\text{r}} \geq \varepsilon_{\text{r}} \\ \mathcal{P}_{\text{dis}}^{\text{JM}}\left(\boldsymbol{J}_{\text{equ}},\boldsymbol{M}_{\text{equ}}\right) \text{ or } \mathcal{P}_{\text{dis}}^{\text{JE}}\left(\boldsymbol{J}_{\text{equ}},\boldsymbol{M}_{\text{equ}}\right) & \text{, if } \sigma \neq 0 \text{ and } \mu_{\text{r}} \leq \varepsilon_{\text{r}} \end{cases} \tag{1-16}$$

For effectively suppressing the spurious modes, some somewhat different schemes were proposed to integrate the dependence relation between $\boldsymbol{J}_{\text{equ}}$ and $\boldsymbol{M}_{\text{equ}}$ into ES-DPO, and the schemes are the dependent variable elimination (DVE) proposed in Refs. [8,30,31,33], the solution domain compression (SDC) proposed in Refs. [14,15], and the intermediate variable method (IVM) proposed in Ref. [34].

As exhibited in Refs. [8,14,15,30,31,33,34], the above schemes can effectively resolve the PROBLEM V mentioned in Sec. 1.2.3.

**ES-WET-Based Resolution for PROBLEM VI**

As shown in Fig. 1-1 and ES-WET (1-2)&(1-3), **the steady working of CM needs a non-zero externally incident field $\boldsymbol{F}_{\text{inc}}$ as the driver for material structure**[8,14]. Thus, <u>CMT (including SM-CMT[5~7], IE-CMT[9~13], and ES-WET-CMT[8,14,15]) is not a source-free modal analysis theory</u>. Specifically, when we do the CM analysis for a EM structure, the related total field is the summation of $\boldsymbol{F}_{\text{inc}}$ and $\boldsymbol{F}_{\text{sca}}$, but not the only $\boldsymbol{F}_{\text{sca}}$.

**The *m*-th modal total field $\boldsymbol{F}^{m}$ as the summation of $\boldsymbol{F}_{\text{inc}}^{m}$ and $\boldsymbol{F}_{\text{sca}}^{m}$ satisfies the homogeneous tangential electric field boundary condition on metallic boundaries, but neither $\boldsymbol{F}_{\text{inc}}^{m}$ nor $\boldsymbol{F}_{\text{sca}}^{m}$ satisfies.** This is just the reason why the modal characteristic field $\boldsymbol{F}_{\text{sca}}^{m}$ cannot satisfy the condition as mentioned in the PROBLEM VI.

## 1.3 Major Problem and Challenge in the Realm of EM Modal Analysis

Some theoretical and numerical problems on IE-CMT have been resolved under IE framework itself or ES-WET framework, and IE-CMT has had some antenna-oriented engineering applications, but it is recently found out that: **for some classical antennas, such as horn antenna and Yagi-Uda antenna, the IE-CMT-based modal analysis fails**.

For the horn antenna shown in Fig. 1-2(a), its commonly used working mode has the radiation pattern shown in Fig. 1-2(b). But unfortunately, the commonly used mode is not contained in the IE-based CM set. The “radiation patterns” of the first several lower-order IE-based CMs are shown in Fig. 1-3, and, evidently, all of them are not consistent with the commonly used one shown in Fig. 1-2(b).

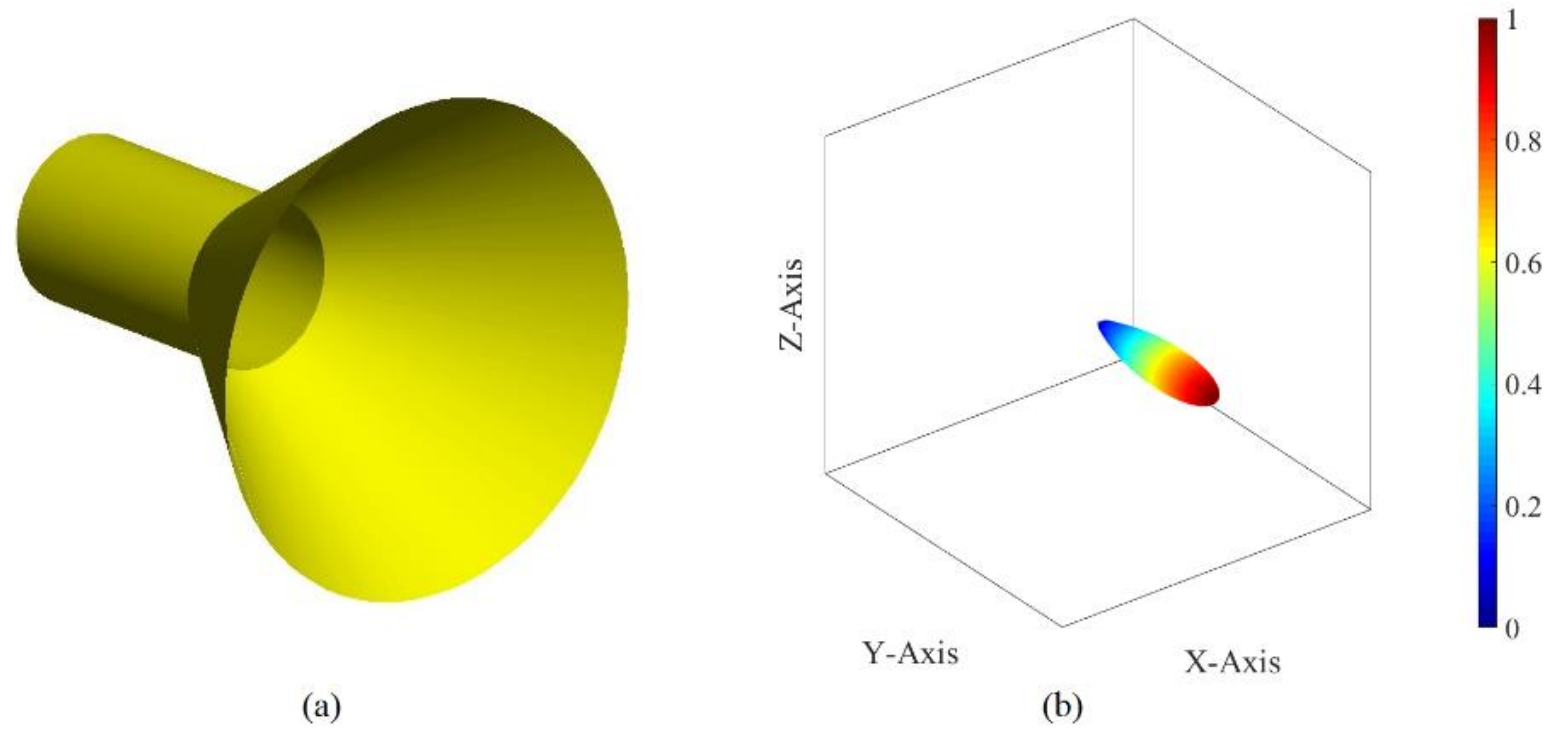

Figure 1-2 (a) Geometry of a horn antenna, whose size is given in Ref. [35]. (b) Radiation pattern of the commonly used mode (working at 11.7 GHz) of the horn antenna.

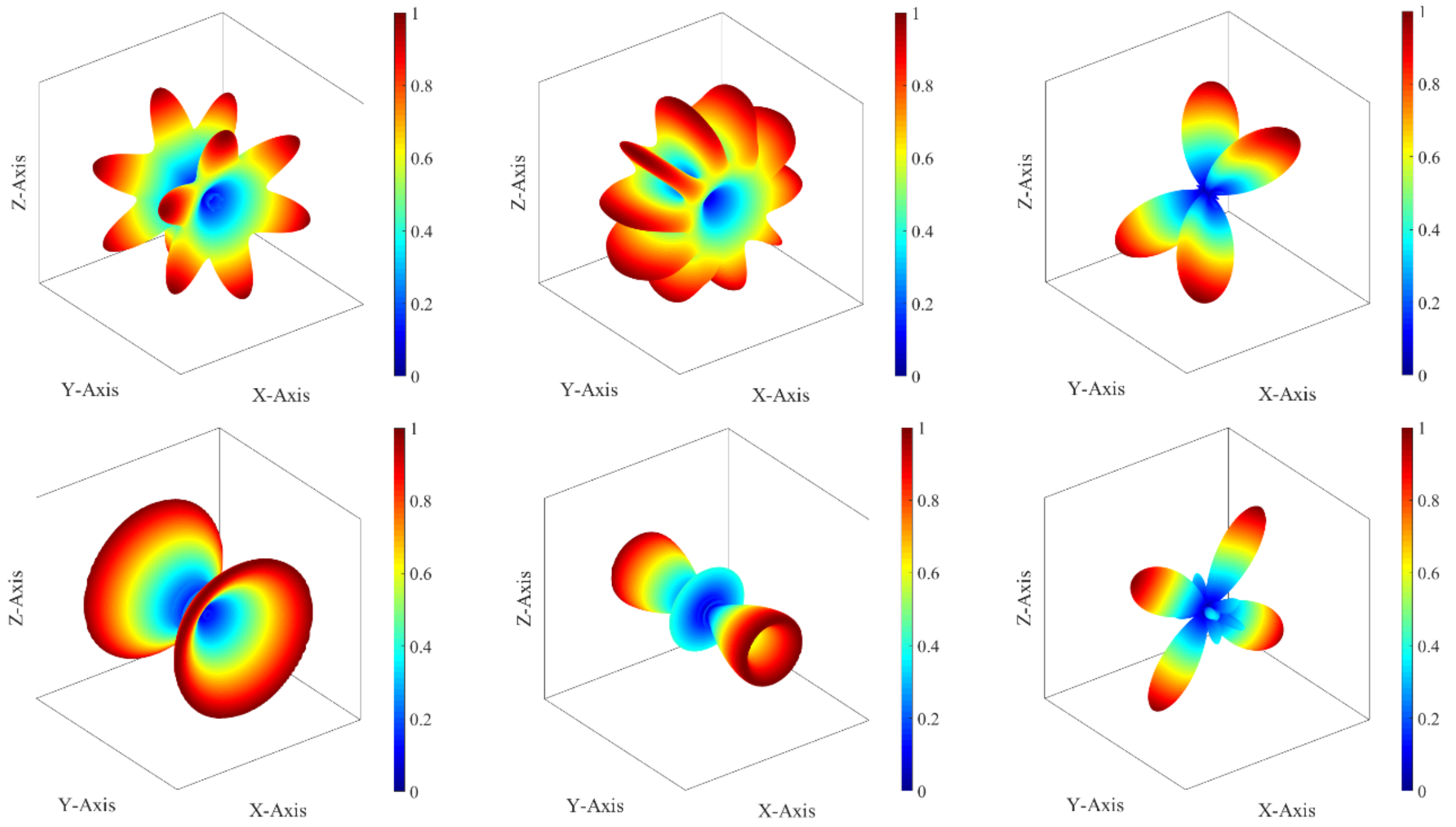

Figure 1-3 “Radiation patterns” of the first several lower-order IE-based CMs working at 11.7 GHz[35]. The CMs are calculated from the formulation proposed in Ref. [10].

For the Yagi-Uda antenna shown in Fig. 1-4(a), its commonly used resonant mode has the radiation pattern shown in Fig. 1-4(b). But unfortunately, the commonly used resonant mode is not contained in the IE-based CM set. The “radiation patterns” of the first several lower-order IE-based resonant CMs are shown in Fig. 1-5, and, evidently, all of them are not consistent with the commonly used one shown in Fig. 1-4(b).

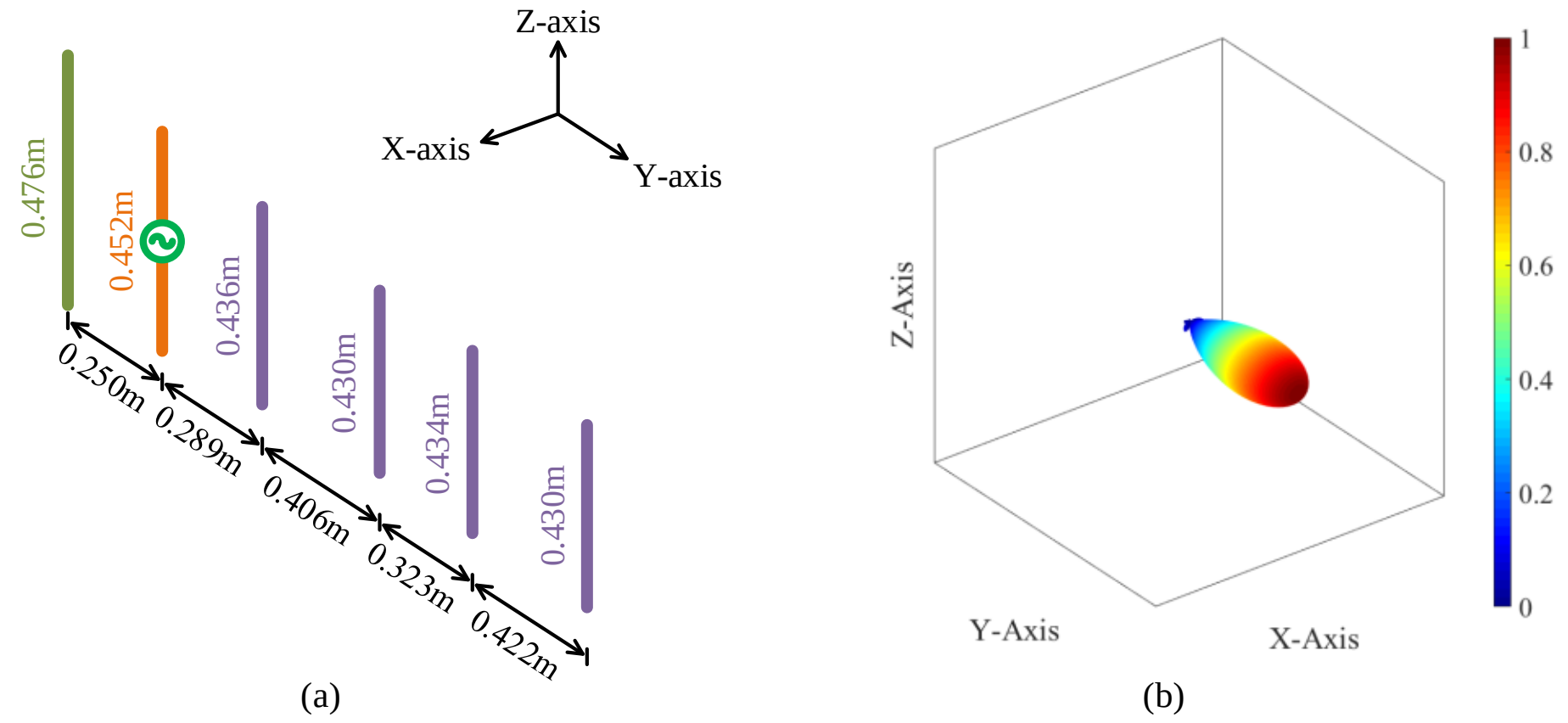

Figure 1-4 (a) Size of a typical 6-element metallic Yagi-Uda antenna. The size is designed according to the formulation proposed in Ref. [36]. (b) Radiation pattern of the commonly used resonant mode (working at 300 MHz) of the Yagi-Uda antenna.

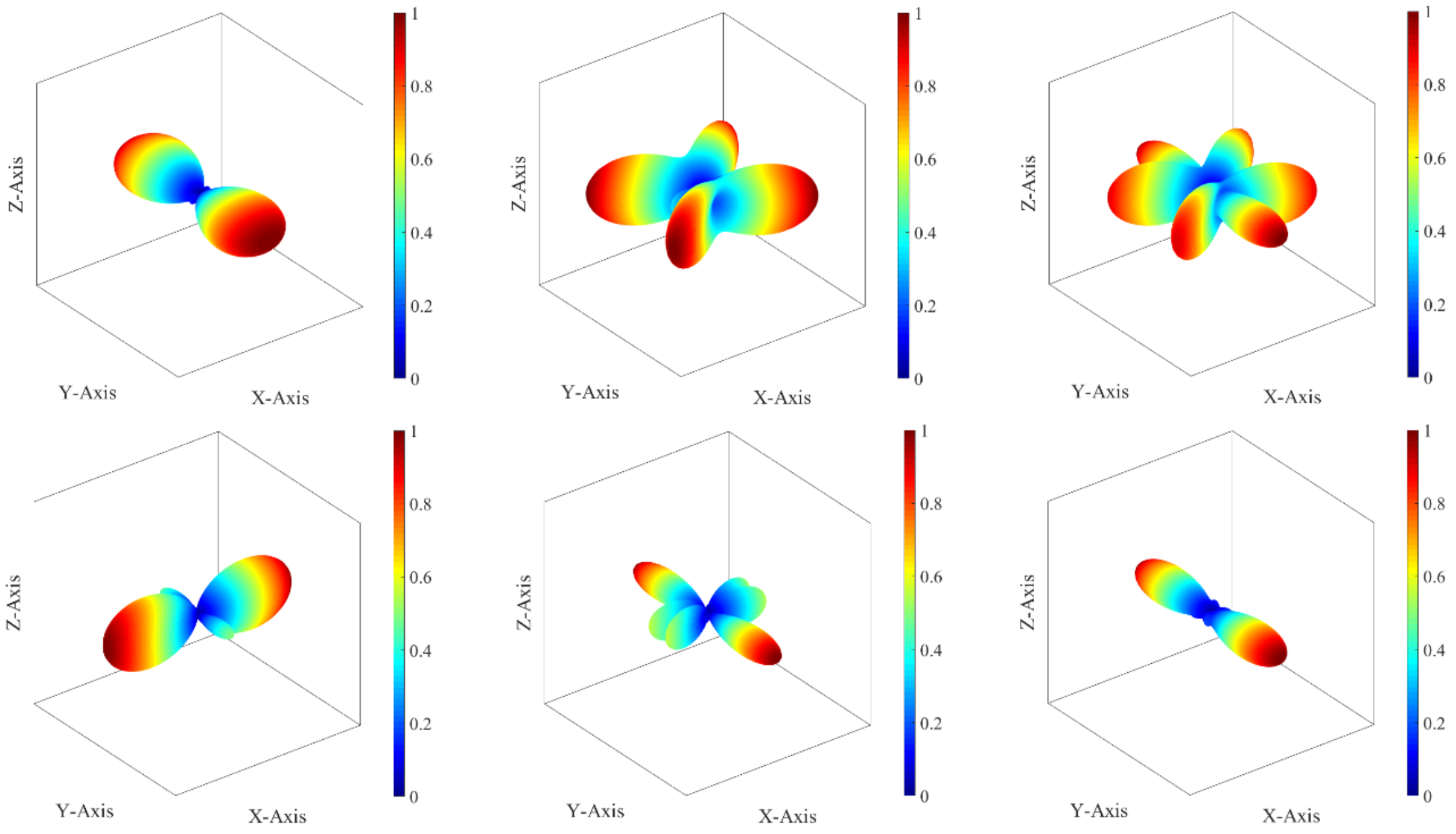

Figure 1-5 “Radiation patterns” of the first several lower-order IE-based resonant CMs[37]. The CMs are calculated from the formulation proposed in Ref. [10].

**The above these expose a major challenge existing in the realm of EM modal analysis: “how to explain the reason leading to the failure of IE-CMT-based antenna modal analysis?” and “how to do an effective modal analysis for the antennas?”**

## 1.4 Main Innovations and Contributions of This Report

This report mainly aims to explain the failure of the IE-CMT-based modal analysis for transmitting antennas and providing effective modal analysis for transmitting antennas.

Besides the above two main aims, this report also aims to generalize the new modal analysis method from transmitting antennas to other kinds of EM structures, such as receiving antennas, wave-guiding structures, and self-oscillating structures etc.

### 1.4.1 Explaining the Failure of the IE-CMT-Based Modal Analysis for Transmitting Antennas

As revealed in ES-WET framework, the physical purpose of IE-CMT is to construct a set of CMs (which can be excited independently) for a pre-selected objective EM structure, and the purpose can be effectively realized by orthogonalizing ES-DPO (which is the operator form of driving power). As exhibited in Fig. 1-1 and Eq. (1-4), the driving power is the power done by an externally incident field on the EM structure, so the externally incident field is just the excitation source used to sustain a steady working of CM.

Evidently, **this kind of excitation manner (i.e., externally-incident-field excitation) is for scattering structures (such as the ones shown in Fig. 1-6(a)) rather than for transmitting antennas (such as the ones shown in Figs. 1-6(b) and 1-6(c))**, as explained by the *IEEE standards terms*[38]:

scattering structure is *a secondary structure generating scattered fields resulted from the scattered currents induced on the structure by some fields incident on the structure from some primary sources*[38-pp.1006];

transmitting antenna is *a device that generates high-frequency electric energy, controlled or modulated, which can be emitted from a finite region in the form of unguided waves*[38-pps.369&1210].

Thus, there exists the conclusion that: **strictly speaking, IE-CMT is a modal analysis theory for scattering structures rather than for transmitting antennas**. This is just the reason why IE-CMT fails to analyze many classical transmitting antennas.

Some more careful discussions on the above conclusion have been given in some literatures, and will also be summarized in the subsequent chapters, from the aspects of excitation manner (Refs. [35,37] & Chap. 2), power transportation process (Ref. [35] & Chap. 3), work-energy transformation process (Ref. [37] & Chaps. 4&5), modal calculation process (Ref. [37]), and modal current distribution (Ref. [37]).

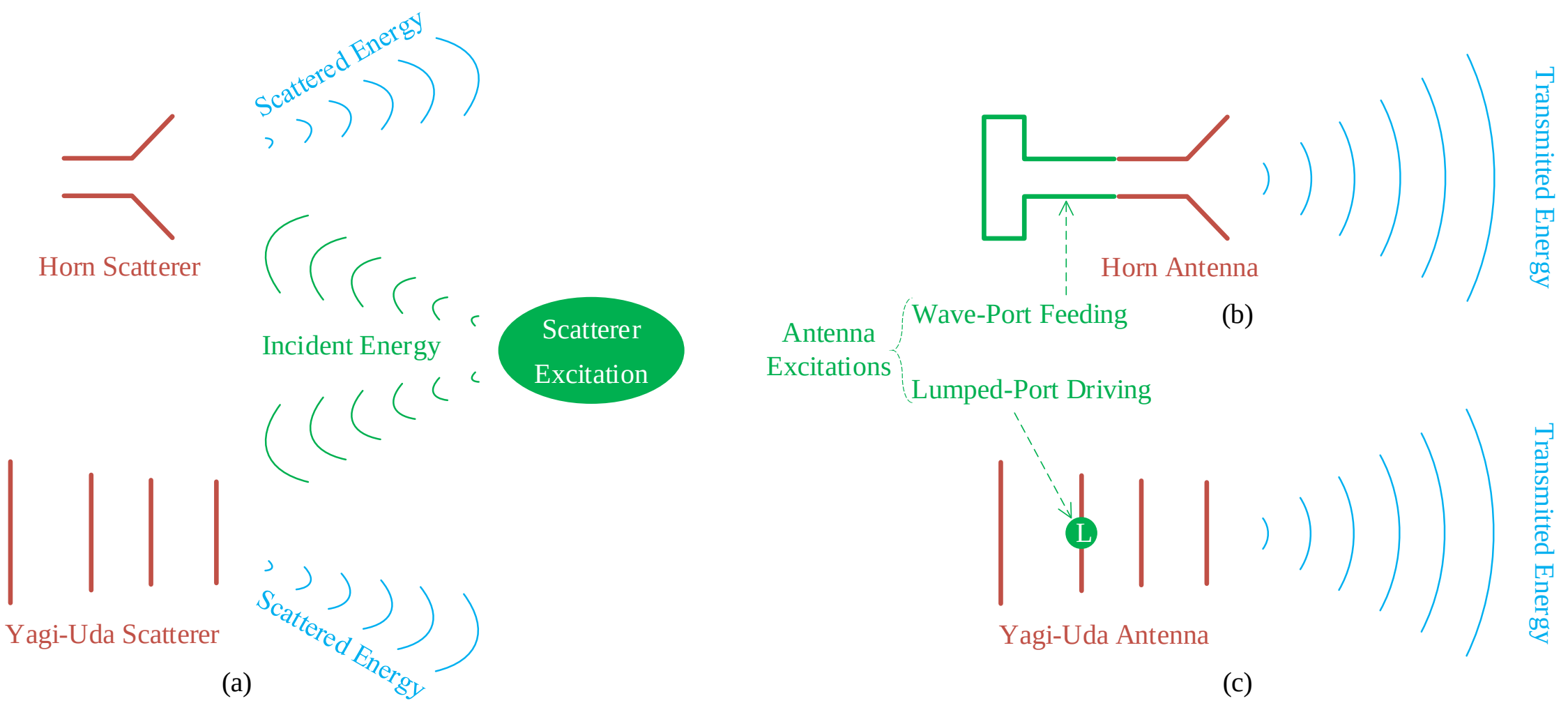

Figure 1-6 (a) Externally-incident-field-driven horn scatterer and Yagi-Uda scatterer, (b) wave-port-fed horn antenna, and (c) lumped-port-driven Yagi-Uda antenna.

Besides the difference between the excitation manners for scattering structures and transmitting antennas, there also exist some differences among the excitation manners for different transmitting antennas. According to the differences of excitation manners, the antennas are usually classified into two categories: wave-port-fed antennas (such as the horn antenna shown in Fig. 1-6(b)) and lumped-port-driven antennas (such as the Yagi-Uda antenna shown in Fig. 1-6(c), where the lumped port Ⓛ can be connected to either voltage source (~) or current source (→)). Due to their different excitation manners, this report discusses the two different kinds of antennas separately.

## 1.4.2 PTT-Based Modal Analysis for Wave-Port-Fed EM Structures

Besides wave-port-fed transmitting antenna, there also exist some other kinds of EM structures fed by wave ports, such as the wave-port-fed feeding waveguide, receiving antenna, and loading waveguide shown in Fig. 1-7.

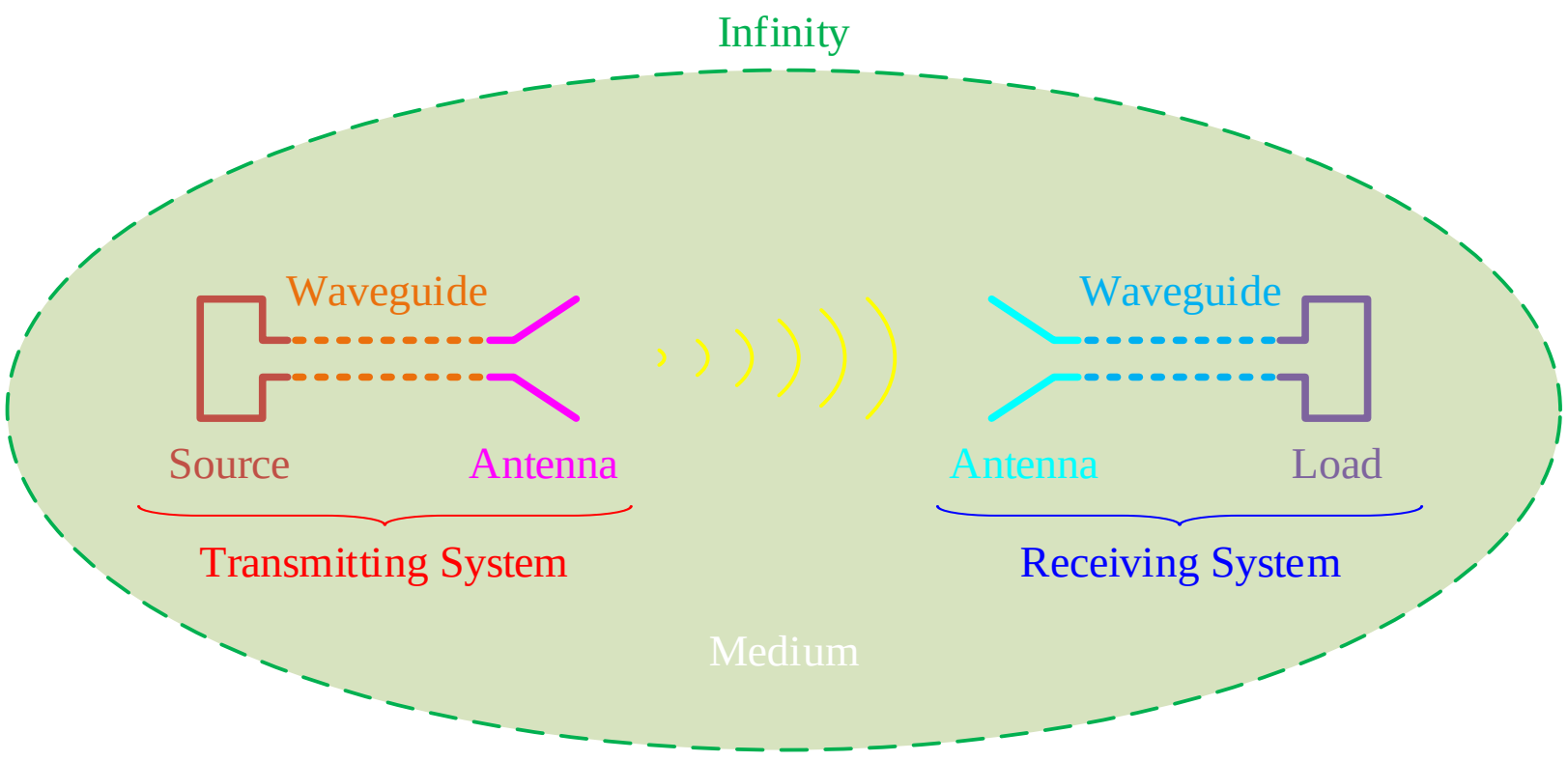

Figure 1-7 Transmitting and receiving systems and their sub-structures.

Qualitatively speaking, along feeding waveguide, EM power flows from power source to transmitting antenna; with the modulation of transmitting antenna, the power is released into surrounding medium, and a part of released power will finally reach receiving system; with the collection of receiving antenna, a part of reached power is received by receiving system (the other part is scattered); along loading waveguide, the received power flows from receiving antenna to power load.

**The above power flow process can be quantitatively formulated in terms of the following POWER TRANSPORT THEOREM (PTT)**[27-Chap.2]

$$P^{\mathrm{S}\rightleftharpoons\mathrm{G}} = \left(P_{\mathrm{dis}}^{\mathrm{G}} + jP_{\mathrm{sto}}^{\mathrm{G}}\right) + \underbrace{\left(P_{\mathrm{dis}}^{\mathrm{A}} + jP_{\mathrm{sto}}^{\mathrm{A}}\right) + \underbrace{\left(P_{\mathrm{Mdis}}^{\mathrm{Mdis}} + jP_{\mathrm{Msto}}^{\mathrm{Msto}}\right) + P_{\mathrm{sca}}^{\mathrm{rad}} + \underbrace{\left(P_{\mathrm{A}}^{\mathrm{dis}} + jP_{\mathrm{A}}^{\mathrm{sto}}\right) + P_{\mathrm{A}\rightleftharpoons\mathrm{G}}}_{P_{\mathrm{M}\rightleftharpoons\mathrm{A}}}}_{P^{\mathrm{A}\rightleftharpoons\mathrm{M}}}}_{P^{\mathrm{G}\rightleftharpoons\mathrm{A}}} \tag{1-17}$$

where the physical meanings of $P^{\mathrm{S}\rightleftharpoons\mathrm{G}}$, $P^{\mathrm{G}\rightleftharpoons\mathrm{A}}$, $P_{\mathrm{M}\rightleftharpoons\mathrm{A}}$, and $P_{\mathrm{A}\rightleftharpoons\mathrm{G}}$ are as follows:

$P^{\mathrm{S}\rightleftharpoons\mathrm{G}}$ is the net power inputted into feeding waveguide, and it is called the input power of feeding waveguide;

$P^{\mathrm{G}\rightleftharpoons\mathrm{A}}$ is the net power inputted into transmitting antenna, and it is called the input power of transmitting antenna;

$P_{\mathrm{M}\rightleftharpoons\mathrm{A}}$ is the net power inputted into receiving antenna, and it is called the input power of receiving antenna;

$P_{\mathrm{A}\rightleftharpoons\mathrm{G}}$ is the net power inputted into loading waveguide, and it is called the input power of loading waveguide.

A rigorous derivation for PTT (1-17) and the physical meanings of the other powers had been carefully discussed in Ref. [27-Chap.2], and will be simply reviewed in Sec. 2.4.

Under a unified PTT framework, this report proposes a universal modal analysis theory — DECOUPLING MODE THEORY (DMT) — for the various wave-port-fed EM structures. **The PTT-based DMT (PTT-DMT) can effectively construct a set of ENERGY-DECOUPLED MODES (DMs) for a pre-selected objective feeding waveguide**[27-Chap.3],[39] **/ transmitting antenna**[27-Chap.6],[35] **/ receiving antenna**[27-Chap.7] **/ loading waveguide**[27-Chap.3] **by orthogonalizing INPUT POWER OPERATOR (IPO) $\boldsymbol{P}^{\mathrm{S}\rightleftharpoons\mathrm{G}}$ / $\boldsymbol{P}^{\mathrm{G}\rightleftharpoons\mathrm{A}}$ / $\boldsymbol{P}_{\mathrm{M}\rightleftharpoons\mathrm{A}}$ / $\boldsymbol{P}_{\mathrm{A}\rightleftharpoons\mathrm{G}}$.** In addition, the PTT-DMT can also be further generalized to constructing the DMs of cascaded systems, such as waveguide-antenna cascaded system[27-Sec.8.2] and antenna-environment-antenna cascaded system[27-Sec.8.3] etc., such that the complicated modal matching process used to analyze cascaded systems can be effectively avoided.

### 1.4.3 PS-WET-Based Modal Analysis for Lumped-Port-Driven EM Structures

**Different from the wave-port-fed EM structures governed by PTT, the energy utilization process of lumped-port-driven transmitting antennas is governed by PARTIAL-STRUCTURE-ORIENTED WORK-ENERGY THEOREM (PS-WET) similar but not identical to the incident-field-driven scattering structures[27-App.H],[37].**

But, the driving power used to sustain a steady work-energy transformation for lumped-port-driven transmitting antenna has a somewhat different operator form from the one for incident-field-driven scattering structure. Taking the Yagi-Uda antenna shown in Fig. 1-6(c) as an example, the antenna-oriented DRIVING POWER OPERATOR (DPO) is[37]

$$P_{\text{driv}} = (1/2)\langle \boldsymbol{J}_{\text{act}}, \boldsymbol{E}_{\text{driv}} \rangle_{\text{active element}} \tag{1-18}$$

where $\boldsymbol{E}_{\text{driv}}$ is the driving field provided by lumped port, and $\boldsymbol{J}_{\text{driv}}$ is the current on the active element (which is directly connected to the lumped port).

Clearly, the main difference between the excitation manners of incident-field-driven scattering structure and lumped-port-driven transmitting antenna is that: **the incident field used to excite the former directly acts on the entire scattering structure (as shown in Fig. 1-6(a)), but the driving field used to excite the latter directly acts on a partial structure of the latter (as shown in Fig. 1-6(c))**[37]. To effectively distinguish the ENTIRE-STRUCTURE-ORIENTED WET/DPO from the PARTIAL-STRUCTURE-ORIENTED WET/DPO, the abbreviation for them are written as ES-WET/ES-DPO and PS-WET/PS-DPO respectively, and <u>their DPOs are also assigned different symbols $P_{\text{DRIV}}$ and $P_{\text{driv}}$</u>.

For an incident-field-driven scattering structure, its energy-decoupled CMs can be effectively constructed by orthogonalizing ES-DPO. Similarly, **this report constructs the DMs of lumped-port-driven transmitting antennas by orthogonalizing PS-DPO**.

In addition, the PS-WET-CMT for lumped-port-driven transmitting antennas can be further generalized to wireless power transfer systems[27-App.G],[40], and a typical two-coil WPT system is shown in Fig. 1-8.

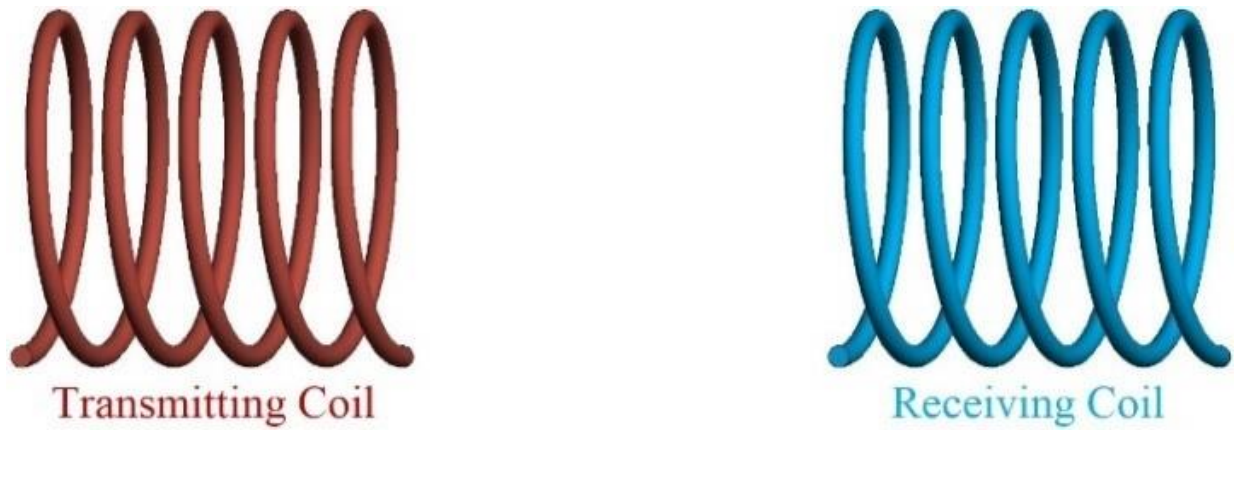

Figure 1-8 Two-coil wireless power transfer system.

### 1.4.4 ES-WET-Based Modal Analysis for Incident-Field-Driven EM Structures

When a scattering structure is placed in non-free-space environment (such as the one shown in Fig. 1-9), the classical scatterer-oriented CMs constructed under ES-WET framework only depend on the inherent characters of the objective scatterer, but depend on neither the external environment nor the external driver. Thus, the ES-WET-based inherent CMs fail to capture the informations of scatterer-environment and scatterer-driver interactions.

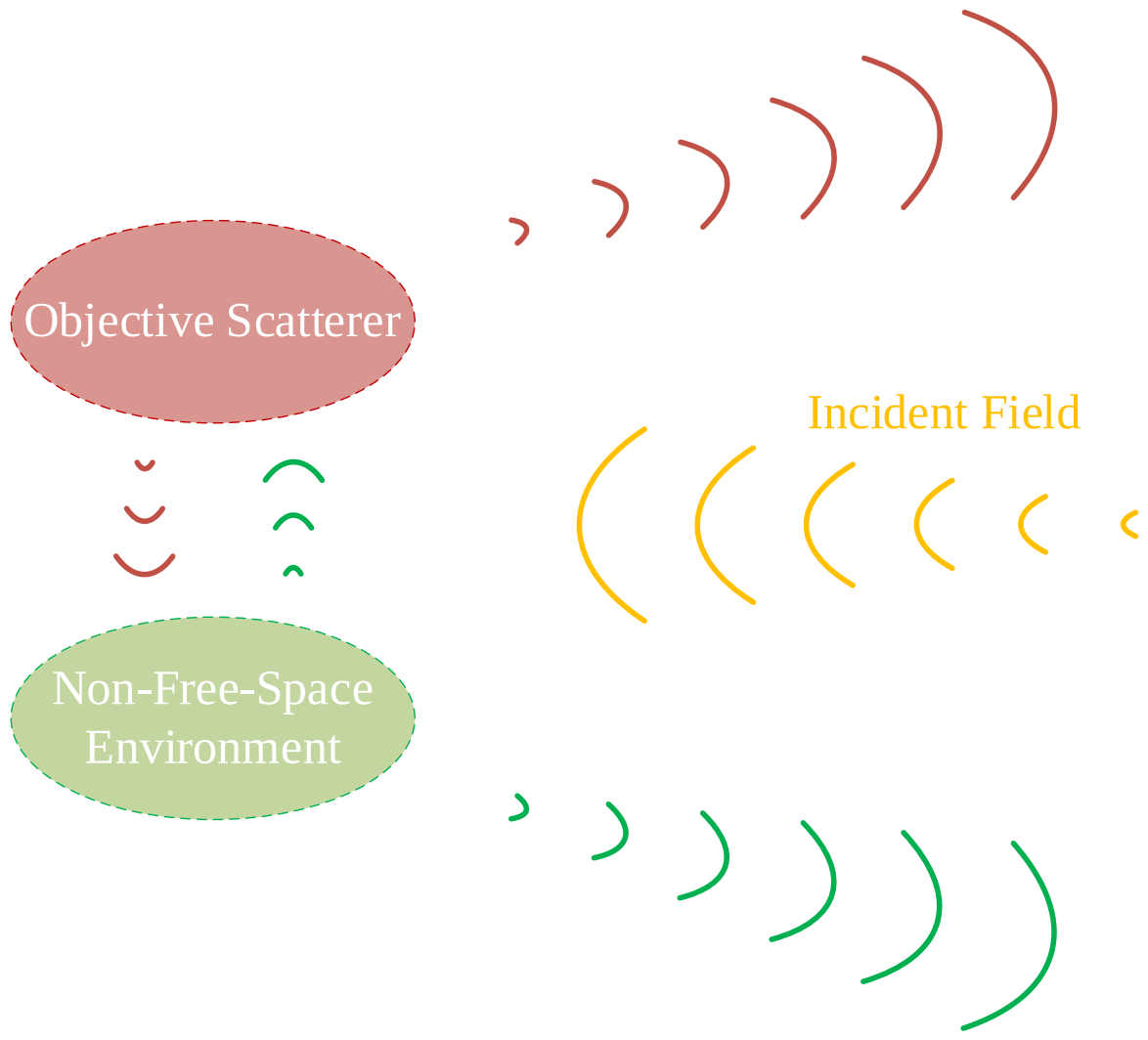

Figure 1-9 An incident-field-driven scattering structure placed in non-free-space environment.

In the last two paragraphs of Ref. [27-Sec.1.2.4.4], a scheme used to incorporate the information of scatterer-environment interaction into CMs was proposed, but Sec. 5.3.1 proves that the CMs calculated from the scheme may not be energy-decoupled. To obtain the energy-decoupled CMs with scatterer-environment interaction information, an alternative ES-WET-based scheme is proposed in the Sec. 5.3.2 of this report for calculating the energy-decoupled environment-dependent CMs of scatterer.

In addition, besides the information of scatterer-environment interaction, the information of scatterer-driver interaction is also valuable sometimes, especially for the near-field scattering applications such as the one shown in Fig. 1-10 (where the scatterer is not placed in the far-field zone of the driver). Based on this observation, the Sec. 5.4 of this report develops an effective scheme used to calculate the CMs which contain the information of scatterer-driver interaction.

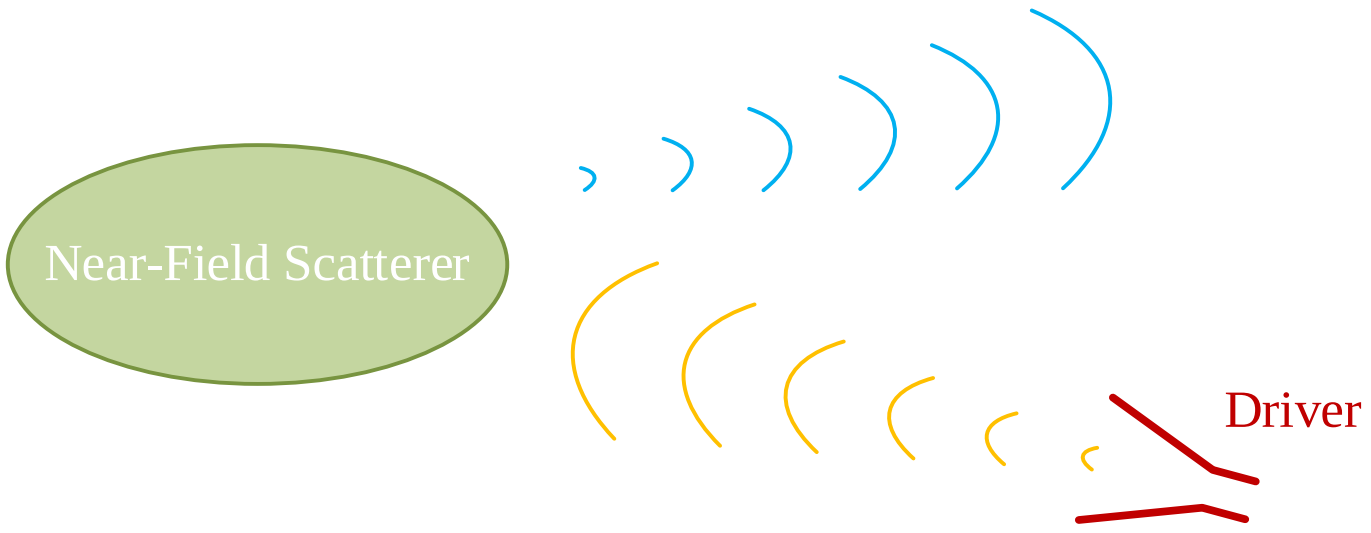

Figure 1-10 A near-field scattering problem. The scatterer is not at the far zone of driver.

### 1.4.5 PtT-Based Modal Analysis for External-Field-Illuminated EM Structures

In Fig. 1-11, the scatterer is a material-coated metallic structure. In stealth and anti-stealth technology[41], we are usually interested in the modes which have relatively small or big radar cross sections (RCSs). Sometimes, we are also interested in the self-oscillating modes (in this case, there is no external field illumination, i.e., external field is 0) of an objective EM structure[42].

In the Secs. 6.2 and 6.3 of this report, some effective schemes will be established for constructing the above-mentioned interested modes of the objective EM structure.

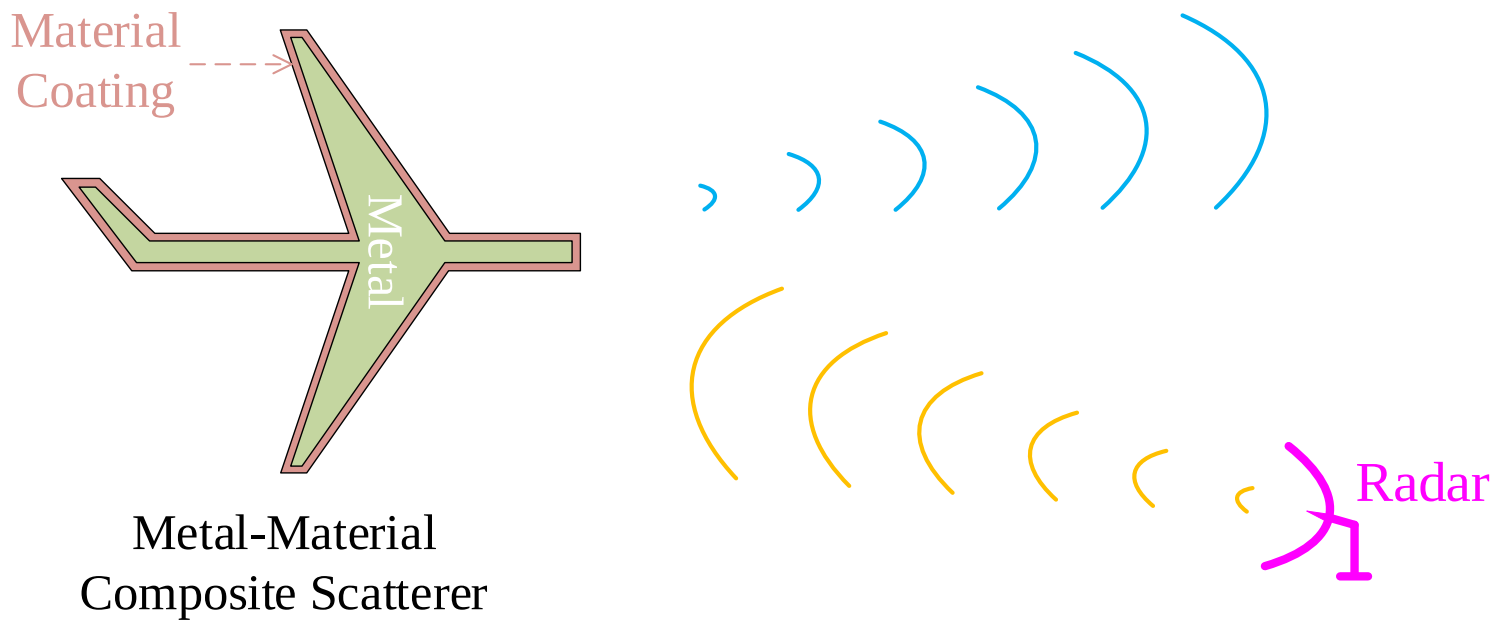

Figure 1-11 A material-coated metallic structure under the illumination of external field.

## 1.5 Comparations Among the Different Modal Analysis Theories

Here, we simply compare the above-mentioned various modal analysis theories from the aspects of objective EM structure, modal excitation source, theoretical framework, modal generating operator, and modal core physical feature, as summarized in Tab. 1-2. The full names of the acronyms used in Tab. 1-2 are listed in the ABBREVIATION LIST after the ABSTRACT of this report.

It is not difficult to find out that: the EMT has a same physical purpose — to construct a set of ENERGY-DECOUPLED MODES — as the ES-WET-CMT, PS-WET-CMT, PTT-DMT, and PtT-DMT. In fact, it had been exhibited in Ref. [27-Chap.3] that the EMT can also be classified into a special case of PTT-DMT, just like having classified IE-CMT into a special case of ES-WET-CMT in Ref. [8].

Table 1-2 Comparisons among various modal analysis theories from the aspects of objective EM structure, modal excitation source, theoretical framework, modal generating operator, and modal core physical feature

| Theory | | EM Structure | | Source | Framework | Operator | Feature | References |
|---|---|---|---|---|---|---|---|---|
| CMT | SM-CMT | lossless | scatterers | externally incident field | SM | PMO | far-field decoupling | [5~7] |
| | | lossy | | | | | | [8-Sec.2.2] |
| | IE-CMT | metallic | | | IE | IMO | not clarified by its founders | [9~11] |
| | | material | | | | | | [12,13,30,34] |
| | | composite | | | | | | [43~48] |
| | ES-WET-CMT | metallic | | | ES-WET | ES-DPO | energy decoupling | [8-Chap.3],[32] (inherent CMs)<br>Sec. 5.2 (inherent CMs)<br>Sec. 5.3 (environment-dependent CMs)<br>Sec. 5.4 (driver-dependent CMs) |
| | | material | | | | | | [8-Chap.4],[14,31,33] (inherent CMs) |
| | | composite | | | | | | [8-Chap.5],[15] (inherent CMs) |
| | PS-WET-CMT | Yagi-Uda antennas | | lumped port / local near field | PS-WET | PS-DPO | | [27-App.H],[37]<br>Secs. 4.2&4.3 |
| | | metallic antennas with passive loads | | | | | | Secs. 4.4&4.5 |
| | | wireless power transfer systems | | | | | | [27-App.G],[40]<br>Sec. 4.6 |
| DMT | PtT-DMT | self-oscillating resonator | | — | PtT | PtFO | | Sec 6.3 |
| | | energy-dissipating material | | wave port | | | | Sec 6.2 |
| | PTT-DMT | metallic | transmitting antennas | | PTT | IPO | | [27-Sec.6.2],[35]<br>Sec. 3.2 |
| | | material | | | | | | [27-Sec.6.3],[35] |
| | | composite | | | | | | [27-Secs.6.4~6.6] |
| | | metallic | receiving antennas | | | | | [27-Secs.7.2&7.3]<br>Sec. 3.3 |
| | | material | | | | | | similar to [27-Sec.7.4] |
| | | composite | | | | | | [27-Sec.7.4] |
| | | metallic | waveguides | | | | | [27-Sec.3.2],[39]<br>Sec. 3.4 |
| | | material | | | | | | [27-Sec.3.3],[39] |
| | | composite | | | | | | [27-Secs.3.4&3.5] |
| | | free-space | | | | | | [27-Sec.3.6]<br>Sec. 3.5 |
| | | waveguide-antenna | cascaded systems | | | | | [27-Sec.8.2]<br>Sec. 3.6.2 |
| | | antenna-antenna | | | | | | [27-Sec.8.3]<br>Sec. 3.6.3 |
| EMT | | waveguides | | | SLT | SLO | | [2,3] |
| | | cavities | | — | | | | |

## 1.6 Research Outline and Roadmap of This Report

In the previous sections, by simply reviewing the research background and significance (Sec. 1.1) and research history and status (Sec. 1.2) related to EM modal analysis, the major problem and challenge (Sec. 1.3) in the realm of EM modal analysis are exposed, and this report's main innovation and contributions (Sec. 1.4) focusing on responding to the problem and challenge are briefly introduced, and the comparations for the modal analysis theories {SM-CMT, IE-CMT, ES-WET-CMT, PS-WET-CMT, PtT-DMT, PTT-DMT, and EMT} involved in this report are provided in Sec. 1.5. In this section, we sketch the research outline and research roadmap (as shown in the following Fig. 1-12) of this report.

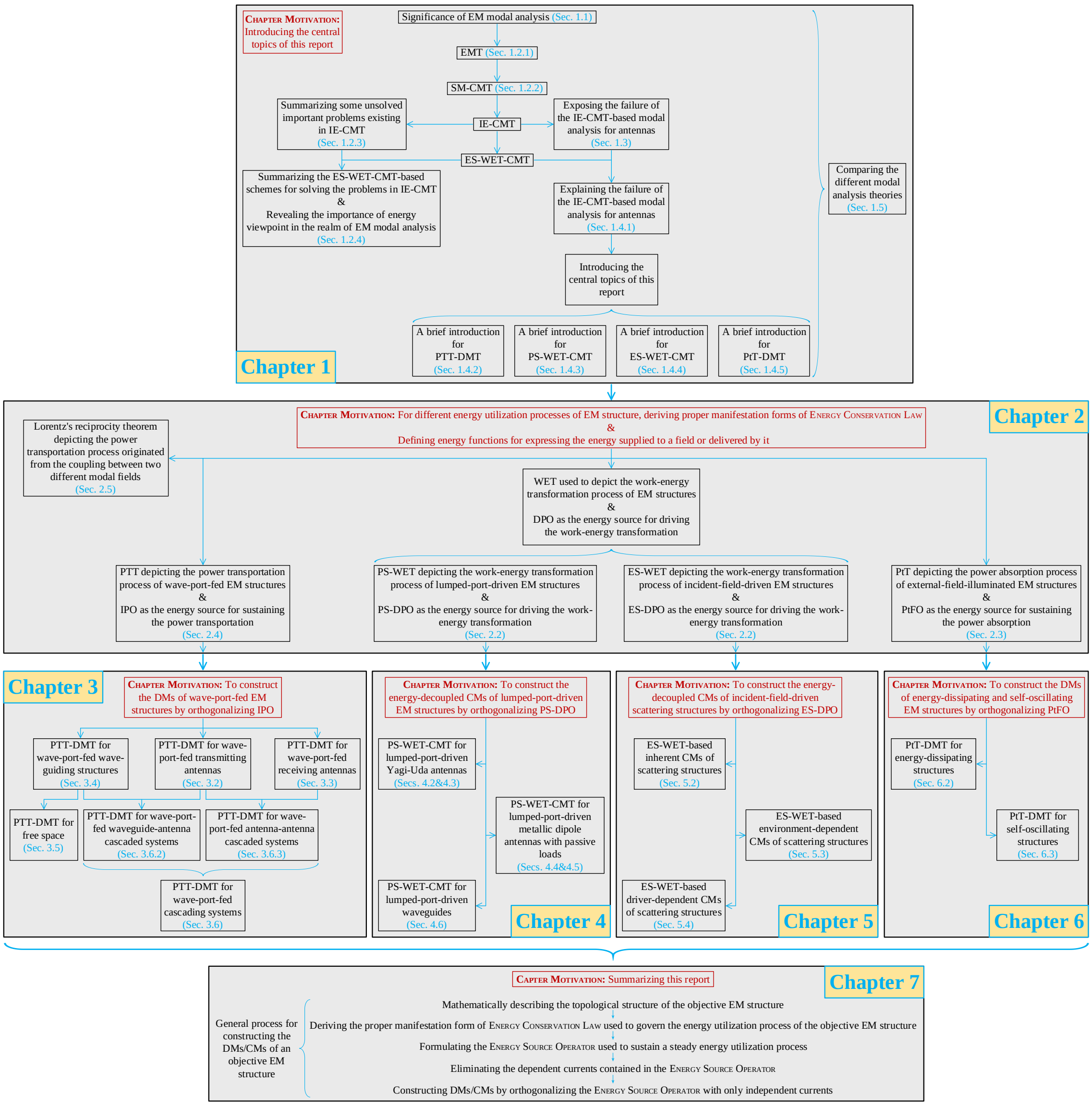

Figure 1-12 Research roadmap of this report.

Based on the research roadmap sketched in Fig. 1-12, the subsequent chapters are organized and dedicated to doing the works summarized as below.

**CHAPTER 2** is dedicated to *introduc*ing *the energy point of view into the study of electromagnetic fields* and *to define energy function for expressing the energy supplied to a field or delivered by it*[49-pp.269].

**CHAPTER 3** is dedicated to establishing an effective modal analysis method — PTT-DMT — for wave-port-fed EM structures, and demonstrating that the PTT-DMT can effectively construct the DMs of wave-port-fed EM structures by orthogonalizing frequency-domain IPO.

**CHAPTER 4** is dedicated to establishing an effective modal analysis method — PS-WET-CMT — for lumped-port-driven EM structures, and demonstrating that the PS-WET-CMT can effectively construct the energy-decoupled CMs of lumped-port-driven EM structures by orthogonalizing frequency-domain PS-DPO.

**CHAPTER 5** is dedicated to (1) generalizing the far-field orthogonality satisfied by the CMs of lossless scatterers to a more general orthogonality relation, (2) comparing the differences between scatterer-oriented modal analysis theory and {antenna, waveguide}-oriented modal analysis theories, and (3) generalizing the conventional scatterer-inherent CMs to environment-dependent CMs and driver-dependent CMs.

**CHAPTER 6** is dedicated to establishing an effective modal analysis method — PtT-DMT — for external-field-illuminated EM structures, and demonstrating that the PtT-DMT can effectively construct the DMs of external-field-illuminated EM structures by orthogonalizing frequency-domain POYNTING'S FLUX OPERATOR (PtFO).

**CHAPTER 7** systematically summarizes the central problems focused on by this report, the fundamental principle established in this report, the main methods used by this report, and the important conclusions and results obtained in this report.

**Appendices** are dedicated to summarizing some important physical quantities, which are frequently used in EM modal analysis and quantitatively depict the modal features in the aspect of utilizing EM energy.

**REFERENCES** list the literatures cited in this report.

# CHAPTER 2 ENERGY CONSERVATION LAW AND ITS DIFFERENT MANIFESTATION FORMS

**CHAPTER MOTIVATION:** *The objective of this chapter is to introduce the energy point of view into the study of electromagnetic fields* and *to define energy functions for expressing the energy supplied to a field or delivered by it*[49-pp.269].

## 2.1 Chapter Introduction

ENERGY CONSERVATION LAW tells us that: energy can neither be created nor destroyed; it can be either transformed from one form into another or transported from one place to another[28,49]. *This chapter introduce*s *the energy point of view into the study of electromagnetic fields*[49-pp.269].

*The* energy *point of view is of great importance to us,* because *many devices are designed to* transform *electric energy into some other form of energy or* transport *electric energy*[49-pp.269]. When an electromagnetic (EM) device works at different working manners/ways, such as scattering, energy-dissipating, transferring, transmitting, receiving, and wave-guiding manners, etc., the ENERGY CONSERVATION LAW governing the energy transformation/transportation process will be manifested in different forms — "ENTIRE-STRUCTURE-ORIENTED WORK-ENERGY THEOREM (ES-WET) form governing the work-energy transformation process of scattering manner"[8,14], "PARTIAL-STRUCTURE-ORIENTED WORK-ENERGY THEOREM (PS-WET) form governing the work-energy transformation process of transmitting and transferring manners"[27-Apps.G&H],[37,40], "POYNTING'S THEOREM (PtT) form governing the energy dissipation process of energy-dissipating manner"[28,49], and "POWER TRANSPORT THEOREM (PTT) form governing the power transportation processes of transmitting, receiving, and wave-guiding manners"[27,35,39].

*Our faith in the validity of* ENERGY CONSERVATION LAW *suggests that it should be possible to define energy functions for expressing the energy supplied to a field or delivered by it*[49-pp.269]. In fact, the above-mentioned energy functions are just the source terms contained in ES-WET, PS-WET, PtT and PTT, and correspond to the energy sources used to sustain the steady work-energy transformation, energy dissipation and power transportation processes, and the ENERGY-DECOUPLED MODES (DMs) of EM structure can be constructed by orthogonalizing the ENERGY SOURCE OPERATORS[8,14,15,27,35,37,39,40].

The central purposes of this chapter are the following three: (1) to provide the mathematical expressions and **physical pictures** of ES-WET, PS-WET, PtT and PTT; (2) to prove the equivalence relation among ES-WET, PS-WET, PtT and PTT; (3) to formulate the energy sources involved in ES-WET, PS-WET, PtT and PTT.

## 2.2 Entire- and Partial-Structure-Oriented Work-Energy Theorems (ES-WET and PS-WET)

This section considers the EM scattering and transmitting problems shown in Figs. 2-1 ~ 2-3 (metallic structure case), 2-4 & 2-5 (material structure case), and 2-6 & 2-7 (composite structure case), and reviews the ES-WET and PS-WET governing the work-energy transformations occurring in the EM scattering and transmitting processes.

### 2.2.1 ES-WET and PS-WET for Metallic Structures

This sub-section focuses on the metallic-structure-related EM problems in Fig. 2-1.

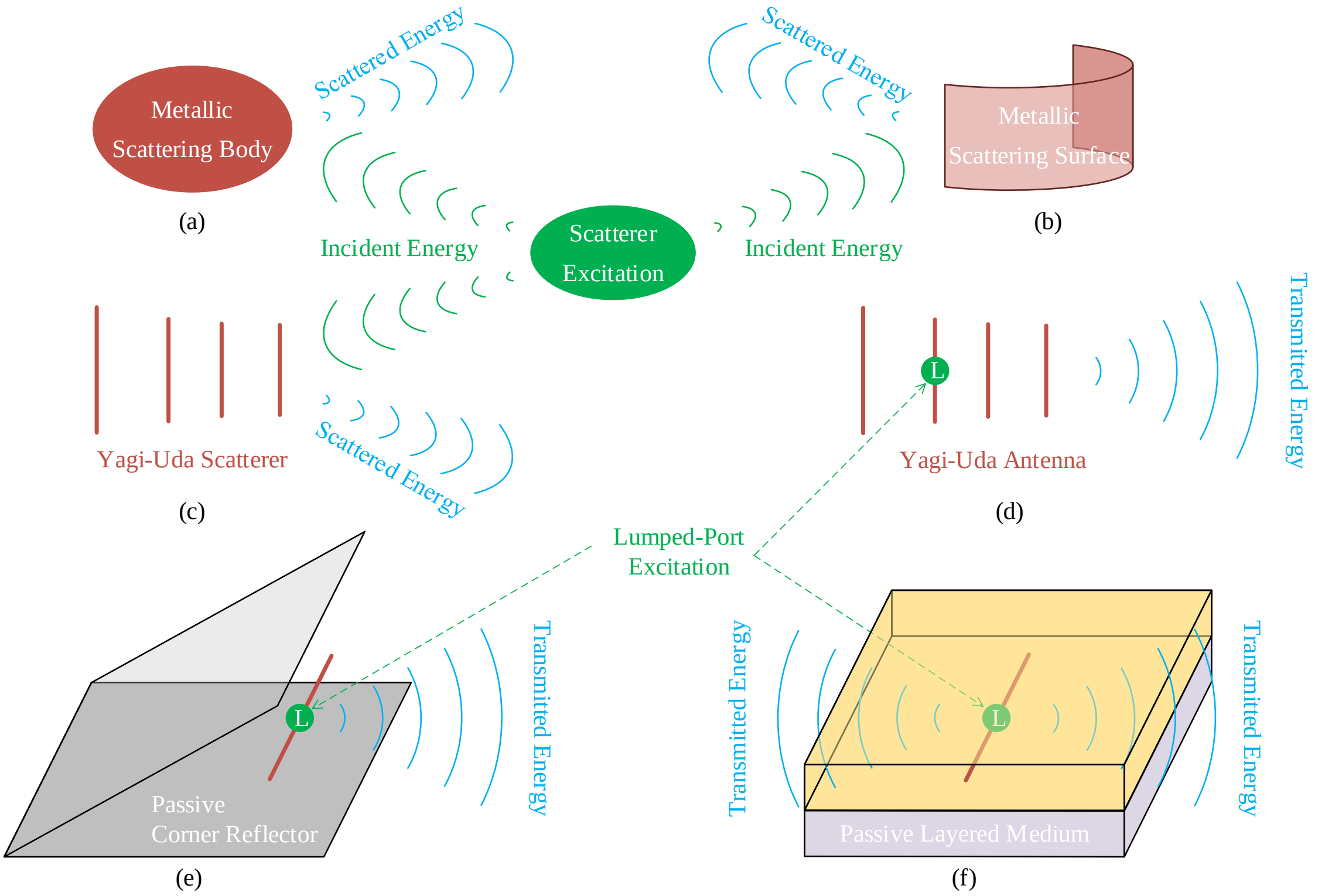

Figure 2-1 External-field-driven single metallic scattering (a) body and (b) surface. (c) External-field-driven metallic Yagi-Uda array scatterer. Lumped-port-driven metallic (d) Yagi-Uda array antenna, (e) dipole antenna loaded by passive corner reflector, and (f) dipole antenna loaded by passive layered medium. In the Figs. 2-1(d~f), the lumped port Ⓛ can be connected to either voltage source or current source .

The EM problems shown in Figs. 2-1(a~c) and Figs. 2-1(d~f) are essentially different from each other, becasue the formers belong to scattering problem but the latters belong to transmitting problem.

As having been frequently mentioned in Refs. [27-Sec.1.3] and [14,35,37], the above-mentioned difference is reflected in the language of the *IEEE standards terms* as follows:

- the formers are *secondary structure*s *generating scattered fields resulted from the scattered currents induced on the structure*s *by some fields incident on the structure*s *from some primary sources*[38-pp.1006];
- the latters are *device*s *that generate high-frequency electric energy, controlled or modulated, which can be emitted from a finite region in the form of unguided waves*[38-pps.369&1210].

In fact, the above-mentioned difference is also reflected in the energy point of view as follows[27-App.H],[37]:

◦ the ENERGY CONSERVATION LAW used to govern the formers' energy utilization processes is manifested in ENTIRE-STRUCTURE-ORIENTED WET (ES-WET) form;

◦ the ENERGY CONSERVATION LAW used to govern the latters' energy utilization process is manifested in PARTIAL-STRUCTURE-ORIENTED WET (PS-WET) form.

Now, we separately discuss the above-mentioned ES-WET and PS-WET forms as below.

**ENTIRE-STRUCTURE-ORIENTED WET (ES-WET)**

For the scattering structure (or simply called scatterer) shown in Fig. 2-1(a), it is a single metallic body which is placed in a non-free-space environment and driven by an externally impressed source. For the convenience of the following discussions, we specially show its topological structure in the following Fig. 2-2.

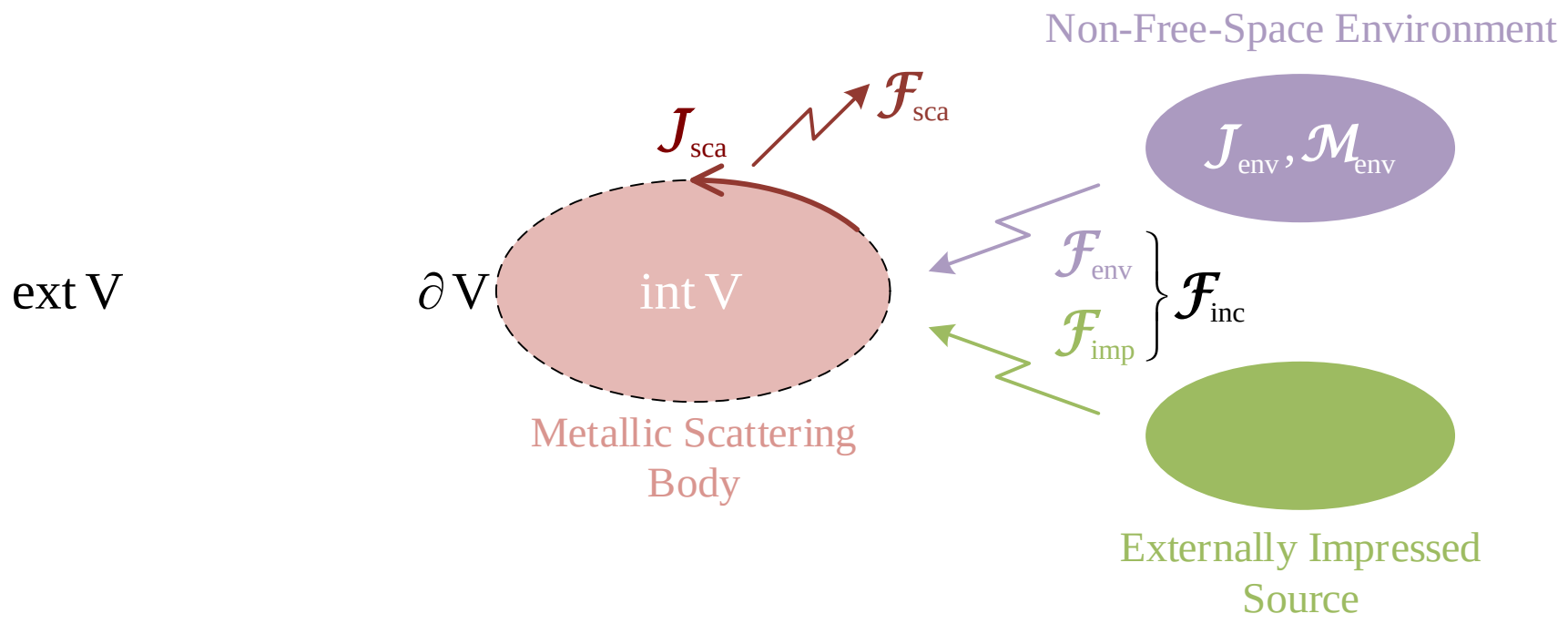

Figure 2-2 Topological structure of the EM scattering problem shown in Fig. 2-1(a).

Here, $\text{int}\,V$, $\partial V$, and $\text{ext}V$ denote the interior, boundary, and exterior of the scatterer.

The field generated by the impressed source is denoted as $\mathcal{F}_{\text{imp}}$. Under the action of $\mathcal{F}_{\text{imp}}$, some currents $\boldsymbol{J}_{\text{sca}}$ and $(\boldsymbol{J}_{\text{env}},\boldsymbol{\mathcal{M}}_{\text{env}})$ will be induced on $\partial \text{V}$ and environment respectively, and then scattered field $\mathcal{F}_{\text{sca}}$ and environmental field $\mathcal{F}_{\text{env}}$ will be generated by $\boldsymbol{J}_{\text{sca}}$ and $(\boldsymbol{J}_{\text{env}},\boldsymbol{\mathcal{M}}_{\text{env}})$ correspondingly. The summation of $\mathcal{F}_{\text{imp}}$ and $\mathcal{F}_{\text{env}}$ is called incident field, and denoted as $\mathcal{F}_{\text{inc}}$, i.e., $\mathcal{F}_{\text{inc}} = \mathcal{F}_{\text{imp}} + \mathcal{F}_{\text{env}}$.

In both $\text{int}\,\text{V}$ and $\text{ext}\text{V}$, which are source-free for $\mathcal{F}_{\text{sca}}$ (because $\boldsymbol{J}_{\text{sca}}$ distributes on $\partial \text{V}$), there exist homogeneous Maxwell's equations $\nabla \times \boldsymbol{\mathcal{H}}_{\text{sca}} = \frac{\partial}{\partial t}\varepsilon_0 \boldsymbol{\mathcal{E}}_{\text{sca}}$ and $\nabla \times \boldsymbol{\mathcal{E}}_{\text{sca}} = -\frac{\partial}{\partial t}\mu_0 \boldsymbol{\mathcal{H}}_{\text{sca}}$, and then exist the following energy relations

$$\oiint_{\partial \text{V}_-} \left(\boldsymbol{\mathcal{E}}_{\text{sca}} \times \boldsymbol{\mathcal{H}}_{\text{sca}}\right) \cdot \boldsymbol{n}_{\partial \text{V}}^- dS = \frac{d}{dt}\left[\frac{1}{2}\left\langle \boldsymbol{\mathcal{H}}_{\text{sca}}, \mu_0 \boldsymbol{\mathcal{H}}_{\text{sca}} \right\rangle_{\text{intV}} + \frac{1}{2}\left\langle \varepsilon_0 \boldsymbol{\mathcal{E}}_{\text{sca}}, \boldsymbol{\mathcal{E}}_{\text{sca}} \right\rangle_{\text{intV}}\right] \quad \text{(2-1a)}$$

$$\begin{aligned}\oiint_{\partial \text{V}_+} \left(\boldsymbol{\mathcal{E}}_{\text{sca}} \times \boldsymbol{\mathcal{H}}_{\text{sca}}\right) \cdot \boldsymbol{n}_{\partial \text{V}}^+ dS = & \oiint_{\text{S}_\infty} \left(\boldsymbol{\mathcal{E}}_{\text{sca}} \times \boldsymbol{\mathcal{H}}_{\text{sca}}\right) \cdot \boldsymbol{n}_{\text{S}_\infty}^+ dS \\ & + \frac{d}{dt}\left[\frac{1}{2}\left\langle \boldsymbol{\mathcal{H}}_{\text{sca}}, \mu_0 \boldsymbol{\mathcal{H}}_{\text{sca}} \right\rangle_{\text{extV}} + \frac{1}{2}\left\langle \varepsilon_0 \boldsymbol{\mathcal{E}}_{\text{sca}}, \boldsymbol{\mathcal{E}}_{\text{sca}} \right\rangle_{\text{extV}}\right]\end{aligned} \quad \text{(2-1b)}$$

Here, $\partial \text{V}_{-/+}$ is the inner/outer surface of $\partial \text{V}$; $\boldsymbol{n}_{\partial \text{V}}^{-/+}$ is the inner/outer normal direction of $\partial \text{V}$; $\text{S}_\infty$, which is a spherical surface with infinite radius, constitutes the boundary of whole three-dimensional Euclidean space $\text{E}_3$; $\boldsymbol{n}_{\text{S}_\infty}^+$ is the outer normal direction of $\text{S}_\infty$. Based on the magnetic field boundary condition $\boldsymbol{n}_{\partial \text{V}}^- \times (\boldsymbol{\mathcal{H}}_{\text{sca}}^- - \boldsymbol{\mathcal{H}}_{\text{sca}}^+) = \boldsymbol{J}_{\text{sca}}$ on $\partial \text{V}$ (where $\boldsymbol{\mathcal{H}}_{\text{sca}}^{-/+}$ distributes on $\partial \text{V}_{-/+}$), the summation of Eqs. (2-1a) and (2-1b) gives the following PtT[28,49]

$$\begin{aligned}-\left\langle \boldsymbol{J}_{\text{sca}}, \boldsymbol{\mathcal{E}}_{\text{sca}} \right\rangle_{\partial \text{V}} = & \oiint_{\text{S}_\infty} \left(\boldsymbol{\mathcal{E}}_{\text{sca}} \times \boldsymbol{\mathcal{H}}_{\text{sca}}\right) \cdot \boldsymbol{n}_{\text{S}_\infty}^+ dS \\ & + \frac{d}{dt}\left[\frac{1}{2}\left\langle \boldsymbol{\mathcal{H}}_{\text{sca}}, \mu_0 \boldsymbol{\mathcal{H}}_{\text{sca}} \right\rangle_{\text{E}_3} + \frac{1}{2}\left\langle \varepsilon_0 \boldsymbol{\mathcal{E}}_{\text{sca}}, \boldsymbol{\mathcal{E}}_{\text{sca}} \right\rangle_{\text{E}_3}\right]\end{aligned} \quad \text{(2-2)}$$

where the relations $(1/2) < \boldsymbol{\mathcal{H}}_{\text{sca}}, \mu_0 \boldsymbol{\mathcal{H}}_{\text{sca}} >_{\partial \text{V}} + (1/2) < \varepsilon_0 \boldsymbol{\mathcal{E}}_{\text{sca}}, \boldsymbol{\mathcal{E}}_{\text{sca}} >_{\partial \text{V}} = 0$ and $\text{E}_3 = \text{int}\,\text{V} \bigcup \partial \text{V} \bigcup \text{ext}\text{V}$ have been utilized.

Substituting the tangential electric field boundary condition $[\boldsymbol{\mathcal{E}}_{\text{inc}} + \boldsymbol{\mathcal{E}}_{\text{sca}}]^{\tan} = 0$ on $\partial \text{V}$ into PtT (2-2), it is immediate to derive the following relation[8-Secs.2.4.1&3.2.1],[32]

$$\begin{aligned}\left\langle \boldsymbol{J}_{\text{sca}}, \boldsymbol{\mathcal{E}}_{\text{inc}} \right\rangle_{\partial \text{V}} = & \oiint_{\text{S}_\infty} \left(\boldsymbol{\mathcal{E}}_{\text{sca}} \times \boldsymbol{\mathcal{H}}_{\text{sca}}\right) \cdot \boldsymbol{n}_{\text{S}_\infty}^+ dS \\ & + \frac{d}{dt}\left[\frac{1}{2}\left\langle \boldsymbol{\mathcal{H}}_{\text{sca}}, \mu_0 \boldsymbol{\mathcal{H}}_{\text{sca}} \right\rangle_{\text{E}_3} + \frac{1}{2}\left\langle \varepsilon_0 \boldsymbol{\mathcal{E}}_{\text{sca}}, \boldsymbol{\mathcal{E}}_{\text{sca}} \right\rangle_{\text{E}_3}\right]\end{aligned} \quad \text{(2-3)}$$

which has a very clear physical picture: **the power done by incident field on scattered current is transformed into two parts — part I (the first term in the right-hand side) is carried by scattered field from scatterer to infinity, and part II (the second term** **in the right-hand side) is used to contribute the energy stored in magnetic and electric fields**. Relation (2-3) is a quantitative description for the work-energy transformation process shown in Fig. 2-1(a), and the $\mathcal{F}_{\rm inc}$ used to deliver energy to $\mathcal{J}_{\rm sca}$ directly acts on entire scattering structure, so it is called ENTIRE-STRUCTURE-ORIENTED WORK-ENERGY THEOREM (ES-WET, which is the time-differential version with power dimension. For the version with energy dimension, please see Ref. [8-Secs.2.4.1]).

In fact, it is also easy to derive PtT (2-2) from ES-WET (2-3) by utilizing the relation $[\mathcal{E}_{\rm inc} + \mathcal{E}_{\rm sca}]^{\rm tan} = 0$ on $\partial V$. Thus, we conclude here that: $\boxed{\text{PtT} \Leftrightarrow \text{ES-WET}}$. In addition, <u>the left-hand side of ES-WET (2-3) is just the energy source used to drive a steady work-energy transformation</u>, so it is particularly called entire-structure-oriented driving power, and the corresponding operator expression is called ENTIRE-STRUCTURE-ORIENTED DRIVING POWER OPERATOR (ES-DPO). As exhibited in Refs. [8-Chap.3] and [32], the frequency-domain ES-DPO is an effective modal generating operator for calculating the energy-decoupled CHARACTERISTIC MODES (CMs) of metallic scattering structures.

In addition, the above-obtained formulations and conclusions are also valid for the EM scattering problems shown in Fig. 2-1(b), whose scattering structure is an open metallic surface, and Fig. 2-1(c), whose scattering structure is constituted by multiple discrete metallic elements.

### PARTIAL-STRUCTURE-ORIENTED WET (PS-WET)

For the Yagi-Uda array antenna shown in Fig. 2-1(d), its all elements can be classified into two categories[50] — active element (with boundary surface $S_{\rm act}$) and passive elements (with boundary surface $S_{\rm pas}$) — as shown in the following Fig. 2-3.

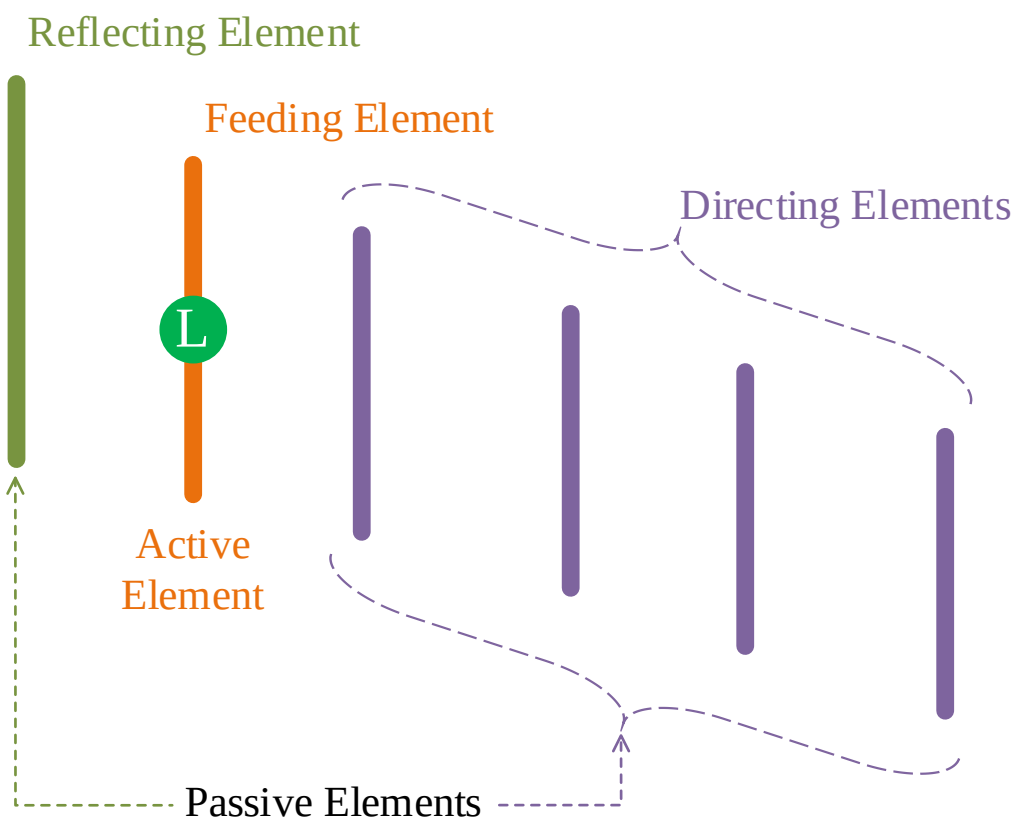

Figure 2-3 Topological structure of the EM transmitting problem shown in Fig. 2-1(d).

The working of the antenna originates from the driving supplied by the lumped port Ⓛ

(which can be connected to either voltage source or current source ) on the active element. The lumped-port driving can be equivalently treated as a field driving, and the equivalent driving field $\mathcal{F}_{\rm driv}$ acts on the active element only[27-App.H],[37].

Field $\mathcal{F}_{\rm driv}$ will induce a current on $\mathrm{S}_{\rm act}$, and then a field will be generated by $\mathrm{S}_{\rm act}$. The $\mathrm{S}_{\rm act}$-generated field will also induce a current on $\mathrm{S}_{\rm pas}$, and then a field will be generated by $\mathrm{S}_{\rm pas}$. In fact, the $\mathrm{S}_{\rm pas}$-generated field will react on $\mathrm{S}_{\rm act}$, and then affect the current distribution on $\mathrm{S}_{\rm act}$. Through a complicated process, the above action and reaction will reach a dynamical equilibrium, and the $\mathrm{S}_{\rm act}$-based and $\mathrm{S}_{\rm pas}$-based induced currents working at the equilibrium state are denoted as $\mathcal{J}_{\rm act}$ and $\mathcal{J}_{\rm pas}$ respectively. The fields generated by $\mathcal{J}_{\rm act}$ and $\mathcal{J}_{\rm pas}$ are denoted as $\mathcal{F}_{\rm act}$ and $\mathcal{F}_{\rm pas}$ correspondingly, and the summation of $\mathcal{F}_{\rm act}$ and $\mathcal{F}_{\rm pas}$ is just the field generated by whole antenna, and it is denoted as $\mathcal{F}_{\rm ant}$, i.e., $\mathcal{F}_{\rm ant} = \mathcal{F}_{\rm act} + \mathcal{F}_{\rm pas}$.

From the Maxwell's equations satisfied by current $\mathcal{J}_{\rm act} + \mathcal{J}_{\rm pas}$ and field $\mathcal{F}_{\rm ant}$, we can derive the following PtT

$$
\begin{aligned}
-\langle \mathcal{J}_{\rm act}, \mathcal{E}_{\rm ant} \rangle_{\mathrm{S}_{\rm act}} &= -< \mathcal{J}_{\rm act} + \mathcal{J}_{\rm pas}, \overbrace{\mathcal{E}_{\rm act} + \mathcal{E}_{\rm pas}}^{\mathcal{E}_{\rm ant}} >_{\mathrm{S}_{\rm act} \cup \mathrm{S}_{\rm pas}} \\
&= \oiint_{\mathrm{S}_\infty} \left( \mathcal{E}_{\rm ant} \times \mathcal{H}_{\rm ant} \right) \cdot \boldsymbol{n}^{+}_{\mathrm{S}_\infty} dS \\
&\quad + \frac{d}{dt} \left[ \frac{1}{2} \langle \mathcal{H}_{\rm ant}, \mu_0 \mathcal{H}_{\rm ant} \rangle_{\mathrm{E}_3} + \frac{1}{2} \langle \varepsilon_0 \mathcal{E}_{\rm ant}, \mathcal{E}_{\rm ant} \rangle_{\mathrm{E}_3} \right]
\end{aligned}
\tag{2-4}
$$

where the first equality is due to that the tangential $\mathcal{E}_{\rm ant}$ is zero on $\mathrm{S}_{\rm pas}$, and the derivation for the second equality is similar to driving PtT (2-2). Substituting the homogeneous tangential electric field boundary condition $[\mathcal{E}_{\rm driv} + \mathcal{E}_{\rm ant}]^{\rm tan} = 0$ on $\mathrm{S}_{\rm act}$ into the above PtT (2-4), it is immediate to derive the following relation[27-App.H],[37]

$$
\begin{aligned}
\langle \mathcal{J}_{\rm act}, \mathcal{E}_{\rm driv} \rangle_{\mathrm{S}_{\rm act}} &= \oiint_{\mathrm{S}_\infty} \left( \mathcal{E}_{\rm ant} \times \mathcal{H}_{\rm ant} \right) \cdot \boldsymbol{n}^{+}_{\mathrm{S}_\infty} dS \\
&\quad + \frac{d}{dt} \left[ \frac{1}{2} \langle \mathcal{H}_{\rm ant}, \mu_0 \mathcal{H}_{\rm ant} \rangle_{\mathrm{E}_3} + \frac{1}{2} \langle \varepsilon_0 \mathcal{E}_{\rm ant}, \mathcal{E}_{\rm ant} \rangle_{\mathrm{E}_3} \right]
\end{aligned}
\tag{2-5}
$$

which has a very similar physical interpretation to the previous ES-WET (2-3).

The main difference between the above relation (2-5) and the previous ES-WET (2-3) is that: the $\mathcal{F}_{\rm driv}$ used to deliver energy to $\mathcal{J}_{\rm act}$ only acts on "a partial structure" rather than "the entire structure like the $\mathcal{F}_{\rm inc}$ in ES-WET (2-3)". Thus, relation (2-5) is particularly called PARTIAL-STRUCTURE-ORIENTED WORK-ENERGY THEOREM (ES-WET), and the ENERGY SOURCE OPERATOR in its left-hand side is correspondingly called PARTIAL-STRUCTURE-ORIENTED DRIVING POWER OPERATOR (PS-DPO). As exhibited in Refs. [27-App.H] and [37], the frequency-domain PS-DPO is an effective operator for calculating the energy-decoupled CMs of lumped-port-driven metallic transmitting antennas.

Obviously, we also have conclusion that $\boxed{\text{PtT} \Leftrightarrow \text{PS-WET}}$ just like the $\boxed{\text{PtT} \Leftrightarrow \text{ES-WET}}$ obtained previously.

In addition, the above-obtained formulations and conclusions are also valid for the EM transmitting problems shown in Fig. 2-1(e), which is a lumped-port driven dipole antenna loaded by a passive corner reflector, and Fig. 2-1(f), which is a lumped-port driven dipole antenna loaded by a passive layered medium.

## 2.2.2 ES-WET and PS-WET for Material Structures

This sub-section will generalize the metallic-structure-oriented ES-WET and PS-WET obtained in the above Sec. 2.2.1 to material structures.

**ENTIRE-STRUCTURE-ORIENTED WET (ES-WET)**

Now, we focus on discussing the ES-WET-governed material scattering problem shown in the following Fig. 2-4.

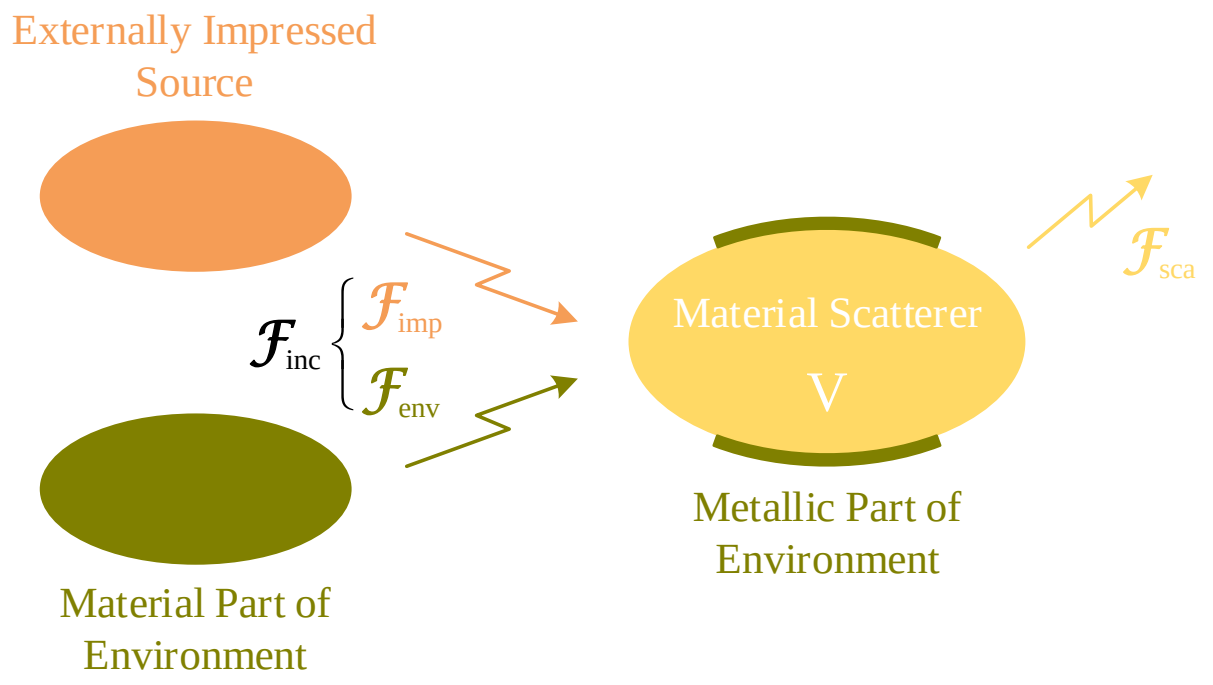

Figure 2-4 External-field-driven material scatterer placed in non-free-space environment.

In Fig. 2-4, the scatterer is a material body which is placed in a non-free-space environment and driven by an externally impressed source. The non-free-space environment is constituted by three parts “metallic part, material part, and vacuum part”, and the impressed source is placed in the vacuum part of environment. The region occupied by the material scatterer is denoted as $V$, and it is with material parameters $\mu$, $\varepsilon$, and $\sigma$, which are real, symmetrical, and time-independent.

The field generated by the impressed source is denoted as $\mathcal{F}_{\text{imp}}$. Under the action of $\mathcal{F}_{\text{imp}}$, some currents $\mathcal{C}_{\text{sca}}$ and $\mathcal{C}_{\text{env}}$ will be induced on the scatterer and environment respectively, and then scattered field $\mathcal{F}_{\text{sca}}$ and environmental field $\mathcal{F}_{\text{env}}$ will be generated by $\mathcal{C}_{\text{sca}}$ and $\mathcal{C}_{\text{env}}$ correspondingly. The summation of $\mathcal{F}_{\text{imp}}$ and $\mathcal{F}_{\text{env}}$ is called incident field, and denoted as $\mathcal{F}_{\rm inc}$, i.e., $\mathcal{F}_{\rm inc} = \mathcal{F}_{\rm imp} + \mathcal{F}_{\rm env}$. The summation of $\mathcal{F}_{\rm inc}$ and $\mathcal{F}_{\rm sca}$ is called total field, and denoted as $\mathcal{F}$, i.e., $\mathcal{F} = \mathcal{F}_{\rm inc} + \mathcal{F}_{\rm sca}$.

The currents $(\mathcal{J}_{\rm sca}, \mathcal{M}_{\rm sca})$ and fields $(\mathcal{E}, \mathcal{H})$ satisfy relations $\mathcal{J}_{\rm sca} = \boldsymbol{\sigma} \cdot \mathcal{E} + \frac{\partial}{\partial t} \Delta\boldsymbol{\varepsilon} \cdot \mathcal{E}$ and $\mathcal{M}_{\rm sca} = \frac{\partial}{\partial t} \Delta\boldsymbol{\mu} \cdot \mathcal{H}$ [8-App.3], where $\Delta\boldsymbol{\mu} = \boldsymbol{\mu} - \mathbf{I}\mu_0$ and $\Delta\boldsymbol{\varepsilon} = \boldsymbol{\varepsilon} - \mathbf{I}\varepsilon_0$. The currents $(\mathcal{J}_{\rm sca}, \mathcal{M}_{\rm sca})$ and fields $(\mathcal{E}_{\rm sca}, \mathcal{H}_{\rm sca})$ satisfy Maxwell's equations $\nabla \times \mathcal{H}_{\rm sca} = \mathcal{J}_{\rm sca} + \frac{\partial}{\partial t} \varepsilon_0 \mathcal{E}_{\rm sca}$ and $\nabla \times \mathcal{E}_{\rm sca} = -\mathcal{M}_{\rm sca} - \frac{\partial}{\partial t} \mu_0 \mathcal{H}_{\rm sca}$ [28], and then satisfy the following relation

$$-\langle \mathcal{J}_{\rm sca}, \mathcal{E}_{\rm sca} \rangle_{\rm V} - \langle \mathcal{M}_{\rm sca}, \mathcal{H}_{\rm sca} \rangle_{\rm V} = \oint\!\!\!\oint_{\rm S_\infty} (\mathcal{E}_{\rm sca} \times \mathcal{H}_{\rm sca}) \cdot \boldsymbol{n}^+_{\rm S_\infty} dS + \frac{d}{dt}\left[\frac{1}{2}\langle \mathcal{H}_{\rm sca}, \mu_0 \mathcal{H}_{\rm sca} \rangle_{\rm E_3} + \frac{1}{2}\langle \varepsilon_0 \mathcal{E}_{\rm sca}, \mathcal{E}_{\rm sca} \rangle_{\rm E_3}\right] \tag{2-6}$$

Employing the above-mentioned relations and relation $\mathcal{F}_{\rm inc} = \mathcal{F} - \mathcal{F}_{\rm sca}$, it is immediate to derive the following relation[8-Sec.4.2.1],[30]

$$\begin{aligned}\langle \mathcal{J}_{\rm sca}, \mathcal{E}_{\rm inc} \rangle_{\rm V} + \langle \mathcal{M}_{\rm sca}, \mathcal{H}_{\rm inc} \rangle_{\rm V} = & \oint\!\!\!\oint_{\rm S_\infty} (\mathcal{E}_{\rm sca} \times \mathcal{H}_{\rm sca}) \cdot \boldsymbol{n}^+_{\rm S_\infty} dS \\ & + \frac{d}{dt}\left[\frac{1}{2}\langle \mathcal{H}_{\rm sca}, \mu_0 \mathcal{H}_{\rm sca} \rangle_{\rm E_3} + \frac{1}{2}\langle \varepsilon_0 \mathcal{E}_{\rm sca}, \mathcal{E}_{\rm sca} \rangle_{\rm E_3}\right] \\ & + \langle \boldsymbol{\sigma} \cdot \mathcal{E}, \mathcal{E} \rangle_{\rm V} \\ & + \frac{d}{dt}\left[\frac{1}{2}\langle \mathcal{H}, \Delta\boldsymbol{\mu} \cdot \mathcal{H} \rangle_{\rm V} + \frac{1}{2}\langle \Delta\boldsymbol{\varepsilon} \cdot \mathcal{E}, \mathcal{E} \rangle_{\rm V}\right]\end{aligned} \tag{2-7}$$

(Annotation: the first two terms on the right-hand side equal $-\langle \mathcal{J}_{\rm sca}, \mathcal{E}_{\rm sca} \rangle_{\rm V} - \langle \mathcal{M}_{\rm sca}, \mathcal{H}_{\rm sca} \rangle_{\rm V}$)

The above relation (2-7) has a very similar (but not identical) physical picture to the previous ES-WET (2-3), and the physical picture is that: **the power done by incident field on scattered current is transformed into three parts — part I (the first and second terms in the right-hand side) is supplied to scattered field by scattered current (mathematically as $-<\mathcal{J}_{\rm sca}, \mathcal{E}_{\rm sca}>_{\rm V} - <\mathcal{M}_{\rm sca}, \mathcal{H}_{\rm sca}>_{\rm V}$), and part II (the third term in the right-hand side) is converted into Joule heat, and part III (the fourth term in the right-hand side) is used to contribute the magnetization and polarization energy**. Obviously, the part I "$-<\mathcal{J}_{\rm sca}, \mathcal{E}_{\rm sca}>_{\rm V} - <\mathcal{M}_{\rm sca}, \mathcal{H}_{\rm sca}>_{\rm V}$" can be further decomposed into two sub-parts — one is carried by scattered field from scatterer to infinity and the other is used to contribute the magnetic and electric field energy, and the part I "$-<\mathcal{J}_{\rm sca}, \mathcal{E}_{\rm sca}>_{\rm V} - <\mathcal{M}_{\rm sca}, \mathcal{H}_{\rm sca}>_{\rm V}$" will be further discussed in Sec. 2.3.2.

Thus, the above relation (2-7) is similarly called ENTIRE-STRUCTURE-ORIENTED WORK-ENERGY THEOREM (ES-WET, which is the time-differential version with power dimension. For the version with energy dimension, please see Refs. [8-Sec.2.4.2] and [14]). In addition, the left-hand side of ES-WET (2-7), which is just the energy source used to drive a steady work-energy transformation, is called ENTIRE-STRUCTURE-ORIENTED DRIVING POWER OPERATOR (ES-DPO). As exhibited in Refs. [8,14,33], the frequency-domain ES-DPO is an effective modal generating operator for calculating the energy-decoupled CMs of material scattering structures.

**PARTIAL-STRUCTURE-ORIENTED WET (PS-WET)**

The PS-WET-governed material transmitting problem shown in the following Fig. 2-5 (which is a local-near-field-driven material Yagi-Uda array antenna) can be similarly discussed as exhibited in Refs. [27-App.H5] and [37-Sec.III].

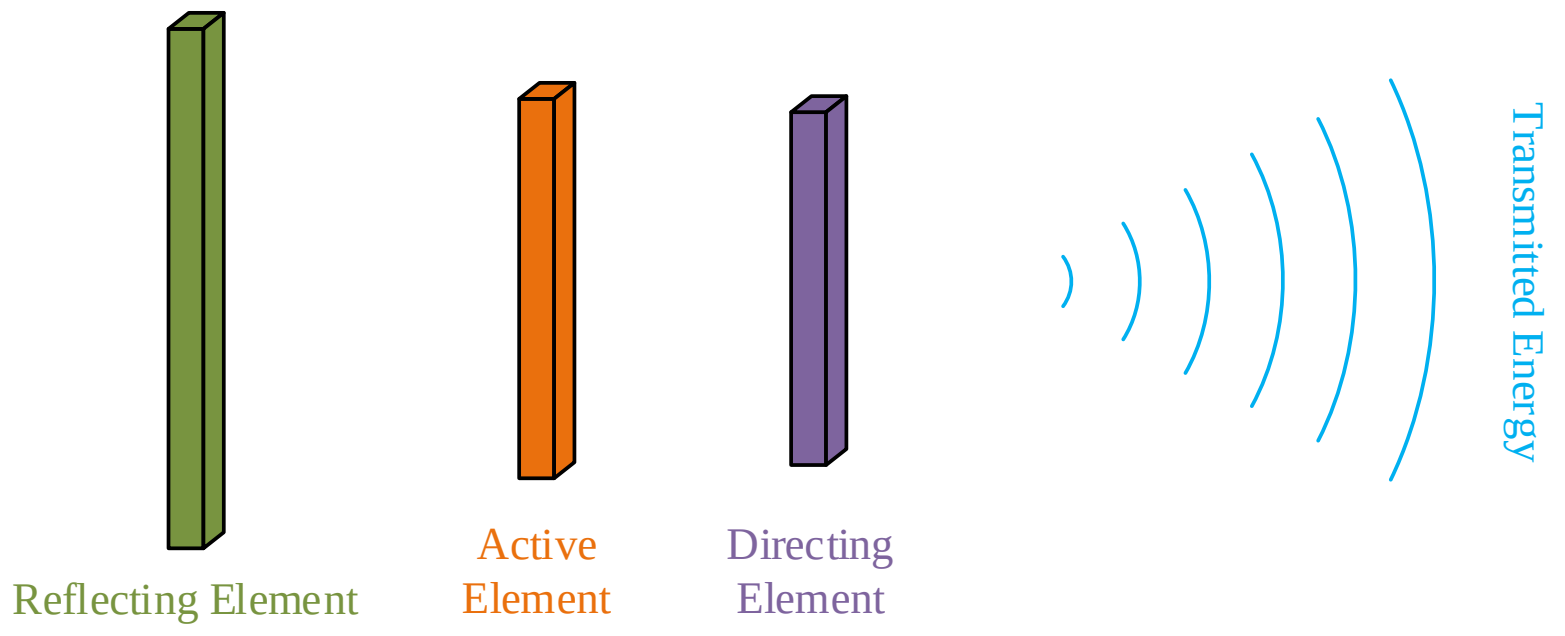

Figure 2-5 Local-near-field-driven material Yagi-Uda array antenna. Here, the local near-field driving directly acts on the active element.

Here, we only provide the material-antenna-oriented PS-WET as follows:

$$\begin{aligned}\left\langle \boldsymbol{J}_{\mathrm{act}},\boldsymbol{\mathcal{E}}_{\mathrm{driv}}\right\rangle_{\mathrm{V}_{\mathrm{act}}}+\left\langle \boldsymbol{\mathcal{M}}_{\mathrm{act}},\boldsymbol{\mathcal{H}}_{\mathrm{driv}}\right\rangle_{\mathrm{V}_{\mathrm{act}}} = & \oiint_{\mathrm{S}_{\infty}}\left(\boldsymbol{\mathcal{E}}_{\mathrm{ant}}\times\boldsymbol{\mathcal{H}}_{\mathrm{ant}}\right)\cdot\boldsymbol{n}_{\mathrm{S}_{\infty}}^{+}\,dS+\left\langle \boldsymbol{\sigma}\cdot\boldsymbol{\mathcal{E}},\boldsymbol{\mathcal{E}}\right\rangle_{\mathrm{V}_{\mathrm{act}}\cup\mathrm{V}_{\mathrm{pas}}}\\ & +\frac{d}{dt}\left[\frac{1}{2}\left\langle \boldsymbol{\mathcal{H}}_{\mathrm{ant}},\mu_{0}\boldsymbol{\mathcal{H}}_{\mathrm{ant}}\right\rangle_{\mathrm{E}_{3}}+\frac{1}{2}\left\langle \varepsilon_{0}\boldsymbol{\mathcal{E}}_{\mathrm{ant}},\boldsymbol{\mathcal{E}}_{\mathrm{ant}}\right\rangle_{\mathrm{E}_{3}}\right]\\ & +\frac{d}{dt}\left[\frac{1}{2}\left\langle \boldsymbol{\mathcal{H}},\Delta\boldsymbol{\mu}\cdot\boldsymbol{\mathcal{H}}\right\rangle_{\mathrm{V}_{\mathrm{act}}\cup\mathrm{V}_{\mathrm{pas}}}+\frac{1}{2}\left\langle \Delta\boldsymbol{\varepsilon}\cdot\boldsymbol{\mathcal{E}},\boldsymbol{\mathcal{E}}\right\rangle_{\mathrm{V}_{\mathrm{act}}\cup\mathrm{V}_{\mathrm{pas}}}\right]\end{aligned} \tag{2-8}$$

In PS-WET (2-8), $\mathcal{F}_{\mathrm{driv}}$ is the near-field driving, which only acts on the active element; $\mathcal{C}_{\mathrm{act}}$ is the induced current on the active element; $\mathcal{F}_{\mathrm{ant}}$ is the field generated by whole material Yagi-Uda antenna; $\mathcal{F}$ is the summation of $\mathcal{F}_{\mathrm{driv}}$ and $\mathcal{F}_{\mathrm{ant}}$; $\mathrm{V}_{\mathrm{act}}$ is the region occupated by the active element, and $\mathrm{V}_{\mathrm{pas}}$ is the region occupated by the reflecting and directing elements. The left-hand side term is the PS-DPO used to sustain a steady work-energy transformation of the local-near-field-driven material Yagi-Uda antenna, and it is an effective operator for calculating the energy-decoupled CMs[27-App.H5],[37-Sec.III].

## 2.2.3 ES-WET and PS-WET for Composite Structures

The above metallic-structure-oriented and material-structure-oriented ES-WET and PS-

WET can be further generalized to metal-material composite structures.

**ENTIRE-STRUCTURE-ORIENTED WET (ES-WET)**

For the metal-material composite scatterer shown in Fig. 2-6, it is constituted by metallic {line, surface, body} and material body. The ES-WET used to govern the energy utilization process occuring in the scattering working manner of the composite scatterer had been carefully discussed in Refs. [8-Chap.5] and [15].

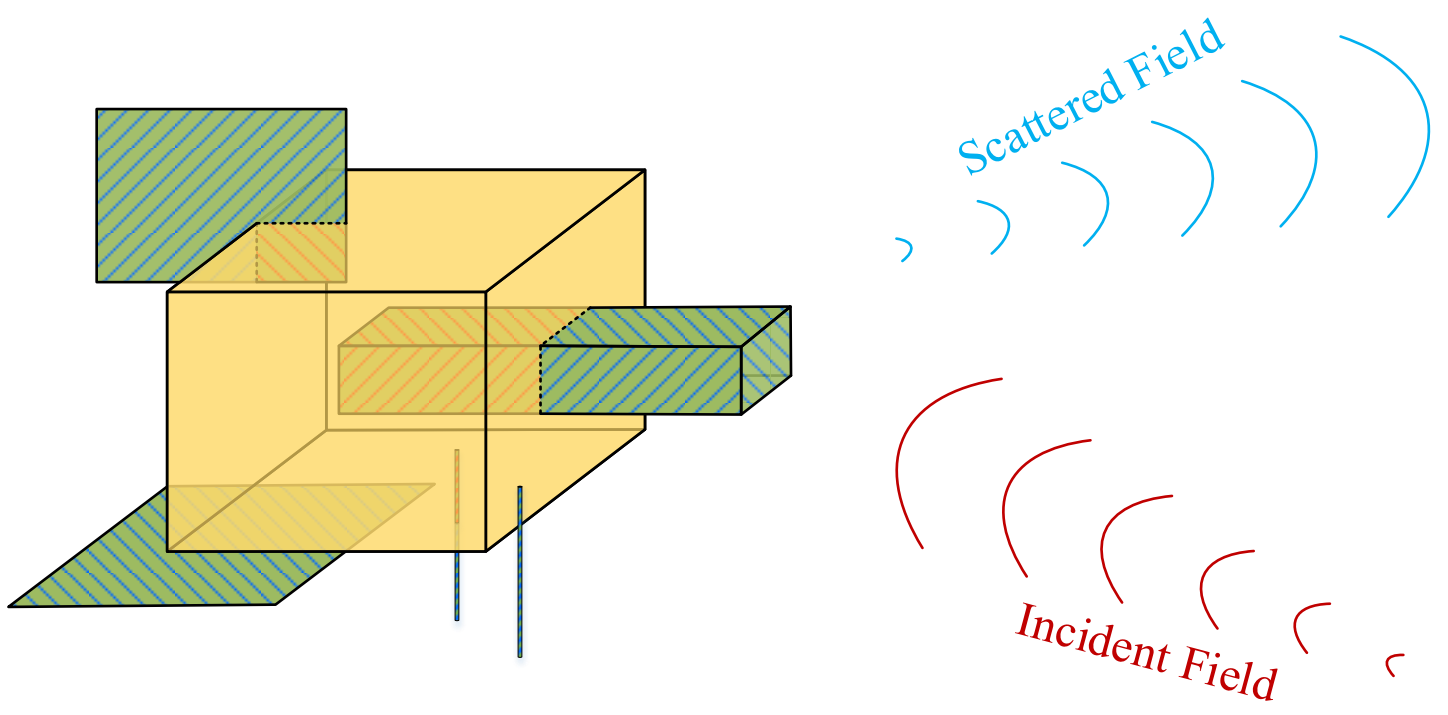

Figure 2-6 External-field-driven metal-material composite scatterer considered in Refs. [8-Chap.5]&[15] and constituted by metallic {line, surface, body} & material body.

The ES-WET for the composite scatterer has the same form as the one given in Eq. (2-7), except that: for the composite scatterer, the $C_{\text{sca}}$ represents the summation of the currents on metallic {line, surface, body} and material body; the integral domain V need to be replaced by the region occupated by whole composite scatterer.

**PARTIAL-STRUCTURE-ORIENTED WET (PS-WET)**

Figure 2-7 shows a typical PS-WET-governed metal-material composite transmitting problem. The transmitting antenna is a lumped-port-driven metallic dipole probe (i.e., voltage probe or current probe) loaded by a passive dielectric resonator antenna (DRA).

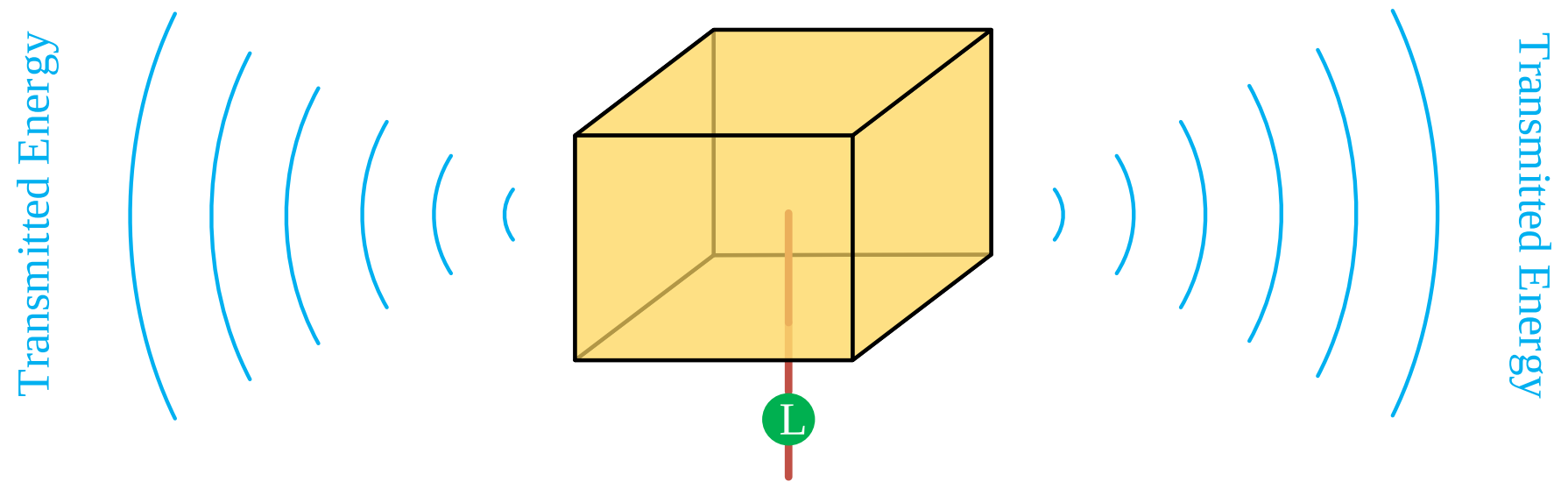

Figure 2-7 DRA which is excited by a lumped-port-driven dipole probe, where the lumped port (L) can be connected to either voltage source (~) or current source (→) .

The PS-WET for the composite structure can be discussed similarly to the previous metallic and material Yagi-Uda antenna cases. Here, we only provide the metal-material-antenna-oriented PS-WET as follows:

$$\begin{aligned}\left\langle \boldsymbol{\mathcal{J}}_{\text{probe}}, \boldsymbol{\mathcal{E}}_{\text{driv}} \right\rangle_{\text{S}_{\text{probe}}} =& \oiint_{\text{S}_\infty} \left( \boldsymbol{\mathcal{E}}_{\text{ant}} \times \boldsymbol{\mathcal{H}}_{\text{ant}} \right) \cdot \boldsymbol{n}_{\text{S}_\infty}^{+} dS + \left\langle \boldsymbol{\sigma} \cdot \boldsymbol{\mathcal{E}}, \boldsymbol{\mathcal{E}} \right\rangle_{\text{V}_{\text{DRA}}} \\ &+ \frac{d}{dt}\left[ (1/2)\left\langle \boldsymbol{\mathcal{H}}_{\text{ant}}, \mu_0 \boldsymbol{\mathcal{H}}_{\text{ant}} \right\rangle_{\text{E}_3} + (1/2)\left\langle \varepsilon_0 \boldsymbol{\mathcal{E}}_{\text{ant}}, \boldsymbol{\mathcal{E}}_{\text{ant}} \right\rangle_{\text{E}_3} \right] \\ &+ \frac{d}{dt}\left[ (1/2)\left\langle \boldsymbol{\mathcal{H}}, \Delta\boldsymbol{\mu} \cdot \boldsymbol{\mathcal{H}} \right\rangle_{\text{V}_{\text{DRA}}} + (1/2)\left\langle \Delta\boldsymbol{\varepsilon} \cdot \boldsymbol{\mathcal{E}}, \boldsymbol{\mathcal{E}} \right\rangle_{\text{V}_{\text{DRA}}} \right]\end{aligned} \tag{2-9}$$

In PS-WET (2-9), $\boldsymbol{\mathcal{F}}_{\text{driv}}$ is the driving field generated by lumped port and only acting on probe; $\boldsymbol{\mathcal{J}}_{\text{probe}}$ is the induced current on probe; $\boldsymbol{\mathcal{F}}_{\text{ant}}$ is the field generated by whole transmitting antenna (including both the active probe and the passive DRA); $\boldsymbol{\mathcal{F}}$ is the summation of $\boldsymbol{\mathcal{F}}_{\text{driv}}$ and $\boldsymbol{\mathcal{F}}_{\text{ant}}$; $\text{V}_{\text{DRA}}$ is the region occupated by DRA. The left-hand side term is the PS-DPO used to sustain a steady work-energy transformation of the lumped-port-driven composite antenna, and it is an effective CM generating operator.

## 2.3 Poynting's Theorem (PtT)

In the previous Sec. 2.2.1, we had proved the conclusions that $\boxed{\text{PtT} \Leftrightarrow \text{ES-WET}}$ in metallic scatterer case and $\boxed{\text{PtT} \Leftrightarrow \text{PS-WET}}$ in passively-loaded metallic antenna case. In this section, we will further generalize the conclusions to material scatterer case, and the {passively-loaded material antenna, composite scatterer, passively-loaded composite antenna} cases can be similarly discussed.

Generally speaking, there are two ways to formulate the PtT of material scatterer. The first way, which we prefer, is to employ the homogeneous Maxwell's equations satisfied by the total fields on scatterer; the second way is to employ the inhomogeneous Maxwell's equations satisfied by the scattered fields on scatterer. The following two sub-sections discuss the two ways separately, and explain why we prefer the first way.

### 2.3.1 PtT Formulated by Total Fields

On region $\text{V}$, scattered fields satisfy inhomogenous Maxwell's equations $\{\nabla \times \boldsymbol{\mathcal{H}}_{\text{sca}} = \boldsymbol{\mathcal{J}}_{\text{sca}} + \frac{\partial}{\partial t}\varepsilon_0 \boldsymbol{\mathcal{E}}_{\text{sca}}, \nabla \times \boldsymbol{\mathcal{E}}_{\text{sca}} = -\boldsymbol{\mathcal{M}}_{\text{sca}} - \frac{\partial}{\partial t}\mu_0 \boldsymbol{\mathcal{H}}_{\text{sca}}\}$, and incident fields satisfy homogeneous Maxwell's equations $\{\nabla \times \boldsymbol{\mathcal{H}}_{\text{inc}} = \frac{\partial}{\partial t}\varepsilon_0 \boldsymbol{\mathcal{E}}_{\text{inc}}, \nabla \times \boldsymbol{\mathcal{E}}_{\text{inc}} = -\frac{\partial}{\partial t}\mu_0 \boldsymbol{\mathcal{H}}_{\text{inc}}\}$ (because $\boldsymbol{\mathcal{F}}_{\text{inc}}$ is source-free on $\text{V}$), so total fields satisfy inhomogenous Maxwell's equations $\{\nabla \times \boldsymbol{\mathcal{H}} = \boldsymbol{\mathcal{J}}_{\text{sca}} + \frac{\partial}{\partial t}\varepsilon_0 \boldsymbol{\mathcal{E}}, \nabla \times \boldsymbol{\mathcal{E}} = -\boldsymbol{\mathcal{M}}_{\text{sca}} - \frac{\partial}{\partial t}\mu_0 \boldsymbol{\mathcal{H}}\}$ [28] (because $\boldsymbol{\mathcal{F}} = \boldsymbol{\mathcal{F}}_{\text{sca}} + \boldsymbol{\mathcal{F}}_{\text{inc}}$).

Employing the above these Maxwell's equations, it is easy to prove the following three relations[27-Sec.1.2.4.4]

$$
\begin{aligned}
\langle \boldsymbol{J}_{\text{sca}}, \boldsymbol{\mathcal{E}}_{\text{inc}} \rangle_{\text{V}} + \langle \boldsymbol{\mathcal{M}}_{\text{sca}}, \boldsymbol{\mathcal{H}}_{\text{inc}} \rangle_{\text{V}} = & \oint\!\!\!\oint_{\partial \text{V}} \left( \boldsymbol{\mathcal{E}} \times \boldsymbol{\mathcal{H}} \right) \cdot \boldsymbol{n}_{\partial \text{V}}^{-} dS \\
& + \oint\!\!\!\oint_{\text{S}_\infty} \left( \boldsymbol{\mathcal{E}}_{\text{sca}} \times \boldsymbol{\mathcal{H}}_{\text{sca}} \right) \cdot \boldsymbol{n}_{\text{S}_\infty}^{+} dS \\
& + \frac{d}{dt} \left[ \frac{1}{2} \langle \boldsymbol{\mathcal{H}}_{\text{sca}}, \mu_0 \boldsymbol{\mathcal{H}}_{\text{sca}} \rangle_{\text{E}_3 \setminus \text{V}} + \frac{1}{2} \langle \varepsilon_0 \boldsymbol{\mathcal{E}}_{\text{sca}}, \boldsymbol{\mathcal{E}}_{\text{sca}} \rangle_{\text{E}_3 \setminus \text{V}} \right] \qquad \text{(2-10a)} \\
& - \frac{d}{dt} \left[ \frac{1}{2} \langle \boldsymbol{\mathcal{H}}, \mu_0 \boldsymbol{\mathcal{H}}_{\text{inc}} \rangle_{\text{V}} + \frac{1}{2} \langle \varepsilon_0 \boldsymbol{\mathcal{E}}, \boldsymbol{\mathcal{E}}_{\text{inc}} \rangle_{\text{V}} \right] \\
& - \frac{d}{dt} \left[ \frac{1}{2} \langle \boldsymbol{\mathcal{H}}_{\text{sca}}, \mu_0 \boldsymbol{\mathcal{H}}_{\text{inc}} \rangle_{\text{V}} + \frac{1}{2} \langle \varepsilon_0 \boldsymbol{\mathcal{E}}_{\text{sca}}, \boldsymbol{\mathcal{E}}_{\text{inc}} \rangle_{\text{V}} \right]
\end{aligned}
$$

and

$$
\begin{aligned}
& \frac{d}{dt} \left[ (1/2) \langle \boldsymbol{\mathcal{H}}, \Delta\boldsymbol{\mu} \cdot \boldsymbol{\mathcal{H}} \rangle_{\text{V}} + (1/2) \langle \Delta\boldsymbol{\varepsilon} \cdot \boldsymbol{\mathcal{E}}, \boldsymbol{\mathcal{E}} \rangle_{\text{V}} \right] \\
& \quad = \frac{d}{dt} \left[ (1/2) \langle \boldsymbol{\mathcal{H}}, \boldsymbol{\mu} \cdot \boldsymbol{\mathcal{H}} \rangle_{\text{V}} + (1/2) \langle \boldsymbol{\varepsilon} \cdot \boldsymbol{\mathcal{E}}, \boldsymbol{\mathcal{E}} \rangle_{\text{V}} \right] \\
& \quad - \frac{d}{dt} \left[ (1/2) \langle \boldsymbol{\mathcal{H}}_{\text{sca}}, \mu_0 \boldsymbol{\mathcal{H}}_{\text{sca}} \rangle_{\text{V}} + (1/2) \langle \varepsilon_0 \boldsymbol{\mathcal{E}}_{\text{sca}}, \boldsymbol{\mathcal{E}}_{\text{sca}} \rangle_{\text{V}} \right] \qquad \text{(2-10b)} \\
& \quad - \frac{d}{dt} \left[ (1/2) \langle \boldsymbol{\mathcal{H}}, \mu_0 \boldsymbol{\mathcal{H}}_{\text{inc}} \rangle_{\text{V}} + (1/2) \langle \varepsilon_0 \boldsymbol{\mathcal{E}}, \boldsymbol{\mathcal{E}}_{\text{inc}} \rangle_{\text{V}} \right] \\
& \quad - \frac{d}{dt} \left[ (1/2) \langle \boldsymbol{\mathcal{H}}_{\text{sca}}, \mu_0 \boldsymbol{\mathcal{H}}_{\text{inc}} \rangle_{\text{V}} + (1/2) \langle \varepsilon_0 \boldsymbol{\mathcal{E}}_{\text{sca}}, \boldsymbol{\mathcal{E}}_{\text{inc}} \rangle_{\text{V}} \right]
\end{aligned}
$$

and

$$
\begin{aligned}
& \frac{d}{dt} \left[ \frac{1}{2} \langle \boldsymbol{\mathcal{H}}_{\text{sca}}, \mu_0 \boldsymbol{\mathcal{H}}_{\text{sca}} \rangle_{\text{E}_3} + \frac{1}{2} \langle \varepsilon_0 \boldsymbol{\mathcal{E}}_{\text{sca}}, \boldsymbol{\mathcal{E}}_{\text{sca}} \rangle_{\text{E}_3} \right] \\
& \quad = \frac{d}{dt} \left[ \frac{1}{2} \langle \boldsymbol{\mathcal{H}}_{\text{sca}}, \mu_0 \boldsymbol{\mathcal{H}}_{\text{sca}} \rangle_{\text{E}_3 \setminus \text{V}} + \frac{1}{2} \langle \varepsilon_0 \boldsymbol{\mathcal{E}}_{\text{sca}}, \boldsymbol{\mathcal{E}}_{\text{sca}} \rangle_{\text{E}_3 \setminus \text{V}} \right] \qquad \text{(2-10c)} \\
& \quad + \frac{d}{dt} \left[ \frac{1}{2} \langle \boldsymbol{\mathcal{H}}_{\text{sca}}, \mu_0 \boldsymbol{\mathcal{H}}_{\text{sca}} \rangle_{\text{V}} + \frac{1}{2} \langle \varepsilon_0 \boldsymbol{\mathcal{E}}_{\text{sca}}, \boldsymbol{\mathcal{E}}_{\text{sca}} \rangle_{\text{V}} \right]
\end{aligned}
$$

Substituting the above relations (2-10a)~(2-10c) into the previous ES-WEP (2-7), it is immediate to derive the following relation[27-Sec.1.2.4.4]

$$
\oint\!\!\!\oint_{\partial \text{V}} \left( \boldsymbol{\mathcal{E}} \times \boldsymbol{\mathcal{H}} \right) \cdot \boldsymbol{n}_{\partial \text{V}}^{-} dS = \langle \boldsymbol{\sigma} \cdot \boldsymbol{\mathcal{E}}, \boldsymbol{\mathcal{E}} \rangle_{\text{V}} + \frac{d}{dt} \left[ \frac{1}{2} \langle \boldsymbol{\mathcal{H}}, \boldsymbol{\mu} \cdot \boldsymbol{\mathcal{H}} \rangle_{\text{V}} + \frac{1}{2} \langle \boldsymbol{\varepsilon} \cdot \boldsymbol{\mathcal{E}}, \boldsymbol{\mathcal{E}} \rangle_{\text{V}} \right] \qquad \text{(2-11)}
$$

and it is just the famous POYNTING'S THEOREM (PtT) satisfied by total fields. In addition, substituting relations (2-10a)~(2-10c) into PtT (2-11), the ES-WEP (2-7) is naturally derived. Thus, we conclude here that: $\boxed{\text{ES-WEP} \Leftrightarrow \text{PtT}}$ also holds in the material scatterer case.

The PtT (2-11) has a clear physical picture[28,49]: **the energy penetrated into a region is converted into two parts — part I (the first term in the right-hand side) corresponds to the Joule heat dissipated in the region, and part II (the second term in the right-hand side) corresponds to the increment of the stored energy in the region**. In addition, the left-hand side of PtT (2-11) is just the source used to sustain the energy dissipation process, and its operator form is particularly called POYNTING'S FLUX OPERATOR (PtFO). To be exhibited in Chap. 6, the frequency-domain PtFO is an effective modal generating operator for calculating the ENERGY-DECOUPLED MODES (DMs) of energy-dissipating structures and self-oscillating structures.

## 2.3.2 PtT Formulated by Scattered Fields

In the previous Sec. 2.2.2, we derive Eq. (2-6) from the inhomogeneous Maxwell's equations $\{\nabla\times\boldsymbol{\mathcal{H}}_{\rm sca}=\boldsymbol{\mathcal{J}}_{\rm sca}+\frac{\partial}{\partial t}\varepsilon_0\boldsymbol{\mathcal{E}}_{\rm sca},\nabla\times\boldsymbol{\mathcal{E}}_{\rm sca}=-\boldsymbol{\mathcal{M}}_{\rm sca}-\frac{\partial}{\partial t}\mu_0\boldsymbol{\mathcal{H}}_{\rm sca}\}$ satisfied by scattered fields. Sometimes, the Eq. (2-6) is also called PtT. Here, we emphasize that this version of PtT is the one satisfied and formulated by scattered fields.

**The field-current interaction term, i.e., the left-hand side of the scattered-field-formed PtT (2-6), quantitatively depicts how the scattered currents supply energy to the scattered fields, but doesn't include any information on where the scattered currents get the energy.** In addition, the polarization and magnetization energies related to the scattering process are also not included in the scattered-field-formed PtT (2-6) (for more details, please see Ref. [51-Sec.VIII-A]).

The above these are the main reasons why "we prefer the total-field-formed PtT (2-11) rather than the scattered-field-formed PtT (2-6), when we do the energy-viewpoint-based modal analysis for material energy-dissipating and self-oscillating structures".

## 2.4 Power Transport Theorem (PTT)

In this section, we rewrite the PtT (2-11) in another equivalent form as POWER TRANSPORT THEOREM (PTT), and then use the PTT to quantitatively depict the power transportation process of wave-port-fed EM structures.

For the convenience of the following discussions, we re-plot the region $\mathbb{V}$ shown in Fig. 2-4 as the following one.

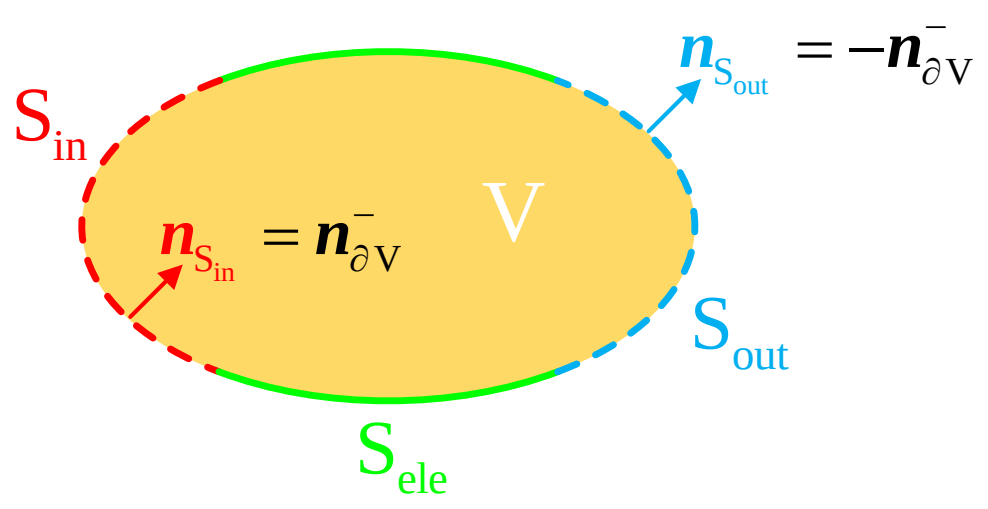

Figure 2-8 Sub-boundaries of the region V shown in Fig. 2-4 and their normal directions.

Here, the boundary $\partial V$ has been divided into three pairwisely disjoint parts $S_{in}$ (a penetrable sub-boundary), $S_{out}$ (a penetrable sub-boundary), and $S_{ele}$ (an impenetrable electric wall), i.e., $\partial V = S_{in} \cup S_{out} \cup S_{ele}$. The normal directions of $S_{in}$ and $S_{out}$ are particularly denoted as $\boldsymbol{n}_{S_{in}}$ and $\boldsymbol{n}_{S_{out}}$ respectively, and point to the interior and exterior of V respectively, so $\boldsymbol{n}_{S_{in}} = \boldsymbol{n}_{\partial V}^{-}$ and $\boldsymbol{n}_{S_{out}} = -\boldsymbol{n}_{\partial V}^{-}$.

Thus, the PtT given in Eq. (2-11) can be equivalently rewritten as the following alternative form[27-Sec.2.2]

$$\begin{aligned}\iint_{S_{in}} (\mathcal{E} \times \mathcal{H}) \cdot \boldsymbol{n}_{S_{in}} dS \;=\; & \iint_{S_{out}} (\mathcal{E} \times \mathcal{H}) \cdot \boldsymbol{n}_{S_{out}} dS + \langle \boldsymbol{\sigma} \cdot \mathcal{E}, \mathcal{E} \rangle_V \\ & + \frac{d}{dt}\Big[ (1/2)\langle \mathcal{H}, \boldsymbol{\mu} \cdot \mathcal{H} \rangle_V + (1/2)\langle \boldsymbol{\varepsilon} \cdot \mathcal{E}, \mathcal{E} \rangle_V \Big]\end{aligned} \tag{2-12}$$

which also has the following frequency-domain version[27-Sec.2.2],[35,39]

$$\begin{aligned}\overbrace{(1/2)\iint_{S_{in}} (\boldsymbol{E} \times \boldsymbol{H}^{\dagger}) \cdot \boldsymbol{n}_{S_{in}} dS}^{P_{in}} \;=\; & (1/2)\iint_{S_{out}} (\boldsymbol{E} \times \boldsymbol{H}^{\dagger}) \cdot \boldsymbol{n}_{S_{out}} dS + (1/2)\langle \boldsymbol{\sigma} \cdot \boldsymbol{E}, \boldsymbol{E} \rangle_V \\ & + j\, 2\omega \Big[ (1/4)\langle \boldsymbol{H}, \boldsymbol{\mu} \cdot \boldsymbol{H} \rangle_V - (1/4)\langle \boldsymbol{\varepsilon} \cdot \boldsymbol{E}, \boldsymbol{E} \rangle_V \Big]\end{aligned} \tag{2-13}$$

for time-harmonic EM problem.

Equations (2-12) and (2-13) have a very clear physical picture[27-Sec.2.2],[35,39]: **the net power-flow passing into V through input port $S_{in}$ is transformed into three parts — part I (the first term in the right-hand side) flows away from V through output port $S_{out}$, and part II (the second term in the right-hand side) is converted into Joule heat, and part III (the third term in the right-hand side) is used to contribute the energy stored in V**. Obviously, Eqs. (2-12) and (2-13) are just a quantitative description for the transportation process of the power-flow passing through the two-port region V, so they are particularly called POWER TRANSPORT THEOREM (PTT).

In addition, the left-hand side term is just the power source used to sustain a steady power transportation[27-Sec.2.2],[35,39], so it is particularly called input power, and its operator form is correspondingly called INPUT POWER OPERATOR (IPO). As exhibited in Refs. [27,35,39], the frequency-domain IPO is an effective modal generating operator for calculating the DMs of wave-port-fed EM structures.

In fact, it is obvious that the above PTT (2-12) is equivalent to the previous PtT (2-11), so we have the following beautiful conclusion

$$\boxed{\left.\begin{matrix}\text{ES-WET}\\ \text{PS-WET}\end{matrix}\right\}\text{WET} \quad \Leftrightarrow \quad \text{PtT} \quad \Leftrightarrow \quad \text{PTT}} \tag{2-14}$$

by employing the relations $\boxed{\text{ES-WET} \Leftrightarrow \text{PtT}}$ and $\boxed{\text{PS-WET} \Leftrightarrow \text{PtT}}$ obtained in the previous Secs. 2.2 and 2.3.

In Ref. [27-Sec.2.3], the transceiving system shown in the following Fig. 2-9 was divided into a series of cascaded two-port EM structures: feeding waveguide $\text{V}^{\text{G}}$, transmitting antenna $\text{V}^{\text{A}}$, propagating medium $\text{M}$, receiving antenna $\text{V}_{\text{A}}$, and loading waveguide $\text{V}_{\text{G}}$.

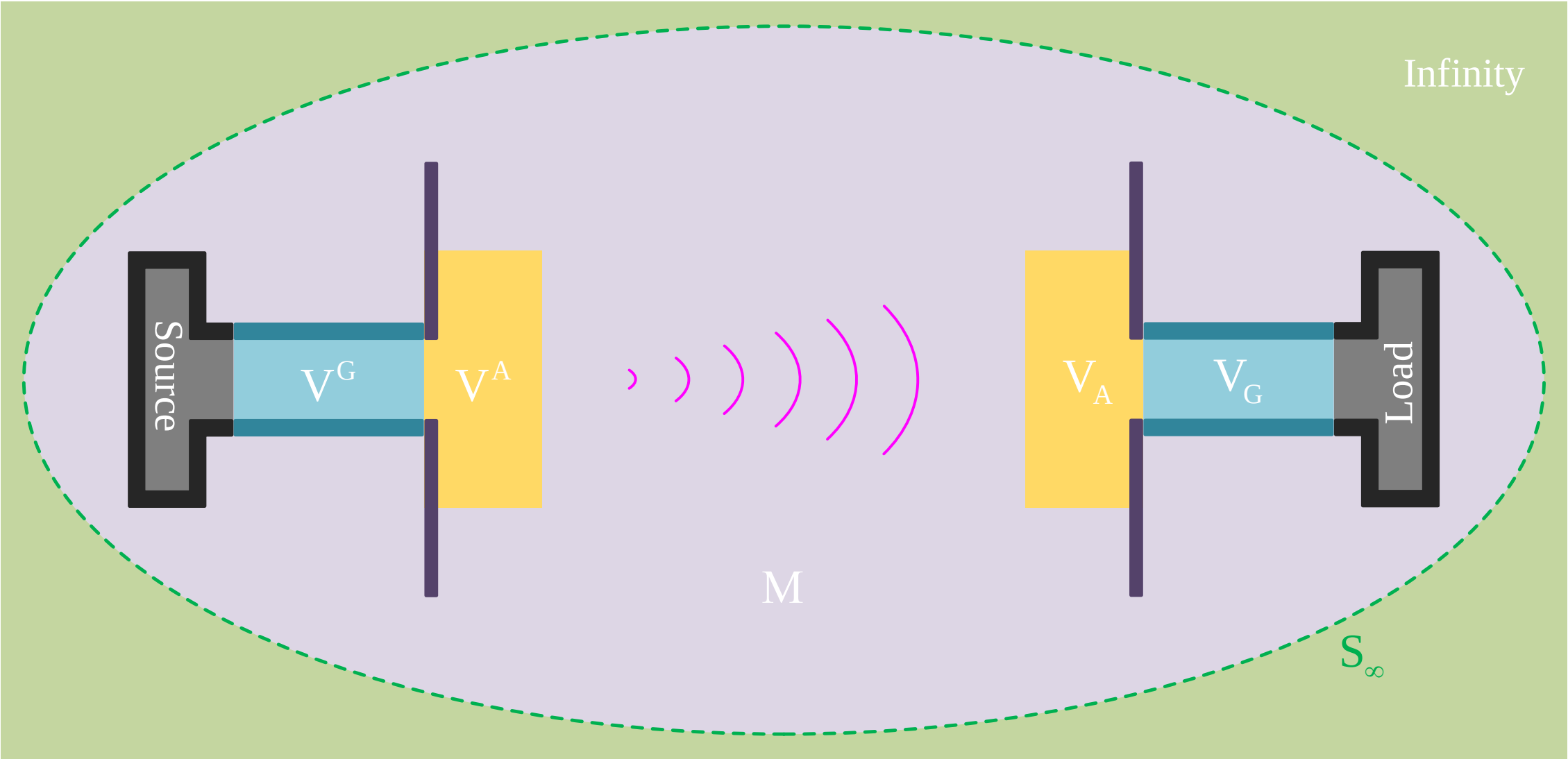

Figure 2-9 Region division for transceiving system.

Applying frequency-domain PTT (2-13) to the two-port EM structures, the following generalized PTT for the cascaded system in Fig. 2-9 is immediately obtained[27-Sec.2.4].

$$P^{\text{S}\rightleftharpoons\text{G}} = \left(P_{\text{dis}}^{\text{G}} + jP_{\text{sto}}^{\text{G}}\right) + \underbrace{\left(P_{\text{dis}}^{\text{A}} + jP_{\text{sto}}^{\text{A}}\right) + \underbrace{\left(P_{\text{Mdis}}^{\text{Mdis}} + jP_{\text{Msto}}^{\text{Msto}}\right) + P_{\text{sca}}^{\text{rad}} + \underbrace{\left(P_{\text{A}}^{\text{dis}} + jP_{\text{A}}^{\text{sto}}\right) + P_{\text{A}\rightleftharpoons\text{G}}}_{P_{\text{M}\rightleftharpoons\text{A}}}}_{P^{\text{A}\rightleftharpoons\text{M}}}}_{P^{\text{G}\rightleftharpoons\text{A}}} \tag{2-15}$$

where the physical meanings of the various powers are as follows:

$P^{\text{S}\rightleftharpoons\text{G}}$ is the net power inputted into $\text{V}^{\text{G}}$, and it is transformed into three parts — a part $P_{\text{dis}}^{\text{G}}$ dissipated in $\text{V}^{\text{G}}$, a part $P_{\text{sto}}^{\text{G}}$ contributing to the energy stored in $\text{V}^{\text{G}}$, and a part $P^{\text{G}\rightleftharpoons\text{A}}$ transported from $\text{V}^{\text{G}}$ to $\text{V}^{\text{A}}$;

$P^{\text{G}\rightleftharpoons\text{A}}$ is the net power inputted into $\text{V}^{\text{A}}$, and it is transformed into three parts — a part $P_{\text{dis}}^{\text{A}}$ dissipated in $\text{V}^{\text{A}}$, a part $P_{\text{sto}}^{\text{A}}$ contributing to the energy stored in $\text{V}^{\text{A}}$, and a part $P^{\text{A}\rightleftharpoons\text{M}}$ transported from $\text{V}^{\text{A}}$ to $\text{M}$;

$P^{\text{A}\rightleftharpoons\text{M}}$ is the net power inputted into $\text{M}$, and it is transformed into four parts — a part $P_{\text{Mdis}}^{\text{Mdis}}$ dissipated in $\text{M}$, a part $P_{\text{Msto}}^{\text{Msto}}$ contributing to the energy stored in $\text{M}$, a part $P_{\text{sca}}^{\text{rad}}$ reaching infinity, and a part $P_{\text{M}\rightleftharpoons\text{A}}$ transported from $\text{M}$ to $\text{V}_{\text{A}}$;

$P_{\text{M}\rightleftharpoons\text{A}}$ is the net power inputted into $\text{V}_{\text{A}}$, and it is transformed into three parts — a part $P_{\text{A}}^{\text{dis}}$ dissipated in $\text{V}_{\text{A}}$, a part $P_{\text{A}}^{\text{sto}}$ contributing to the energy stored in $\text{V}_{\text{A}}$, and a part $P_{\text{A}\rightleftharpoons\text{G}}$ transported from $\text{V}_{\text{A}}$ to $\text{V}_{\text{G}}$;

$P_{\text{A}\rightleftharpoons\text{G}}$ is the net power inputted into $\text{V}_{\text{G}}$, and it is transformed into three parts — a part dissipated in $\text{V}_{\text{G}}$, a part contributing to the energy stored in $\text{V}_{\text{G}}$, and a part transported from $\text{V}_{\text{G}}$ to load.

The specific mathematical expressions for the above-mentioned powers can be found in Ref. [27-Sec.2.4]. In the future Chap. 3, we will employ the IPOs $P^{\text{G}\rightleftharpoons\text{A}}$, $P_{\text{M}\rightleftharpoons\text{A}}$, $P^{\text{O}\rightleftharpoons\text{G}}$ / $P_{\text{A}\rightleftharpoons\text{G}}$, and $P^{\text{A}\rightleftharpoons\text{M}}$ to calculate the DMs of wave-port-fed transmitting antennas, receiving antennas, wave-guiding structures, and free space respectively.

## 2.5 Lorentz's Reciprocity Theorem

In the previous sections, we respectively discussed the different manifestation forms of ENERGY CONSERVATION LAW used to govern the energy couplings "between currents $(\boldsymbol{J}_{\text{sca/act}},\boldsymbol{M}_{\text{sca/act}})$ and the fields $(\boldsymbol{E}_{\text{inc/driv}},\boldsymbol{H}_{\text{inc/driv}})$ of other currents" (Sec. 2.2), "between currents $(\boldsymbol{J}_{\text{sca}},\boldsymbol{M}_{\text{sca}})$ and the fields $(\boldsymbol{E}_{\text{sca}},\boldsymbol{H}_{\text{sca}})$ of themselves" (Sec. 2.3.2), and "between electric/magnetic field $\boldsymbol{E}$ / $\boldsymbol{H}$ and the magnetic/electric field $\boldsymbol{H}$ / $\boldsymbol{E}$ of itself" (Secs. 2.3.1 and 2.4). In this section, we discuss another manifestation form of ENERGY CONSERVATION LAW — LORENTZ'S RECIPROCITY THEOREM form — used to govern the energy coupling between fields $(\boldsymbol{E}_m,\boldsymbol{H}_m)$ and other fields $(\boldsymbol{H}_n,\boldsymbol{E}_n)$, where $m \neq n$.

Two different sets of fields $(\boldsymbol{E}_m,\boldsymbol{H}_m)$ and $(\boldsymbol{E}_n,\boldsymbol{H}_n)$ satisfy the following Maxwell's equations

$$\begin{cases}\nabla\times\boldsymbol{H}_{m/n} = \boldsymbol{\sigma}\cdot\boldsymbol{E}_{m/n} + j\omega\boldsymbol{\varepsilon}\cdot\boldsymbol{E}_{m/n}\\ \nabla\times\boldsymbol{E}_{m/n} = -j\omega\boldsymbol{\mu}\cdot\boldsymbol{H}_{m/n}\end{cases} \quad\text{and}\quad \begin{cases}\nabla\times\boldsymbol{H}_{m/n}^{\dagger} = \boldsymbol{\sigma}\cdot\boldsymbol{E}_{m/n}^{\dagger} - j\omega\boldsymbol{\varepsilon}\cdot\boldsymbol{E}_{m/n}^{\dagger}\\ \nabla\times\boldsymbol{E}_{m/n}^{\dagger} = j\omega\boldsymbol{\mu}\cdot\boldsymbol{H}_{m/n}^{\dagger}\end{cases} \tag{2-16}$$

on the region V (shown in Fig. 2-8) with material parameters $\boldsymbol{\mu}$, $\boldsymbol{\varepsilon}$, and $\boldsymbol{\sigma}$, which are real, symmetrical, and time-independent. From the above Maxwell's equations, the following relations

$$\nabla\times\left[\boldsymbol{\mu}^{-1}\cdot\left(\nabla\times\boldsymbol{E}_{m/n}\right)\right]=-j\omega\left(\boldsymbol{\sigma}\cdot\boldsymbol{E}_{m/n}+j\omega\boldsymbol{\varepsilon}\cdot\boldsymbol{E}_{m/n}\right) \tag{2-17a}$$

$$\nabla\times\left[\boldsymbol{\mu}^{-1}\cdot\left(\nabla\times\boldsymbol{E}_{m/n}^{\dagger}\right)\right]=j\omega\left(\boldsymbol{\sigma}\cdot\boldsymbol{E}_{m/n}^{\dagger}-j\omega\boldsymbol{\varepsilon}\cdot\boldsymbol{E}_{m/n}^{\dagger}\right) \tag{2-17b}$$

can be easily derived, where $\boldsymbol{\mu}^{-1}$ is the inverse of $\boldsymbol{\mu}$.

Based on Gauss's divergence Theorem[52] and some simple vectorial operations, it is not difficult to obtain the following generalized vector-vector Green's second theorem[8-App.C2]

$$\begin{aligned}&\iiint_{\Omega}\left(\boldsymbol{P}\cdot\left\{\nabla\times\left[\boldsymbol{\chi}^{-1}\cdot(\nabla\times\boldsymbol{Q})\right]\right\}-\left\{\nabla\times\left[\boldsymbol{\chi}^{-1}\cdot(\nabla\times\boldsymbol{P})\right]\right\}\cdot\boldsymbol{Q}\right)dV\\&=\oiint_{\partial\Omega}\left\{\boldsymbol{P}\times\left[\boldsymbol{\chi}^{-1}\cdot\left(\nabla\times\boldsymbol{Q}\right)\right]+\left[\boldsymbol{\chi}^{-1}\cdot\left(\nabla\times\boldsymbol{P}\right)\right]\times\boldsymbol{Q}\right\}\cdot\boldsymbol{n}_{\partial\Omega}^{-}dS\end{aligned} \tag{2-18}$$

In the above Eq. (2-18), $\boldsymbol{P}$ and $\boldsymbol{Q}$ are two differentiable vectors distributing on three-dimensional region $\Omega$; $\boldsymbol{\chi}$ is a two-order symmetrical dyad, and its elements are real; $\boldsymbol{\chi}^{-1}$ represents the inverse of $\boldsymbol{\chi}$; $\partial\Omega$ is the boundary surface of $\Omega$; $\boldsymbol{n}_{\partial\Omega}^{-}$ is the inner normal direction of $\partial\Omega$, and it points to the interior of $\Omega$.

Substituting $\{\boldsymbol{P}=\boldsymbol{E}_m;\boldsymbol{Q}=\boldsymbol{E}_n^{\dagger};\boldsymbol{\chi}=\boldsymbol{\mu};\Omega=\mathrm{V};\partial\Omega=\partial\mathrm{V};\boldsymbol{n}_{\partial\Omega}^{-}=\boldsymbol{n}_{\partial\mathrm{V}}^{-}\}$ into Eq. (2-18), and employing Eqs. (2-16) and (2-17), we immediately obtain the following relation

$$\begin{aligned}&(1/2)\oiint_{\partial\mathrm{V}}\left(\boldsymbol{E}_m\times\boldsymbol{H}_n^{\dagger}\right)\cdot\boldsymbol{n}_{\partial\mathrm{V}}^{-}dS+(1/2)\oiint_{\partial\mathrm{V}}\left(\boldsymbol{E}_n^{\dagger}\times\boldsymbol{H}_m\right)\cdot\boldsymbol{n}_{\partial\mathrm{V}}^{-}dS\\&=\frac{1}{2}\langle\boldsymbol{\sigma}\cdot\boldsymbol{E}_n,\boldsymbol{E}_m\rangle_{\mathrm{V}}+\frac{1}{2}\langle\boldsymbol{\sigma}\cdot\boldsymbol{E}_m,\boldsymbol{E}_n\rangle_{\mathrm{V}}-j\,2\omega\iiint_{\mathrm{V}}\left[\frac{1}{4}\boldsymbol{E}_m\cdot\left(\boldsymbol{\varepsilon}\cdot\boldsymbol{E}_n^{\dagger}\right)-\frac{1}{4}\left(\boldsymbol{\varepsilon}\cdot\boldsymbol{E}_m\right)\cdot\boldsymbol{E}_n^{\dagger}\right]dV\\&=(1/2)\langle\boldsymbol{\sigma}\cdot\boldsymbol{E}_n,\boldsymbol{E}_m\rangle_{\mathrm{V}}+(1/2)\langle\boldsymbol{\sigma}\cdot\boldsymbol{E}_m,\boldsymbol{E}_n\rangle_{\mathrm{V}}\end{aligned} \tag{2-19}$$

where the second equality is because of that $\boldsymbol{E}_m\cdot(\boldsymbol{\varepsilon}\cdot\boldsymbol{E}_n^{\dagger})=(\boldsymbol{\varepsilon}\cdot\boldsymbol{E}_m)\cdot\boldsymbol{E}_n^{\dagger}$, which can be proven by employing the method used in Ref. [8-App.C2]. When the region V is lossless, i.e., $\boldsymbol{\sigma}=0$, the above relation (2-19) is further simplified to the following COMPLEX LORENTZ'S RECIPROCITY THEOREM

$$(1/2)\oiint_{\partial\mathrm{V}}\left(\boldsymbol{E}_m\times\boldsymbol{H}_n^{\dagger}\right)\cdot\boldsymbol{n}_{\partial\mathrm{V}}^{-}dS+(1/2)\oiint_{\partial\mathrm{V}}\left(\boldsymbol{E}_n^{\dagger}\times\boldsymbol{H}_m\right)\cdot\boldsymbol{n}_{\partial\mathrm{V}}^{-}dS=0 \tag{2-20}$$

which is different from (but not contradictory to) the conventional LORENTZ'S RECIPROCITY THEOREM[53]. From this sense, the relation (2-19) can be viewed as a generalized version of COMPLEX LORENTZ'S RECIPROCITY THEOREM. The above original and generalized COMPLEX LORENTZ'S RECIPROCITY THEOREMS have the following time-domain forms

$$
\begin{aligned}
&(1/T)\int_{t_0}^{t_0+T}\Big[\oiint_{\partial \mathrm{V}}\big(\boldsymbol{\mathcal{E}}_m\times\boldsymbol{\mathcal{H}}_n+\boldsymbol{\mathcal{E}}_n\times\boldsymbol{\mathcal{H}}_m\big)\cdot\boldsymbol{n}_{\partial \mathrm{V}}^{-}dS\Big]dt\\
=\!=\!=\!=\!=\;&(1/T)\int_{t_0}^{t_0+T}\Big[\langle\boldsymbol{\sigma}\cdot\boldsymbol{\mathcal{E}}_n,\boldsymbol{\mathcal{E}}_m\rangle_{\mathrm{V}}+\langle\boldsymbol{\sigma}\cdot\boldsymbol{\mathcal{E}}_m,\boldsymbol{\mathcal{E}}_n\rangle_{\mathrm{V}}\Big]dt \qquad (2\text{-}21)\\
\xrightarrow{\boldsymbol{\sigma}=0}\;&0
\end{aligned}
$$

where $T$ is the time period of time-harmonic EM field.

Decomposing boundary $\partial \mathrm{V}$ in terms of $\partial \mathrm{V}=\mathrm{S}_{\mathrm{in}}\cup\mathrm{S}_{\mathrm{ele}}\cup\mathrm{S}_{\mathrm{out}}$ as shown in Fig. 2-8, and employing the homogeneous tangential electric field boundary condition on $\mathrm{S}_{\mathrm{ele}}$, the time-domain LORENTZ'S RECIPROCITY THEOREM (2-21) can be expressed as the following more inspired form

$$
\begin{aligned}
&\frac{1}{T}\int_{t_0}^{t_0+T}\left[\iint_{\mathrm{S}_{\mathrm{in}}}\big(\boldsymbol{\mathcal{E}}_m\times\boldsymbol{\mathcal{H}}_n+\boldsymbol{\mathcal{E}}_n\times\boldsymbol{\mathcal{H}}_m\big)\cdot\boldsymbol{n}_{\mathrm{S}_{\mathrm{in}}}dS\right]dt\\
=\!=\!=\!=\!=\;&\frac{1}{T}\int_{t_0}^{t_0+T}\left[\iint_{\mathrm{S}_{\mathrm{out}}}\big(\boldsymbol{\mathcal{E}}_m\times\boldsymbol{\mathcal{H}}_n+\boldsymbol{\mathcal{E}}_n\times\boldsymbol{\mathcal{H}}_m\big)\cdot\boldsymbol{n}_{\mathrm{S}_{\mathrm{out}}}dS\right]dt+\frac{1}{T}\int_{t_0}^{t_0+T}\Big[\langle\boldsymbol{\sigma}\cdot\boldsymbol{\mathcal{E}}_n,\boldsymbol{\mathcal{E}}_m\rangle_{\mathrm{V}}+\langle\boldsymbol{\sigma}\cdot\boldsymbol{\mathcal{E}}_m,\boldsymbol{\mathcal{E}}_n\rangle_{\mathrm{V}}\Big]dt\\
\xrightarrow{\boldsymbol{\sigma}=0}\;&\frac{1}{T}\int_{t_0}^{t_0+T}\left[\iint_{\mathrm{S}_{\mathrm{out}}}\big(\boldsymbol{\mathcal{E}}_m\times\boldsymbol{\mathcal{H}}_n+\boldsymbol{\mathcal{E}}_n\times\boldsymbol{\mathcal{H}}_m\big)\cdot\boldsymbol{n}_{\mathrm{S}_{\mathrm{out}}}dS\right]dt \qquad (2\text{-}22)
\end{aligned}
$$

where the relations $\boldsymbol{n}_{\partial \mathrm{V}}^{-}=\boldsymbol{n}_{\mathrm{S}_{\mathrm{in}}}$ and $\boldsymbol{n}_{\partial \mathrm{V}}^{-}=-\boldsymbol{n}_{\mathrm{S}_{\mathrm{out}}}$ have been utilized. The above time-domain LORENTZ'S RECIPROCITY THEOREM (2-22) has the following physical interpretation: **the coupled power-flow inputted into region V by passing through input port $\mathrm{S}_{\mathrm{in}}$ is transformed into two parts — part I is the coupled power-flow outputted from region V by passing through output port $\mathrm{S}_{\mathrm{out}}$ and part II is the coupled power-dissipation converted into Joule heat**. Obviously, the above-obtained LORENTZ'S RECIPROCITY THEOREM for the two-port region V shown in Fig. 2-8 can be easily generalized to the seperated regions and cascaded systems shown in Fig. 2-9.

To be exhibited in the subsequent chapters, the above-discussed LORENTZ'S RECIPROCITY THEOREM is very useful to prove many beautiful energy features of the DMs (Chaps. 3 and 6) and energy-decoupled CMs (Chaps. 4 and 5) of EM structures. In addition, the conventional RAYLEIGH-CARSON RECIPROCITY THEOREM[8-App.C3] can also be generalized to complex version, and energy-viewpoint-based physical interpretations for the complex version can be similarly discussed, and they are not explicitly provided here.

## 2.6 Chapter Summary

For a certain EM structure, it has many different working manners, such as scattering manner, energy-dissipating/self-oscillating manner, and transceiving manner. In different working manners, the ENERGY CONSERVATION LAW always holds, but it has different manifestation forms.

From Maxwell's equations, this chapter derives the mathematical expressions of some different manifestation forms of ENERGY CONSERVATION LAW, and then formulates the energy sources used to sustain the steady energy utilization processes of the different manners. Specifically speaking:

- When EM structure works at scattering manner, a work-energy transformation process occurs, and the ES-WET form of ENERGY CONSERVATION LAW is an effective quantitative depiction for the work-energy transformation. ES-DPO is the source term contained in ES-WET, and it is just the energy source for sustaining a steady work-energy transformation.
- When lumped-port-driven EM structure works at transmitting and transfering manners, a work-energy transformation process occurs, and the PS-WET form of ENERGY CONSERVATION LAW is an effective quantitative depiction for the work-energy transformation process. PS-DPO is the source term contained in PS-WET, and it is just the energy source for sustaining a steady work-energy transformation.
- When EM structure works at energy-dissipating/self-oscillating manner, a energy dissipation / self-oscillation process occurs, and the PtT form of ENERGY CONSERVATION LAW is a quantitative depiction for the energy dissipation / self-oscillation. PtFO is the source term contained in PtT, and the non-zero PtFO is just the energy source for sustaining a steady energy dissipation, and the self-oscillation process doesn't need a non-zero PtFO.
- When wave-port-fed EM structure works at transmitting, receiving, and wave-guiding manners, a power transportation process occurs, and the PTT form of ENERGY CONSERVATION LAW is an effective quantitative depiction for the power transportation process. IPO is the source term contained in PTT, and it is just the energy source for sustaining a steady power transportation.

In addition, during the process to discuss ES-WET, PS-WET, PtT and PTT, this chapter also highlights their **physical pictures** and equivalence relation, besides their energy sources ES-DPO, PS-DPO, PtFO and IPO.

By orthogonalizing the above-mentioned energy source terms IPO, PS-DPO, ES-DPO, and PtFO, the subsequent chapters focus on constructing the energy-decoupled modes of wave-port-fed EM structures (Chap. 3), lumped-port-driven EM structures (Chap. 4), incident-field-driven EM structures (Chap. 5), and external-field-illuminated EM structures (Chap. 6). When the energy-decoupled modes have been constructed, we will further prove some fascinating features of the modes by employing the LORENTZ'S RECIPROCITY THEOREM obtained in this chapter.

# CHAPTER 3 PTT-BASED MODAL ANALYSIS FOR WAVE-PORT-FED EM STRUCTURES

**CHAPTER MOTIVATION:** This chapter focuses on establishing an effective modal analysis method — DECOUPLING MODE THEORY (DMT) — for wave-port-fed electromagnetic (EM) structures in POWER TRANSPORT THEOREM (PTT) framework. The PTT-based DMT (PTT-DMT) can effectively construct the ENERGY-DECOUPLED MODES (DMs) of wave-port-fed EM structures by orthogonalizing frequency-domain INPUT POWER OPERATOR (IPO).

## 3.1 Chapter Introduction

Figure 3-1 illustrates a typical transceiving system, which is constituted by a series of cascaded wave-port-fed EM structures — wave-port-fed feeding waveguide, transmitting antenna, receiving antenna, and loading waveguide, etc.[27-Chap.2].

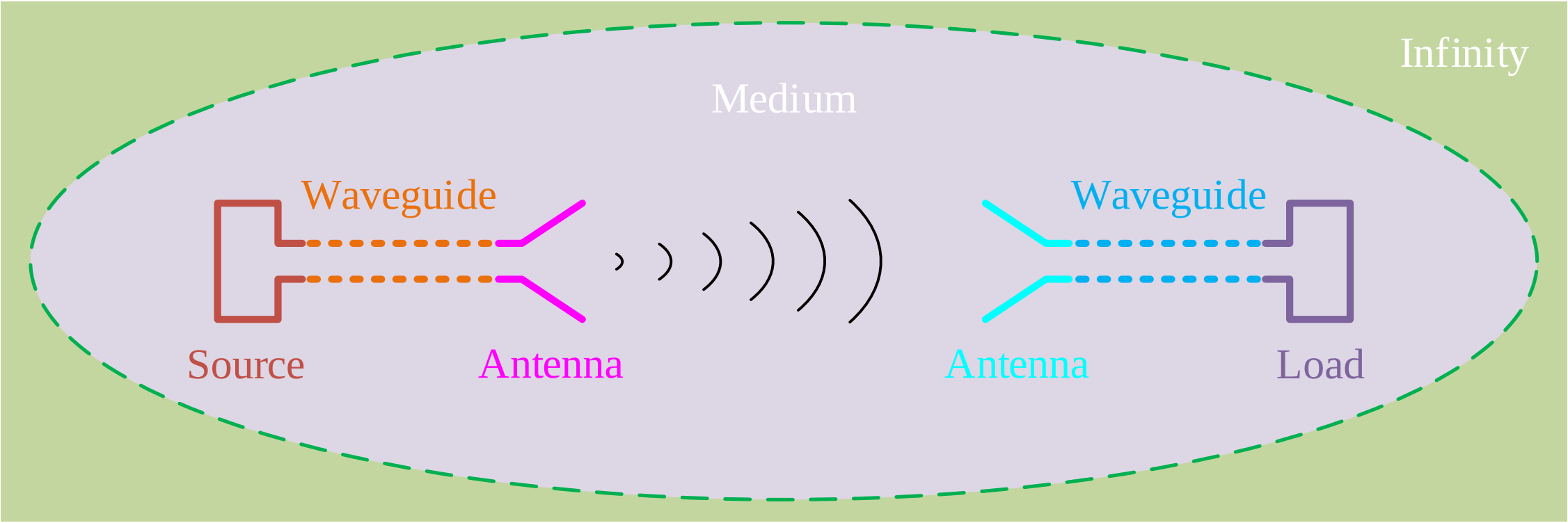

Figure 3-1 Transceiving system constituted by a series of cascaded wave-port-fed structures.

During the working process of whole transceiving system, there exists a strong inter-action among the EM fields generated by the EM structures[54,55]. The inter-action will lead to a complicated inter-transformation/inter-excitation among the fundamental modes of the EM structures[54,55]. Modal matching method[54~56] is an effective one for quantitatively describing and analyzing the modal inter-transformation/inter-excitation. However, before utilizing the modal matching method, it is indispensable to separately calculate the fundamental modes of the EM structures (i.e., to do the modal analysis for the EM structures separately) beforehand.

Under a unified PTT framework, this chapter is devoted to establishing an universal modal analysis method — PTT-DMT — for the various wave-port-fed EM structures.

## 3.2 PTT-Based DMs of Wave-Port-Fed Transmitting Antennas

Taking the metallic horn antenna shown in Fig. 3-2 as a typical example, this section establishes PTT-DMT and constructs DMs for wave-port-fed transmitting antennas.

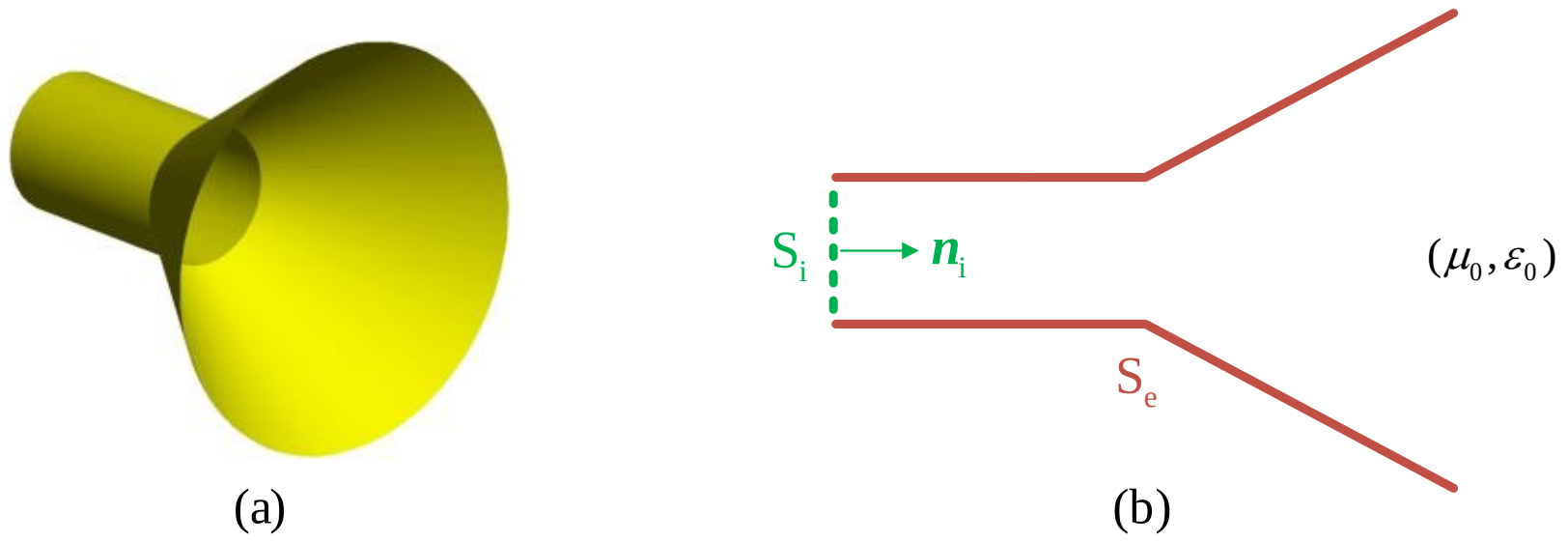

Figure 3-2 (a) Geometry and (b) topology of a metallic horn antenna.

For the horn shown in Fig. 3-2, its surrounding environment is free space (which has parameters $\mu_0$ and $\varepsilon_0$), and its input port (i.e., feeding port) and electric wall (i.e., horn wall) are $\mathrm{S}_\mathrm{i}$ and $\mathrm{S}_\mathrm{e}$. In addition, the normal direction of $\mathrm{S}_\mathrm{i}$ is $\boldsymbol{n}_\mathrm{i}$.

Based on the conclusions given in Sec. 2.4, the source used to sustain a steady power transmitting of the horn is input power $P_\mathrm{in} = (1/2)\iint_{\mathrm{S}_\mathrm{i}} (\boldsymbol{E}\times\boldsymbol{H}^\dagger)\cdot\boldsymbol{n}_\mathrm{i} dS$ [27-Chaps.2&6], and it has the following operator expression[27-Sec.6.2],[35]

$$\begin{aligned} P_\mathrm{in} &= -(1/2)\left\langle \boldsymbol{J}_\mathrm{i}, -j\omega\mu_0\mathcal{L}_0(\boldsymbol{J}_\mathrm{i}+\boldsymbol{J}_\mathrm{e}) - \mathcal{K}_0(\boldsymbol{M}_\mathrm{i})\right\rangle_{\mathrm{S}_\mathrm{i}^+} \\ &= -(1/2)\left\langle \boldsymbol{M}_\mathrm{i}, \mathcal{K}_0(\boldsymbol{J}_\mathrm{i}+\boldsymbol{J}_\mathrm{e}) - j\omega\varepsilon_0\mathcal{L}_0(\boldsymbol{M}_\mathrm{i})\right\rangle_{\mathrm{S}_\mathrm{i}^+}^\dagger \end{aligned} \tag{3-1}$$

called INPUT POWER OPERATOR (IPO). In IPO (3-1), $\boldsymbol{J}_\mathrm{i} = \boldsymbol{n}_\mathrm{i}\times\boldsymbol{H}$ and $\boldsymbol{M}_\mathrm{i} = \boldsymbol{E}\times\boldsymbol{n}_\mathrm{i}$ are the equivalent currents on $\mathrm{S}_\mathrm{i}$; $\boldsymbol{J}_\mathrm{e}$ is the induced current on $\mathrm{S}_\mathrm{e}$; $\mathrm{S}_\mathrm{i}^+$ is the right-side surface of $\mathrm{S}_\mathrm{i}$. In addition, the right-hand sides of the first and second equalities in IPO (3-1) are called JE and HM interaction forms respectively.

Employing the basis function expansions for the currents involved in IPO (3-1), the integral operators are immediately discretized into the following matrix forms

$$P_\mathrm{in} = \mathbb{J}_\mathrm{i}^\dagger \cdot \mathbb{P}_\mathrm{JE} \cdot \begin{bmatrix} \mathbb{J}_\mathrm{i} \\ \mathbb{J}_\mathrm{e} \\ \mathbb{M}_\mathrm{i} \end{bmatrix} = \begin{bmatrix} \mathbb{J}_\mathrm{i} \\ \mathbb{J}_\mathrm{e} \\ \mathbb{M}_\mathrm{i} \end{bmatrix}^\dagger \cdot \mathbb{P}_\mathrm{HM} \cdot \mathbb{M}_\mathrm{i} \tag{3-2}$$

where $\mathbb{J}_\mathrm{i}$, $\mathbb{J}_\mathrm{e}$ and $\mathbb{M}_\mathrm{i}$ are the basis function expansion coefficient vectors of $\boldsymbol{J}_\mathrm{i}$, $\boldsymbol{J}_\mathrm{e}$ and $\boldsymbol{M}_\mathrm{i}$.

In fact, the above-mentioned currents are not independent of each other, because of the following integral equations

$$\left[\mathcal{K}_0\left(\boldsymbol{J}_{\rm i}+\boldsymbol{J}_{\rm e}\right)-j\omega\varepsilon_0\mathcal{L}_0\left(\boldsymbol{M}_{\rm i}\right)\right]_{{\rm S}_{\rm i}^+}^{\tan} = \boldsymbol{J}_{\rm i}\times\boldsymbol{n}_{\rm i} \tag{3-3a}$$

$$\left[-j\omega\mu_0\mathcal{L}_0\left(\boldsymbol{J}_{\rm i}+\boldsymbol{J}_{\rm e}\right)-\mathcal{K}_0\left(\boldsymbol{M}_{\rm i}\right)\right]_{{\rm S}_{\rm i}^+}^{\tan} = \boldsymbol{n}_{\rm i}\times\boldsymbol{M}_{\rm i} \tag{3-3b}$$

$$\left[-j\omega\mu_0\mathcal{L}_0\left(\boldsymbol{J}_{\rm i}+\boldsymbol{J}_{\rm e}\right)-\mathcal{K}_0\left(\boldsymbol{M}_{\rm i}\right)\right]_{{\rm S}_{\rm e}}^{\tan} = 0 \tag{3-4}$$

Here, the first two equations originate from the definitions of $\boldsymbol{J}_{\rm i}$ (DoJ) and $\boldsymbol{M}_{\rm i}$ (DoM) respectively, and the last equation is due to that the tangential $\boldsymbol{E}$ is 0 on ${\rm S}_{\rm e}$.

The method of moments (MoM) can easily discretize the integral equations (3-3a)~(3-4) into matrix equations, and the matrix equations can give the the following transformations

$$\mathbb{T}_{\rm DoJ}\cdot\mathbb{J}_{\rm i} = \begin{bmatrix}\mathbb{J}_{\rm i}\\ \mathbb{J}_{\rm e}\\ \mathbb{M}_{\rm i}\end{bmatrix} = \mathbb{T}_{\rm DoM}\cdot\mathbb{M}_{\rm i} \tag{3-5}$$

from independent current $\mathbb{J}_{\rm i}/\mathbb{M}_{\rm i}$ into all currents $(\mathbb{J}_{\rm i},\mathbb{J}_{\rm e},\mathbb{M}_{\rm i})$.

Employing transformation (3-5), the dependent currents involved in matrix forms (3-2) can be eliminated as follows[27-Sec.6.2],[35]:

$$P_{\rm in} = \mathbb{C}_{\rm i}^\dagger\cdot\mathbb{P}_{\rm in}\cdot\mathbb{C}_{\rm i} = \begin{cases}\mathbb{J}_{\rm i}^\dagger\cdot\overbrace{\mathbb{P}_{\rm JE}\cdot\mathbb{T}_{\rm DoJ}}^{\mathbb{P}_{\rm JE\text{-}DoJ}}\cdot\mathbb{J}_{\rm i}\\ \mathbb{M}_{\rm i}^\dagger\cdot\underbrace{\mathbb{T}_{\rm DoM}^\dagger\cdot\mathbb{P}_{\rm HM}}_{\mathbb{P}_{\rm HM\text{-}DoM}}\cdot\mathbb{M}_{\rm i}\end{cases} \tag{3-6}$$

In Eq. (3-6), $\mathbb{C}_{\rm i}$ is the independent current, which is either $\mathbb{J}_{\rm i}$ or $\mathbb{M}_{\rm i}$.

Using the above $\mathbb{P}_{\rm in}$, the DMs can be derived from solving the following modal decoupling equation[27-Sec.6.2],[35]

$$\mathbb{P}_{\rm in}^-\cdot\mathbb{C}_{\rm i} = \theta\,\mathbb{P}_{\rm in}^+\cdot\mathbb{C}_{\rm i} \tag{3-7}$$

where $\mathbb{P}_{\rm in}^+ = (\mathbb{P}_{\rm in}+\mathbb{P}_{\rm in}^\dagger)/2$ and $\mathbb{P}_{\rm in}^- = (\mathbb{P}_{\rm in}-\mathbb{P}_{\rm in}^\dagger)/2j$ are the positive and negative Hermitian parts of $\mathbb{P}_{\rm in}$. The above-obtained modes satisfy the following frequency-domain power-decoupling relation

$$(1/2)\iint_{{\rm S}_{\rm i}}\left(\boldsymbol{E}_n\times\boldsymbol{H}_m^\dagger\right)\cdot\boldsymbol{n}_{\rm i}dS = \left(1+j\theta_m\right)\delta_{mn} \tag{3-8}$$

and then the following time-domain energy-decoupling relation (or alternatively called time-averaged power-decoupling relation)

$$(1/T)\int_{t_0}^{t_0+T}\left[\iint_{{\rm S}_{\rm i}}\left(\boldsymbol{\mathcal{E}}_n\times\boldsymbol{\mathcal{H}}_m\right)\cdot\boldsymbol{n}_{\rm i}dS\right]dt = \delta_{mn} \tag{3-9}$$

where $T$ is the time period of the time-harmonic EM field, and all modal real powers have been normalized to 1 following the convention used in Ref. [10] (the physical explanation for this kind of modal normalization had been carefully discussed in Refs. [27-Sec.1.2.4.7] and [14]). Evidently, the energy-decoupling relation (3-9) has a very clear physical interpretation: **in any integral period, there doesn't exist net energy exchange between any two different modes**[27-Sec.6.2],[35]. Thus, the modes derived from Eq. (3-7) are energy-decoupled.

Here, we provide a simple but important corollary of energy-decoupling relation (3-9) as follows:

$$(1/T)\int_{t_0}^{t_0+T}\left[\iint_{\mathrm{S_i}}\left(\boldsymbol{\mathcal{E}}_n\times\boldsymbol{\mathcal{H}}_m+\boldsymbol{\mathcal{E}}_m\times\boldsymbol{\mathcal{H}}_n\right)\cdot\boldsymbol{n}_{\mathrm{i}}dS\right]dt=2\delta_{mn} \tag{3-10}$$

which is consistent with the energy-decoupling relation satisfied by the DMs on the input port of receiving antennas. In addition, utilizing time-domain Lorentz's Reciprocity Theorem (2-22) and the homogeneous tangential electric field boundary condition on $\mathrm{S_e}$, we have the following more general energy-decoupling relation

$$(1/T)\int_{t_0}^{t_0+T}\left[\oint\!\!\!\oint_{\mathrm{S}}\left(\boldsymbol{\mathcal{E}}_n\times\boldsymbol{\mathcal{H}}_m+\boldsymbol{\mathcal{E}}_m\times\boldsymbol{\mathcal{H}}_n\right)\cdot\boldsymbol{n}dS\right]dt=2\delta_{mn} \tag{3-11}$$

where the closed integral surface $\mathrm{S}$ is an arbitrary surface enclosing the whole transmitting antenna as shown in Fig. 3-3, and $\boldsymbol{n}$ is the outer normal direction of $\mathrm{S}$.

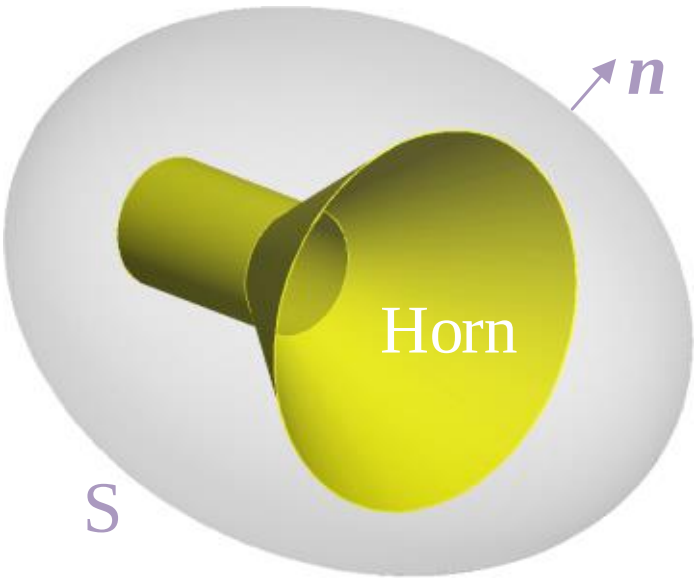

Figure 3-3 A closed surface enclosing whole transmitting horn antenna.

Obviously, the most special case of $\mathrm{S}$ is just $\mathrm{S}_\infty$, which is a spherical surface with infinite radius.

For the DMs obtained above, we propose the following field-based definitions for modal input "impedance $Z_{\mathrm{in}}$, resistance $R_{\mathrm{in}}$, reactance $X_{\mathrm{in}}$" and "admittance $Y_{\mathrm{in}}$, conductance $G_{\mathrm{in}}$, susceptance $B_{\mathrm{in}}$"[27-Sec.6.2.4.3],[35]

$$Z_{\mathrm{in}} = \frac{(1/2)\iint_{\mathrm{S_i}}\left(\boldsymbol{E}\times\boldsymbol{H}^{\dagger}\right)\cdot\boldsymbol{n}_{\mathrm{i}}dS}{(1/2)\langle\boldsymbol{J}_{\mathrm{i}},\boldsymbol{J}_{\mathrm{i}}\rangle_{\mathrm{S_i}}} = R_{\mathrm{in}}+j\,X_{\mathrm{in}} \tag{3-12a}$$

$$Y_{\text{in}} = \frac{(1/2)\iint_{S_i} \left(\boldsymbol{E} \times \boldsymbol{H}^{\dagger}\right) \cdot \boldsymbol{n}_i dS}{(1/2)\langle \boldsymbol{M}_i, \boldsymbol{M}_i \rangle_{S_i}} = G_{\text{in}} + j\,B_{\text{in}} \tag{3-12b}$$

where $R_{\text{in}} = \text{Re}\,Z_{\text{in}}$, $X_{\text{in}} = \text{Im}\,Z_{\text{in}}$, $G_{\text{in}} = \text{Re}\,Y_{\text{in}}$, and $B_{\text{in}} = \text{Im}\,Y_{\text{in}}$. The modal resistance $R_{\text{in}}$ and conductance $G_{\text{in}}$ will be utilized to recognize the resonant DMs in the following numerical examples.

Here, we provide a PTT-DMT-based modal analysis for the circular metallic horn shown in Fig. 3-4 for verifying the validity of the above formulations.

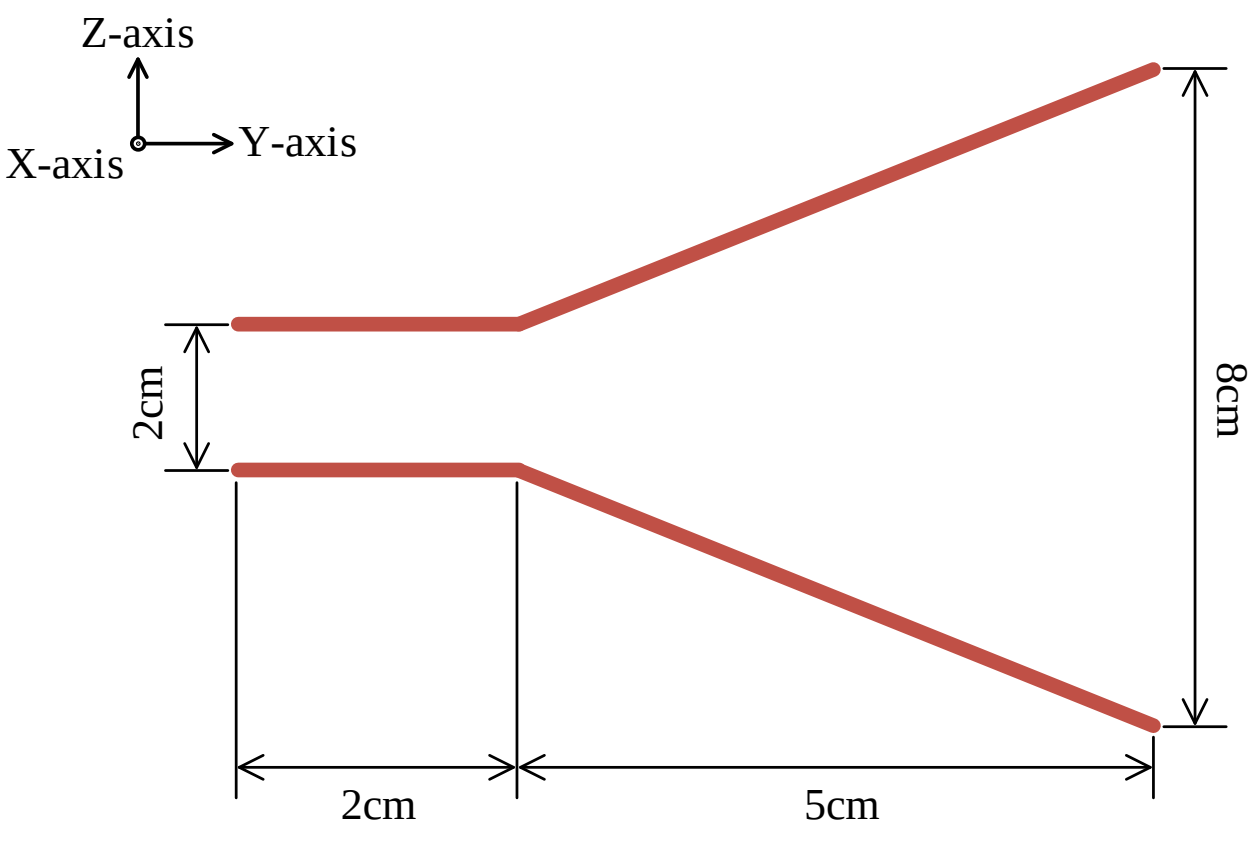

Figure 3-4 Size of a circular metallic horn.

We use the JE-DoJ-based IPO $\mathbb{P}_{\text{JE-DoJ}}$ to calculate the DMs of the horn, and show the associated modal resistance curves in Fig. 3-5. If *"resonance" is defined as the working state at which* $R_{\text{in}}$ *attains its maximum*[57-pp.440], then Fig. 3-5 implies that DM 1 and DM 2 are resonant at 9.3 GHz, and DM 3 and DM 4 are resonant at 14.9 GHz.

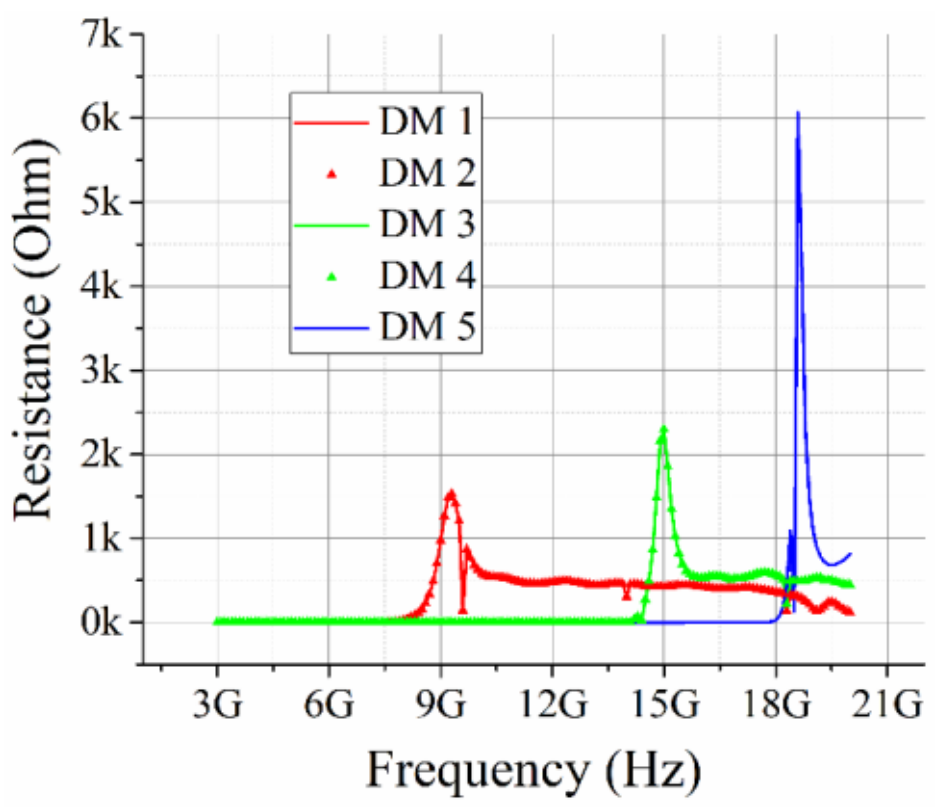

Figure 3-5 Modal input resistances of the first several lower-order DMs.

For the resonant DM 1 and DM 2 (working at 9.3 GHz), their modal port currents are shown in the following Fig. 3-6.

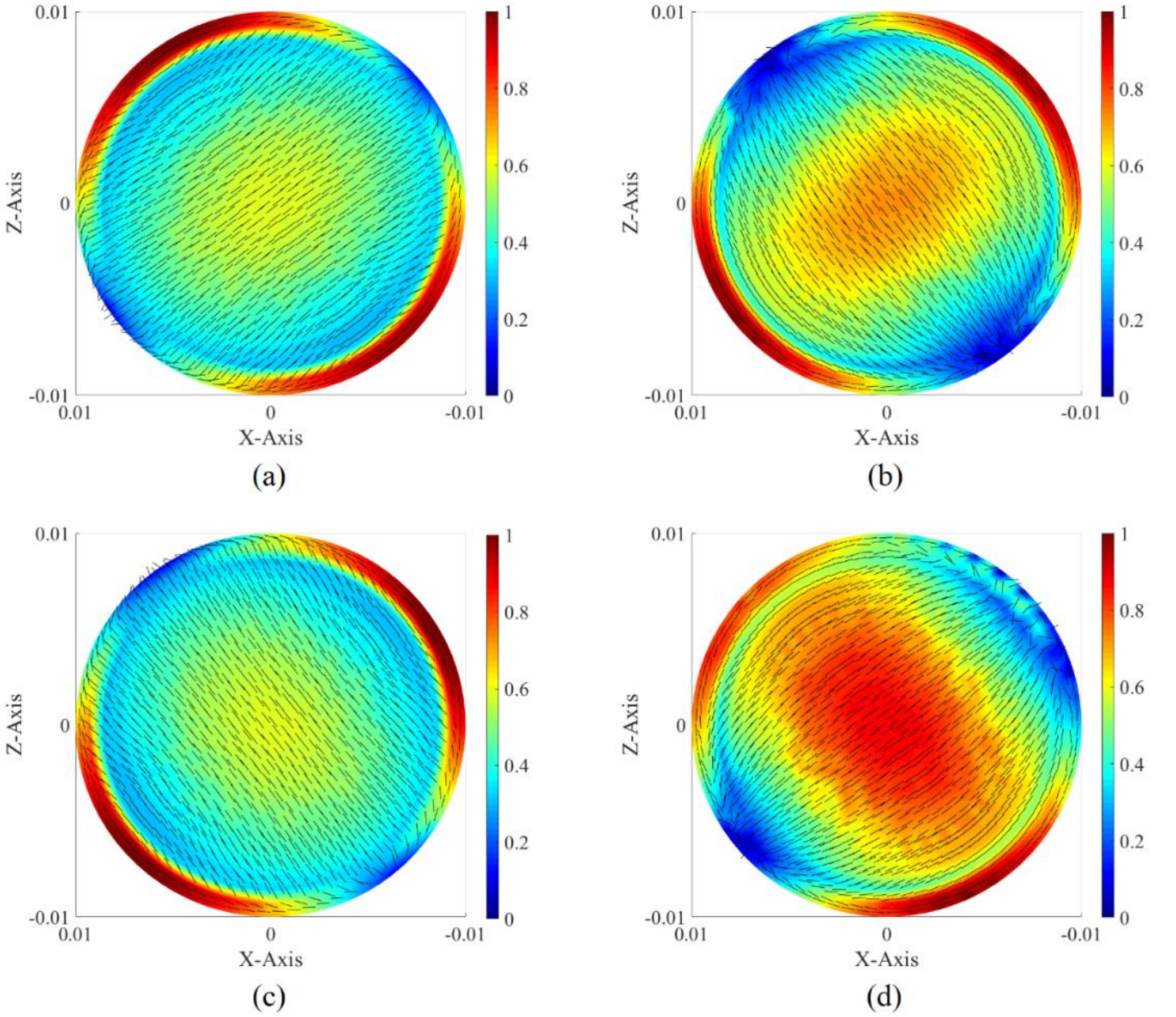

Figure 3-6 Distributions of the modal port currents at 9.3 GHz. (a) Modal port electric current $\boldsymbol{J}_{\mathrm{i}}$ of DM 1, (b) modal port magnetic current $\boldsymbol{M}_{\mathrm{i}}$ of DM 1, (c) modal port electric current $\boldsymbol{J}_{\mathrm{i}}$ of DM 2, and (d) modal port magnetic current $\boldsymbol{M}_{\mathrm{i}}$ of DM 2.

Evidently, the DM 1 and DM 2 are a pair of spatially degenerate states (due to Y-axis rotational symmetry of the horn). For the first degenerate state (at 9.3 GHz), its wall current distribution, radiation pattern, and field distributions are shown in Fig. 3-7.

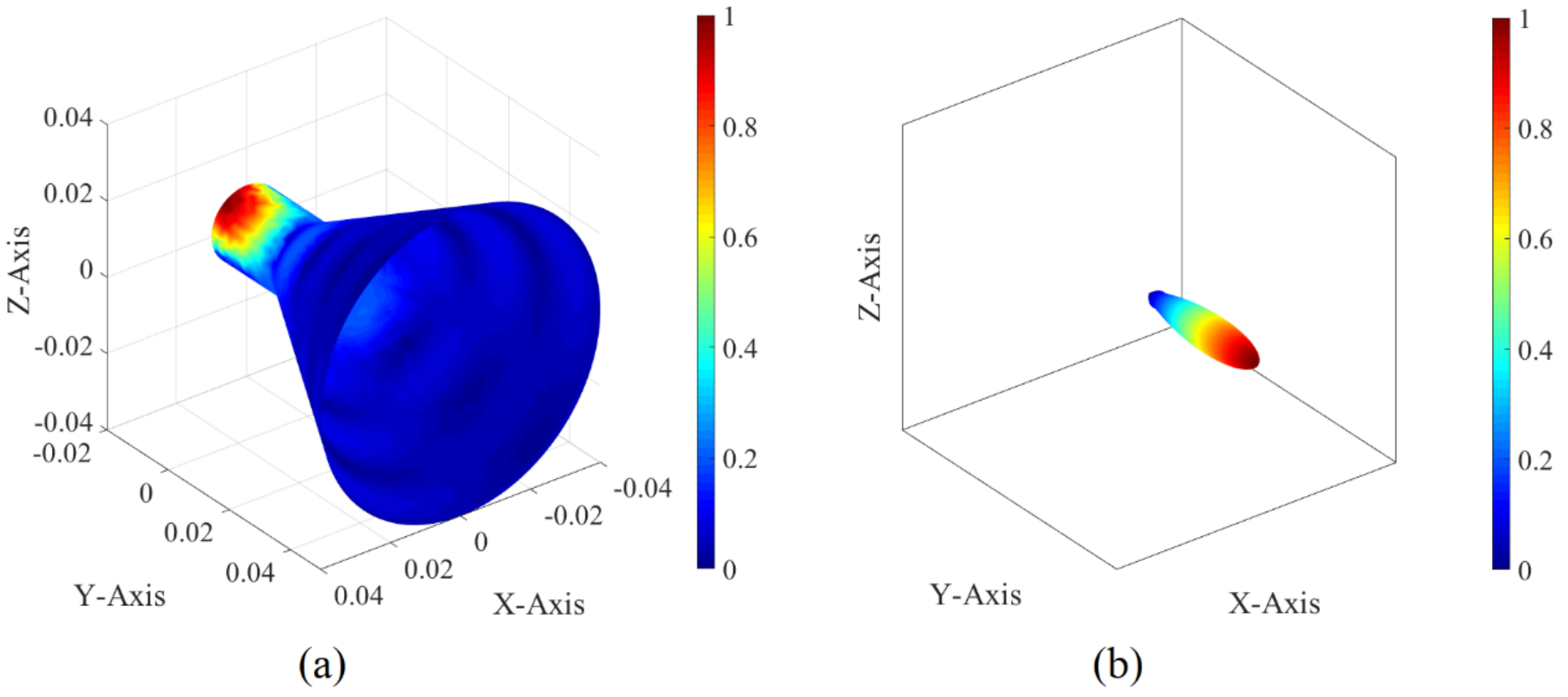

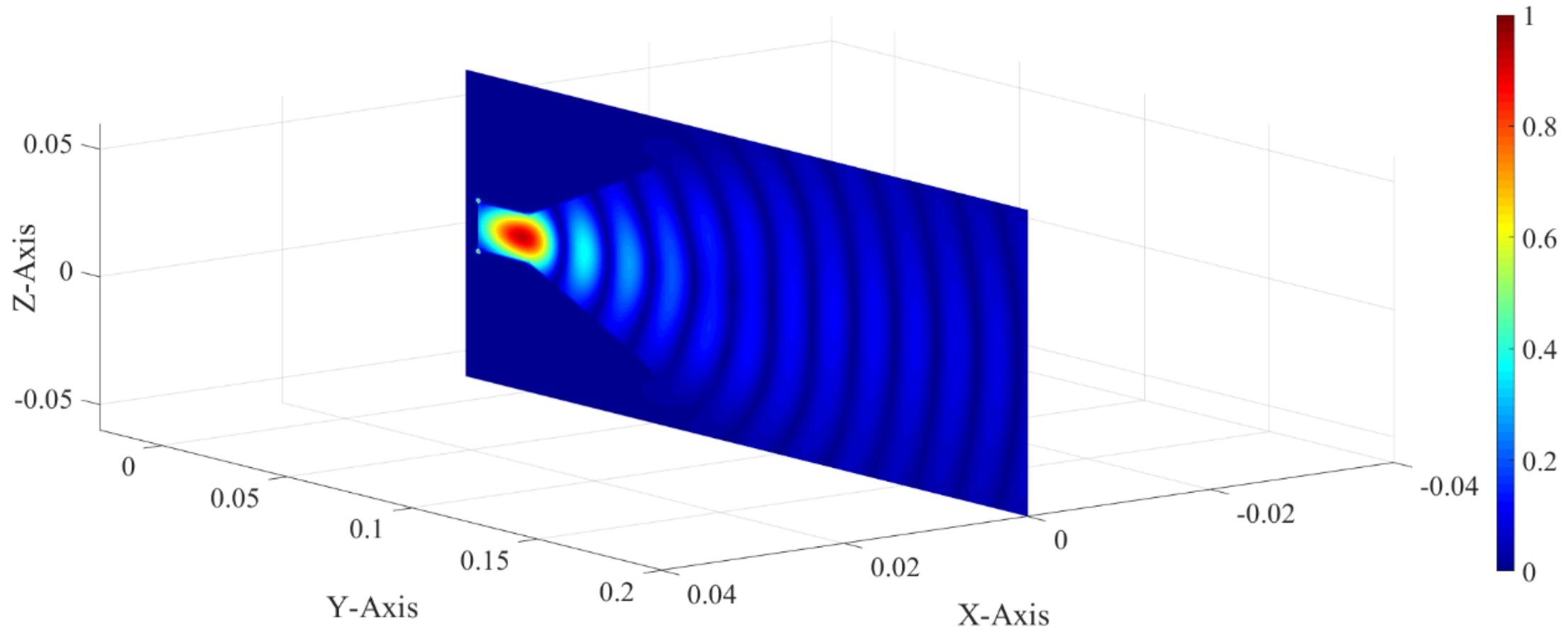

(c)

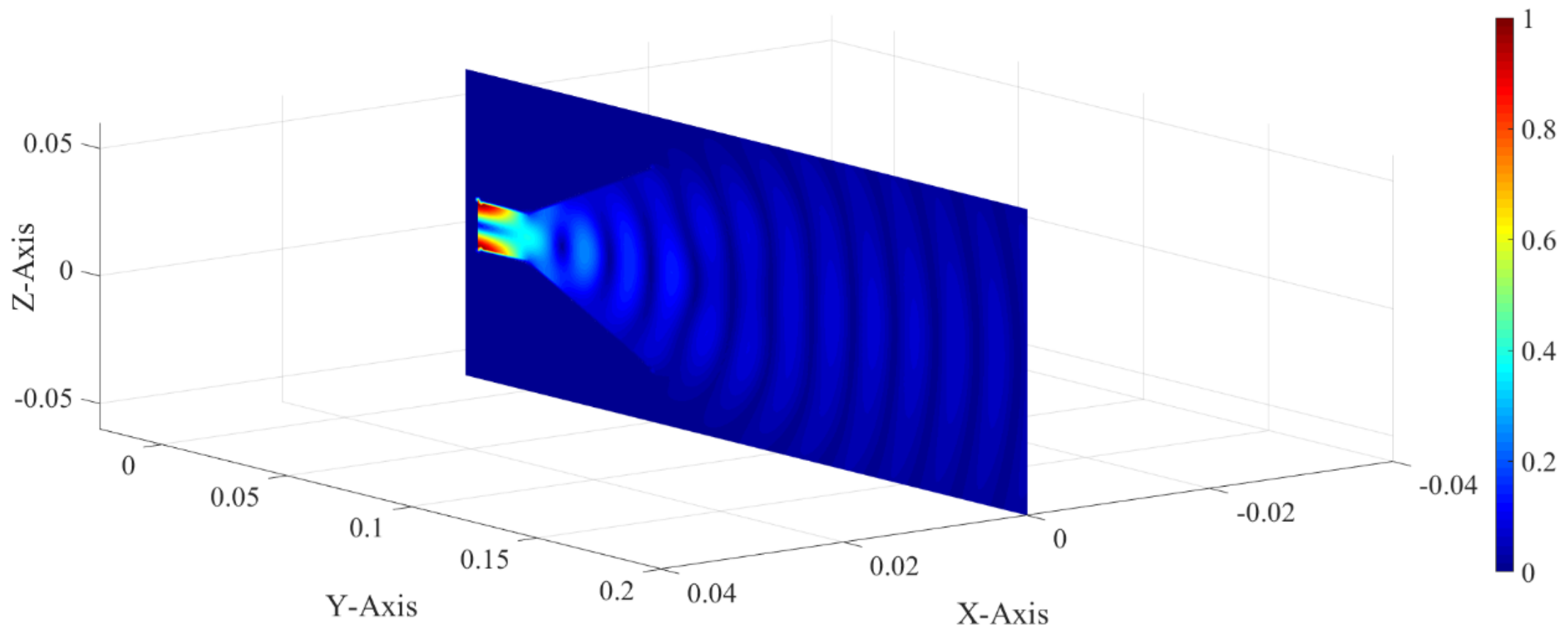

(d)

Figure 3-7 (a) Wall electric current $\boldsymbol{J}_{\rm e}$, (b) far-field radiation pattern, (c) electric field magnitude $|\boldsymbol{\mathcal{E}}|$, and (d) magnetic field magnitude $|\boldsymbol{\mathcal{H}}|$ of the resonant DM 1 (working at 9.3 GHz).

Because the DM 1 and DM 2 are spatially degenerate as illustrated by their modal port electric and magnetic currents shown in Fig. 3-6, then the wall electric current distribution, far-field radiation pattern, electric field magnitude distribution, and magnetic field magnitude distribution of the second degenerate state are completely similar to the figures shown in Fig. 3-7 (except a spatial rotation around Y-axis). Thus, we don't explicitly provide the distributions here.

In addition, as exhibited in Refs. [27-Secs.6.2.5.2&6.2.5.3] and [35], the above formulations can also be directly utilized to do the modal analysis for the aperture-fed parabolic reflector antenna shown in Fig. 3-8(a) and the horn-fed parabolic reflector antenna shown in Fig. 3-8(b).

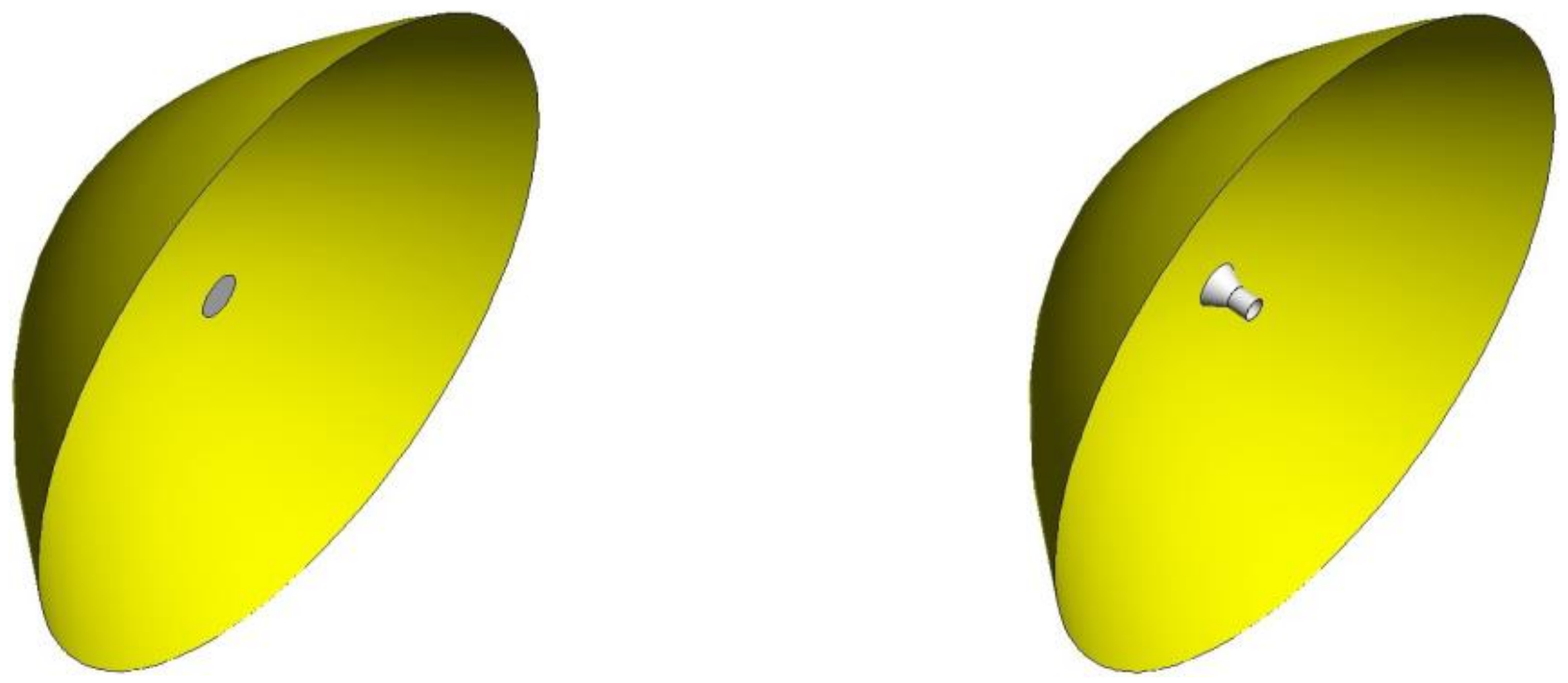
(a) (b)

Figure 3-8 Geometries of (a) aperture-fed and (b) horn-fed parabolic reflector antennas.

As exhibited in Refs. [27-Sec.6.3] and [35], the above metallic-antenna-oriented PTT-DMT can be easily generalized to wave-port-fed material transmitting antennas, and two typical material antennas are shown in the following Fig. 3-9.

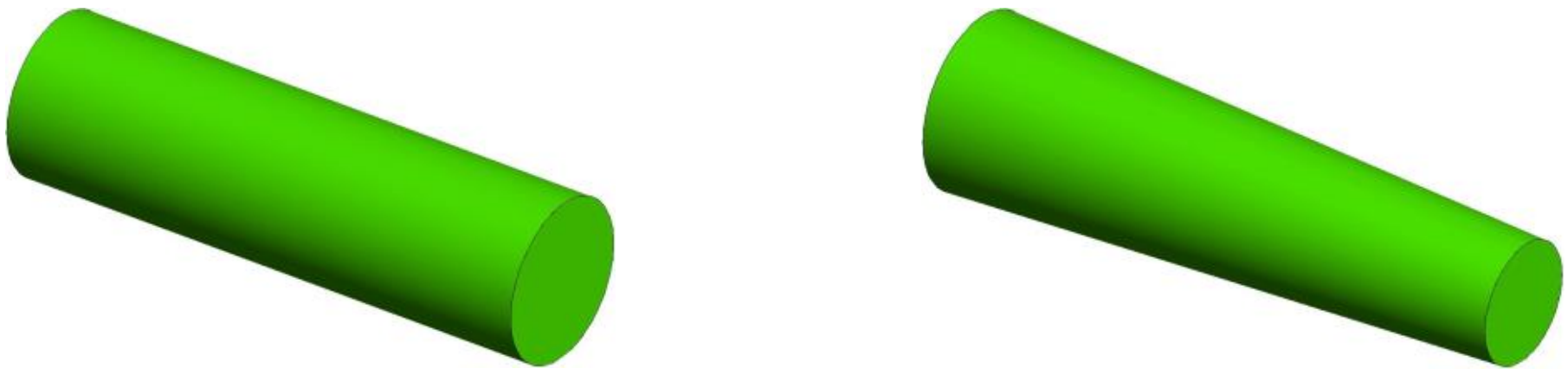
(a) (b)

Figure 3-9 Geometries of (a) cylindrical and (b) conical dielectric rod antennas.

As exhibited in Ref. [27-Secs.6.4~6.6], the PTT-DMT for wave-port-fed metallic and material transmitting antennas can be further generalized to wave-port-fed metal-material composite transmitting antennas, such as the coaxial-fed dielectric resonator antenna mounted on metallic ground plane shown in Fig. 3-10(a) and the metallic-horn-fed dielectric lens antenna shown in Fig. 3-10(b).

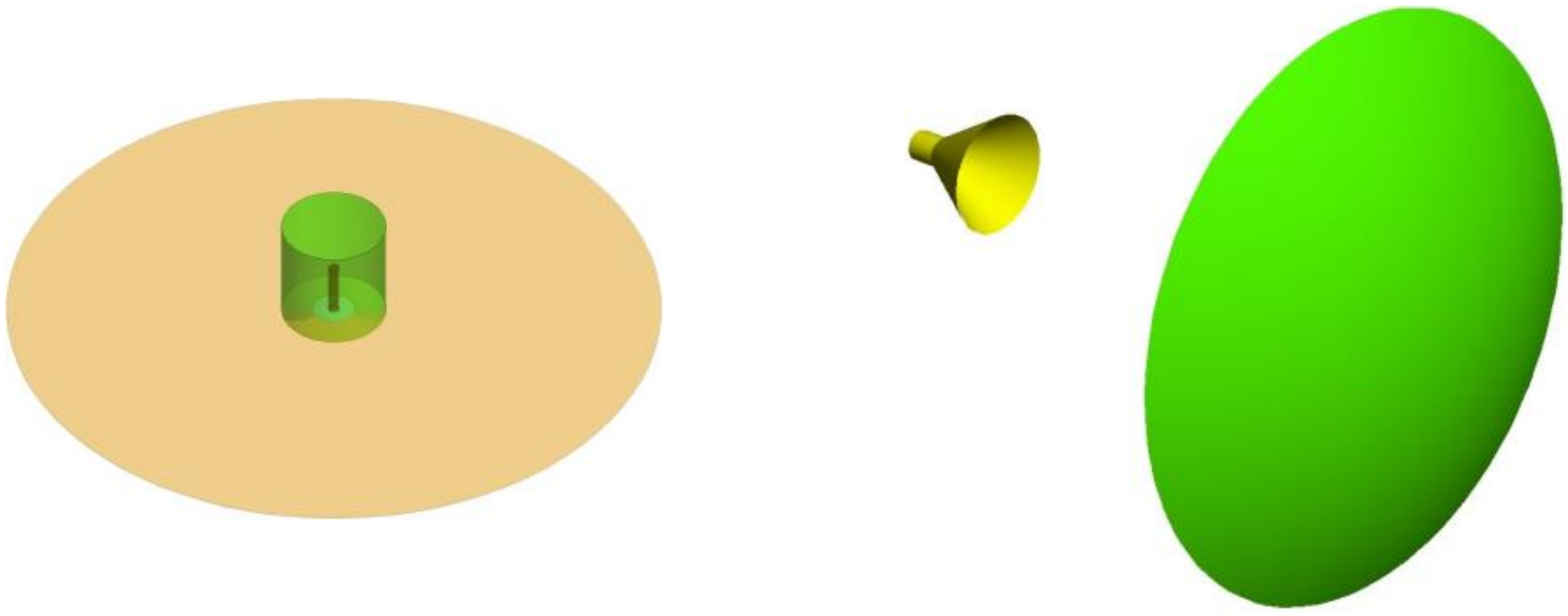
(a) (b)

Figure 3-10 Geometries of (a) coaxial-fed dielectric resonator antenna mounted on metallic ground plane and (b) metallic-horn-fed dielectric lens antenna.

In addition, taking the horn array shown in Fig. 3-11 as a typical example, Ref. [27-Sec.6.7] also generalized the PTT-DMT for single transmitting antenna to transmitting antenna array.

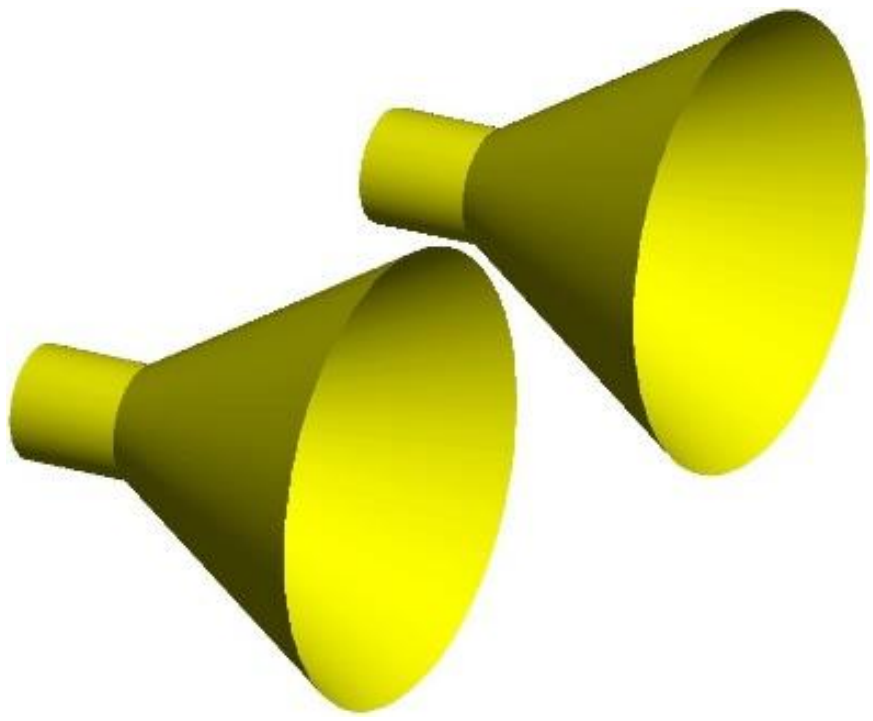

Figure 3-11 Geometry of a transmitting antenna array constituted by two metallic horns.

## 3.3 PTT-Based DMs of Wave-Port-Fed Receiving Antennas

This section is devoted to generalizing the above PTT-DMT from transmitting antennas to receiving antennas, by focusing on a typical metallic receiving horn shown in the following Fig. 3-12.

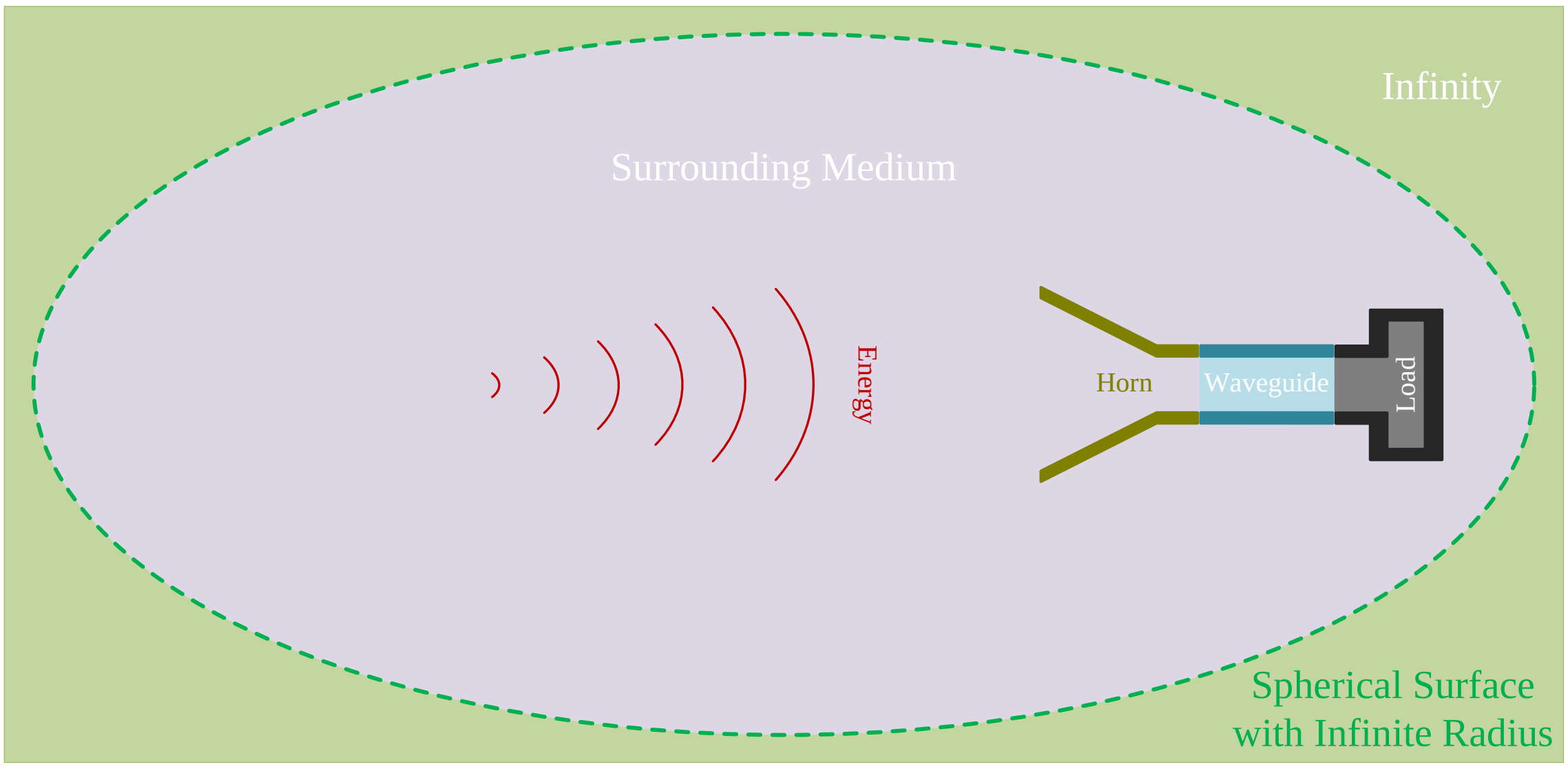

Figure 3-12 Geometry of the receiving problem considered in Sec. 3.3.

The horn is loaded by a metallic loading waveguide, and fed by an arbitrary power source from surrounding medium.

The topological structure corresponding to the above receiving problem is illustrated in the following Fig. 3-13.

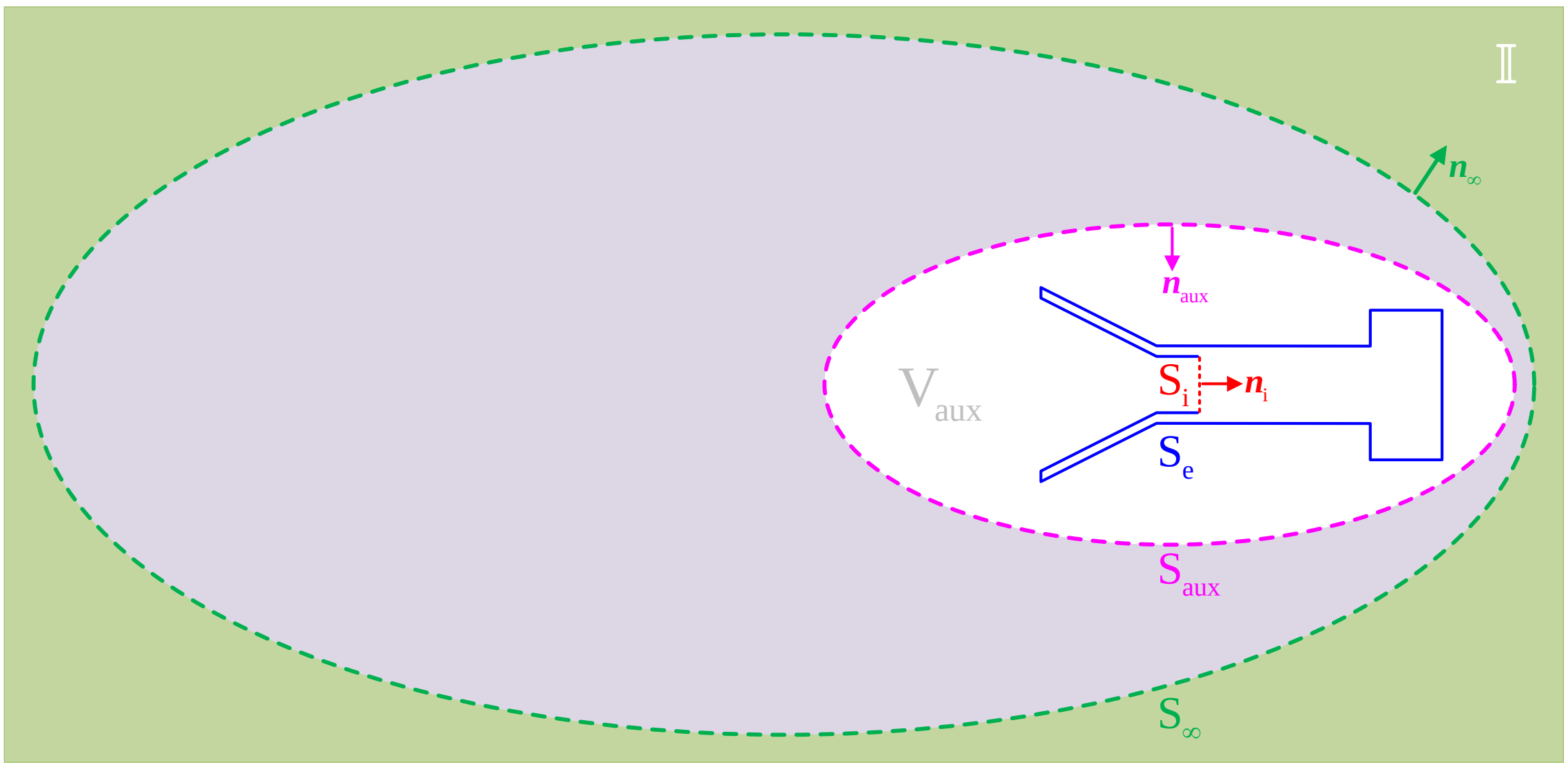

Figure 3-13 Topological structure of the receiving problem shown in Fig. 3-12, where $\mathrm{S}_{\mathrm{aux}}$ is an auxiliary closed surface enclosing whole receiving system.

In the above Fig. 3-13, $\mathrm{S}_{\mathrm{aux}}$ is an **arbitrary** penetrable auxiliary surface enclosing whole receiving system (horn + waveguide + load), and it is employed to model the power source used to feed the receiving horn. The region sandwiched between the auxiliary surface and receiving system is denoted as $\mathrm{V}_{\mathrm{aux}}$, and there doesn't exist any source distributing in $\mathrm{V}_{\mathrm{aux}}$, i.e., the permeability and permittivity of $\mathrm{V}_{\mathrm{aux}}$ are $\mu_0$ and $\varepsilon_0$. The electric wall used to separate receiving system from surrounding environment is denoted as $\mathrm{S}_{\mathrm{e}}$. The penetrable inferface between surrounding environment and horn is just the input port of the receiving antenna, and it is denoted as $\mathrm{S}_{\mathrm{i}}$. Clearly, for the specific receiving problem shown in Figs. 3-12 and 3-13, the penetrable surface $\mathrm{S}_{\mathrm{i}}$ is also "the output port of receiving antenna" / "the input port of loading waveguide" (but, this is not a general conclusion for other kinds of receiving problems, such as the one considered in Ref. [27-Sec.7.4]). The normal directions of $\mathrm{S}_{\mathrm{aux}}$ and $\mathrm{S}_{\mathrm{i}}$ are denoted as $\boldsymbol{n}_{\mathrm{aux}}$ and $\boldsymbol{n}_{\mathrm{i}}$, and they are defined as Fig. 3-13.

If the equivalent surface electric and magnetic currents on $\mathrm{S}_{\mathrm{aux/i}}$ are defined as $\boldsymbol{J}_{\mathrm{aux/i}} = \boldsymbol{n}_{\mathrm{aux/i}} \times \boldsymbol{H}$ and $\boldsymbol{M}_{\mathrm{aux/i}} = \boldsymbol{E} \times \boldsymbol{n}_{\mathrm{aux/i}}$ respectively, and the induced surface electric current on $\mathrm{S}_{\mathrm{e}}$ is denoted as $\boldsymbol{J}_{\mathrm{e}}$, then the input power $P_{\mathrm{in}} = (1/2)\iint_{\mathrm{S}_{\mathrm{i}}} (\boldsymbol{E} \times \boldsymbol{H}^{\dagger}) \cdot \boldsymbol{n}_{\mathrm{i}} dS$ as the power source used to feed the horn can be expressed in terms of the following IPO[27-Sec.7.3]

$$
\begin{aligned}
P_{\mathrm{in}} &= -(1/2)\left\langle \boldsymbol{J}_{\mathrm{i}}, \mathcal{E}_0\left(\boldsymbol{J}_{\mathrm{aux}} + \boldsymbol{J}_{\mathrm{e}} - \boldsymbol{J}_{\mathrm{i}}, \boldsymbol{M}_{\mathrm{aux}} - \boldsymbol{M}_{\mathrm{i}}\right)\right\rangle_{\mathrm{S}_{\mathrm{i}}^-} \\
&= -(1/2)\left\langle \boldsymbol{M}_{\mathrm{i}}, \mathcal{H}_0\left(\boldsymbol{J}_{\mathrm{aux}} + \boldsymbol{J}_{\mathrm{e}} - \boldsymbol{J}_{\mathrm{i}}, \boldsymbol{M}_{\mathrm{aux}} - \boldsymbol{M}_{\mathrm{i}}\right)\right\rangle_{\mathrm{S}_{\mathrm{i}}^-}^{\dagger}
\end{aligned}
\qquad (3\text{-}13)
$$

where $\mathrm{S}_{\mathrm{i}}^{-}$ is the left-side surface of $\mathrm{S}_{\mathrm{i}}$, and the operators $\mathcal{E}_0$ and $\mathcal{H}_0$ are defined as that $\mathcal{E}_0(\boldsymbol{J},\boldsymbol{M}) = -j\omega\mu_0\mathcal{L}_0(\boldsymbol{J}) - \mathcal{K}_0(\boldsymbol{M})$ and $\mathcal{H}_0(\boldsymbol{J},\boldsymbol{M}) = \mathcal{K}_0(\boldsymbol{J}) - j\omega\varepsilon_0\mathcal{L}_0(\boldsymbol{M})$. The basis function expansion for the currents involved in IPO (3-13) makes the integral operators be discretized into the following matrix operators

$$P_{\mathrm{in}} = \mathbb{J}_{\mathrm{i}}^{\dagger} \cdot \mathbb{P}_{\mathrm{JE}} \cdot \begin{bmatrix} \mathbb{J}_{\mathrm{aux}} \\ \mathbb{J}_{\mathrm{e}} \\ \mathbb{J}_{\mathrm{i}} \\ \mathbb{M}_{\mathrm{aux}} \\ \mathbb{M}_{\mathrm{i}} \end{bmatrix} = \begin{bmatrix} \mathbb{J}_{\mathrm{aux}} \\ \mathbb{J}_{\mathrm{e}} \\ \mathbb{J}_{\mathrm{i}} \\ \mathbb{M}_{\mathrm{aux}} \\ \mathbb{M}_{\mathrm{i}} \end{bmatrix}^{\dagger} \cdot \mathbb{P}_{\mathrm{HM}} \cdot \mathbb{M}_{\mathrm{i}} \tag{3-14}$$

where $\mathbb{J}_{\mathrm{aux}}$, $\mathbb{J}_{\mathrm{e}}$, $\mathbb{J}_{\mathrm{i}}$, $\mathbb{M}_{\mathrm{aux}}$, and $\mathbb{M}_{\mathrm{i}}$ are the basis function expansion coefficient vectors of $\boldsymbol{J}_{\mathrm{aux}}$, $\boldsymbol{J}_{\mathrm{e}}$, $\boldsymbol{J}_{\mathrm{i}}$, $\boldsymbol{M}_{\mathrm{aux}}$, and $\boldsymbol{M}_{\mathrm{i}}$ respectively.

The currents appeared in the above IPO are not independent, because they satisfy the following integral equations

$$\left[\mathcal{H}_0\left(\boldsymbol{J}_{\mathrm{aux}} + \boldsymbol{J}_{\mathrm{e}} - \boldsymbol{J}_{\mathrm{i}}, \boldsymbol{M}_{\mathrm{aux}} - \boldsymbol{M}_{\mathrm{i}}\right)\right]_{\mathrm{S}_{\mathrm{aux}}^{-}}^{\tan} = \boldsymbol{J}_{\mathrm{aux}} \times \boldsymbol{n}_{\mathrm{aux}} \tag{3-15a}$$

$$\left[\mathcal{E}_0\left(\boldsymbol{J}_{\mathrm{aux}} + \boldsymbol{J}_{\mathrm{e}} - \boldsymbol{J}_{\mathrm{i}}, \boldsymbol{M}_{\mathrm{aux}} - \boldsymbol{M}_{\mathrm{i}}\right)\right]_{\mathrm{S}_{\mathrm{aux}}^{-}}^{\tan} = \boldsymbol{n}_{\mathrm{aux}} \times \boldsymbol{M}_{\mathrm{aux}} \tag{3-15b}$$

$$\left[\mathcal{E}_0\left(\boldsymbol{J}_{\mathrm{aux}} + \boldsymbol{J}_{\mathrm{e}} - \boldsymbol{J}_{\mathrm{i}}, \boldsymbol{M}_{\mathrm{aux}} - \boldsymbol{M}_{\mathrm{i}}\right)\right]_{\mathrm{S}_{\mathrm{e}}}^{\tan} = 0 \tag{3-16}$$

$$\left[\mathcal{E}_0\left(\boldsymbol{J}_{\mathrm{aux}} + \boldsymbol{J}_{\mathrm{e}} - \boldsymbol{J}_{\mathrm{i}}, \boldsymbol{M}_{\mathrm{aux}} - \boldsymbol{M}_{\mathrm{i}}\right)\right]_{\mathrm{S}_{\mathrm{i}}^{-}}^{\tan} = \left[\mathcal{E}_0\left(\boldsymbol{J}_{\mathrm{i}}, \boldsymbol{M}_{\mathrm{i}}\right)\right]_{\mathrm{S}_{\mathrm{i}}^{+}}^{\tan} \tag{3-17a}$$

$$\left[\mathcal{H}_0\left(\boldsymbol{J}_{\mathrm{aux}} + \boldsymbol{J}_{\mathrm{e}} - \boldsymbol{J}_{\mathrm{i}}, \boldsymbol{M}_{\mathrm{aux}} - \boldsymbol{M}_{\mathrm{i}}\right)\right]_{\mathrm{S}_{\mathrm{i}}^{-}}^{\tan} = \left[\mathcal{H}_0\left(\boldsymbol{J}_{\mathrm{i}}, \boldsymbol{M}_{\mathrm{i}}\right)\right]_{\mathrm{S}_{\mathrm{i}}^{+}}^{\tan} \tag{3-17b}$$

Here, Eqs. (3-15a) and (3-15b) are because of the definitions of $\boldsymbol{J}_{\mathrm{aux}}$ (DoJ) and $\boldsymbol{M}_{\mathrm{aux}}$ (DoM) where $\mathrm{S}_{\mathrm{aux}}^{-}$ is the inner surface of $\mathrm{S}_{\mathrm{aux}}$, and Eq. (3-16) is due to the homogeneous tangential electric field boundary condition on $\mathrm{S}_{\mathrm{e}}$, and Eqs. (3-17a) and (3-17b) are originated from the perfectly matching condition of $\boldsymbol{E}$ and $\boldsymbol{H}$ used on $\mathrm{S}_{\mathrm{i}}$ where $\mathrm{S}_{\mathrm{i}}^{-}$ and $\mathrm{S}_{\mathrm{i}}^{+}$ are the left-side and right-side surfaces of $\mathrm{S}_{\mathrm{i}}$ (in fact, **the perfectly matching condition can be viewed as a counterpart of the famous Sommerfeld's radiation condition at infinity**, and the theoretical foundation supporting us to utilize the perfectly matching condition is that we are now doing the modal analysis for the receiving horn[27-Chap.7]). Applying MoM to Eqs. (3-15a)~(3-17b), the integral equations are immediately transformed into some matrix equations, and the latters imply some different matrix transformations from the independent current into the other currents. Utilizing the matrix equations corresponding to integral equations (3-15a)/(3-15b), (3-16), (3-17a), and (3-17b), we obtain the following transformations

$$\mathbb{T}_{\text{DoJ}} \cdot \mathbb{J}_{\text{aux}} = \begin{bmatrix} \mathbb{J}_{\text{aux}} \\ \mathbb{J}_{\text{e}} \\ \mathbb{J}_{\text{i}} \\ \mathbb{M}_{\text{aux}} \\ \mathbb{M}_{\text{i}} \end{bmatrix} = \mathbb{T}_{\text{DoM}} \cdot \mathbb{M}_{\text{aux}} \quad \text{and} \quad \begin{cases} \mathbb{J}_{\text{i}} = \mathbb{t}_{\text{DoJ}} \cdot \mathbb{J}_{\text{aux}} \\ \mathbb{M}_{\text{i}} = \mathbb{t}_{\text{DoM}} \cdot \mathbb{M}_{\text{aux}} \end{cases} \tag{3-18}$$

from the independent current $\mathbb{J}_{\text{aux}} / \mathbb{M}_{\text{aux}}$ into the other currents.

Inserting the above matrix transformations into the previous matrix-formed IPOs (3-14), the following IPO[27-Sec.7.3]

$$P_{\text{in}} = \mathbb{C}_{\text{aux}}^{\dagger} \cdot \mathbb{P}_{\text{in}} \cdot \mathbb{C}_{\text{aux}} = \begin{cases} \mathbb{J}_{\text{aux}}^{\dagger} \cdot \overbrace{\mathbb{t}_{\text{DoJ}}^{\dagger} \cdot \mathbb{P}_{\text{JE}} \cdot \mathbb{T}_{\text{DoJ}}}^{\mathbb{P}_{\text{JE-DoJ}}} \cdot \mathbb{J}_{\text{aux}} \\ \mathbb{M}_{\text{aux}}^{\dagger} \cdot \underbrace{\mathbb{T}_{\text{DoM}}^{\dagger} \cdot \mathbb{P}_{\text{HM}} \cdot \mathbb{t}_{\text{DoM}}}_{\mathbb{P}_{\text{HM-DoM}}} \cdot \mathbb{M}_{\text{aux}} \end{cases} \tag{3-19}$$

with only independent current $\mathbb{C}_{\text{aux}}$ (where $\mathbb{C}_{\text{aux}} = \mathbb{J}_{\text{aux}} / \mathbb{M}_{\text{aux}}$) is obtained. **In the first paragraph of Ref. [27-Sec.7.3.5], it was pointed out that: $\mathbb{P}_{\text{in}}^{+}$ is usually not positive definite. Thus, the DMs of the receiving horn cannot be effectively derived from solving equation $\mathbb{P}_{\text{in}}^{-} \cdot \mathbb{C}_{\text{aux}} = \theta\, \mathbb{P}_{\text{in}}^{+} \cdot \mathbb{C}_{\text{aux}}$ [27-Sec.7.3.5], where $\mathbb{P}_{\text{in}}^{+} = (\mathbb{P}_{\text{in}} + \mathbb{P}_{\text{in}}^{\dagger}) / 2$ and $\mathbb{P}_{\text{in}}^{-} = (\mathbb{P}_{\text{in}} - \mathbb{P}_{\text{in}}^{\dagger}) / 2j$. As a compromise scheme, Ref. [27-Sec.7.3.5] proposed an alternatively auxiliary power $P_{\text{aux}} = (1/2) \oiint_{S_{\text{aux}}} (\boldsymbol{E} \times \boldsymbol{H}^{\dagger}) \cdot \boldsymbol{n}_{\text{aux}} dS$ used to calculate DMs for the receiving horn.** Obviously, $P_{\text{aux}}$ has the following integral operator and matrix operator forms

$$\begin{aligned} P_{\text{aux}} &= -\frac{1}{2} \left\langle \boldsymbol{J}_{\text{aux}}, \mathcal{E}_0 \left( \boldsymbol{J}_{\text{aux}} + \boldsymbol{J}_{\text{e}} - \boldsymbol{J}_{\text{i}}, \boldsymbol{M}_{\text{aux}} - \boldsymbol{M}_{\text{i}} \right) \right\rangle_{S_{\text{aux}}^{-}} = -\frac{1}{2} \left\langle \boldsymbol{M}_{\text{aux}}, \mathcal{H}_0 \left( \boldsymbol{J}_{\text{aux}} + \boldsymbol{J}_{\text{e}} - \boldsymbol{J}_{\text{i}}, \boldsymbol{M}_{\text{aux}} - \boldsymbol{M}_{\text{i}} \right) \right\rangle_{S_{\text{aux}}^{-}}^{\dagger} \\ &= \mathbb{J}_{\text{aux}}^{\dagger} \cdot \mathbb{P}_{\text{je}} \cdot \begin{bmatrix} \mathbb{J}_{\text{aux}} \\ \mathbb{J}_{\text{e}} \\ \mathbb{J}_{\text{i}} \\ \mathbb{M}_{\text{aux}} \\ \mathbb{M}_{\text{i}} \end{bmatrix} = \begin{bmatrix} \mathbb{J}_{\text{aux}} \\ \mathbb{J}_{\text{e}} \\ \mathbb{J}_{\text{i}} \\ \mathbb{M}_{\text{aux}} \\ \mathbb{M}_{\text{i}} \end{bmatrix}^{\dagger} \cdot \mathbb{P}_{\text{hm}} \cdot \mathbb{M}_{\text{aux}} \end{aligned} \tag{3–20}$$

where the $\mathbb{P}_{\text{je}}$ and $\mathbb{P}_{\text{hm}}$ are different from the $\mathbb{P}_{\text{JE}}$ and $\mathbb{P}_{\text{HM}}$ used in Eqs. (3-14) and (3-19). Substituting Eq. (3-18) into Eqs. (3-20), the following auxiliary power operator

$$P_{\text{aux}} = \mathbb{C}_{\text{aux}}^{\dagger} \cdot \mathbb{P}_{\text{aux}} \cdot \mathbb{C}_{\text{aux}} = \begin{cases} \mathbb{J}_{\text{aux}}^{\dagger} \cdot \overbrace{\mathbb{P}_{\text{je}} \cdot \mathbb{T}_{\text{DoJ}}}^{\mathbb{P}_{\text{je-DoJ}}} \cdot \mathbb{J}_{\text{aux}} \\ \mathbb{M}_{\text{aux}}^{\dagger} \cdot \underbrace{\mathbb{T}_{\text{DoM}}^{\dagger} \cdot \mathbb{P}_{\text{hm}}}_{\mathbb{P}_{\text{hm-DoM}}} \cdot \mathbb{M}_{\text{aux}} \end{cases} \tag{3-21}$$

with only independent current $\mathbb{C}_{\text{aux}}$ (where $\mathbb{C}_{\text{aux}} = \mathbb{J}_{\text{aux}} / \mathbb{M}_{\text{aux}}$) is obtained, where the $\mathbb{P}_{\text{je-DoJ}}$ and $\mathbb{P}_{\text{hm-DoM}}$ are different from the $\mathbb{P}_{\text{JE-DoJ}}$ and $\mathbb{P}_{\text{HM-DoM}}$ used in Eq. (3-19).

Employing the above auxiliary power matrix $\mathbb{P}_{\text{aux}}$, a series of modes can be derived from solving the following auxiliary modal decoupling equation

$$\mathbb{P}_{\text{aux}}^{-} \cdot \mathbb{C}_{\text{aux}} = \alpha \, \mathbb{P}_{\text{aux}}^{+} \cdot \mathbb{C}_{\text{aux}} \tag{3-22}$$

where $\mathbb{P}_{\text{aux}}^{+} = (\mathbb{P}_{\text{aux}} + \mathbb{P}_{\text{aux}}^{\dagger})/2$ and $\mathbb{P}_{\text{aux}}^{-} = (\mathbb{P}_{\text{aux}} - \mathbb{P}_{\text{aux}}^{\dagger})/2j$ are the positive and negative Hermitian parts of $\mathbb{P}_{\text{aux}}$. Obviousy, the modes derived from Eq. (3-22) satisfy power-decoupling relation

$$(1/2)\oint\!\!\!\oint_{\text{S}_{\text{aux}}} \left(\boldsymbol{E}_n \times \boldsymbol{H}_m^{\dagger}\right) \cdot \boldsymbol{n}_{\text{aux}} dS = \left(1 + j\alpha_m\right)\delta_{mn} \tag{3-23}$$

and then the following energy-decoupling relation

$$(1/T)\int_{t_0}^{t_0+T} \left[ \oint\!\!\!\oint_{\text{S}_{\text{aux}}} \left(\boldsymbol{\mathcal{E}}_n \times \boldsymbol{\mathcal{H}}_m + \boldsymbol{\mathcal{E}}_m \times \boldsymbol{\mathcal{H}}_n\right) \cdot \boldsymbol{n}_{\text{aux}} dS \right] dt = 2\delta_{mn} \tag{3-24}$$

Employing time-domain LORENTZ'S RECIPROCITY THEOREM (2-22), the above relation (3-24) immediately leads to the following energy-decoupling relation

$$(1/T)\int_{t_0}^{t_0+T} \left[ \iint_{\text{S}_{\text{i}}} \left(\boldsymbol{\mathcal{E}}_n \times \boldsymbol{\mathcal{H}}_m + \boldsymbol{\mathcal{E}}_m \times \boldsymbol{\mathcal{H}}_n\right) \cdot \boldsymbol{n}_{\text{i}} dS \right] dt = 2\delta_{mn} \tag{3-25}$$

on input port $\text{S}_{\text{i}}$, and the proof process for "Eq. (3-24) → Eq. (3-25)" doesn't depend on the specific choice of auxiliary surface $\text{S}_{\text{aux}}$. **Thus, the modes derived from Eq. (3-22) are indeed energy-decoupled on the input port $\text{S}_{\text{i}}$ of the receiving horn.**

Following the convention of Sec. 3.2 (for transmitting antennas), the modal input "impedance $Z_{\text{in}}$, resistance $R_{\text{in}}$, reactance $X_{\text{in}}$" and "admittance $Y_{\text{in}}$, conductance $G_{\text{in}}$, susceptance $B_{\text{in}}$" of the receiving antenna shown in Figs. 3-12 and 3-13 can be defined as the ones given in Eqs. (3-12a) and (3-12b). Here, we propose another physical quantity "modal power/energy transport coefficient (TC) from $\text{S}_{\text{aux}}$ to $\text{S}_{\text{i}}$" as follows:

$$\begin{aligned}
\text{TC}_{\text{S}_{\text{aux}} \to \text{S}_{\text{i}}} &= \frac{(1/T)\int_{t_0}^{t_0+T} \left[ \iint_{\text{S}_{\text{i}}} \left(\boldsymbol{\mathcal{E}} \times \boldsymbol{\mathcal{H}}\right) \cdot \boldsymbol{n}_{\text{i}} dS \right] dt}{(1/T)\int_{t_0}^{t_0+T} \left[ \oint\!\!\!\oint_{\text{S}_{\text{aux}}} \left(\boldsymbol{\mathcal{E}} \times \boldsymbol{\mathcal{H}}\right) \cdot \boldsymbol{n}_{\text{aux}} dS \right] dt} \\
&= \frac{\text{Re}\left\{(1/2)\iint_{\text{S}_{\text{i}}} \left(\boldsymbol{E} \times \boldsymbol{H}^{\dagger}\right) \cdot \boldsymbol{n}_{\text{i}} dS\right\}}{\text{Re}\left\{(1/2)\oint\!\!\!\oint_{\text{S}_{\text{aux}}} \left(\boldsymbol{E} \times \boldsymbol{H}^{\dagger}\right) \cdot \boldsymbol{n}_{\text{aux}} dS\right\}} \\
&= \frac{\mathbb{C}_{\text{aux}} \cdot \mathbb{P}_{\text{in}}^{+} \cdot \mathbb{C}_{\text{aux}}}{\mathbb{C}_{\text{aux}} \cdot \mathbb{P}_{\text{aux}}^{+} \cdot \mathbb{C}_{\text{aux}}}
\end{aligned} \tag{3-26}$$

In the above Eq. (3-26), the second equality is due to the time-harmonic property of the fields; the third equality is because of Eqs. (3-19) and (3-21).

Here, we consider a concrete example — a metallic receiving horn fed by an auxiliary spherical surface. The geometries of the metallic horn and auxiliary feeding surface are shown in Fig. 3-14.

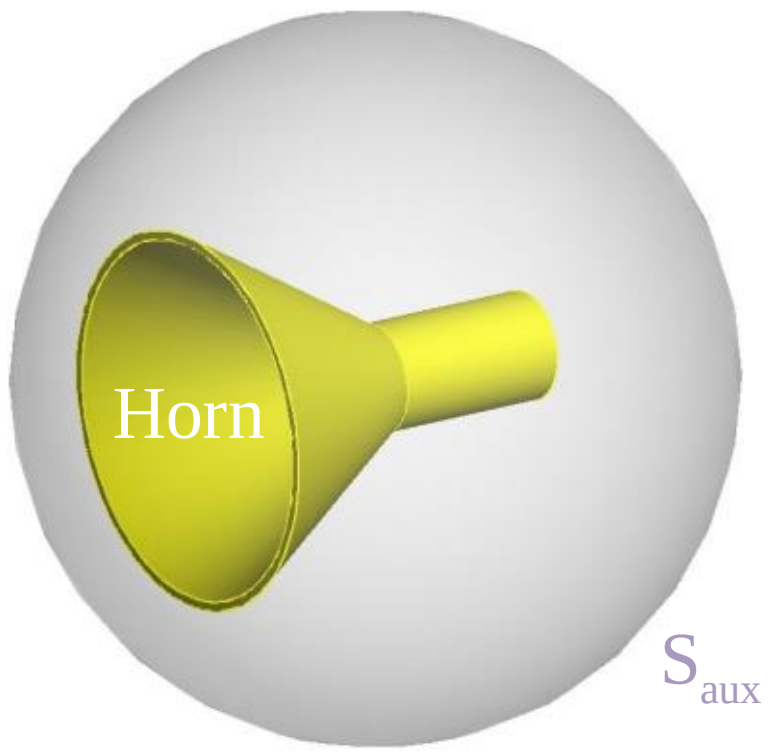

Figure 3-14 Geometry of a metallic receiving horn fed by an auxiliary spherical surface, where the spherical surface encloses whole horn.

The metallic receiving horn antenna is the one with a 1/4 size of the metallic transmitting horn antenna considered in the previous Sec. 3.2. The auxiliary feeding surface is with radius 2 cm.

By orthogonalizing the je-DoJ and hm-DoM based formulations of $P_{aux}$, we derive the DMs of the horn, and plot the associated modal resistance $R_{aux}$ curves in following Fig. 3-15.

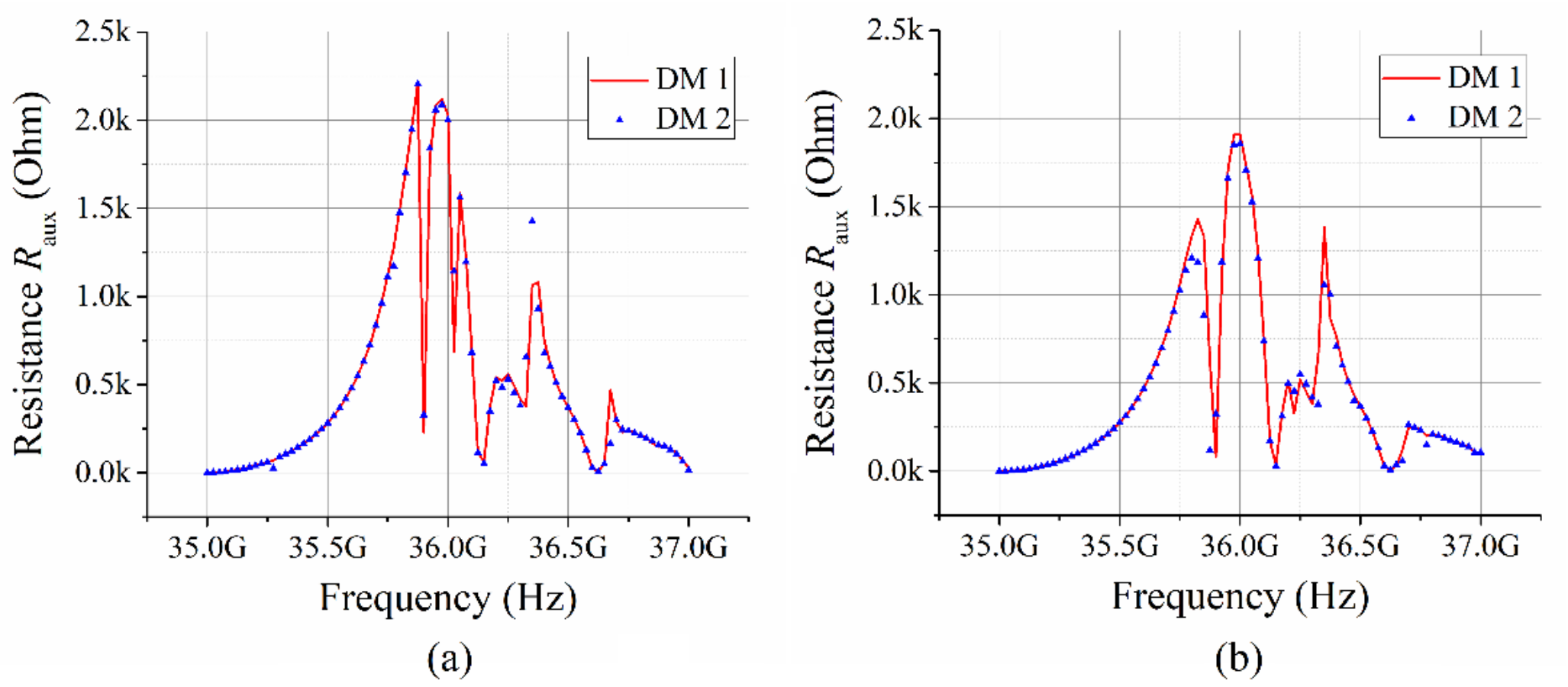

Figure 3-15 The first several lower-order modal resistance $R_{aux}$ curves calculated from the (a) je-DoJ-based and (b) hm-DoM-based formulations of $P_{aux}$.

Taking the DM 1 shown in Fig. 3-15(a) as an example, its modal electric and magnetic currents distributing on antenna input port $S_i$ are illustrated in the following Fig. 3-16.

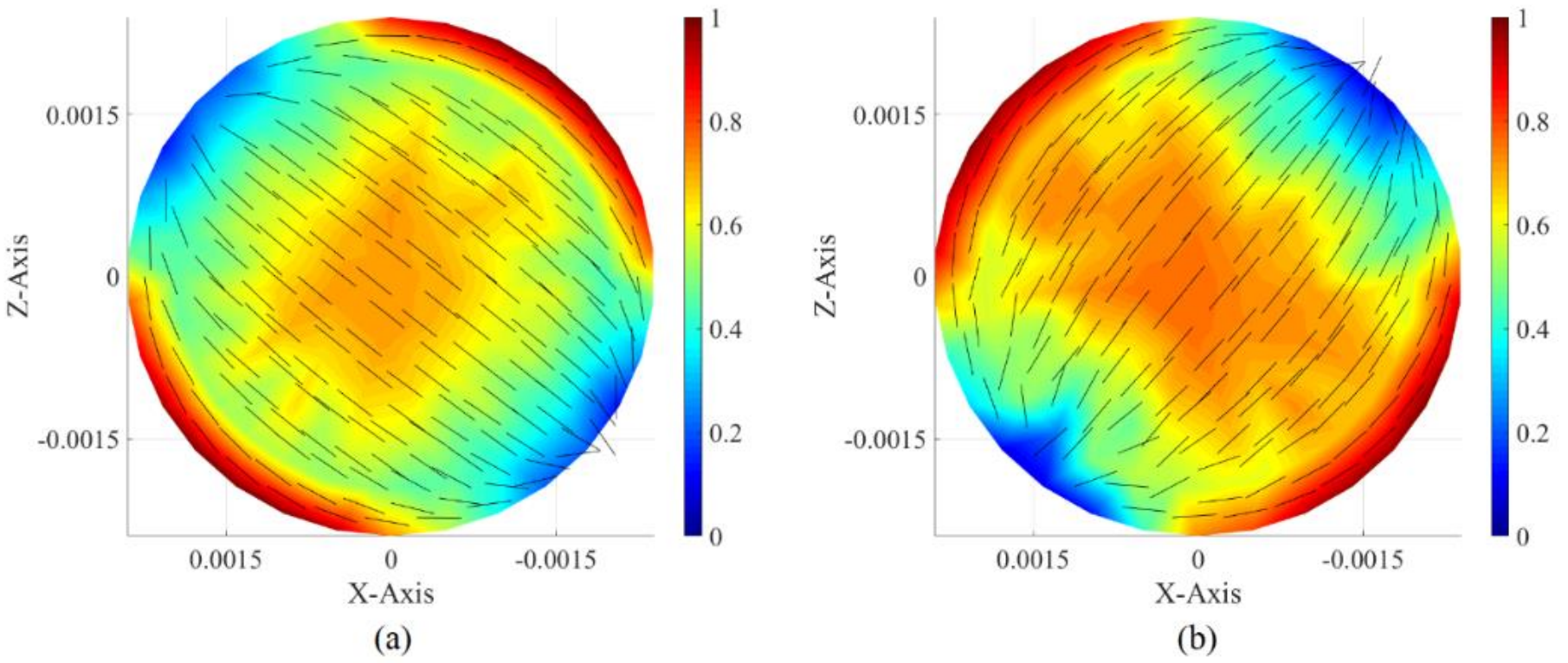

Figure 3-16 Modal (a) electric and (b) magnetic currents on antenna input port $S_i$ .

For the above-mentioned mode working at 35.875 GHz and 35.975 GHz, we also plot their modal electric fields with a series of time points $t = \{ 0.10T, 0.20T, 0.30T, 0.40T, 0.50T \}$ in the following Fig. 3-17, where $T$ is the time period of the time-harmonic field.

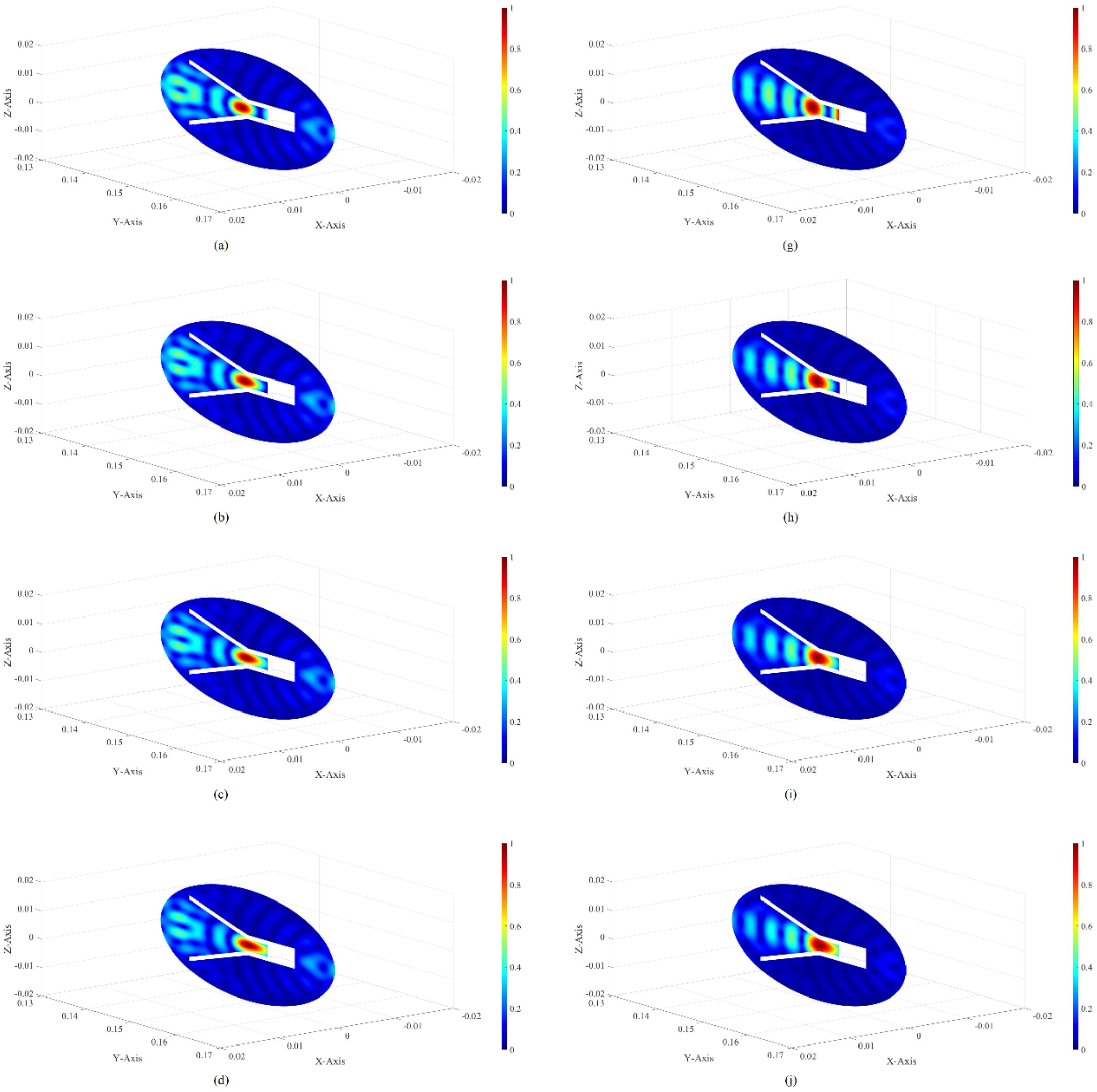

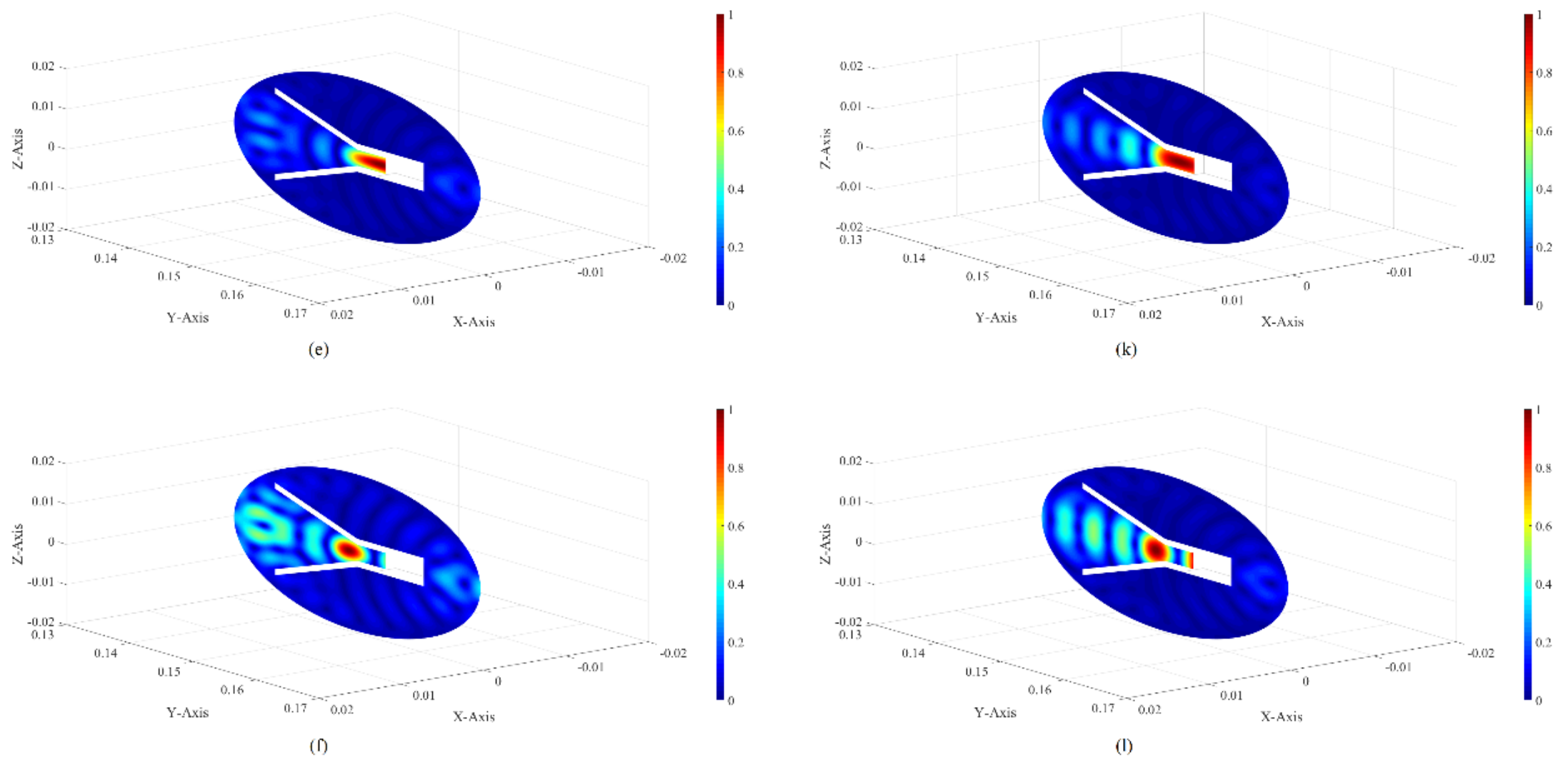

Figure 3-17 Modal electric field working at 35.875 GHz with a series of time points (a) 0.0$T$, (b) 0.1$T$, (c) 0.2$T$, (d) 0.3$T$, (e) 0.4$T$, and (f) 0.5$T$. Modal electric field working at 35.975 GHz with a series of time points (g) 0.0$T$, (h) 0.1$T$, (i) 0.2$T$, (j) 0.3$T$, (k) 0.4$T$, and (l) 0.5$T$.

In fact, as exhibited in Ref. [27-Sec.7.4], the above formulations can be further generalized to some more complicated receiving antennas, such as the one shown in the following Fig. 3-18.

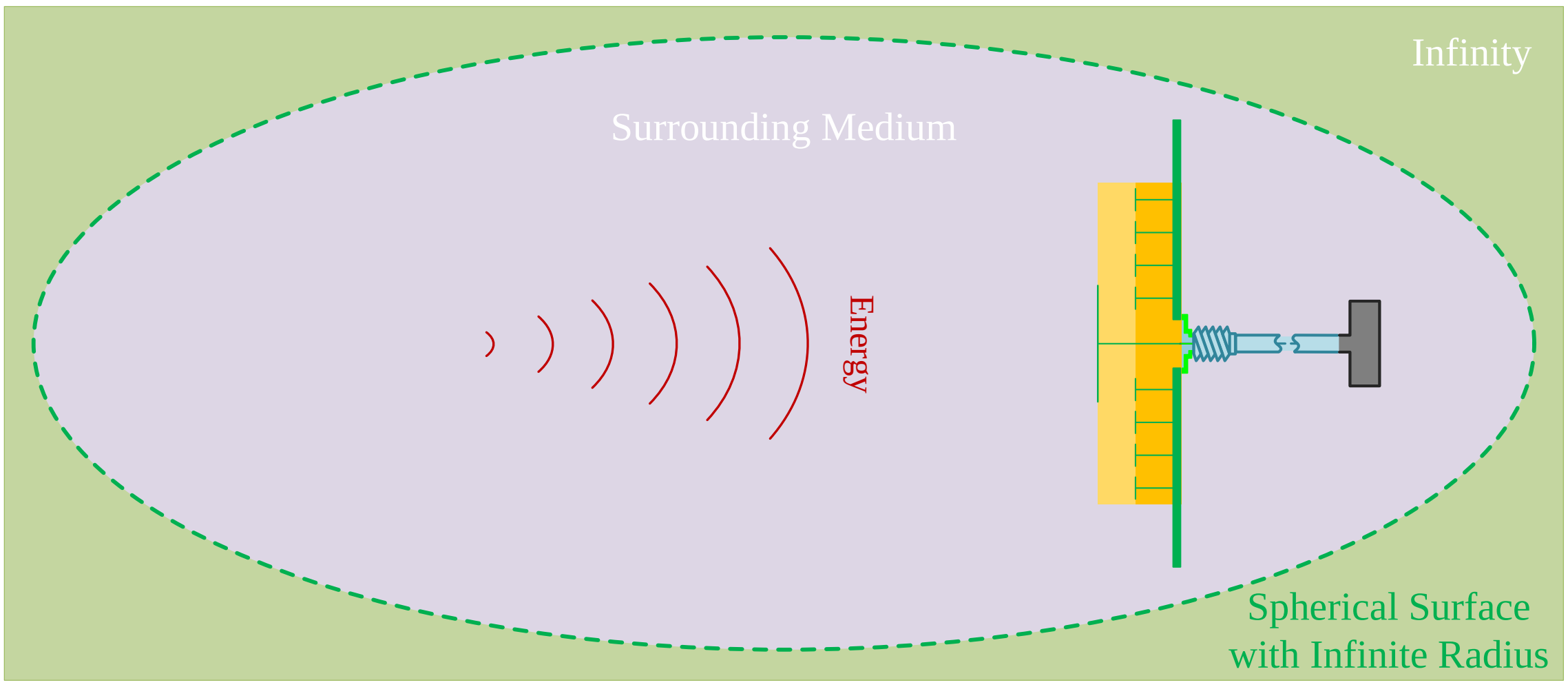

Figure 3-18 Geometry of a coaxial-loaded meta-surface receiving antenna.

## 3.4 PTT-Based DMs of Wave-Port-Fed Wave-Guiding Structures

Taking the metallic tube waveguide shown in the following Fig. 3-19 as a typical example, this section establishes PTT-DMT and constructs DMs for wave-port-fed wave-guiding structures.

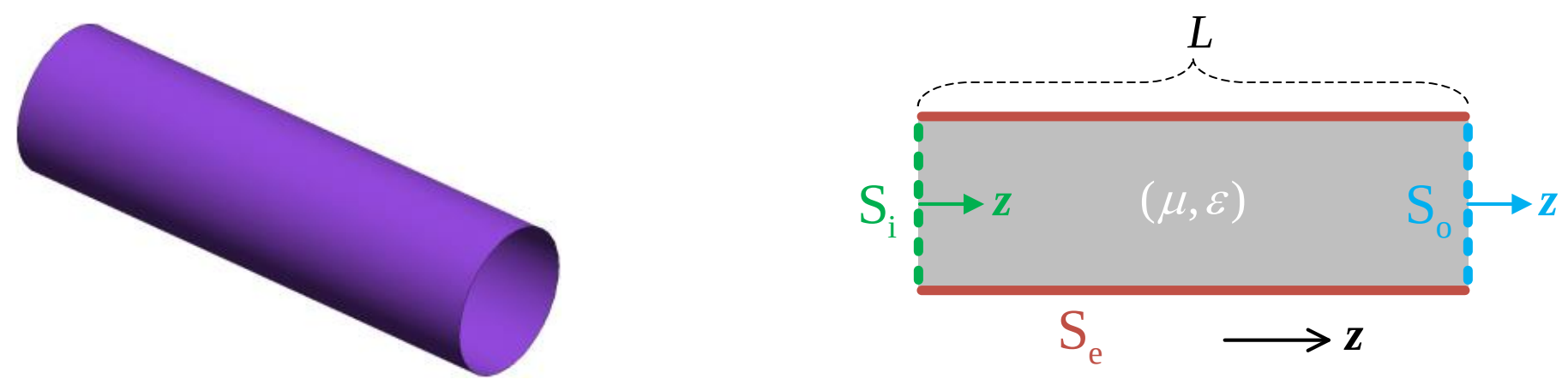

(a) (b)

Figure 3-19 (a) Geometry and (b) topology of a metallic tube waveguide.

The waveguide is filled by homogeneous isotropic lossless material with parameters $(\mu,\varepsilon)$. The longitudinal direction of the waveguide is selected as Z-axis with unit vector $\boldsymbol{z}$, and its lateral direction has an arbitrary cross section. In the figure, only the part (with longitudinal length $L$) sandwiched between two cross sections $\mathrm{S_i}$ and $\mathrm{S_o}$ is illustrated, and the electric wall of the part is denoted as $\mathrm{S_e}$.

As explained in the previous Sec. 2.4, input power $P_{\mathrm{in}} = (1/2)\iint_{\mathrm{S_i}} (\boldsymbol{E} \times \boldsymbol{H}^{\dagger}) \cdot \boldsymbol{z} dS$ is just the source to sustain a steady power transportation of the waveguide section[27-Chaps.2&3]. If the equivalent electric and magnetic currents on $\mathrm{S_{i/o}}$ are defined as $\boldsymbol{J}_{\mathrm{i/o}} = \boldsymbol{z} \times \boldsymbol{H}$ and $\boldsymbol{M}_{\mathrm{i/o}} = \boldsymbol{E} \times \boldsymbol{z}$ respectively, and the induced current on $\mathrm{S_e}$ is denoted as $\boldsymbol{J}_{\mathrm{e}}$, then the input power can be rewritten as the following JE-formed and HM-formed operator expressions[27-Sec.3.2],[39]

$$
\begin{aligned}
P_{\mathrm{in}} &= -(1/2)\left\langle \boldsymbol{J}_{\mathrm{i}}, \mathcal{E}\left(\boldsymbol{J}_{\mathrm{i}} + \boldsymbol{J}_{\mathrm{e}} - \boldsymbol{J}_{\mathrm{o}}, \boldsymbol{M}_{\mathrm{i}} - \boldsymbol{M}_{\mathrm{o}}\right)\right\rangle_{\mathrm{S_i^+}} \\
&= -(1/2)\left\langle \boldsymbol{M}_{\mathrm{i}}, \mathcal{H}\left(\boldsymbol{J}_{\mathrm{i}} + \boldsymbol{J}_{\mathrm{e}} - \boldsymbol{J}_{\mathrm{o}}, \boldsymbol{M}_{\mathrm{i}} - \boldsymbol{M}_{\mathrm{o}}\right)\right\rangle_{\mathrm{S_i^+}}^{\dagger}
\end{aligned}
\tag{3-27}
$$

where $\mathrm{S_i^+}$ is the right-side surface of $\mathrm{S_i}$, and the operators $\mathcal{E}$ and $\mathcal{H}$ are defined as that $\mathcal{E}(\boldsymbol{J},\boldsymbol{M}) = -j\omega\mu\mathcal{L}(\boldsymbol{J}) - \mathcal{K}(\boldsymbol{M})$ and $\mathcal{H}(\boldsymbol{J},\boldsymbol{M}) = \mathcal{K}(\boldsymbol{J}) - j\omega\varepsilon\mathcal{L}(\boldsymbol{M})$.

By expanding the currents in terms of some proper basis functions, the integral-operator-formed IPOs (3-27) can be discretized into the following matrix operator forms

$$
P_{\mathrm{in}} = \mathbb{J}_{\mathrm{i}}^{\dagger} \cdot \mathbb{P}_{\mathrm{JE}} \cdot \begin{bmatrix} \mathbb{J}_{\mathrm{i}} \\ \mathbb{J}_{\mathrm{e}} \\ \mathbb{J}_{\mathrm{o}} \\ \mathbb{M}_{\mathrm{i}} \\ \mathbb{M}_{\mathrm{o}} \end{bmatrix} = \begin{bmatrix} \mathbb{J}_{\mathrm{i}} \\ \mathbb{J}_{\mathrm{e}} \\ \mathbb{J}_{\mathrm{o}} \\ \mathbb{M}_{\mathrm{i}} \\ \mathbb{M}_{\mathrm{o}} \end{bmatrix}^{\dagger} \cdot \mathbb{P}_{\mathrm{HM}} \cdot \mathbb{M}_{\mathrm{i}}
\tag{3-28}
$$

where $\mathbb{J}_{\mathrm{i}}$, $\mathbb{J}_{\mathrm{e}}$, $\mathbb{J}_{\mathrm{o}}$, $\mathbb{M}_{\mathrm{i}}$, and $\mathbb{M}_{\mathrm{o}}$ are the basis function expansion coefficient vectors of $\boldsymbol{J}_{\mathrm{i}}$, $\boldsymbol{J}_{\mathrm{e}}$, $\boldsymbol{J}_{\mathrm{o}}$, $\boldsymbol{M}_{\mathrm{i}}$, and $\boldsymbol{M}_{\mathrm{o}}$ respectively.

In fact, the currents mentioned above are not independent of each other, because they satisfy the following integral equations

$$\left[\mathcal{H}\left(\boldsymbol{J}_{\mathrm{i}}+\boldsymbol{J}_{\mathrm{e}}-\boldsymbol{J}_{\mathrm{o}},\boldsymbol{M}_{\mathrm{i}}-\boldsymbol{M}_{\mathrm{o}}\right)\right]_{\mathrm{S}_{\mathrm{i}}^{+}}^{\tan} = \boldsymbol{J}_{\mathrm{i}}\times\boldsymbol{z} \tag{3-29a}$$

$$\left[\mathcal{E}\left(\boldsymbol{J}_{\mathrm{i}}+\boldsymbol{J}_{\mathrm{e}}-\boldsymbol{J}_{\mathrm{o}},\boldsymbol{M}_{\mathrm{i}}-\boldsymbol{M}_{\mathrm{o}}\right)\right]_{\mathrm{S}_{\mathrm{i}}^{+}}^{\tan} = \boldsymbol{z}\times\boldsymbol{M}_{\mathrm{i}} \tag{3-29b}$$

$$\left[\mathcal{E}\left(\boldsymbol{J}_{\mathrm{i}}+\boldsymbol{J}_{\mathrm{e}}-\boldsymbol{J}_{\mathrm{o}},\boldsymbol{M}_{\mathrm{i}}-\boldsymbol{M}_{\mathrm{o}}\right)\right]_{\mathrm{S}_{\mathrm{e}}}^{\tan} = 0 \tag{3-30}$$

$$\left[\mathcal{E}\left(\boldsymbol{J}_{\mathrm{i}}+\boldsymbol{J}_{\mathrm{e}}-\boldsymbol{J}_{\mathrm{o}},\boldsymbol{M}_{\mathrm{i}}-\boldsymbol{M}_{\mathrm{o}}\right)\right]_{\mathrm{S}_{\mathrm{o}}^{-}}^{\tan} = \left[\mathcal{E}\left(\boldsymbol{J}_{\mathrm{o}},\boldsymbol{M}_{\mathrm{o}}\right)\right]_{\mathrm{S}_{\mathrm{o}}^{+}}^{\tan} \tag{3-31a}$$

$$\left[\mathcal{H}\left(\boldsymbol{J}_{\mathrm{i}}+\boldsymbol{J}_{\mathrm{e}}-\boldsymbol{J}_{\mathrm{o}},\boldsymbol{M}_{\mathrm{i}}-\boldsymbol{M}_{\mathrm{o}}\right)\right]_{\mathrm{S}_{\mathrm{o}}^{-}}^{\tan} = \left[\mathcal{H}\left(\boldsymbol{J}_{\mathrm{o}},\boldsymbol{M}_{\mathrm{o}}\right)\right]_{\mathrm{S}_{\mathrm{o}}^{+}}^{\tan} \tag{3-31b}$$

Here, Eqs. (3-29a) and (3-29b) are based on the definitions of $\boldsymbol{J}_{\mathrm{i}}$ (DoJ) and $\boldsymbol{M}_{\mathrm{i}}$ (DoM), and Eq. (3-30) is based on the homogeneous tangential electric field boundary condition on $\mathrm{S}_{\mathrm{e}}$, and Eqs. (3-31a) and (3-31b) are based on the traveling-wave condition on $\mathrm{S}_{\mathrm{o}}$ where $\mathrm{S}_{\mathrm{o}}^{-}$ and $\mathrm{S}_{\mathrm{o}}^{+}$ are the left-side and right-side surfaces of $\mathrm{S}_{\mathrm{o}}$ (in fact, the traveling-wave condition can be viewed as a counterpart of the famous Sommerfeld's radiation condition at infinity and also a counterpart of the perfectly matching condition used on the output port of receiving antenna, and a careful discussion for it can be found in Ref. [27-Secs.3.2.1.3&3.2.1.4]).

Similarly to the previous cases of antennas, the integral equations (3-29a)~(3-31b) can be discretized into some matrix equations by using MoM, and the matrix equations lead to the following transformations

$$\mathbb{T}_{\mathrm{DoJ}}\cdot\mathbb{J}_{\mathrm{i}} = \begin{bmatrix}\mathbb{J}_{\mathrm{i}}\\ \mathbb{J}_{\mathrm{e}}\\ \mathbb{J}_{\mathrm{o}}\\ \mathbb{M}_{\mathrm{i}}\\ \mathbb{M}_{\mathrm{o}}\end{bmatrix} = \mathbb{T}_{\mathrm{DoM}}\cdot\mathbb{M}_{\mathrm{i}} \tag{3-32}$$

from independent current $\mathbb{J}_{\mathrm{i}}/\mathbb{M}_{\mathrm{i}}$ into all currents $(\mathbb{J}_{\mathrm{i}},\mathbb{J}_{\mathrm{e}},\mathbb{J}_{\mathrm{o}},\mathbb{M}_{\mathrm{i}},\mathbb{M}_{\mathrm{o}})$.

Substituting transformations (3-32) into matrix forms (3-28), the matrix-formed IPO $P_{\mathrm{in}} = \mathbb{C}_{\mathrm{i}}^{\dagger}\cdot\mathbb{P}_{\mathrm{in}}\cdot\mathbb{C}_{\mathrm{i}}$ with only independent current $\mathbb{C}_{\mathrm{i}}$ (which is either $\mathbb{J}_{\mathrm{i}}$ or $\mathbb{M}_{\mathrm{i}}$) is obtained, where the specific expression for $\mathbb{P}_{\mathrm{in}}$ is similar to the one given in the previous Eq. (3-6). Employing the $\mathbb{P}_{\mathrm{in}}$, the DMs can be derived from solving modal decoupling equation $\mathbb{P}_{\mathrm{in}}^{-}\cdot\mathbb{C}_{\mathrm{i}} = \theta\,\mathbb{P}_{\mathrm{in}}^{+}\cdot\mathbb{C}_{\mathrm{i}}$ with $\mathbb{P}_{\mathrm{in}}^{+} = (\mathbb{P}_{\mathrm{in}}+\mathbb{P}_{\mathrm{in}}^{\dagger})/2$ and $\mathbb{P}_{\mathrm{in}}^{-} = (\mathbb{P}_{\mathrm{in}}-\mathbb{P}_{\mathrm{in}}^{\dagger})/2j$ [27-Sec.3.2],[39]. The obtained DMs satisfy the following energy-decoupling relation

$$\left(1/T\right)\int_{t_0}^{t_0+T}\left[\iint_{\mathrm{S}_{\mathrm{i}}}\left(\boldsymbol{\mathcal{E}}_n\times\boldsymbol{\mathcal{H}}_m+\boldsymbol{\mathcal{E}}_m\times\boldsymbol{\mathcal{H}}_n\right)\cdot\boldsymbol{n}_{\mathrm{i}}dS\right]dt = 2\delta_{mn} \tag{3-33}$$

and then time-domain LORENTZ'S RECIPROCITY THEOREM (2-22) implies that

$$(1/T)\int_{t_0}^{t_0+T}\left[\iint_{S}\left(\boldsymbol{\mathcal{E}}_n\times\boldsymbol{\mathcal{H}}_m+\boldsymbol{\mathcal{E}}_m\times\boldsymbol{\mathcal{H}}_n\right)\cdot\boldsymbol{n}dS\right]dt=2\delta_{mn} \tag{3-34}$$

where S is an arbitrary waveguide cross-section, and $\boldsymbol{n}$ is the normal direction of S.

Following the convention of Sec. 3.2 (for transmitting antennas), the modal input "impedance $Z_{\text{in}}$, resistance $R_{\text{in}}$, reactance $X_{\text{in}}$" and "admittance $Y_{\text{in}}$, conductance $G_{\text{in}}$, susceptance $B_{\text{in}}$" of the waveguide section shown in Fig. 3-19 can be defined as the ones given in Eq. (3-12). Obviously, both the $R_{\text{in}}$ and $G_{\text{in}}$ are the real functions about working frequency $f$, so both the $R_{\text{in}}(f)$ and $G_{\text{in}}(f)$ curves can be easily obtained by utilizing a frequency-sweep calculation. If the following two conditions

Condition 1. frequencies $\{f_1, f_2, \cdots, f_\xi, \cdots\}$ are all the local maximum points of the *m*-th modal $R_{\text{in}}(f)$ or $G_{\text{in}}(f)$ curve, and

Condition 2. frequencies $\{f_1, f_2, \cdots, f_\xi, \cdots\}$ satisfy monotonously increasing relation $f_1 < f_2 < \cdots < f_\xi < \cdots$ are satisfied simultaneously, then the cut-off frequency $f_{\text{cut}}$ of the *m*-th DM can be calculated as the following explicit expression[27-Sec.3.2],[39]

$$f_{\text{cut}}=\sqrt{f_\xi^2-\left(\frac{\xi}{2L\sqrt{\mu\varepsilon}}\right)^2} \tag{3-35}$$

Among all the $\{f_{\text{cut}}\}$, the smallest one corresponds to the dominant mode.

Here, we use the above formulations to calculate DMs of a circular metallic waveguide. The waveguide is with cross-section radius 1 cm and infinite longitudinal length, and its tube is filled by the material with $\mu_{\text{r}}=1$ and $\varepsilon_{\text{r}}=5$. The following calculation only covers a section with longitudinal length 5 cm, and the longitudinal infinity feature of the waveguide is modeled by the travelling-wave condition on output port. The following calculation is based on the HM-DoM formulation, and the associated resistance and conductance curves are shown in Fig. 3-20.

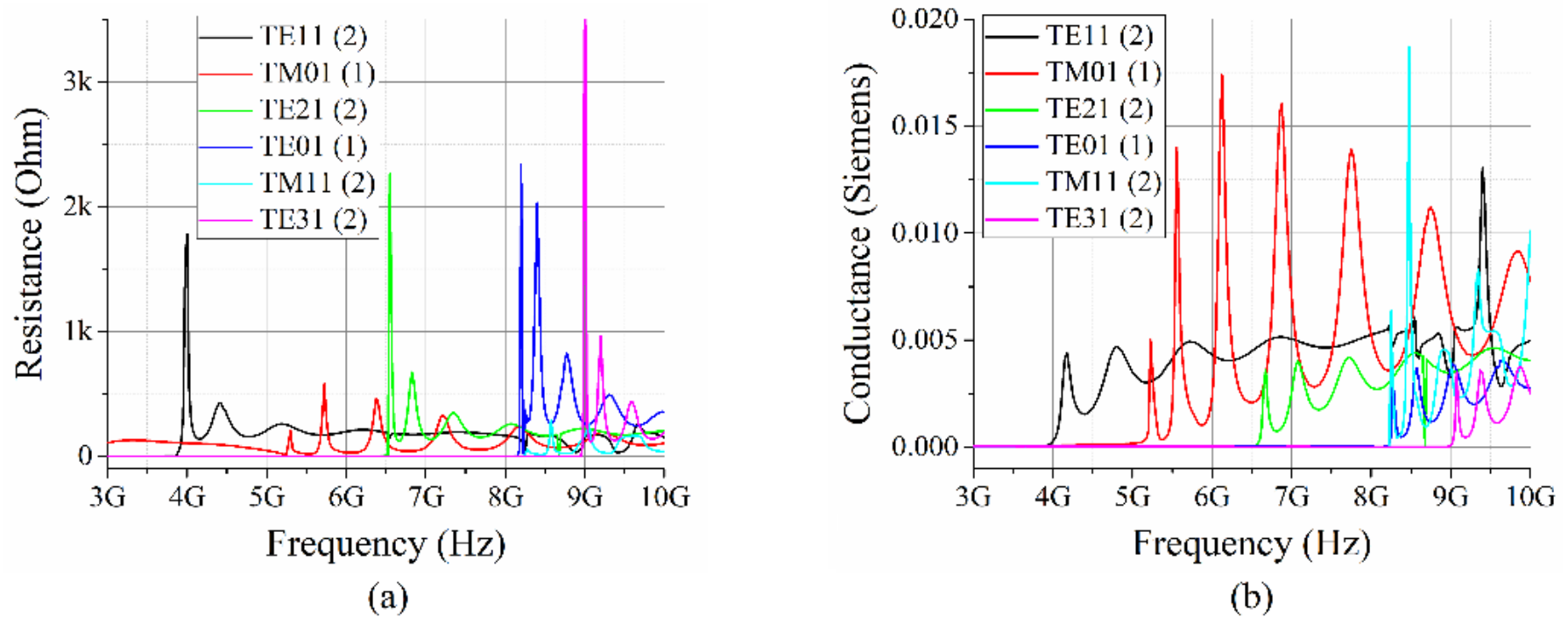

Figure 3-20 Modal (a) resistance and (b) conductance calculated from the HM-DoM formula.

Taking the first mode working at 4.175 GHz as an example (corresponding to the first local maximum of the conductance curve), we show its modal magnetic current $\boldsymbol{M}_{\rm i}$ on input port $\rm S_i$ in the following Fig. 3-21, and the corresponding tangential modal electric field distributing on $\rm S_i$ can be determined as that $[\boldsymbol{E}]^{\rm tan} = \boldsymbol{z} \times \boldsymbol{M}_{\rm i}$ on $\rm S_i$.

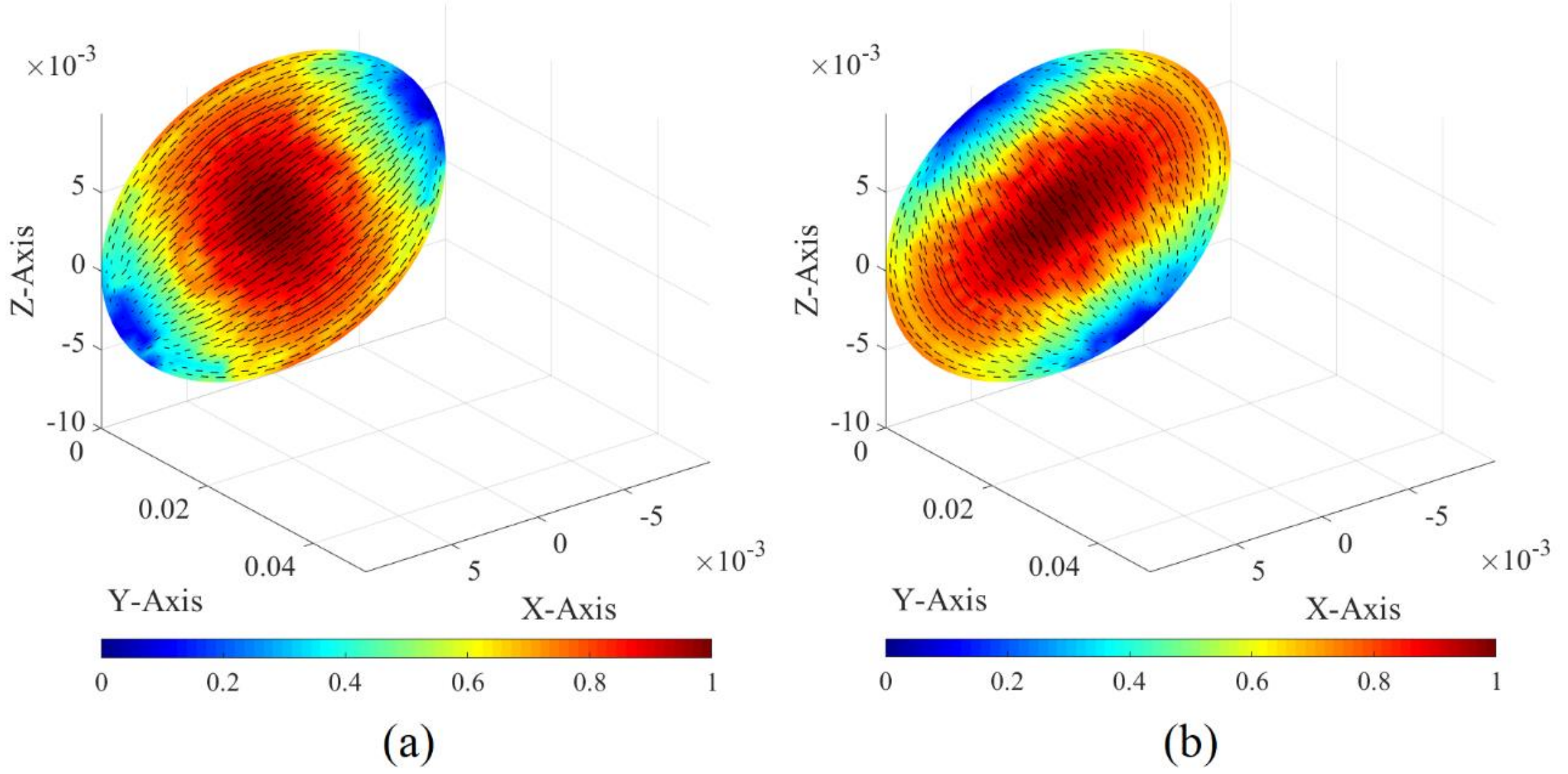

Figure 3-21 Modal magnetic current $\boldsymbol{M}_{\rm i}$ of the (a) first and (b) second degenerate states of the first mode shown in Fig. 3-20(b).

From Fig. 3-21, it is easy to recognize that the mode is just the classical TE11 eigen-mode of circular metallic waveguide. Now, we separately plot the modal conductance curve of the TE11 mode in the following Fig. 3-22, and mark a series of critical points in the conductance curve.

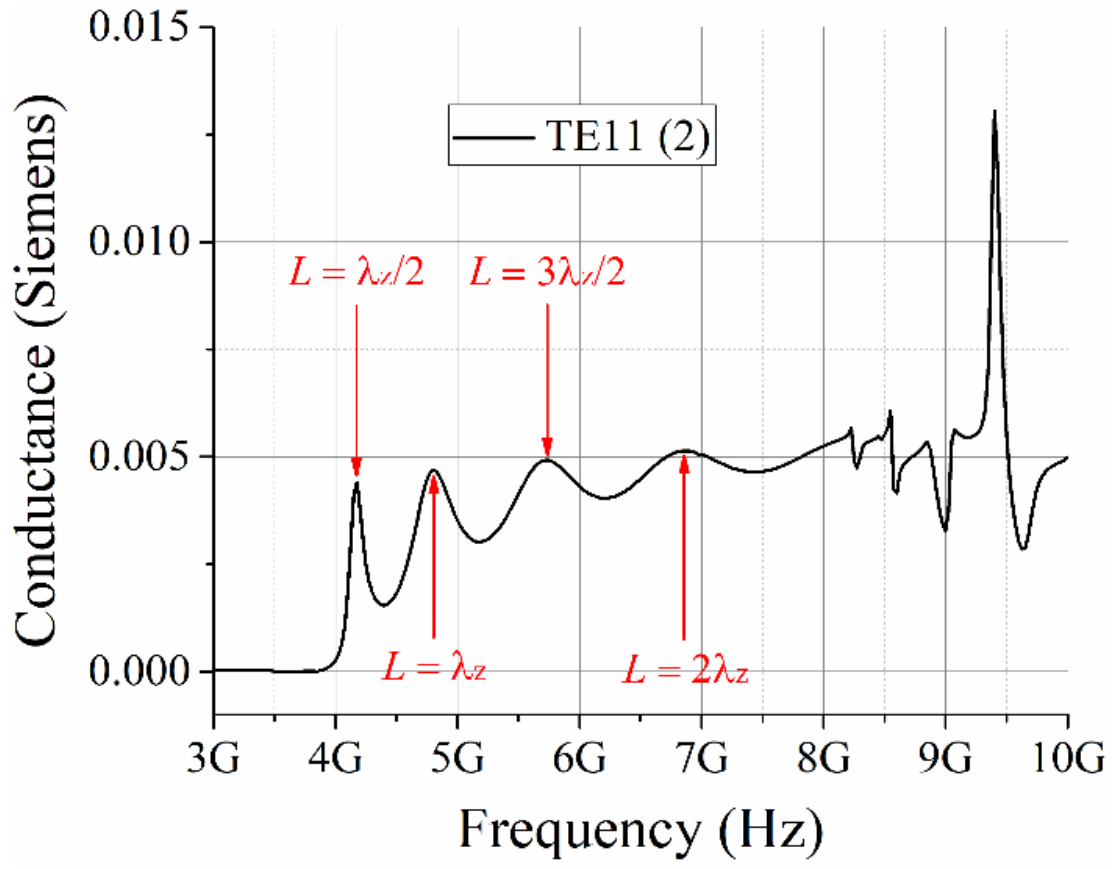

Figure 3-22 Some critical points in the conductance curve of the TE11 mode in Fig. 3-20(b).

The modal wall electric current and electric energy density distributions corresponding to the critical points are shown in the following Figs. 3-23 and 3-24 respectively.

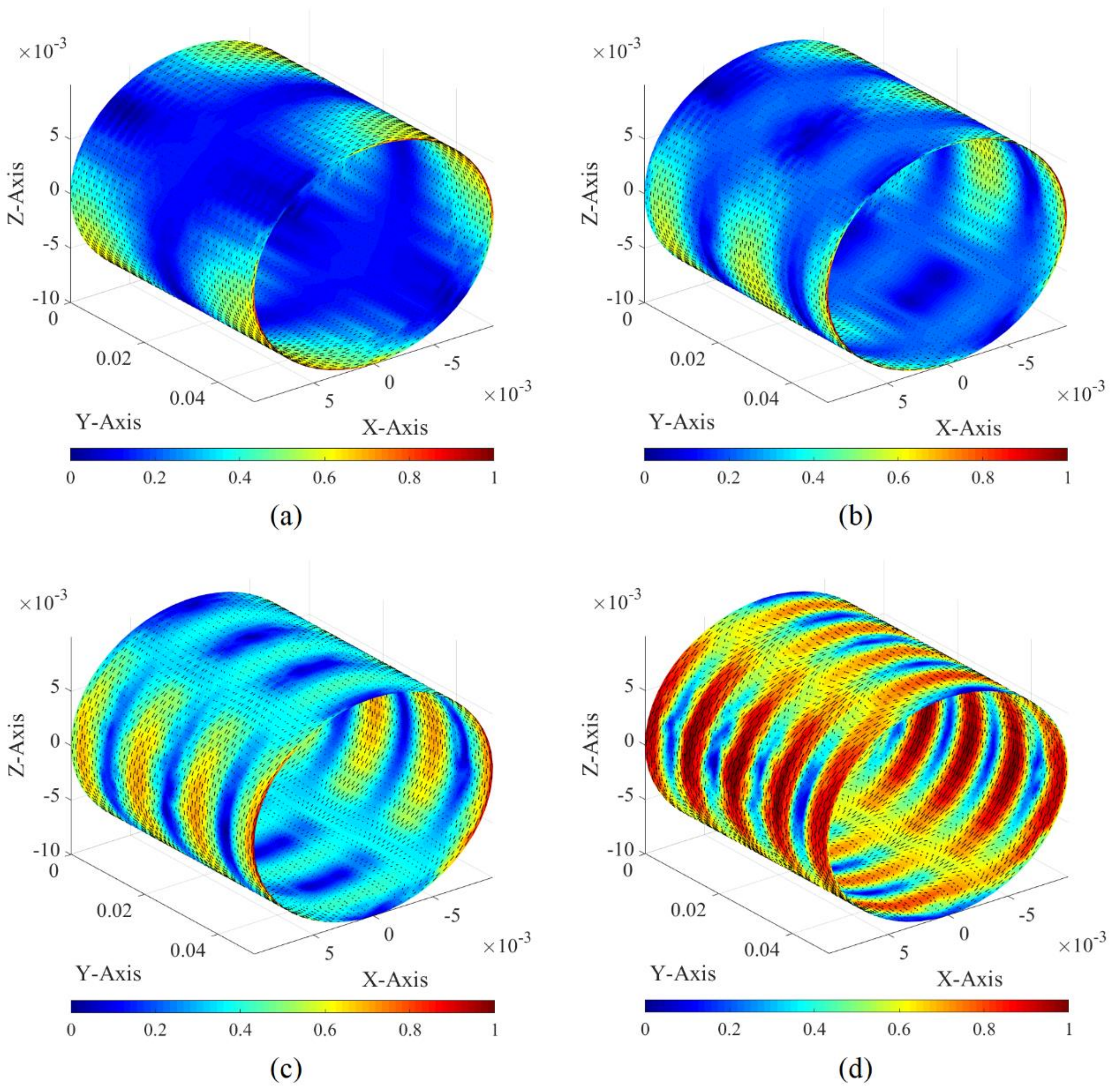

Figure 3-23 Waveguide wall electric currents of the TE11 mode working at (a) 4.175 GHz, (b) 4.800 GHz, (c) 5.725 GHz, and (d) 6.875 GHz.

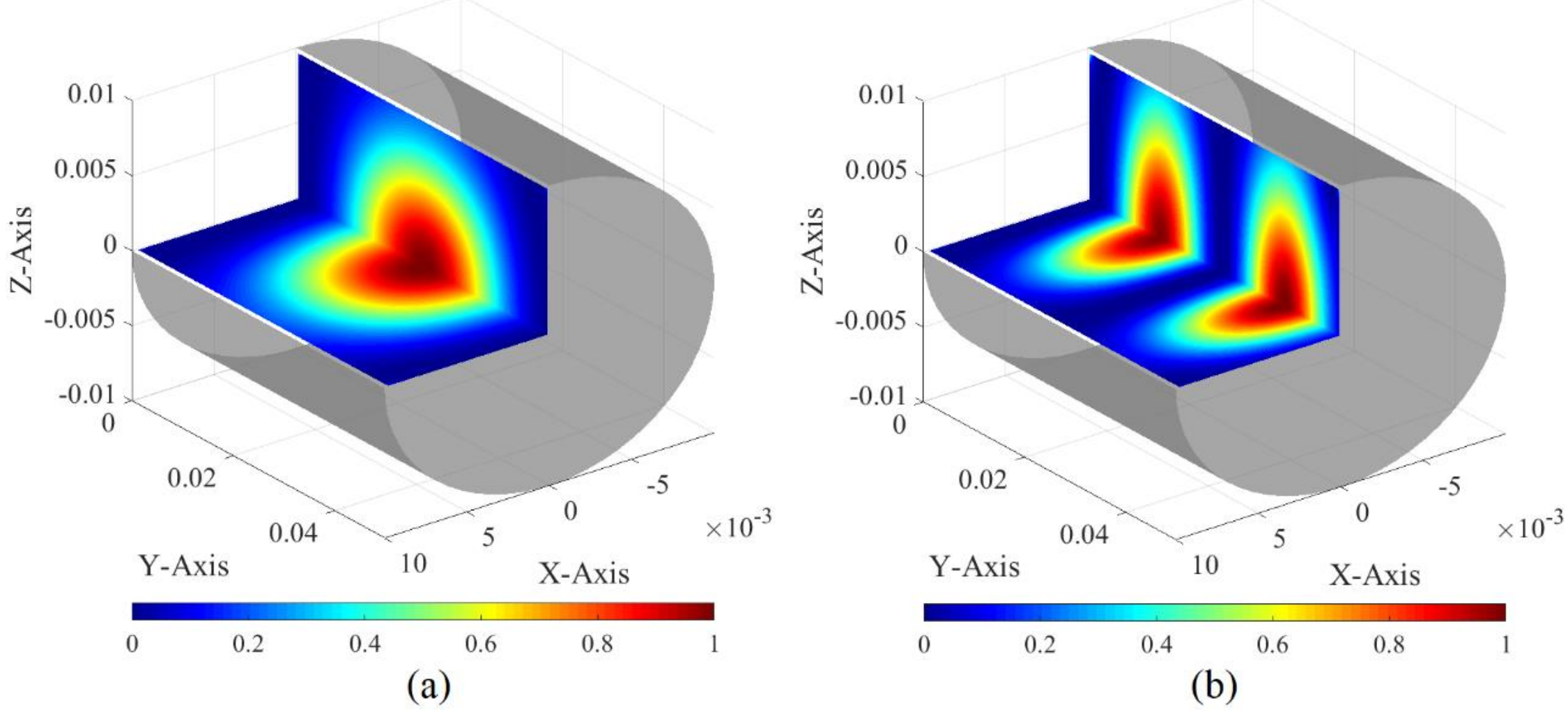

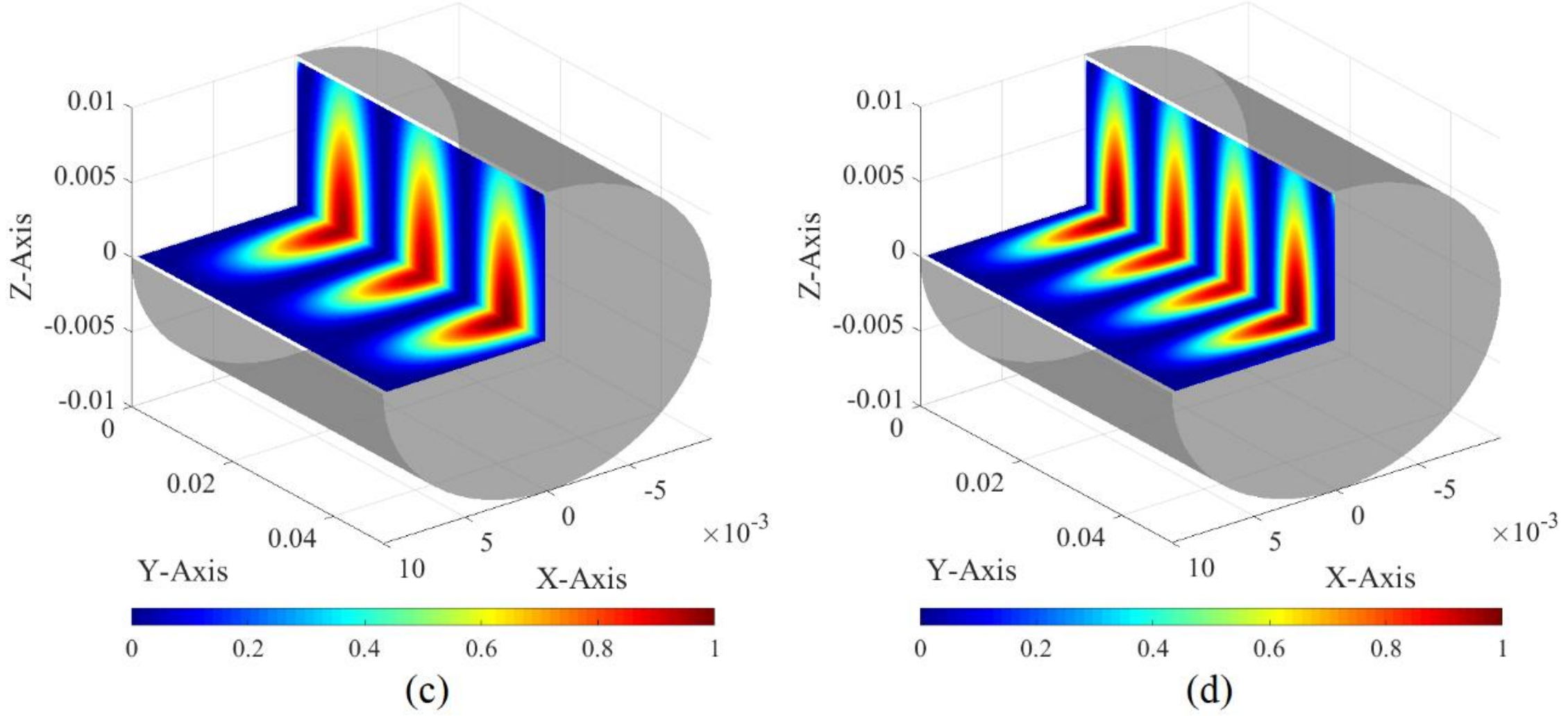

Figure 3-24 Electric energy density distributions (on xOy and yOz planes) of the TE11 mode working at a series of resonance frequencies (a) 4.175 GHz, (b) 4.800 GHz, (c) 5.725 GHz, and (d) 6.875 GHz.

The figures imply that: 5 cm is equal to the $\lambda_z/2$, $\lambda_z$, $3\lambda_z/2$, and $2\lambda_z$ of the TE11 mode working at 4.175 GHz, 4.800 GHz, 5.725 GHz, and 6.875 GHz respectively, where $\lambda_z$ represents the waveguide wavelength along Z-axis direction. Based on the conclusions given in Ref. [27-Sec.3.2], it is easy to explain why the modal conductance curve achieves the local maximums at 4.800 GHz (corresponding to $\lambda_z = 5\,\text{cm}$) and 6.875 GHz (corresponding to $2\lambda_z = 5\,\text{cm}$). Now, we focus on explaining the reasons leading to the local maximums at 4.175 GHz (corresponding to $\lambda_z/2 = 5\,\text{cm}$) and 5.725 GHz (corresponding to $3\lambda_z/2 = 5\,\text{cm}$).

Because of the time-harmonic distributions of the modal fields along Z-axis direction, the travelling-wave modal fields must satisfy the relation that $\boldsymbol{F}(z) = -\boldsymbol{F}(z+\lambda_z/2) = \boldsymbol{F}(z+\lambda_z)$. Based on this observation, we can conclude here that: the Eqs. (3-31a) and (3-31b) are also applicable to the case that the distance between $\text{S}_\text{i}$ and $\text{S}_\text{o}$ is $\lambda_z/2$. The case that the distance between $\text{S}_\text{i}$ and $\text{S}_\text{o}$ is $3\lambda_z/2$ can be similarly explained.

Because of these above, we can calculate the cutoff frequency of the TE11 mode from substituting $\{\xi = 1;\ f_\xi = 4.175\text{GHz};\ L = 5\,\text{cm};\ \mu = \mu_0,\ \varepsilon = 5\varepsilon_0\}$ into Eq. (3-35), and then the derived cutoff frequency is 3.9539 GHz. Similarly, we can also obtain the cutoff frequencies of the other modes shown in Fig. 3-20, and we list the obtained cutoff frequencies in the following Tab. 3-1.

Table 3-1 Cutoff frequencies (GHz) and degeneracy degrees (in the brackets) of the lower-order travelling-wave modes derived from novel PTT-DMT and classical eigen-mode theory[2,3]

| | **Novel PTT-DMP** | | **Classical Eigen-Mode Theory[2,3]** |
|---|---|---|---|
| | **Recognized from the Local Maximum of $R_{\text{in}}$ Curve** | **Recognized from the Local Maximum of $G_{\text{in}}$ Curve** | |
| **TE11** | 3.7686 (2) | **3.9539 (2)** | 3.9288 (2) |
| **TM01** | **5.1276 (1)** | 5.0501 (1) | 5.1316 (1) |
| **TE21** | 6.4113 (2) | **6.5390 (2)** | 6.5171 (2) |
| **TE01** | 8.0897 (1) | **8.1657 (1)** | 8.1763 (1) |
| **TM11** | **8.1657 (2)** | 8.1403 (2) | 8.1763 (2) |
| **TE31** | 8.8996 (2) | **8.9754 (2)** | 8.9646 (2) |

By comparing the results derived from the novel PTT-DMT and the classical eigen-mode theory, it is not difficult to find out that the results are agreed well with each other. At the same time, it is easy to observe that: for the TE modes, the results derived from the conductance curves are more desirable; for the TM modes, the results derived from the resistance curves are more desirable.

In addition, we also plot their modal equivalent electric current or modal equivalent magnetic current distributions on input port $S_{\text{i}}$ in the following Fig. 3-25 for readers' reference. The currents are consistent with the port currents of the classical eigen-modes given in Ref. [3-pps.492&493].

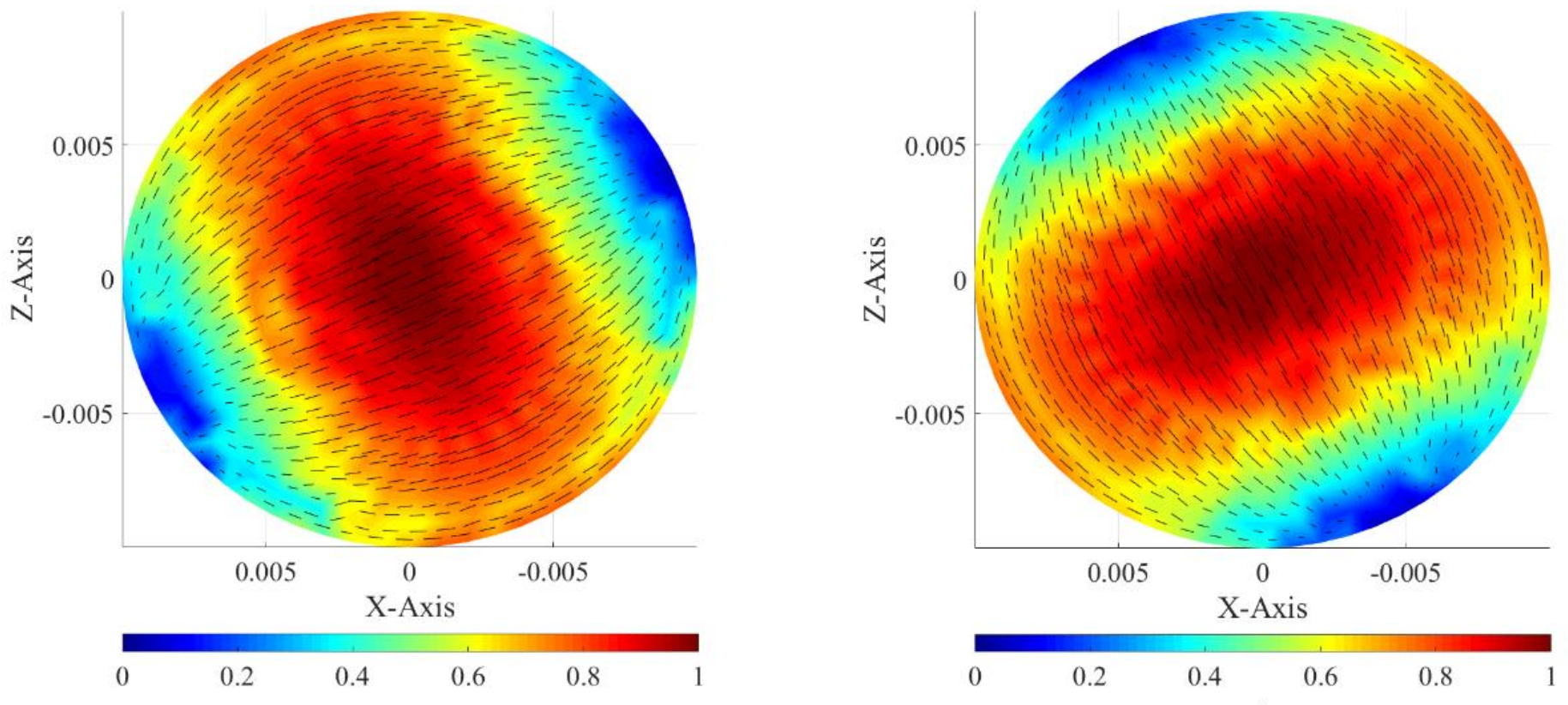

(a) (b)

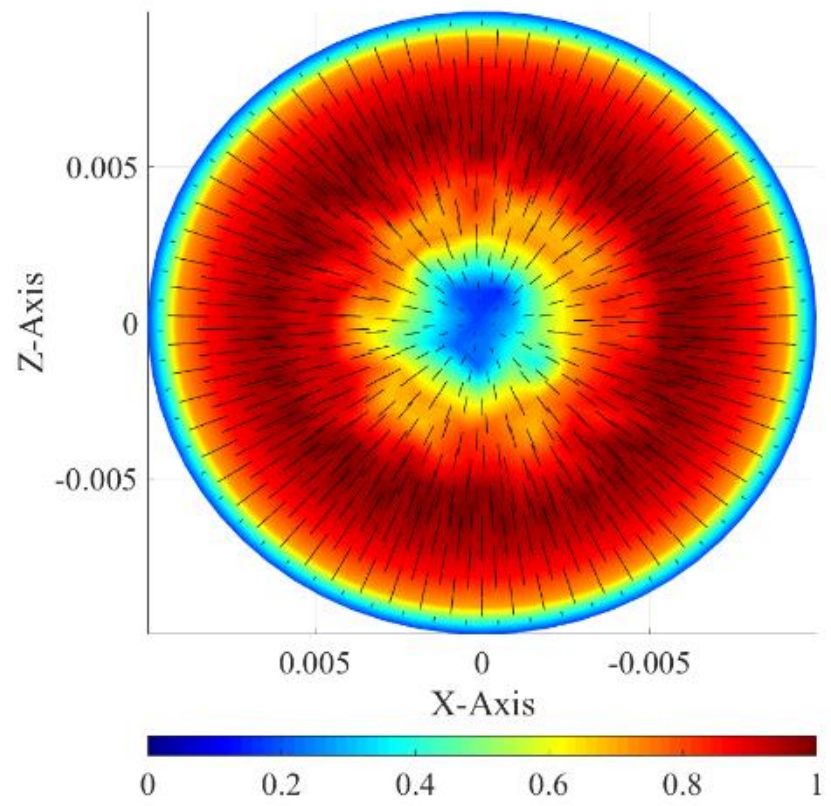

(c)

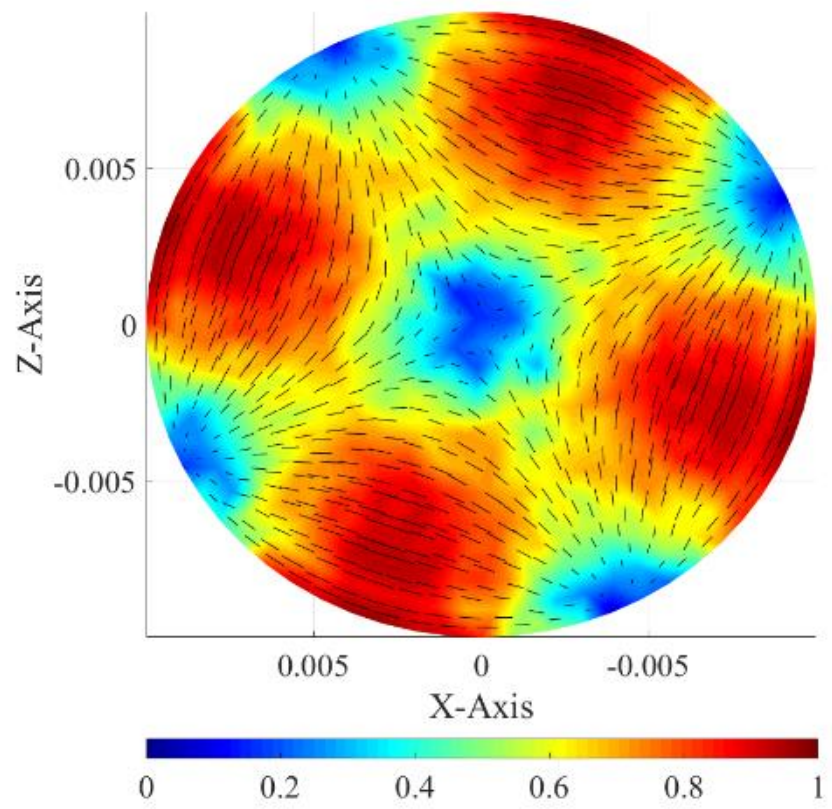

(d)

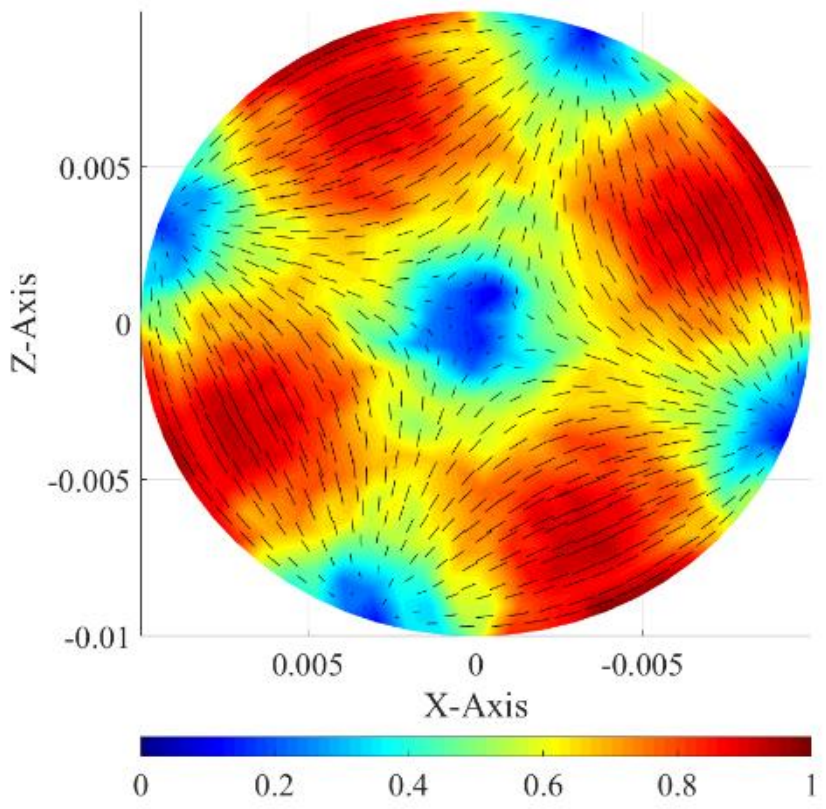

(e)

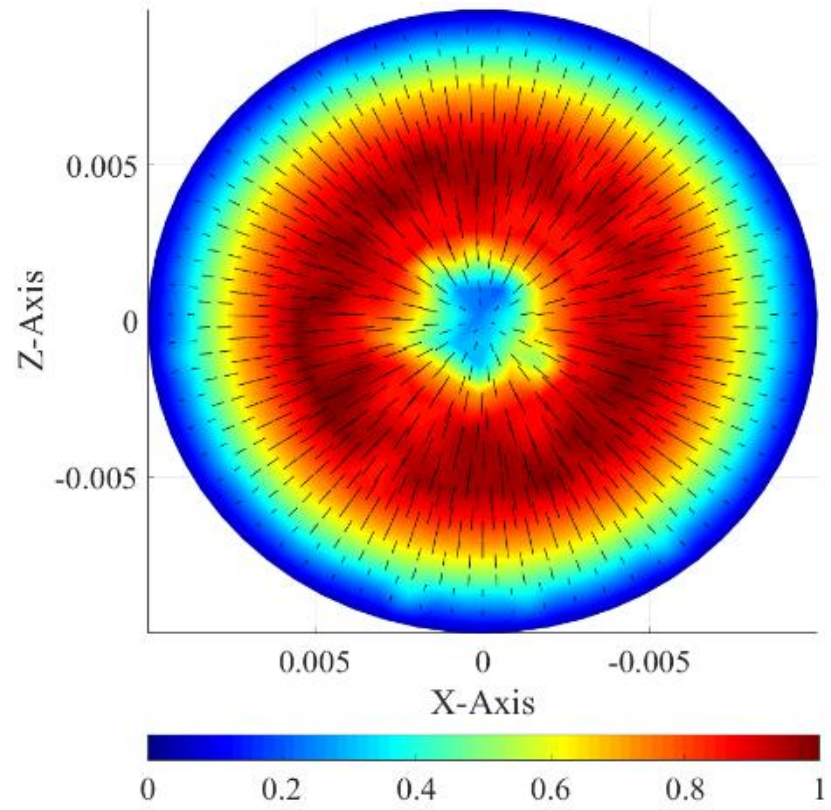

(f)

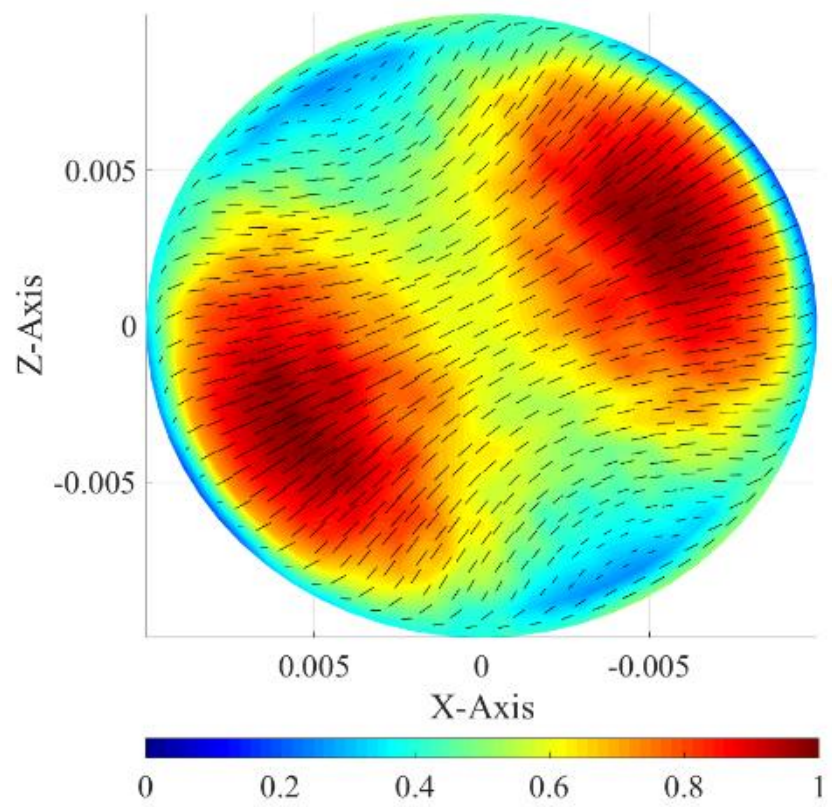

(g)

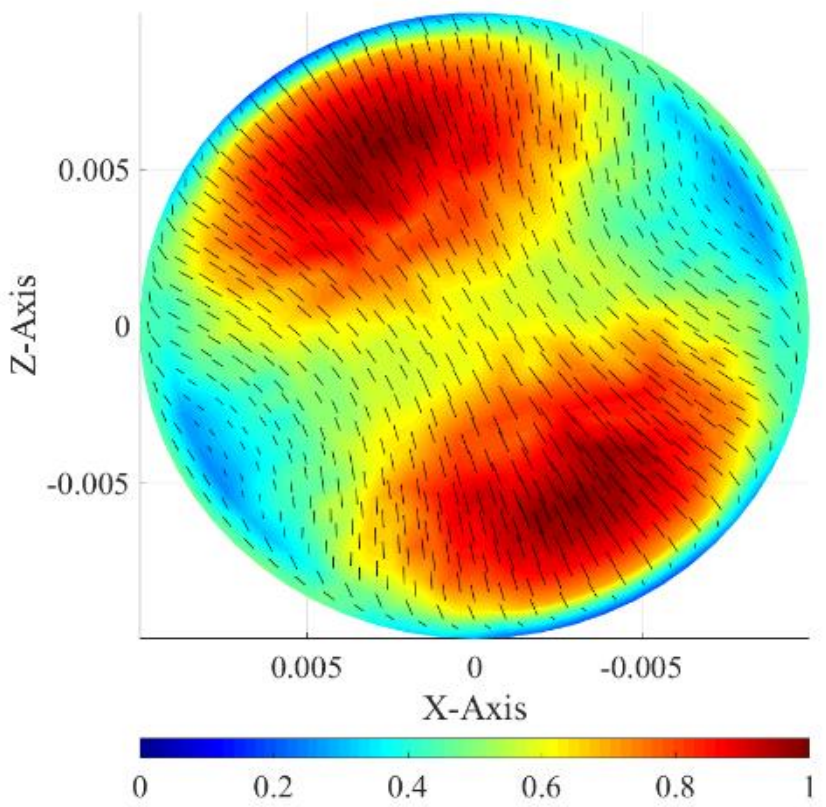

(h)

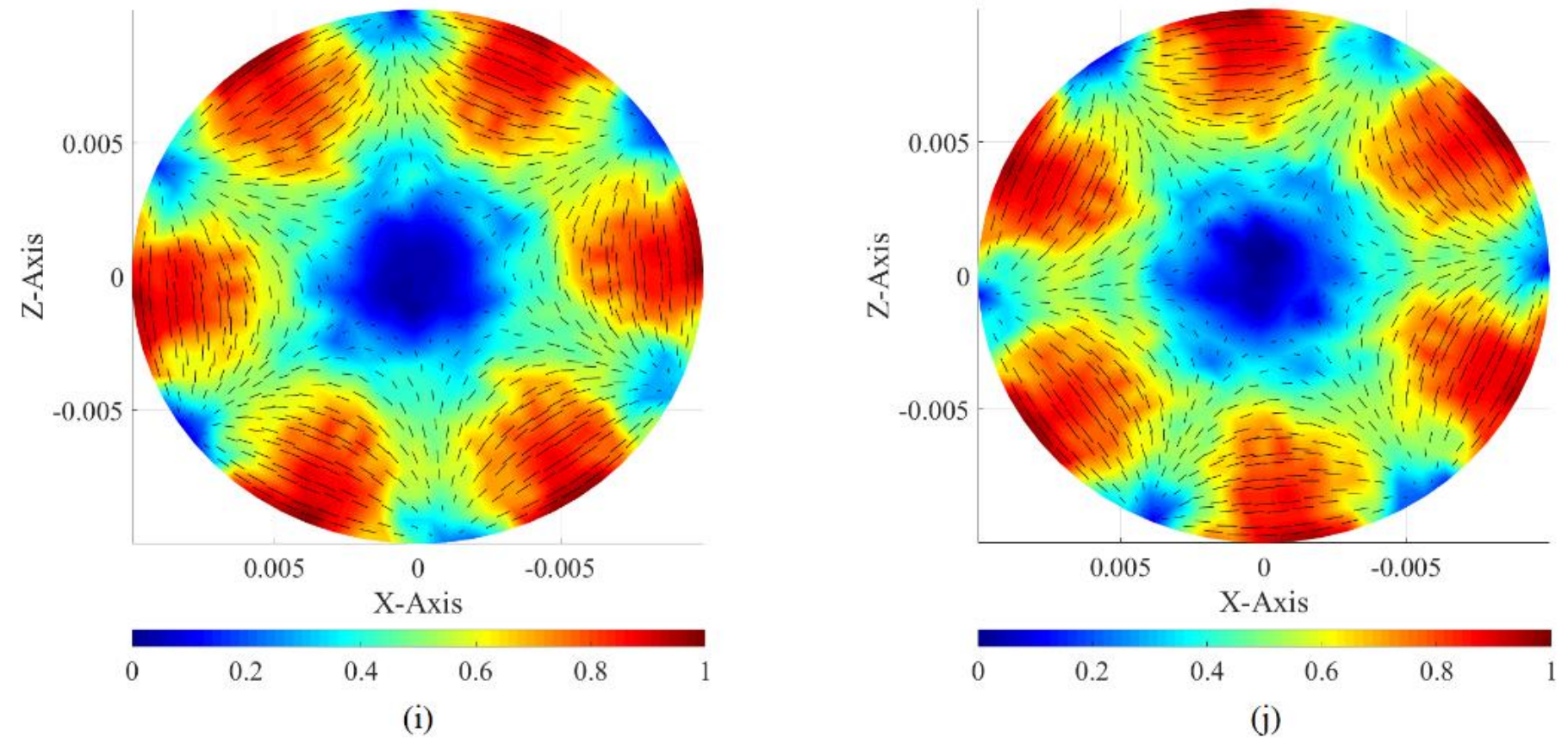

Figure 3-25 Modal equivalent currents on input port $S_i$. (a) $\boldsymbol{M}_i$ of the 1st degenerate state of TE11; (b) $\boldsymbol{M}_i$ of the 2nd degenerate state of TE11; (c) $\boldsymbol{J}_i$ of TM01; (d) $\boldsymbol{M}_i$ of the 1st degenerate state of TE21; (e) $\boldsymbol{M}_i$ of the 2nd degenerate state of TE21; (f) $\boldsymbol{M}_i$ of TE01; (g) $\boldsymbol{J}_i$ of the 1st degenerate state of TM11; (h) $\boldsymbol{J}_i$ of the 2nd degenerate state of TM11; (i) $\boldsymbol{M}_i$ of the 1st degenerate state of TE31; (j) $\boldsymbol{M}_i$ of the 2nd degenerate state of TE31.

The above these imply that the novel PTT-DMT indeed has ability to construct the travelling-wave modes of the circular metallic waveguide.

Besides the above circular metallic waveguide, the PTT-DMT is also directly applicable to the following rectangular[39] and coaxial[27-Sec.3.2.5.2],[39] metallic waveguides.

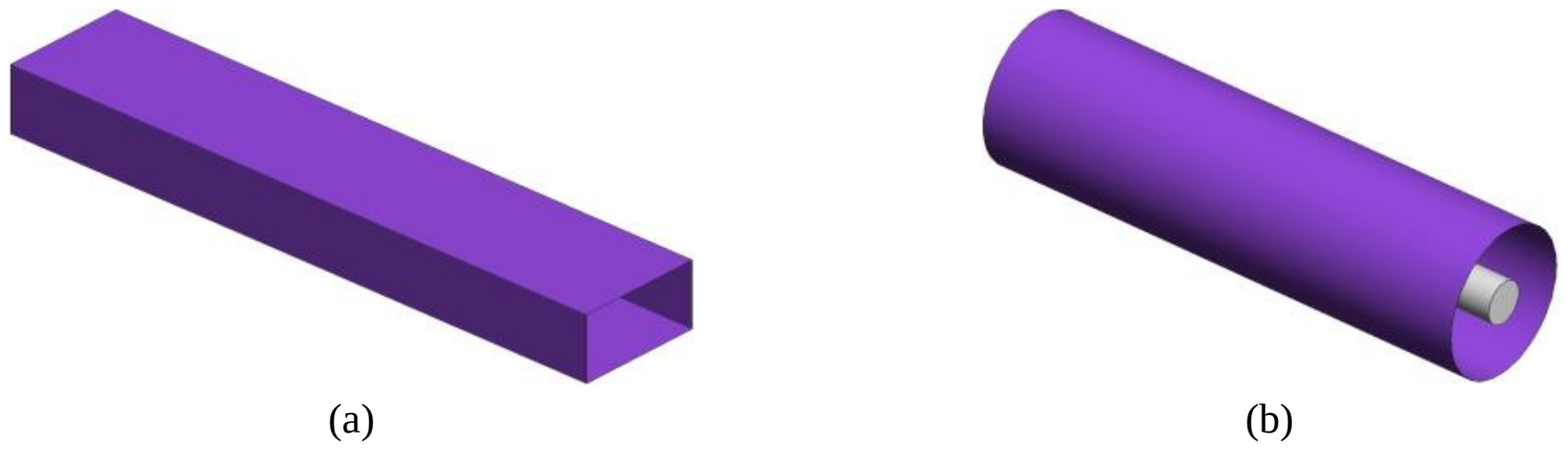

Figure 3-26 Geometries of (a) rectangular and (b) coaxial metallic waveguides.

As exhibited in Refs. [27-Secs.3.3&3.4] and [39], the above metallic-waveguide-oriented PTT-DMT can be easily generalized to material and composite waveguides, and a typical material waveguide and a typical composite waveguide are shown in following Fig. 3-27.

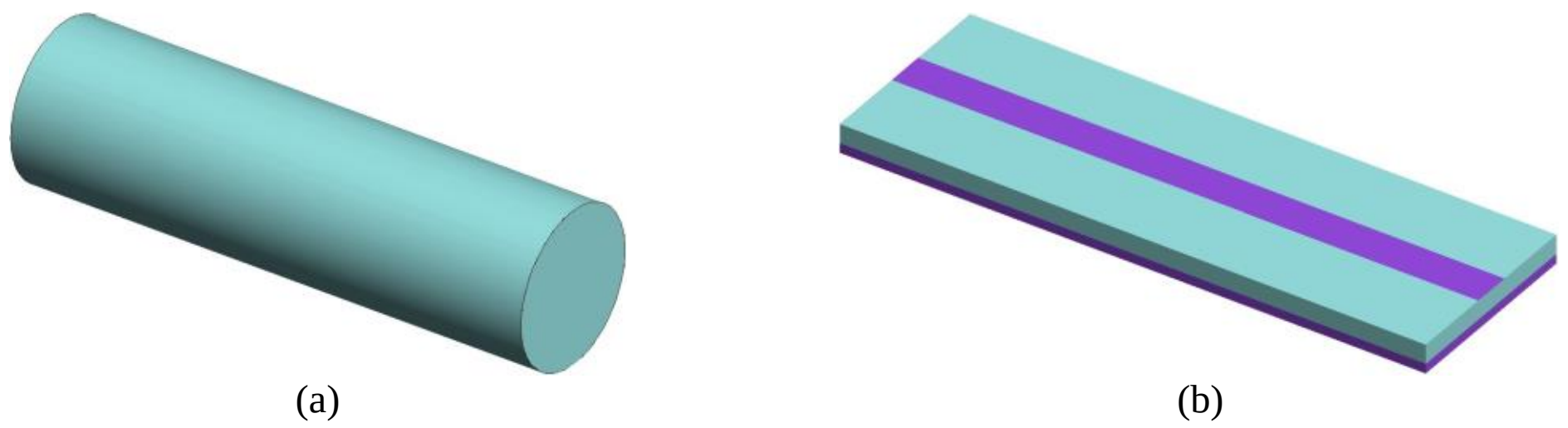

Figure 3-27 Geometries of (a) dielectric waveguide and (b) microstrip line.

In addition, as exhibited in Ref. [27-Sec.3.5], the PTT-DMT for the above-mentioned standard longitudinally homogeneous waveguides can also be further generalized to the longitudinally inhomogeneous waveguides, such as the ones shown in Fig. 3-28. The validity of the travelling-wave condition on the output ports of the non-standard waveguides is originated from that we are doing modal analysis for the waveguides.

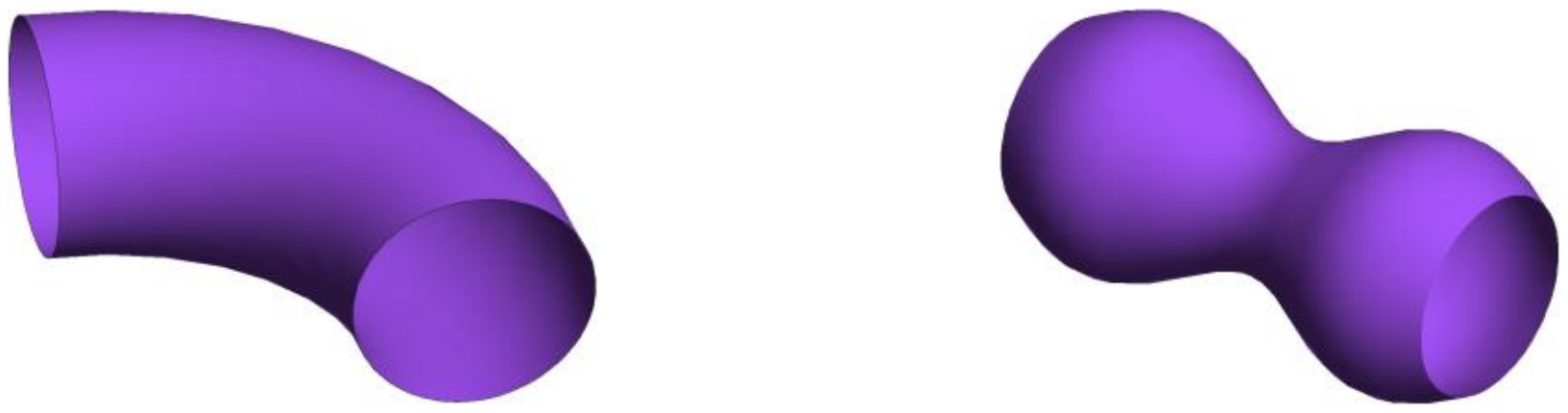

(a) (b)

Figure 3-28 Geometries of two typical longitudinally inhomogeneous waveguides.

## 3.5 PTT-Based DMs of Free Space (Free-Space Waveguide)

Now we consider some currents $(\boldsymbol{J},\boldsymbol{M})$ distributing in source region, and the source region is surrounded by free space, as shown in Fig. 3-29. The currents $(\boldsymbol{J},\boldsymbol{M})$ generate some fields $(\boldsymbol{E},\boldsymbol{H})$ in whole three-dimensional Euclidean space $\mathbb{E}_3$.

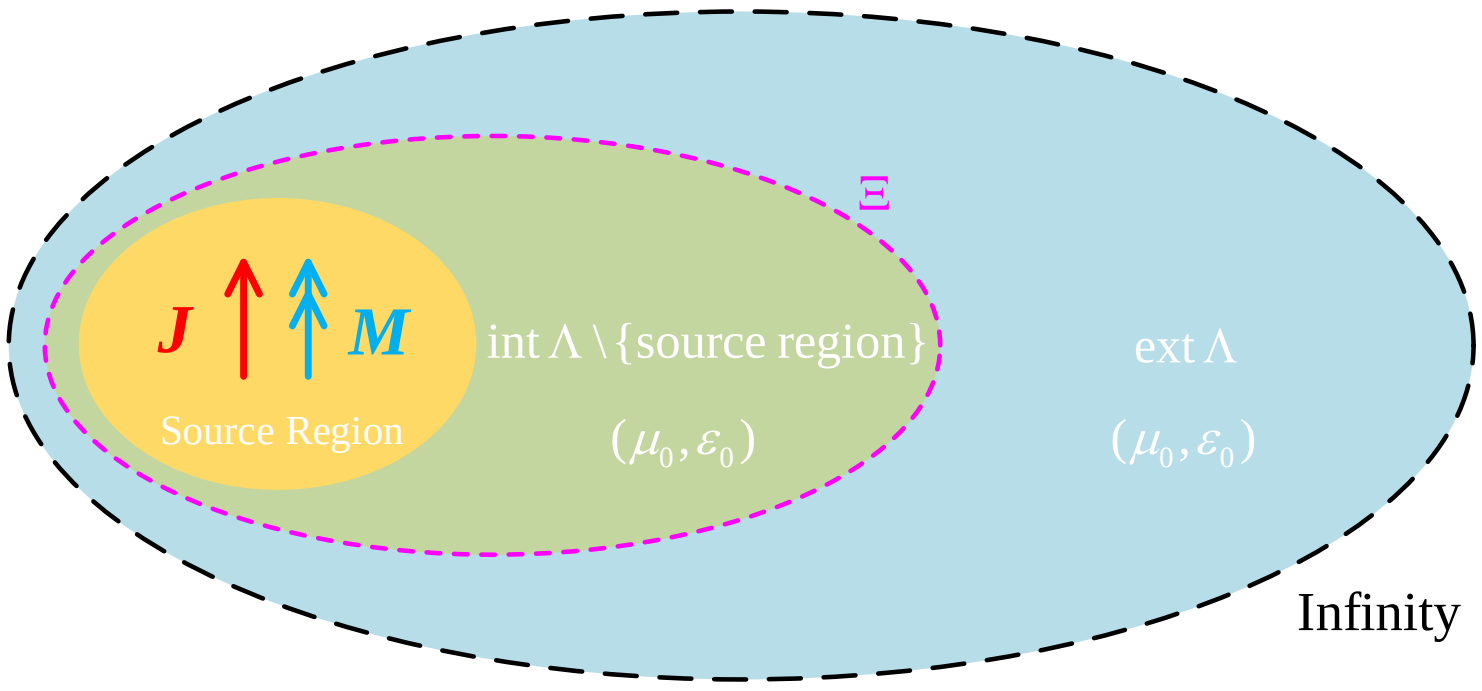

Figure 3-29 Surface $\Xi$ divides whole $\mathbb{E}_3$ into two parts $\text{int}\Lambda$ and $\text{ext}\Lambda$. Whole source region belongs to $\text{int}\Lambda$. $\text{ext}\Lambda$ and $\text{int}\Lambda \setminus \{\text{source region}\}$ are with free-space material parameters $(\mu_0,\varepsilon_0)$.

If $\Xi$ is a closed surface enclosing the whole source region, and the region enclosed by $\Xi$ is denoted as $\Lambda$ (where the source region belongs to $\Lambda$, but it is not necessarily identical to $\Lambda$), then there exists the following Huygens-Fresnel principle[8-App.C]:

$$\mathbf{G}_0^{JF} * (\boldsymbol{n}\times\boldsymbol{H}) + \mathbf{G}_0^{MF} * (\boldsymbol{E}\times\boldsymbol{n}) = \begin{cases} \boldsymbol{F}(\boldsymbol{r}) &, \quad \boldsymbol{r}\in\text{ext}\Lambda \\ 0 &, \quad \boldsymbol{r}\in\text{int}\Lambda \end{cases} \tag{3-36}$$

In Huygens-Fresnel principle (3-36), $\text{int}\Lambda$ and $\text{ext}\Lambda$ are the interior and exterior of $\Lambda$, and $\text{int}\Lambda \bigcup \Xi \bigcup \text{ext}\Lambda = \mathbb{E}_3$; $\boldsymbol{n}$ is the normal direction of $\Xi$, and it points to $\text{ext}\Lambda$.

The above Huygens-Fresnel principle implies that: in free space, the fields $(\boldsymbol{E},\boldsymbol{H})$ propagate from the source region to infinity, and there doesn't exist any reflection. This feature is very similar to the one satisfied by the travelling-wave modes working in the various wave-guiding structures discussed in the previous Sec. 3.4[27-Sec.3.6]. Thus, **this report treats the free space as a wave-guiding structure used to guide EM energy from the source region to infinity**[27-Sec.3.6]. This section focuses on constructing the DMs of the free-space waveguide, and has a similar organization as the previous Sec. 3.4.

The region occupied by free space is denoted as $\mathrm{V}$. The boundary of $\mathrm{V}$ is constituted by two closed surfaces $\mathrm{S_i}\bigcup\mathrm{S_e}$ and $\mathrm{S_\infty}$, and $\mathrm{S_i}\bigcup\mathrm{S_e}$ encloses whole source region, and $\mathrm{S_\infty}$ is a spherical surface with infinite radius. Surface $\mathrm{S_\infty}$ is just the output port of $\mathrm{V}$, and the fields automatically satisfy Sommerfeld's radiation condition on the output port. Usually, $\mathrm{S_i}\bigcup\mathrm{S_e}$ includes two parts — a penetrable part $\mathrm{S_i}$ and an impenetrable part $\mathrm{S_e}$, and the penetrable part $\mathrm{S_i}$ is just the input port of $\mathrm{V}$, and the impenetrable part $\mathrm{S_e}$ is usually electric wall, and this general case had been carefully discussed in Ref. [27-Sec.3.6]. Here, we focus on a more special case that the whole $\mathrm{S_i}\bigcup\mathrm{S_e}$ is penetrable, i.e., $\mathrm{S_e}=\varnothing$, and **this special case can be used to establish the connection between DMs and many classical modes, such as spherical modes**.

If the equivalent currents on $\mathrm{S_i}$ are defined as $\boldsymbol{J}_\mathrm{i}=\boldsymbol{n}_\mathrm{i}\times\boldsymbol{E}$ and $\boldsymbol{M}_\mathrm{i}=\boldsymbol{H}\times\boldsymbol{n}_\mathrm{i}$ (where $\boldsymbol{n}_\mathrm{i}$ is the outer normal of $\mathrm{S_i}$), the input power $P_\mathrm{in}=(1/2)\iint_{\mathrm{S_i}}(\boldsymbol{E}\times\boldsymbol{H}^\dagger)\cdot\boldsymbol{n}_\mathrm{i}dS$ passing through $\mathrm{S_i}$ can be written as the following integral operators

$$\begin{aligned}P_\mathrm{in}&=-(1/2)\langle\boldsymbol{J}_\mathrm{i},-j\omega\mu_0\mathcal{L}_0(\boldsymbol{J}_\mathrm{i})-\mathcal{K}_0(\boldsymbol{M}_\mathrm{i})\rangle_{\mathrm{S_i^+}}\\&=-(1/2)\langle\boldsymbol{M}_\mathrm{i},\mathcal{K}_0(\boldsymbol{J}_\mathrm{i})-j\omega\varepsilon_0\mathcal{L}_0(\boldsymbol{M}_\mathrm{i})\rangle_{\mathrm{S_i^+}}^\dagger\end{aligned}\tag{3-37}$$

where $\mathrm{S_i^+}$ is the outer-side surface of $\mathrm{S_i}$. Similarly to the pervasive sections, the integral operators can be easily discretized into the following matrix operators

$$P_\mathrm{in}=\mathbb{J}_\mathrm{i}^\dagger\cdot\mathbb{P}_\mathrm{JE}\cdot\begin{bmatrix}\mathbb{J}_\mathrm{i}\\\mathbb{M}_\mathrm{i}\end{bmatrix}=\begin{bmatrix}\mathbb{J}_\mathrm{i}\\\mathbb{M}_\mathrm{i}\end{bmatrix}^\dagger\cdot\mathbb{P}_\mathrm{HM}\cdot\mathbb{M}_\mathrm{i}\tag{3-38}$$

The above $\mathbb{J}_\mathrm{i}$ and $\mathbb{M}_\mathrm{i}$ are not independent, and they satisfy the transformations $\mathbb{M}_\mathrm{i}=\mathbb{T}_\mathrm{DoJ}\cdot\mathbb{J}_\mathrm{i}$ and $\mathbb{J}_\mathrm{i}=\mathbb{T}_\mathrm{DoM}\cdot\mathbb{M}_\mathrm{i}$ which respectively originate from integral equations $[\mathcal{K}_0(\boldsymbol{J}_\mathrm{i})-j\omega\varepsilon_0\mathcal{L}_0(\boldsymbol{M}_\mathrm{i})]_{\mathrm{S_i^+}}^\mathrm{tan}=\boldsymbol{J}_\mathrm{i}\times\boldsymbol{n}_\mathrm{i}$ and $[-j\omega\mu_0\mathcal{L}_0(\boldsymbol{J}_\mathrm{i})-\mathcal{K}_0(\boldsymbol{M}_\mathrm{i})]_{\mathrm{S_i^+}}^\mathrm{tan}=\boldsymbol{n}_\mathrm{i}\times\boldsymbol{M}_\mathrm{i}$. Substituting the transformations into matrix operators (3-38), the matrix-formed IPO $P_\mathrm{in}=\mathbb{C}_\mathrm{i}^\dagger\cdot\mathbb{P}_\mathrm{in}\cdot\mathbb{C}_\mathrm{i}$ with only independent current $\mathbb{C}_\mathrm{i}$ (either $\mathbb{J}_\mathrm{i}$ or $\mathbb{M}_\mathrm{i}$) is obtained, where the specific expression for $\mathbb{P}_\mathrm{in}$ is similar to the one given in Eq. (3-6). Using the $\mathbb{P}_{\text{in}}$, the DMs can be derived from solving modal decoupling equation $\mathbb{P}_{\text{in}}^{-}\cdot\mathbb{C}_{\text{i}}=\theta\,\mathbb{P}_{\text{in}}^{+}\cdot\mathbb{C}_{\text{i}}$ with $\mathbb{P}_{\text{in}}^{+}=(\mathbb{P}_{\text{in}}+\mathbb{P}_{\text{in}}^{\dagger})/2$ and $\mathbb{P}_{\text{in}}^{-}=(\mathbb{P}_{\text{in}}-\mathbb{P}_{\text{in}}^{\dagger})/2j$, and the obtained DMs satisfy that

$$\left(1/T\right)\int_{t_0}^{t_0+T}\left[\oiint_{S}\left(\boldsymbol{\mathcal{E}}_n\times\boldsymbol{\mathcal{H}}_m+\boldsymbol{\mathcal{E}}_m\times\boldsymbol{\mathcal{H}}_n\right)\cdot\boldsymbol{n}dS\right]dt=2\delta_{mn} \tag{3-39}$$

where S is an arbitrary closed surface enclosing $S_i$, and ***n*** is the normal direction of S.

Here, we select the input port $S_{\text{i}}$ of free-space waveguide as a spherical surface with radius 3cm. By orthogonalizing the JE-DoJ-based IPO, we obtain the DMs of free-space waveguide. Modal resistance curves of the obtained DMs are shown in Fig. 3-30.

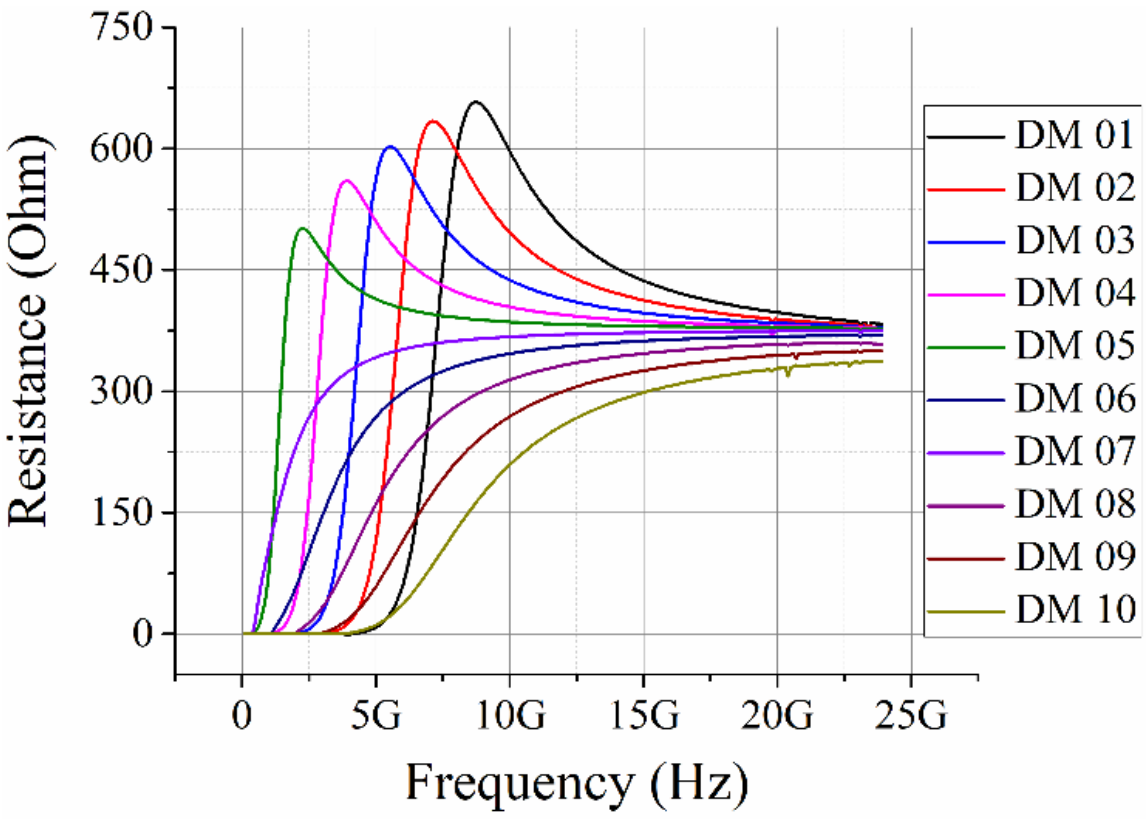

Figure 3-30 Modal resistance curves of some typical DMs.

The equivalent electric currents and radiation patterns of 10 typical DMs are shown in the following Fig. 3-31.

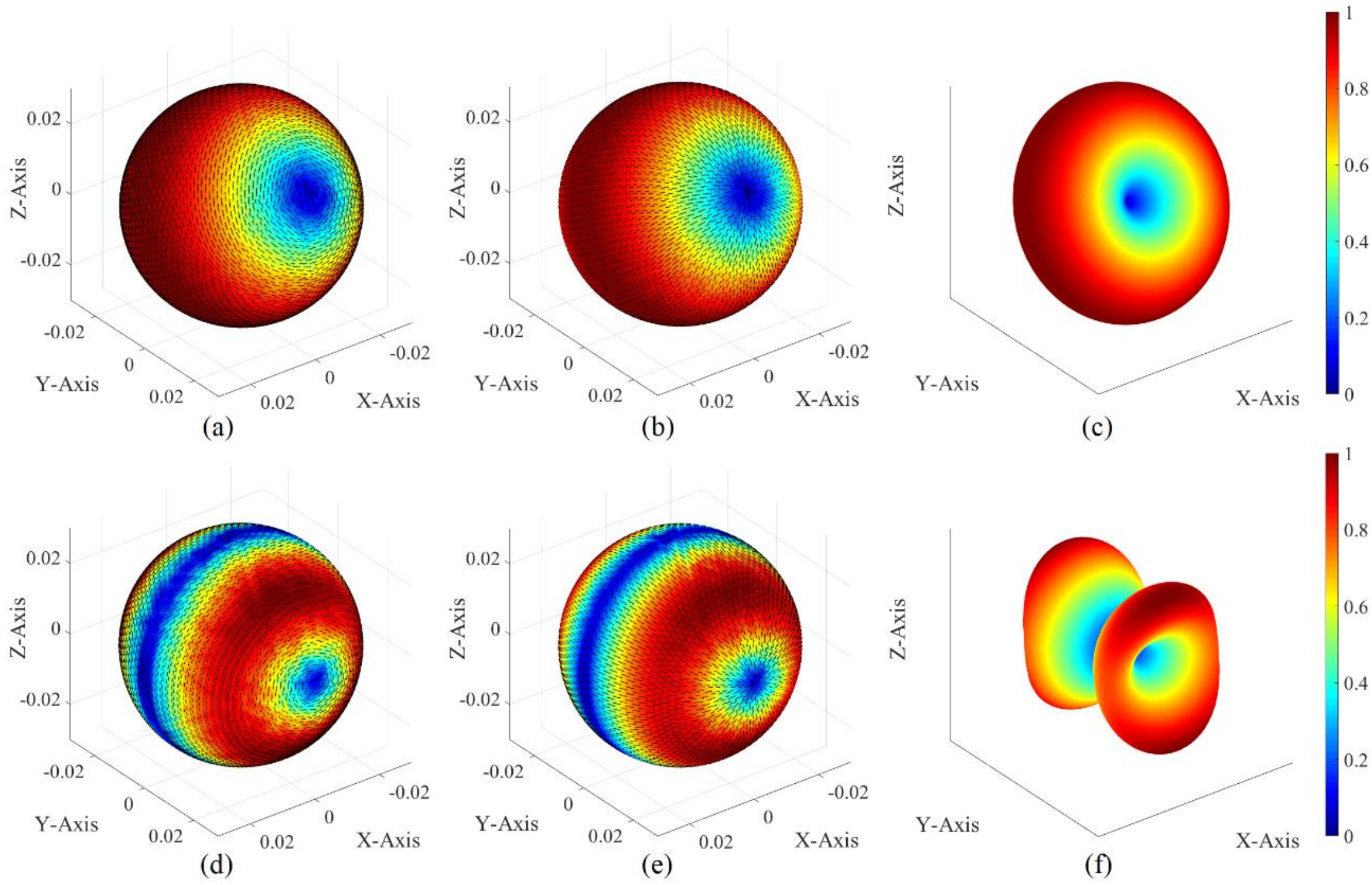

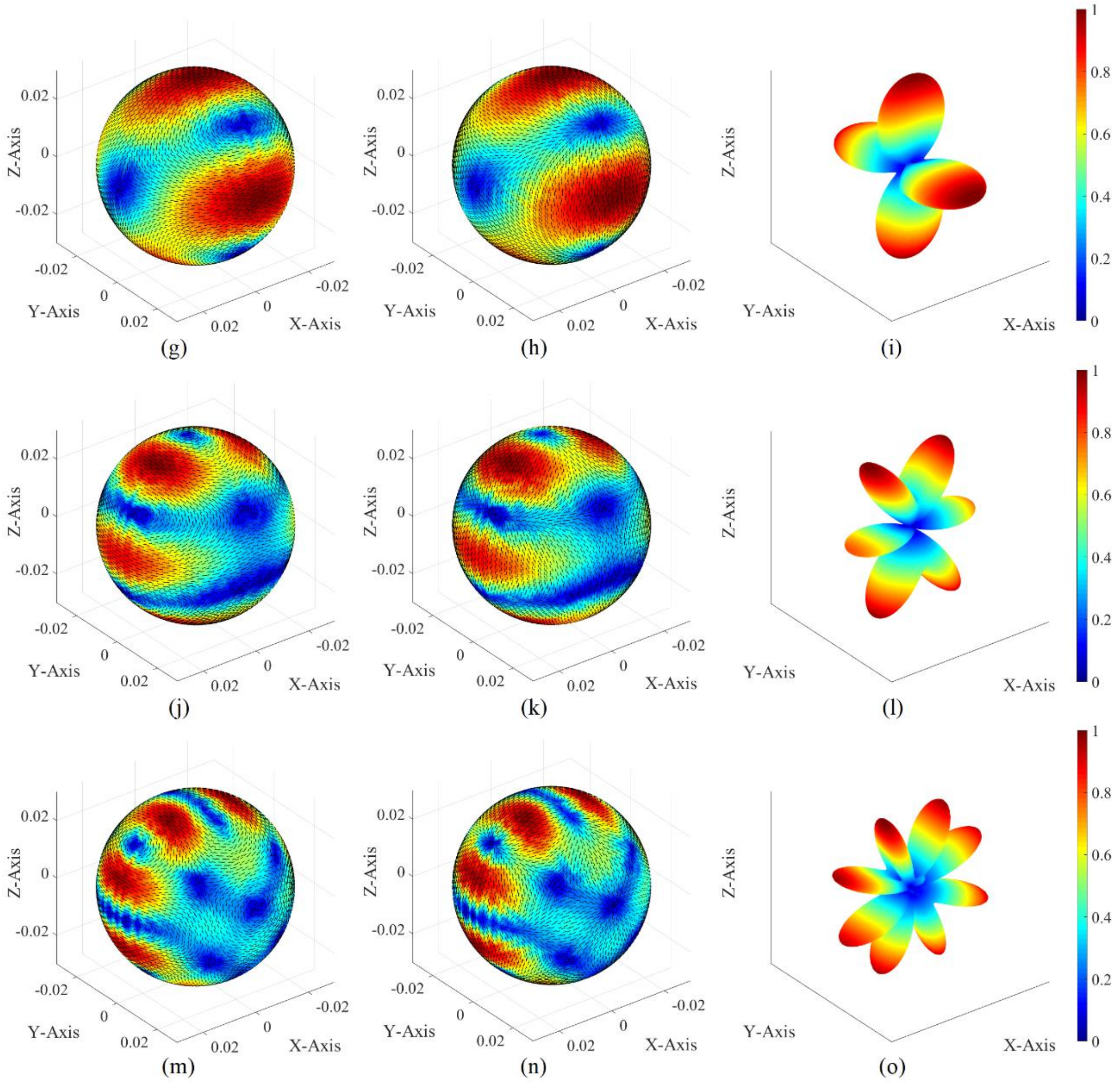

Figure 3-31 Modal equivalent electric currents and radiation patterns of 10 typical DMs.

Clearly, besides the above spherical-surface-fed free-space waveguide, PTT-DMT is also applicable to the following rectangular-surface-fed and tetrahedral-surface-fed free-space waveguides, and then leads to “rectangular modes” and “tetrahedral modes”.

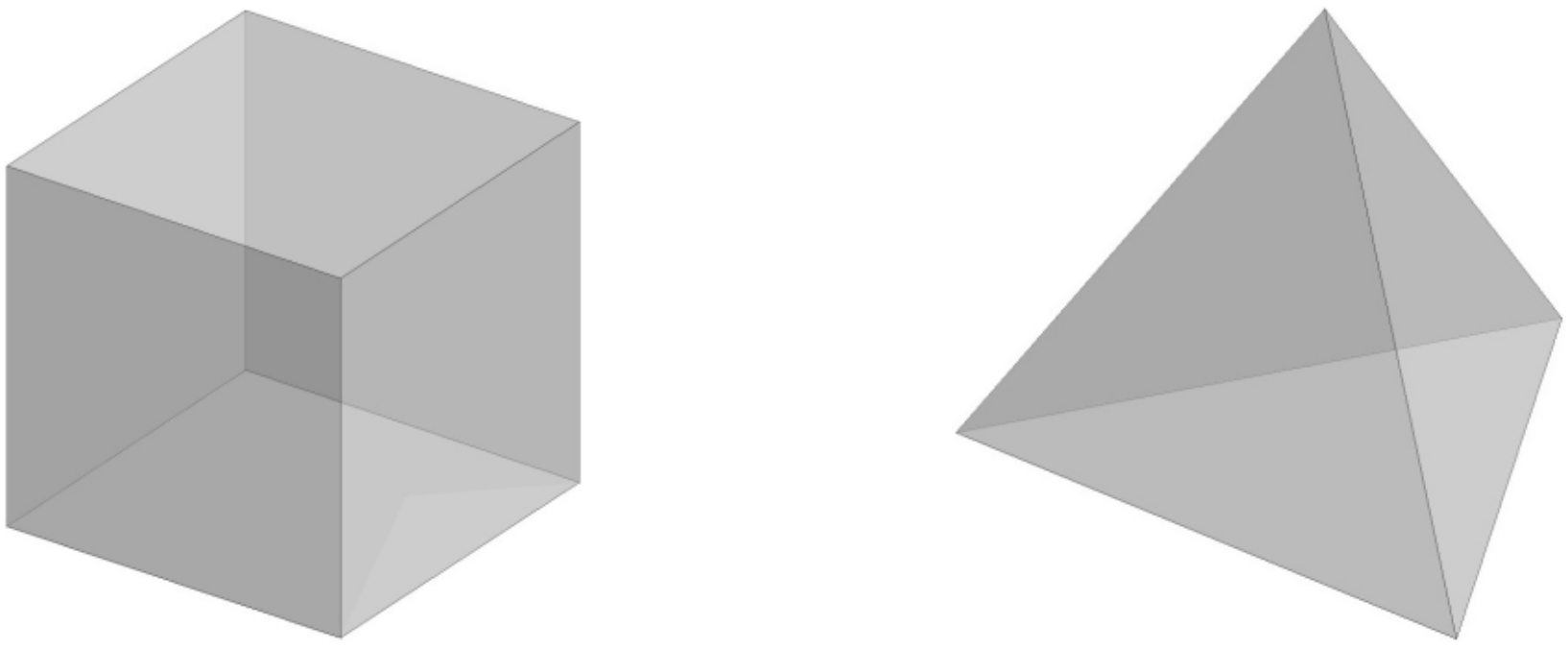

(a) (b)

Figure 3-32 (a) Rectangular and (b) tetrahedral input ports for feeding free-space waveguide.

## 3.6 PTT-Based DMs of Wave-Port-Fed Combined Systems

The one shown in the following Fig. 3-33 is a complete transceiving system, which is constituted by a series of cascaded structures — power source, feeding waveguide, transmitting antenna, surrounding medium, receiving antenna, loading waveguide, and power load.

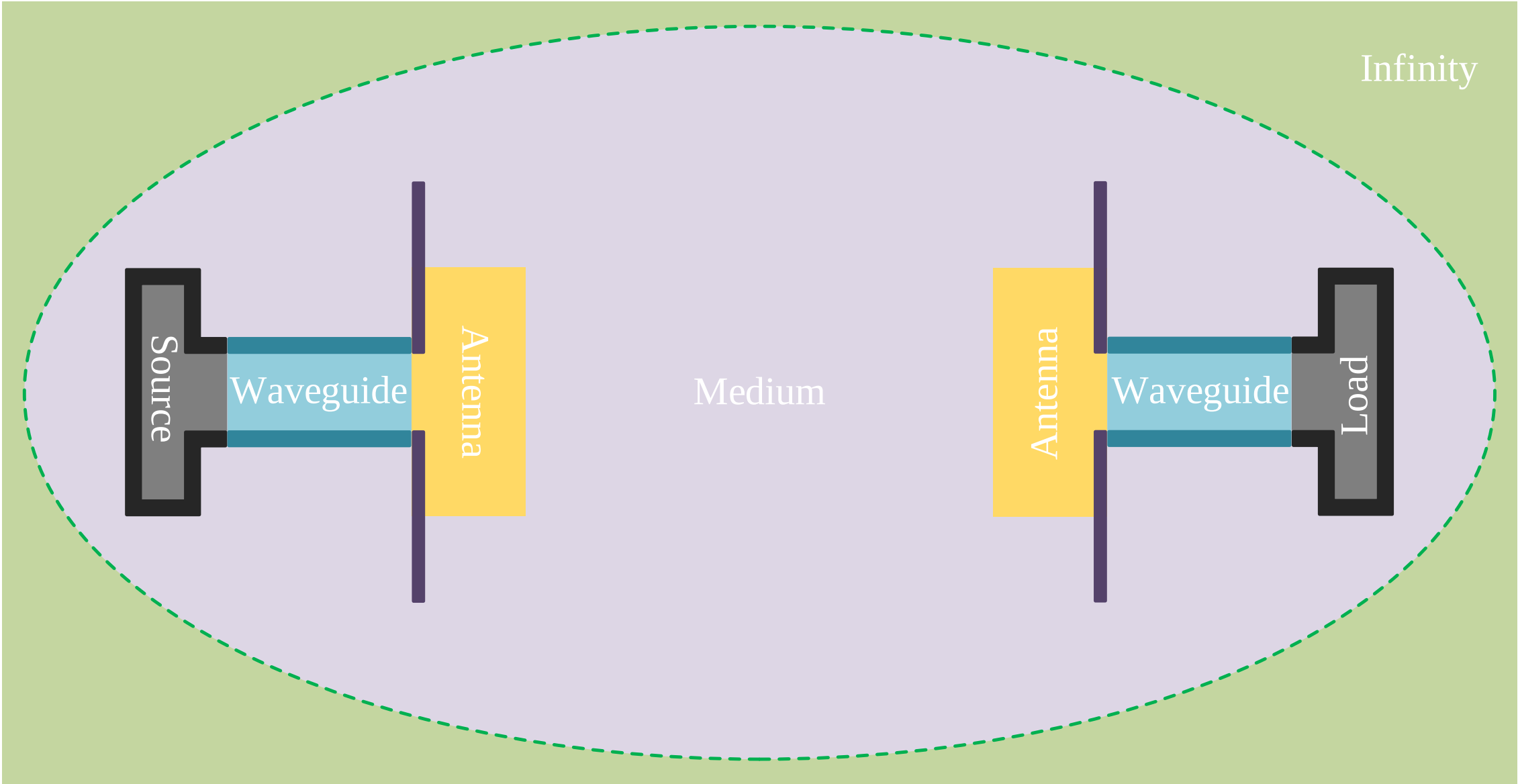

Figure 3-33 Transceiving system and its sub-structures.

During the working process of whole transceiving system, there exists a strong inter-action among the EM fields generated by the structures[54~56]. The inter-action will lead to a complicated inter-transformation/inter-excitation among the fundamental modes of the structures[54~56]. This section focuses on discussing the modal inter-transformation/inter-excitation.

### 3.6.1 Modal Matching Process for Multiple Cascaded Structures

Modal matching method[54~56] is a scheme for quantitatively describing and analyzing the modal inter-transformation/excitation. Some careful discussions for the eigen-mode-based modal matching method can be found in Refs. [54~56], and a detailed discussion for the DM-based modal matching method can be found in Ref. [27-Chap.5].

In Ref. [27-Chap.5], it was also exposed that: the modal matching process is very cumbersome, and the reason leading to the indispensability for modal matching process originates from seperately treating the regions occupied by the structures. Based on this observation, we propose some schemes for avoiding the modal matching process as below.

### 3.6.2 PTT-Based DMs of Waveguide-Antenna Cascaded Systems

Taking the one shown in Fig. 3-34 (which is surrounded by free space and constituted by a metallic tube waveguide and a metallic horn transmitting antenna) as an example, this section focuses on establishig the PTT-DMT for waveguide-antenna cascaded systems.

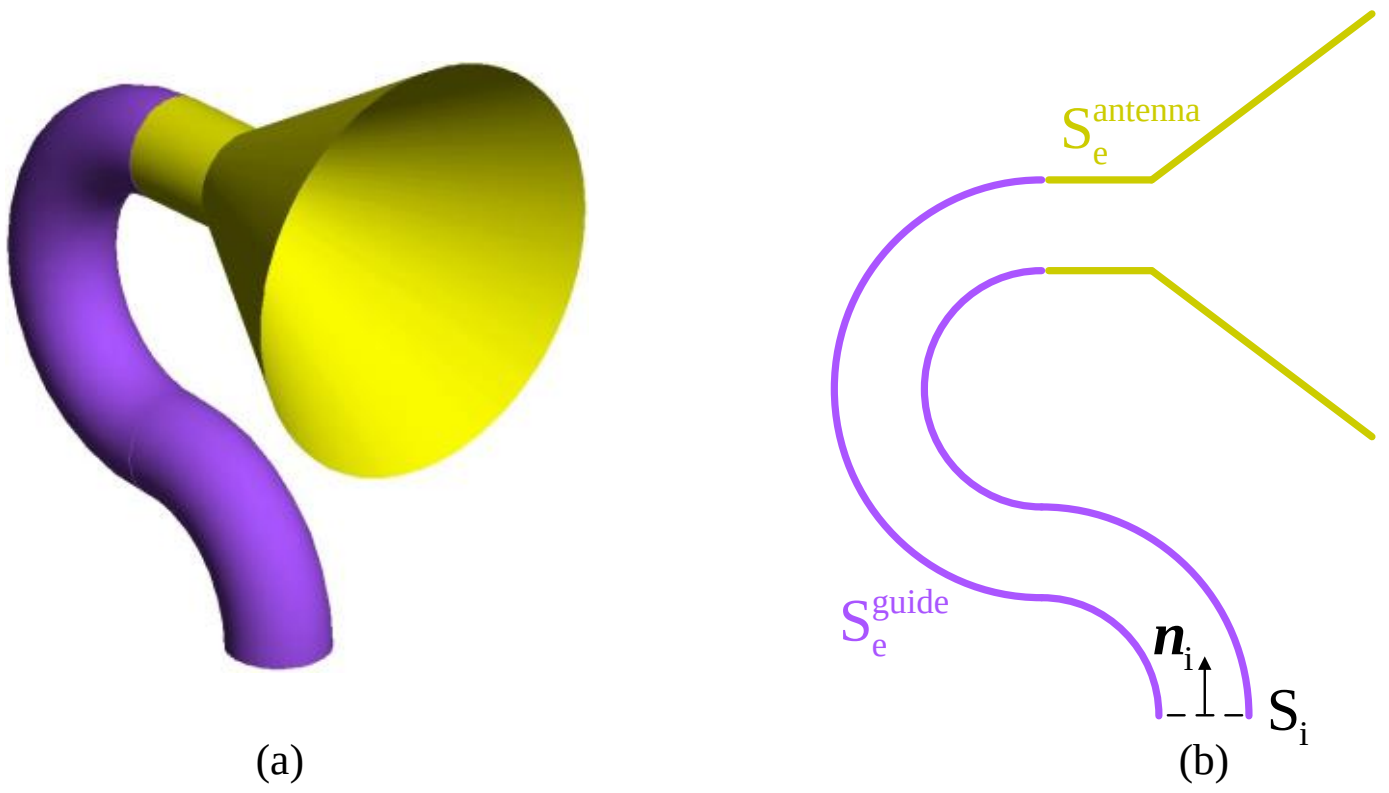

Figure 3-34 (a) Geometry and (b) topology of a typical waveguide-antenna cascaded system.

As shown in Fig. 3-34(b), the input port of waveguide is denoted as $S_i$, and its normal direction is denoted as $\boldsymbol{n}_i$; the electric walls of waveguide and antenna are denoted as $S_e^{guide}$ and $S_e^{antenna}$ respectively. Obviously, if we denote the union of $S_e^{guide}$ and $S_e^{antenna}$ as $S_e$, i.e., $S_e = S_e^{guide} \bigcup S_e^{antenna}$, then the topological structure of the above cascaded system is identical to the topological structure of the horn discussed in Sec. 3.2. Thus, the formulations established in Sec. 3.2 can be directly applied to calculating the DMs of the cascaded system shown in Fig. 3-34.

Now, we let the specific size of the cascaded system in Fig. 3-34 be as the one shown in following Fig. 3-35, and employ the JE-DoJ-based formulation to calculate the DMs.

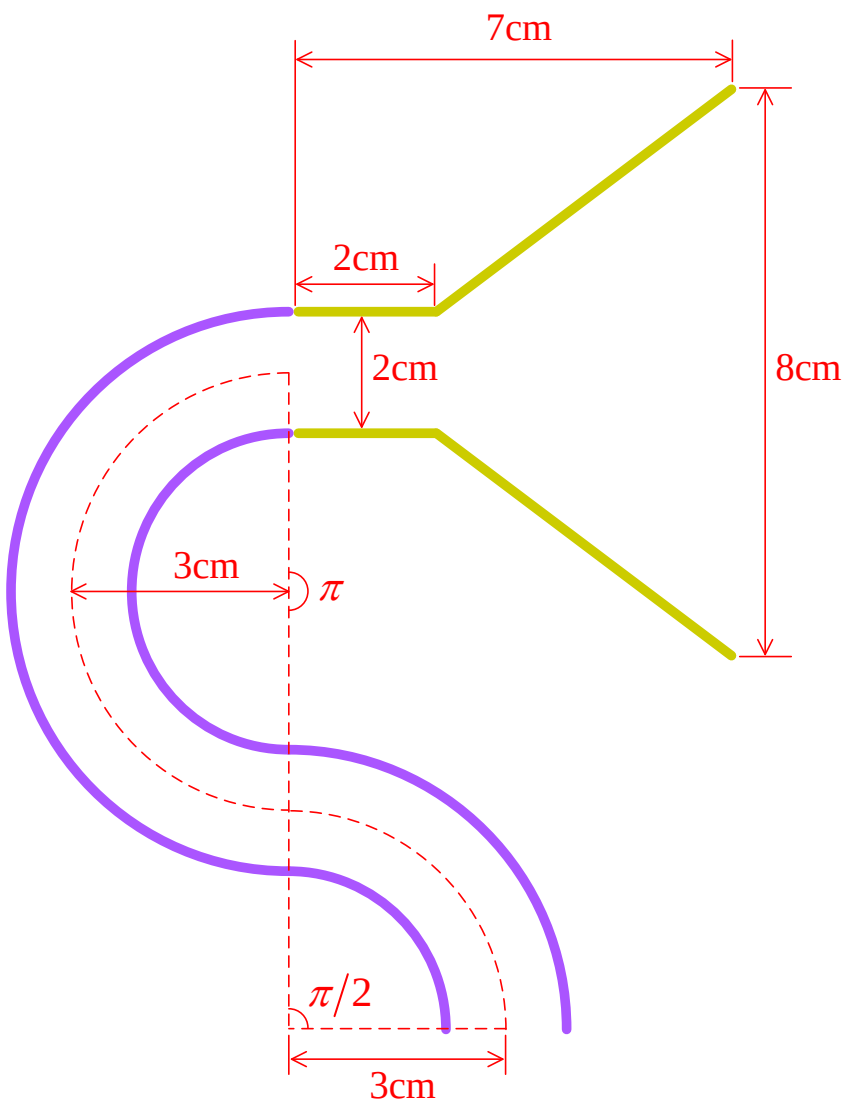

Figure 3-35 Size of a specific waveguide-antenna cascaded system.

The modal resistance curves corresponding to the first two lower-order DMs are shown in the following Fig. 3-36.

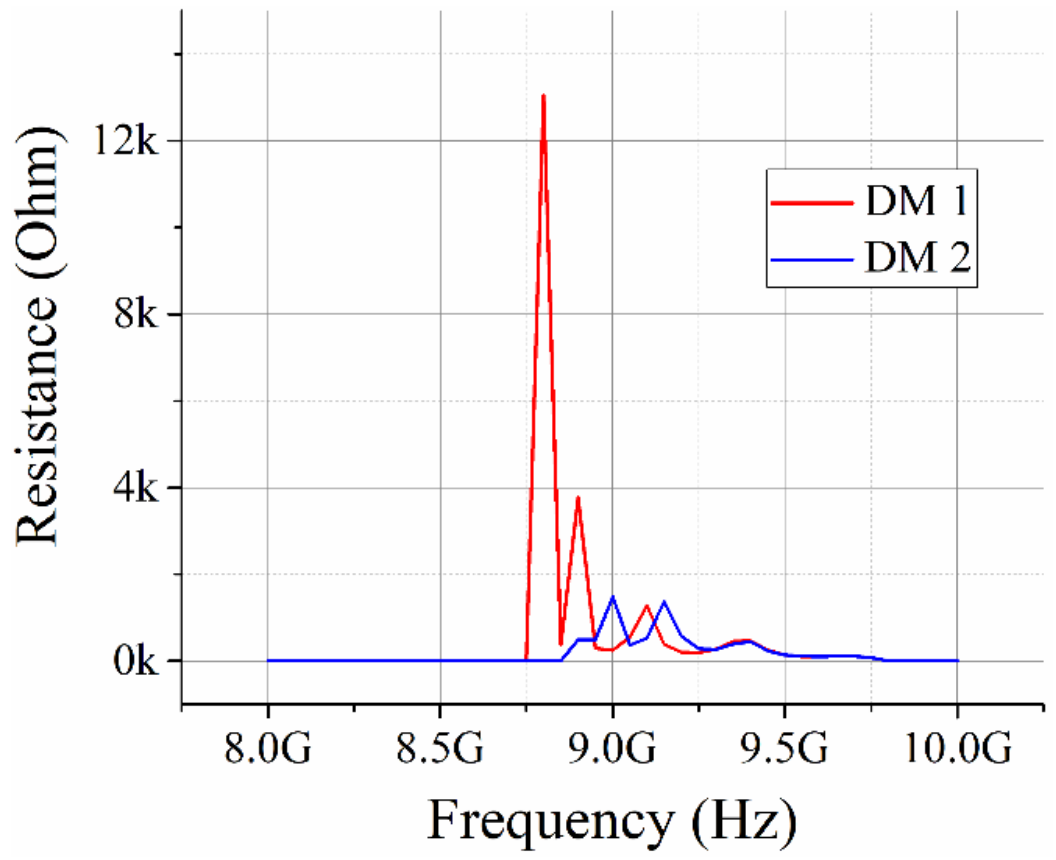

Figure 3-36 Modal resistance curves of the first several lower-order DMs derived from the JE-DoJ-based formulation.

Taking the DM 1 as an example, its equivalent surface electric current and equivalent surface magnetic current distributing on the input port of whole cascaded system are shown in the following Fig. 3-37.

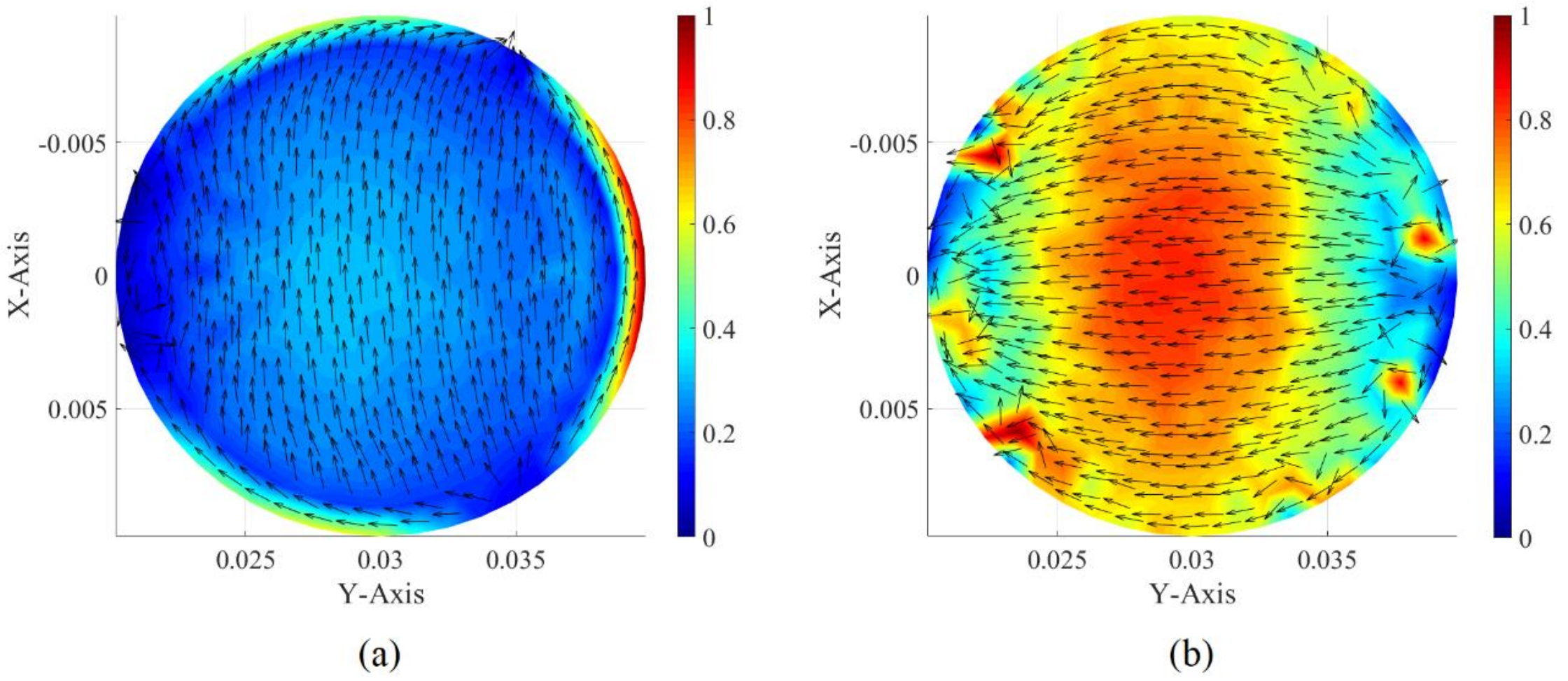

Figure 3-37 Modal equivalent (a) electric and (b) magnetic currents of the JE-DoJ-based DM 1 shown in Fig. 3-36.

From the Fig. 3-36, it is easy to find out that the DM 1 curve reaches the local peaks at 8.8 GHz, 8.9 GHz, 9.1 GHz, and 9.4 GHz. The modal induced electric currents (on the metallic electric wall) corresponding to the four peak/resonance frequencies are shown in the following Fig. 3-38.

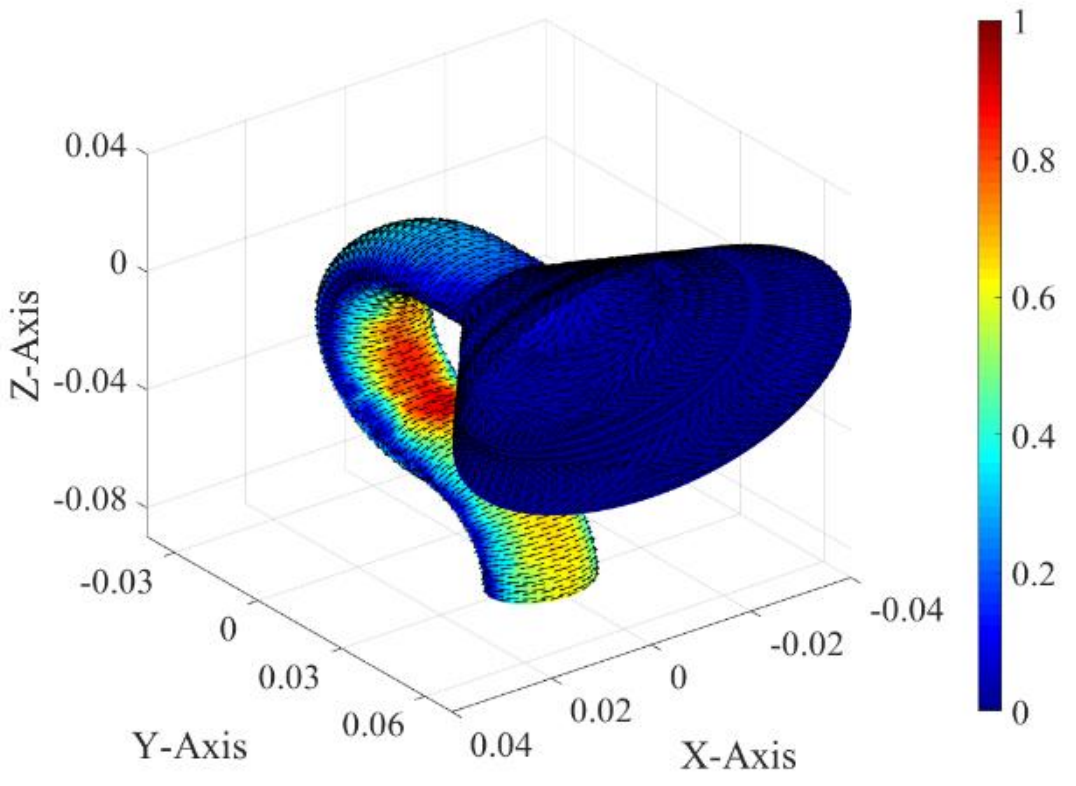

(a)

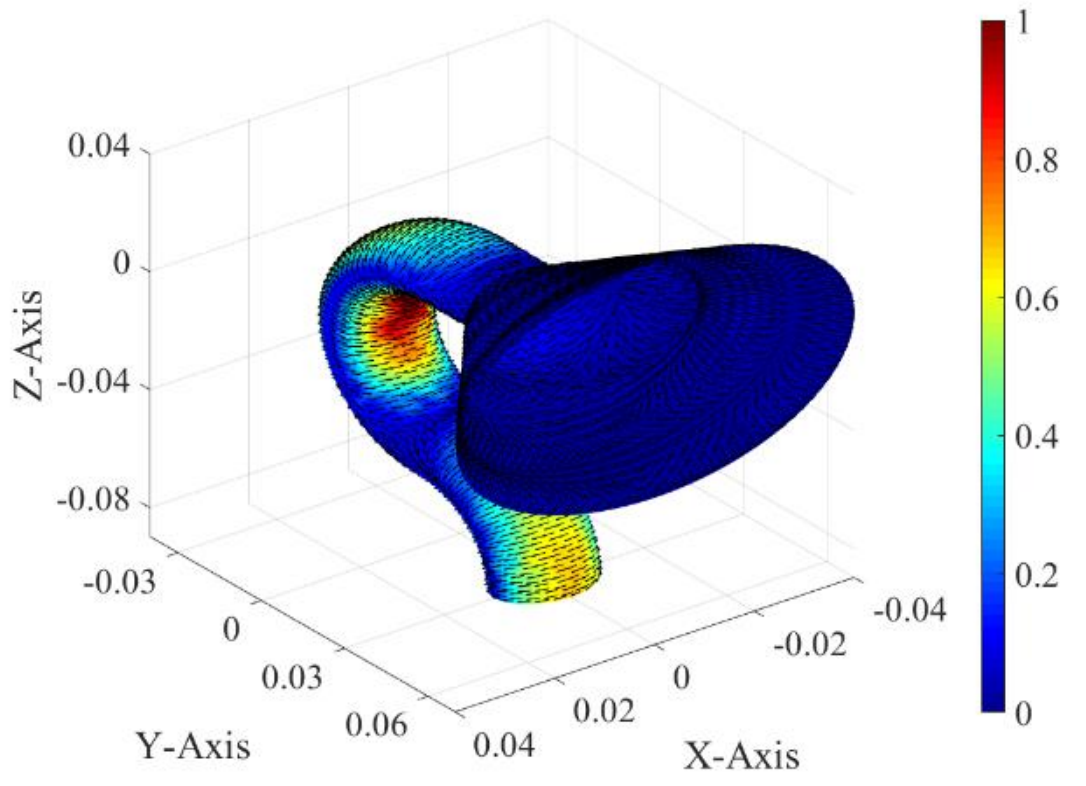

(b)

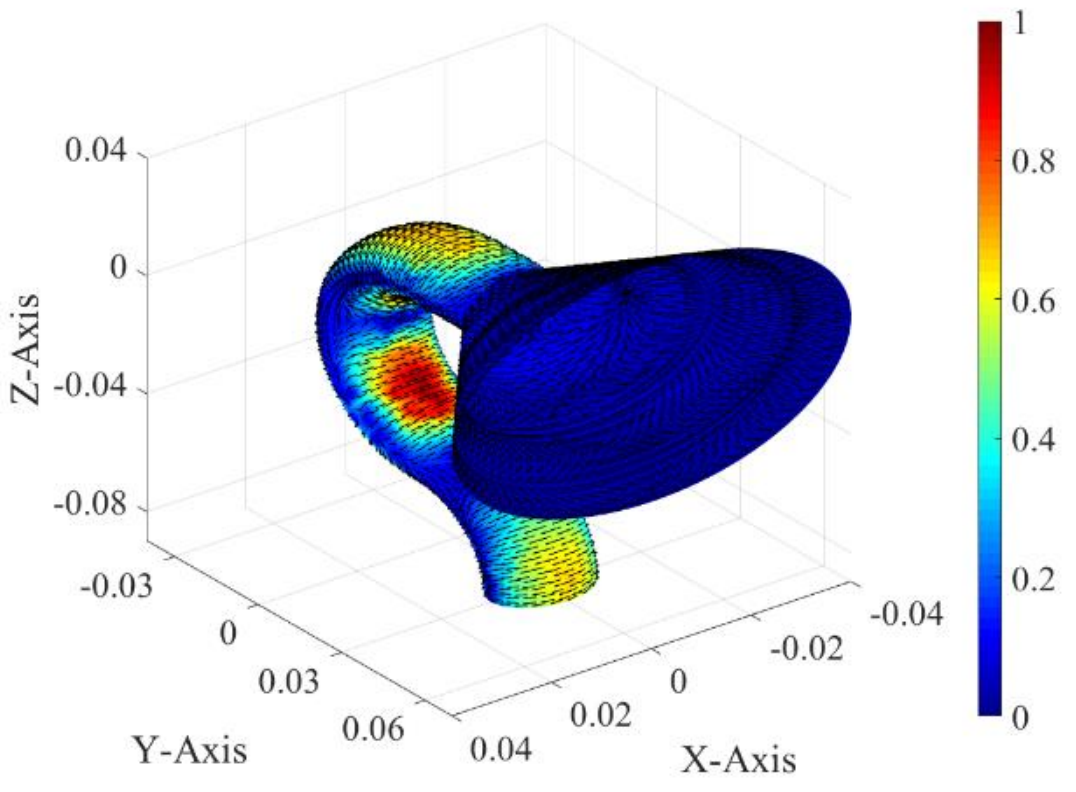

(c)

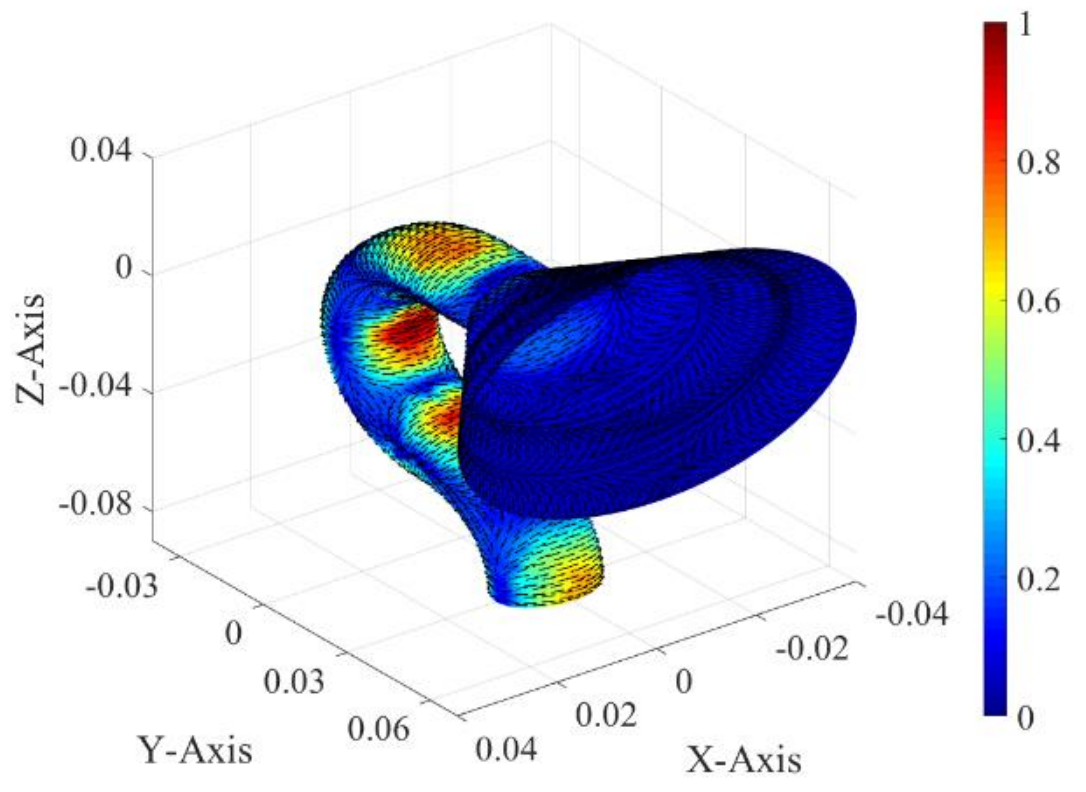

(d)

Figure 3-38 Modal induced electric current (on metallic electric wall) of the JE-DoJ-based DM 1 shown in Fig. 3-36. (a) 8.8 GHz; (b) 8.9 GHz; (c) 9.1 GHz; (d) 9.4 GHz.

In addition, we also plot the magnitude distributions of the modal electric field (on yOz plane) of the JE-DoJ-based DM 1 (shown in the previous Fig. 3-36) working at the four resonance frequencies 8.8 GHz, 8.9 GHz, 9.1 GHz, and 9.4 GHz in the following Fig. 3-39.

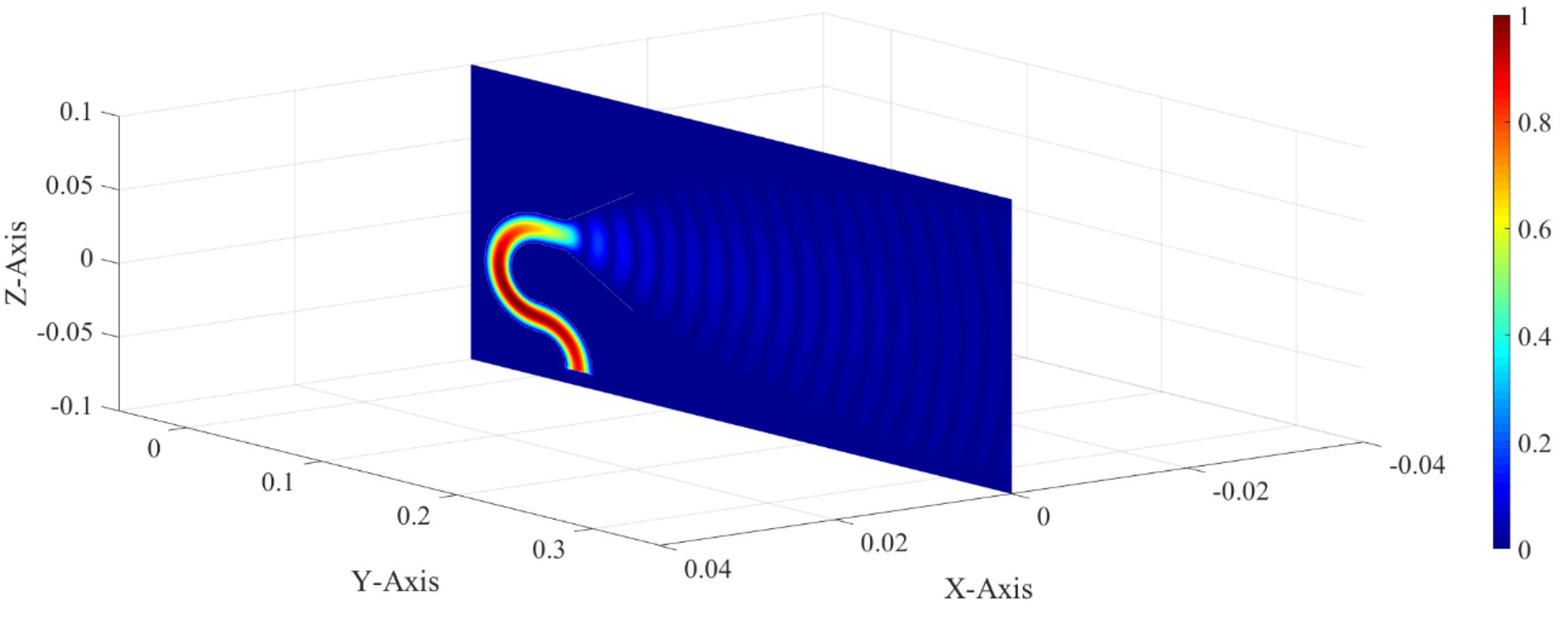

(a)

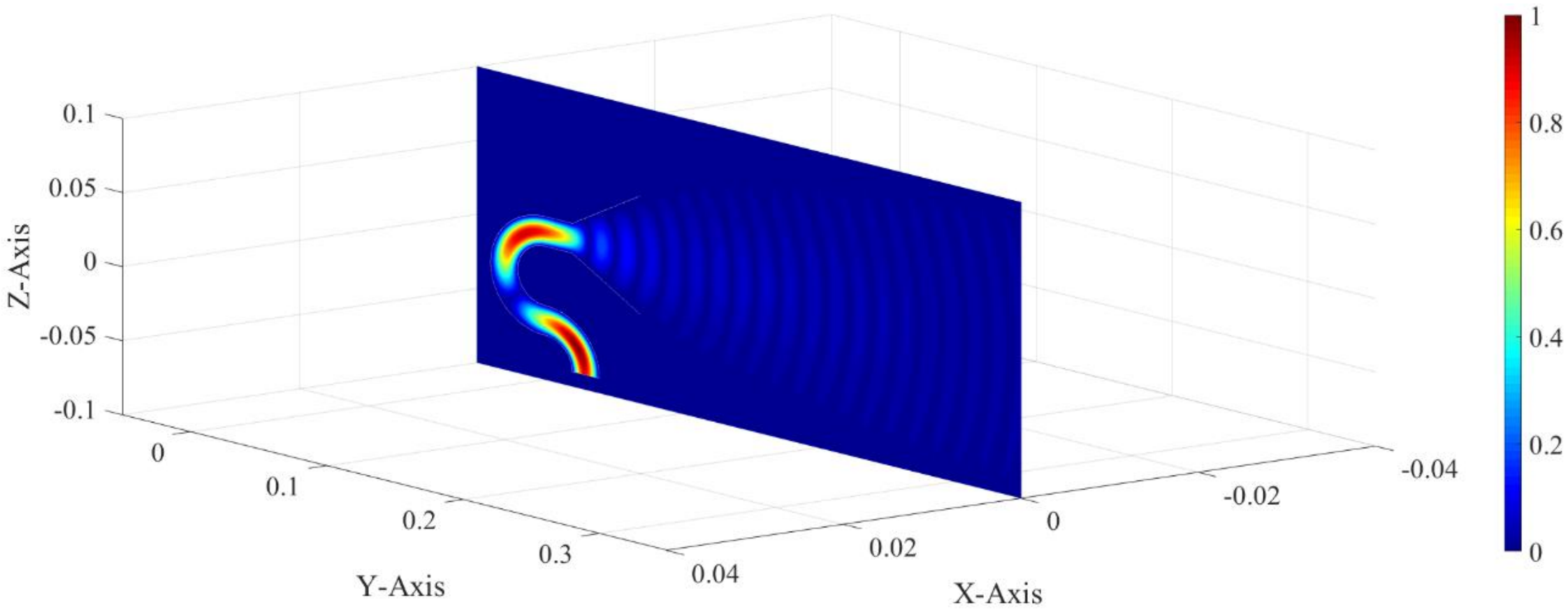

(b)

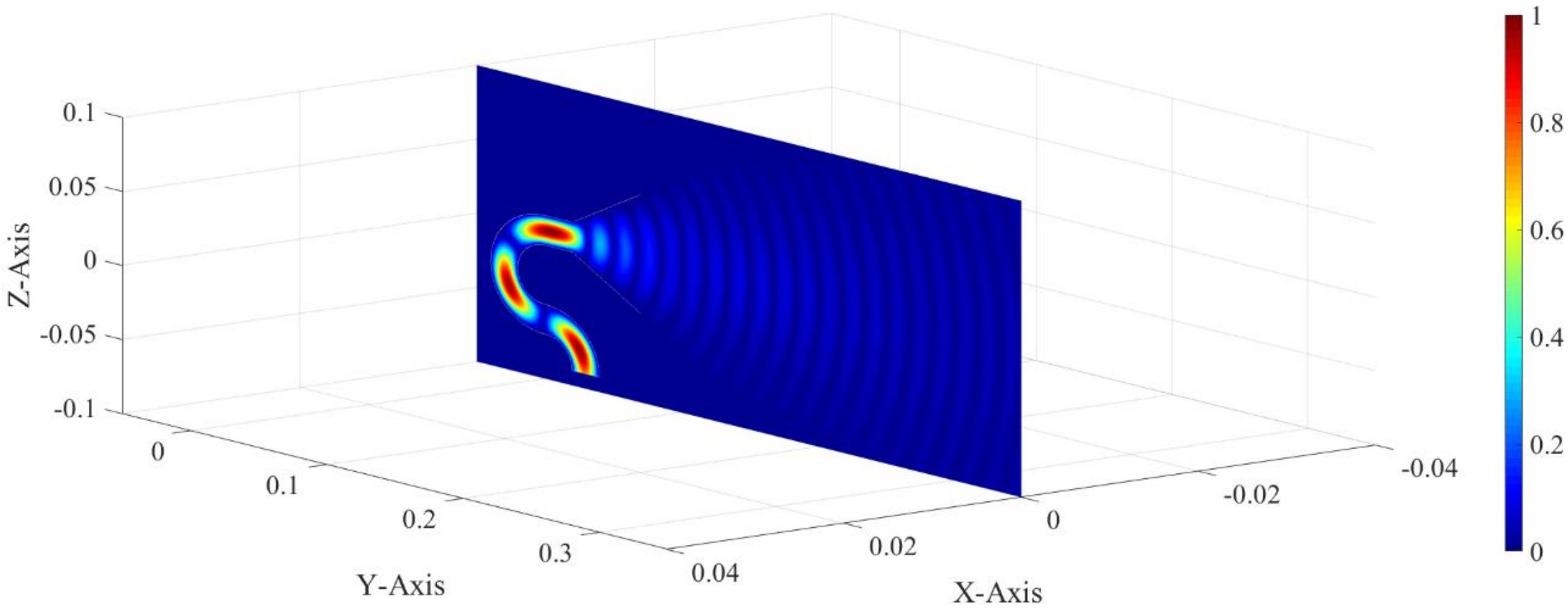

(c)

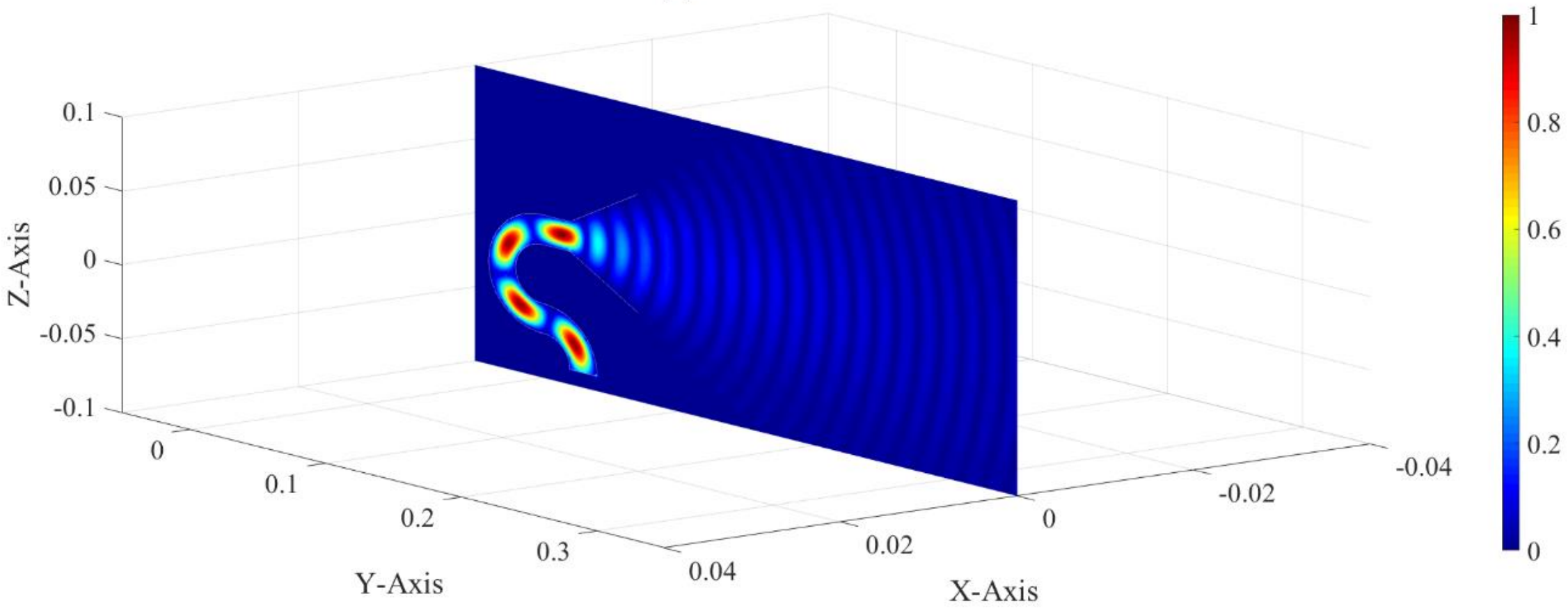

(d)

Figure 3-39 Modal electric field distributions (on yOz plane) of the JE-DoJ-based DM 1 shown in Fig. 3-36 at a series of resonance frequencies. (a) 8.8 GHz; (b) 8.9 GHz; (c) 9.1 GHz; (d) 9.4 GHz.

At last, we visually plot the radiation pattern of the JE-DoJ-based DM 1 in the following Fig. 3-40.

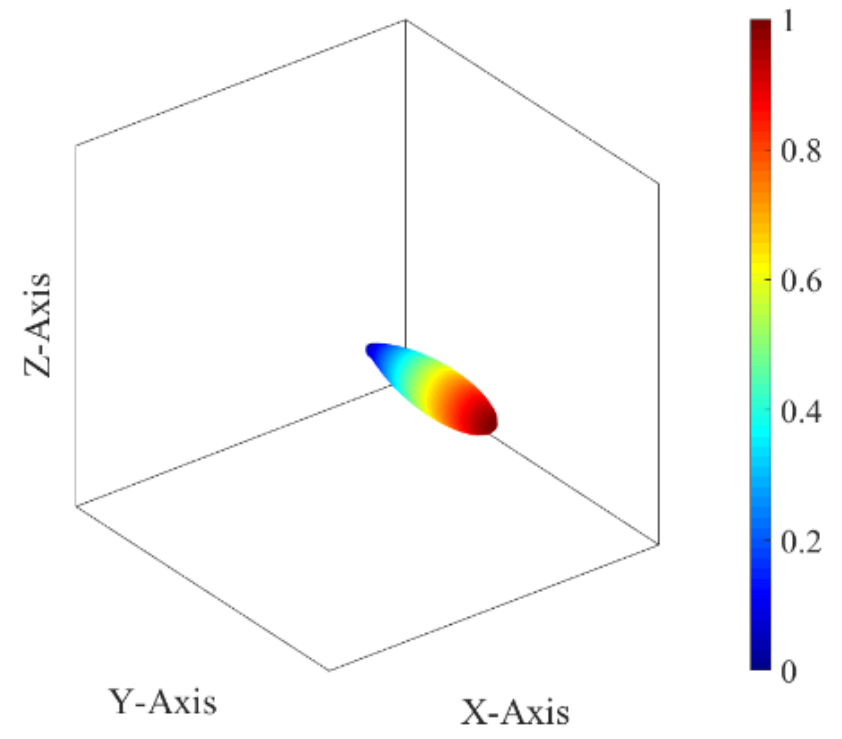

Figure 3-40 Modal radiation pattern of the JE-DoJ-based DM 1 shown in Fig. 3-36.

In fact, as exhibited in Ref. [27-Sec.8.2.6.2], the above method can also be directly used to do the modal analysis for the waveguide-fed 2-element horn antenna array shown in the following Fig. 3-41.

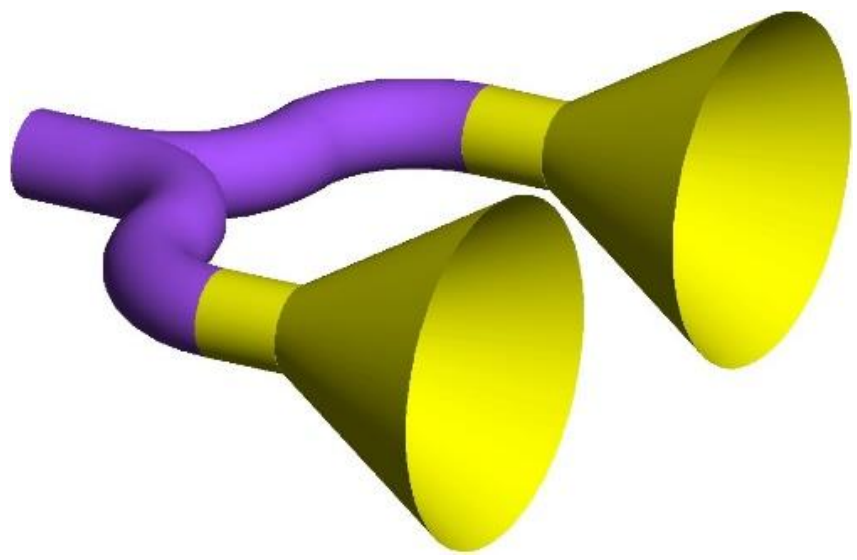

Figure 3-41 Geometry of a waveguide-fed horn antenna array.

In addition, as exhibited in Ref. [27-Sec.8.2], the PTT-DMT for the above-mentioned relatively simple waveguide-antenna cascaded system can be generalized to some more complicated waveguide-antenna cascaded systems, such as the one shown in Fig. 3-42.

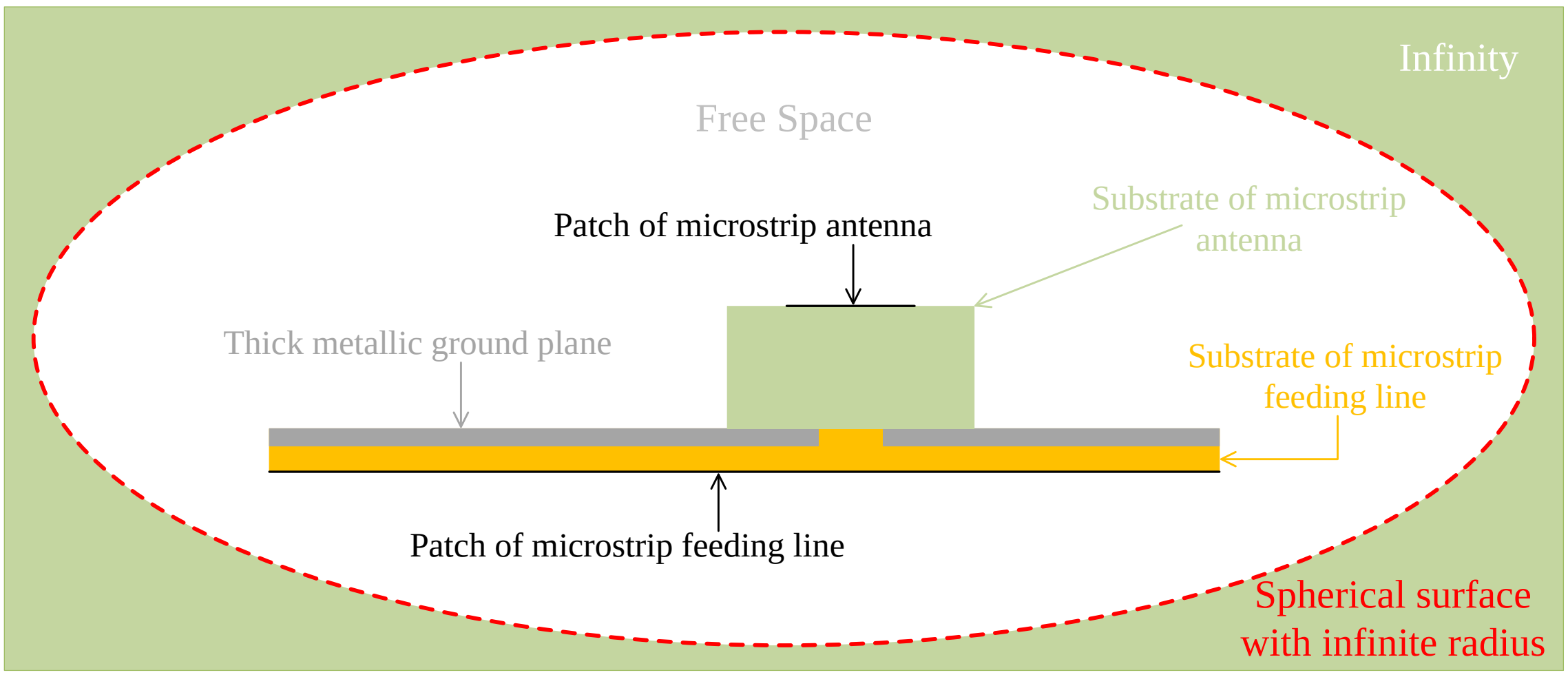

Figure 3-42 Geometry of a somewhat complicated waveguide-antenna cascaded system, which is constituted by a microstrip feeding line and a microstrip patch antenna.

### 3.6.3 PTT-Based DMs of Waveguide-Antenna-Medium-Antenna-Waveguide Cascaded Systems

In this section, we further generalize the PTT-DMT to a kind of more complicated multi-structure cascaded system — waveguide-antenna-medium-antenna-waveguide cascaded system, and a typical one is shown in the following Fig. 3-43.

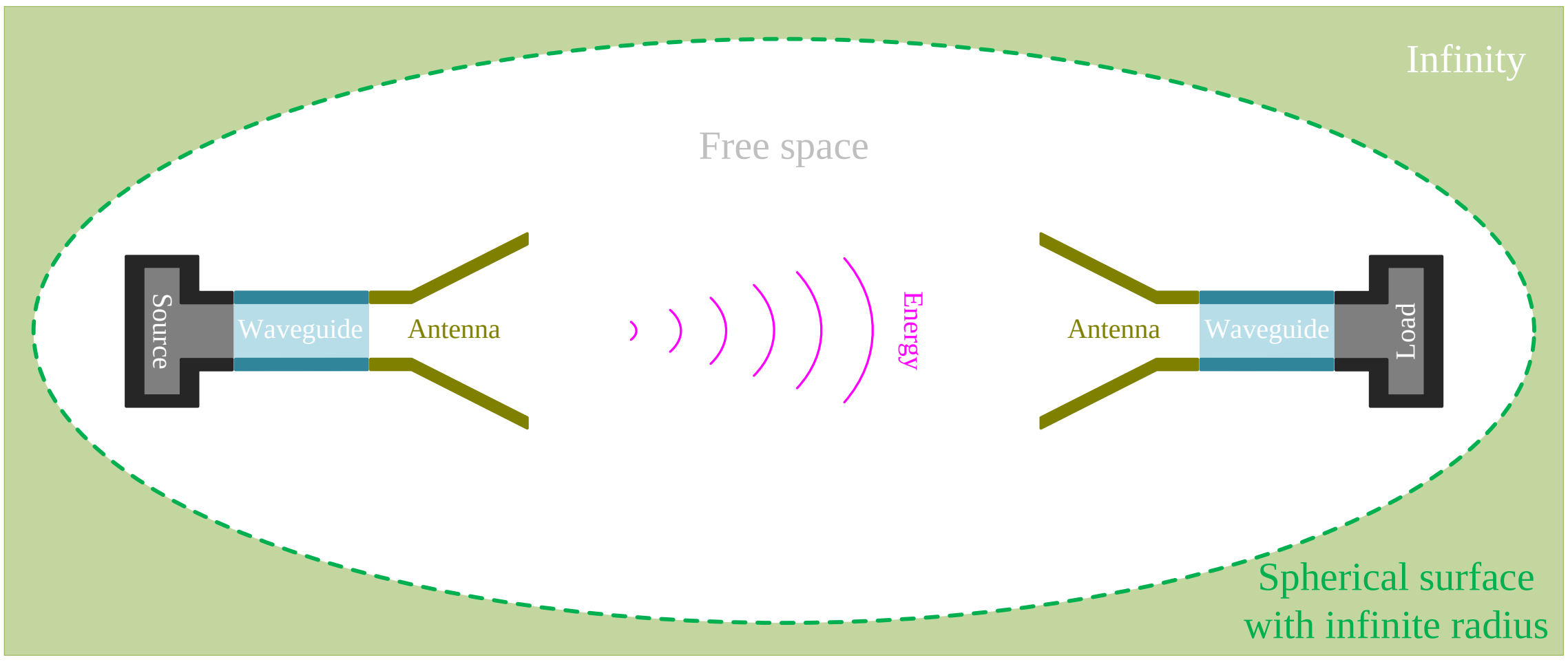

Figure 3-43 Geometry of a multi-structure cascaded system constituted by feeding waveguide, transmitting horn, free-space medium, receiving horn, and loading waveguide.

The topological structure of the EM problem in Fig. 3-43 is illustrated in Fig. 3-44.

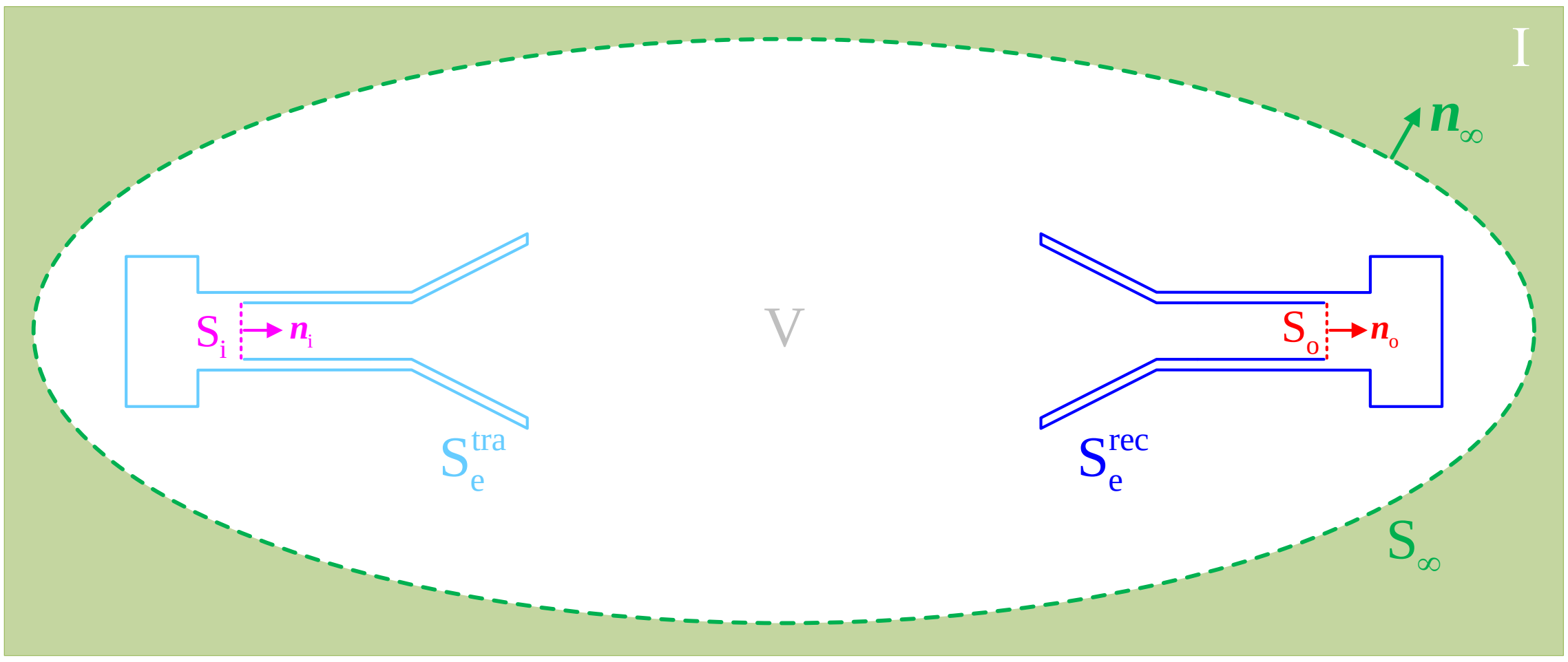

Figure 3-44 Topological structure of the EM problem shown in Fig. 3-43.

Here, $S_i$ and $S_o$ are the input and output ports of the cascaded system, and their normal directions are $\boldsymbol{n}_i$ and $\boldsymbol{n}_o$ respectively. The surrounding medium is free space, and is denoted as $V$. The electric walls used to separate $V$ from the transmitting and receiving systems are denoted as $S_e^{tra}$ and $S_e^{rec}$ respectively. For simplifying the symbolic system, the $S_e^{tra}$ and $S_e^{rec}$ are treated as a whole $S_e$, i.e., $S_e = S_e^{tra} \bigcup S_e^{rec}$.

If the equivalent currents on $\mathrm{S_{i/o}}$ are defined as that $\boldsymbol{J}_{\mathrm{i/o}} = \boldsymbol{n}_{\mathrm{i/o}} \times \boldsymbol{H}$ and $\boldsymbol{M}_{\mathrm{i/o}} = \boldsymbol{E} \times \boldsymbol{n}_{\mathrm{i/o}}$, and the induced current on $\mathrm{S_e}$ is denoted as $\boldsymbol{J}_{\mathrm{e}}$, then input power $P_{\mathrm{in}} = (1/2)\iint_{\mathrm{S_i}} (\boldsymbol{E} \times \boldsymbol{H}^{\dagger}) \cdot \boldsymbol{n}_{\mathrm{i}} dS$ can be expressed as the following integral operators

$$\begin{aligned} P_{\mathrm{in}} &= -(1/2)\left\langle \boldsymbol{J}_{\mathrm{i}}, \mathcal{E}_0\left(\boldsymbol{J}_{\mathrm{i}} + \boldsymbol{J}_{\mathrm{e}} - \boldsymbol{J}_{\mathrm{o}}, \boldsymbol{M}_{\mathrm{i}} - \boldsymbol{M}_{\mathrm{o}}\right)\right\rangle_{\mathrm{S_i^+}} \\ &= -(1/2)\left\langle \boldsymbol{M}_{\mathrm{i}}, \mathcal{H}_0\left(\boldsymbol{J}_{\mathrm{i}} + \boldsymbol{J}_{\mathrm{e}} - \boldsymbol{J}_{\mathrm{o}}, \boldsymbol{M}_{\mathrm{i}} - \boldsymbol{M}_{\mathrm{o}}\right)\right\rangle_{\mathrm{S_i^+}}^{\dagger} \end{aligned} \tag{3-40}$$

where $\mathrm{S_i^+}$ is the right-side surface of $\mathrm{S_i}$. Obviously, these IPOs have the same mathematical structure as the ones given in Eq. (3-27) (metallic waveguide case). In fact, the integral equations used to establish the transformations among the currents involved in the above IPOs also have the same mathematical structure as the ones given in Eqs. (3-29a)~(3-31b) (metallic waveguide case). Thus, the modal decoupling equation used to calculate the DMs of the above cascaded system has also identical mathematical structure to the one used for the metallic waveguide case, and the obtained DMs satisfy

$$(1/T)\int_{t_0}^{t_0+T}\left[\iint_{\mathrm{S_i}}\left(\boldsymbol{\mathcal{E}}_n \times \boldsymbol{\mathcal{H}}_m\right)\cdot \boldsymbol{n}_{\mathrm{i}} dS\right]dt = \delta_{mn} = (1/T)\int_{t_0}^{t_0+T}\left[\iint_{\mathrm{S_i}}\left(\boldsymbol{\mathcal{E}}_m \times \boldsymbol{\mathcal{H}}_n\right)\cdot \boldsymbol{n}_{\mathrm{i}} dS\right]dt \tag{3-41}$$

and then LORENTZ'S RECIPROCITY THEOREM (2-22) implies energy-decoupling relation

$$(1/T)\int_{t_0}^{t_0+T}\left[\iint_{\mathrm{S_o}}\left(\boldsymbol{\mathcal{E}}_n \times \boldsymbol{\mathcal{H}}_m + \boldsymbol{\mathcal{E}}_m \times \boldsymbol{\mathcal{H}}_n\right)\cdot \boldsymbol{n}_{\mathrm{o}} dS\right]dt = 2\delta_{mn} \tag{3-42}$$

and also implies the following more general energy-decoupling relation

$$(1/T)\int_{t_0}^{t_0+T}\left[\iint_{\mathrm{S}}\left(\boldsymbol{\mathcal{E}}_n \times \boldsymbol{\mathcal{H}}_m + \boldsymbol{\mathcal{E}}_m \times \boldsymbol{\mathcal{H}}_n\right)\cdot \boldsymbol{n} dS\right]dt = 2\delta_{mn} \tag{3-43}$$

where S is an arbitrary closed surface seperating the transmitting antenna from the receiving antenna as shown in Fig. 3-45, and $\boldsymbol{n}$ is the outer normal direction of $\mathrm{S}$.

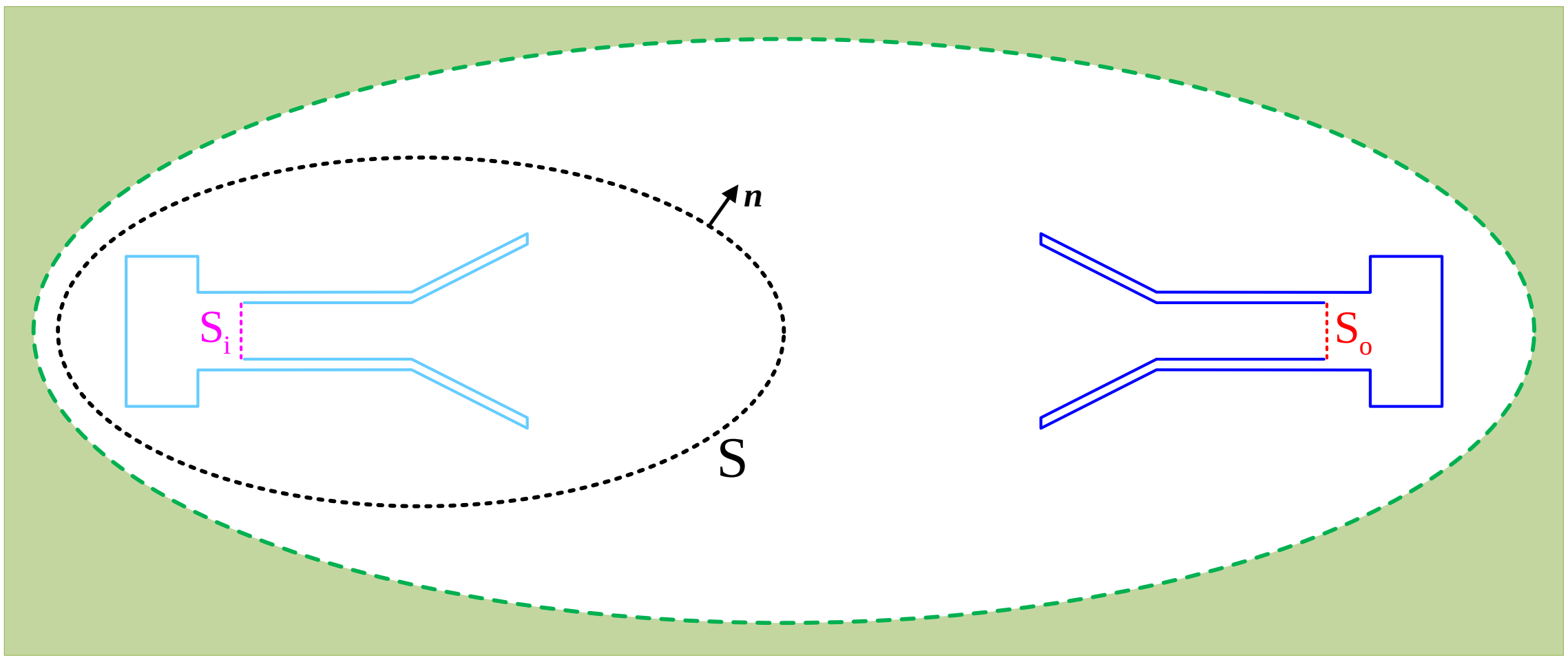

Figure 3-45 A closed surface seperating transmitting and receiving antennas.

For a specific waveguide-antenna-medium-antenna-waveguide cascaded system shown in Fig. 3-46, we do the PTT-DMT-based modal analysis here.

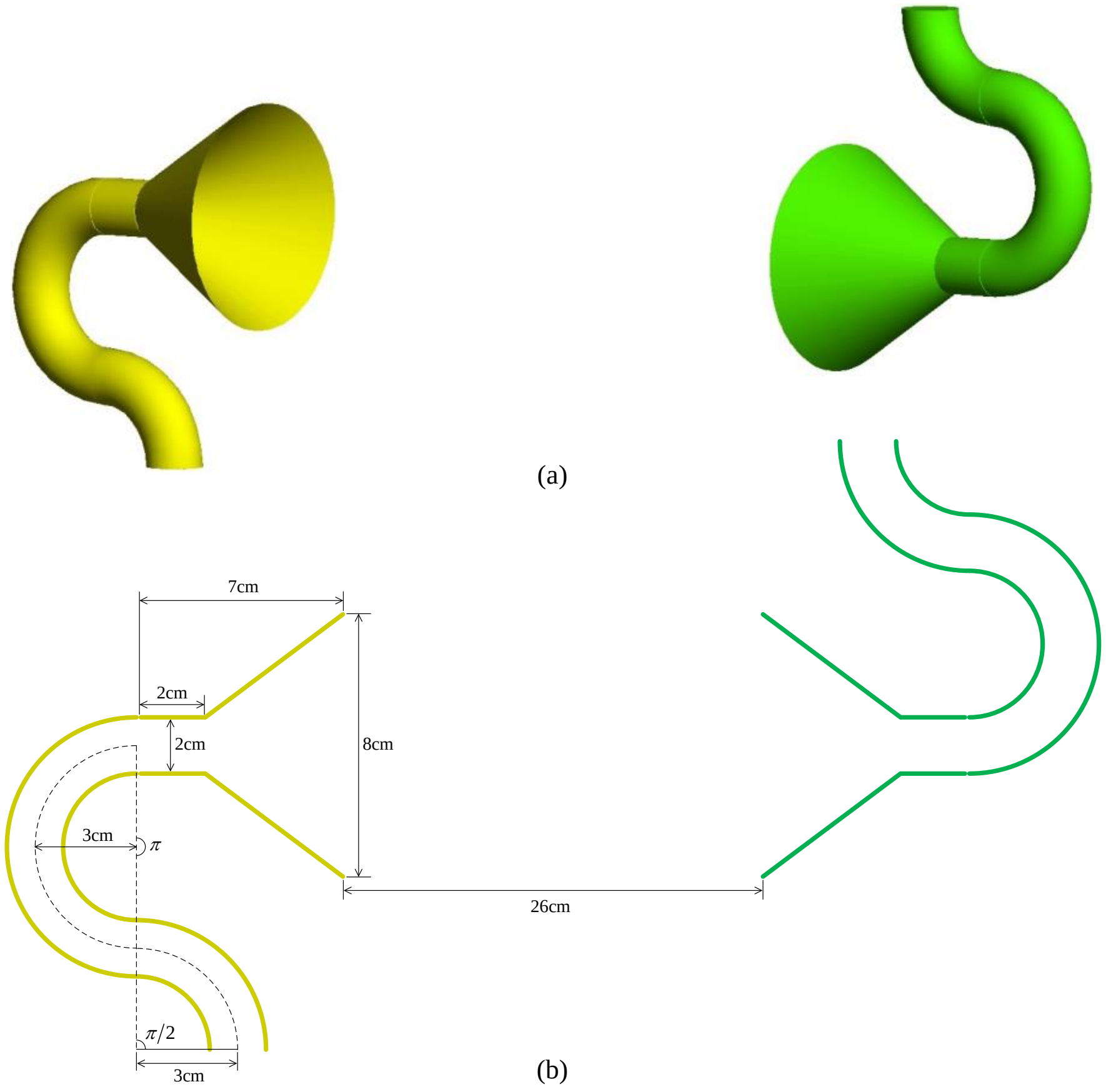

Figure 3-46 (a) Geometry and (b) size of a specific cascaded system.

The JE-DoJ-based modal resistance curves are shown in Fig. 3-47.

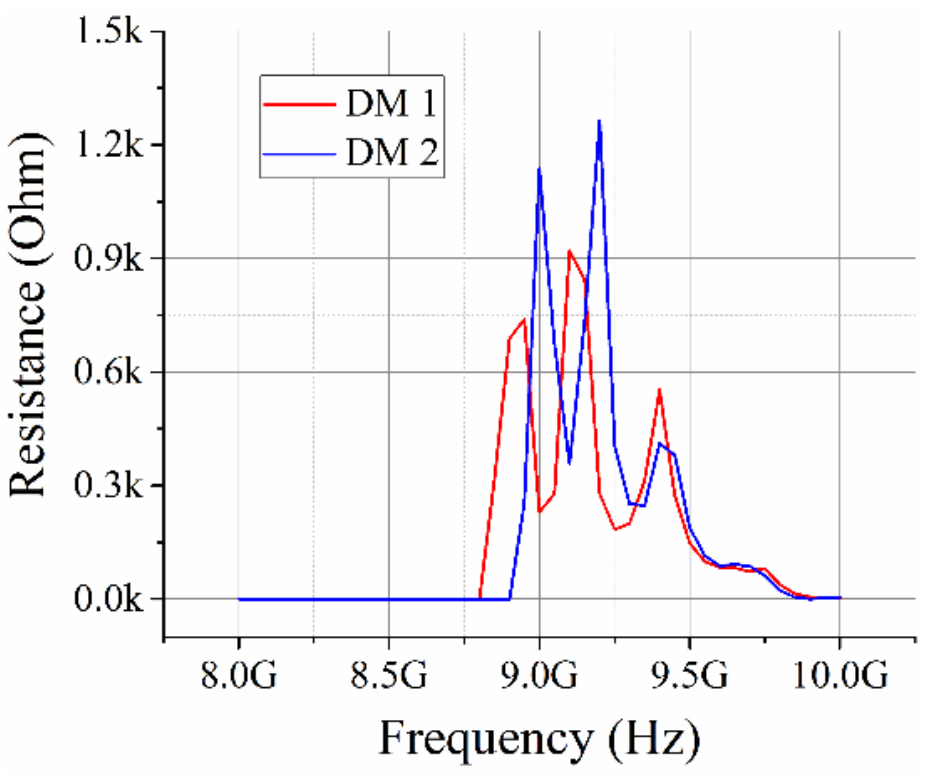

Figure 3-47 Modal resistance curves of the first several lower-order JE-DoJ-based DMs.

For the DM 1, it is easy to find out that the curve reaches the local peaks at 8.95 GHz, 9.10 GHz, and 9.40 GHz. The modal electric field distributions (on yOz plane) corresponding to the three peak/resonance frequencies are shown in Fig. 3-48.

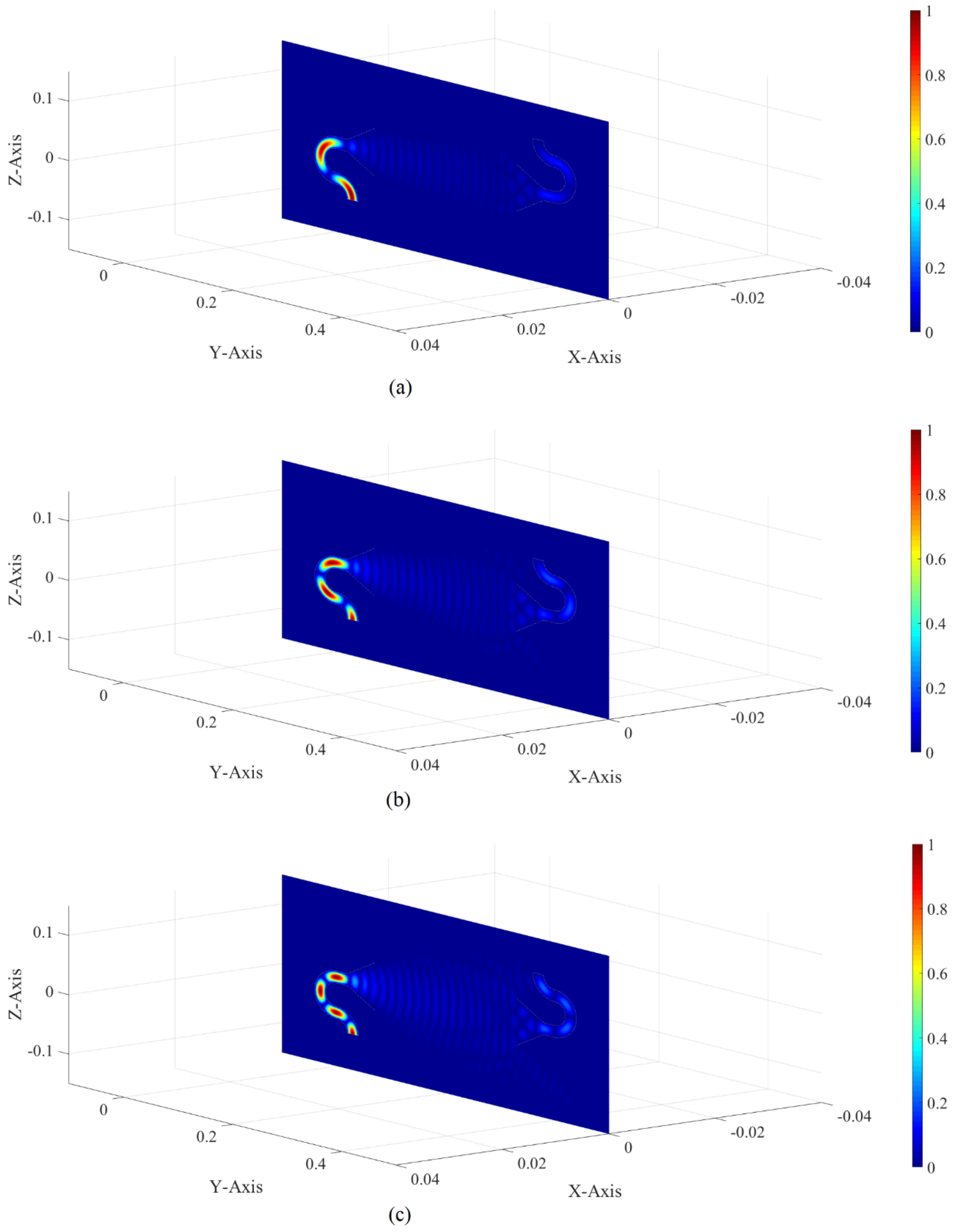

Figure 3-48 Modal electric field distributions (on yOz plane) of the JE-DoJ-based DM 1 at a series of resonance frequencies. (a) 8.95 GHz; (b) 9.10 GHz; (c) 9.40 GHz.

In fact, as illustrated in Ref. [27-Sec.8.3], the above modal analysis method can also be further generalized to some more complicated cascaded systems, such as the one shown in the following Fig. 3-49. In the figure, the transmitting and receiving antennas are two dielectric resonantor antennas mounted on metallic ground planes, and the transmitting/receiving antenna is fed/loaded by metallic tube waveguide, and there is a scatterer placed in the environment surrounding the transmitting and receiving systems.

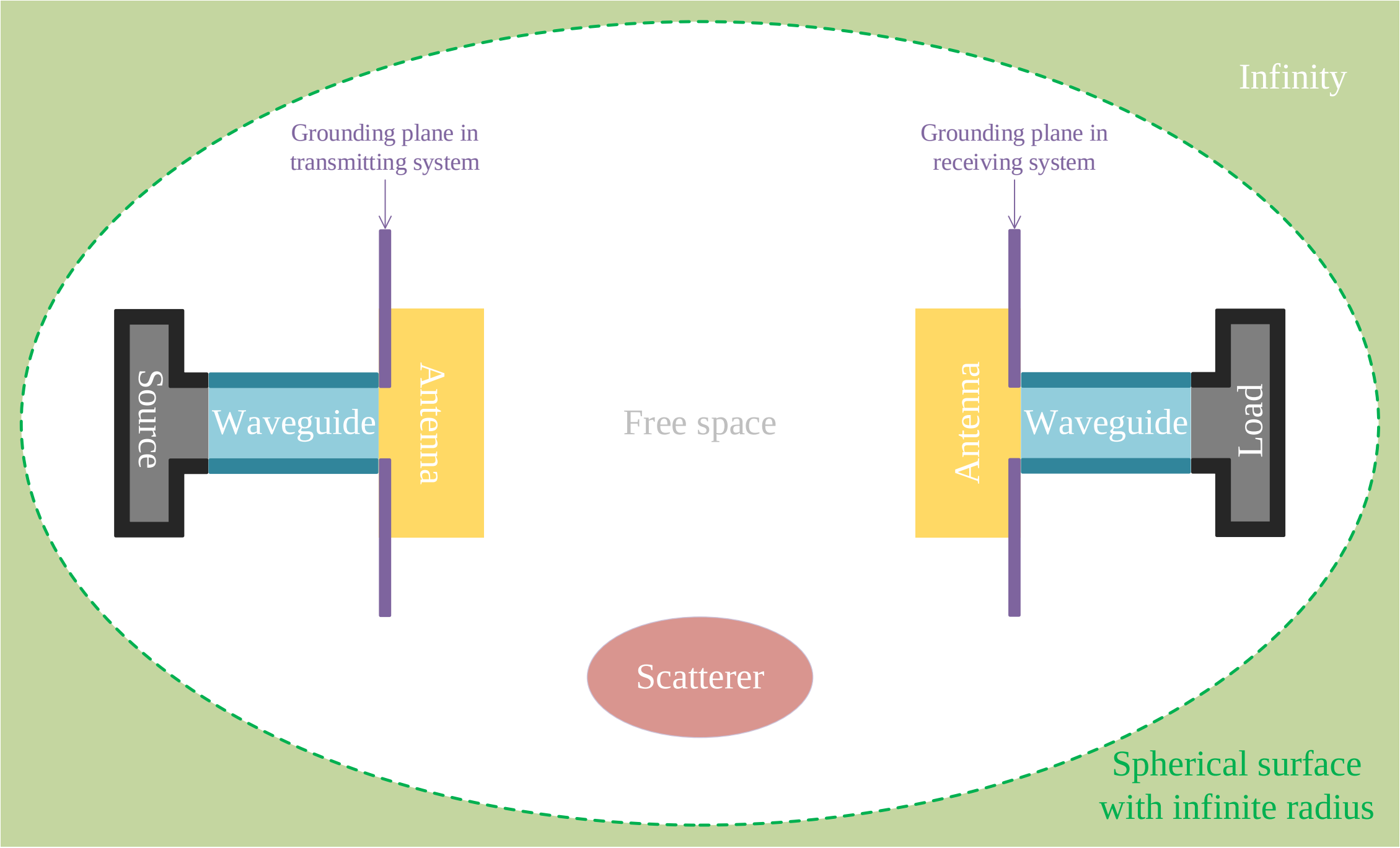

Figure 3-49 A more general waveguide-antenna-medium-antenna-waveguide cascaded system discussed in Ref. [27-Sec.8.3].

## 3.7 Chapter Summary

This chapter focuses on establish an effective energy-viewpoint-based modal analysis method — PTT-DMT — for wave-port-fed EM structures.

The main novel works done in this chapter are reflected in the following several aspects.

1) The whole wave-port-fed transceiving system is divided into a series of cascaded wave-port-fed EM structures — feeding waveguide, transmitting antenna, receiving antenna, and loading waveguide, etc.
2) Taking metallic transmitting horn as example, Sec. 3.2 establishes the PTT-DMT for wave-port-fed transmitting antennas.
3) Taking metallic receiving horn as example, Sec. 3.3 establishes the PTT-DMT for wave-port-fed receiving antennas.
4) Taking metallic tube waveguide as example, Sec. 3.4 establishes the PTT-DMT for wave-port-fed wave-guiding structures.

5) Following the idea of Sec. 3.4, the PTT-DMT for wave-port-fed wave-guiding structures is generalized to free-space waveguide in Sec. 3.5.
6) Due to the unified framework of the PTT-DMTs for various wave-port-fed EM structures, Sec. 3.6 establishes the PTT-DMT for the combined systems constituted by some cascaded EM structures (such as waveguide-antenna cascaded system and waveguide-antenna-medium-antenna-waveguide cascaded system), such that the complicated modal matching process used to analyze cascaded EM structures is successfully avoided.

# CHAPTER 4 PS-WET-Based Modal Analysis for Lumped-Port-Driven EM Structures

**Chapter Motivation:** This chapter is dedicated to providing an effective energy-based modal analysis method to lumped-port-driven electromagnetic (EM) structures.

## 4.1 Chapter Introduction

Yagi-Uda antenna was first introduced by Uda and Yagi in the 1920s[58,59]. In 1984, the *Proceedings of the IEEE* reprinted several classical articles for celebrating the centennial year of IEEE (1884-1984), and Yagi's article[59] became the only reprinted one in the realm of EM antenna. This fact clearly illustrates the great significance of Yagi-Uda antenna in *Antennas & Propagation Society*. A typical Yagi-Uda antenna is shown in Fig. 4-1(a), and it is constituted by a row of discrete metallic linear elements, one of which is driven by a lumped port while the others act as parasitic radiators (or called passive radiators) whose currents are induced by near-field mutual coupling.

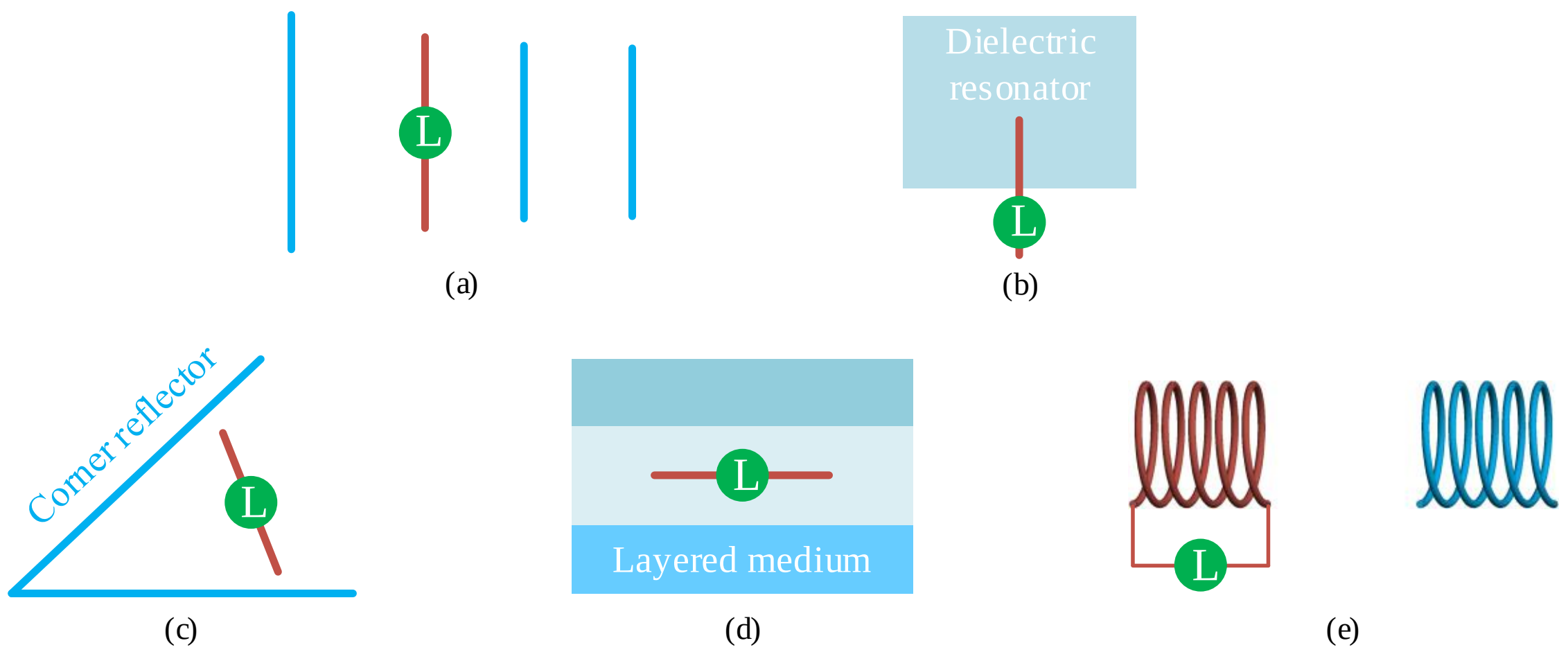

Figure 4-1 Lumped-port-driven metallic (a) Yagi-Uda array antenna (b) dipole antenna loaded by passive dielectric resonator, (c) dipole antenna loaded by passive metallic corner reflector, (d) dipole antenna loaded by passive layered medium, and (e) wireless power transfer system. Here, the lumped port L can be connected to either voltage source or current source .

The Yagi-Uda antenna shown in Fig. 4-1(a) is a typical lumped-port-driven EM structure, and Figs. 4-1(b~e) exhibit another some typical lumped-port-driven EM structures. Figure 4-1(b) is a lumped-port-driven dipole antenna loaded by a passive dielectric resonator; Fig. 4-1(c)/(d) is a lumped-port-driven dipole antenna loaded by a passive corner reflector / layered medium; Fig. 4-1(e) is a lumped-port-driven two-coil wireless power transfer system, which is designed for wirelessly transferring EM power from the transmitting coil to the receiving coil. Here, the lumped port Ⓛ can be connected to either voltage source ⓒ or current source ⊕ .

As exhibited in Refs. [27-Apps.G&H] and [37,40], the conventional scatterer-oriented CHARACTERISTIC MODE THEORY (CMT) fails to analyze the lumped-port-driven EM structures. This chapter is dedicated to generalizing the conventional CMT from scattering structures to lumped-port-driven structures, and the generalized CMT is established under PARTIAL-STRUCTURE-ORIENTED WORK-ENERGY THEOREM (PS-WET) framework. The PS-WET-based CMT (PS-WET-CMT) can effectively construct the energy-decoupled CHARACTERISTIC MODES (CMs) of lumped-port-driven EM structures by orthogonalizing PARTIAL-STRUCTURE-ORIENTED DRIVING POWER OPERATOR (PS-DPO).

## 4.2 PS-WET-Based Energy-Decoupled CMs of Lumped-Port-Driven Metallic Transmitting Antennas

Taking the lumped-port-driven metallic Yagi-Uda antenna shown in Fig. 4-2 as example, this section focuses on establishing the PS-WET-CMT for lumped-port-driven metallic transmitting antennas.

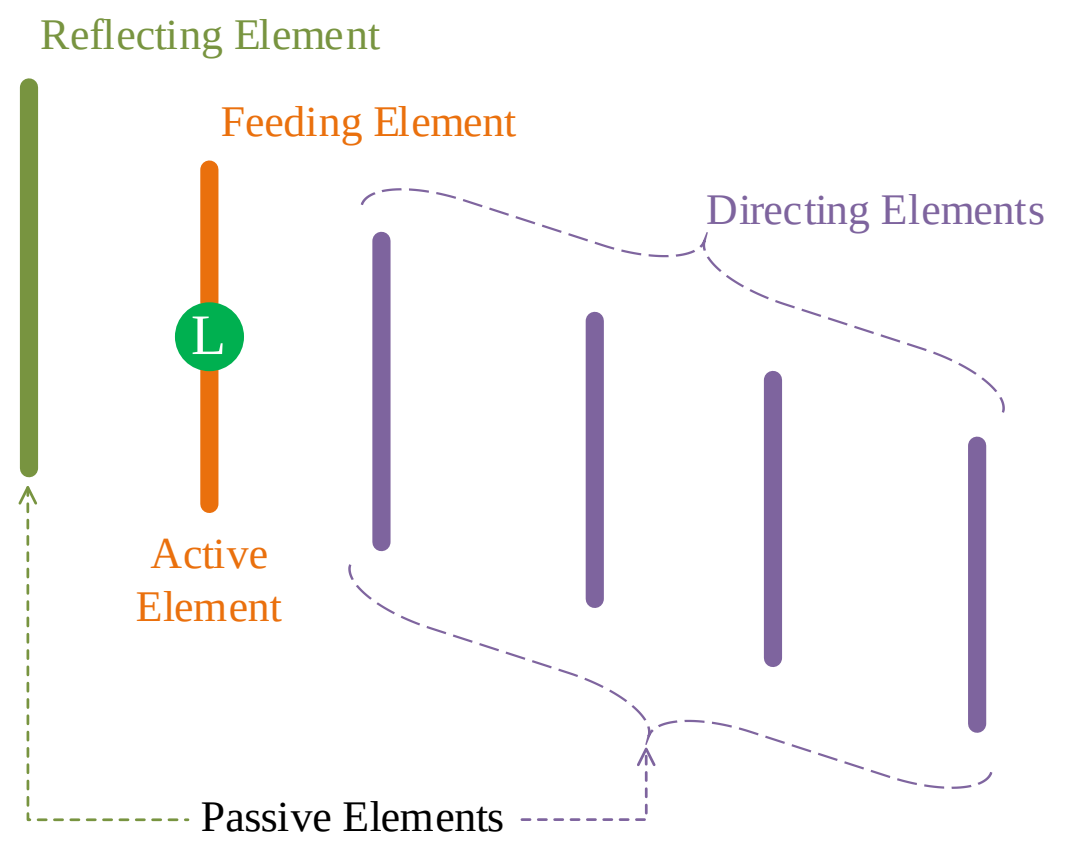

Figure 4-2 Geometry of a 6-element metallic Yagi-Uda antenna driven by a lumped port.

In Fig. 4-2, the Yagi-Uda antenna is placed in free space, which has parameters $(\mu_0, \varepsilon_0)$. In addition, the antenna elements are classified into two groups[50] — active element (with boundary surface $\mathbb{S}_a$) and passive elements (with boundary surface $\mathbb{S}_p$), where $\mathbb{S}_a$ is the feeding element directly connected with lumped port, and $\mathbb{S}_p$ is the union of reflecting element and directing elements.

The lumped port has an equivalent field effect, and the equivalent field is called driving field, and denoted as $\boldsymbol{F}_{\text{driv}}$. As explained in Refs. [27-App.H] and [37], the $\boldsymbol{F}_{\text{driv}}$ only drives $\text{S}_{\text{a}}$, but doesn't drive $\text{S}_{\text{p}}$ directly. The driver for $\text{S}_{\text{p}}$ is the field generated by $\text{S}_{\text{a}}$, i.e., **$\text{S}_{\text{a}}$ acts as an energy relay between the lumped port and $\text{S}_{\text{p}}$**. When the whole Yagi-Uda antenna works at stationary state, the currents on $\text{S}_{\text{a}}$ and $\text{S}_{\text{p}}$ are denoted as $\boldsymbol{J}_{\text{a}}$ and $\boldsymbol{J}_{\text{p}}$ respectively. The field generated by $\boldsymbol{J}_{\text{a}}+\boldsymbol{J}_{\text{p}}$ is just the field generated by whole Yagi-Uda antenna, and it is denoted as $\boldsymbol{F}$.

ENERGY CONSERVATION LAW implies that the interactions among lumped port and antenna elements will result in a work-energy transformation. As proved in Sec. 2.2, the work-energy transformation is quantified by PS-WET, and the source used to sustain the work-energy transformation, i.e., the source term in PS-WET, is following driving power

$$P_{\text{driv}}=\left(1/2\right)\left\langle \boldsymbol{J}_{\text{a}},\boldsymbol{E}_{\text{driv}}\right\rangle_{\text{S}_{\text{a}}} \tag{4-1}$$

Here, it is necessary to emphasize that the "driven current" and "integral domain" are $\boldsymbol{J}_{\text{a}}$ and $\text{S}_{\text{a}}$ rather than $\boldsymbol{J}_{\text{a}}+\boldsymbol{J}_{\text{p}}$ and $\text{S}_{\text{a}}\bigcup\text{S}_{\text{p}}$, because $\boldsymbol{E}_{\text{driv}}$ doesn't directly act on $\boldsymbol{J}_{\text{p}}$ (i.e., $\boldsymbol{E}_{\text{driv}}$ is restricted on $\text{S}_{\text{a}}$)[27-App.H],[37].

On the active element boundary $\text{S}_{\text{a}}$, there exists relation $[\boldsymbol{E}_{\text{driv}}]^{\text{tan}}=-[\boldsymbol{E}]^{\text{tan}}$, due to the homogeneous tangential electric field boundary condition on metallic boundary. Thus, the tangential $\boldsymbol{E}_{\text{driv}}$ on $\text{S}_{\text{a}}$ can be expressed in terms of the function of $(\boldsymbol{J}_{\text{a}},\boldsymbol{J}_{\text{p}})$. Then, driving power can be expressed as the following integral operator[27-App.H],[37]

$$P_{\text{driv}}=-\left(1/2\right)\left\langle \boldsymbol{J}_{\text{a}},-j\omega\mu_0\mathcal{L}_0\left(\boldsymbol{J}_{\text{a}}+\boldsymbol{J}_{\text{p}}\right)\right\rangle_{\text{S}_{\text{a}}} \tag{4-2}$$

called PS-DPO. If the current basis function expansions are applied to PS-DPO (4-2), the integral operator is discretized into the following matrix operator

$$P_{\text{driv}}=\mathbb{J}_{\text{a}}^{\dagger}\cdot\begin{bmatrix}\mathbb{P}_{\text{aa}} & \mathbb{P}_{\text{ap}}\end{bmatrix}\cdot\begin{bmatrix}\mathbb{J}_{\text{a}}\\ \mathbb{J}_{\text{p}}\end{bmatrix} \tag{4-3}$$

where $\mathbb{J}_{\text{a}}$ and $\mathbb{J}_{\text{p}}$ are the basis function expansion coefficient vectors for $\boldsymbol{J}_{\text{a}}$ and $\boldsymbol{J}_{\text{p}}$, and $\mathbb{P}_{\text{aa}}$ and $\mathbb{P}_{\text{ap}}$ are the power quadratic matrices corresponding to power terms $-(1/2)<\boldsymbol{J}_{\text{a}},-j\omega\mu_0\mathcal{L}_0(\boldsymbol{J}_{\text{a}})>_{\text{S}_{\text{a}}}$ and $-(1/2)<\boldsymbol{J}_{\text{a}},-j\omega\mu_0\mathcal{L}_0(\boldsymbol{J}_{\text{p}})>_{\text{S}_{\text{a}}}$.

In fact, currents $\boldsymbol{J}_{\text{a}}$ and $\boldsymbol{J}_{\text{p}}$ are not independent of each other, and they satisfy the following integral equation

$$\left[-j\omega\mu_0\mathcal{L}_0\left(\boldsymbol{J}_{\text{a}}+\boldsymbol{J}_{\text{p}}\right)\right]_{\text{S}_{\text{p}}}^{\text{tan}}=0 \tag{4-4}$$

due to the homogeneous tangential electric field boundary condition on metallic boundary

$\mathrm{S_p}$. Applying the method of moments to Eq. (4-4), the integral equation is discretized into a matrix equation. By solving the matrix equation, we obtain the transformation

$$\begin{bmatrix} \mathbb{J}_\mathrm{a} \\ \mathbb{J}_\mathrm{p} \end{bmatrix} = \mathbb{T} \cdot \mathbb{J}_\mathrm{a} \tag{4-5}$$

from independent current $\mathbb{J}_\mathrm{a}$ into all currents $(\mathbb{J}_\mathrm{a}, \mathbb{J}_\mathrm{p})$.

Substituting transformation (4-5) into matrix operator (4-3), the following matrix operator[27-App.H],[37]

$$P_\mathrm{driv} = \mathbb{J}_\mathrm{a}^\dagger \cdot \mathbb{P}_\mathrm{driv} \cdot \mathbb{J}_\mathrm{a} \tag{4-6}$$

with only independent current $\mathbb{J}_\mathrm{a}$ is obtained, where $\mathbb{P}_\mathrm{driv} = [\mathbb{P}_\mathrm{aa}\ \mathbb{P}_\mathrm{ap}] \cdot \mathbb{T}$. The CMs of the Yagi-Uda antenna can be calculated from solving the following characteristic equation

$$\mathbb{P}_\mathrm{driv}^- \cdot \mathbb{J}_\mathrm{a} = \theta\, \mathbb{P}_\mathrm{driv}^+ \cdot \mathbb{J}_\mathrm{a} \tag{4-7}$$

where $\mathbb{P}_\mathrm{driv}^+$ and $\mathbb{P}_\mathrm{driv}^-$ are the positive and negative Hermitian parts of $\mathbb{P}_\mathrm{driv}$, i.e., $\mathbb{P}_\mathrm{driv}^+ = (\mathbb{P}_\mathrm{driv} + \mathbb{P}_\mathrm{driv}^\dagger)/2$ and $\mathbb{P}_\mathrm{driv}^- = (\mathbb{P}_\mathrm{driv} - \mathbb{P}_\mathrm{driv}^\dagger)/2j$ [60-Sec.0.2.5].

It is easy to prove that the above-obtained CMs satisfy the following frequency-domain power-decoupling relation[27-App.H],[37]

$$(1/2)\left\langle \boldsymbol{J}_\mathrm{a}^m, \boldsymbol{E}_\mathrm{driv}^n \right\rangle_{\mathrm{S_a}} = (1 + j\theta_m)\delta_{mn} \tag{4-8}$$

and then the following time-domain energy-decoupling relation (or alternatively called time-averaged power-decoupling relation)

$$(1/T)\int_{t_0}^{t_0+T} \left\langle \boldsymbol{J}_\mathrm{a}^m, \boldsymbol{\mathcal{E}}_\mathrm{driv}^n \right\rangle_{\mathrm{S_a}} dt = \delta_{mn} \tag{4-9}$$

where $T$ is the time period of the time-harmonic EM field, and all modal real powers are normalized to 1 according to the convention used in Ref. [10] (for the physical reason of the normalization, please see Refs. [27-Sec.1.2.4.7] and [14]). Evidently, energy-decoupling relation (4-9) has a very clear physical interpretation: **in any integral period, there doesn't exist net energy delivery from the *n*-th modal driving field $\boldsymbol{\mathcal{E}}_\mathbf{driv}^n$ to the *m*-th modal induced current $\boldsymbol{J}_\mathbf{a}^m$ (i.e., from lumped port to Yagi-Uda antenna) if $m \neq n$**. Thus, the above-obtained CMs are energy-decoupled.

Now, we discuss another orthogonality satisfied by the above-obtained energy-decoupled CMs as below. Because of the homogeneous tangential electric field boundary conditions $[\boldsymbol{\mathcal{E}}_\mathrm{driv} + \boldsymbol{\mathcal{E}}]^\mathrm{tan} = 0$ and $[\boldsymbol{\mathcal{E}}]^\mathrm{tan} = 0$ on $\mathrm{S_a}$ and $\mathrm{S_p}$ respectively, energy-decoupling relation (4-9) implies the following far-field orthogonality

$$\begin{aligned}\delta_{mn} &= (1/T)\int_{t_0}^{t_0+T}\left\langle \boldsymbol{J}_{\rm a}^m+\boldsymbol{J}_{\rm p}^m,-\boldsymbol{\mathcal{E}}^n\right\rangle_{\rm S_a\cup S_p} dt\\ &= (1/T)\int_{t_0}^{t_0+T}\left[\oiint_{\rm S_\infty}\left(\boldsymbol{\mathcal{E}}^m\times\boldsymbol{\mathcal{H}}^n\right)\cdot\boldsymbol{n}_\infty^+ dS\right]dt\\ &\quad+(1/T)\int_{t_0}^{t_0+T}\left\{\frac{d}{dt}\left[(1/2)\left\langle\boldsymbol{\mathcal{H}}^n,\mu_0\boldsymbol{\mathcal{H}}^m\right\rangle_{\rm E_3}+(1/2)\left\langle\varepsilon_0\boldsymbol{\mathcal{E}}^n,\boldsymbol{\mathcal{E}}^m\right\rangle_{\rm E_3}\right]\right\}dt\\ &= (1/T)\int_{t_0}^{t_0+T}\left[\oiint_{\rm S_\infty}\left(\boldsymbol{\mathcal{E}}^m\times\boldsymbol{\mathcal{H}}^n\right)\cdot\boldsymbol{n}_\infty^+ dS\right]dt\end{aligned} \tag{4-10}$$

In the above orthogonality (4-10), the derivation for the second equality is similar to deriving POYNTING'S THEOREM (2-6); the third equality is due to the periodicity of the time-harmonic EM field; the integral domain $\rm E_3$ is the whole three-dimensional Euclidean space; integral surface $\rm S_\infty$ is the outer boundary of $\rm E_3$, and it is usually selected as a spherical surface with infinite radius; vector $\boldsymbol{n}_\infty^+$ is the outer normal direction of $\rm S_\infty$, and points to infinity. Obviously, far-field orthogonality (4-10) implies that

$$(1/T)\int_{t_0}^{t_0+T}\left[\oiint_{\rm S_\infty}\left(\boldsymbol{\mathcal{E}}^m\times\boldsymbol{\mathcal{H}}^n+\boldsymbol{\mathcal{E}}^n\times\boldsymbol{\mathcal{H}}^m\right)\cdot\boldsymbol{n}_\infty^+ dS\right]dt=2\delta_{mn} \tag{4-11}$$

and then LORENTZ'S RECIPROCITY THEOREM (2-22) implies energy-decoupling relation

$$(1/T)\int_{t_0}^{t_0+T}\left[\oiint_{\rm S}\left(\boldsymbol{\mathcal{E}}^m\times\boldsymbol{\mathcal{H}}^n+\boldsymbol{\mathcal{E}}^n\times\boldsymbol{\mathcal{H}}^m\right)\cdot\boldsymbol{n}dS\right]dt=2\delta_{mn} \tag{4-12}$$

where S is an arbitrary closed surface enclosing whole Yagi-Uda antenna as shown in Fig. 4-3, and $\boldsymbol{n}$ is the normal direction of $\rm S$.

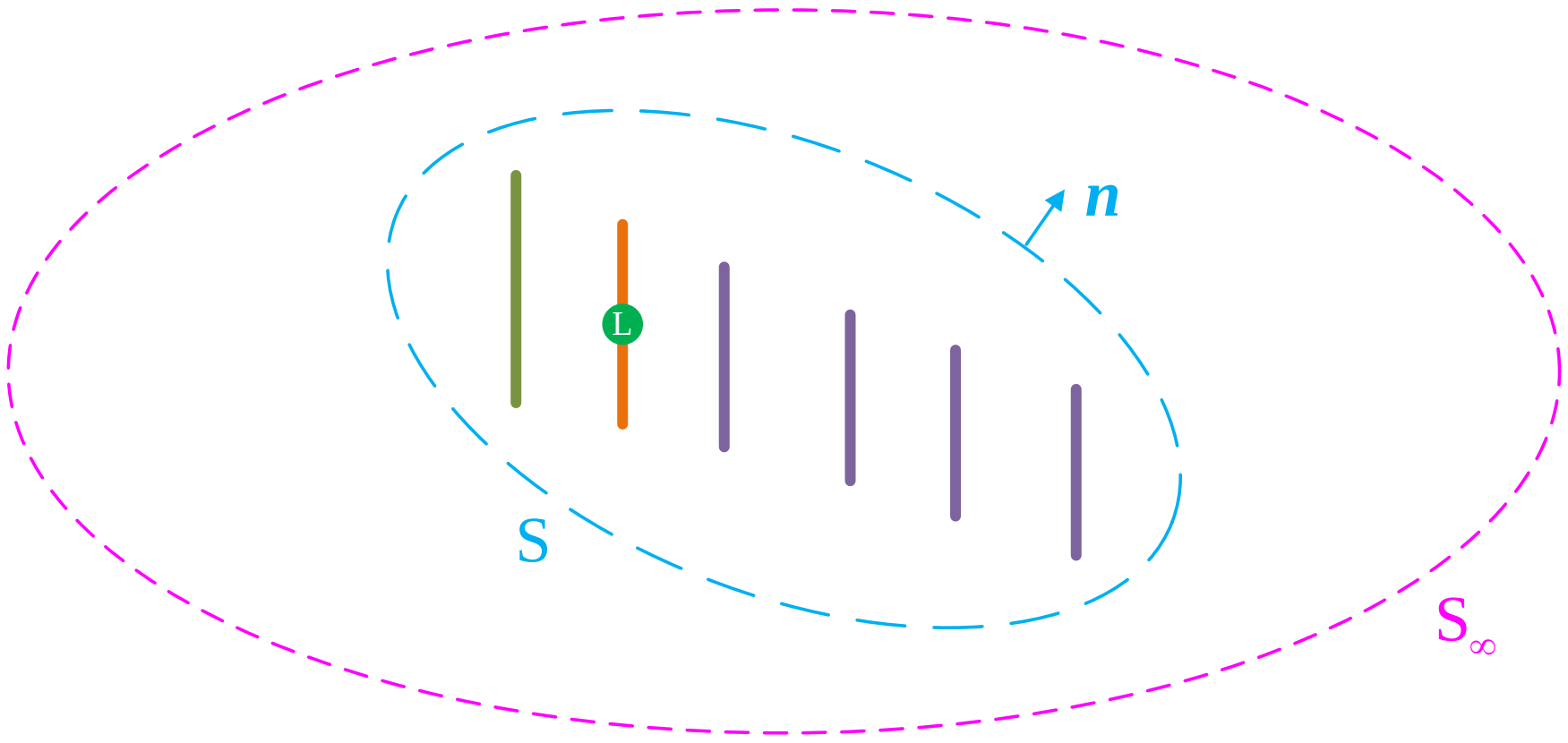

Figure 4-3 A closed surface $\rm S$ enclosing whole Yagi-Uda antenna.

The modal significance (MS) $\text{MS}=1/|1+j\theta|$ is usually employed to depict the modal feature in the aspect of utilizing EM energy. Some careful interpretations for the physical meaning of MS had been provided in Refs. [27-Sec.9.4] and [14], and will be simply summarized in the App. A2 of this report.

Here, we use the above PS-WET-CMT to do the modal analysis for a specific metallic Yagi-Uda antenna, whose size is shown in Fig. 4-4.

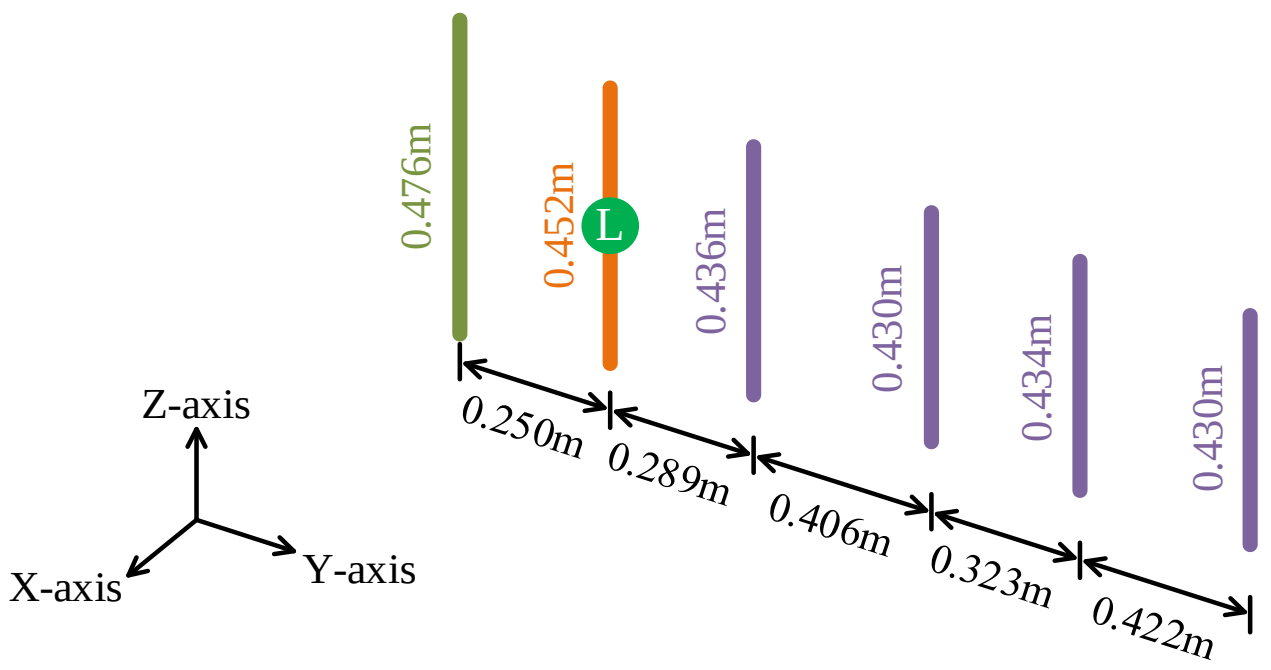

Figure 4-4 A Yagi-Uda antenna designed by using the formulations proposed in Ref. [36].

The above Yagi-Uda antenna is designed by using the method proposed in Ref. [36]. The MSs associated to the first 4 lower-order energy-decoupled CMs are shown in Fig. 4-5.

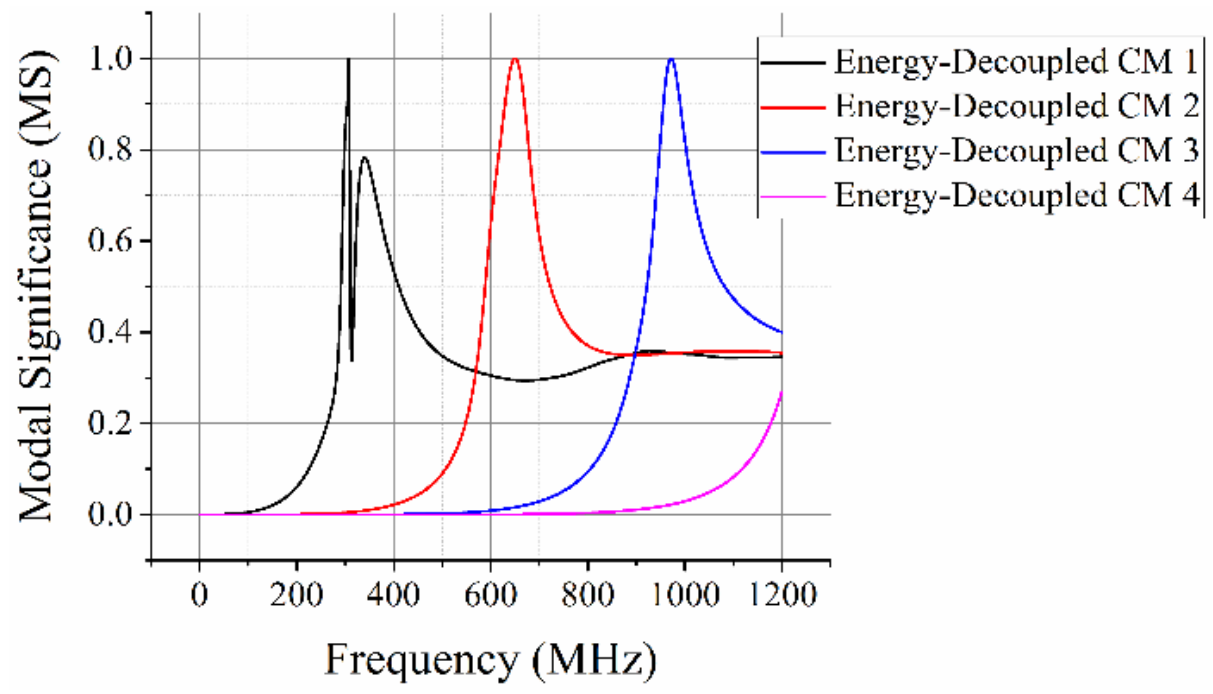

Figure 4-5 MSs of the first 4 lower-order energy-decoupled CMs.

The figure implies that: the CM 1 is resonant at 307.3 MHz, which frequency is consistent with the one calculated from the formulation proposed in Ref. [36] except a 2% numerical error. Besides the resonant CM 1, there also exist two higher-order resonant CMs, the resonant CM 2 at 649.8 MHz and the resonant CM 3 at 971.8 MHz. The radiation patterns of the above-mentioned three resonant CMs are shown in Fig. 4-6.

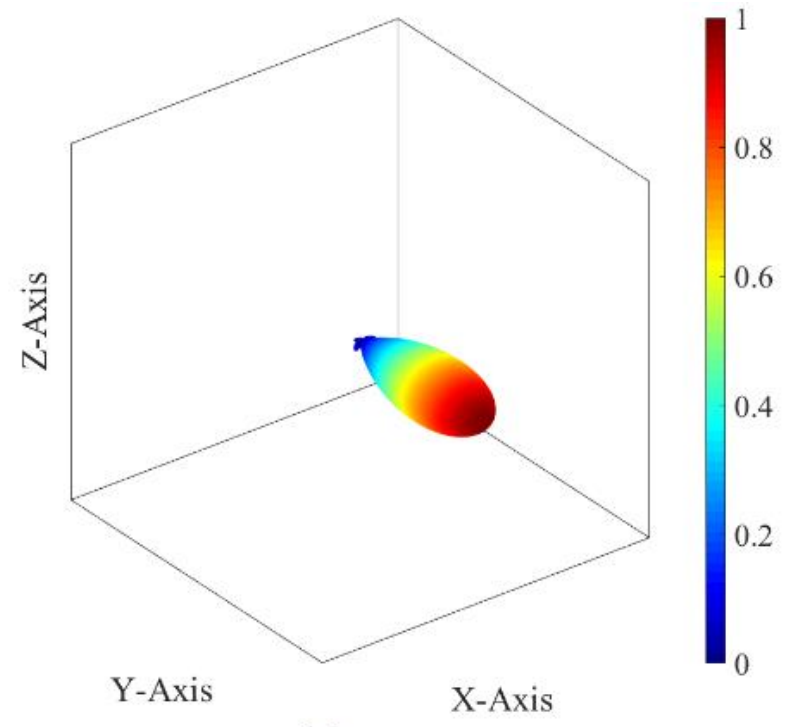

(a)

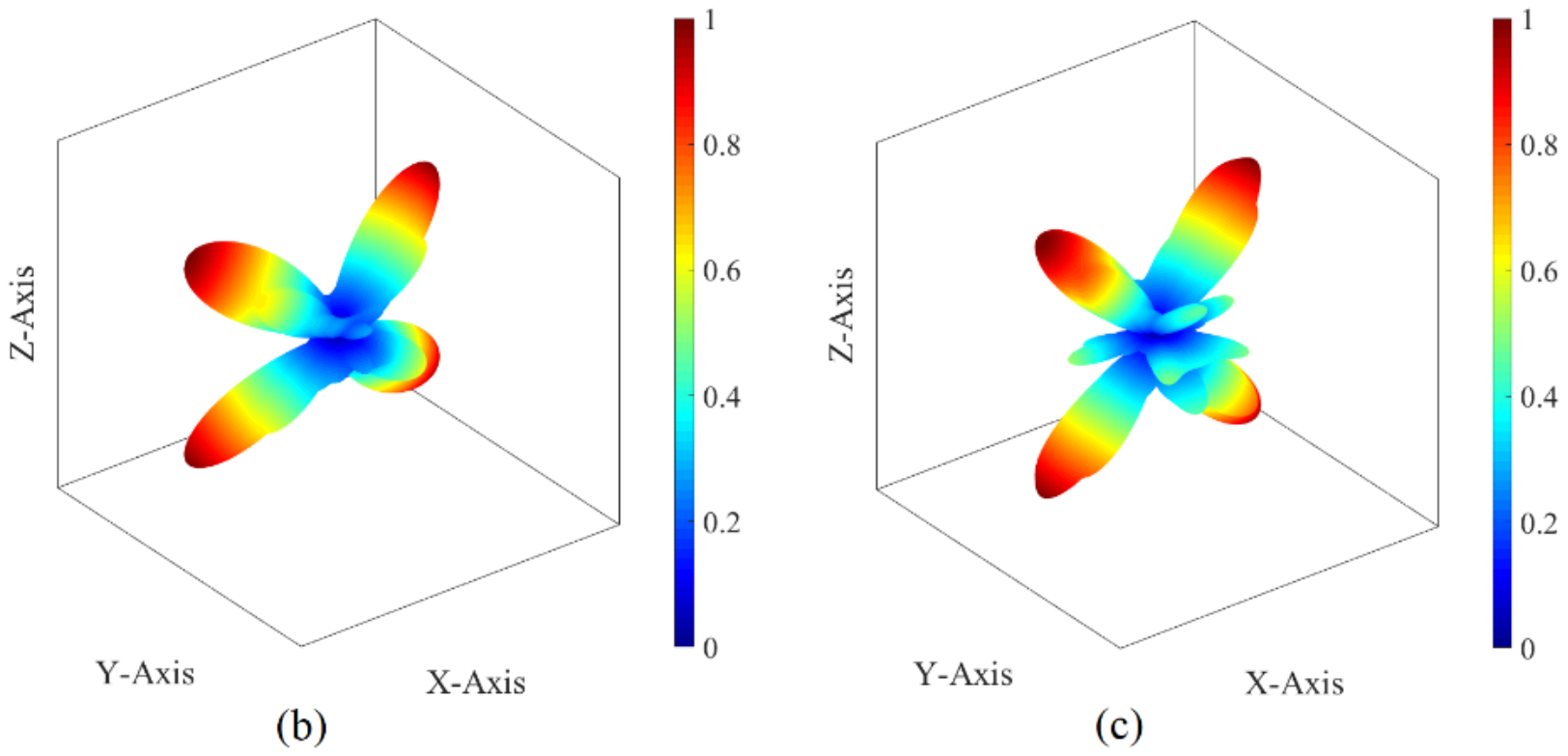

Figure 4-6 Radiation patterns of the three resonant CMs in Fig. 4-5. (a) Resonant CM 1 at 307.3 MHz; (b) resonant CM 2 at 649.8 MHz; (c) resonant CM 3 at 971.8 MHz.

Obviously, only the dominant resonant CM 1 works at end-fire state, but the higher-order resonant CMs don't. In fact, this is just the reason why *higher resonances are available near lengths of* $\lambda$*,* $3\lambda/2$*, and so forth, but are seldom used*[61-pp.562]. Here, we also illustrate the field distribution of the dominant resonant CM 1 in the following Fig. 4-7. The figure clearly exhibits that the EM power of the end-fire state indeed propagates along the direction from reflecting element to directing elements.

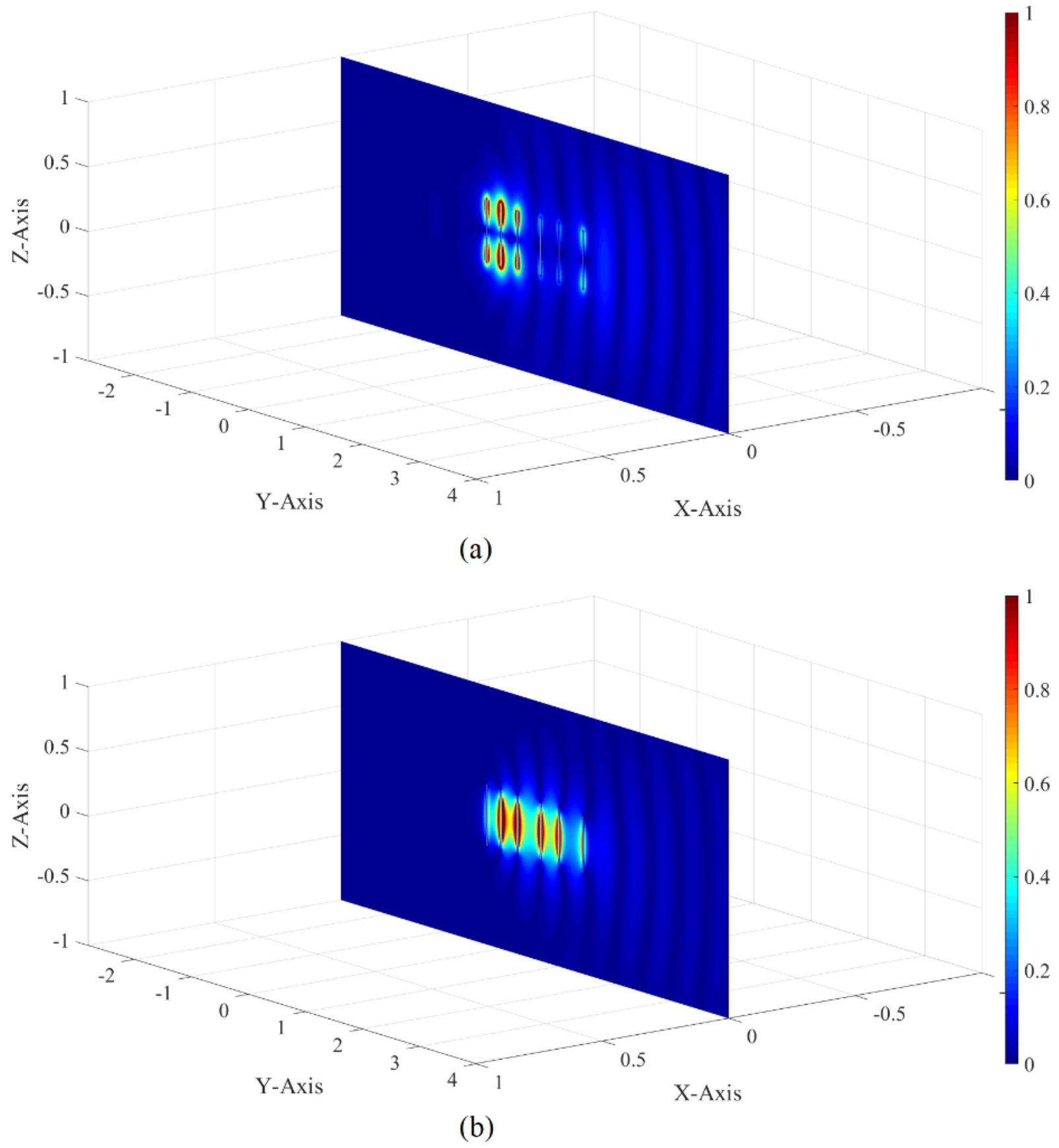

Figure 4-7 Distributions of the (a) electric and (b) magnetic fields of the dominant resonant CM working at 307.3 MHz (end-fire state).

## 4.3 PS-WET-Based Energy-Decoupled CMs of Local-Near-Field-Driven Material Transmitting Antennas

Generally speaking, the lumped port is usually used to directly drive a metallic part of antenna. Thus, it is not easy to provide a typical purely material transmitting antenna driven by lumped port. In this section, we discuss a material-antenna-oriented driving manner — local near field driving — which has many similarities to lumped port driving.

Taking the 3-element material Yagi-Uda array antenna shown in Fig. 4-8 as a typical example, this section establishes the PS-WET-based CM analysis for local-near-field-driven material transmitting antennas. For simplying the following discussions, the array elements are restricted to being non-magnetic, and their complex permittivities are denoted as $\boldsymbol{\mu}_{\rm a}^{\rm c}$, $\boldsymbol{\mu}_{\rm p1}^{\rm c}$, and $\boldsymbol{\mu}_{\rm p2}^{\rm c}$, and the purely magnetic case and magneto-dielectric case can be similarly discussed.

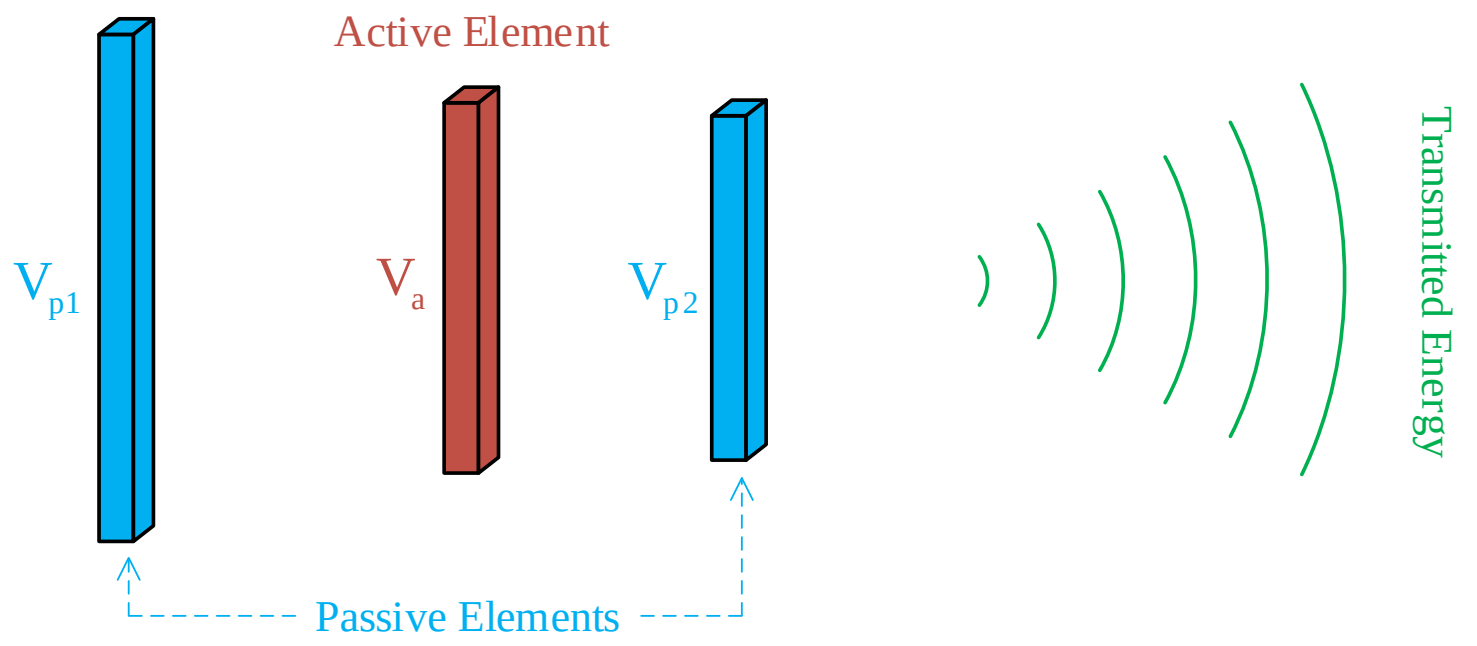

Figure 4-8 Geometry of a local near-field-driven 3-element material Yagi-Uda transmitting antenna, which was proposed in Ref. [62].

In the following parts of this section, we will provide two different CM calculation formulations — volume formulation[27-Sec.H5.1],[37-Sec.III-A] and surface formulation[27-Sec.H5.2],[37-Sec.III-B].

### 4.3.1 Volume CM Formulation

For the array antenna shown in Fig. 4-8, the regions occupied by its array elements are denoted as $\rm V_a$, $\rm V_{p1}$, and $\rm V_{p2}$. If the induced volume electric currents distributing on the elements are denoted as $\boldsymbol{j}_{\rm a}$, $\boldsymbol{j}_{\rm p1}$, and $\boldsymbol{j}_{\rm p2}$ respectively, then the corresponding PS-DPO is as follows[27-Sec.H5.1],[37-Sec.III-A]:

$$P_{\rm driv} = (1/2)\left\langle \boldsymbol{j}_{\rm a}, \left(j\omega\Delta\boldsymbol{\varepsilon}_{\rm a}^{\rm c}\right)^{-1}\cdot\boldsymbol{j}_{\rm a} + j\omega\mu_0\mathcal{L}_0\left(\boldsymbol{j}_{\rm a}+\boldsymbol{j}_{\rm p1}+\boldsymbol{j}_{\rm p2}\right)\right\rangle_{\rm V_a} \tag{4-13}$$

where $\Delta\boldsymbol{\varepsilon}_{\rm a}^{\rm c} = \boldsymbol{\varepsilon}_{\rm a}^{\rm c} - \mathbf{I}\varepsilon_0$ and $\mathbf{I}$ is the unit dyad. In fact, the currents involved in PS-DPO

(4-13) are not independent of each other, and their dependence relations are governed by integral equations

$$\boldsymbol{j}_{\text{p1}} = j\omega\Delta\boldsymbol{\varepsilon}_{\text{p1}}^{\text{c}} \cdot \left[-j\omega\mu_0\mathcal{L}_0\left(\boldsymbol{j}_{\text{a}}+\boldsymbol{j}_{\text{p1}}+\boldsymbol{j}_{\text{p2}}\right)\right] \quad \text{on} \quad \text{V}_{\text{p1}} \tag{4-14a}$$

$$\boldsymbol{j}_{\text{p2}} = j\omega\Delta\boldsymbol{\varepsilon}_{\text{p2}}^{\text{c}} \cdot \left[-j\omega\mu_0\mathcal{L}_0\left(\boldsymbol{j}_{\text{a}}+\boldsymbol{j}_{\text{p1}}+\boldsymbol{j}_{\text{p2}}\right)\right] \quad \text{on} \quad \text{V}_{\text{p2}} \tag{4-14b}$$

because of volume equivalence principle[8-App.A] and that driving field is 0 on $\text{V}_{\text{p1}} \cup \text{V}_{\text{p2}}$.

Expanding the currents in PS-DPO (4-13) in terms of some proper basis functions, the integral-operator-formed PS-DPO (4-13) can be discretized into matrix operator form. Applying the method of moments to integral equations (4-14), they will be discretized into matrix equations, and the matrix equations implies a matrix transformation from independent current into the other currents. Substituting the matrix transformation into the matrix-operator-formed PS-DPO, we immediately obtain the following

$$P_{\text{driv}} = \text{j}_{\text{a}}^{\dagger} \cdot \text{p}_{\text{driv}} \cdot \text{j}_{\text{a}} \tag{4-15}$$

with only $\text{j}_{\text{a}}$, which is the basis function expansion coefficient vector of independent current $\boldsymbol{j}_{\text{a}}$.

The same as the metallic Yagi-Uda antenna case discussed in Sec. 4.2, the CMs of the material Yagi-Uda antenna can be calculated from solving characteristic equation $\text{p}_{\text{driv}}^{-} \cdot \text{j}_{\text{a}} = \theta\, \text{p}_{\text{driv}}^{+} \cdot \text{j}_{\text{a}}$, where $\text{p}_{\text{driv}}^{+}$ and $\text{p}_{\text{driv}}^{-}$ are the positive and negative Hermitian parts of $\text{p}_{\text{driv}}$, and $\text{p}_{\text{driv}}^{+} = (\text{p}_{\text{driv}} + \text{p}_{\text{driv}}^{\dagger})/2$ and $\text{p}_{\text{driv}}^{-} = (\text{p}_{\text{driv}} - \text{p}_{\text{driv}}^{\dagger})/2j$.

### 4.3.2 Surface CM Formulation

For establishing the surface CM formulation, we denote the boundary surfaces of $\text{V}_{\text{a}}$, $\text{V}_{\text{p1}}$, and $\text{V}_{\text{p2}}$ as $\text{S}_{\text{a}}$, $\text{S}_{\text{p1}}$, and $\text{S}_{\text{p2}}$ respectively. If the equivalent surface currents distributing on the boundary surfaces are denoted as $(\boldsymbol{J}_{\text{a}},\boldsymbol{M}_{\text{a}})$, $(\boldsymbol{J}_{\text{p1}},\boldsymbol{M}_{\text{p1}})$, and $(\boldsymbol{J}_{\text{p2}},\boldsymbol{M}_{\text{p2}})$, which are defined by employing the inner normal directions of the boundary surfaces, then the PS-DPO given in Eq. (4-13) can be alternatively written as the following surface-current version[27-Sec.H5.1],[37-Sec.III-A]

$$\begin{aligned} P_{\text{driv}} = & -(1/2)\left\langle \boldsymbol{J}_{\text{a}}, \mathcal{E}_0\left(\boldsymbol{J}_{\text{a}}+\boldsymbol{J}_{\text{p1}}+\boldsymbol{J}_{\text{p2}}, \boldsymbol{M}_{\text{a}}+\boldsymbol{M}_{\text{p1}}+\boldsymbol{M}_{\text{p2}}\right)\right\rangle_{\text{S}_{\text{a}}^{-}} \\ & -(1/2)\left\langle \boldsymbol{M}_{\text{a}}, \mathcal{H}_0\left(\boldsymbol{J}_{\text{a}}+\boldsymbol{J}_{\text{p1}}+\boldsymbol{J}_{\text{p2}}, \boldsymbol{M}_{\text{a}}+\boldsymbol{M}_{\text{p1}}+\boldsymbol{M}_{\text{p2}}\right)\right\rangle_{\text{S}_{\text{a}}^{-}} \end{aligned} \tag{4-16}$$

where integral surface $\text{S}_{\text{a}}^{-}$ is the inner boundary surface of $\text{V}_{\text{a}}$. The currents involved in PS-DPO (4-16) are also not independent of each other, and they satisfy the following integrate equations

$$\left[\mathcal{E}_{\rm a}\left(\boldsymbol{J}_{\rm a},\boldsymbol{M}_{\rm a}\right)\right]_{S_{\rm a}^-}^{\tan} = \boldsymbol{n}_{\rm a}^- \times \boldsymbol{M}_{\rm a} \tag{4-17}$$

$$\left[\mathcal{E}_{\rm p1}\left(\boldsymbol{J}_{\rm p1},\boldsymbol{M}_{\rm p1}\right)\right]_{S_{\rm p1}^-}^{\tan} = -\left[\mathcal{E}_0\left(\boldsymbol{J}_{\rm a}+\boldsymbol{J}_{\rm p1}+\boldsymbol{J}_{\rm p2},\boldsymbol{M}_{\rm a}+\boldsymbol{M}_{\rm p1}+\boldsymbol{M}_{\rm p2}\right)\right]_{S_{\rm p1}^+}^{\tan} \tag{4-18a}$$

$$\left[\mathcal{H}_{\rm p1}\left(\boldsymbol{J}_{\rm p1},\boldsymbol{M}_{\rm p1}\right)\right]_{S_{\rm p1}^-}^{\tan} = -\left[\mathcal{H}_0\left(\boldsymbol{J}_{\rm a}+\boldsymbol{J}_{\rm p1}+\boldsymbol{J}_{\rm p2},\boldsymbol{M}_{\rm a}+\boldsymbol{M}_{\rm p1}+\boldsymbol{M}_{\rm p2}\right)\right]_{S_{\rm p1}^+}^{\tan} \tag{4-18b}$$

$$\left[\mathcal{E}_{\rm p2}\left(\boldsymbol{J}_{\rm p2},\boldsymbol{M}_{\rm p2}\right)\right]_{S_{\rm p2}^-}^{\tan} = -\left[\mathcal{E}_0\left(\boldsymbol{J}_{\rm a}+\boldsymbol{J}_{\rm p1}+\boldsymbol{J}_{\rm p2},\boldsymbol{M}_{\rm a}+\boldsymbol{M}_{\rm p1}+\boldsymbol{M}_{\rm p2}\right)\right]_{S_{\rm p2}^+}^{\tan} \tag{4-19a}$$

$$\left[\mathcal{H}_{\rm p2}\left(\boldsymbol{J}_{\rm p2},\boldsymbol{M}_{\rm p2}\right)\right]_{S_{\rm p2}^-}^{\tan} = -\left[\mathcal{H}_0\left(\boldsymbol{J}_{\rm a}+\boldsymbol{J}_{\rm p1}+\boldsymbol{J}_{\rm p2},\boldsymbol{M}_{\rm a}+\boldsymbol{M}_{\rm p1}+\boldsymbol{M}_{\rm p2}\right)\right]_{S_{\rm p2}^+}^{\tan} \tag{4-19b}$$

Here, Eq. (4-17) is based on the definition of $\boldsymbol{M}_{\rm a}$; Eqs. (4-18a) and (4-18b) are based on the electric and magnetic field tangential continuation conditions on $\rm S_{p1}$; Eqs. (4-19a) and (4-19b) are based on the electric and magnetic field tangential continuation conditions on $\rm S_{p2}$. Operators $\mathcal{E}_0$ and $\mathcal{H}_0$ are defined as that $\mathcal{E}_0(\boldsymbol{J},\boldsymbol{M}) = -j\omega\mu_0\mathcal{L}_0(\boldsymbol{J}) - \mathcal{K}_0(\boldsymbol{M})$ and $\mathcal{H}_0(\boldsymbol{J},\boldsymbol{M}) = \mathcal{K}_0(\boldsymbol{J}) - j\omega\varepsilon_0\mathcal{L}_0(\boldsymbol{M})$; operators $\mathcal{E}_{\rm a/p1/p2}$ and $\mathcal{H}_{\rm a/p1/p2}$ transform the currents $(\boldsymbol{J}_{\rm a/p1/p2},\boldsymbol{M}_{\rm a/p1/p2})$ into the electric and magnetic fields distributing on $\rm V_{a/p1/p2}$. In addition, $\rm S_{p1/p2}^-$ and $\rm S_{p1/p2}^+$ are the inner and outer boundary surfaces of $\rm V_{p1/p2}$.

Similary to deriving PS-DPO (4-15) from Eqs. (4-13)~(4-14b), the following PS-DPO (4-20) can be derived from Eqs. (4-17)~(4-19b).

$$P_{\rm driv} = \mathbb{M}_{\rm a}^\dagger \cdot \mathbb{P}_{\rm driv} \cdot \mathbb{M}_{\rm a} \tag{4-20}$$

with only $\mathbb{M}_{\rm a}$, which is the basis function expansion coefficient vector of independent current $\boldsymbol{M}_{\rm a}$. The $\mathbb{P}_{\rm driv}$ is different from the $\mathbb{p}_{\rm driv}$ used in Eq. (4-15). The CMs of the material Yagi-Uda antenna can be derived from solving equation $\mathbb{P}_{\rm driv}^- \cdot \mathbb{M}_{\rm a} = \theta\, \mathbb{P}_{\rm driv}^+ \cdot \mathbb{M}_{\rm a}$, where $\mathbb{P}_{\rm driv}^+$ and $\mathbb{P}_{\rm driv}^-$ are the positive and negative Hermitian parts of $\mathbb{P}_{\rm driv}$.

### 4.3.3 Numerical Verification

Here, we use the above volume and surface formulations to do the modal analysis for a specific material Yagi-Uda antenna, which is reported in Ref. [62]. For the specific antenna, its geometry is shown in Fig. 4-8; its all elements are with $4.0\,\text{mm} \times 4.0\,\text{mm}$ cross section; its elements $\rm V_a$, $\rm V_{p1}$, and $\rm V_{p2}$ have lengths 46.35 mm, 77.6 mm, and 44.4 mm respectively; the side-to-side distance between $\rm V_a$ and $\rm V_{p1}$ is 23.0 mm, and the side-to-side distance between $\rm V_a$ and $\rm V_{p2}$ is 10.7 mm. The complex permittivities of the elements are that $\boldsymbol{\varepsilon}_{\rm a}^{\rm c} = \boldsymbol{\varepsilon}_{\rm p1}^{\rm c} = \boldsymbol{\varepsilon}_{\rm p2}^{\rm c} = \mathbf{I}34\varepsilon_0$.

The characteristic values (in decibel) of the dominant CM calculated from the above volume and surface formulations and the modal $S_{11}$ parameter (in decibel) reported in Ref. [62] are shown in Fig. 4-9 simultaneously for comparison.

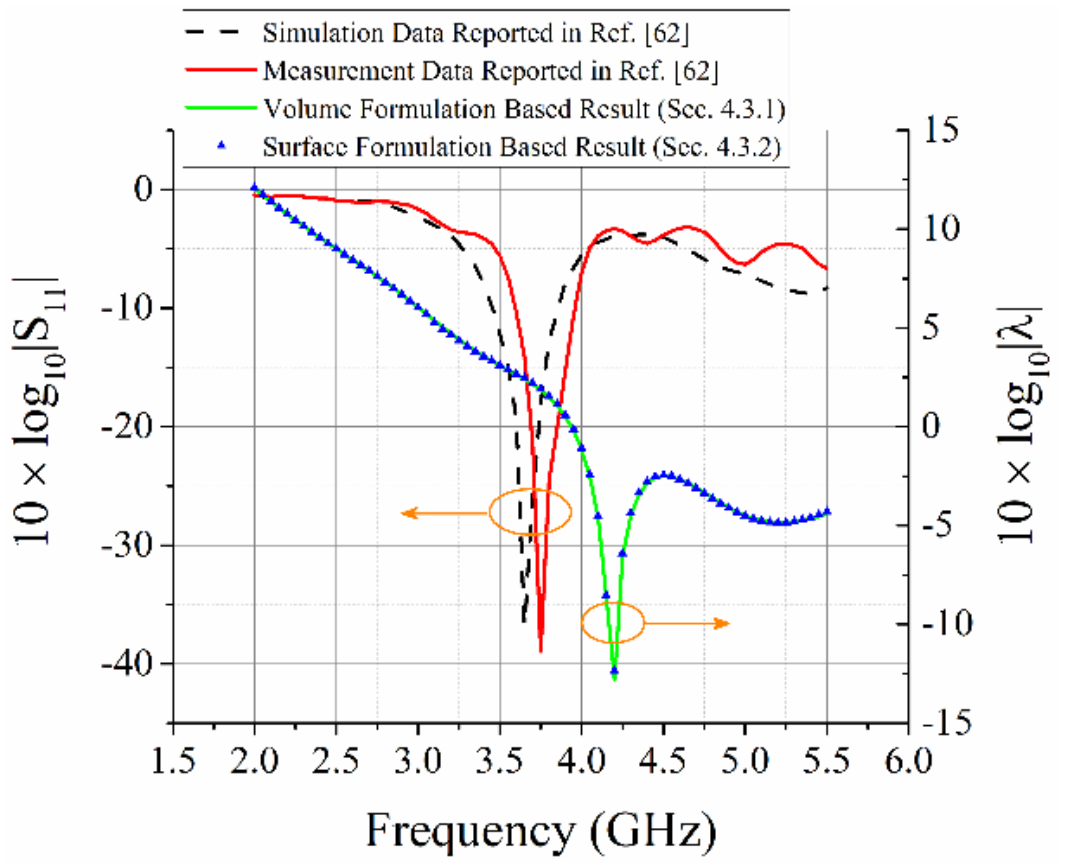

Figure 4-9 Modal parameters of the dominant CM of the material Yagi-Uda antenna reported in Ref. [62].

Clearly, the volume-formulation-based and surface-formulation-based results are consistent with each other; the PS-WET-based resonance frequency is basically consistent with the data reported in Ref. [62], and the slight discrepancy is mainly originated from ignoring the feeding structure.

For the PS-WET-based resonant CM working at 4.2 GHz, its modal radiation pattern and field distributions are shown in Fig. 4-10 and Fig. 4-11.

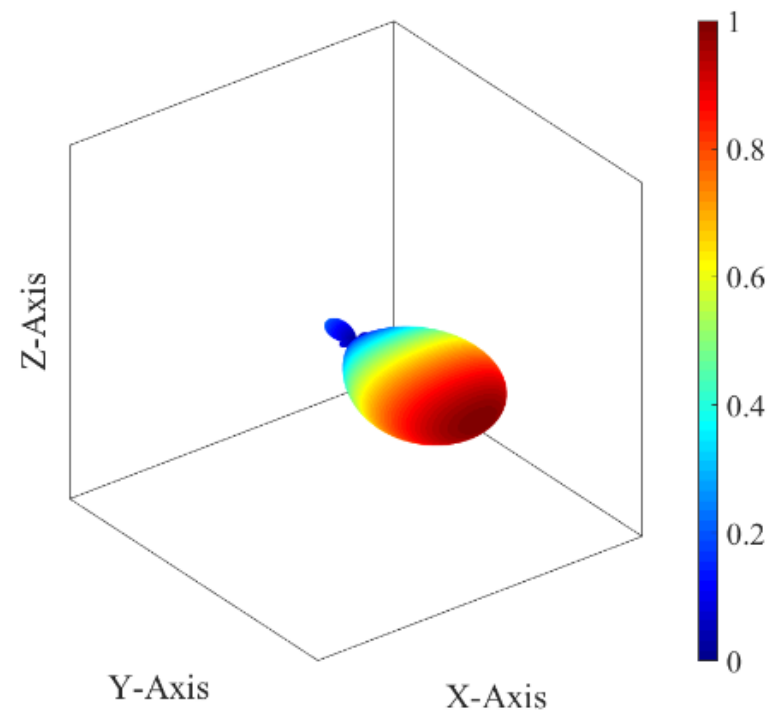

Figure 4-10 Radiation pattern of the PS-WET-based resonant CM.

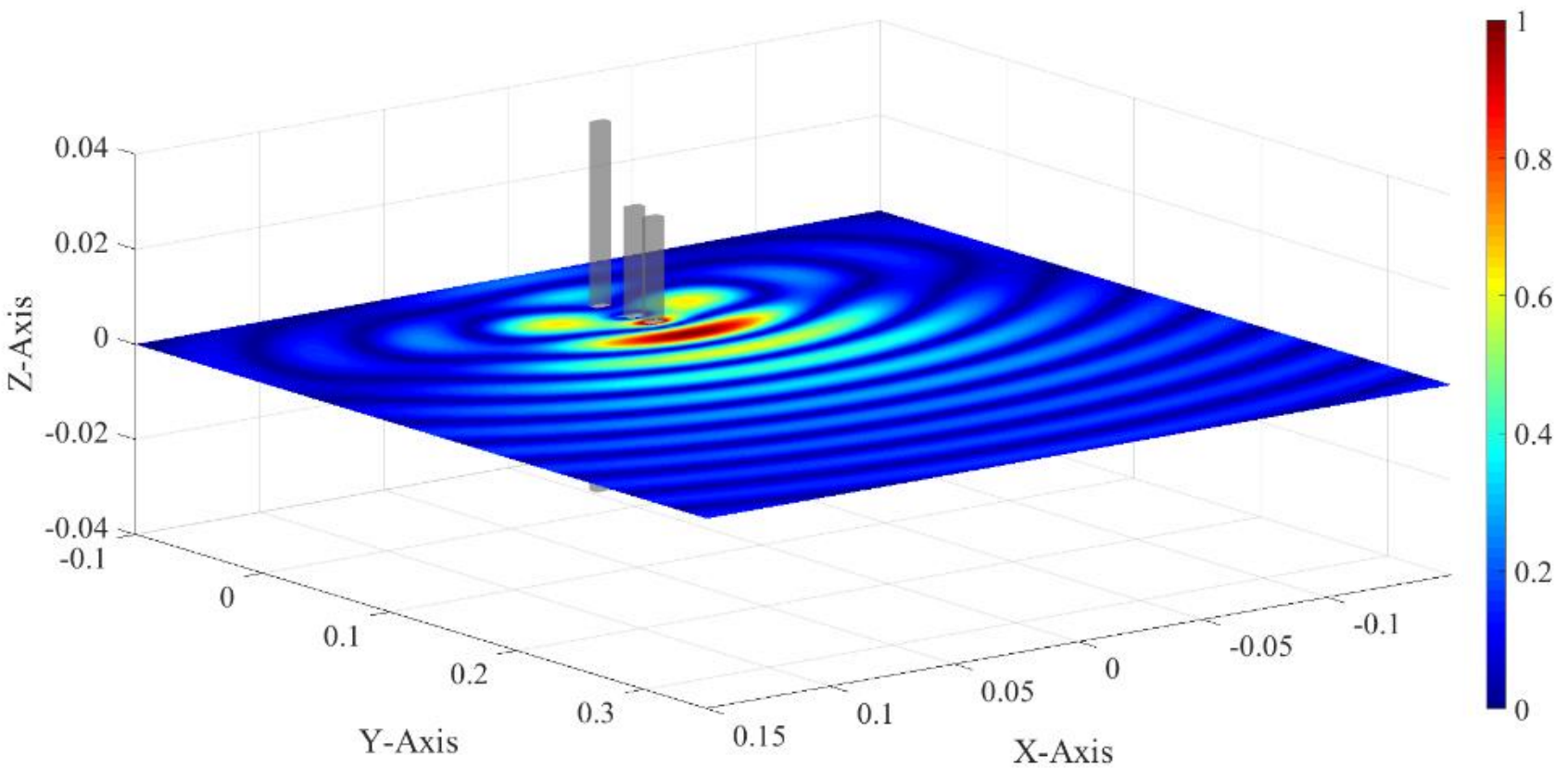

(a)

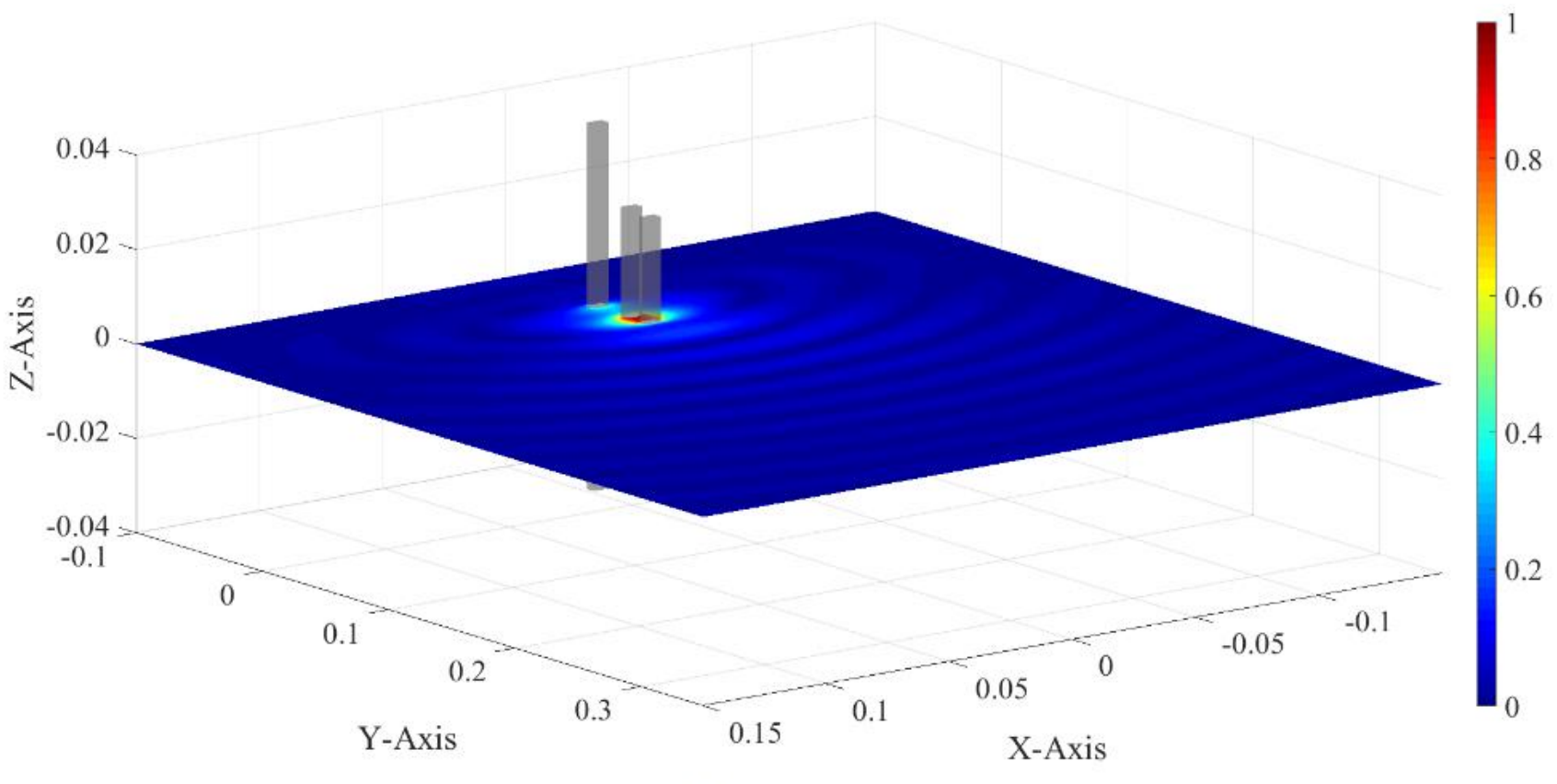

(b)

Figure 4-11 Modal (a) electric and (b) magnetic field distributions of the PS-WET-based resonant CM.

Evidently, Fig. 4-10 and Fig. 4-11 satisfy the well-known end-fire feature of linear Yagi-Uda antenna — the radiative power propagates along the direction from reflecting element to directing element.

## 4.4 PS-WET-Based Energy-Decoupled CMs of Lumped-Port-Driven Metallic Dipole Antennas Loaded by Passive Dielectric Resonator Antennas

Taking the lumped-port-driven dipole antenna loaded by dielectric resonantor shown in Fig. 4-1(b) as an example, this section discuss how to calculate the PS-WET-based energy-decoupled CMs of lumped-port-driven metal-material composite transmitting antennas. In this section, we provide two different CM calculation formulations — surface-volume formulation and surface formulation.

### 4.4.1 Surface-Volume CM Formulation

For establishing the surface-volume CM formulation of the antenna shown Fig. 4-1(b), we plot the topological structure of the antenna as follows:

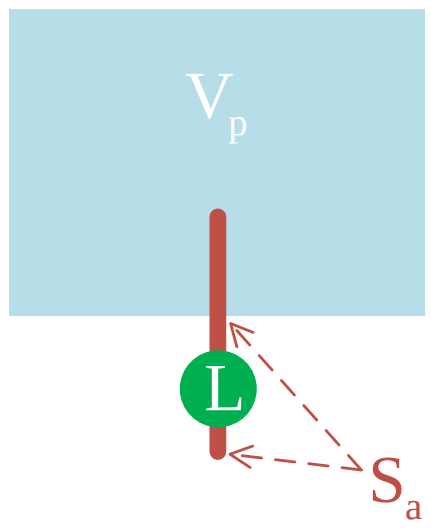

Figure 4-12 Topological structure of the antenna shown in Fig. 4-1(b).

In the figure, the boundary surface of the metallic dipole is denoted as $\mathrm{S_a}$; the region occupied by the passive dielectric resonator is denoted as $\mathrm{V_p}$, which is with material parameters $(\boldsymbol{\mu}_\mathrm{p}, \boldsymbol{\varepsilon}_\mathrm{p}, \boldsymbol{\sigma}_\mathrm{p})$, and the corresponding complex permittivity is $\boldsymbol{\varepsilon}_\mathrm{p}^\mathrm{c} = \boldsymbol{\varepsilon}_\mathrm{p} - j\boldsymbol{\sigma}_\mathrm{p} / \omega$.

When the lumped-port-driven composite antenna works at time-harmonic stationary state, the induced surface electric current on $\mathrm{S_a}$ is denoted as $\boldsymbol{J}_\mathrm{a}$, and the induced volume electric and magnetic currents on $\mathrm{V_p}$ are denoted as $(\boldsymbol{j}_\mathrm{p}, \boldsymbol{m}_\mathrm{p})$. Thus, the PS-DPO corresponding to the antenna can be written as follows:

$$\begin{aligned} P_\mathrm{driv} &= (1/2)\langle \boldsymbol{J}_\mathrm{a}, \boldsymbol{E}_\mathrm{driv} \rangle_{\mathrm{S_a}} \\ &= -(1/2)\left\langle \boldsymbol{J}_\mathrm{a}, -j\omega\mu_0\mathcal{L}_0(\boldsymbol{J}_\mathrm{a}) - j\omega\mu_0\mathcal{L}_0(\boldsymbol{j}_\mathrm{p}) - \mathcal{K}_0(\boldsymbol{m}_\mathrm{p}) \right\rangle_{\mathrm{S_a}} \end{aligned} \tag{4-21}$$

In the above PS-DPO (4-21), $\boldsymbol{E}_\mathrm{driv}$ is the driving field generated by the lumped port; the second equality is because of the homogeneous tangential electric field boundary condition on $\mathrm{S_a}$; the $-j\omega\mu_0\mathcal{L}_0(\boldsymbol{J}_\mathrm{a})$ and $-j\omega\mu_0\mathcal{L}_0(\boldsymbol{j}_\mathrm{p}) - \mathcal{K}_0(\boldsymbol{m}_\mathrm{p})$ in the right-hand side of the second equality are the electric fields generated by $\boldsymbol{J}_\mathrm{a}$ and $(\boldsymbol{j}_\mathrm{p}, \boldsymbol{m}_\mathrm{p})$ respectively.

Due to the volume equivalence principle[8-App.A] on the DRA, there exist the following integral equations

$$\boldsymbol{j}_\mathrm{p} = j\omega\Delta\boldsymbol{\varepsilon}_\mathrm{p}^\mathrm{c} \cdot \left[ -j\omega\mu_0\mathcal{L}_0(\boldsymbol{J}_\mathrm{a} + \boldsymbol{j}_\mathrm{p}) - \mathcal{K}_0(\boldsymbol{m}_\mathrm{p}) \right] \quad \text{on} \quad \mathrm{V_p} \tag{4-22a}$$

$$\boldsymbol{m}_\mathrm{p} = j\omega\Delta\boldsymbol{\mu}_\mathrm{p} \cdot \left[ \mathcal{K}_0(\boldsymbol{J}_\mathrm{a} + \boldsymbol{j}_\mathrm{p}) - j\omega\varepsilon_0\mathcal{L}_0(\boldsymbol{m}_\mathrm{p}) \right] \quad \text{on} \quad \mathrm{V_p} \tag{4-22b}$$

satisfied by currents $\boldsymbol{J}_\mathrm{a}$ and $(\boldsymbol{j}_\mathrm{p}, \boldsymbol{m}_\mathrm{p})$.

Similary to deriving PS-DPO (4-15) from Eqs. (4-13)~(4-14b), the following PS-DPO (4-23) can be derived from Eqs. (4-21)~(4-22b).

$$P_\mathrm{driv} = \mathbb{J}_\mathrm{a}^\dagger \cdot \mathrm{p}_\mathrm{driv} \cdot \mathbb{J}_\mathrm{a} \tag{4-23}$$

with only $\mathbb{J}_\mathrm{a}$, which is the basis function expansion coefficient vector of independent current $\boldsymbol{J}_\mathrm{a}$. The CMs of the lumped-port-driven composite antenna can be calculated from solving characteristic equation $\mathrm{p}_\mathrm{driv}^- \cdot \mathbb{J}_\mathrm{a} = \theta\, \mathrm{p}_\mathrm{driv}^+ \cdot \mathbb{J}_\mathrm{a}$, where $\mathrm{p}_\mathrm{driv}^+$ and $\mathrm{p}_\mathrm{driv}^-$ are the positive and negative Hermitian parts of $\mathrm{p}_\mathrm{driv}$.

### 4.4.2 Surface CM Formulation

For establishing the surface CM formulation of the antenna shown Fig. 4-1(b), we plot another topological structure of the antenna in Fig. 4-13. In the figure, the whole $\mathrm{S_a}$ is decomposed into two pairwisely disjoint parts $\mathrm{S_{a0}}$ and $\mathrm{S_{ap}}$, where $\mathrm{S_{a0/ap}}$ is the interface between probe and environment/resonator; the whole boundary surface of $\mathrm{V_p}$ is also decomposed into two pairwisely disjoint parts $S_{p0}$ and $S_{pa}$, where $S_{p0/pa}$ is the interface between dielectric resonator and environment/probe. It is obvious that $S_{ap} = S_{pa}$.

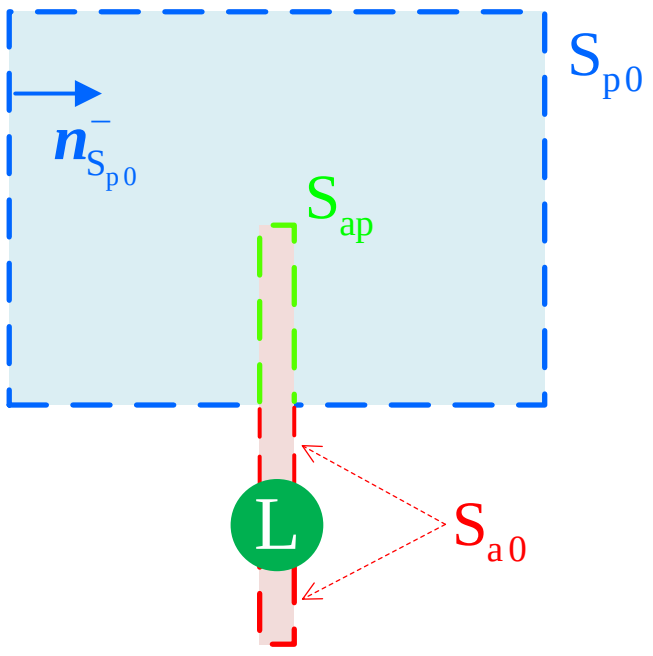

Figure 4-13 Another topological structure of the antenna shown in Fig. 4-1(b).

When the lumped-port-driven composite antenna works at time-harmonic stationary state, the induced surface electric currents on $S_{a0}$ and $S_{ap}$ are denoted as $\boldsymbol{J}_{a0}$ and $\boldsymbol{J}_{ap}$ respectively; the equivalent surface currents on $S_{p0}$ are defined as that $\boldsymbol{J}_{p0} = \boldsymbol{n}^{-}_{S_{p0}} \times \boldsymbol{H}_{-}$ and $\boldsymbol{M}_{p0} = \boldsymbol{E}_{-} \times \boldsymbol{n}^{-}_{S_{p0}}$, where $(\boldsymbol{E}_{-}, \boldsymbol{H}_{-})$ are the field in the interior of $V_p$ and $\boldsymbol{n}^{-}_{S_{p0}}$ is the inner normal direction of $S_{p0}$. Thus, the PS-DPO corresponding to the antenna can be written as follows:

$$P_{driv} = -(1/2)\left\langle \boldsymbol{J}_{a0} + \boldsymbol{J}_{ap}, -j\omega\mu_0 \mathcal{L}_0\left(\boldsymbol{J}_{a0} - \boldsymbol{J}_{p0}\right) - \mathcal{K}_0\left(-\boldsymbol{M}_{p0}\right)\right\rangle_{S_{a0} \cup S_{ap}} \tag{4-24}$$

Due to the tangential electric and magnetic field continuation conditions on $S_{p0}$, there exist the following integral equations

$$\left[\mathcal{E}_p\left(\boldsymbol{J}_{ap} + \boldsymbol{J}_{p0}, \boldsymbol{M}_{p0}\right)\right]^{tan}_{S^{-}_{p0}} = \left[\mathcal{E}_0\left(\boldsymbol{J}_{a0} - \boldsymbol{J}_{p0}, -\boldsymbol{M}_{p0}\right)\right]^{tan}_{S^{+}_{p0}} \tag{4-25a}$$

$$\left[\mathcal{H}_p\left(\boldsymbol{J}_{ap} + \boldsymbol{J}_{p0}, \boldsymbol{M}_{p0}\right)\right]^{tan}_{S^{-}_{p0}} = \left[\mathcal{H}_0\left(\boldsymbol{J}_{a0} - \boldsymbol{J}_{p0}, -\boldsymbol{M}_{p0}\right)\right]^{tan}_{S^{+}_{p0}} \tag{4-25b}$$

satisfied by the currents involved in PS-DPO (4-24). Here, operators $\mathcal{E}_0$ and $\mathcal{H}_0$ are the same as the ones used in Sec. 4.3.2; operators $\mathcal{E}_p$ and $\mathcal{H}_p$ transform the currents $(\boldsymbol{J}_{ap}, \boldsymbol{J}_{p0}, \boldsymbol{M}_{p0})$ into the electric and magnetic fields distributing on $V_p$. In addition, $S^{-}_{p0}$ and $S^{+}_{p0}$ belong to the inner and outer boundary surfaces of $V_p$.

Similary to deriving PS-DPO (4-15) from Eqs. (4-13)~(4-14b), the following PS-DPO (4-26) can be derived from Eqs. (4-24)~(4-25b).,

$$P_{driv} = \mathbb{J}^{\dagger}_{a} \cdot \mathbb{P}_{driv} \cdot \mathbb{J}_{a}, \text{ where } \mathbb{J}_{a} = \begin{bmatrix} \mathbb{J}_{a0} \\ \mathbb{J}_{ap} \end{bmatrix} \tag{4-26}$$

with only independent current $\mathbb{J}_a$. The CMs of the lumped-port-driven composite antenna can be calculated from solving characteristic equation $\mathbb{P}_{\text{driv}}^{-} \cdot \mathbb{J}_{\text{a}} = \theta\, \mathbb{P}_{\text{driv}}^{+} \cdot \mathbb{J}_{\text{a}}$, where $\mathbb{P}_{\text{driv}}^{+}$ and $\mathbb{P}_{\text{driv}}^{-}$ are the positive and negative Hermitian parts of $\mathbb{P}_{\text{driv}}$.

In fact, the above PS-WET-based surface-volume and surface CM formulations can be further generalized to metal-material composite Yagi-Uda antennas (such as the one shown in Fig. 4-14, which had been discussed in Refs. [27-Sec.H6] and [37-Sec.IV]).

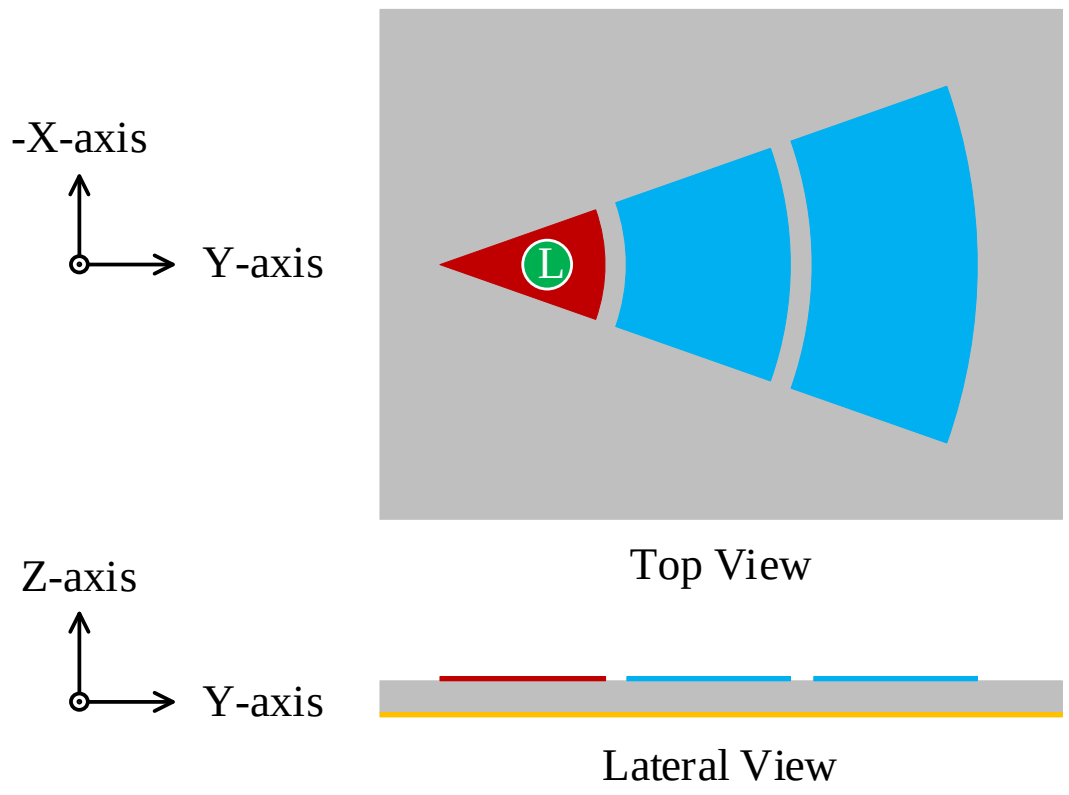

Figure 4-14 A typical metal-material composite quasi Yagi-Uda antenna reported in Ref. [63].

In addition, we want to emphasize here that: for the antenna shown in Fig. 4-14, if the active patch is driven by a wave-port-fed coaxial probe, a more reasonable modal analysis is the PTT-DMT given in the previous Sec. 3.2.

## 4.5 PS-WET-Based Energy-Decoupled CMs of Lumped-Port-Driven Metallic Dipole Antennas Loaded by Passive Corner Reflectors / Layered Mediums

In this section, we discuss how to calculate the PS-WET-based energy-decoupled CMs of the EM structures in Fig. 4-1(c) and Fig. 4-1(d). The one in Fig. 4-1(c) is a lumped-port-driven dipole antenna loaded by a passive corner reflector, and its topological structure is shown in Fig. 4-15(a). The one in Fig. 4-1(d) is a lumped-port-driven dipole antenna loaded by a passive layered medium, and its topological structure is shown in Fig. 4-15(b).

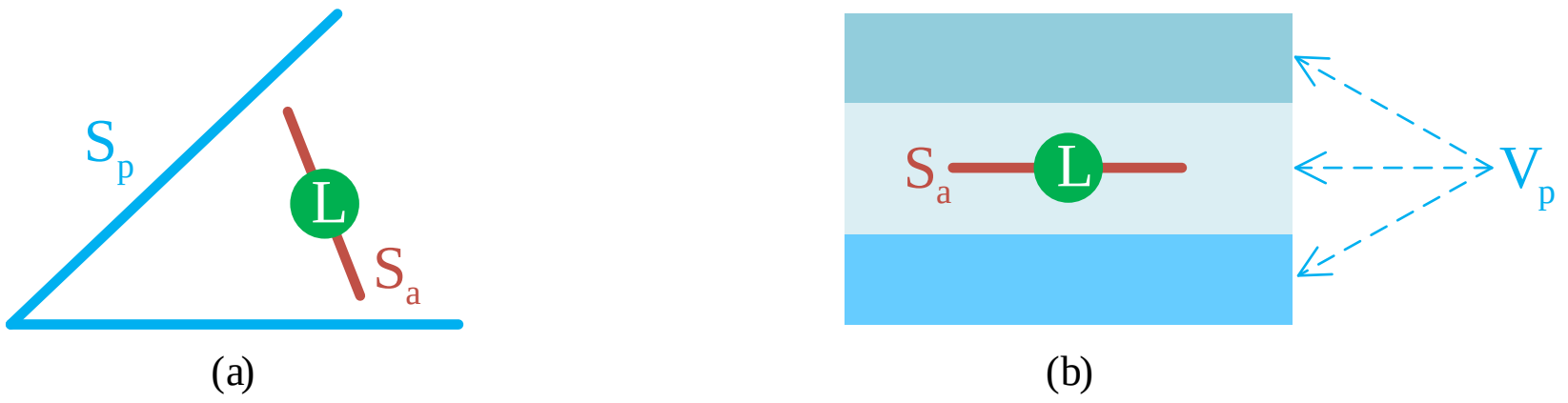

Figure 4-15 Topological structures of lumped-port-driven metallic dipole antennas loaded by passive (a) corner reflector and (b) layered medium.

In the following two sub-sections, we separately discuss two cases: (1) the passively loaded structures are with finite sizes; (2) the passively loaded structures are with infinite sizes.

## 4.5.1 Corner Reflectors / Layered Mediums with Finite Sizes

When the passively loaded corner reflector shown in Fig. 4-15(a) is with finite size, the PS-WET-based CM calculation formulation is the same as the one provided in Sec. 4.2 (lumped-port-driven metallic Yagi-Uda antenna case).

When the passively loaded layered medium shown in Fig. 4-15(b) is with finite size, the PS-WET-based CM calculating formulation can be established by using a similar method to the one given in Sec. 4.4 (lumped-port-driven dipole antenna loaded by a passive dielectric resonator case).

## 4.5.2 Corner Reflectors / Layered Mediums with Infinite Sizes

When the passively loaded corner reflector and layered medium are with infinite sizes, the above-proposed schemes are not applicable, because the schemes will generate infinite unknowns (if some sub-domain basis functions are used to discretize the PS-DPO). Here, we propose an effective scheme to overcome the problem.

For the EM problems shown in Fig. 4-15, the corresponding PS-DPOs can be uniformly written as follows:

$$P_{\rm driv} = \frac{1}{2}\langle \boldsymbol{J}_{\rm a}, \boldsymbol{E}_{\rm driv}\rangle_{\rm S_a} = -\frac{1}{2}\langle \boldsymbol{J}_{\rm a}, \boldsymbol{E}_{\rm a} + \boldsymbol{E}_{\rm p}\rangle_{\rm S_a} \tag{4-27}$$

In Eq. (4-27), $\boldsymbol{E}_{\rm a}$ and $\boldsymbol{E}_{\rm p}$ are the fields generated by the currents induced on the active dipole and passive load, and the second equality is based on the homogeneous tangential electric field boundary condition $[\boldsymbol{E}_{\rm driv} + \boldsymbol{E}_{\rm a} + \boldsymbol{E}_{\rm p}]^{\rm tan} = 0$ on $\rm S_a$.

If the Green's functions $\mathbf{G}$ of the regions defined by the corner reflector[53] and the layered medium[54] exist, then the $P_{\rm driv}$ can be simplified as follows:

$$P_{\rm driv} = -(1/2)\langle \boldsymbol{J}_{\rm a}, \mathbf{G} * \boldsymbol{J}_{\rm a}\rangle_{\rm S_a} \tag{4-28}$$

which involves independent current $\boldsymbol{J}_{\rm a}$ only, where $\mathbf{G} * \boldsymbol{J}_{\rm a} = \boldsymbol{E}_{\rm a} + \boldsymbol{E}_{\rm p}$.①

By orthogonalizing the above PS-DPO, the energy-decoupled CMs of the lumped-port-driven dipole antennas with infinite-sized passive loads can be obtained.

① A similar result in the case of "a finite metallic object passively loaded by an infinite (or a very large) metallic ground plane" was also obtained by some other researchers, such as Yubo Wen (文宇波).

## 4.6 PS-WET-Based Energy-Decoupled CMs of Lumped-Port-Driven Wave-Guiding Structures

Figure 4-16 shows a two-coil wireless power transfer (WPT) system designed for wirelessly transferring EM power from the transmitting coil to the receiving coil. The earliest researches on WPT can be dated back to the pioneers Hutin&Leblanc[64] and Tesla[65]. Taking the lumped-port-driven metallic WPT system as example, this section focuses on generalizing the above PS-WET-CMT from lumped-port-driven transmitting antennas to lumped-port-driven wave-guiding structures.

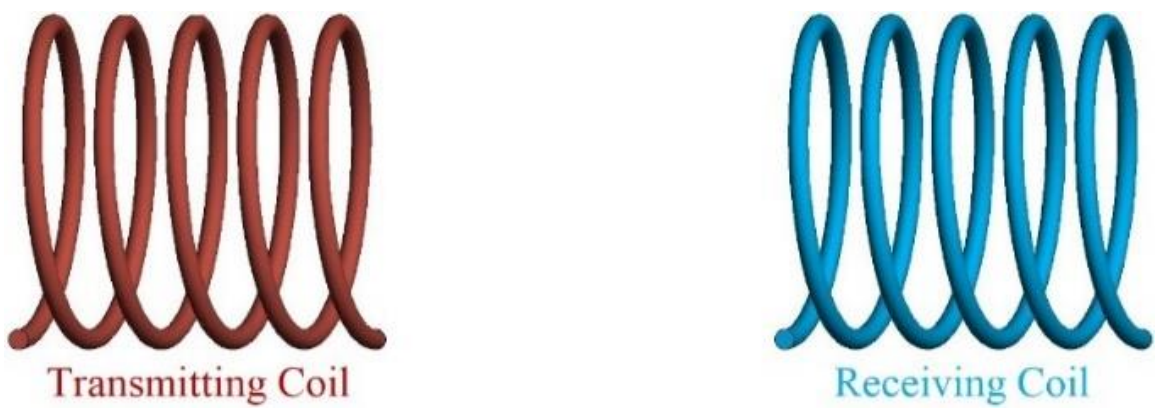

Figure 4-16 Geometry of a typical two-coil wireless power transfer (WPT) system reported in Ref. [67].

In Fig. 4-16, the WPT system is placed in free space, which has parameters $(\mu_0, \varepsilon_0)$ and outer boundary $\mathrm{S}_\infty$, and is constituted by a transmitting coil $\mathrm{S}_\mathrm{t}$ and a receiving coil $\mathrm{S}_\mathrm{r}$. Usually, the $\mathrm{S}_\mathrm{t}$ is driven by a voltage source, and the $\mathrm{S}_\mathrm{r}$ is loaded by a load, as described in Ref. [66]. When doing the modal analysis for the WPT system, the load is usually assumed to be perfectly matched[27-App.G],[40].

As explained in Refs. [27-App.G] and [40], the working mechanism of the above WPT system is similar to the previously discussed Yagi-Uda antenna in Sec. 4.2. At the stationary working state, the driving field generated by lumped port is denoted as $\boldsymbol{\mathcal{E}}_\mathrm{driv}$; the currents induced on $\mathrm{S}_\mathrm{t}$ and $\mathrm{S}_\mathrm{r}$ are denoted as $\boldsymbol{J}_\mathrm{t}$ and $\boldsymbol{J}_\mathrm{r}$; the fields generated by $\boldsymbol{J}_\mathrm{t}$ and $\boldsymbol{J}_\mathrm{r}$ are denoted as $\boldsymbol{\mathcal{F}}_\mathrm{t}$ and $\boldsymbol{\mathcal{F}}_\mathrm{r}$, and their summation is denoted as $\boldsymbol{\mathcal{F}}$, i.e., $\boldsymbol{\mathcal{F}} = \boldsymbol{\mathcal{F}}_\mathrm{t} + \boldsymbol{\mathcal{F}}_\mathrm{r}$, which is just the field generated by whole WPT system (except the port).

Because of the homogeneous tangential electric field boundary condition $[\boldsymbol{\mathcal{E}}_\mathrm{driv} + \boldsymbol{\mathcal{E}}]^\mathrm{tan} = 0$ on $\mathrm{S}_\mathrm{t}$, the driving power $\mathcal{P}_\mathrm{driv} = <\boldsymbol{J}_\mathrm{t}, \boldsymbol{\mathcal{E}}_\mathrm{driv}>_{\mathrm{S}_\mathrm{t}}$ used to drive the WPT system can be decomposed as follows[27-App.G],[40]:

$$\begin{aligned}
\overbrace{\langle \boldsymbol{J}_\mathrm{t}, \boldsymbol{\mathcal{E}}_\mathrm{driv} \rangle_{\mathrm{S}_\mathrm{t}}}^{\mathcal{P}_\mathrm{driv}} = \langle \boldsymbol{J}_\mathrm{t}, -\boldsymbol{\mathcal{E}} \rangle_{\mathrm{S}_\mathrm{t}} &= \langle \boldsymbol{J}_\mathrm{t}, -\boldsymbol{\mathcal{E}}_\mathrm{t} - \boldsymbol{\mathcal{E}}_\mathrm{r} \rangle_{\mathrm{S}_\mathrm{t}} + \overbrace{\langle \boldsymbol{J}_\mathrm{t}, -\boldsymbol{\mathcal{E}}_\mathrm{t} - \boldsymbol{\mathcal{E}}_\mathrm{r} \rangle_{\mathrm{S}_\mathrm{r}}}^{0} \\
&= \langle \boldsymbol{J}_\mathrm{t}, -\boldsymbol{\mathcal{E}}_\mathrm{t} \rangle_{\mathrm{S}_\mathrm{t}} + \underbrace{\overbrace{\langle \boldsymbol{J}_\mathrm{t}, -\boldsymbol{\mathcal{E}}_\mathrm{r} \rangle_{\mathrm{S}_\mathrm{t}}}^{\mathcal{P}_\mathrm{tra}} + \overbrace{\langle \boldsymbol{J}_\mathrm{t}, -\boldsymbol{\mathcal{E}}_\mathrm{t} \rangle_{\mathrm{S}_\mathrm{r}} + \langle \boldsymbol{J}_\mathrm{t}, -\boldsymbol{\mathcal{E}}_\mathrm{r} \rangle_{\mathrm{S}_\mathrm{r}}}^{0}}_{\mathcal{P}_\mathrm{tra}}
\end{aligned} \tag{4-29}$$

where the second equality is due to that $\mathcal{E} = \mathcal{E}_{\rm t} + \mathcal{E}_{\rm r}$ and that tangential $\mathcal{E}$ is 0 on $\rm S_r$. Based on the Maxwell's equations satisfied by $\mathcal{J}_{\rm t}$ and $\mathcal{F}_{\rm t}$, the first term $< \mathcal{J}_{\rm t}, -\mathcal{E}_{\rm t} >_{\rm S_t}$ in the right-hand side of Eq. (4-29) can be alternatively expressed as the following POYNTING'S THEOREM (PtT)

$$\langle \mathcal{J}_{\rm t}, -\mathcal{E}_{\rm t} \rangle_{\rm S_t} = \oiint_{\rm S_\infty} \left( \mathcal{E}_{\rm t} \times \mathcal{H}_{\rm t} \right) \cdot \boldsymbol{n}_\infty dS + \frac{d}{dt}\left[ \frac{1}{2} \langle \mathcal{H}_{\rm t}, \mu_0 \mathcal{H}_{\rm t} \rangle_{\rm E_3} + \frac{1}{2} \langle \varepsilon_0 \mathcal{E}_{\rm t}, \mathcal{E}_{\rm t} \rangle_{\rm E_3} \right] \tag{4-30}$$

The above PtT (4-30) is a quantitative description for the way how $\mathcal{J}_{\rm t}$ supplies power to $\mathcal{F}_{\rm t}$. It will be proved in the following discussions that $\mathcal{J}_{\rm r}$ is uniquely determined by $\mathcal{J}_{\rm t}$, so <u>$\mathcal{J}_{\rm t}$ can also be viewed as the source for supplying power to $\mathcal{F}_{\rm r}$</u>, and the supplied power can be expressed in terms of $< \mathcal{J}_{\rm t}, -\mathcal{E}_{\rm r} >_{\rm S_t}$, i.e., the $\mathcal{P}_{\rm tra}$ in Eq. (4-29), and then $\mathcal{P}_{\rm tra}$ is called transferred power from $\rm S_t$ to $\rm S_r$ [27-App.G],[40].

For WPT applications, the transferred power $\mathcal{P}_{\rm tra}$ is desired, and the power $< \mathcal{J}_{\rm t}, -\mathcal{E}_{\rm t} >_{\rm S_t}$ is unwanted and expected to be as small as possible. Based on this, we introduce the following concept of transferring coefficient (TC)[27-App.G],[40]

$$\mathrm{TC} = \frac{\left(1/T\right)\int_{t_0}^{t_0+T} \mathcal{P}_{\rm tra} dt}{\left(1/T\right)\int_{t_0}^{t_0+T} \mathcal{P}_{\rm driv} dt} \tag{4-31}$$

to quantify the transferring efficiency of the WPT system. From a relatively mathematical point of view, the central aim of designing transferring system is to search for a physically realizable working mode (or working state) such that TC is maximized. In this section, the central aim is realized by applying a PS-WET-CMT-based modal analysis to the WPT system.

In the following, we simply summarize the mathematical formulations used to establish the PS-WET-CMT for the WPT system, and the details for the formulations can be found in Refs. [27-App.G] and [40]. Because there exist relations $(1/T)\int_{t_0}^{t_0+T} \mathcal{P}_{\rm tra} dt = \mathrm{Re}\, P_{\rm tra}$ and $(1/T)\int_{t_0}^{t_0+T} \mathcal{P}_{\rm driv} dt = \mathrm{Re}\, P_{\rm driv}$, then the following is discussed in frequency domain.

The frequency-domain driving power $P_{\rm driv}$ has the following integral operator expression[27-App.G],[40]

$$P_{\rm driv} = -\left(1/2\right) \langle \boldsymbol{J}_{\rm t}, -j\omega\mu_0 \mathcal{L}_0 \left( \boldsymbol{J}_{\rm t} + \boldsymbol{J}_{\rm r} \right) \rangle_{\rm S_t} \tag{4-32}$$

called frequency-domain PS-DPO. The currents $\boldsymbol{J}_{\rm t}$ and $\boldsymbol{J}_{\rm r}$ involved in PS-DPO are not independent, and they satisfy the following integral equation

$$\left[-j\omega\mu_0\mathcal{L}_0\left(\boldsymbol{J}_{\mathrm{t}}+\boldsymbol{J}_{\mathrm{r}}\right)\right]_{\mathrm{S_r}}^{\tan}=0 \tag{4-33}$$

Similarly to the previous sections, it is easy to discretize PS-DPO (4-32) and integral equation (4-33) into matrix forms. Employing the matrix forms, we can obtain $P_{\mathrm{driv}}=\mathbb{J}_{\mathrm{t}}^{\dagger}\cdot\mathbb{P}_{\mathrm{driv}}\cdot\mathbb{J}_{\mathrm{t}}$, which involves independent current $\mathbb{J}_{\mathrm{t}}$ only. Using the positive and negative Hermitian parts of $\mathbb{P}_{\mathrm{driv}}$, the characteristic equation $\mathbb{P}_{\mathrm{driv}}^{-}\cdot\mathbb{J}_{\mathrm{t}}=\theta\,\mathbb{P}_{\mathrm{driv}}^{+}\cdot\mathbb{J}_{\mathrm{t}}$ can be formulated, and it gives the energy-decoupled CMs of the WPT system.

The WPT system considered in Ref. [66] is constituted by two metallic coils as shown in Fig. 4-16. The coils have the same radius 30 cm, height 20 cm, and turns 5.25. The coils are placed coaxially, and their distance is 2 m. The optimally transferring frequency (i.e. the working frequency of the optimally transferring mode) calculated from the coupled-mode theory used in Ref. [66] is $10.56\pm0.3\,\mathrm{MHz}$, and the optimally transferring frequency obtained from the measurement done in Ref. [66] is 9.90 MHz. The reason leading to a 5% discrepancy between the theoretical and measured values was explained in Ref. [66].

We use the PS-WET-CMT established in this section to calculate the energy-decoupled CMs of the WPT system, and show the TC curves of the first 5 CMs in the following Fig. 4-17.

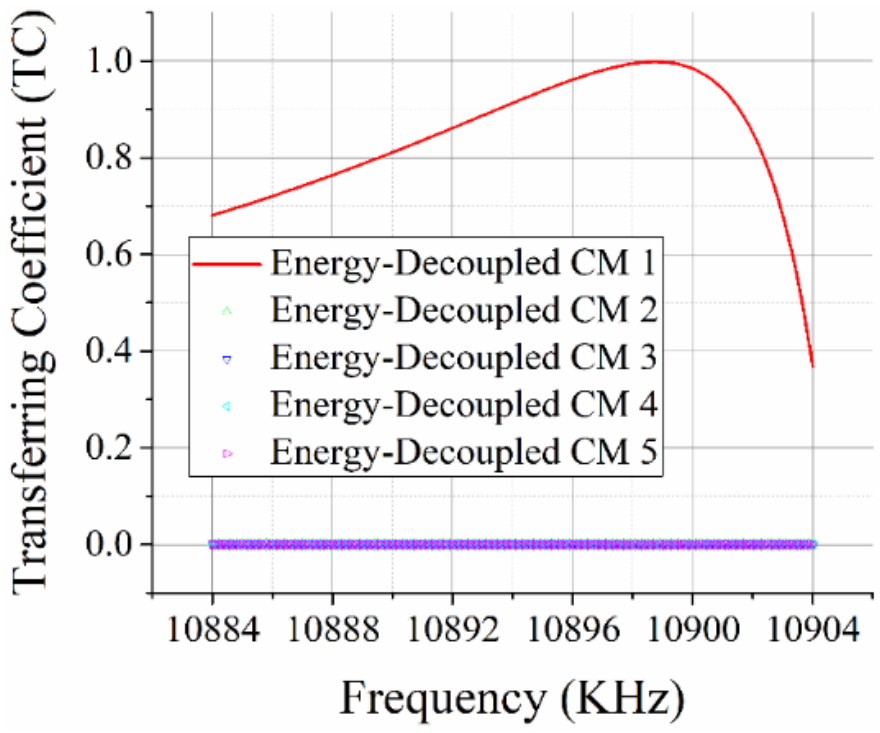

Figure 4-17 TC curves of the first 5 lower-order energy-decoupled CMs.

It is not difficult to observe that the CM 1 at 10.8988 MHz (which corresponds to the local maximum of the TC curve) works at the optimally transferring state. The coil current distribution and time-averaged magnetic energy density distribution of the optimally transferring mode are shown in Fig. 4-18 and Fig. 4-19 respectively. Evidently, the CM 1 working at 10.8988 MHz corresponds to a half-wave current distribution for both coil T and coil R as shown in Fig. 4-18, and it indeed can efficiently transfer EM power from coil T to coil R in a wireless manner as shown in Fig. 4-17 and Fig. 4-19.

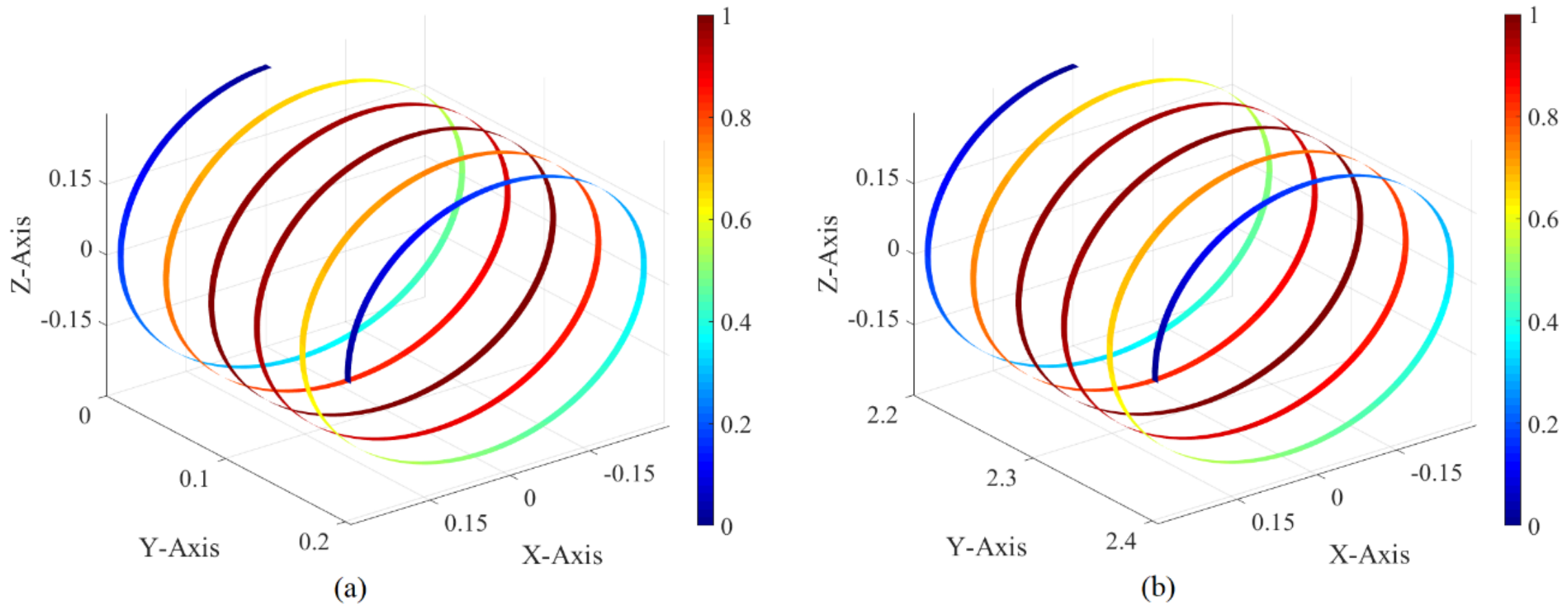

Figure 4-18 For the CM 1 working at 10.8988 MHz, its current magnitudes distributing on (a) coil T and (b) coil R.

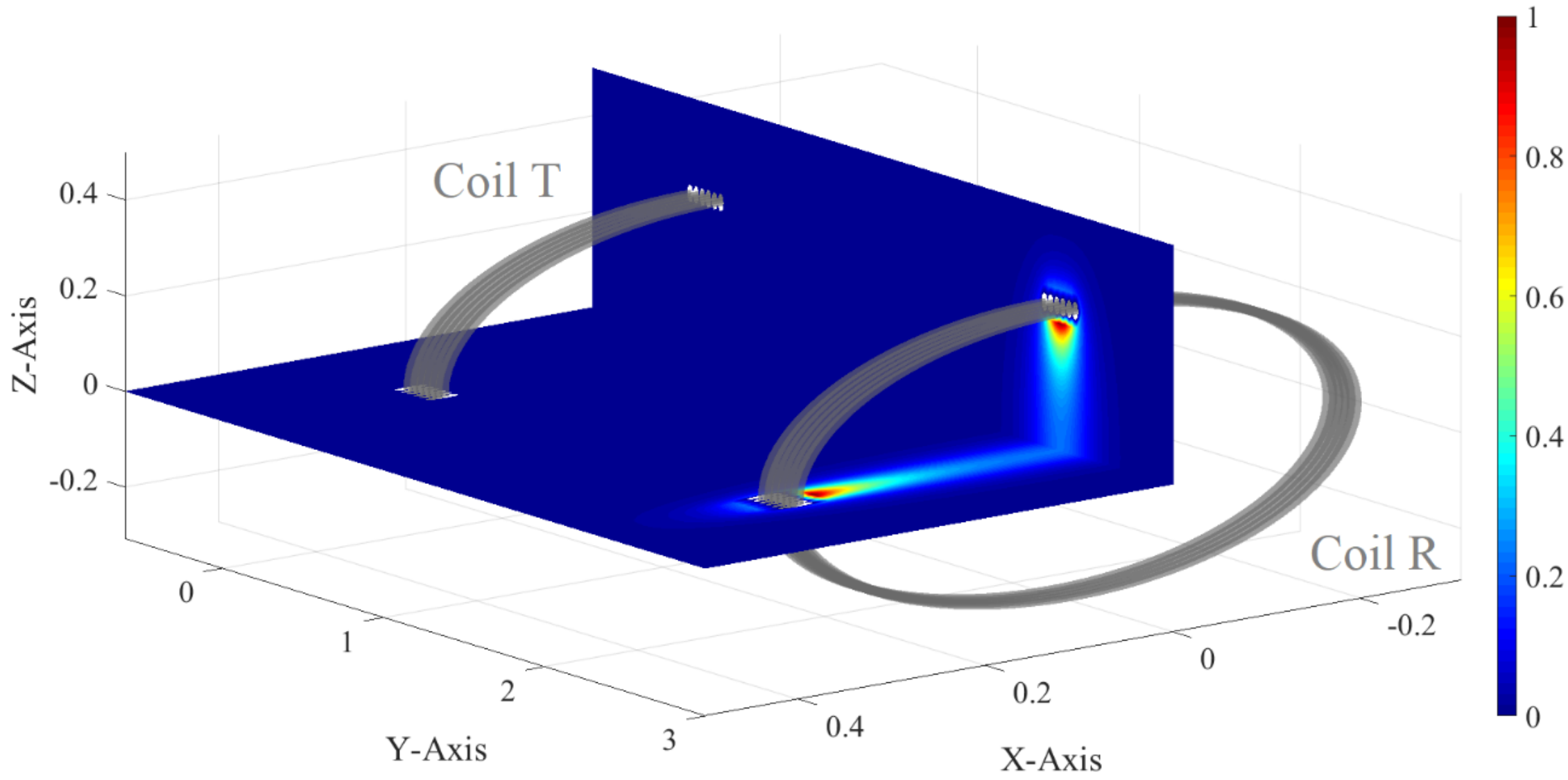

Figure 4-19 For the CM 1 working at 10.8988 MHz, its time-averaged magnetic energy density distribution on xOy and yOz planes.

In addition, the optimally transferring frequencies calculated from the classical coupled-mode theory ($10.56 \pm 0.3\,\mathrm{MHz}$) proposed in Ref. [67] and the PS-WET-CMT (10.8988 MHz) used in this section are consistent with each other. The advantage of PS-WET-CMT over coupled-mode theory is reflected as follows:

Aspect 1. The PS-WET-CMT is a field-based modal analysis method, which is directly derived from Maxwell's equations and doesn't use any approximation; the coupled-mode theory is a circuit-model-based modal analysis method, which employs some circuit-model-based approximate quantities (such as scalar voltage, scalar current, effective inductance, and effective capacitance, etc.).

Aspect 2. The PS-WET-CMT is applicable to the coils working at arbitrary frequency; the coupled-mode theory is only applicable to the coils working at low frequency at which the circuit model exists.

Aspect 3. The PS-WET-CMT is applicable to the coils with arbitrary geometries; the coupled-mode theory is only applicable to the coils with simple geometries, such that the coils can support sinusoidal scalar currents.

By an alternative field-based modal analysis, we verify that the CM 1 at 10.8988 MHz is indeed the most efficient mode for WPT as below, and then exhibit that the optimally transferring mode is indeed included in the CM set constructed by PS-WET-CMT.

Similarly to obtaining $P_{\text{driv}} = \mathbb{J}_{\text{t}}^{\dagger} \cdot \mathbb{P}_{\text{driv}} \cdot \mathbb{J}_{\text{t}}$, the transferred power $\mathcal{P}_{\text{tra}}$ can be formulated as $P_{\text{tra}} = \mathbb{J}_{\text{t}}^{\dagger} \cdot \mathbb{P}_{\text{tra}} \cdot \mathbb{J}_{\text{t}}$, where $\mathbb{P}_{\text{tra}}$ can be calculated as the formulations provided in Refs. [27-App.G] and [40]. Then, the mode maximizing TC (4-31) can be obtained from solving the following equation[67]

$$\mathbb{P}_{\text{tra}}^{+} \cdot \mathbb{J}_{\text{t}} = \tau \, \mathbb{P}_{\text{driv}}^{+} \cdot \mathbb{J}_{\text{t}} \tag{4-34}$$

where $\mathbb{P}_{\text{driv}}^{+}$ and $\mathbb{P}_{\text{tra}}^{+}$ are the positive Hermitian parts of $\mathbb{P}_{\text{driv}}$ and $\mathbb{P}_{\text{tra}}$ respectively. Using the equation, we calculate the optimally transferring mode, and show the associated TC curve in Fig. 4-20.

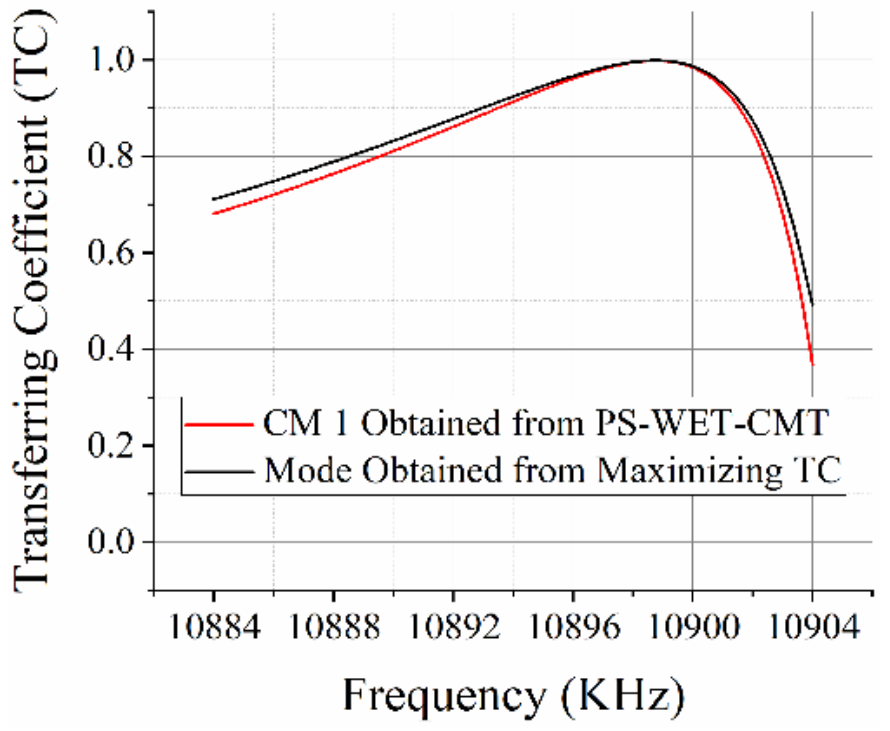

Figure 4-20 TC curves of the optimally transferring modes obtained from two somewhat different modal analysis methods provided in this section.

Obviously, both the obtained optimally transferring frequency and optimally transferring coefficient are consistent with the ones obtained from the PS-WET-CMT-based modal analysis method.

In fact, the above PS-WET-CMT for the two-coil WPT system shown in Fig. 4-16 can be directly applied to some more complicated metallic WPT systems, such as the coaxial $N$-coil WPT system ( $N > 2$ )[27-App.G7.6],[40-Sec.V] shown in Fig. 4-21(a) and the non-coaxial $N$-coil WPT system ( $N \geq 2$ )[27-App.G7.6],[40-Sec.V] shown in Fig. 4-21(b) and can also be applied to the two-coil WPT system with obstacle between the coils, such as the one shown in Fig. 4-22[27-App.G7.7],[40-Sec.V].

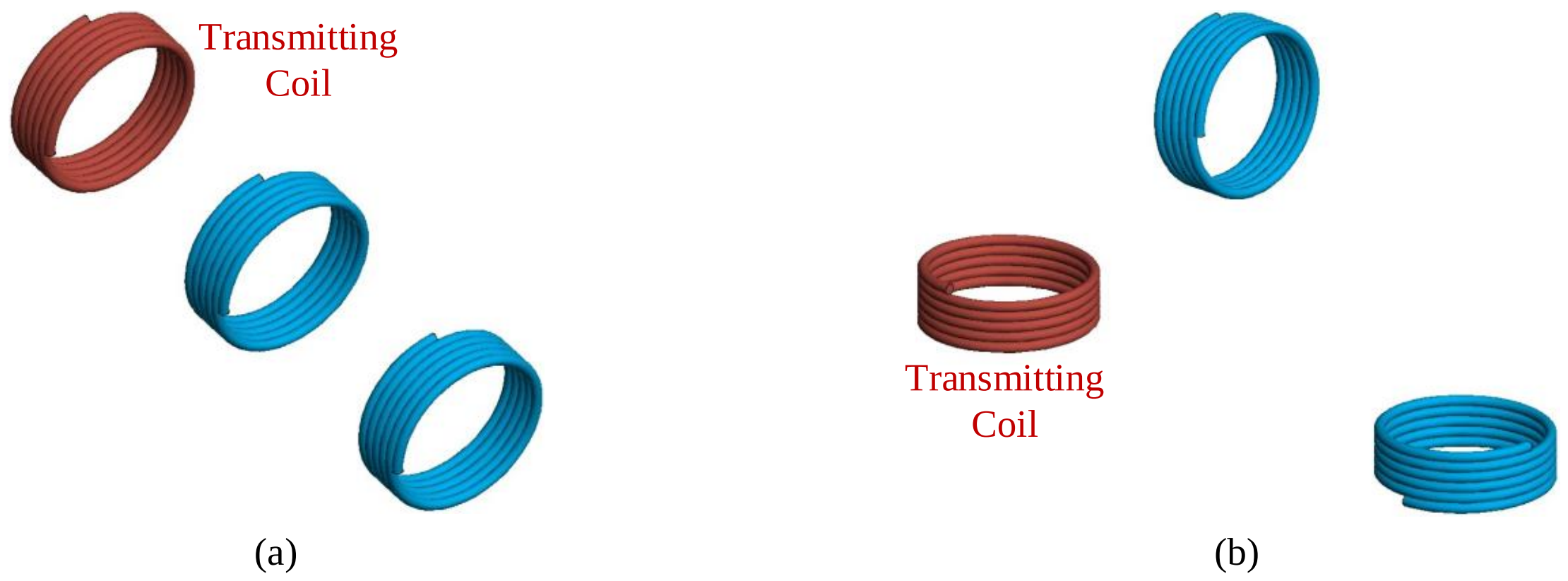

Figure 4-21 Two typical three-coil WPT systems. (a) Three coils are placed coaxially; (b) three coils are placed non-coaxially.

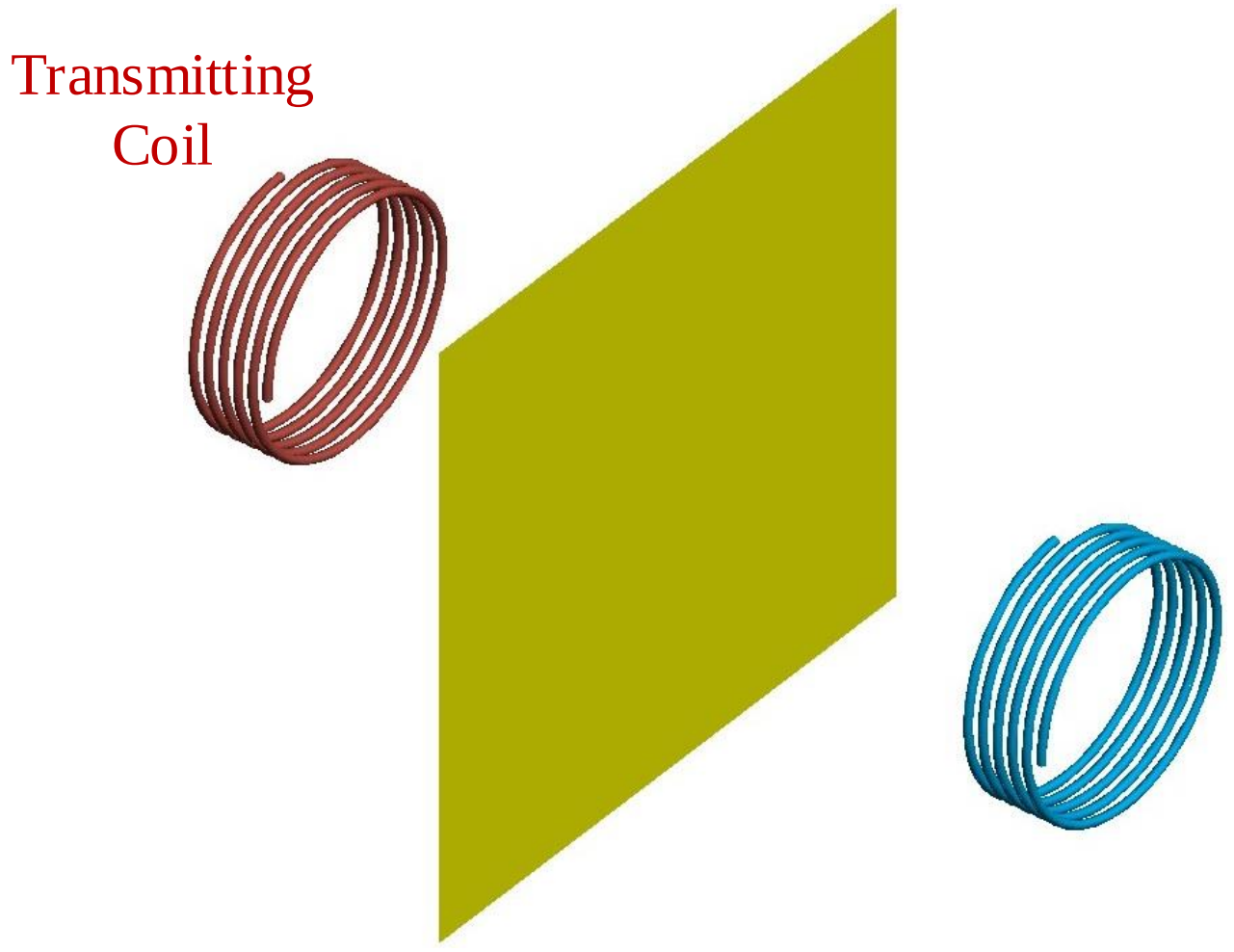

Figure 4-22 A typical two-coil WPT system with a metallic plane obstacle.

In addition, the PS-WET-CMT for the WPT systems constituted by metallic coils can also be further generalized to the WPT systems constituted by material coils.

## 4.7 Chapter Summary

For lumped-port-driven EM structures, their energy utilization process (especially energy source) is different from the energy utilization processes of both incident-field-driven scattering structures and wave-port-fed transceiving systems. Thus, the conventional scatterer-oriented ENTIRE-STRUCTURE-ORIENTED WORK-ENERGY THEOREM based CMT (ES-WET-CMT, to be discussed in the following Chap. 5) and the novel transceiver-oriented Power Transport Theorem based Decoupling Mode Theory (PTT-DMT, had been discussed in the previous Chap. 3) fail to analyze the lumped-port-driven EM structures.

The central purpose of this chapter is to establish an effective energy-viewpoint-based modal analysis method for lumped-port-driven EM structures. Taking "metallic and material Yagi-Uda antennas", "metallic dipole antennas with passive loads", and "two-coil WPT system" as examples, this chapter exhibits that: by properly generalizing the conventional scatterer-oriented ES-WET-CMT, an effective modal analysis method — PS-WET-CMT — for lumped-port-driven EM structures can be established under PS-WET framework. By orthogonalizing frequency-domain PS-DPO, the PS-WET-CMT can effectively construct the energy-decoupled CMs of lumped-port-driven EM structures. Here, the PS-DPO is just the Energy Source Operator contained in PS-WET.

The essential difference between PS-WET-CMT and ES-WET-CMT is that: the modal generating operator PS-DPO used by the former is the power done by the driving field acting on a part of the objective EM structure, but the modal generating operator ES-DPO (Entire-Structure-Oriented Driving Power Operator) used by the latter is the power done by the incident field acting on the entire objective EM structure.

# CHAPTER 5 ES-WET-BASED MODAL ANALYSIS FOR INCIDENT-FIELD-DRIVEN EM STRUCTURES

**CHAPTER MOTIVATION:** The main destinations of this chapter are the following three: (1) to generalize the far-field orthogonality satisfied by the characteristic modes of lossless scatterers to a more general orthogonality relation; (2) to compare the differences between scatterer-oriented modal analysis theory and {antenna, waveguide}-oriented modal analysis theories; (3) to generalize scatterer-oriented modal analysis theory from the conventional scattering problem to some more complicated scattering problems.

## 5.1 Chapter Introduction

Scatterer-oriented CHARACTERISTIC MODE THEORY (CMT) exists some different versions, such as scattering matrix based CMT (SM-CMT)[5~7], integral equation based CMT (IE-CMT)[9~13], and ENTIRE-STRUCTURE-ORIENTED WORK-ENERGY THEOREM based CMT (ES-WET-CMT)[8,14,15].

The ES-WET-CMT treats all external fields (including the fields generated by externally impressed source and external environment) of objective scatterer as a whole — external incident field, so the CHARACTERISTIC MODES (CMs) calculated from ES-WET-CMT depend only on the inherent physical characters of the objective scatterer[27-Sec.1.2.4.4],[8,14]. For a relatively general metal-material composite scatterer shown in Fig. 5-1, Refs. [8-Chap.5] and [15] carefully discussed the method to calculate its ENTIRE-STRUCTURE-ORIENTED WORK-ENERGY THEOREM (ES-WET) based inherent CMs.

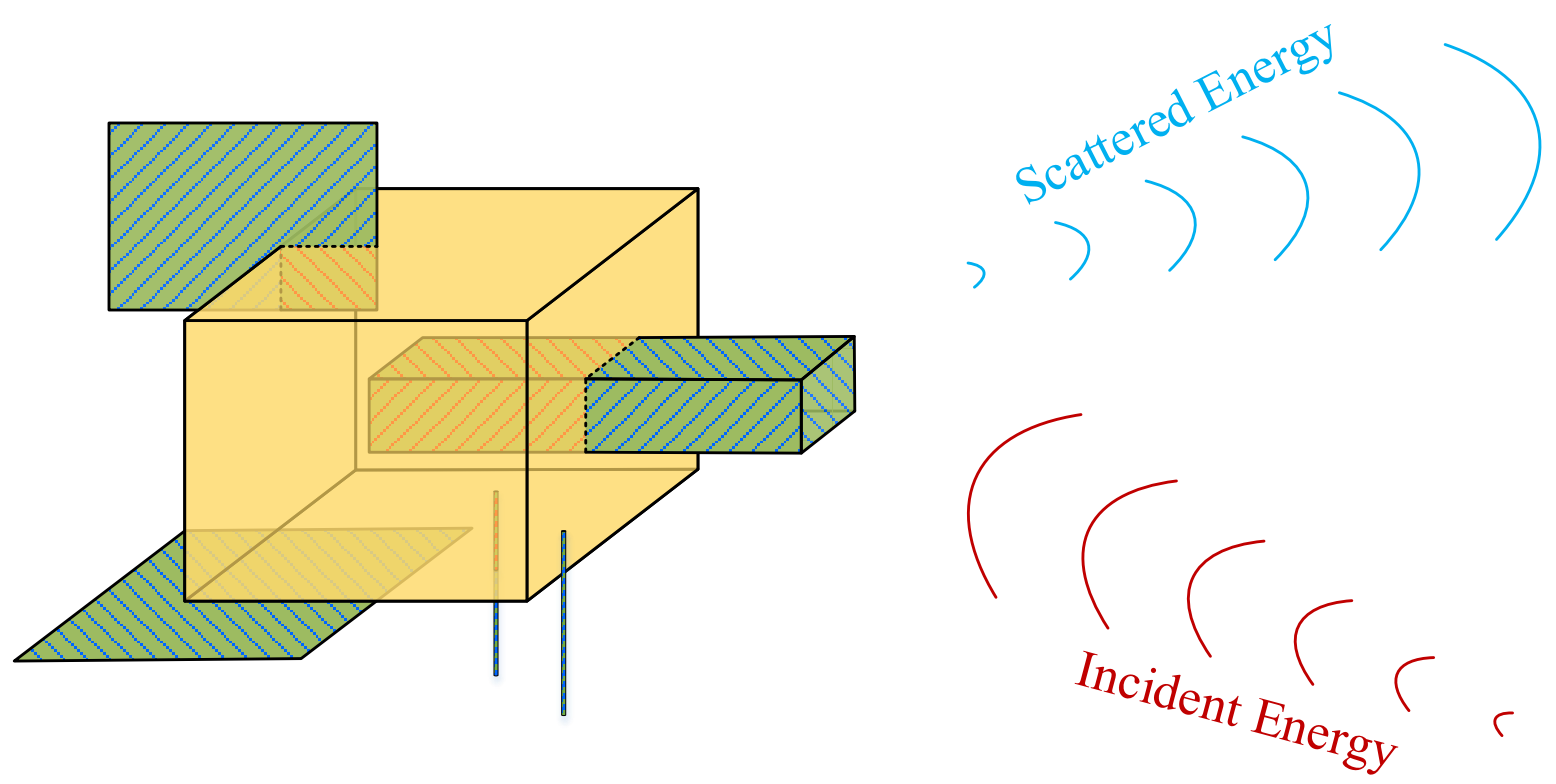

Figure 5-1 Geometry of a metal-material composite scatterer considered in Refs. [8-Chap.5] and [15] and constituted by metallic {line, surface, body} and material body.

Taking three typical electromagnetic scattering structures (metallic horn scatterer, metallic Yagi-Uda array scatterer, and metallic two-coil scatterer) as examples, the subsequent Sec. 5.2 focuses on two main purposes: (1) to obtain some further results about the modal orthogonalities satisfied by the scattering CMs and (2) to compare the differences between the scatterer-oriented ES-WET-CMT established in Refs. [8,14,15] and the {antenna, waveguide}-oriented modal analysis theories established in Refs. [27-Chap.6], [27-Apps.G&H], and [35,37,40].

When we want to calculate the scattering CMs containing the information of scatterer-environment interaction, the obtained CMs must be environment-dependent, so they must be different from the ES-WET-based inherent CMs calculated in Refs. [8,14,15]. To calculate the environment-dependent CMs, Ref. [27-Sec.1.2.4.4] proposed a scheme: to treate the original scatterer and the non-free-space environment as a whole — augmented scattering system, and then to calculate the CMs of the augmented scattering system, and the obtained modal currents distributing on the original scatterer are just the environment-dependent CM currents[27-Sec.1.2.4.4]. But in fact, the above-obtained environment-dependent CMs are not energy-decoupled as explained in the Sec. 5.3.1 of this report. To effectively calculate the environment-dependent energy-decoupled CMs, we propose an alternative scheme in the Sec. 5.3.2 of this report.

Similarly to generalizing the classical inherent CMs to environment-dependent CMs, Sec. 5.4 further generalizes them to driver-dependent CMs, which contain the information of scatterer-driver interaction. Just like the classical inherent CMs, both the generalized environment-dependent and driver-dependent CMs are constructed from orthogonalizing proper Energy Source Operators.

## 5.2 ES-WET-Based Inherent CMs of Scattering Structures

By three relatively simple but very typical examples (metallic horn, Yagi-Uda array, two-coil system), this section derives some further results satisfied by scattering CMs, and exhibits the differences between the scatterer-oriented CMT established in Refs. [8~15] and the {antenna, waveguide}-oriented modal analysis theories established in Refs. [27-Chap.6], [27-Apps.G&H], and [35,37,40].

### 5.2.1 Example I: Metallic Horn Treated as Scatterer

The following figure illustrates an electromagnetic (EM) scattering problem, and the scatterer is a metallic horn.

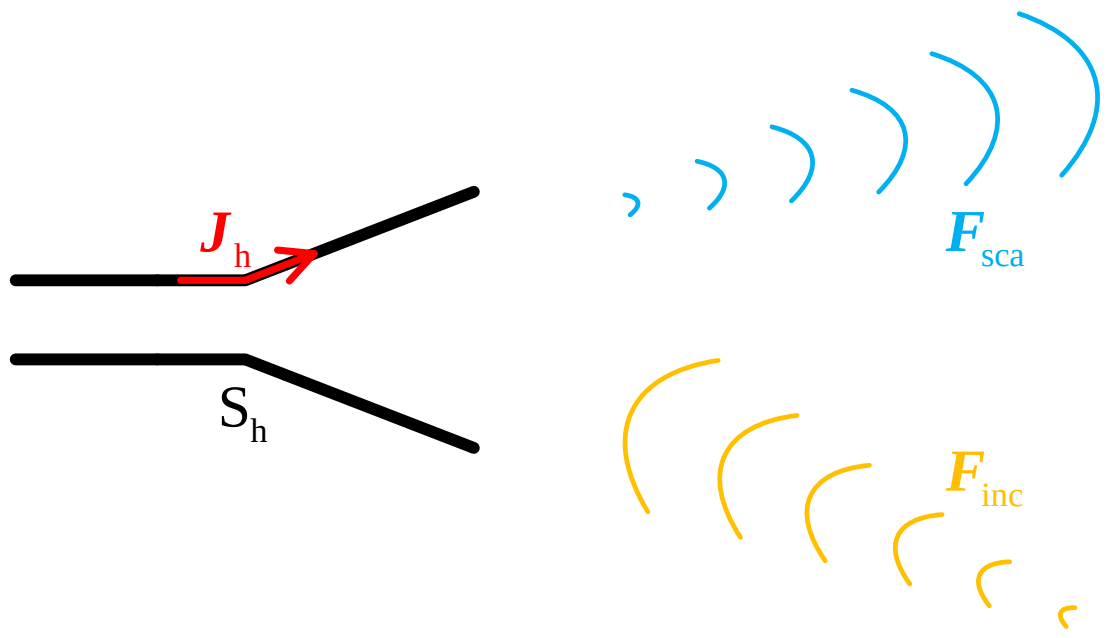

Figure 5-2 EM scattering problem considered in Sec. 5.2.1, where the scatterer is a metallic horn placed in free space.

Taking the scattering problem shown in Fig. 5-2 as a typical example, this subsection compares the differences between the conventional CMT for scattering structures and the Power Transport Theorem based Decoupling Mode Theory (PTT-DMT) for wave-port-fed transmitting antennas.

In the figure, the horn has boundary surface $S_h$, and is placed in free space with material parameters $(\mu_0, \varepsilon_0)$, and is driven by an externally incident field $\boldsymbol{F}_{inc}$. Under the driving of $\boldsymbol{F}_{inc}$, a current $\boldsymbol{J}_h$ will be induced on $S_h$, and a field $\boldsymbol{F}_{sca}$ is generated by the current correspondingly.

The ES-WET implies that the source used to sustain a steady EM scattering is the entire-structure-oriented driving power $P_{DRIVE} = (1/2) < \boldsymbol{J}_h, \boldsymbol{E}_{inc} >_{S_h}$, and the $P_{DRIVE}$ has the following integral and matrix operator forms

$$P_{DRIVE} = -(1/2)\left\langle \boldsymbol{J}_h, -j\omega\mu_0\mathcal{L}_0(\boldsymbol{J}_h)\right\rangle_{S_h} = \mathbb{J}_h^\dagger \cdot \mathbb{P}_{DRIVE} \cdot \mathbb{J}_h \tag{5-1}$$

called Entire-Structure-Oriented Driving Power Operator (ES-DPO), where $\mathbb{J}_h$ is the basis function expansion coefficient vector of $\boldsymbol{J}_h$. By orthogonalizing the ES-DPO, i.e., solving characteristic equation $\mathbb{P}_{DRIV}^- \cdot \mathbb{J}_h = \theta\, \mathbb{P}_{DRIV}^+ \cdot \mathbb{J}_h$ (where $\mathbb{P}_{DRIV}^+$ and $\mathbb{P}_{DRIV}^-$ are the positive and negative Hermitian parts of $\mathbb{P}_{DRIV}$), the CMs satisfying the following relations can be obtained.

$$\begin{aligned}\frac{1}{2}\left\langle \boldsymbol{J}_h^m, \boldsymbol{E}_{inc}^n \right\rangle_{S_h} &= (1 + j\,\theta_m)\delta_{mn} \\ &= \underbrace{\overbrace{\frac{1}{2}\oint\!\!\!\oint_{S_\infty}\left[\boldsymbol{E}_{sca}^n \times \left(\boldsymbol{H}_{sca}^m\right)^\dagger\right]\cdot \boldsymbol{n}_\infty dS}^{\delta_{mn}} + \overbrace{j\,2\omega\left[\frac{1}{4}\left\langle \boldsymbol{H}_{sca}^m, \mu_0 \boldsymbol{H}_{sca}^n\right\rangle_{E_3} - \frac{1}{4}\left\langle \varepsilon_0 \boldsymbol{E}_{sca}^m, \boldsymbol{E}_{sca}^n\right\rangle_{E_3}\right]}^{\theta_m\delta_{mn}}}_{\text{whose real part is } (1/T)\int_{t_0}^{t_0+T}\left\langle \mathcal{J}_h^m, \mathcal{E}_{inc}^n\right\rangle_{S_h} dt}\end{aligned} \tag{5-2}$$

which implies that the obtained CMs are energy-decoupled. In relation (5-2), $\mathrm{E}_3$ is the whole three-dimensional Euclidean space, and $\mathrm{S}_\infty$ is the boundary of $\mathrm{E}_3$, and $\boldsymbol{n}_\infty$ is the outer normal direction of $\mathrm{S}_\infty$, and $T$ is the time period of the time-harmonic field. Obviously, the far-field orthogonality $(1/2)\oiint_{\mathrm{S}_\infty}[\boldsymbol{E}_{\mathrm{sca}}^n \times (\boldsymbol{H}_{\mathrm{sca}}^m)^\dagger]\cdot \boldsymbol{n}_\infty dS = \delta_{mn}$ implies the following orthogonality

$$(1/T)\int_{t_0}^{t_0+T}\left[\oiint_{\mathrm{S}_\infty}\left(\boldsymbol{\mathcal{E}}_{\mathrm{sca}}^n \times \boldsymbol{\mathcal{H}}_{\mathrm{sca}}^m + \boldsymbol{\mathcal{E}}_{\mathrm{sca}}^m \times \boldsymbol{\mathcal{H}}_{\mathrm{sca}}^n\right)\cdot \boldsymbol{n}_\infty dS\right]dt = 2\delta_{mn} \tag{5-3}$$

and then time-domain LORENTZ'S RECIPROCITY THEOREM (2-22) implies the following orthogonality

$$(1/T)\int_{t_0}^{t_0+T}\left[\oiint_{\mathrm{S}}\left(\boldsymbol{\mathcal{E}}_{\mathrm{sca}}^n \times \boldsymbol{\mathcal{H}}_{\mathrm{sca}}^m + \boldsymbol{\mathcal{E}}_{\mathrm{sca}}^m \times \boldsymbol{\mathcal{H}}_{\mathrm{sca}}^n\right)\cdot \boldsymbol{n} dS\right]dt = 2\delta_{mn} \tag{5-4}$$

where S is an arbitrary closed surface enclosing whole horn scatterer as shown in Fig. 5-3, and $\boldsymbol{n}$ is the outer normal direction of S. Here, we want to emphasize that the fields satisfying orthogonalities (5-3) and (5-4) are the modal **scattered** fields. In addition, the orthogonalities (5-3) and (5-4) are also valid for any **lossless** scatterer, such as the metallic Yagi-Uda scatterer to be discussed in Sec. 5.2.2 and the metallic two-coil scatterer to be discussed in Sec. 5.2.3.

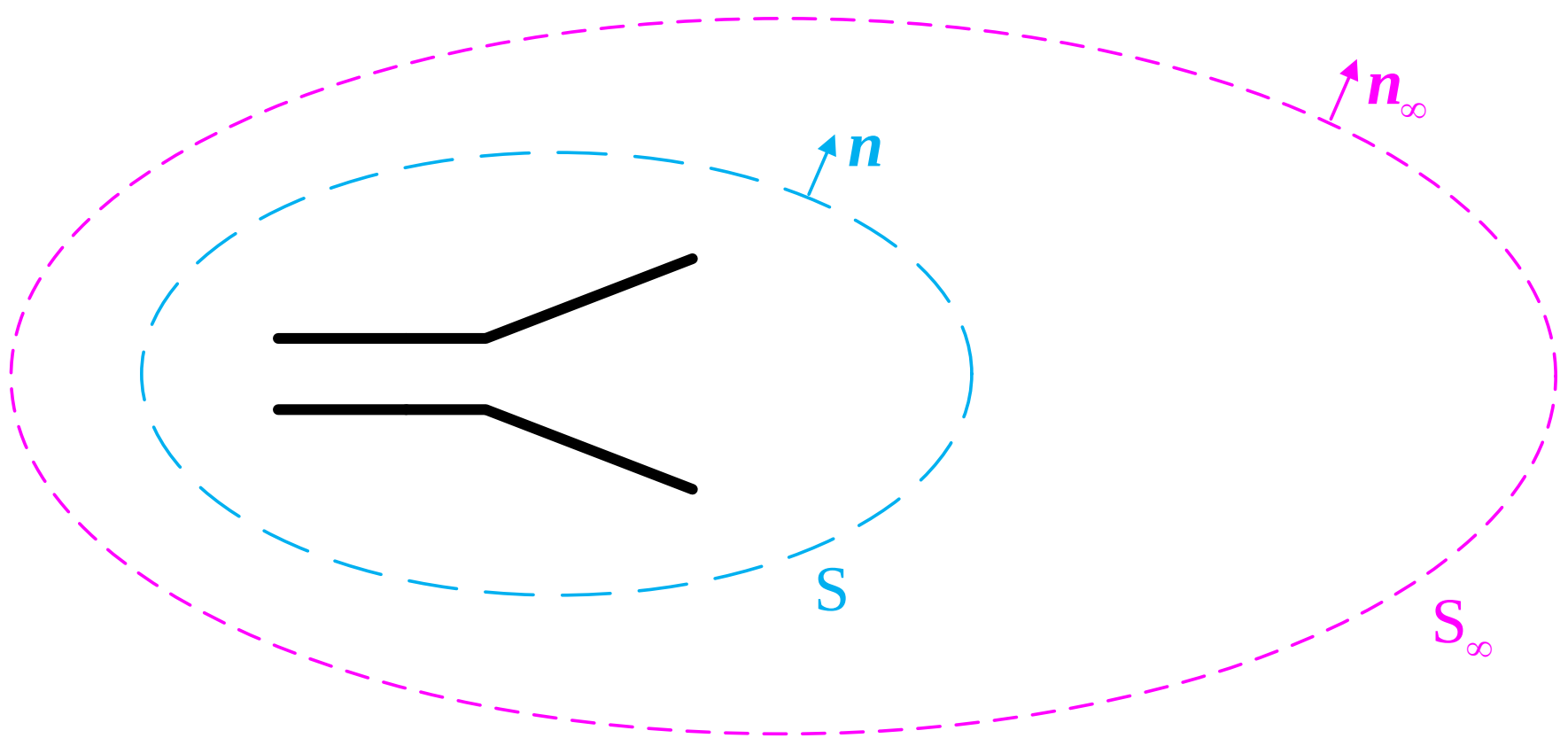

Figure 5-3 A closed surface S enclosing whole horn scatterer.

Because the above modal calculation process doesn't consider the specific form of the external environment and excitation, then the obtained CMs only depend on the inherent characters of the scattering object[8,14], so the CMs are particularly called inherent CMs. For a specific horn scatterer having the size shown in Fig. 3-4, its ES-WET-based inherent CMs are calculated, and the associated modal significances (MSs) are shown in Fig. 5-4.

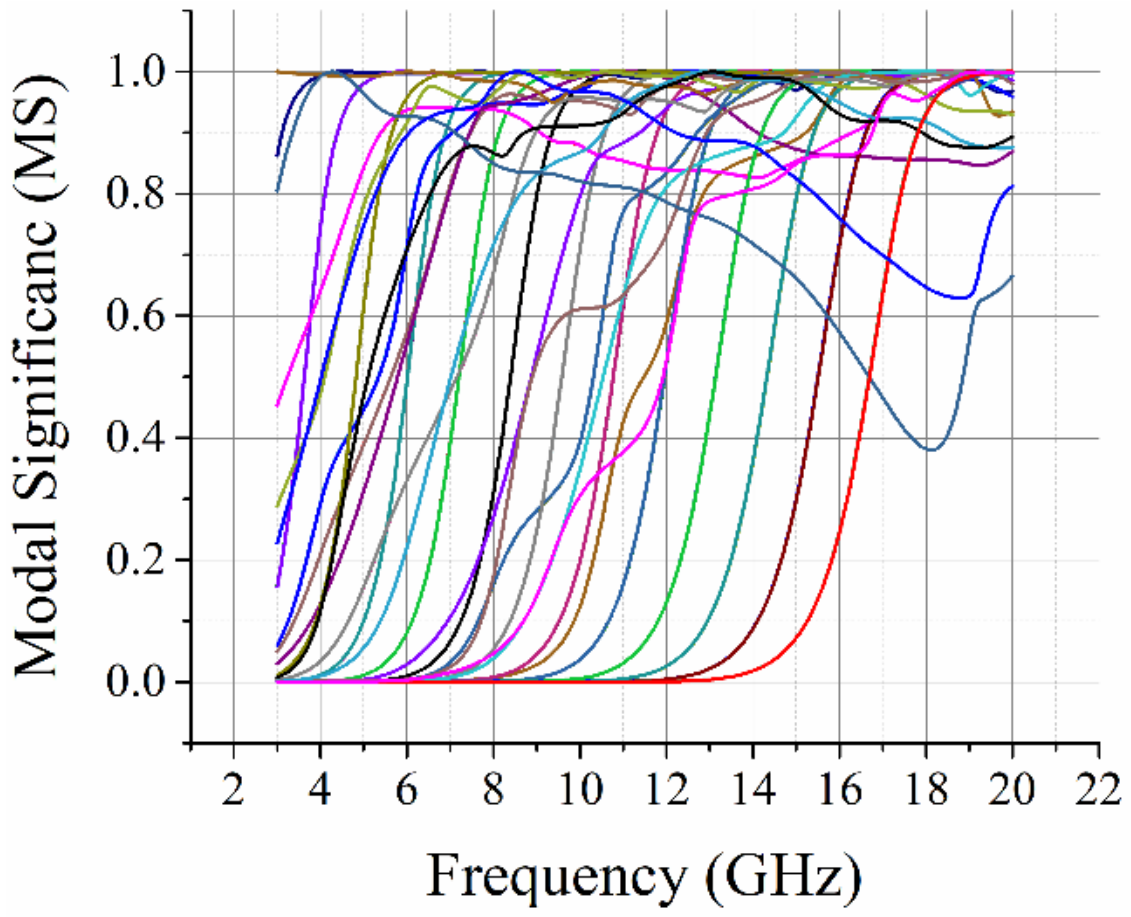

Figure 5-4 MSs of the first several lower-order ES-WET-based inherent CMs of the horn scatterer having the size shown in Fig. 3-4.

The radiation patterns of the first several lower-order ES-WET-based inherent CMs working at 9.3 GHz are shown in the following Fig. 5-5.

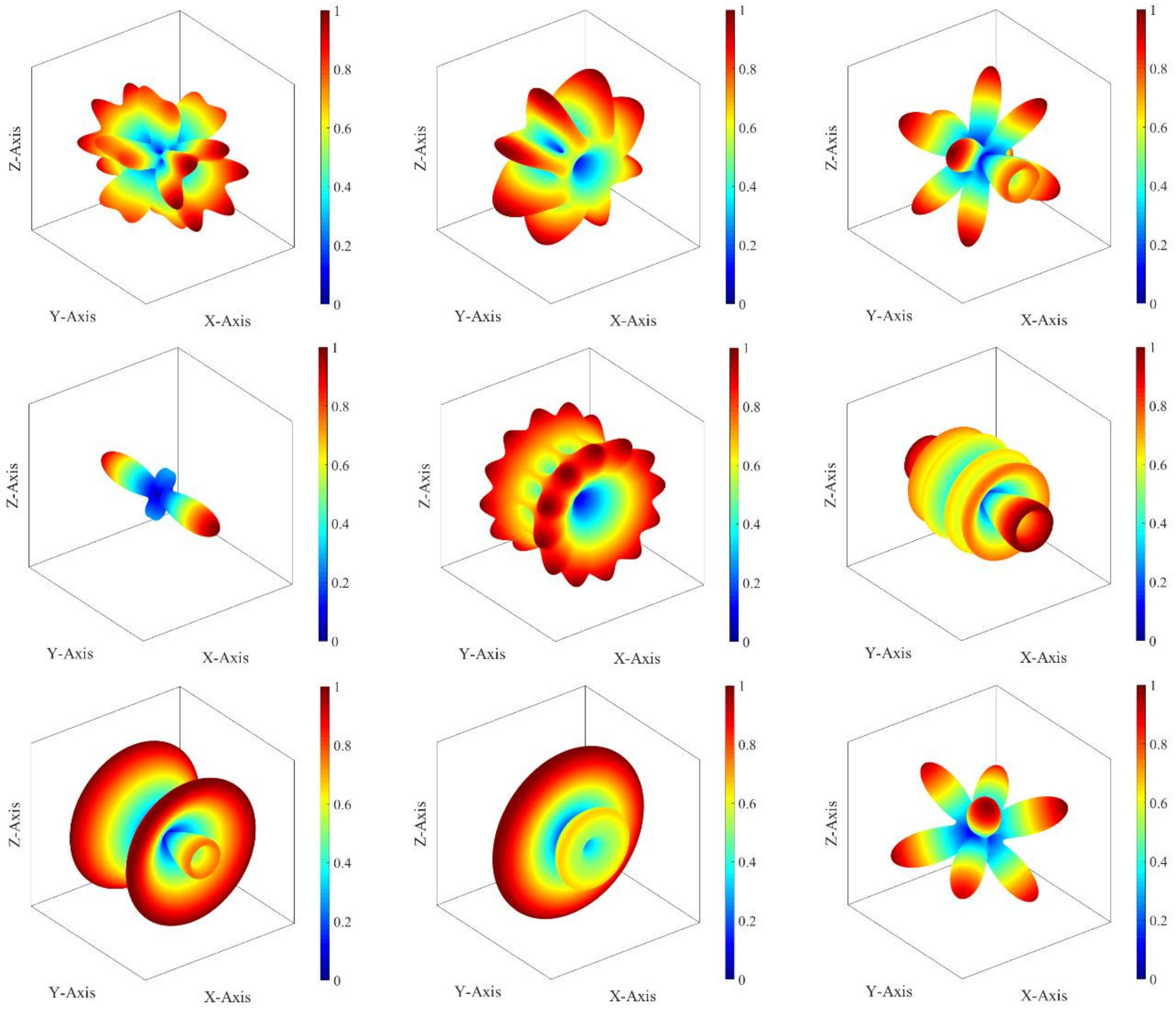

Figure 5-5 Radiation patterns of the first several lower-order ES-WET-based inherent CMs working at 9.3 GHz.

Evidently, the commonly used end-fire mode with radiation pattern Fig. 3-7(b) is not contained in the above-obtained ES-WET-based inherent CM set. This implies that the ES-WET-CMT fails to analyze the wave-port-fed transmitting antennas[35]. The reason leading to the failure of the ES-WET-CMT-based modal analysis for the horn antenna is that: **the ES-DPO used to calculate the ES-WET-based CMs is the source term contained in ES-WET (which governs the scattering process of horn scatterer), but not the source term contained in PTT (which governs the transmitting process of horn antenna)**.

## 5.2.2 Example II: Metallic Yagi-Uda Array Treated as Scatterer

The following figure illustrates a EM scattering problem, and the scatterer is a metallic Yagi-Uda array.

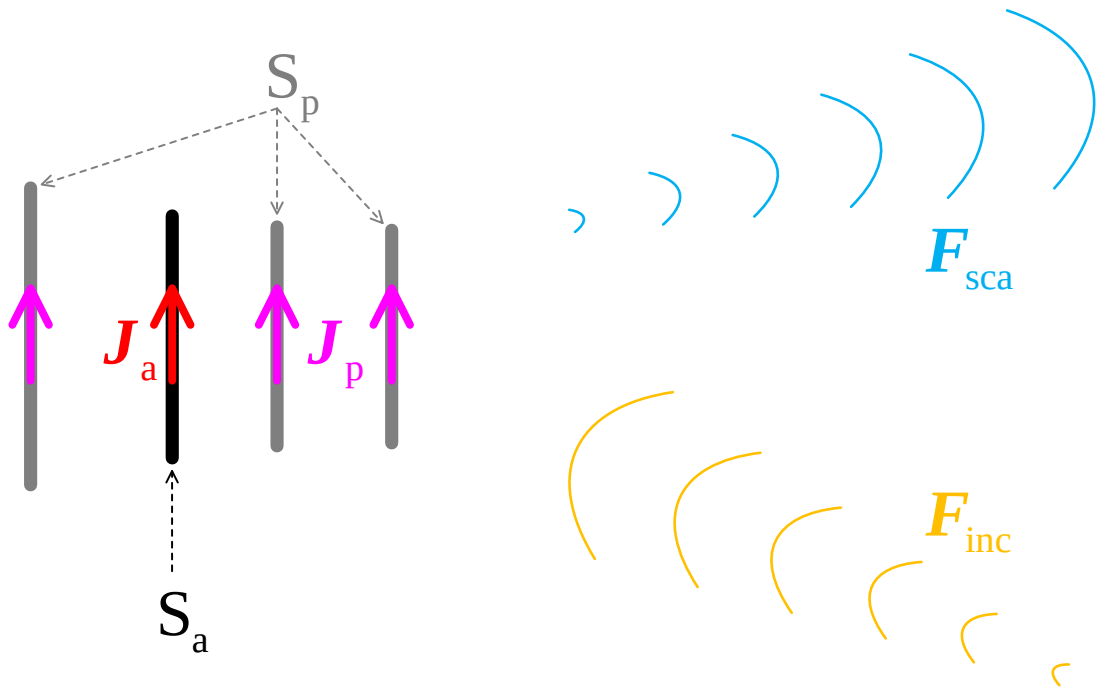

Figure 5-6 EM scattering problem considered in Sec. 5.2.2, where the scatterer is a metallic Yagi-Uda array placed in free space.

Taking the scattering problem shown in Fig. 5-6 as a typical example, this subsection compares the differences between the conventional CMT for scattering structures and the Partial-Structure-Oriented Work-Energy Theorem based CMT (PS-WET-CMT) for lumped-port-driven transmitting antennas.

In the figure, the Yagi-Uda array is placed in free space with material parameters $(\mu_0, \varepsilon_0)$, and is driven by an externally incident field $\boldsymbol{F}_{\text{inc}}$; the active element and passive elements of the Yagi-Uda array are denoted as $\text{S}_{\text{a}}$ and $\text{S}_{\text{p}}$ respectively the same as Sec. 4.2. Under the driving of $\boldsymbol{F}_{\text{inc}}$, currents $\boldsymbol{J}_{\text{a}}$ and $\boldsymbol{J}_{\text{p}}$ will be induced on $\text{S}_{\text{a}}$ and $\text{S}_{\text{p}}$, and field $\boldsymbol{F}_{\text{sca}}$ is generated by current $\boldsymbol{J}_{\text{a}} + \boldsymbol{J}_{\text{p}}$ correspondingly.

Entire-structure-oriented driving power $P_{\text{DRIV}} = (1/2) < \boldsymbol{J}_{\text{a}} + \boldsymbol{J}_{\text{p}}, \boldsymbol{E}_{\text{inc}} >_{\text{S}_{\text{a}} \cup \text{S}_{\text{p}}}$, which is the source to result in an entire-structure-oriented EM scattering process, has the following integral and matrix operator forms

$$P_{\rm DRIV} = -(1/2)\left\langle \boldsymbol{J}_{\rm a} + \boldsymbol{J}_{\rm p}, -j\omega\mu_0\mathcal{L}_0\left(\boldsymbol{J}_{\rm a} + \boldsymbol{J}_{\rm p}\right)\right\rangle_{\rm S_a \cup S_p} = \mathbb{J}_{\rm ap}^{\dagger} \cdot \mathbb{P}_{\rm DRIV} \cdot \mathbb{J}_{\rm ap}, \text{ where } \mathbb{J}_{\rm ap} = \begin{bmatrix} \mathbb{J}_{\rm a} \\ \mathbb{J}_{\rm p} \end{bmatrix} \quad (5\text{-}5)$$

called ES-DPO, where $\mathbb{J}_{\rm a}$ and $\mathbb{J}_{\rm p}$ are the basis function expansion coefficient vectors of $\boldsymbol{J}_{\rm a}$ and $\boldsymbol{J}_{\rm p}$ respectively. By solving characteristic equation $\mathbb{P}_{\rm DRIV}^{-} \cdot \mathbb{J}_{\rm ap} = \theta\, \mathbb{P}_{\rm DRIV}^{+} \cdot \mathbb{J}_{\rm ap}$, the CMs satisfying the following relations

$$\begin{aligned}\frac{1}{2}\left\langle \boldsymbol{J}_{\rm a}^{m} + \boldsymbol{J}_{\rm p}^{m}, \boldsymbol{E}_{\rm inc}^{n}\right\rangle_{\rm S_a \cup S_p} &= \left(1 + j\,\theta_m\right)\delta_{mn} \\ &= \underbrace{\frac{1}{2}\oint\!\!\!\oint_{\rm S_\infty}\left[\boldsymbol{E}_{\rm sca}^{n} \times \left(\boldsymbol{H}_{\rm sca}^{m}\right)^{\dagger}\right]\cdot \boldsymbol{n}_\infty dS}_{\delta_{mn}} + \underbrace{j\,2\omega\left[\frac{1}{4}\left\langle \boldsymbol{H}_{\rm sca}^{m}, \mu_0 \boldsymbol{H}_{\rm sca}^{n}\right\rangle_{\rm E_3} - \frac{1}{4}\left\langle \varepsilon_0 \boldsymbol{E}_{\rm sca}^{m}, \boldsymbol{E}_{\rm sca}^{n}\right\rangle_{\rm E_3}\right]}_{\theta_m\delta_{mn}}\end{aligned} \quad (5\text{-}6)$$

can be obtained. Just like the CMs obtained in the previous Sec. 5.2.1 (focusing on horn scatterer), the above CMs of the Yagi-Uda array scatterer are also both inherent and energy-decoupled. In addition, the CMs of the Yagi-Uda scatterer also satisfy orthogonality $(1/T)\int_{t_0}^{t_0+T}[\oint\!\!\!\oint_{\rm S}(\boldsymbol{\mathcal{E}}_{\rm sca}^{n} \times \boldsymbol{\mathcal{H}}_{\rm sca}^{m} + \boldsymbol{\mathcal{E}}_{\rm sca}^{m} \times \boldsymbol{\mathcal{H}}_{\rm sca}^{n})\cdot \boldsymbol{n}dS]dt = 2\delta_{mn}$, where S is an arbitrary closed surface enclosing whole horn scatterer.

For a specific Yagi-Uda array scatterer having the size shown in Fig. 4-4 (whose size is designed by using the formulations proposed in Ref. [36], and it is the same as the one considered in Refs. [27-App.H] and [37]), its ES-WET-based inherent CMs are calculated, and the associated MSs are shown in Fig. 5-7.

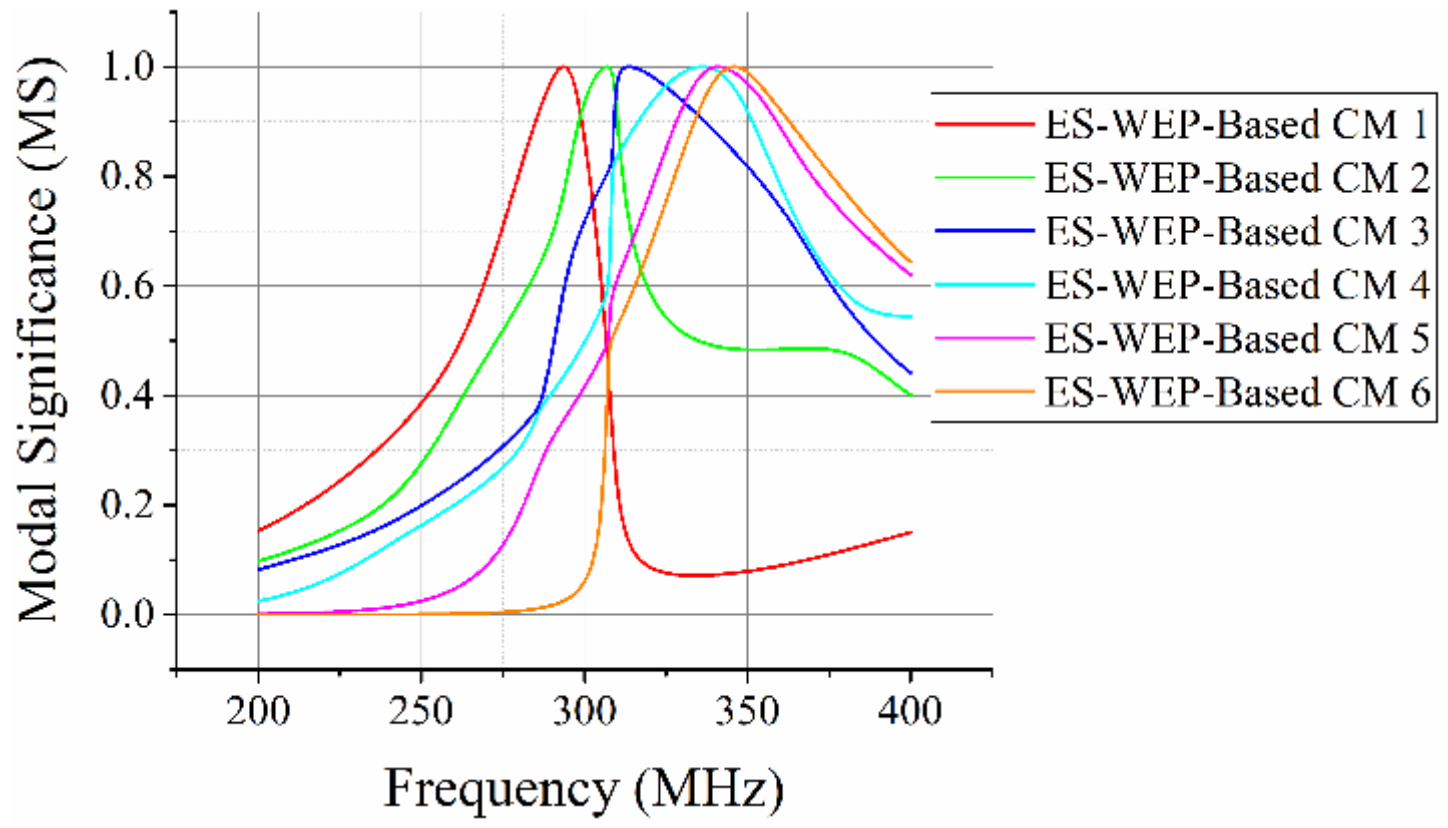

Figure 5-7 MS curves[27-App.H],[37] of the first several lower-order ES-WET-based inherent CMs of the Yagi-Uda array scatterer having the size shown in Fig. 4-4.

From the above Fig. 5-7, it is not difficult to find out that the Yagi-Uda scatterer is resonant at frequencies 293.6 MHz, 307.0 MHz, 313.6 MHz, 336.2 MHz, 340.9 MHz, and 345.9 MHz. The radiation patterns of the ES-WET-based resonant CMs are shown in the following Fig. 5-8.

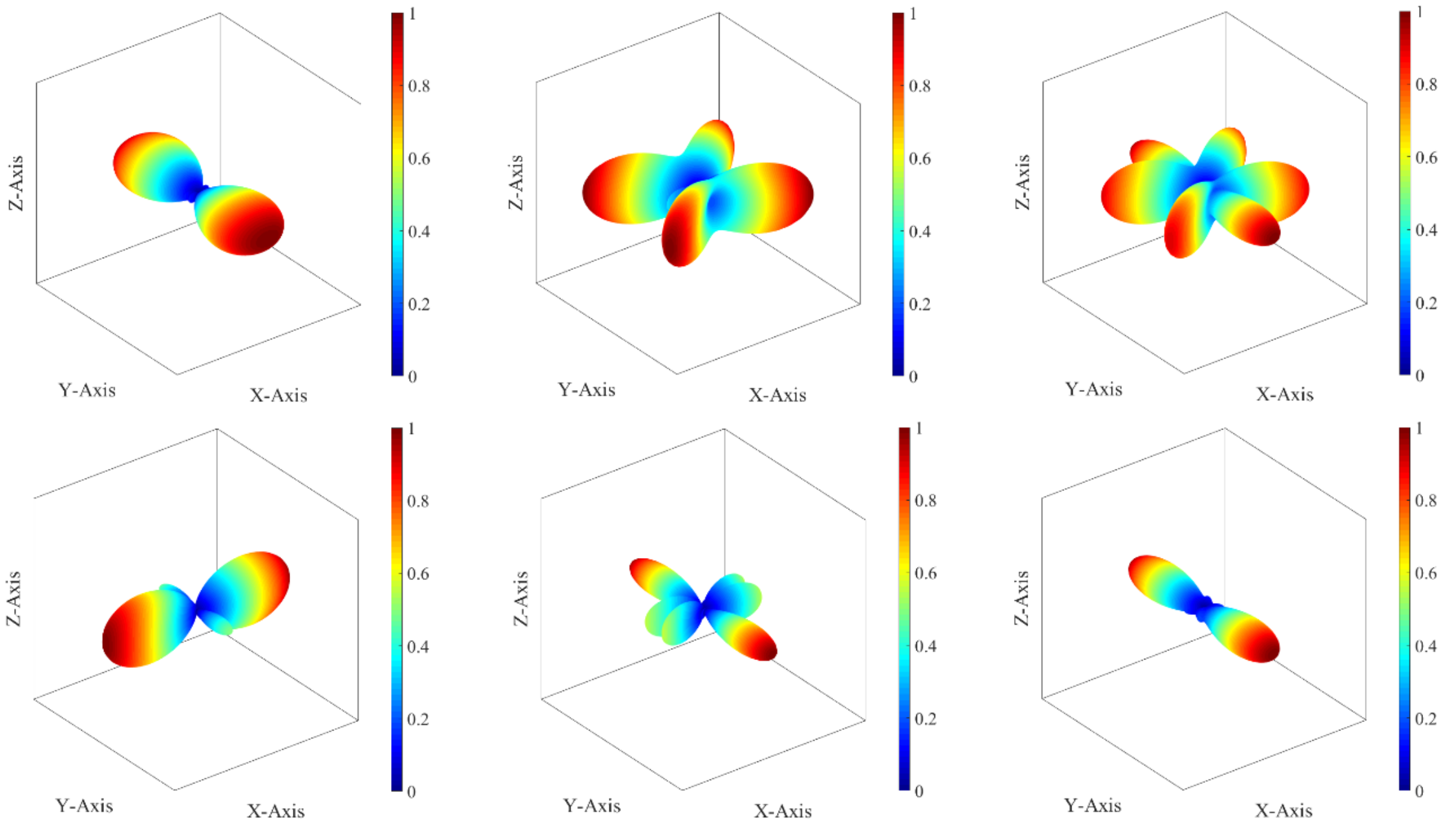

Figure 5-8 Radiation patterns of the ES-WET-based resonant CMs shown in Fig. 5-7.

Evidently, the commonly used end-fire mode with radiation pattern Fig. 4-6(a) is not contained in the above-obtained ES-WET-based inherent CM set. This implies that the ES-WET-CMT fails to analyze the lumped-port-driven transmitting antennas[27-App.H],[37]. In fact, the reason leading to the failure of the ES-WET-CMT-based modal analysis for the Yagi-Uda antenna is that: **the ES-DPO used to calculate the ES-WET-based CMs is the source term contained in ES-WET (which governs the scattering process of Yagi-Uda array scatterer), but not the source term contained in PS-WET (which governs the transmitting process of Yagi-Uda array antenna)**.

## 5.2.3 Example III: Metallic Multi-Coil System Treated as Scatterer

The following figure illustrates a EM scattering problem, and the scatterer is a metallic two-coil system.

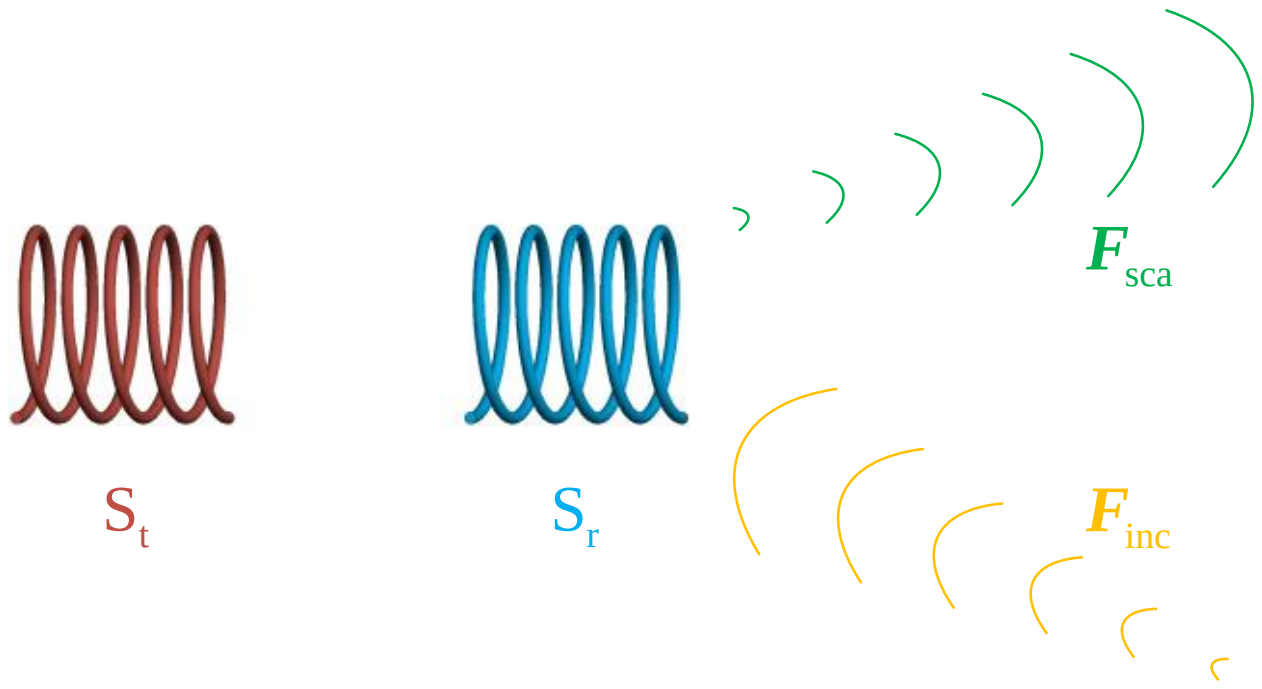

Figure 5-9 EM scattering problem considered in Sec. 5.2.3, where the scatterer is a metallic two-coil system placed in free space.

Taking the scattering problem shown in Fig. 5-9 as a typical example, this subsection compares the differences between the ES-WET-CMT for scattering structures and the PS-WET-CMT for lumped-port-driven wireless power transfer (WPT) systems.

In the figure, the two-coil system is placed in free space with material parameters $(\mu_0, \varepsilon_0)$, and is driven by an externally incident field $\boldsymbol{F}_{\text{inc}}$; the transmitting and receiving coils are denoted as $\text{S}_{\text{t}}$ and $\text{S}_{\text{r}}$ respectively like Sec. 4.6 did. Under the driving of $\boldsymbol{F}_{\text{inc}}$, currents $\boldsymbol{J}_{\text{t}}$ and $\boldsymbol{J}_{\text{r}}$ will be induced on $\text{S}_{\text{t}}$ and $\text{S}_{\text{r}}$, and field $\boldsymbol{F}_{\text{sca}}$ is generated by current $\boldsymbol{J}_{\text{t}} + \boldsymbol{J}_{\text{r}}$ correspondingly.

For the two-coil system, its entire-structure-oriented driving power $P_{\text{DRIV}} = (1/2) < \boldsymbol{J}_{\text{t}} + \boldsymbol{J}_{\text{r}}, \boldsymbol{E}_{\text{inc}} >_{\text{S}_{\text{t}} \cup \text{S}_{\text{r}}}$ has the following integral and matrix operator forms

$$P_{\text{DRIV}} = -(1/2)\left\langle \boldsymbol{J}_{\text{t}} + \boldsymbol{J}_{\text{r}}, -j\omega\mu_0\mathcal{L}_0\left(\boldsymbol{J}_{\text{t}} + \boldsymbol{J}_{\text{r}}\right)\right\rangle_{\text{S}_{\text{t}} \cup \text{S}_{\text{r}}} = \mathbb{J}_{\text{tr}}^{\dagger} \cdot \mathbb{P}_{\text{DRIV}} \cdot \mathbb{J}_{\text{tr}}, \text{ where } \mathbb{J}_{\text{tr}} = \begin{bmatrix} \mathbb{J}_{\text{t}} \\ \mathbb{J}_{\text{r}} \end{bmatrix} \tag{5-7}$$

called ES-DPO, where $\mathbb{J}_{\text{t}}$ and $\mathbb{J}_{\text{r}}$ are the basis function expansion coefficient vectors of $\boldsymbol{J}_{\text{t}}$ and $\boldsymbol{J}_{\text{r}}$ respectively. By solving characteristic equation $\mathbb{P}_{\text{DRIV}}^{-} \cdot \mathbb{J}_{\text{tr}} = \theta\, \mathbb{P}_{\text{DRIV}}^{+} \cdot \mathbb{J}_{\text{tr}}$, the CMs satisfying the following relations

$$\begin{aligned} \frac{1}{2}\left\langle \boldsymbol{J}_{\text{t}}^{m} + \boldsymbol{J}_{\text{r}}^{m}, \boldsymbol{E}_{\text{inc}}^{n} \right\rangle_{\text{S}_{\text{t}} \cup \text{S}_{\text{r}}} &= \left(1 + j\,\theta_m\right)\delta_{mn} \\ &= \underbrace{\frac{1}{2}\oint\!\!\!\oint_{\text{S}_\infty}\left[\boldsymbol{E}_{\text{sca}}^{n} \times \left(\boldsymbol{H}_{\text{sca}}^{m}\right)^{\dagger}\right] \cdot \boldsymbol{n}_\infty dS}_{\delta_{mn}} + \underbrace{j\,2\omega\left[\frac{1}{4}\left\langle \boldsymbol{H}_{\text{sca}}^{m}, \mu_0 \boldsymbol{H}_{\text{sca}}^{n} \right\rangle_{\text{E}_3} - \frac{1}{4}\left\langle \varepsilon_0 \boldsymbol{E}_{\text{sca}}^{m}, \boldsymbol{E}_{\text{sca}}^{n} \right\rangle_{\text{E}_3}\right]}_{\theta_m \delta_{mn}} \end{aligned} \tag{5-8}$$

can be obtained. Just like the ES-WET-based CMs obtained in the previous Secs. 5.2.1 and 5.2.2, the above CMs of two-coil scatterer are also energy-decoupled and inherent.

For a specific two-coil scatterer having the size as the one considered in Sec. 4.6 (which is the same as the one considered in Refs. [27-App.G] and [40,66]), its ES-WET-based inherent CMs are calculated, and the associated MSs are shown in Fig. 5-10.

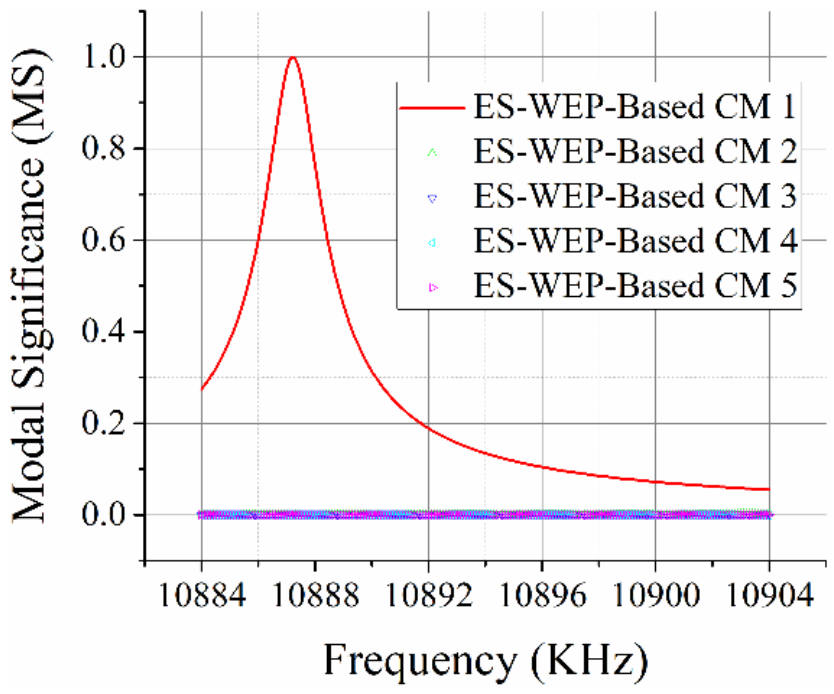

Figure 5-10 MSs of the first several lower-order ES-WET-based inherent CMs of the two-coil scatterer which is the same as the one considered in Ref. [66] and Sec. 4.6.

From Fig. 5-10, it is easy to find out that the two-coil scatterer is resonant at 108.872 MHz. The time-averaged magnetic energy density distribution of the resonant CM is shown in the following Fig. 5-11.

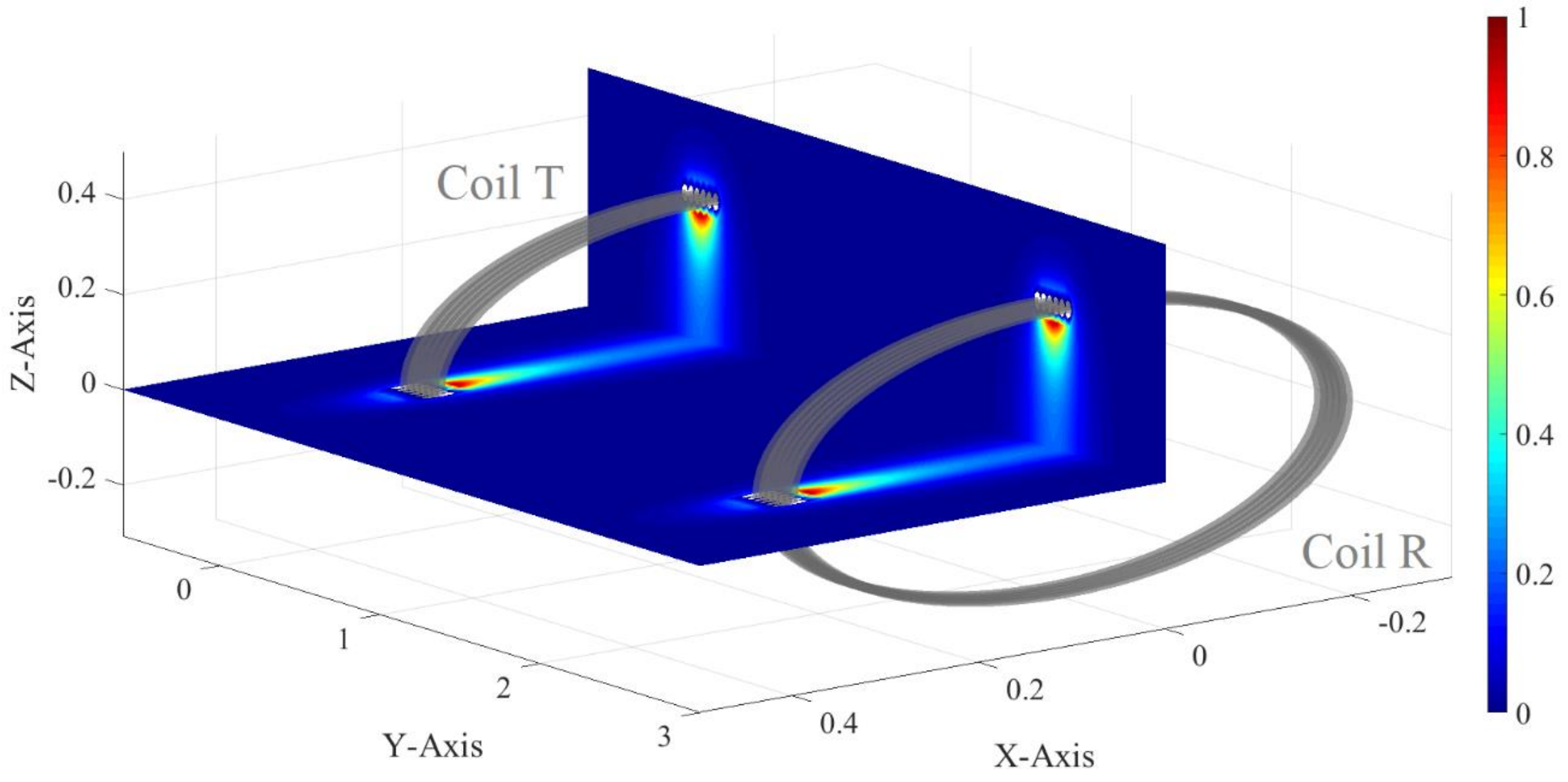

Figure 5-11 Time-averaged magnetic energy density distribution of the ES-WET-based resonant CM shown in Fig. 5-10.

Evidently, the optimally transferring mode given in Sec. 4.6 is not contained in the above-obtained ES-WET-based inherent CM set. This implies that the ES-WET-CMT fails to analyze the lumped-port-driven WPT system[27-App.G],[40]. The reason leading to the failure of the ES-WET-CMT-based modal analysis for the two-coil WPT system is that: **the ES-DPO used to calculate the ES-WET-based CMs is the source term contained in ES-WET (which governs the scattering process of two-coil scattering system), but not the source term contained in PS-WET (which governs the power transferring process of two-coil WPT system)**.

## 5.3 ES-WET-Based Environment-Dependent CMs of Scatterers

In this section, we consider the EM scattering problem shown in Fig. 5-12. Taking the problem as a typical example, this section discusses two different schemes used to construct the environment-dependent CMs under ES-WET framework, and the ES-WET-based environment-dependent CMs contain the informations of scatterer-environment interaction. Section 5.3.1 reviews an old scheme proposed in Ref. [27-Sec.1.2.4.4] which outputs energy-coupled CMs, and Sec. 5.3.2 proposes a new scheme which outputs energy-decoupled CMs.

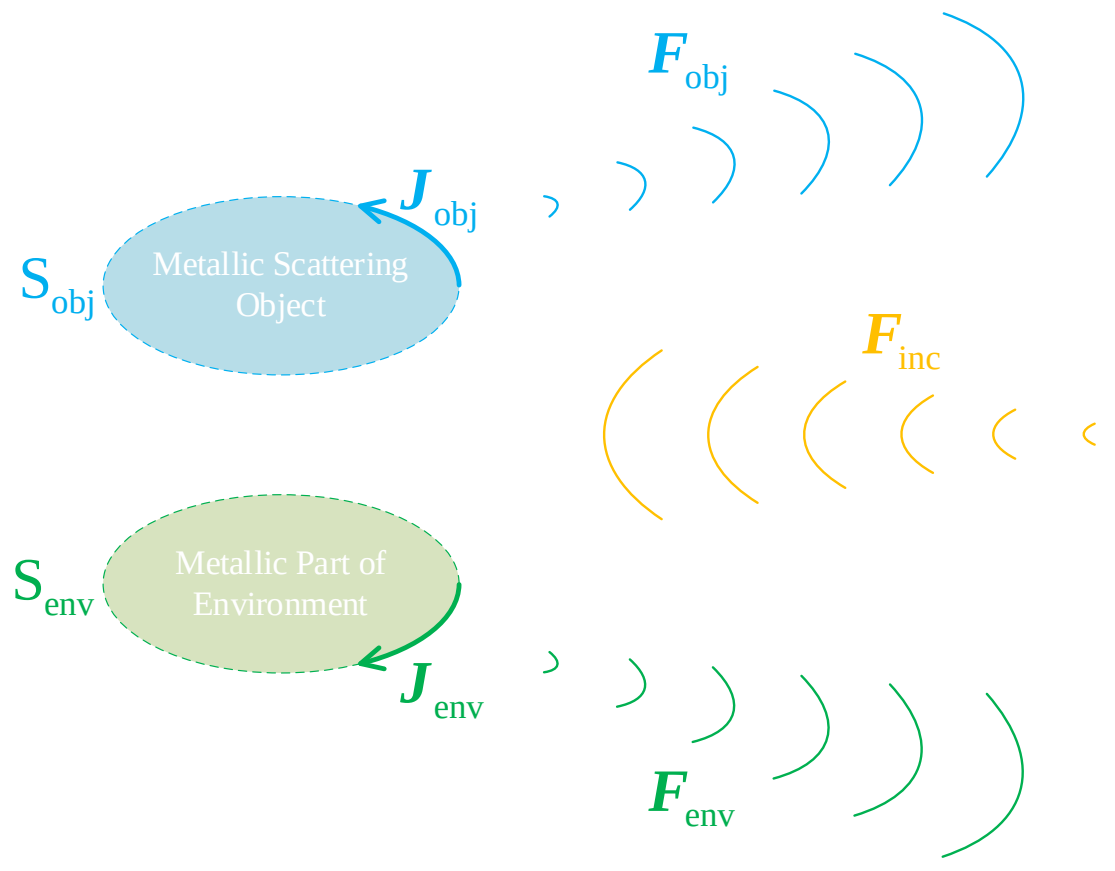

Figure 5-12 EM scattering problem considered in Sec. 5.3. The metallic scatterer is placed in a non-free-space environment. The environment includes a metallic body.

In the above figure, the boundaries of scattering structure and metallic environment are denoted as $S_{obj}$ and $S_{env}$ respectively. The externally impressed field used to drive a steady scattering process is $\boldsymbol{F}_{imp}$. Under the driving of $\boldsymbol{F}_{imp}$, currents $\boldsymbol{J}_{obj}$ and $\boldsymbol{J}_{env}$ will be induced on $S_{obj}$ and $S_{env}$, and they generate fields $\boldsymbol{F}_{obj}$ and $\boldsymbol{F}_{env}$ respectively.

## 5.3.1 An Old Scheme

The scheme proposed in the last two paragraphs of Ref. [27-Sec.1.2.4.4] constructs the CMs which can orthogonalize entire-structure-oriented driving power $P_{DRIV} = (1/2) < \boldsymbol{J}_{obj} + \boldsymbol{J}_{env}, \boldsymbol{E}_{imp} >_{S_{obj} \cup S_{env}}$.

The above entire-structure-oriented driving power has the following integral and matrix operator expressions

$$\begin{aligned} P_{DRIV} &= -(1/2)\left\langle \boldsymbol{J}_{obj} + \boldsymbol{J}_{env}, -j\omega\mu_0 \mathcal{L}_0\left(\boldsymbol{J}_{obj} + \boldsymbol{J}_{env}\right)\right\rangle_{S_{obj} \cup S_{env}} \\ &= \underbrace{\begin{bmatrix} \mathbb{J}_{obj} \\ \mathbb{J}_{env} \end{bmatrix}^{\dagger}}_{\mathbb{J}} \cdot \underbrace{\begin{bmatrix} \mathbb{P}_{oo} & \mathbb{P}_{oe} \\ \mathbb{P}_{eo} & \mathbb{P}_{ee} \end{bmatrix}}_{\mathbb{P}_{DRIV}} \cdot \underbrace{\begin{bmatrix} \mathbb{J}_{obj} \\ \mathbb{J}_{env} \end{bmatrix}}_{\mathbb{J}} \end{aligned} \tag{5-9}$$

where $\mathbb{J}_{obj}$ and $\mathbb{J}_{env}$ are the basis function expansion coefficient vectors of $\boldsymbol{J}_{obj}$ and $\boldsymbol{J}_{env}$. By solving characteristic equation $\mathbb{P}_{DRIV}^{-} \cdot \mathbb{J} = \theta\, \mathbb{P}_{DRIV}^{+} \cdot \mathbb{J}$ (where $\mathbb{P}_{DRIV}^{+}$ and $\mathbb{P}_{DRIV}^{-}$ are the positive and negative Hermitian parts of $\mathbb{P}_{DRIV}$), the CMs satisfying the following relation can be obtained.

$$(1/2)\left\langle \boldsymbol{J}_{obj}^{m} + \boldsymbol{J}_{env}^{m}, \boldsymbol{E}_{imp}^{n}\right\rangle_{S_{obj} \cup S_{env}} = \left(1 + j\,\theta_m\right)\delta_{mn} \tag{5-10}$$

Thus, the modal impressed fields $\{\boldsymbol{E}_{imp}^{n}\}$ and environment-dependent modal currents $\{\boldsymbol{J}_{obj}^{m}\}$ satisfy the following relation

$$(1/2)\left\langle \boldsymbol{J}_{\text{obj}}^{m}, \boldsymbol{E}_{\text{imp}}^{n} \right\rangle_{S_{\text{obj}}} \neq 0, \text{ if } \; m \neq n \tag{5-11}$$

This implies that the above CMs are not energy-decoupled, though they are indeed environment-dependent (i.e., contain the scatterer-environment interaction information).

For a specific example having the size shown in Fig. 5-13, the ES-WET-based environment-dependent energy-coupled CMs of the scattering structure working at 5 GHz have the environment-dependent modal currents shown in Fig. 5-14.

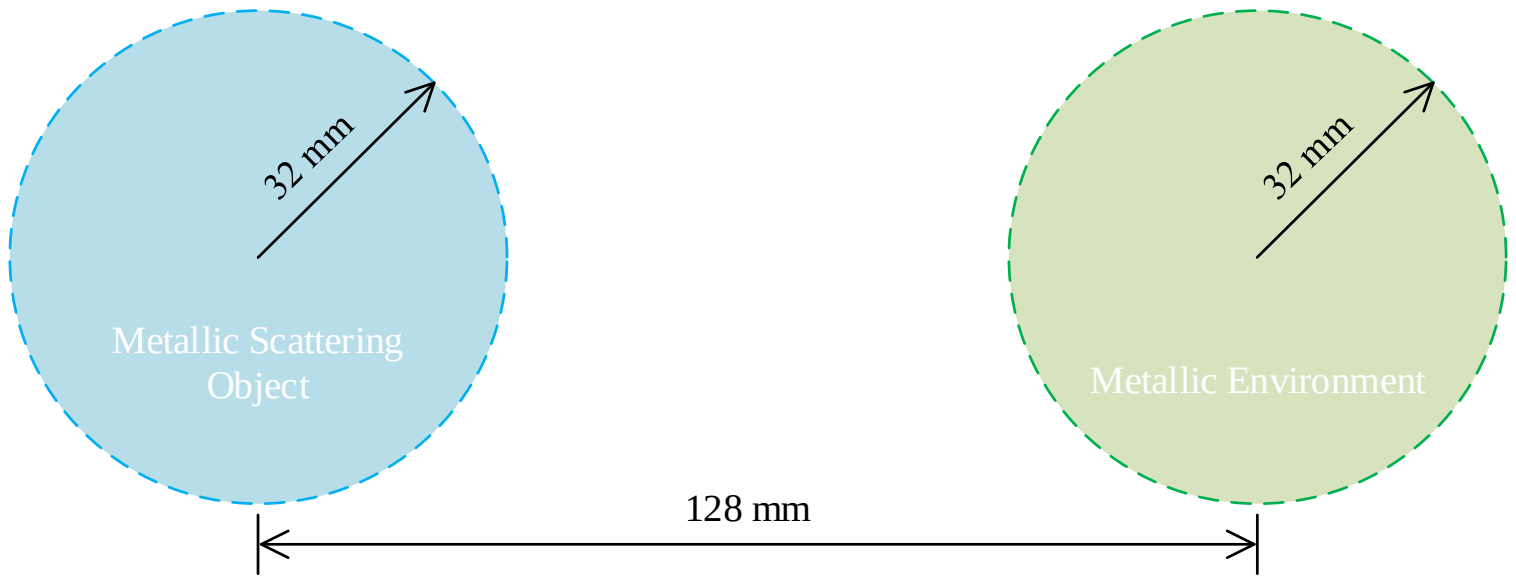

Figure 5-13 Geometry and size of a metallic scatterer placed in a metallic environment.

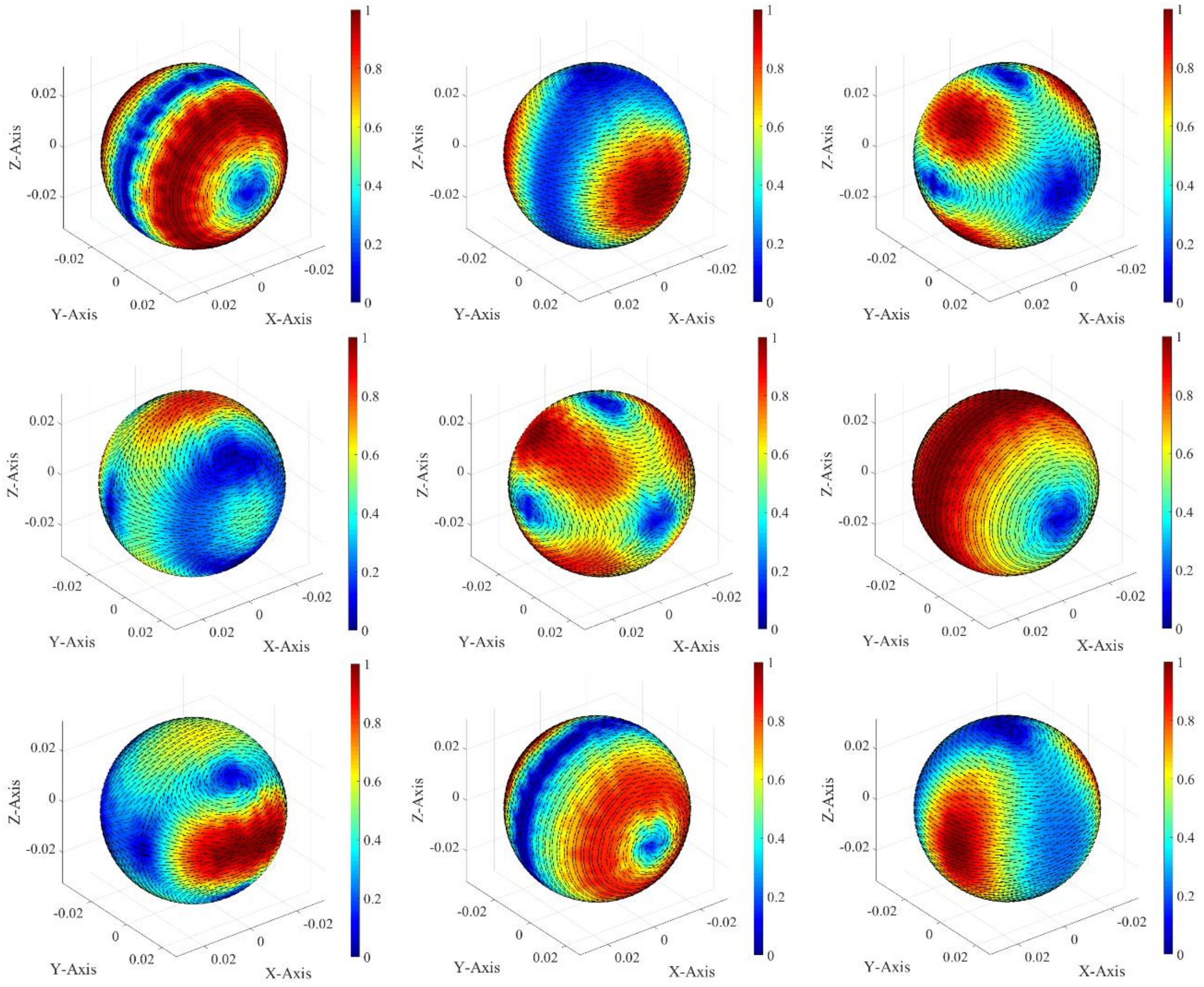

Figure 5-14 Modal currents of the first several lower-order ES-WET-based environment-dependent energy-coupled CMs calculated from the scheme proposed in the last paragraphs of Ref. [27-Sec.1.2.4.4].

For comparation, we also calculate the inherent CMs of the scatterer shown in Fig. 5-13, and show the associated inherent CM currents at 5 GHz in the following Fig. 5-15.

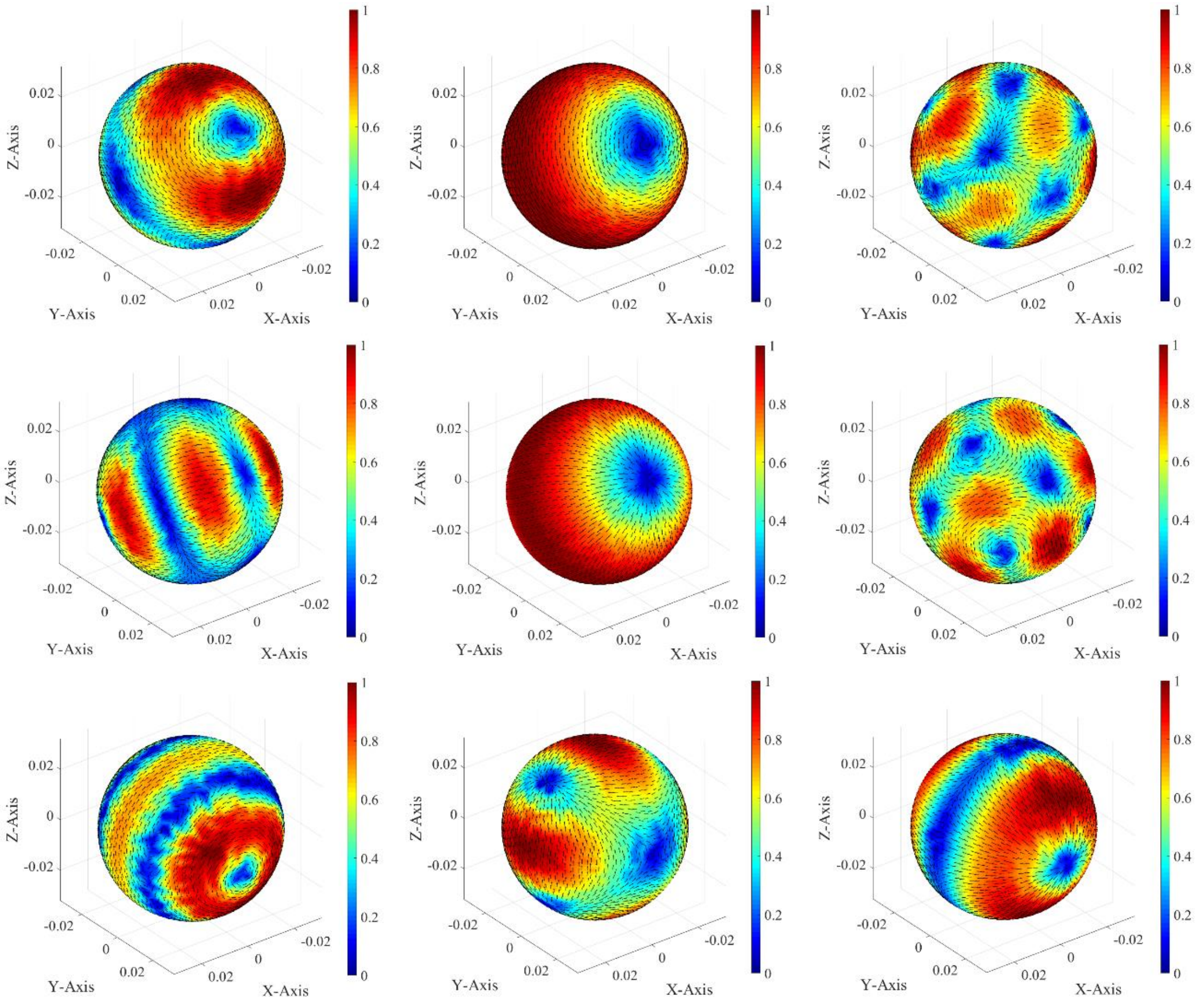

Figure 5-15 Modal currents of the first several lower-order ES-WET-based inherent CM currents (at 5 GHz) of the scattering metallic sphere shown in Fig. 5-13.

Evidently, the environment-dependent CMs shown in Fig. 5-14 are different from the classical inherent CMs shown in Fig. 5-15. **The reason leading to the difference is that the former incorporates the information of scatterer-environment interaction, but the latter doesn't.**

## 5.3.2 An Alternative Scheme

Equation (5-11) explicitly exhibits that the ES-WET-based environment-dependent CMs derived from orthogonalizing ES-DPO (5-9) are not energy-decoupled.

To resolve this problem, this sub-section proposes an alternative CM generating operator as follows:

$$
\begin{aligned}
P_{\text{imp}\to\text{obj}} = (1/2)\langle \boldsymbol{J}_{\text{obj}}, \boldsymbol{E}_{\text{imp}} \rangle_{\text{S}_{\text{obj}}} &= -(1/2)\langle \boldsymbol{J}_{\text{obj}}, \boldsymbol{E}_{\text{obj}} + \boldsymbol{E}_{\text{env}} \rangle_{\text{S}_{\text{obj}}} \\
&= -(1/2)\langle \boldsymbol{J}_{\text{obj}}, -j\omega\mu_0\mathcal{L}_0\left(\boldsymbol{J}_{\text{obj}} + \boldsymbol{J}_{\text{env}}\right)\rangle_{\text{S}_{\text{obj}}} \\
&= \underbrace{\begin{bmatrix} \mathbb{J}_{\text{obj}} \\ \mathbb{J}_{\text{env}} \end{bmatrix}^{\dagger}}_{\mathbb{J}} \cdot \underbrace{\begin{bmatrix} \mathbb{P}_{\text{oo}} & \mathbb{P}_{\text{oe}} \\ \mathbb{O}_{\text{eo}} & \mathbb{O}_{\text{ee}} \end{bmatrix}}_{\mathbb{P}_{\text{imp}\to\text{obj}}} \cdot \underbrace{\begin{bmatrix} \mathbb{J}_{\text{obj}} \\ \mathbb{J}_{\text{env}} \end{bmatrix}}_{\mathbb{J}}
\end{aligned}
\tag{5-12}
$$

Here, the first equality is for decoupling the obtained $\{\boldsymbol{E}_{\text{imp}}^{n}\}$ and $\{\boldsymbol{J}_{\text{obj}}^{m}\}$; the second equality is due to the homogeneous tangential electric field boundary condition on metallic boundary $\text{S}_{\text{obj}}$; the third equality is because of that $\boldsymbol{E}_{\text{obj/env}} = -j\omega\mu_0\mathcal{L}_0(\boldsymbol{J}_{\text{obj/env}})$; the fourth equality originates from expanding the involved currents in terms of some proper current basis functions. In the right-hand side of the last equality of Eq. (5-12), the sub-vectors $\mathbb{J}_{\text{obj}}$ and $\mathbb{J}_{\text{env}}$ have the same meanings as the ones in ES-DPO (5-9); the sub-matrices $\mathbb{P}_{\text{oo}}$ and $\mathbb{P}_{\text{oe}}$ are the same as the ones in ES-DPO (5-9); the sub-matrices $\mathbb{O}_{\text{eo}}$ and $\mathbb{O}_{\text{ee}}$ are zero matrices.

The environment-dependent energy-decoupled CMs, which satisfy the following decoupling relation

$$
(1/2)\langle \boldsymbol{J}_{\text{obj}}^{m}, \boldsymbol{E}_{\text{imp}}^{n} \rangle_{\text{S}_{\text{obj}}} = (1 + j\,\theta_m)\delta_{mn} \tag{5-13}
$$

can be derived from orthogonalizing power operator $\mathbb{P}_{\text{imp}\to\text{obj}}$, i.e., solving the following characteristic equation

$$
\mathbb{P}_{\text{imp}\to\text{obj}}^{-} \cdot \mathbb{J} = \theta\, \mathbb{P}_{\text{imp}\to\text{obj}}^{+} \cdot \mathbb{J} \tag{5-14}
$$

where $\mathbb{P}_{\text{imp}\to\text{obj}}^{+}$ and $\mathbb{P}_{\text{imp}\to\text{obj}}^{-}$ are the positive and negative Hermitian parts of $\mathbb{P}_{\text{imp}\to\text{obj}}$.

## 5.4 ES-WET-Based Driver-Dependent CMs of Scatterers

In this section, we consider the EM scattering problem shown in Fig. 5-16. The scattering object is a metallic body, and it is driven by a metallic transmitting horn.

As shown in Fig. 5-16, the boundaries of scatterer and horn are denoted as $\text{S}_{\text{s}}$ and $\text{S}_{\text{h}}$ respectively; the input port of horn is denoted as $\text{S}_{\text{i}}$. When the whole scattering problem works at steady state, the currents distributing on $\text{S}_{\text{h}}$ and $\text{S}_{\text{i}}$ are denoted as $\boldsymbol{J}_{\text{h}}$ and $(\boldsymbol{J}_{\text{i}}, \boldsymbol{M}_{\text{i}})$ respectively. The field transmitted from horn, which can be expressed in terms of the function of $\boldsymbol{J}_{\text{h}}$ and $(\boldsymbol{J}_{\text{i}}, \boldsymbol{M}_{\text{i}})$[27-Sec.6.2],[35], is just the externally incident field $\boldsymbol{F}_{\text{inc}}$ used to drive the steady scattering of scatterer. $\boldsymbol{F}_{\text{inc}}$ will result in an induced current $\boldsymbol{J}_{\text{s}}$ on $\text{S}_{\text{s}}$, and $\boldsymbol{J}_{\text{s}}$ will generate a scattered field $\boldsymbol{F}_{\text{sca}}$ correspondingly. The summation of $\boldsymbol{F}_{\text{inc}}$ and $\boldsymbol{F}_{\text{sca}}$ is just the total field $\boldsymbol{F}$, i.e., $\boldsymbol{F} = \boldsymbol{F}_{\text{inc}} + \boldsymbol{F}_{\text{sca}}$.

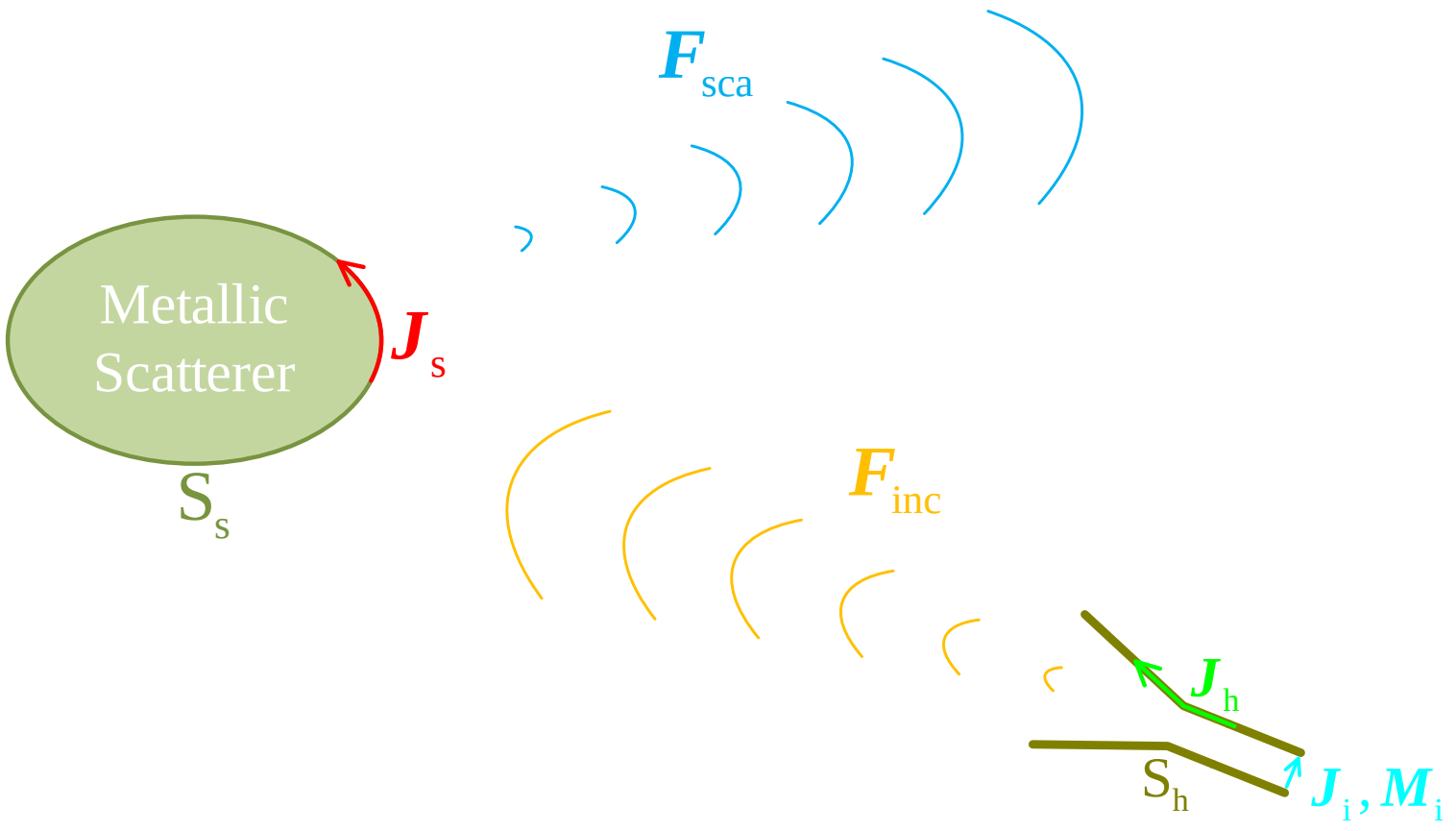

Figure 5-16 EM scattering problem considered in Sec. 5.4, where the scatterer is a metallic body and is driven by a metallic horn.

Following the idea of Sec. 5.3, this section also proposes two somewhat different schemes for constructing the driver-dependent CMs as below.

### 5.4.1 Scheme I

This scheme is based on the antenna-oriented POWER TRANSPORT THEOREM based DECOUPLING MODE THEORY (PTT-DMT) established in Sec. 3.2 and Refs. [27-Chap.6]&[35]. This scheme treats the union of the transmitting horn and the scattering structure as a whole — augmented transmitting antenna (for details, please see Refs. [27-Sec.2.3.4&Chap.6]), and constructs the PTT-DMT-based ENERGY-DECOUPLED MODES (DMs) of the augmented antenna. The obtained DM currents on the scatterer are just the driver-dependent characteristic currents of the scatterer under the driving of the horn.

The PTT implies that the energy source used to sustain a steady EM transmitting process is the input power $P_{\text{in}} = (1/2)\iint_{\text{S}_{\text{i}}} (\boldsymbol{E} \times \boldsymbol{H}^{\dagger}) \cdot \boldsymbol{n}_{\text{i}} dS$ (where $\boldsymbol{n}_{\text{i}}$ is the normal direction of $\text{S}_{\text{i}}$), and it has the following operator forms

$$P_{\text{in}} = -(1/2)\left\langle \boldsymbol{J}_{\text{i}}, \mathcal{E}_0\left(\boldsymbol{J}_{\text{i}} + \boldsymbol{J}_{\text{h}} + \boldsymbol{J}_{\text{s}}, \boldsymbol{M}_{\text{i}}\right)\right\rangle_{\text{S}_{\text{i}}^{+}} = -(1/2)\left\langle \boldsymbol{M}_{\text{i}}, \mathcal{H}_0\left(\boldsymbol{J}_{\text{i}} + \boldsymbol{J}_{\text{h}} + \boldsymbol{J}_{\text{s}}, \boldsymbol{M}_{\text{i}}\right)\right\rangle_{\text{S}_{\text{i}}^{+}}^{\dagger}$$
$$= \mathbb{J}_{\text{i}}^{\dagger} \cdot \mathbb{P}_{\text{JE}} \cdot \begin{bmatrix} \mathbb{J}_{\text{i}} \\ \mathbb{J}_{\text{h}} \\ \mathbb{J}_{\text{s}} \\ \mathbb{M}_{\text{i}} \end{bmatrix} = \begin{bmatrix} \mathbb{J}_{\text{i}} \\ \mathbb{J}_{\text{h}} \\ \mathbb{J}_{\text{s}} \\ \mathbb{M}_{\text{i}} \end{bmatrix}^{\dagger} \cdot \mathbb{P}_{\text{HM}} \cdot \mathbb{M}_{\text{i}} \tag{5-15}$$

where $\mathcal{E}_0(\boldsymbol{J}, \boldsymbol{M}) = -j\omega\mu_0\mathcal{L}_0(\boldsymbol{J}) - \mathcal{K}_0(\boldsymbol{M})$ and $\mathcal{H}_0(\boldsymbol{J}, \boldsymbol{M}) = \mathcal{K}_0(\boldsymbol{J}) - j\omega\varepsilon_0\mathcal{L}_0(\boldsymbol{M})$. In fact, the above-mentioned currents are not independent of each other, and they satisfy some integral equations, and the integral equations imply the transformations from independent current $\mathbb{J}_{\text{i}} / \mathbb{M}_{\text{i}}$ into the other currents, as follows:

$$\left.\begin{array}{r}\left[\mathcal{H}_0\left(\boldsymbol{J}_{\rm i}+\boldsymbol{J}_{\rm h}+\boldsymbol{J}_{\rm s},\boldsymbol{M}_{\rm i}\right)\right]_{{\rm S}_{\rm i}^+}^{\tan}=\boldsymbol{J}_{\rm i}\times\boldsymbol{n}_{\rm i}\\ \left[\mathcal{E}_0\left(\boldsymbol{J}_{\rm i}+\boldsymbol{J}_{\rm h}+\boldsymbol{J}_{\rm s},\boldsymbol{M}_{\rm i}\right)\right]_{{\rm S}_{\rm i}^+}^{\tan}=\boldsymbol{n}_{\rm i}\times\boldsymbol{M}_{\rm i}\\ \left[\mathcal{E}_0\left(\boldsymbol{J}_{\rm i}+\boldsymbol{J}_{\rm h}+\boldsymbol{J}_{\rm s},\boldsymbol{M}_{\rm i}\right)\right]_{{\rm S}_{\rm h}\cup{\rm S}_{\rm s}}^{\tan}=0\end{array}\right\}\Rightarrow\mathbb{T}_{\rm DoJ}\cdot\mathbb{J}_{\rm i}=\begin{bmatrix}\mathbb{J}_{\rm i}\\ \mathbb{J}_{\rm h}\\ \mathbb{J}_{\rm s}\\ \mathbb{M}_{\rm i}\end{bmatrix}=\mathbb{T}_{\rm DoM}\cdot\mathbb{M}_{\rm i} \tag{5-16}$$

In Eq. (5-16), the first and second integral equations originate from the definitions for currents $\boldsymbol{J}_{\rm i}$ (DoJ) and $\boldsymbol{M}_{\rm i}$ (DoM); the third integral equation is based on the homogeneous tangential electric field boundary condition on metallic boundary ${\rm S}_{\rm h}\cup{\rm S}_{\rm s}$.

Substituting the transformations from independent current $\mathbb{J}_{\rm i}/\mathbb{M}_{\rm i}$ to the other currents into IPOs (5-15), the following IPO

$$P_{\rm in}=\mathbb{C}_{\rm i}^{\dagger}\cdot\mathbb{P}_{\rm in}\cdot\mathbb{C}_{\rm i} \tag{5-17}$$

with only independent current $\mathbb{C}_{\rm i}$ (which is either $\mathbb{J}_{\rm i}$ or $\mathbb{M}_{\rm i}$) can be obtained. By solving the following modal decoupling equation

$$\mathbb{P}_{\rm in}^{-}\cdot\mathbb{C}_{\rm i}=\theta\,\mathbb{P}_{\rm in}^{+}\cdot\mathbb{C}_{\rm i} \tag{5-18}$$

defined by matrix $\mathbb{P}_{\rm in}$, the PTT-DMT-based DMs satisfying the following power-decoupling relation

$$(1/2)\iint_{{\rm S}_{\rm i}}\left(\boldsymbol{E}_n\times\boldsymbol{H}_m^{\dagger}\right)\cdot\boldsymbol{n}_{\rm i}dS=\left(1+j\,\theta_m\right)\delta_{mn} \tag{5-19}$$

are obtained, but the relation $(1/2)<\boldsymbol{J}_{\rm s}^m,\boldsymbol{E}_{\rm inc}^n>_{{\rm S}_{\rm s}}=0$ cannot be guaranteed for $m\neq n$ case. In addition, the obtained modal currents on scatterer are just the driver-dependent characteristic scattering currents, because their calculation process incorporates the informations of scatterer-driver interaction mathematically reflected as Eq. (5-16).

For a specific driver-dependent scattering problem having the size shown in Fig. 5-17, its PTT-DMT-based DMs are calculated, and the associated driver-dependent characteristic scattering currents at 5 GHz are shown in Fig. 5-18.

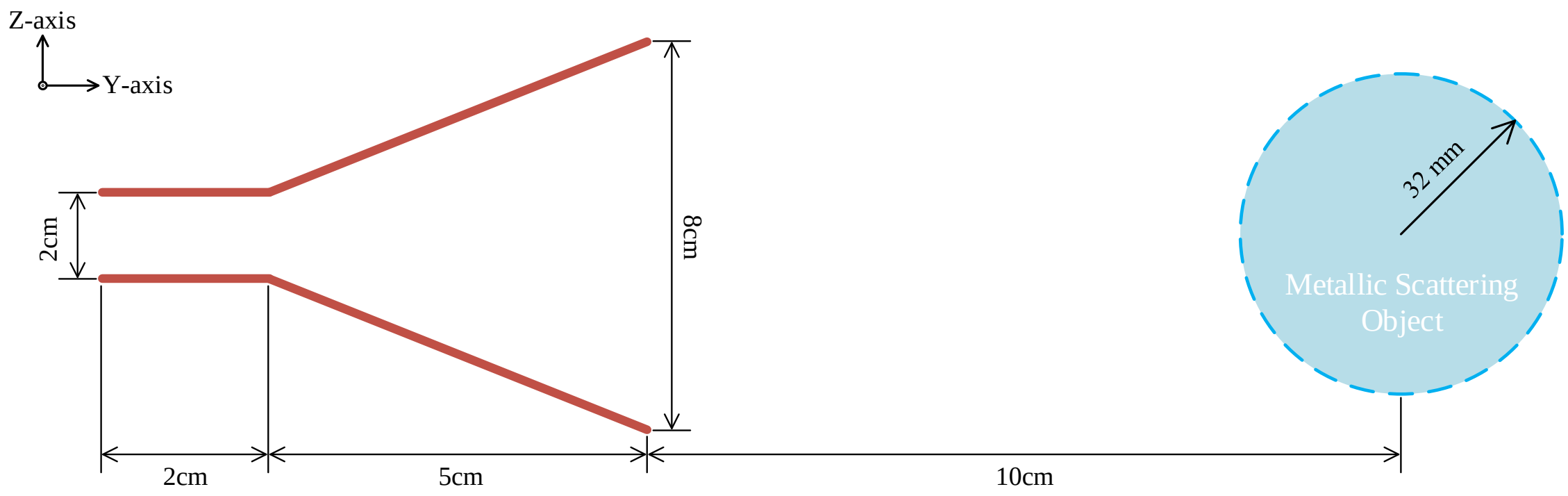

Figure 5-17 A specific driver-dependent EM scattering problem and its geometry and size, where the scatterer is a metallic sphere and the driver is a metallic horn.

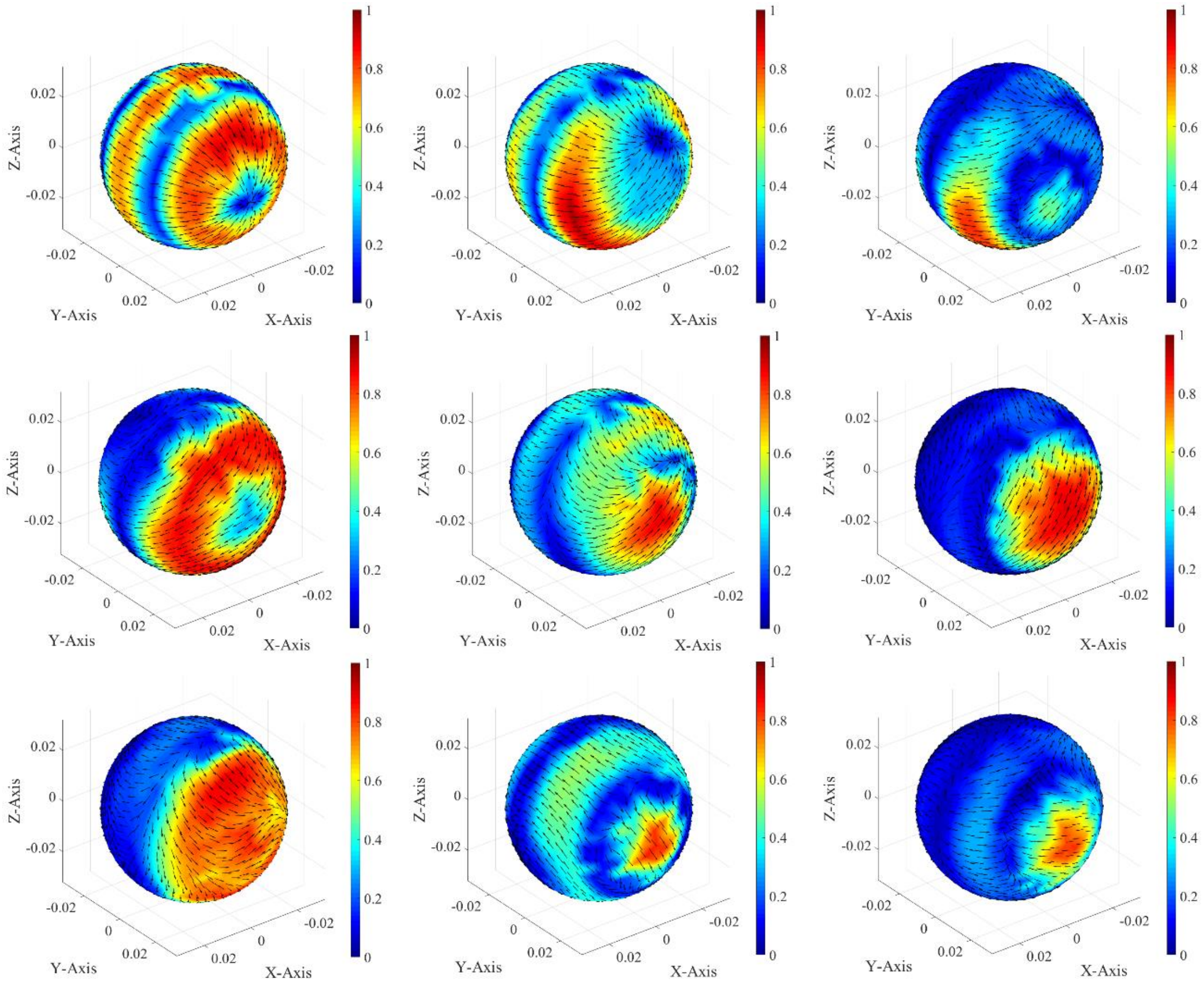

Figure 5-18 The first several lower-order driver-dependent characteristic scattered currents (at 5 GHz) calculated from modal decoupling equation (5-18).

Evidently, the driver-dependent characteristic scattering currents shown in Fig. 5-18 are different from the classical inherent characteristic scattering currents shown in Fig. 5-15. **The reason leading to the difference is that the former incorporates the information of scatterer-driver interaction, but the latter doesn't.**

## 5.4.2 Scheme II

This scheme is based on ES-WET, and focuses on calculating the driver-dependent CMs satisfying energy-decoupling relation $(1/2) < \boldsymbol{J}_{\rm s}^{m}, \boldsymbol{E}_{\rm inc}^{n} >_{S_{\rm s}} = 0$ (if $m \neq n$).

The ES-WET implies that the energy source used to sustain a steady EM scattering process is the entire-structure-oriented driving power $P_{\rm DRIV} = (1/2) < \boldsymbol{J}_{\rm s}, \boldsymbol{E}_{\rm inc} >_{S_{\rm s}}$, and it has the following operator forms

$$P_{\rm DRIV} = (1/2)\left\langle \boldsymbol{J}_{\rm s}, \mathcal{E}_0\left(\boldsymbol{J}_{\rm i} + \boldsymbol{J}_{\rm h}, \boldsymbol{M}_{\rm i}\right)\right\rangle_{S_{\rm s}} = \mathbb{J}_{\rm s}^{\dagger} \cdot \mathbb{P}_{\rm DRIV} \cdot \begin{bmatrix} \mathbb{J}_{\rm i} \\ \mathbb{J}_{\rm h} \\ \mathbb{M}_{\rm i} \end{bmatrix} \tag{5-20}$$

The above-mentioned currents are not independent of each other, because they satisfy transformations (5-16). Substituting transformations (5-16) into ES-DPO (5-20), the following ES-DPO

$$P_{\text{DRIV}} = \mathbb{C}_{\text{i}}^{\dagger} \cdot \mathbb{P}_{\text{DRIV}} \cdot \mathbb{C}_{\text{i}} \tag{5-21}$$

with only independent current $\mathbb{C}_{\text{i}}$ (which is either $\mathbb{J}_{\text{i}}$ or $\mathbb{M}_{\text{i}}$) can be obtained. By solving the following characteristic equation

$$\mathbb{P}_{\text{DRIV}}^{-} \cdot \mathbb{C}_{\text{i}} = \theta\, \mathbb{P}_{\text{DRIV}}^{+} \cdot \mathbb{C}_{\text{i}} \tag{5-22}$$

defined by matrix $\mathbb{P}_{\text{DRIV}}$, the ES-WET-based CMs satisfying the following power-decoupling relation

$$\begin{aligned}\frac{1}{2}\left\langle \boldsymbol{J}_{\text{s}}^{m}, \boldsymbol{E}_{\text{inc}}^{n} \right\rangle_{\text{S}_{\text{s}}} &= \left(1 + j\,\theta_m\right)\delta_{mn} \\ &= \underbrace{\frac{1}{2}\oiint_{\text{S}_\infty}\left[\boldsymbol{E}_{\text{sca}}^{n} \times \left(\boldsymbol{H}_{\text{sca}}^{m}\right)^{\dagger}\right]\cdot \boldsymbol{n}_\infty dS}_{\delta_{mn}} + \underbrace{j\,2\omega\left[\frac{1}{4}\left\langle \boldsymbol{H}_{\text{sca}}^{m}, \mu_0 \boldsymbol{H}_{\text{sca}}^{n}\right\rangle_{\text{E}_3} - \frac{1}{4}\left\langle \varepsilon_0 \boldsymbol{E}_{\text{sca}}^{m}, \boldsymbol{E}_{\text{sca}}^{n}\right\rangle_{\text{E}_3}\right]}_{\theta_m \delta_{mn}}\end{aligned} \tag{5-23}$$

are obtained, and they are driver-dependent because their calculation process incorporates the informations of scatterer-driver interaction mathematically reflected as Eq. (5-16).

## 5.5 Chapter Summary

This chapter focuses on discussing the various ENERGY-DECOUPLED MODES of scattering structures under ENTIRE-STRUCTURE-ORIENTED WORK-ENERGY THEOREM (ES-WET) framework. The main studies and conclusions of this chapter are summarized as below.

Employing three typical EM structures (metallic horn, metallic Yagi-Uda array, and metallic two-coil system), Sec. 5.2 exhibits a fact that: for a certain EM structure, when it works at different working manners, such as scattering manner, transmitting manner, and WPT manner, it has different ENERGY-DECOUPLED MODE sets. The differences are mainly originated from the following reasons.

1) The different working manners have different energy utilization processes, and the different energy utilization processes are governed by different manifestation forms of ENERGY CONSERVATION LAW. Specifically, the scattering manner is governed by ES-WET, and the wave-port-fed transmitting manner is governed by POWER TRANSPORT THEOREM (PTT), and the lumped-port-driven transmitting and WPT manners are governed by PARTIAL-STRUCTURE-ORIENTED WORK-ENERGY THEOREM (PS-WET).

2) For any EM structure, its ENERGY-DECOUPLED MODES orthogonalize the ENERGY SOURCE OPERATOR involved in ENERGY CONSERVATION LAW, but the different manifestation forms of ENERGY CONSERVATION LAW have different ENERGY SOURCE OPERATORS. Specifically, the energy source terms of ES-WET, PTT, and PS-WET are ENTIRE-STRUCTURE-ORIENTED DRIVING POWER OPERATOR (ES-DPO), INPUT POWER OPERATOR (IPO), and PARTIAL-STRUCTURE-ORIENTED DRIVING POWER OPERATOR (PS-DPO) respectively.

The classical ES-WET-based CMs depend only on the inherent EM scattering characters of objective scatterer, and independent of external environment and driver. Taking some typical EM scattering problems as examples, Secs. 5.3 and 5.4 generalize the inherent CMs to some different kinds of non-inherent CMs. The generalizations are mainly concentrated in the following aspects.

i) Section 5.3 reviews an old scheme and proposes a new scheme used to construct the environment-dependent CMs under ES-WET framework, and both the old and new CMs involve the interaction informations between scattering structure and external environment. The main difference between old and new environment-dependent CMs is that the old ones are energy-coupled but the new ones are energy-decoupled.

ii) Section 5.4 proposes two somewhat different driver-dependent CM sets, and they involve the interaction informations between scattering object and external driver. The main difference between the different driver-dependent CM sets is that one CM set is energy-coupled but the other CM set is energy-decoupled.

# CHAPTER 6 PtT-Based Modal Analysis for Energy-Dissipating and Self-Oscillating EM Structures

**Chapter Motivation:** The main destination of this chapter is to further develop energy-viewpoint-based electromagnetic (EM) modal analysis method by employing another manifestation form of Energy Conservation Law — Poynting's Theorem (PtT) form.

## 6.1 Chapter Introduction

In Chap. 2, we discussed five different manifestation forms of Energy Conservation Law, that are Power Transport Theorem (PTT) form, Partial-Structure-Oriented Work-Energy Theorem (PS-WET) form, Entire-Structure-Oriented Work-Energy Theorem (ES-WET) form, Poynting's Theorem (PtT) form, and Lorentz's Reciprocity Theorem form, and exhibited that the different manifestation forms govern the different energy utilization processes of EM structures.

PTT governs the power transportation process of wave-port-fed EM structures. Chapter 3 discussed an effective modal analysis theory — PTT-based Decoupling Mode Theory (PTT-DMT) — for wave-port-fed EM structures, and exhibited an effective modal calculation method — orthogonalizing Input Power Operator (IPO) method — for constructing the Energy-Decoupled Modes (DMs) of wave-port-fed EM structures.

PS-WET governs the work-energy transformation process of lumped-port-driven EM structures. Chapter 4 discussed an effective modal analysis theory — PS-WET-based Characteristic Mode Theory (PS-WET-CMT) — for lumped-port-driven EM structures, and exhibited an effective modal calculation method — orthogonalizing Partial-Structure-Oriented Driving Power Operator (PS-DPO) method — for constructing the energy-decoupled Characteristic Modes (CMs) of lumped-port-driven EM structures.

ES-WET governs the work-energy transformation process of incident-field-driven EM structures. Chapter 5 simply reviewed an effective modal analysis theory — ES-WET-based Characteristic Mode Theory (ES-WET-CMT) — for incident-field-driven EM structures, and exhibited an effective modal calculation method — orthogonalizing Entire-Structure-Oriented Driving Power Operator (ES-DPO) method — for constructing the energy-decoupled CMs of incident-field-driven EM structures.

PtT governs the energy dissipation process and self-oscillation process of lossy and lossless penetrable EM structures. This chapter focuses on establishing an effective modal analysis theory — PtT-based Decoupling Mode Theory (PtT-DMT) — for penetrable EM structures, and proposing an effective modal calculation method — orthogonalizing Poynting's Flux Operator (PtFO) method — for constructing the optimally-energy-dissipating and self-oscillating DMs of penetrable EM structures.

## 6.2 PtT-Based DMs of Energy-Dissipating Structures

Taking the material-coated metallic structure shown in Fig. 6-1 as an example, this section focuses on establishing the PtT-DMT for energy-dissipating structures, and a main function of the PtT-DMT is to construct the DMs and to find the optimally energy-dissipating modes of the objective structure.

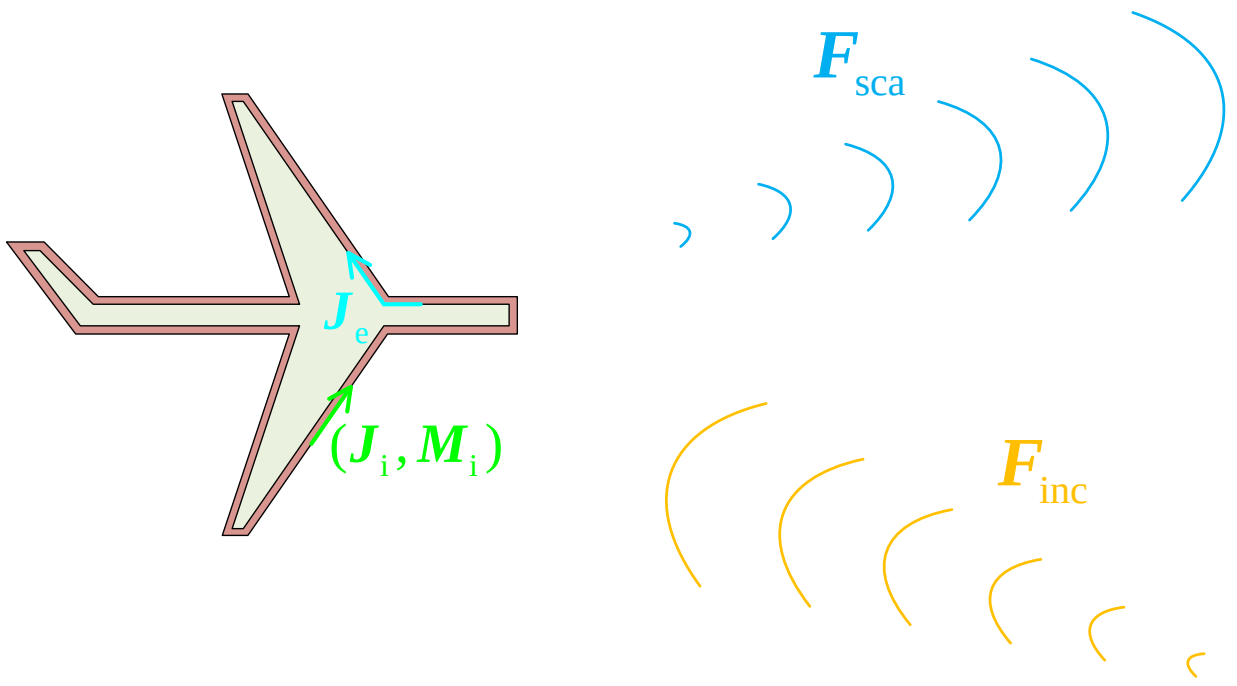

Figure 6-1 A material-coated metallic structure under the illumination of external field.

In the above figure, the whole metal-material composite structure is placed in free space; the region occupied by the material coating is denoted as $V$; the environment-material and material-metal boundaries are denoted as $S_i$ and $S_e$ respectively. Now, there is an external field $\boldsymbol{F}_{inc}$ illuminating on the composite structure. Under the illumination of $\boldsymbol{F}_{inc}$, some currents are induced on the composite structure, and the induced currents generate a scattered field $\boldsymbol{F}_{sca}$. The summation of $\boldsymbol{F}_{inc}$ and $\boldsymbol{F}_{sca}$ is denoted as $\boldsymbol{F}$ (i.e., $\boldsymbol{F} = \boldsymbol{F}_{inc} + \boldsymbol{F}_{sca}$) called total field.

The PtT manifestation form of Energy Conservation Law tells us that: the energy source used to sustain a steady energy dissipation process is the following Poynting's flux

$$\begin{aligned} P_{flux} &= (1/2)\oint\!\!\!\oint_{S_i} \left(\boldsymbol{E}\times\boldsymbol{H}^\dagger\right)\cdot \boldsymbol{n}_i^- dS + (1/2)\oint\!\!\!\oint_{S_e} \left(\boldsymbol{E}\times\boldsymbol{H}^\dagger\right)\cdot \boldsymbol{n}_e^+ dS \\ &= (1/2)\oint\!\!\!\oint_{S_i} \left(\boldsymbol{E}\times\boldsymbol{H}^\dagger\right)\cdot \boldsymbol{n}_i^- dS \\ &= -(1/2)\langle \boldsymbol{J}_i, \boldsymbol{E}\rangle_{S_i} \\ &= -(1/2)\langle \boldsymbol{M}_i, \boldsymbol{H}\rangle_{S_i}^\dagger \end{aligned} \tag{6-1}$$

Here, the first equality directly originates from PtT; the second equality is due to the homogeneous tangential electric field boundary condition on $\mathrm{S_e}$; the third and fourth equalities are based on the following current definitions $\boldsymbol{J}_\mathrm{i} = \boldsymbol{n}_\mathrm{i}^- \times \boldsymbol{H}$ and $\boldsymbol{M}_\mathrm{i} = \boldsymbol{E} \times \boldsymbol{n}_\mathrm{i}^-$ on $\mathrm{S_i}$, where $\boldsymbol{n}_\mathrm{i}^-$ is the normal direction of $\mathrm{S_i}$ and points to the interior of $\mathrm{V}$. For distinguishing the various expressions of the $P_\mathrm{flux}$, the right-hand sides of the second, third, and fourth equalities are called EM, JE, and HM interaction forms respectively. In fact, the $P_\mathrm{flux}$ can also be expressed as the following JM interaction form $P_\mathrm{flux} = (1/2) < \boldsymbol{n}_\mathrm{i}^- \times \boldsymbol{J}_\mathrm{i}, \boldsymbol{M}_\mathrm{i} >_{\mathrm{S_i}}$.

Using surface equivalence principle, the filed $\boldsymbol{F}$ in $\mathrm{V}$ can be expressed in terms of the function of the above-mentioned currents as $\boldsymbol{F} = \mathcal{F}(\boldsymbol{J}_\mathrm{i} + \boldsymbol{J}_\mathrm{e}, \boldsymbol{M}_\mathrm{i})$ where $\boldsymbol{J}_\mathrm{e}$ is the induced electric current on $\mathrm{S_e}$, so the above $P_\mathrm{flux}$ can be re-formulated as the following integral operator forms

$$\begin{aligned} P_\mathrm{flux} &= -(1/2)\left\langle \boldsymbol{J}_\mathrm{i}, \mathcal{E}\left(\boldsymbol{J}_\mathrm{i} + \boldsymbol{J}_\mathrm{e}, \boldsymbol{M}_\mathrm{i}\right)\right\rangle_{S_\mathrm{i}^-} \\ &= -(1/2)\left\langle \boldsymbol{M}_\mathrm{i}, \mathcal{H}\left(\boldsymbol{J}_\mathrm{i} + \boldsymbol{J}_\mathrm{e}, \boldsymbol{M}_\mathrm{i}\right)\right\rangle_{S_\mathrm{i}^-}^\dagger \end{aligned} \tag{6-2}$$

called POYNTING'S FLUX OPERATORS (PtFOs), where $S_\mathrm{i}^-$ is the inner surface of $\mathrm{S_i}$, and then $P_\mathrm{flux}$ can be further discretized into the following matrix operator forms

$$P_\mathrm{flux} = \mathbb{J}_\mathrm{i}^\dagger \cdot \mathbb{P}_\mathrm{JE} \cdot \begin{bmatrix} \mathbb{J}_\mathrm{i} \\ \mathbb{J}_\mathrm{e} \\ \mathbb{M}_\mathrm{i} \end{bmatrix} = \begin{bmatrix} \mathbb{J}_\mathrm{i} \\ \mathbb{J}_\mathrm{e} \\ \mathbb{M}_\mathrm{i} \end{bmatrix}^\dagger \cdot \mathbb{P}_\mathrm{HM} \cdot \mathbb{M}_\mathrm{i} \tag{6-3}$$

if the currents are expanded in terms of some proper basis functions.

Because of the definitions of $\boldsymbol{J}_\mathrm{i}$ (DoJ) and $\boldsymbol{M}_\mathrm{i}$ (DoM) and the homogeneous tangential electric field boundary condition on $\mathrm{S_e}$, the above-mentioned currents satisfy some integral equations, and some transformations from the independent current to the other currents can be derived from discretizing the integral equations into matrix equations as follows:

$$\left.\begin{aligned} \left[\mathcal{H}\left(\boldsymbol{J}_\mathrm{i} + \boldsymbol{J}_\mathrm{e}, \boldsymbol{M}_\mathrm{i}\right)\right]_{S_\mathrm{i}^-}^\mathrm{tan} &= \boldsymbol{J}_\mathrm{i} \times \boldsymbol{n}_\mathrm{i}^- \\ \left[\mathcal{E}\left(\boldsymbol{J}_\mathrm{i} + \boldsymbol{J}_\mathrm{e}, \boldsymbol{M}_\mathrm{i}\right)\right]_{S_\mathrm{i}^-}^\mathrm{tan} &= \boldsymbol{n}_\mathrm{i}^- \times \boldsymbol{M}_\mathrm{i} \\ \left[\mathcal{E}\left(\boldsymbol{J}_\mathrm{i} + \boldsymbol{J}_\mathrm{e}, \boldsymbol{M}_\mathrm{i}\right)\right]_{\mathrm{S_e}}^\mathrm{tan} &= 0 \end{aligned}\right\} \Rightarrow \mathbb{T}_\mathrm{DoJ} \cdot \mathbb{J}_\mathrm{i} = \begin{bmatrix} \mathbb{J}_\mathrm{i} \\ \mathbb{J}_\mathrm{e} \\ \mathbb{M}_\mathrm{i} \end{bmatrix} = \mathbb{T}_\mathrm{DoM} \cdot \mathbb{M}_\mathrm{i} \tag{6-4}$$

Substituting matrix transformations (6-4) into matrix-formed PtFOs (6-3), the following matrix-formed PtFO

$$P_{\text{flux}} = \mathbb{C}_{\text{i}}^{\dagger} \cdot \mathbb{P}_{\text{flux}} \cdot \mathbb{C}_{\text{i}} = \begin{cases} \mathbb{J}_{\text{i}}^{\dagger} \cdot \overbrace{\mathbb{P}_{\text{JE}} \cdot \mathbb{T}_{\text{DoJ}}}^{\mathbb{P}_{\text{JE-DoJ}}} \cdot \mathbb{J}_{\text{i}} \\ \mathbb{M}_{\text{i}}^{\dagger} \cdot \underbrace{\mathbb{T}_{\text{DoM}}^{\dagger} \cdot \mathbb{P}_{\text{HM}}}_{\mathbb{P}_{\text{HM-DoM}}} \cdot \mathbb{M}_{\text{i}} \end{cases} \tag{6-5}$$

with only independent current $\mathbb{C}_{\text{i}}$ (which is either $\mathbb{J}_{\text{i}}$ or $\mathbb{M}_{\text{i}}$ corresponding to $\mathbb{P}_{\text{flux}} = \mathbb{P}_{\text{JE-DoJ}}$ or $\mathbb{P}_{\text{flux}} = \mathbb{P}_{\text{HM-DoM}}$ respectively) is immediately obtained.

Using the above-obtained $\mathbb{P}_{\text{flux}}$, we construct the following modal decoupling equation

$$\mathbb{P}_{\text{flux}}^{-} \cdot \mathbb{C}_{\text{i}} = \theta\, \mathbb{P}_{\text{flux}}^{+} \cdot \mathbb{C}_{\text{i}} \tag{6-6}$$

By solving the above equation, the DMs satisfying the following power-decoupling relations

$$\begin{aligned} (1 + j\,\theta_m)\delta_{mn} &= (1/2)\oint\!\!\!\oint_{S_{\text{i}}} \left(\boldsymbol{E}_n \times \boldsymbol{H}_m^{\dagger}\right) \cdot \boldsymbol{n}_{\text{i}}^{-} dS + (1/2)\oint\!\!\!\oint_{S_{\text{e}}} \left(\boldsymbol{E}_n \times \boldsymbol{H}_m^{\dagger}\right) \cdot \boldsymbol{n}_{\text{e}}^{+} dS \\ &= (1/2)\oint\!\!\!\oint_{S_{\text{i}}} \left(\boldsymbol{E}_n \times \boldsymbol{H}_m^{\dagger}\right) \cdot \boldsymbol{n}_{\text{i}}^{-} dS \\ &= \underbrace{(1/2)\langle \boldsymbol{\sigma} \cdot \boldsymbol{E}_m, \boldsymbol{E}_n \rangle_{\text{V}}}_{\delta_{mn}} + j\underbrace{2\omega\left[(1/4)\langle \boldsymbol{H}_m, \boldsymbol{\mu} \cdot \boldsymbol{H}_n \rangle_{\text{V}} - (1/4)\langle \boldsymbol{\varepsilon} \cdot \boldsymbol{E}_m, \boldsymbol{E}_n \rangle_{\text{V}}\right]}_{\theta_m \delta_{mn}} \end{aligned} \tag{6-7}$$

can be derived. To effectively recognize the optimally energy-dissipating modes contained in the obtained DM set, we introduce the modal "impedance $Z_{\text{flux}}$, resistance $R_{\text{flux}}$, reactance $X_{\text{flux}}$" and "admittance $Y_{\text{flux}}$, conductance $G_{\text{flux}}$, susceptance $B_{\text{flux}}$" as follows:

$$Z_{\text{flux}} = \frac{(1/2)\iint_{S_{\text{i}}} \left(\boldsymbol{E} \times \boldsymbol{H}^{\dagger}\right) \cdot \boldsymbol{n}_{\text{i}}^{-} dS}{(1/2)\langle \boldsymbol{J}_{\text{i}}, \boldsymbol{J}_{\text{i}} \rangle_{S_{\text{i}}}} = R_{\text{flux}} + j\, X_{\text{flux}} \tag{6-8a}$$

$$Y_{\text{flux}} = \frac{(1/2)\iint_{S_{\text{i}}} \left(\boldsymbol{E} \times \boldsymbol{H}^{\dagger}\right) \cdot \boldsymbol{n}_{\text{i}}^{-} dS}{(1/2)\langle \boldsymbol{M}_{\text{i}}, \boldsymbol{M}_{\text{i}} \rangle_{S_{\text{i}}}} = G_{\text{flux}} + j\, B_{\text{flux}} \tag{6-8b}$$

In the above Eqs. (6-8a) and (6-8b), $R_{\text{flux}} = \text{Re}\, Z_{\text{flux}}$, and $X_{\text{flux}} = \text{Im}\, Z_{\text{flux}}$, and $G_{\text{flux}} = \text{Re}\, Y_{\text{flux}}$, and $B_{\text{flux}} = \text{Im}\, Y_{\text{flux}}$.

Here, we consider a specific example. For a lossy dielectric sphere, whose radius is 7 mm and material parameters are $\{\mu_{\text{r}} = 1, \varepsilon_{\text{r}} = 10, \sigma = 0.3\}$, its DMs determined by Eq. (6-6) are calculated, and the modal resistances associated to the first several typical modes are shown in Fig. 6-2. Figure 6-2 implies that: the resistances of DM 1, DM 2, DM 3, and DM 4 achieve their local maximums at 5.775 GHz, 6.000 GHz, 6.275 GHz, and 8.475 GHz respectively. The modal electric and magnetic currents corresponding to these locally maximal resistances are shown in Fig. 6-3.

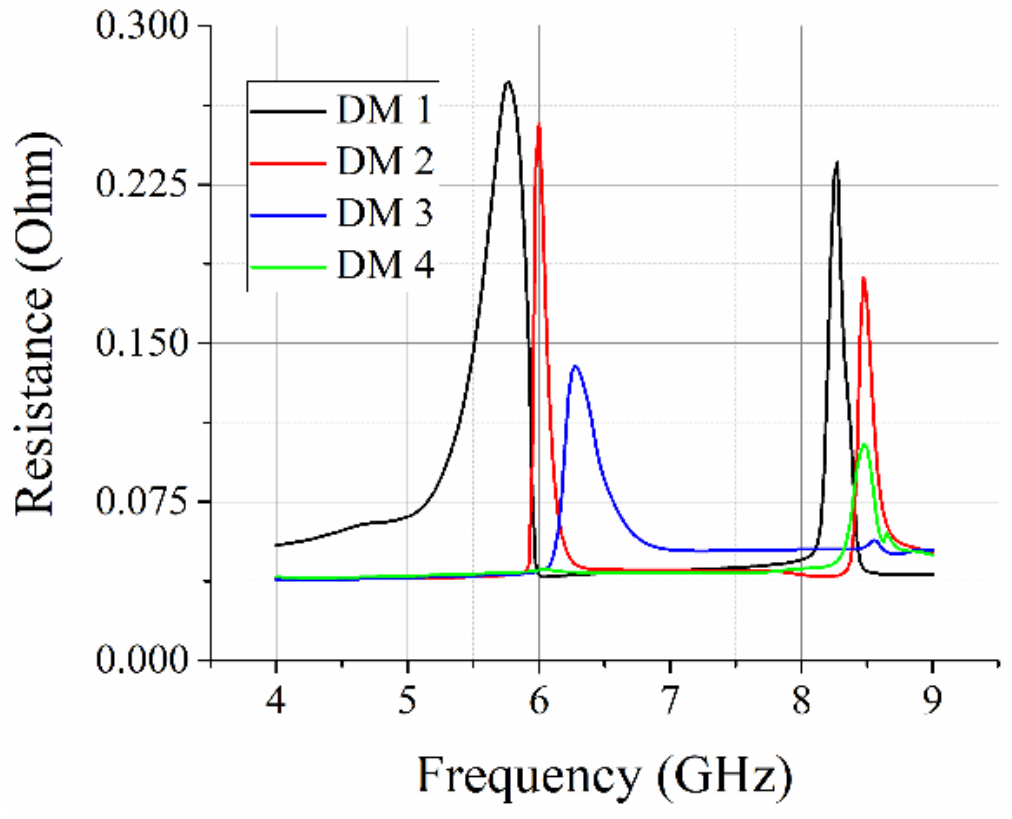

Figure 6-2 Resistance curves of the first several typical DMs.

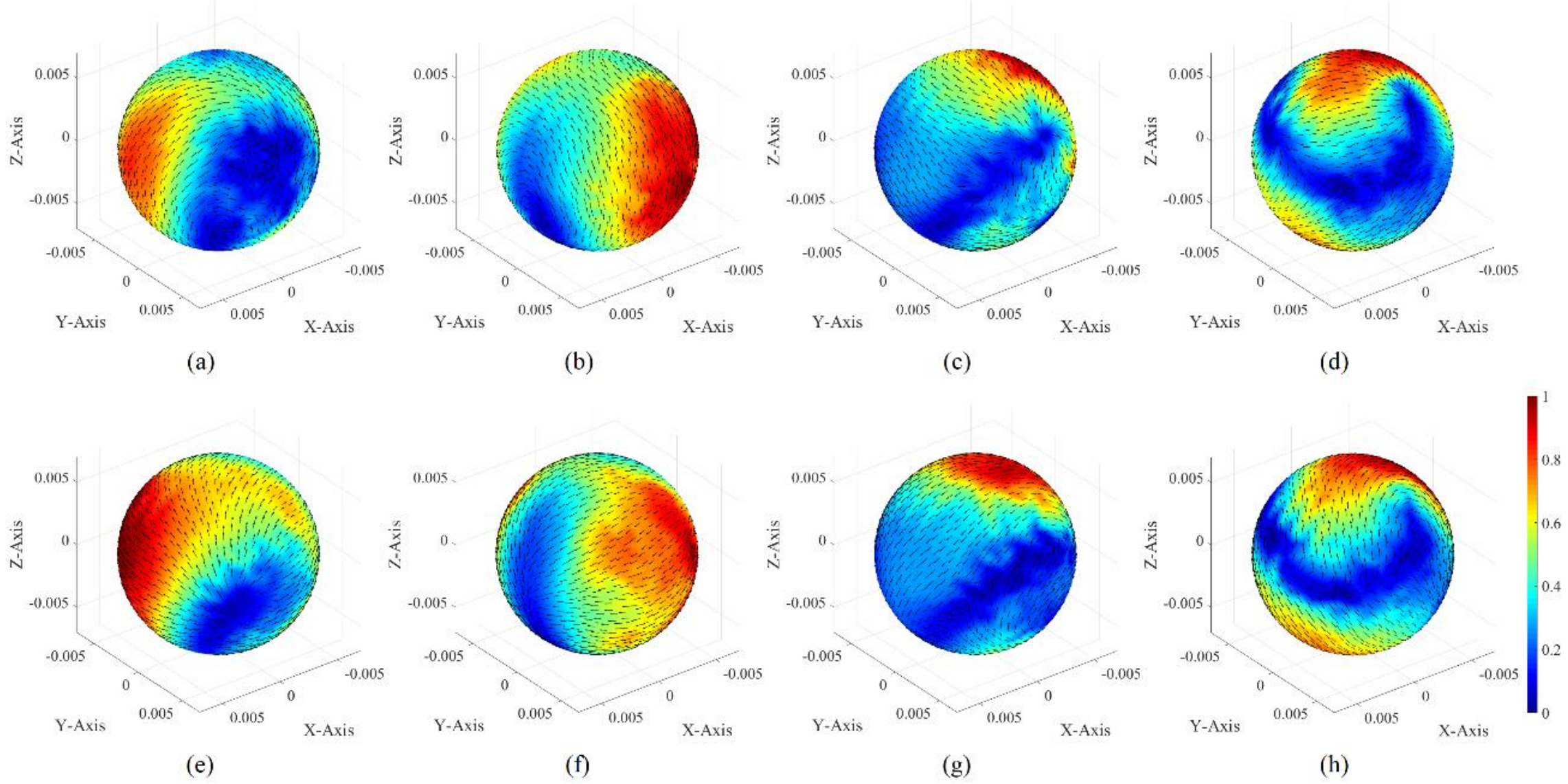

Figure 6-3 (a) Modal $\boldsymbol{J}_{\mathrm{i}}$ of the DM 1 at 5.775 GHz, (b) modal $\boldsymbol{J}_{\mathrm{i}}$ of the DM 2 at 6.000 GHz, (c) modal $\boldsymbol{J}_{\mathrm{i}}$ of the DM 3 at 6.275 GHz, (d) modal $\boldsymbol{J}_{\mathrm{i}}$ of the DM 4 at 8.475 GHz, (e) modal $\boldsymbol{M}_{\mathrm{i}}$ of the DM 1 at 5.775 GHz, (f) modal $\boldsymbol{M}_{\mathrm{i}}$ of the DM 2 at 6.000 GHz, (g) modal $\boldsymbol{M}_{\mathrm{i}}$ of the DM 3 at 6.275 GHz, and (h) modal $\boldsymbol{M}_{\mathrm{i}}$ of the DM 4 at 8.475 GHz.

## 6.3 PtT-Based DMs of Self-Oscillating Structures

Taking the two-body lossless material structure shown in Fig. 6-4 as an example, this section focuses on establishing the PtT-DMT for self-oscillating structures, and **a main function of the PtT-DMT is to find the self-oscillating modes of the lossless structure**. It must be clearly emphasize here that: <u>the EM structures considered in this section are restricted to being lossless</u>.

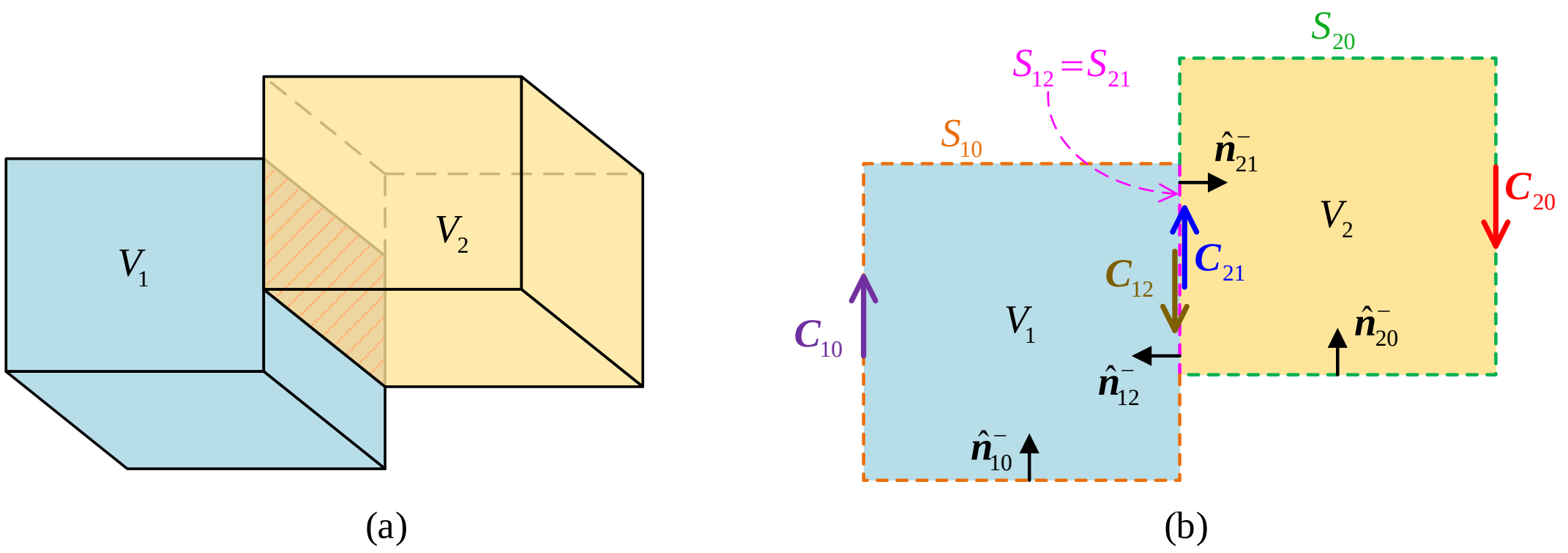

Figure 6-4 (a) Geometry and (b) topology of a two-body material structure.

In the above figure, the whole two-body material structure is placed in free space; the regions occupied by the two material bodies are denoted as $V_1$ and $V_2$; the environment-material boundaries are denoted as $S_{10}$ and $S_{20}$, and the material-material boundary is denoted as $S_{12}$. Now, we suppose there is an external field $\boldsymbol{F}_{\text{inc}}$ illuminating on the material structure. Under the illumination of $\boldsymbol{F}_{\text{inc}}$, some currents are induced on the material structure, and the induced currents generate a scattered field $\boldsymbol{F}_{\text{sca}}$. The summation of $\boldsymbol{F}_{\text{inc}}$ and $\boldsymbol{F}_{\text{sca}}$ is denoted as $\boldsymbol{F}$ called total field, i.e., $\boldsymbol{F}=\boldsymbol{F}_{\text{inc}}+\boldsymbol{F}_{\text{sca}}$.

The PtT manifestation form of ENERGY CONSERVATION LAW tells us that: the Poynting's flux penetrated into the whole material structure is

$$\begin{aligned} P_{\text{flux}} &= (1/2)\iint_{S_{10}}\left(\boldsymbol{E}\times\boldsymbol{H}^{\dagger}\right)\cdot\boldsymbol{n}_{10}^{-}dS+(1/2)\iint_{S_{20}}\left(\boldsymbol{E}\times\boldsymbol{H}^{\dagger}\right)\cdot\boldsymbol{n}_{20}^{-}dS \\ &=-(1/2)\langle\boldsymbol{J}_{10},\boldsymbol{E}\rangle_{S_{10}}-(1/2)\langle\boldsymbol{J}_{20},\boldsymbol{E}\rangle_{S_{20}} \\ &=-(1/2)\langle\boldsymbol{J}_{10},\boldsymbol{E}\rangle_{S_{10}}-(1/2)\langle\boldsymbol{M}_{20},\boldsymbol{H}\rangle_{S_{20}}^{\dagger} \\ &=-(1/2)\langle\boldsymbol{M}_{10},\boldsymbol{H}\rangle_{S_{10}}^{\dagger}-(1/2)\langle\boldsymbol{J}_{20},\boldsymbol{E}\rangle_{S_{20}} \\ &=-(1/2)\langle\boldsymbol{M}_{10},\boldsymbol{H}\rangle_{S_{10}}^{\dagger}-(1/2)\langle\boldsymbol{M}_{20},\boldsymbol{H}\rangle_{S_{20}}^{\dagger} \end{aligned} \tag{6-9}$$

Here, the first equality directly originates from PtT; the last four equalities are based on the following definitions $\boldsymbol{J}_{10/20}=\boldsymbol{n}_{10/20}^{-}\times\boldsymbol{H}$ and $\boldsymbol{M}_{10/20}=\boldsymbol{E}\times\boldsymbol{n}_{10/20}^{-}$ on $S_{10/20}$, where $\boldsymbol{n}_{10/20}^{-}$ is the normal direction of $S_{10/20}$ and points to the interior of $V_{1/2}$. Similarly to the previous Sec. 6.2, the right-hand sides of the equalities are called EHEH, JEJE, JEHM, HMJE, and HMHM forms respectively. Obviously, the $P_{\text{flux}}$ can also be expressed as the following JMJM form $P_{\text{flux}}=(1/2)<\boldsymbol{n}_{10}^{-}\times\boldsymbol{J}_{10},\boldsymbol{M}_{10}>_{S_{10}}+(1/2)<\boldsymbol{n}_{20}^{-}\times\boldsymbol{J}_{20},\boldsymbol{M}_{20}>_{S_{20}}$.

Besides the previously defined equivalent currents $\boldsymbol{J}_{10/20}$ and $\boldsymbol{M}_{10/20}$, we also define the equivalent currents on $S_{12}$ as that $\boldsymbol{J}_{12}=\boldsymbol{n}_{12}^{-}\times\boldsymbol{H}$ and $\boldsymbol{M}_{12}=\boldsymbol{E}\times\boldsymbol{n}_{12}^{-}$, where $\boldsymbol{n}_{12}^{-}$ is the normal direction of $S_{12}$ and points to the interior of $V_1$. Using surface equivalence principle, the filed $\boldsymbol{F}$ in $\mathrm{V}_1$ and $\mathrm{V}_2$ can be expressed in terms of the function of the above-mentioned currents as that $\boldsymbol{F}=\mathcal{F}_1(\boldsymbol{J}_{10}+\boldsymbol{J}_{12},\boldsymbol{M}_{10}+\boldsymbol{M}_{12})$ on $\mathrm{V}_1$ and $\boldsymbol{F}=\mathcal{F}_2(\boldsymbol{J}_{20}-\boldsymbol{J}_{12},\boldsymbol{M}_{20}-\boldsymbol{M}_{12})$ on $\mathrm{V}_2$, so the above $P_{\text{flux}}$ can be re-formulated as the following integral operator forms

$$\begin{aligned}P_{\text{flux}}&=-\frac{1}{2}\left\langle\boldsymbol{J}_{10},\mathcal{E}_1\left(\boldsymbol{J}_{10}+\boldsymbol{J}_{12},\boldsymbol{M}_{10}+\boldsymbol{M}_{12}\right)\right\rangle_{S_{10}^-}-\frac{1}{2}\left\langle\boldsymbol{J}_{20},\mathcal{E}_2\left(\boldsymbol{J}_{20}-\boldsymbol{J}_{12},\boldsymbol{M}_{20}-\boldsymbol{M}_{12}\right)\right\rangle_{S_{20}^-}\\&=-\frac{1}{2}\left\langle\boldsymbol{J}_{10},\mathcal{E}_1\left(\boldsymbol{J}_{10}+\boldsymbol{J}_{12},\boldsymbol{M}_{10}+\boldsymbol{M}_{12}\right)\right\rangle_{S_{10}^-}-\frac{1}{2}\left\langle\boldsymbol{M}_{20},\mathcal{H}_2\left(\boldsymbol{J}_{20}-\boldsymbol{J}_{12},\boldsymbol{M}_{20}-\boldsymbol{M}_{12}\right)\right\rangle_{S_{20}^-}^{\dagger}\\&=-\frac{1}{2}\left\langle\boldsymbol{M}_{10},\mathcal{H}_1\left(\boldsymbol{J}_{10}+\boldsymbol{J}_{12},\boldsymbol{M}_{10}+\boldsymbol{M}_{12}\right)\right\rangle_{S_{10}^-}^{\dagger}-\frac{1}{2}\left\langle\boldsymbol{J}_{20},\mathcal{E}_2\left(\boldsymbol{J}_{20}-\boldsymbol{J}_{12},\boldsymbol{M}_{20}-\boldsymbol{M}_{12}\right)\right\rangle_{S_{20}^-}\\&=-\frac{1}{2}\left\langle\boldsymbol{M}_{10},\mathcal{H}_1\left(\boldsymbol{J}_{10}+\boldsymbol{J}_{12},\boldsymbol{M}_{10}+\boldsymbol{M}_{12}\right)\right\rangle_{S_{10}^-}^{\dagger}-\frac{1}{2}\left\langle\boldsymbol{M}_{20},\mathcal{H}_2\left(\boldsymbol{J}_{20}-\boldsymbol{J}_{12},\boldsymbol{M}_{20}-\boldsymbol{M}_{12}\right)\right\rangle_{S_{20}^-}^{\dagger}\end{aligned}\tag{6-10}$$

where $S_{10/20}^-$ is the inner surface of $\mathrm{S}_{10/20}$, and then $P_{\text{flux}}$ can be further discretized into the following matrix operator forms

$$P_{\text{flux}}=\begin{bmatrix}\mathbb{J}_{10}\\\mathbb{J}_{12}\\\mathbb{J}_{20}\\\mathbb{M}_{10}\\\mathbb{M}_{12}\\\mathbb{M}_{20}\end{bmatrix}^{\dagger}\cdot\mathbb{P}_{\text{JEJE/JEHM/HMJE/HMHM}}\cdot\begin{bmatrix}\mathbb{J}_{10}\\\mathbb{J}_{12}\\\mathbb{J}_{20}\\\mathbb{M}_{10}\\\mathbb{M}_{12}\\\mathbb{M}_{20}\end{bmatrix}\tag{6-11}$$

if the currents are expanded in terms of some proper basis functions.

Because of the definitions of $\{\boldsymbol{J}_{10},\boldsymbol{J}_{20}\}$ (DoJ) and $\{\boldsymbol{M}_{10},\boldsymbol{M}_{20}\}$ (DoM) and the tangential field continuation conditions on $\mathrm{S}_{12}$, the above-mentioned currents satisfy the following integral equations

$$\left[\mathcal{H}_1\left(\boldsymbol{J}_{10}+\boldsymbol{J}_{12},\boldsymbol{M}_{10}+\boldsymbol{M}_{12}\right)\right]_{S_{10}^-}^{\tan}=\boldsymbol{J}_{10}\times\boldsymbol{n}_{10}^-\tag{6-12a}$$

$$\left[\mathcal{E}_1\left(\boldsymbol{J}_{10}+\boldsymbol{J}_{12},\boldsymbol{M}_{10}+\boldsymbol{M}_{12}\right)\right]_{S_{10}^-}^{\tan}=\boldsymbol{n}_{10}^-\times\boldsymbol{M}_{10}\tag{6-12b}$$

$$\left[\mathcal{H}_2\left(\boldsymbol{J}_{20}-\boldsymbol{J}_{12},\boldsymbol{M}_{20}-\boldsymbol{M}_{12}\right)\right]_{S_{20}^-}^{\tan}=\boldsymbol{J}_{20}\times\boldsymbol{n}_{20}^-\tag{6-13a}$$

$$\left[\mathcal{E}_2\left(\boldsymbol{J}_{20}-\boldsymbol{J}_{12},\boldsymbol{M}_{20}-\boldsymbol{M}_{12}\right)\right]_{S_{20}^-}^{\tan}=\boldsymbol{n}_{20}^-\times\boldsymbol{M}_{20}\tag{6-13b}$$

$$\left[\mathcal{E}_1\left(\boldsymbol{J}_{10}+\boldsymbol{J}_{12},\boldsymbol{M}_{10}+\boldsymbol{M}_{12}\right)\right]_{S_{12}^-}^{\tan}=\left[\mathcal{E}_2\left(\boldsymbol{J}_{20}-\boldsymbol{J}_{12},\boldsymbol{M}_{20}-\boldsymbol{M}_{12}\right)\right]_{S_{12}^+}^{\tan}\tag{6-14a}$$

$$\left[\mathcal{H}_1\left(\boldsymbol{J}_{10}+\boldsymbol{J}_{12},\boldsymbol{M}_{10}+\boldsymbol{M}_{12}\right)\right]_{S_{12}^-}^{\tan}=\left[\mathcal{H}_2\left(\boldsymbol{J}_{20}-\boldsymbol{J}_{12},\boldsymbol{M}_{20}-\boldsymbol{M}_{12}\right)\right]_{S_{12}^+}^{\tan}\tag{6-14b}$$

By discretizing the above integral equations into matrix equations and solving the matrix equations, we have the following transformations

$$\begin{bmatrix} \mathbb{J}_{10} \\ \mathbb{J}_{12} \\ \mathbb{J}_{20} \\ \mathbb{M}_{10} \\ \mathbb{M}_{12} \\ \mathbb{M}_{20} \end{bmatrix} = \mathbb{T}_{\text{DoJJ}} \cdot \begin{bmatrix} \mathbb{J}_{10} \\ \mathbb{J}_{20} \end{bmatrix} = \mathbb{T}_{\text{DoJM}} \cdot \begin{bmatrix} \mathbb{J}_{10} \\ \mathbb{M}_{20} \end{bmatrix} = \mathbb{T}_{\text{DoMJ}} \cdot \begin{bmatrix} \mathbb{M}_{10} \\ \mathbb{J}_{20} \end{bmatrix} = \mathbb{T}_{\text{DoMM}} \cdot \begin{bmatrix} \mathbb{M}_{10} \\ \mathbb{M}_{20} \end{bmatrix} \tag{6-15}$$

Substituting matrix transformations (6-15) into matrix-formed PtFOs (6-11), the following matrix-formed PtFO

$$P_{\text{flux}} = \mathbb{C}^{\dagger} \cdot \mathbb{P}_{\text{flux}} \cdot \mathbb{C} = \begin{cases} \begin{bmatrix} \mathbb{J}_{10} \\ \mathbb{J}_{20} \end{bmatrix}^{\dagger} \cdot \overbrace{\mathbb{T}_{\text{DoJJ}}^{\dagger} \cdot \mathbb{P}_{\text{JEJE}} \cdot \mathbb{T}_{\text{DoJJ}}}^{\mathbb{P}_{\text{JEJE-DoJJ}}} \cdot \begin{bmatrix} \mathbb{J}_{10} \\ \mathbb{J}_{20} \end{bmatrix} &, \text{ where } \mathbb{C} = \begin{bmatrix} \mathbb{J}_{10} \\ \mathbb{J}_{20} \end{bmatrix} \\ \begin{bmatrix} \mathbb{J}_{10} \\ \mathbb{M}_{20} \end{bmatrix}^{\dagger} \cdot \overbrace{\mathbb{T}_{\text{DoJM}}^{\dagger} \cdot \mathbb{P}_{\text{JEHM}} \cdot \mathbb{T}_{\text{DoJM}}}^{\mathbb{P}_{\text{JEHM-DoJM}}} \cdot \begin{bmatrix} \mathbb{J}_{10} \\ \mathbb{M}_{20} \end{bmatrix} &, \text{ where } \mathbb{C} = \begin{bmatrix} \mathbb{J}_{10} \\ \mathbb{M}_{20} \end{bmatrix} \\ \begin{bmatrix} \mathbb{M}_{10} \\ \mathbb{J}_{20} \end{bmatrix}^{\dagger} \cdot \underbrace{\mathbb{T}_{\text{DoMJ}}^{\dagger} \cdot \mathbb{P}_{\text{HMJE}} \cdot \mathbb{T}_{\text{DoMJ}}}_{\mathbb{P}_{\text{HMJE-DoMJ}}} \cdot \begin{bmatrix} \mathbb{M}_{10} \\ \mathbb{J}_{20} \end{bmatrix} &, \text{ where } \mathbb{C} = \begin{bmatrix} \mathbb{M}_{10} \\ \mathbb{J}_{20} \end{bmatrix} \\ \begin{bmatrix} \mathbb{M}_{10} \\ \mathbb{M}_{20} \end{bmatrix}^{\dagger} \cdot \underbrace{\mathbb{T}_{\text{DoMM}}^{\dagger} \cdot \mathbb{P}_{\text{HMHM}} \cdot \mathbb{T}_{\text{DoMM}}}_{\mathbb{P}_{\text{HMHM-DoMM}}} \cdot \begin{bmatrix} \mathbb{M}_{10} \\ \mathbb{M}_{20} \end{bmatrix} &, \text{ where } \mathbb{C} = \begin{bmatrix} \mathbb{M}_{10} \\ \mathbb{M}_{20} \end{bmatrix} \end{cases} \tag{6-16}$$

with only independent current $\mathbb{C}$ is immediately obtained.

The positive and negative Hermitian parts of $\mathbb{P}_{\text{flux}}$ are denoted as $\mathbb{P}_{\text{flux}}^{+}$ and $\mathbb{P}_{\text{flux}}^{-}$ respectively. **Based on a similar method used in Ref. [8-Sec.6.4.4], it is easy to prove that $\mathbb{P}_{\text{flux}}^{+} = 0$ theoretically, because the EM structure is restricted to being lossless.** Using the $\mathbb{P}_{\text{flux}}^{-}$, we construct the following equation

$$\mathbb{P}_{\text{flux}}^{-} \cdot \mathbb{C} = \gamma\, \mathbb{C} \tag{6-17}$$

used to determine the DMs. By solving the equation, the DMs satisfying the following power-decoupling relations

$$\begin{aligned} (0 + j\,\theta_m)\delta_{mn} &= (1/2) \iint_{S_{10}} \left( \boldsymbol{E}_n \times \boldsymbol{H}_m^{\dagger} \right) \cdot \boldsymbol{n}_{10}^{-} dS + (1/2) \iint_{S_{20}} \left( \boldsymbol{E}_n \times \boldsymbol{H}_m^{\dagger} \right) \cdot \boldsymbol{n}_{20}^{-} dS \\ &= \underbrace{\frac{1}{2} \langle \boldsymbol{\sigma} \cdot \boldsymbol{E}_m, \boldsymbol{E}_n \rangle_{\text{V}_1 \cup \text{V}_2}}_{0} + \underbrace{j\, 2\omega \left[ \frac{1}{4} \langle \boldsymbol{H}_m, \boldsymbol{\mu} \cdot \boldsymbol{H}_n \rangle_{\text{V}_1 \cup \text{V}_2} - \frac{1}{4} \langle \boldsymbol{\varepsilon} \cdot \boldsymbol{E}_m, \boldsymbol{E}_n \rangle_{\text{V}_1 \cup \text{V}_2} \right]}_{\theta_m \delta_{mn}} \end{aligned} \tag{6-18}$$

can be obtained. **Here, the conclusion $\mathbb{C}_m^{\dagger} \cdot \mathbb{P}_{\text{flux}}^{+} \cdot \mathbb{C}_n = 0$ (because $\mathbb{P}_{\text{flux}}^{+} = 0$) has been utilized to derive relations (6-18).** In addition, we emphasize that: <u>it is better not to replace Eq. (6-17) with Eq. (6-6)</u>, because $\mathbb{P}_{\text{flux}}^{+} = 0$ **theoretically**. The first line of relation (6-18) implies the following energy-decoupling relation

$$(1/T)\int_{t_0}^{t_0+T}\left[\oint\!\!\!\oint_{S_{10}\cup S_{20}}\left(\boldsymbol{\mathcal{E}}_n\times\boldsymbol{\mathcal{H}}_m+\boldsymbol{\mathcal{E}}_m\times\boldsymbol{\mathcal{H}}_n\right)\cdot d\boldsymbol{S}\right]dt=0 \tag{6-19}$$

and then time-domain Lorentz's Reciprocity Theorem (2-22) implies the following orthogonality

$$(1/T)\int_{t_0}^{t_0+T}\left[\oint\!\!\!\oint_{\mathrm{S}}\left(\boldsymbol{\mathcal{E}}_n\times\boldsymbol{\mathcal{H}}_m+\boldsymbol{\mathcal{E}}_m\times\boldsymbol{\mathcal{H}}_n\right)\cdot d\boldsymbol{S}\right]dt=0 \tag{6-20}$$

where <u>S is an arbitrary closed surface completely contained in the union of V1 and V2</u> as shown in Fig. 6-5.

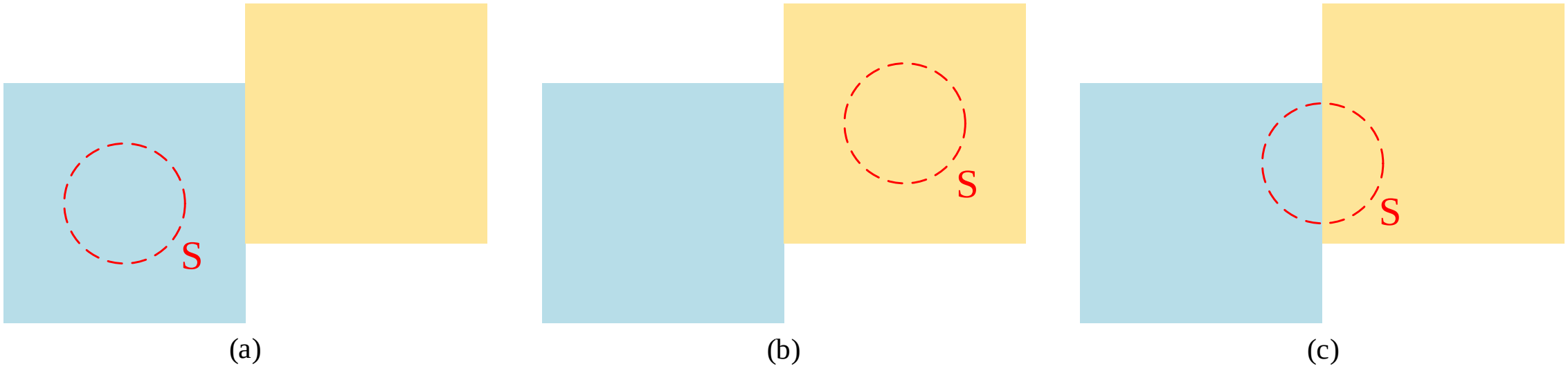

Figure 6-5 A closed surface S completely contained in (a) $\mathrm{V}_1$, (b) $\mathrm{V}_2$, and (c) $\mathrm{V}_1\cup\mathrm{V}_2$.

Here, we want to emphasize again that both the material body $\mathrm{V}_1$ and the material body $\mathrm{V}_2$ are lossless.

Similarly to the Eqs. (6-8a) and (6-8b) given in the previous Sec. 6.2, the modal "impedance $Z_{\text{flux}}$, resistance $R_{\text{flux}}$, reactance $X_{\text{flux}}$" and "admittance $Y_{\text{flux}}$, conductance $G_{\text{flux}}$, susceptance $B_{\text{flux}}$" can also be defined for the DMs obtained in this section. For a DM, if its $X_{\text{flux}}$ is equal to zero, or its $B_{\text{flux}}$ is equal to zero, then it works at self-oscillating state. The resonance frequency $f_{\text{self-osc}}$ satisfying $R_{\text{flux}}(f_{\text{self-osc}})=0=X_{\text{flux}}(f_{\text{self-osc}})$ or $G_{\text{flux}}(f_{\text{self-osc}})=0=B_{\text{flux}}(f_{\text{self-osc}})$ is called self-oscillation frequency.

Here, we consider a specific example. For a lossless dielectric sphere, whose radius is 7 mm and material parameters are $\{\mu_{\mathrm{r}}=1,\varepsilon_{\mathrm{r}}=10,\sigma=0\}$, its DMs determined by Eq. (6-17) are calculated, and the modal reactances associated to the first several typical modes are shown in Fig. 6-6. Figure 6-6 implies that: the reactance of DM 1 is zero at 6.16 GHz. Thus, the DM 1 is self-oscillating at 6.16 GHz as explained previously. The modal electric and magnetic currents corresponding to the self-oscillating mode are shown in Fig. 6-7.

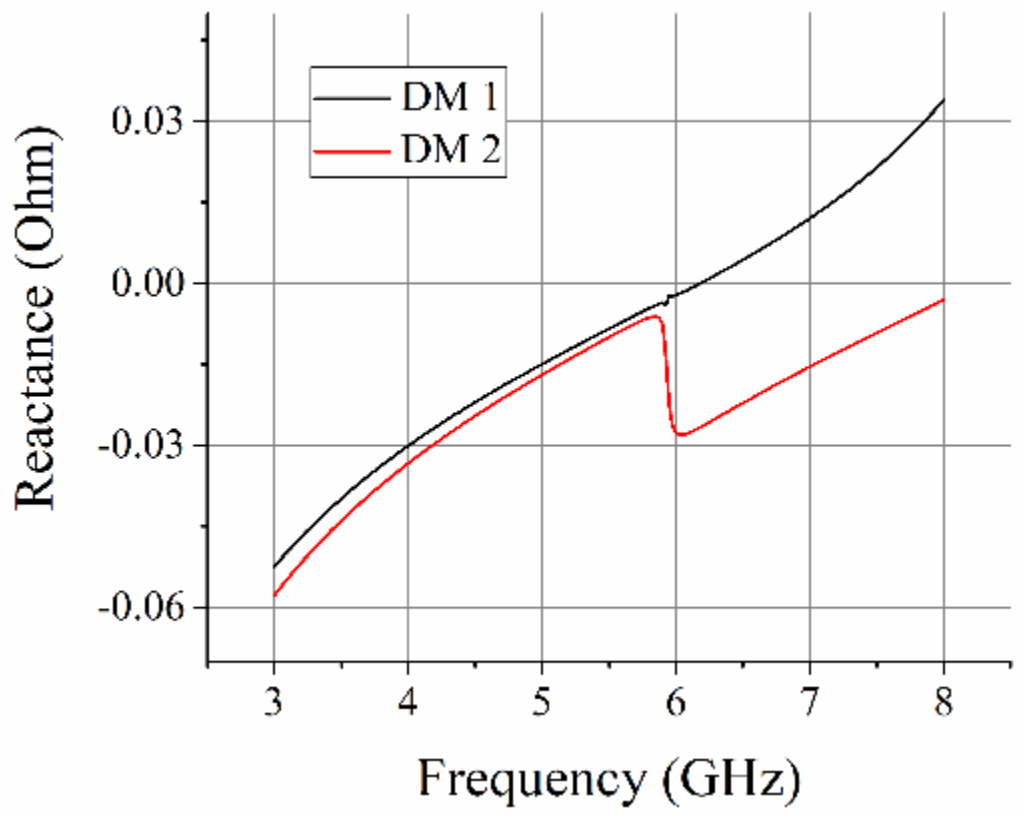

Figure 6-6 Reactance curves of the first several typical DMs.

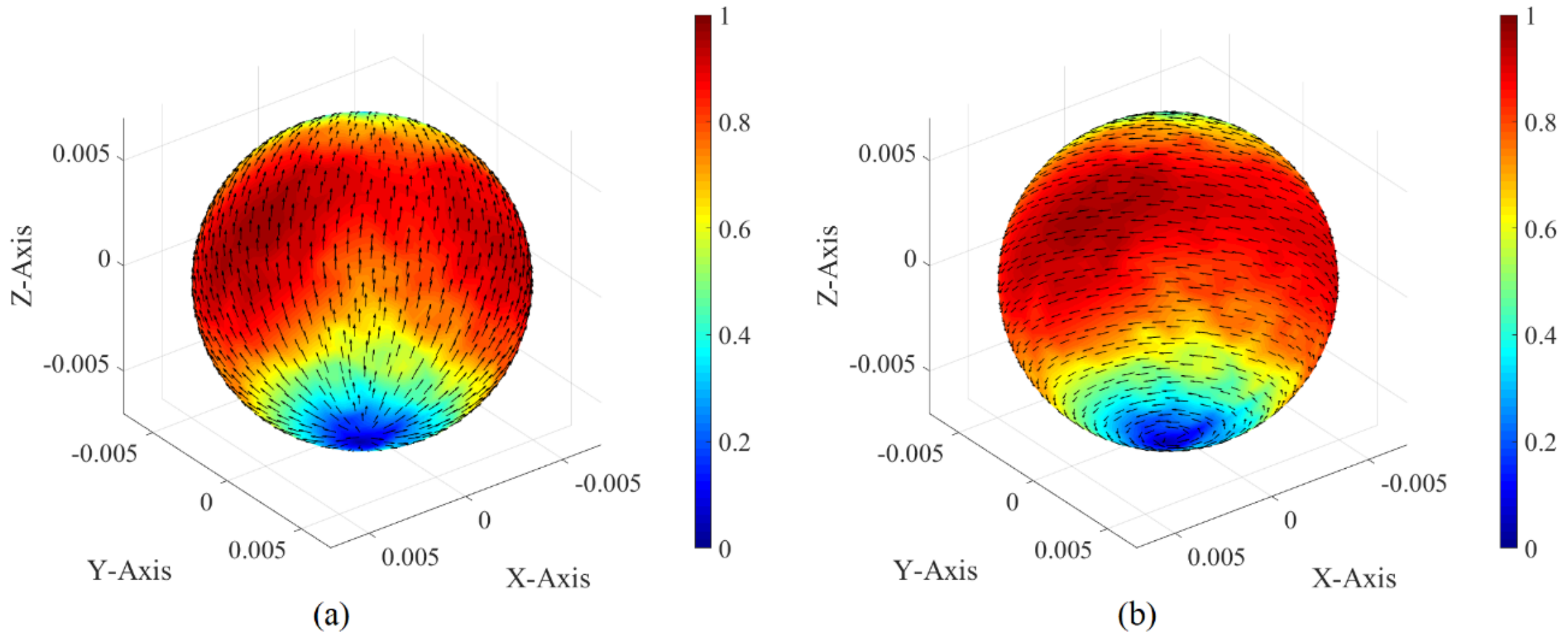

Figure 6-7 Modal (a) $\boldsymbol{J}_{\rm i}$ and (b) $\boldsymbol{M}_{\rm i}$ of the DM 1 at 6.16 GHz.

## 6.4 Chapter Summary

This chapter proposes a PtT-DMT used to do the modal analysis for energy-dissipating and self-oscillating structures.

The PtT-DMT can construct the DMs of energy-dissipating and self-oscillating structures by orthogonalizing PtFO. The optimally energy-dissipating modes of lossy structures and the self-oscillating modes of lossless structures can be effectively found in the obtained DM sets, if the modes are indeed existed.

# CHAPTER 7 CONCLUSIONS

The central purposes of this Post-Doctoral Concluding Report are (1) to reveal the core position of energy viewpoint in the realm of electromagnetic (EM) modal analysis; (2) to show how to do energy-viewpoint-based modal analysis for various EM structures.

The major conclusions related to this report are that: ENERGY CONSERVATION LAW governs the energy utilization process of EM structure, and its energy source term sustains the steady energy utilization process; the whole modal space of a EM structure is spanned by a series of ENERGY-DECOUPLED MODES (DMs), which don't have net energy exchange in any integral period; the DMs can be effectively constructed by orthogonalizing ENERGY SOURCE OPERATOR, which is just the operator form of the energy source term.

Some specific results obtained in this report are summarized as below.

For a certain EM structure, it has many different working manners, such as transmitting, receiving, wave-guiding, power-transferring, scattering, and self-oscillating manners, etc. For example, for the dielectric rod structure shown in Fig. 7-1, transmitting[27-Sec.6.3.5],[35], receiving[27-Sec.7.4.6], wave-guiding[27-Sec.3.3],[39], scattering[8-Secs.4.2.4&4.3.6],[14,30,31,33,34], and self-oscillating manners (Sec. 6.3 of this report) are all its physically realizable working manners.

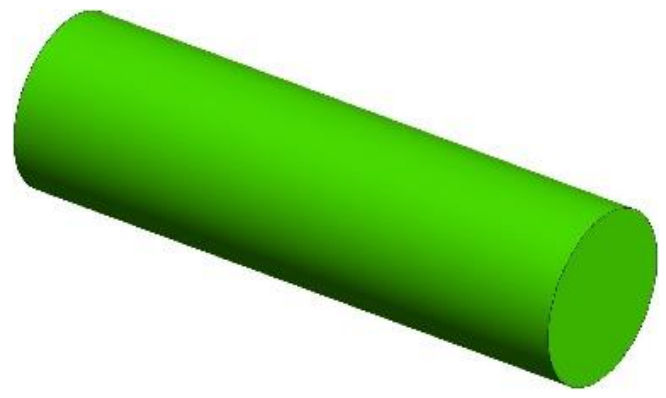

Figure 7-1 Geometry of a dielectric rod structure.

Different working manners correspond to different energy utilization processes. ENERGY CONSERVATION LAW governs all the energy utilization processes, but it has different manifestation forms in the different processes. Different manifestation forms of ENERGY CONSERVATION LAW have different energy source terms, and different energy source terms imply different excitation ways. Thus, different excitation ways result in different working manners, and, at the same time, different working manners also need to be sustained by different excitation ways. Specifically speaking:

i) wave-port-fed EM structures (such as the sub-structures shown in Fig. 7-2) result in a power transportation process, and the process is governed by Power Transport Theorem (PTT) form, and the source term in PTT form is input power, which is used to sustain a steady power transportation;

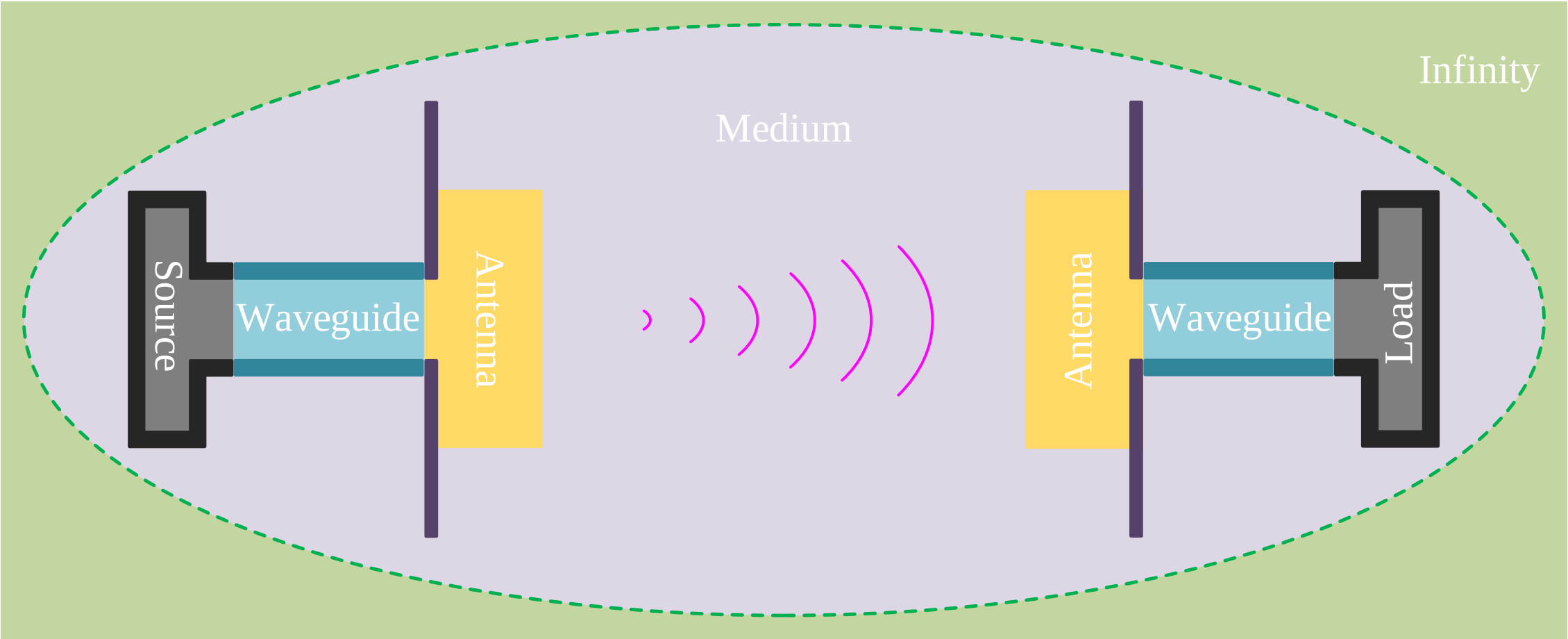

Figure 7-2 Geometries of a transceiving system and its sub-structures, which are a series of cascaded wave-port-fed EM structures (feeding waveguide, transmitting antenna, propagating medium, receiving antenna, and loading waveguide).

ii) lumped-port-driven/local-near-field-driven EM structures (such as the ones shown in Fig. 7-3) result in a work-energy transformation process, and the process is governed by Partial-Structure-Oriented Work-Energy Theorem (PS-WET) form, and the source term in PS-WET form is partial-structure-oriented driving power, which is used to sustain a steady work-energy transformation;

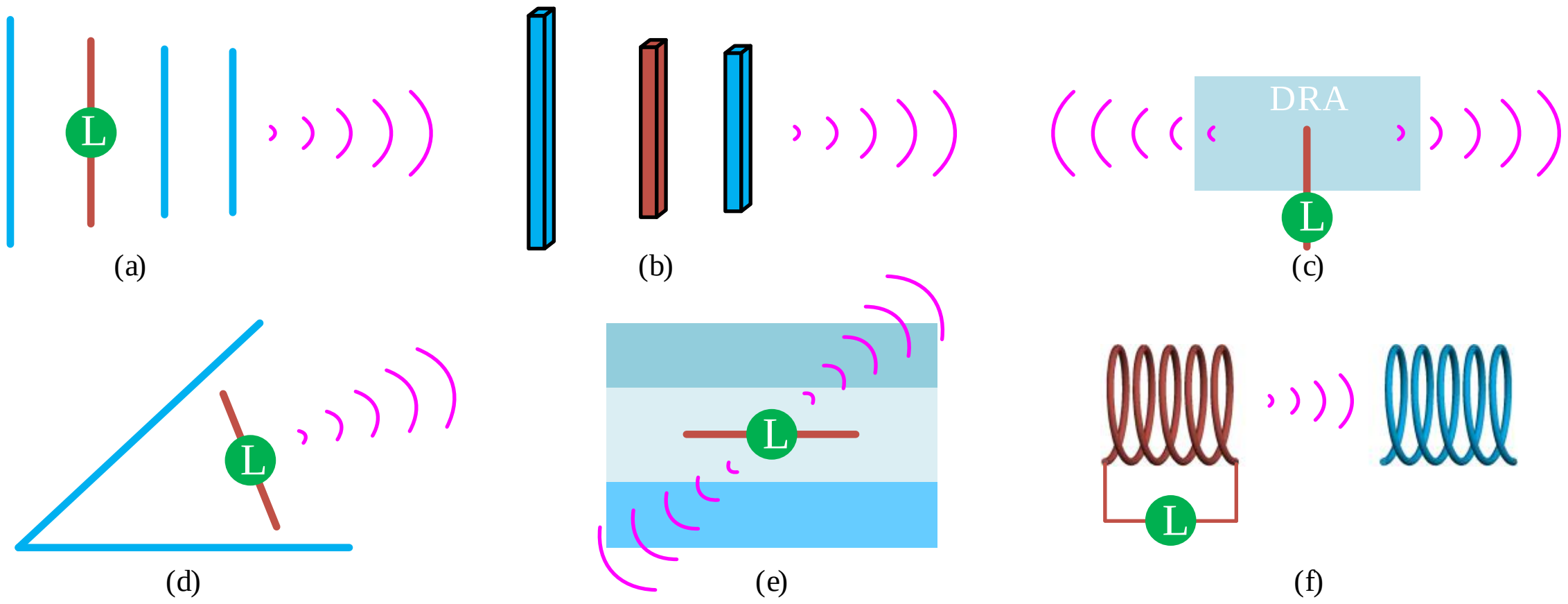

Figure 7-3 Geometries of lumped-port-driven (a) metallic Yagi-Uda antenna, (b) material Yagi-Uda antenna, (c) metallic dipole antenna passively loaded by DRA, (d) metallic dipole antenna passively loaded by corner reflector, (e) metallic dipole antenna passively loaded by layered medium, and (f) two-coil WPT system.

iii) incident-field-driven EM structures (such as the one shown in Fig. 7-4) result in a work-energy transformation process, and the process is governed by ENTIRE-STRUCTURE-ORIENTED WORK-ENERGY THEOREM (ES-WET) form, and the source term in ES-WET form is entire-structure-oriented driving power, which is used to sustain a steady work-energy transformation;

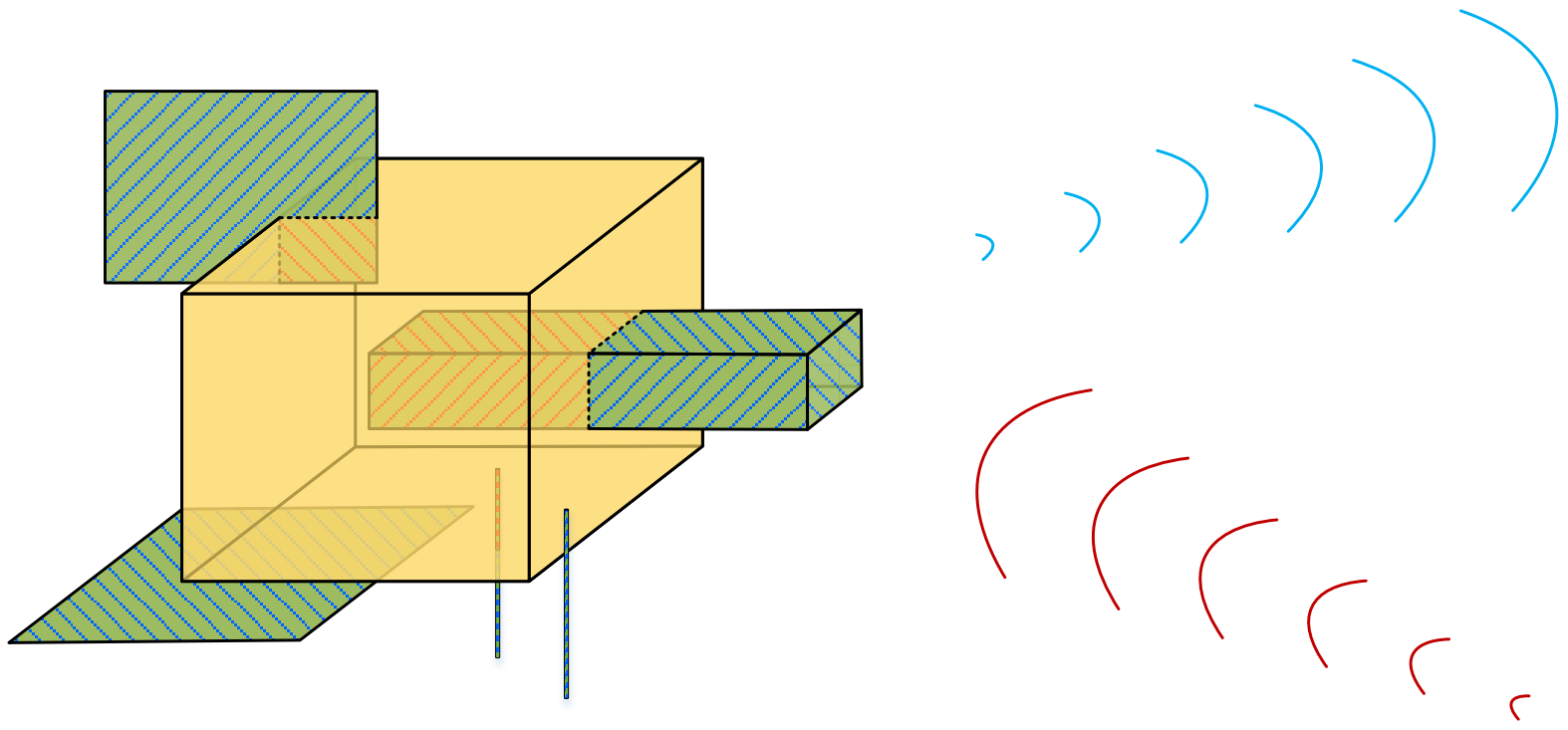

Figure 7-4 Geometry of an incident-field-driven metal-material composite scatterer, which is constituted by metallic {line, surface, body} and material body.

iv) penetrable EM structures (such as the one shown in Fig. 7-5) have energy dissipation (if structures are lossy) and self-oscillation (if structures are lossless) processes, and the processes are governed by POYNTING'S THEOREM (PtT) form, and the source term in PtT form is Poynting's flux, which is used to sustain a steady energy utilization;

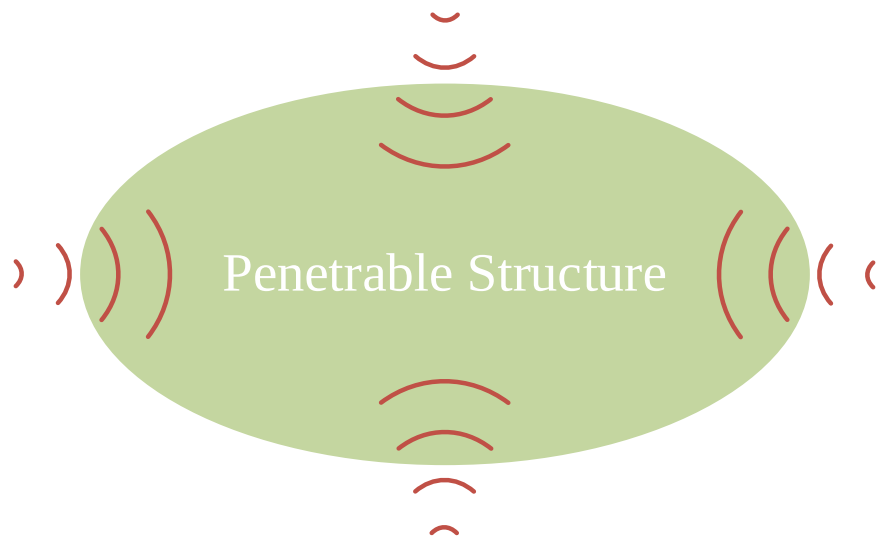

Figure 7-5 Geometry of a external-field-illuminated penetrable material structure.

v) besides the energy coupling between the electric and magnetic fields of a single mode, there also exists the energy coupling between the fields of two different modes, and the energy coupling process between the two different modal fields is governed by LORENTZ'S RECIPROCITY THEOREM (LRT) form.

The above energy principles constitute the theoretical foundation of this report. Using the principles, this report proposes some energy-based EM modal analysis methods as below.

When the EM structure works at a certain manner (transmitting or receiving or scattering or others), it has many different physically realizable modes, and all of the modes constitute a linear space called modal space. Usually, modal analysis method is dedicated to constructing a set of Energy-Decoupled Modes (DMs, which have no net energy exchange in integral period) in modal space. The DMs can be effectively derived from orthogonalizing the corresponding Energy Source Operator. Specifically speaking:

**a) PTT-DMT for Wave-Port-Fed EM Structures**

For wave-port-fed transmitting/receiving antennas (such as the ones shown in Fig. 7-6), their DMs can be effectively constructed by orthogonalizing the Input Power Operator (IPO) contained in PTT, and the corresponding modal analysis method is called PTT-based Decoupling Mode Theory (PTT-DMT).

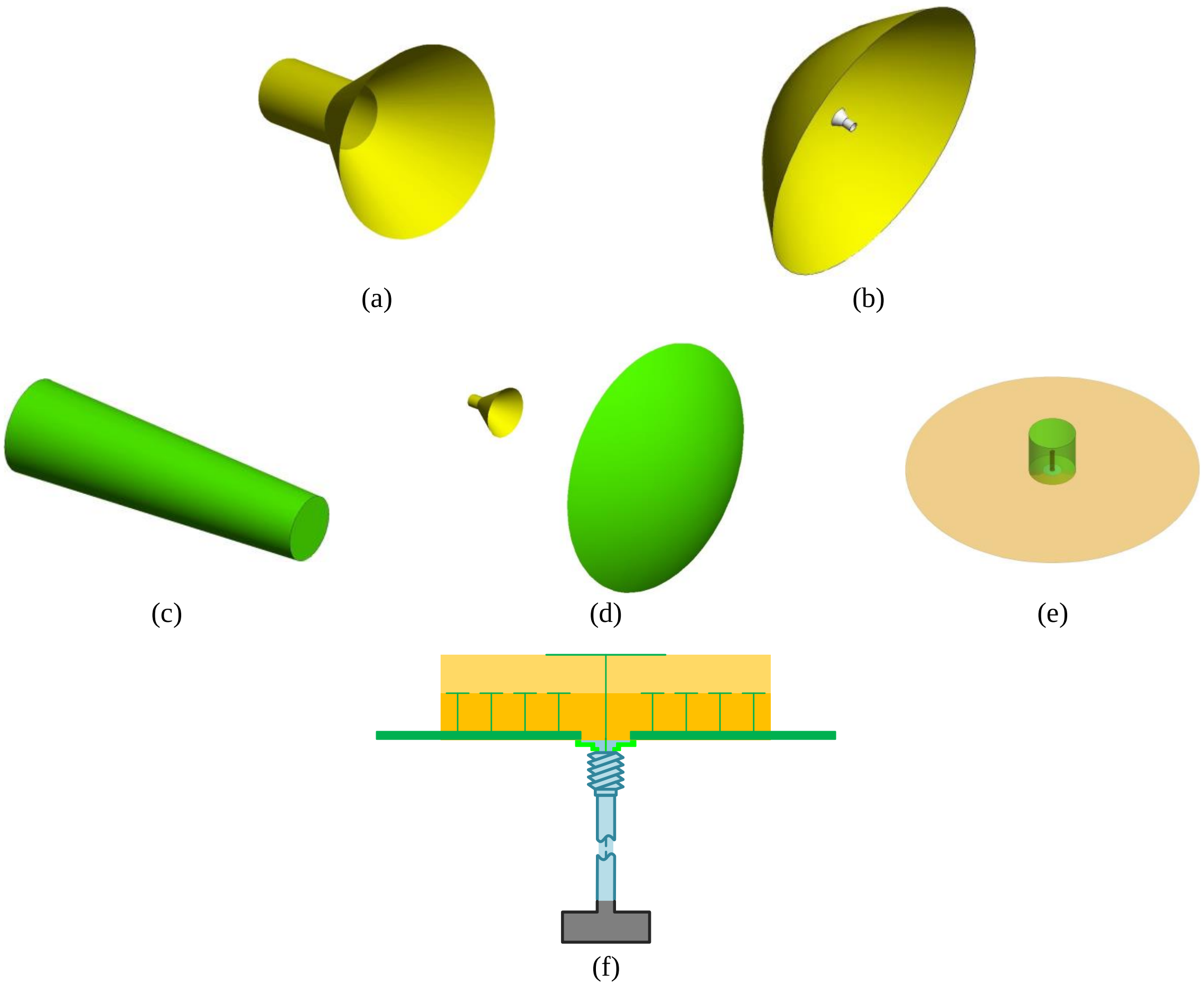

Figure 7-6 Geometries of some typical wave-port-fed transmitting/receiving antennas. (a) Metallic horn antenna, (b) horn-excited metallic parabolic reflector antenna, (c) dielectric rod antenna, (d) horn-excited dielectric lens antenna, (e) dielectric resonator antenna placed on metallic ground plane, and (f) meta-material antenna placed on metallic ground plane.

Besides the single antennas, the PTT-DMT and orthogonalizing IPO method are also suitable for wave-port-fed array antennas, such as the one shown in Fig. 7-7.

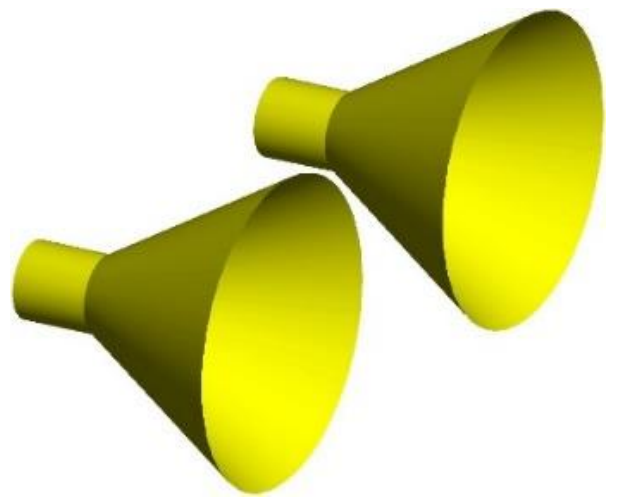

Figure 7-7 Geometry of a transmitting antenna array constituted by two metallic horns.

For wave-port-fed wave-guiding structures (such as the ones in Fig. 7-8), their DMs can also be effectively constructed by the PTT-DMT-based orthogonalizing IPO method.

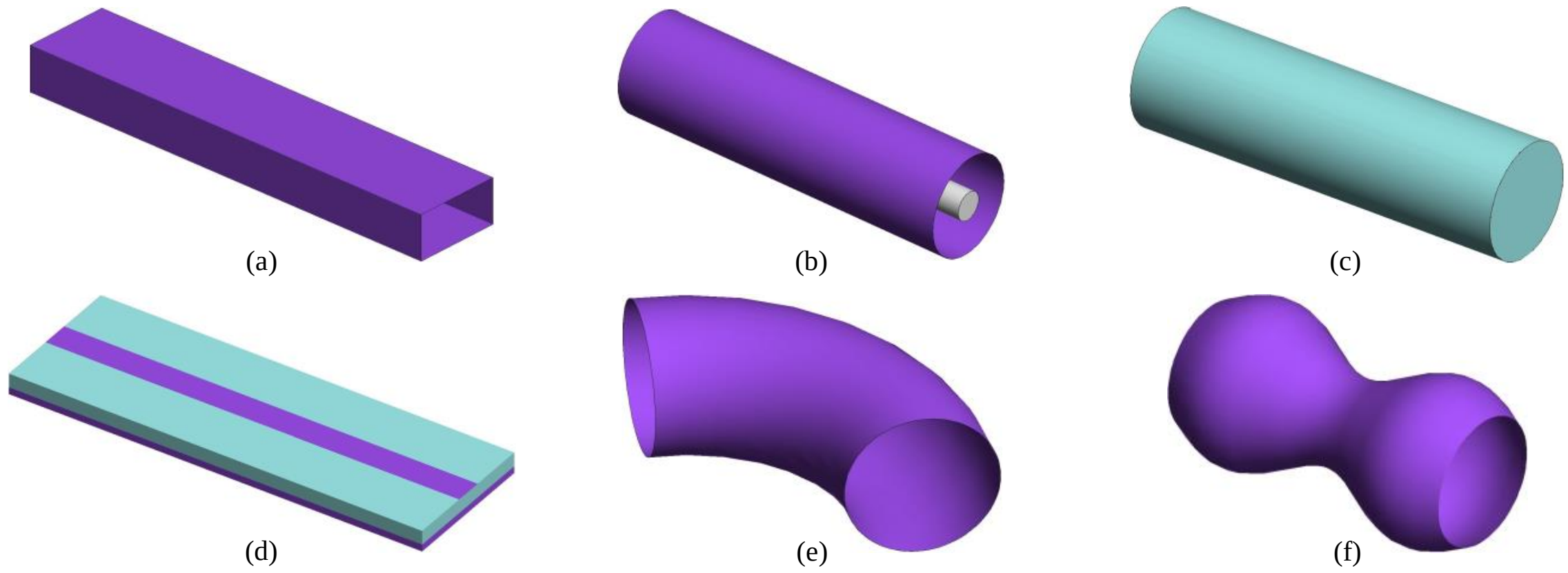

Figure 7-8 Geometries of some typical wave-port-fed wave-guiding structures. (a) Metallic tube waveguide, (b) metallic coaxial waveguide, (c) dielectric waveguide, (d) microstrip line, and (e&f) two non-standard metallic tube waveguides.

Because the modal analysis frameworks and modal construction methods for the wave-port-fed transmitting/receiving antennas and wave-guiding structures are universal, then they are directly applicable to the waveguide-antenna (such as the ones shown in Fig. 7-9) and antenna-antenna (such as the one shown in Fig. 7-10) cascaded systems.

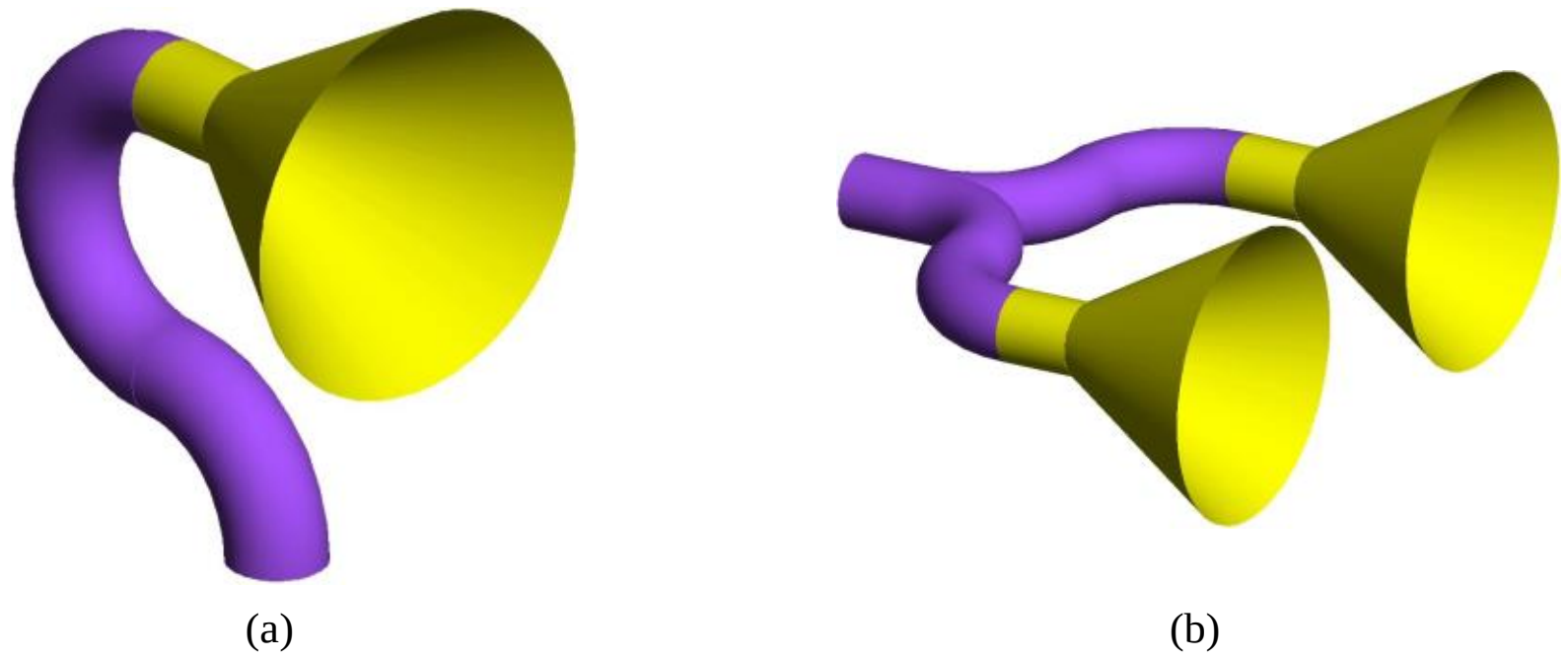

Figure 7-9 Geometries of two typical waveguide-antenna cascaded systems.

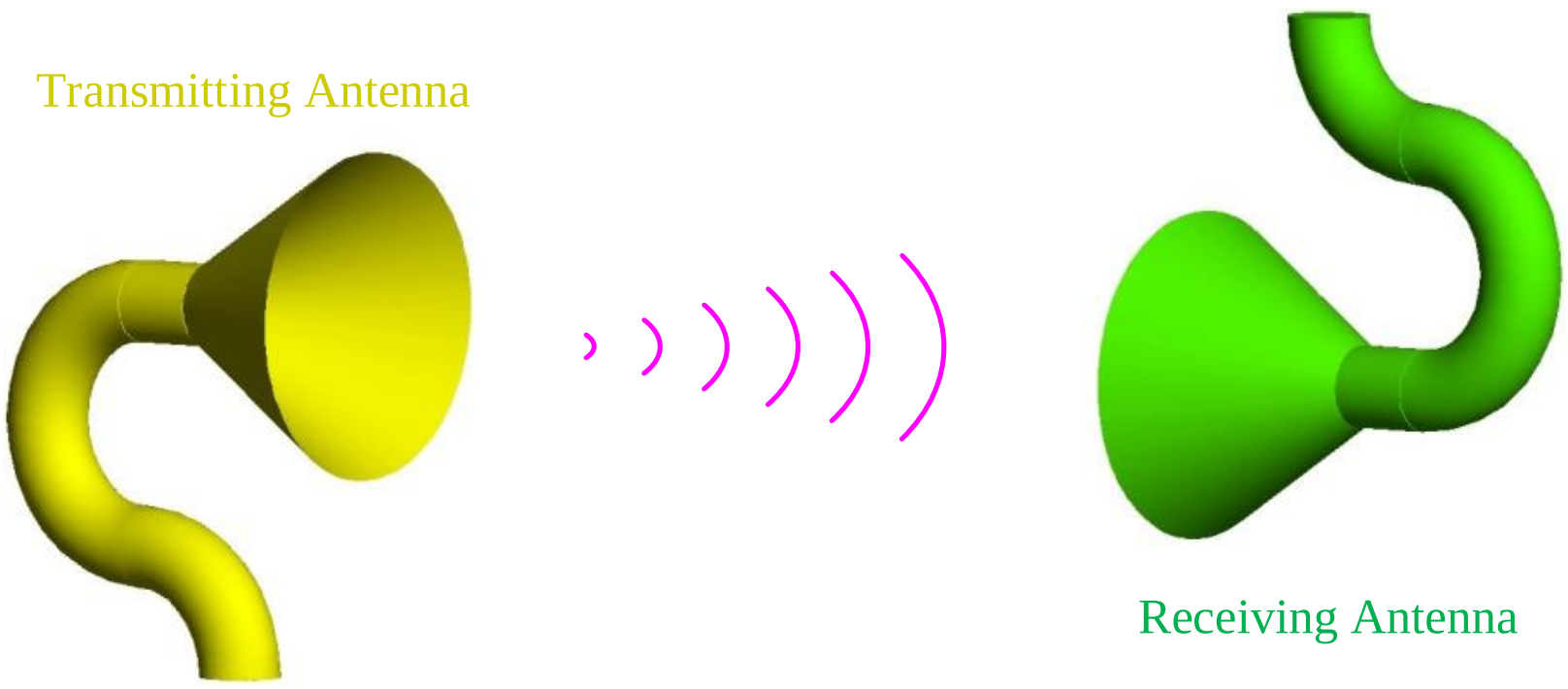

Figure 7-10 Geometry of a typical antenna-antenna cascaded system.

### b) PS-WET-CMT for Lumped-Port-Driven EM Structures

For lumped-port-driven/local-near-field-driven transmitting antennas (such as the ones shown in the previous Figs. 7-3(a~e)), their energy-decoupled CHARACTERISTIC MODES (CMs) can be effectively constructed by orthogonalizing the PARTIAL-STRUCTURE-ORIENTED DRIVING POWER OPERATOR (PS-DPO) contained in PS-WET, and the corresponding modal analysis method is called PS-WET-based CHARACTERISTIC MODE THEORY (CMT). For lumped-port-driven wave-guiding structures / wireless power transfer systems (such as the ones shown in Figs. 7-3(f) and 7-11), their energy-decoupled CMs can also be effectively constructed by using the orthogonalizing PS-DPO method under the PS-WET-based CMT (PS-WET-CMT) framework.

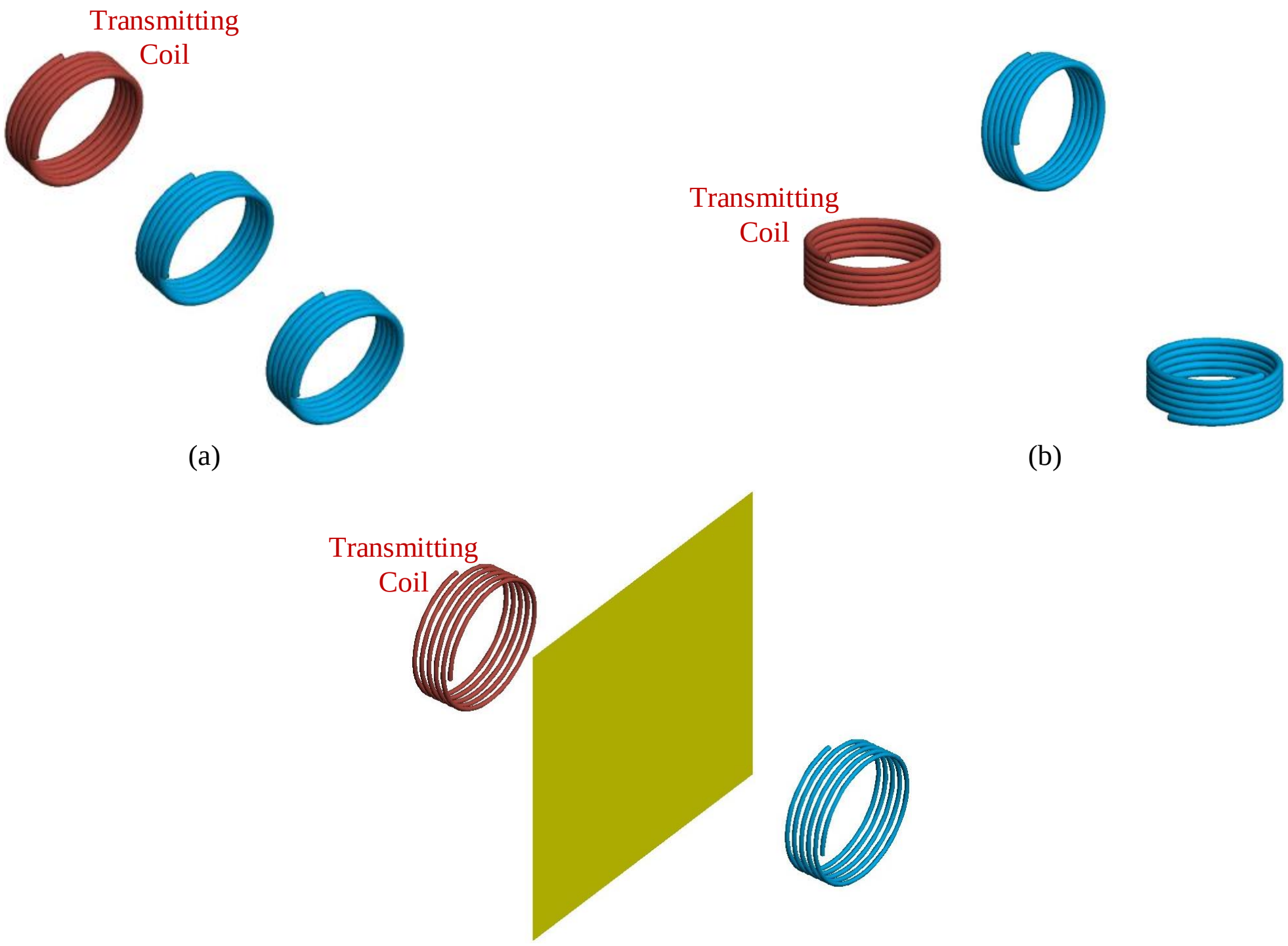

Figure 7-11 Geometries of some typical wireless power transfer systems.

**c) ES-WET-CMT for Incident-Field-Driven EM Structures**

For incident-field-driven scattering structures (such as the one shown in Fig. 7-4), their conventional INTEGRAL-EQUATION BASED CMT (IE-CMT) can also be effectively established under ES-WET framework, and their conventional CMs calculated from orthogonalizing IMPEDANCE MATRIX OPERATOR (IMO) can also be effectively calculated from orthogonalizing ENTIRE-STRUCTURE-ORIENTED DRIVING POWER OPERATOR (ES-DPO). A series of unsolved problems existing in IE-CMT can be successfully resolved under the ES-WET-based CMT (ES-WET-CMT) framework, for example:

On PROBLEM I. Under ES-WET framework, externally impressed and environmental fields are treated as a whole — externally incident field, such that the finally obtained CMs only depend on the inherent scattering characters of the scattering object, and the theoretical foundation of this treatment is ES-WET and its energy source term ES-DPO.

On PROBLEM II. Under ES-WET framework, it is revealed that: in the aspect of generating energy-decoupled CMs, the symmetric IMO in IE-CMT and the ES-DPO in ES-WET-CMT are equivalent to each other; the CMs generated by the asymmetric IMO in IE-CMT cannot guarantee the energy-decoupling feature. Thus, the symmetric IMO is more reasonable in the aspect of generating energy-decoupled CMs.

On PROBLEM III. Under ES-WET framework, it is revealed that energy decoupling is a more essential and indispensable feature than far-field orthogonality.

On PROBLEM IV. Employing the concept of driving power and modal decomposition introduced under ES-WET framework, it is revealed that $|\lambda_n|$ is a quantitative depiction for the energy-decoupling degree between modal incident field and modal scattered current of the $n$-th CM.

On PROBLEM V. Under ES-WET framework, it is revealed that: (1) the unwanted modes outputted from the conventional IE-CMT-based formulations originate from using inproper modal generating operators; (2) the spurious modes outputted from the conventional IE-CMT-based formulations originate from overlooking the dependence relations among the currents involved in modal generating operator. Under ES-WET framework, some desired modal generating operators and spurious mode suppression schemes are proposed to suppress the unwanted and spurious modes.

On Problem VI. Under ES-WET framework, it is revealed that: the steady working of a CM (except the internally resonant mode) need to be sustained by a non-zero modal incident field. Thus, CMT (both IE-CMT and ES-WET-CMT) is not a source-free modal analysis theory. Then, the total field related to the EM problem analyzed by CMT is the summation of modal scattered field and non-zero modal incident field. The modal total field indeed satisfies the boundary condition on metallic boundaries, but the modal scattered field (i.e., usually so-called modal field) is not necessary to satisfy the boundary condition on metallic boundaries.

Besides the above-mentioned ES-WET-CMT-based resolutions for the problems, some generalizations for the conventional scatterer-oriented CMT are also done under ES-WET framework. Specifically, for the scatterer shown in Fig. 7-12(a), its surrounding environment is not free space; for the scatterer shown in Fig. 7-12(b), its driver is not placed in the far zone of the scatterer.

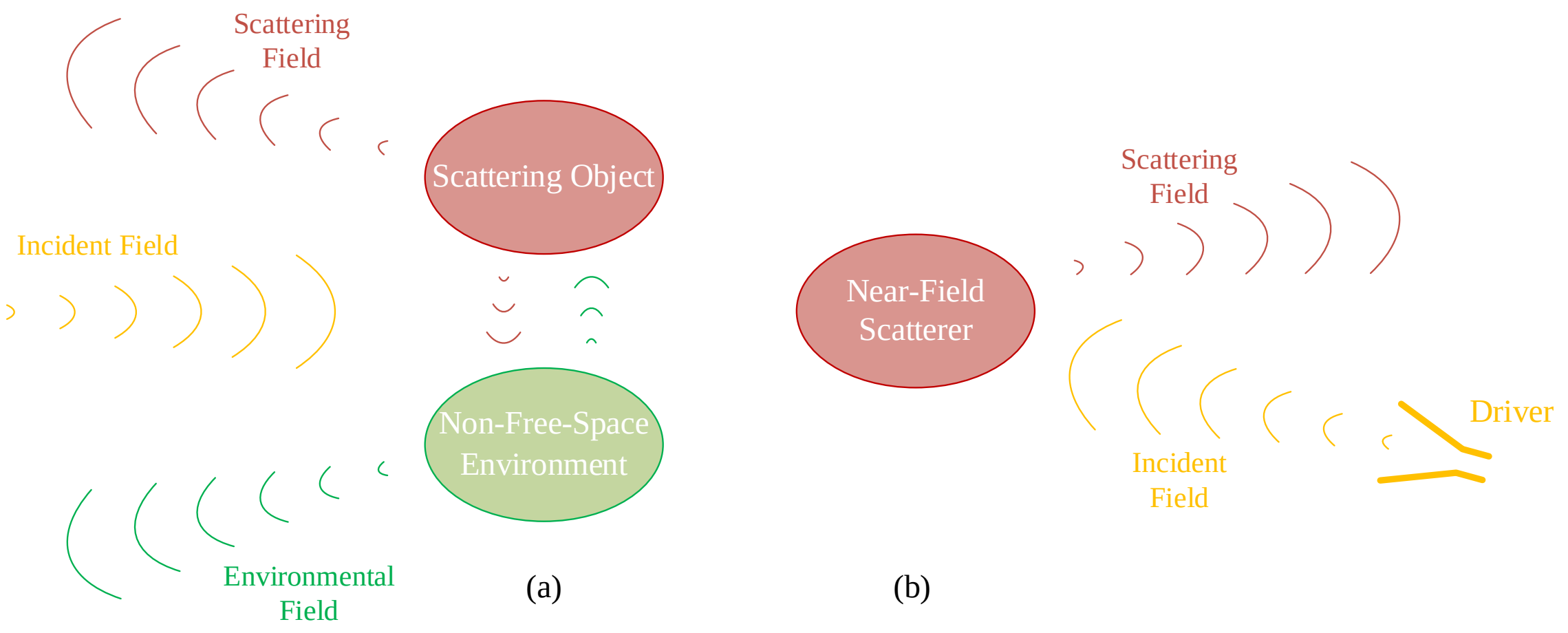

Figure 7-12 (a) An incident-field-driven scatterer surrounded by a non-free-space environment and (b) a scatterer driven by the incident field originating from a horn not placed in far zone.

The CMs derived from the conventional CMT only depend on the inherent scattering characters of the objective scatterer, but are independent of the environment surrounding the scatterer and the driver used to excite the scatterer. This report proposes some methods for constructing the environment-dependent CMs with scatterer-environment interaction informations and the driver-dependent CMs with scatterer-driver interaction informations.

In fact, compared with the above-mentioned results achieved by transforming IE-CMT into ES-WET-CMT, a more important achievement obtained in ES-WET framework is to reveal a fact that: both the IE-CMT and ES-WET-CMT are the modal analysis theories for scattering structures rather than for {transmitting/receiving antennas, wave-guiding structures, transfering structures}; many transmitting-problem-oriented IE-CMT-based engineering applications are approximate but not rigorous.

**d) PtT-DMT for External-Field-Illuminated EM Structures**

For external-field-illuminated EM structures (an extreme case is null-field illumination) such as the one shown in Fig. 7-13, this report proposes a modal analysis theory — PtT-Based Decoupling Mode Theory (PtT-DMT) — to construct the optimally energy-dissipating modes and self-oscillating modes (null-field illumination case), by orthogonalizing Poynting's Flux Operator (PtFO).

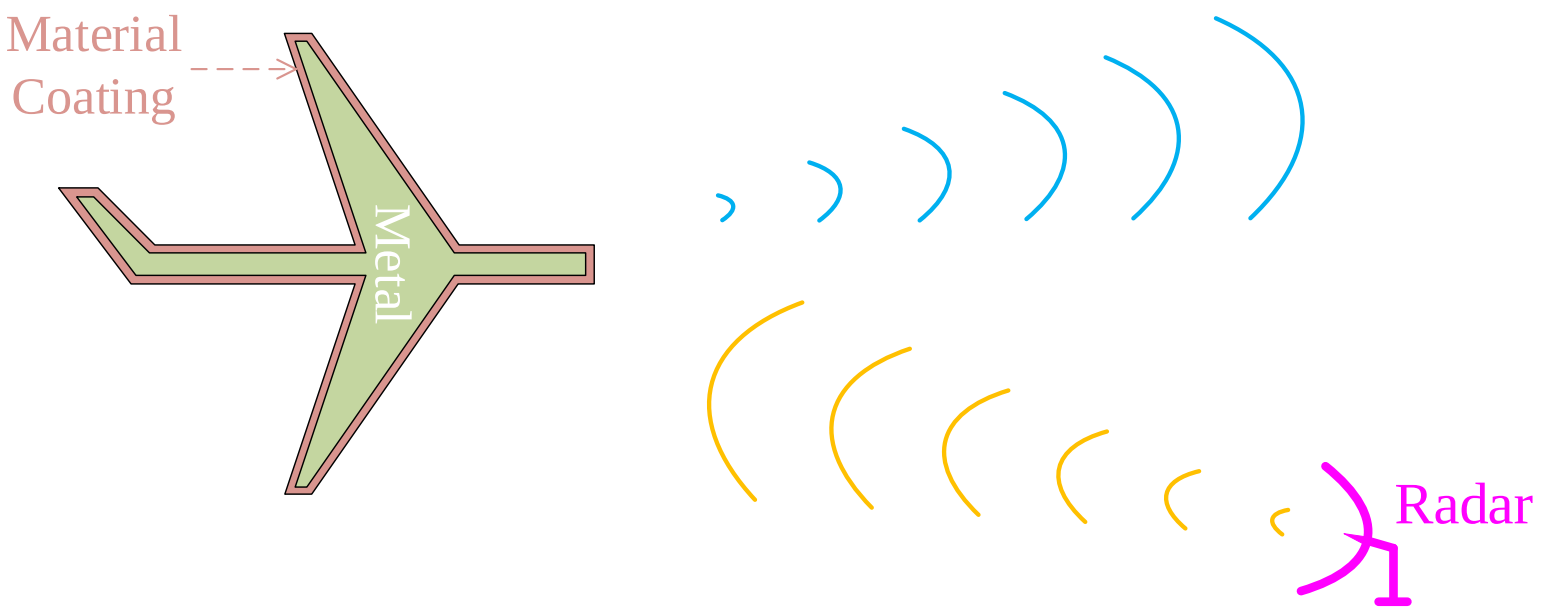

Figure 7-13 External-field-illuminated metallic object with material coating.

**e) LRT-Based Further Conclusions on Modal Energy Decoupling**

Based on LRT, this report derives some beautiful conclusions on the energy-decoupling features satisfied by the DMs and CMs.

In addition, some important physical quantities frequently used in EM modal analysis (such as, Θ-factor, MS, impedance and TC, which have abilities to quantitatively depict the modal features in the aspect of utilizing EM energy) are also summarized. Specifically:

Θ-factor. In the cases of PTT-DMT (Chap. 3) and PtT-DMT (Chap. 6), the Θ-factor quantitatively depicts the decoupling degree between the electric field energy and magnetic field energy carried by working mode. In the case of PS-WET-CMT (Chap. 4) / ES-WET-CMT (Chap. 5), the Θ-factor quantitatively depicts the decoupling degree between the energies carried by the driving/incident field and induced/scattered current of working mode.

In fact, $|\theta_n| \ / \ |\lambda_n|$ is just the Θ-factor of the *n*-th DM itself, and this fact gives $|\theta_n| \ / \ |\lambda_n|$ a very clear physical meaning — "electric-magnetic / field-current energy-decoupling degree" of the *n*-th DM/CM.

MS. $MS_n$ (modal significance of the *n*-th DM/CM) has the following two noteworthy physical interpretations (1) modal weight of a DM/CM in whole modal expansion formulation; (2) "electric-magnetic energy-coupling degree / field-current energy-coupling degree" of the *n*-th DM/CM.

Impedance. For wave-port-fed EM structures, the conventional circuit-based definitions for modal "impedance, resistance, reactance" and "admittance, conductance, susceptance" can also be effectively defined by employing the language of field. The field-based definitions are very useful for recognizing the resonant DMs derived from PTT-DMT (Chap. 3), the optimally energy-dissipating modes derived from PtT-DMT (Sec. 6.2), and the self-oscillating modes derived from PtT-DMT (Sec. 6.3).

TC. For wave-port-fed EM structures, a concept of energy transport coefficient is introduced to quantify the energy transporting efficiency from one port to another port, and it is valuable for recognizing the optimally transporting modes.

For lumped-port-driven wireless power transfer systems, a concept of energy transfer coefficient is introduced to quantify the transferring efficiency of wireless power transfer systems, and it is valuable for recognizing the optimally transferring modes.

In fact, this Post-Doctoral Concluding Report "Energy-Viewpoint-Based Electromagnetic Modal Analysis" is not only a systematical integration but also a further sublimation for the author's Doctoral Dissertation "Research on the Work-Energy Principle Based Characteristic Mode Theory for Scattering Systems"[8] and Post-Doctoral Research Report "Research on the Power Transport Theorem Based Decoupling Mode Theory for Transceiving Systems"[27], which can be downloaded from the following links.

Doctoral Dissertation: https://arxiv.org/abs/1907.11787

Post-Doctoral Research Report: https://arxiv.org/abs/2103.01853

# APPENDICES

The following appendices focus on summarizing some physical quantities, which are frequently used in electromagnetic (EM) modal analysis and can quantitatively depict the modal features in the aspect of utilizing EM energy.

## Appendix A Universal Physical Quantities

In this appendix, we summarize some universal physical quantities, which are universally applicable to the DMs derived from PTT-DMT (Chap. 3), PS-WET-CMT (Chap. 4), ES-WET-CMT (Chap. 5), and PtT-DMT (Chap. 6).

Due to the completeness of DM set $\{\mathbb{C}_n\}$, any working mode $\mathbb{C}$ can be expanded as that $\mathbb{C} = \sum_n c_n \mathbb{C}_n$, where $\{c_n\}$ are the expansion coefficients. Because of the orthogonality of DMs, the time-average power of $\mathbb{C}$ can be expanded as follows:

$$(1/T)\int_{t_0}^{t_0+T} \mathcal{P} dt = \operatorname{Re} P = \sum\nolimits_n |c_n|^2 \operatorname{Re} P_n \tag{A-1}$$

In Eq. (A-1), $T$ is the time period of time-harmonic field; $\mathcal{P}$ and $P$ are the time-domain and frequency-domain power sources used to deliver energy to $\mathbb{C}$; $P_n$ is the frequency-domain power source used to deliver energy to $\mathbb{C}_n$.

Equation (A-1) implies that: if $\{\operatorname{Re} P_n\}$ are normalized to 1, we obtain the following famous Parseval's identity

$$(1/T)\int_{t_0}^{t_0+T} \mathcal{P} dt = \sum\nolimits_n |c_n|^2 \tag{A-2}$$

where

$$c_n = \frac{1}{1+j\theta_n} \cdot \begin{cases} \left.\begin{matrix} (1/2)\iint_{\text{input port}} \left(\boldsymbol{E}_n \times \boldsymbol{H}^\dagger\right) \cdot d\boldsymbol{S} \\ (1/2)\iint_{\text{input port}} \left(\boldsymbol{E} \times \boldsymbol{H}_n^\dagger\right) \cdot d\boldsymbol{S} \end{matrix}\right\} & \text{PTT-DMT} \\ (1/2)\langle \boldsymbol{J}_n, \boldsymbol{E}_{\text{driv}} \rangle_{\text{partial structure}} + (1/2)\langle \boldsymbol{M}_n, \boldsymbol{H}_{\text{driv}} \rangle_{\text{partial structure}} & \text{PS-WET-CMT} \\ (1/2)\langle \boldsymbol{J}_n, \boldsymbol{E}_{\text{inc}} \rangle_{\text{entire structure}} + (1/2)\langle \boldsymbol{M}_n, \boldsymbol{H}_{\text{inc}} \rangle_{\text{entire structure}} & \text{ES-WET-CMT} \\ \left.\begin{matrix} (1/2)\iint_{\text{material-environment boundary}} \left(\boldsymbol{E}_n \times \boldsymbol{H}^\dagger\right) \cdot d\boldsymbol{S} \\ (1/2)\iint_{\text{material-environment boundary}} \left(\boldsymbol{E} \times \boldsymbol{H}_n^\dagger\right) \cdot d\boldsymbol{S} \end{matrix}\right\} & \text{PtT-DMT} \end{cases} \tag{A-3}$$

for the DMs/CMs derived from PTT-DMT (Chap. 3), PS-WET-CMT (Chap. 4), ES-WET-

CMT (Chap. 5), and PtT-DMT (Chap. 6). In Eq. (A-3), $\theta_n$ is obtained from solving modal decoupling equation (PTT-DMT and PtT-DMT cases) or characteristic equation (PS-WET-CMT and ES-WET-CMT cases).

## A1 Θ-Factor (Generalized Q-Factor)

This App. A1 is devoted to introducing a novel physical quantity "electric-magnetic energy-decoupling factor / field-current energy-decoupling factor" (Θ-factor), which can be viewed as a generalization for the conventional quality factor (Q-factor).

Following the ideas proposed in Refs. [8-Sec.3.3] and [32], the whole DM set $\{\mathbb{C}_n\}$ can be decomposed into three sub-sets — purely inductive DM set $\{\mathbb{C}_\zeta^{\rm ind}\}$ (constituted by all inductive DMs), purely resonant DM set $\{\mathbb{C}_\varsigma^{\rm res}\}$ (constituted by all resonant DMs), and purely capacitive DM set $\{\mathbb{C}_\xi^{\rm cap}\}$ (constituted by all capacitive DMs). Thus the modal expansion formulation $\mathbb{C} = \sum_n c_n \mathbb{C}_n$ can be alternatively written as the following more illuminating form

$$\mathbb{C} = \underbrace{\sum\nolimits_\zeta c_\zeta^{\rm ind}\mathbb{C}_\zeta^{\rm ind}}_{\mathbb{B}^{\rm ind}} + \underbrace{\sum\nolimits_\varsigma c_\varsigma^{\rm res}\mathbb{C}_\varsigma^{\rm res}}_{\mathbb{B}^{\rm res}} + \underbrace{\sum\nolimits_\xi c_\xi^{\rm cap}\mathbb{C}_\xi^{\rm cap}}_{\mathbb{B}^{\rm cap}} \tag{A-4}$$

where the building block components $\mathbb{B}^{\rm ind}$, $\mathbb{B}^{\rm res}$, and $\mathbb{B}^{\rm cap}$ used to constitute whole working mode $\mathbb{C}$ are called purely inductive term, purely resonant term, and purely capacitive term respectively. Clearly, the three components are energy-decoupled.

Based on the above modal decomposition formulation (A-4), we introduce a novel concept of Θ-factor for any working mode $\mathbb{C}$ as follows[14,35,37],[27-Sec.9.3]:

$$\Theta\left(\mathbb{C}\right) = \frac{{\rm Im}\left(1\mathbb{B}^{\rm ind} + 0\mathbb{B}^{\rm res} + j\mathbb{B}^{\rm cap}\right)\cdot\mathbb{P}\cdot\left(1\mathbb{B}^{\rm ind} + 0\mathbb{B}^{\rm res} + j\mathbb{B}^{\rm cap}\right)}{{\rm Re}\,\mathbb{C}\cdot\mathbb{P}\cdot\mathbb{C}} \tag{A-5}$$

Here, $\mathbb{P}$ is the matrix form of the ENERGY SOURCE OPERATOR with only independent currents; specifically, $\mathbb{P} = \mathbb{P}_{\rm in}$ for PTT-DMT case (Chap. 3), and $\mathbb{P} = \mathbb{P}_{\rm driv}$ for PS-WET-CMT case (Chap. 4), and $\mathbb{P} = \mathbb{P}_{\rm DRIV}$ for ES-WET-CMT case (Chap. 5), and $\mathbb{P} = \mathbb{P}_{\rm flux}$ for PtT-DMT case (Chap. 6). As explained previously, the Θ-factor quantitatively depicts the decoupling degree between the electric field energy and magnetic field energy carried by the mode in the cases of PTT-DMT (Chap. 3) and PtT-DMT (Chap. 6)[35],[27-Sec.9.3]; the Θ-factor quantitatively depicts the decoupling degree between the energies carried by driving field and scattered current in the cases of PS-WET-CMT (Chap. 4) and ES-WET-CMT (Chap. 5)[14,37]. Thus, the Θ-factor is usually called "electric-magnetic energy-decoupling factor / field-current energy-decoupling factor".

Obviously, the above novel Θ-factor can be viewed as a generalization for the classical Q-factor. In addition, for any single DM $\mathbb{C}_n$, there exists the following more simplified relation[14,35,37],[27-Sec.9.3]

$$\Theta\left(\mathbb{C}_n\right)=\begin{cases}|\theta_n| & \text{in the cases of PTT-DMT and PtT-DMT}\\ |\lambda_n| & \text{in the cases of PS-WET-CMT and ES-WET-CMT}\end{cases} \tag{A-6}$$

and this relation (A-6) clearly reveals the physical meaning of $|\theta_n|$ — the "electric-magnetic energy-decoupling factor / field-current energy-decoupling factor" of the *n*-th DM itself.

## A2 Modal Significance (MS)

This App. A2 is devoted to summarizing two somewhat different (but not contradictory) physical meanings of modal significance (MS).

Based on Parseval's identity (A-2) and expansion coefficient (A-3), the following modal significance (MS)[25]

$$\text{MS}_n=\frac{1}{|1+j\theta_n|} \tag{A-7}$$

is usually used to quantitatively depict the weight of a DM in whole modal expansion formulation $\mathbb{C}=\sum_n c_n\mathbb{C}_n$.

Obviously, $\text{MS}_n=1/|1+j|\theta_n||$, because $\theta_n$ is a purely real number, so $\text{MS}_n$ is a monotonically decreasing function about $|\theta_n|$. It is thus clear that, besides the above physical interpretation — modal weight of a DM in whole modal expansion formulation, $\text{MS}_n$ has another noteworthy physical interpretation — "electric-magnetic energy-coupling degree / field-current energy-coupling degree" of the *n*-th DM[14],[27-Sec.9.4].

# Appendix B Physical Quantities Defined on Wave Port

In this App. B, we discuss some physical quantities, which are defined on wave port and then applicable to the DMs derived from PTT-DMT (Chap. 3) and PtT-DMT (Chap. 6).

## B1 Modal Input Impedance and Admittance

For the DMs of the wave-port-fed EM structures discussed in Chap. 3, we proposed the following field-based definitions for modal input "impedance $Z_{\text{in}}$, resistance $R_{\text{in}}$, reactance $X_{\text{in}}$" and "admittance $Y_{\text{in}}$, conductance $G_{\text{in}}$, susceptance $B_{\text{in}}$"[27,35,39]

$$Z_{\text{in}} = \frac{(1/2)\iint_{\text{wave port}} \left(\boldsymbol{E}\times\boldsymbol{H}^{\dagger}\right)\cdot d\boldsymbol{S}}{(1/2)\langle \boldsymbol{J},\boldsymbol{J}\rangle_{\text{wave port}}} = R_{\text{in}} + j\,X_{\text{in}} \tag{A-8a}$$

$$Y_{\text{in}} = \frac{(1/2)\iint_{\text{wave port}} \left(\boldsymbol{E}\times\boldsymbol{H}^{\dagger}\right)\cdot d\boldsymbol{S}}{(1/2)\langle \boldsymbol{M},\boldsymbol{M}\rangle_{\text{wave port}}} = G_{\text{in}} + j\,B_{\text{in}} \tag{A-8b}$$

In Eqs. (A-8a) and (A-8b), $R_{\text{in}} = \text{Re}\,Z_{\text{in}}$, $X_{\text{in}} = \text{Im}\,Z_{\text{in}}$, $G_{\text{in}} = \text{Re}\,Y_{\text{in}}$, and $B_{\text{in}} = \text{Im}\,Y_{\text{in}}$. The modal $R_{\text{in}}$ and $G_{\text{in}}$ have been systematically utilized to recognize the resonant DMs of wave-port-fed EM structures in Refs. [27,35,39].

Obviously, the above field-based definitions for the DMs derived from PTT-DMT (established in Chap. 3 and Refs. [27,35,39]) are also suitable for the DMs derived from PtT-DMT (established in Chap. 6).

## B2 Modal Energy Transport Coefficient from Port 1 to Port 2

In the Sec. 3.3 (receiving antenna case) of this report, we introduced a novel parameter “modal energy transport coefficient (TC) from $\text{S}_{\text{aux}}$ to $\text{S}_{\text{i}}$” as defined in Eq. (3-26), where $\text{S}_{\text{aux}}$ is an auxiliary port to model the excitation for the receiving antenna and $\text{S}_{\text{i}}$ is the input port of the receiving antenna.

In fact, the receiving-antenna-oriented TC defined in Eq. (3-26) can be further generalized to an arbitrary region with two wave ports “$\text{S}_1$ and $\text{S}_2$” as the following expression

$$\text{TC}_{\text{S}_1\rightarrow\text{S}_2} = \frac{(1/T)\int_{t_0}^{t_0+T}\left[\iint_{\text{S}_2}\left(\boldsymbol{\mathcal{E}}\times\boldsymbol{\mathcal{H}}\right)\cdot d\boldsymbol{S}\right]dt}{(1/T)\int_{t_0}^{t_0+T}\left[\iint_{\text{S}_1}\left(\boldsymbol{\mathcal{E}}\times\boldsymbol{\mathcal{H}}\right)\cdot d\boldsymbol{S}\right]dt} = \frac{\text{Re}\left\{(1/2)\iint_{\text{S}_2}\left(\boldsymbol{E}\times\boldsymbol{H}^{\dagger}\right)\cdot d\boldsymbol{S}\right\}}{\text{Re}\left\{(1/2)\iint_{\text{S}_1}\left(\boldsymbol{E}\times\boldsymbol{H}^{\dagger}\right)\cdot d\boldsymbol{S}\right\}} \tag{A-9}$$

In the above definition (A-9), ports $\text{S}_1$ and $\text{S}_2$ can be either open surfaces or closed surfaces.

# Appendix C Physical Quantities Defined on Lumped Port

In this appendix, we introduce some physical quantities, which are defined on lumped port and then applicable to the energy-decoupled CMs derived from PS-WET-CMT (Chap. 4).

For wireless power transfer (WPT) applications, the transferred power $\mathcal{P}_{\text{tra}}$, which is only a part of whole driving power $\mathcal{P}_{\text{driv}}$, is desired. Based on this, we introduce the following concept of energy transfer coefficient (TC)[27-App.G],[40]

$$\mathrm{TC} = \frac{(1/T)\int_{t_0}^{t_0+T} \mathcal{P}_{\mathrm{tra}} dt}{(1/T)\int_{t_0}^{t_0+T} \mathcal{P}_{\mathrm{driv}} dt} \tag{A-10}$$

to quantify the transferring efficiency of the WPT system. In fact, the concept of TC can be further generalized to the passively-loaded transmitting antennas discussed in Chap. 4.

# ACKNOWLEDGEMENTS

This work is dedicated to my mother for her constant understanding, support, and encouragement.

In the above-listed references, some references were submitted to *Institute of Electrical and Electronics Engineers* (IEEE), for example:

Ref. [14] with Manuscript DOI [AP2004-0708] was submitted to IEEE-TAP on 2020-04-12;

Ref. [15] with Manuscript DOI [AP2004-0709] was submitted to IEEE-TAP on 2020-04-12;

Ref. [35] with Manuscript DOI [AP2104-0777] was submitted to IEEE-TAP on 2021-04-14;

Ref. [37] with Manuscript DOI [AP2101-0036] was submitted to IEEE-TAP on 2021-01-06;

Ref. [40] with Manuscript DOI [AP2101-0035] was submitted to IEEE-TAP on 2021-01-06;

Ref. [39] with Manuscript DOI [TMTT-2021-04-0464] was submitted to IEEE-TMTT on 2021-04-14.

The Refs. [14,15,37,40] had been attached to the back of the Ref. [27] being available from link https://www.researchgate.net/publication/349728281_Research_on_the_Power_Transport_Theorem_Based_Decoupling_Mode_Theory_for_Transceiving_Systems.